\documentclass[12pt]{report}
\input epsf
\usepackage{graphics,linde2003}
\begin{document}
\pagenumbering{roman} \setcounter{page}{5} \pagestyle{myheadings}
\thispagestyle{empty} \hskip 1cm \vskip 3cm
\begin{center}
{\Large\bf PARTICLE PHYSICS\\
\vskip 1cm AND INFLATIONARY COSMOLOGY\footnote{This is the LaTeX
version of my book ``Particle Physics and Inflationary Cosmology''
(Harwood, Chur, Switzerland,
1990).}}\\
\vskip 1.4cm \vskip 2.2cm {\bf Andrei Linde} \vskip 0.7cm
Department of Physics, Stanford University, Stanford CA
94305-4060,
 USA
\end{center}
\newpage
\hskip 1 cm
\vskip 1 cm
{\centerline{\large Abstract}}

\

This is the LaTeX version of my book ``Particle Physics and Inflationary Cosmology'' (Harwood, Chur, Switzerland, 1990). I decided to put it to hep-th, to make it easily available. Many things happened during the 15 years since the time when it
was written. In particular, we have learned a lot about the high temperature behavior in the electroweak theory and about baryogenesis. A discovery of the acceleration of the universe has changed the way we are thinking about the problem of the vacuum energy: Instead of trying to explain why it is zero, we are trying to understand why it is anomalously small. Recent cosmological observations have shown that the universe is flat, or almost exactly flat, and confirmed many other predictions of inflationary theory.  Many new versions of this theory have been developed, including hybrid inflation and inflationary models based on string theory. There was a substantial progress in the theory of reheating of the universe after inflation, and in the theory of eternal inflation. 

\

It s clear, therefore, that some parts of the book should be updated, which I might do sometimes in the future. I hope, however, that this book may be of some interest even in its original form. I am  using it in my lectures on inflationary cosmology at Stanford, supplementing it with the discussion of the subjects mentioned above. I would suggest to read this book in parallel with the book by Liddle and Lyth ``Cosmological Inflation and Large Scale Structure,'' with the book by Mukhanov  ``Physical Foundations of Cosmology,'' which is to be published soon, and with my review article hep-th/0503195, which contains a discussion of some (but certainly not all) of the recent developments in inflationary theory.

 \newpage

\tableofcontents
\cleardoublepage
\vspace*{6pc}
\thispagestyle{empty}
\markboth{PREFACE TO THE SERIES}{\bktit}
\vbox{\parindent 0pt\raggedright\LARGE \bf
Preface to the Series\vspace{3pc}}%
\addcontentsline{toc}{mystart}{Preface to the Series}
\noindent{}%
The series of volumes, {\it Contemporary Concepts in Physics},
is addressed to the professional physicist and to the serious
graduate student of physics.  The subjects to be covered will
include those at the forefront of current research. It is anticipated
that the various volumes in the series will be rigorous and complete
in their treatment, supplying the intellectual tools necessary for the
appreciation of the present status of the areas under consideration  and
providing the framework upon which future developments may be based.
\cleardoublepage
\thispagestyle{empty}
\markboth{INTRODUCTION}{\bktit}%
\vspace*{6pc}
\vbox{\parindent 0pt\raggedright\LARGE \bf Introduction\vspace{3pc}}%
\addcontentsline{toc}{mystart}{Introduction}
\noindent{}%
With the invention and development of unified
\index{Gauge theories}%
gauge theories of weak and electromagnetic interactions, a
genuine revolution has taken place in elementary particle
physics in the last 15 years.  One of the basic underlying
ideas of these theories is that of spontaneous symmetry
breaking between different types of interactions due to the
appearance of constant classical scalar fields $\varphi$ over
all space (the so-called
\index{Higgs fields}%
Higgs fields).  Prior to the
appearance of these fields, there is no fundamental
difference between strong, weak, and electromagnetic
interactions.  Their spontaneous appearance over all space
essentially signifies a restructuring of the vacuum, with
certain vector (gauge) fields acquiring high mass as a
result.  The interactions mediated by these vector fields
then become short-range, and this leads to symmetry
breaking between the various interactions described by the
unified theories.

The first consistent description of strong and weak
interactions was obtained within the scope of gauge
theories with spontaneous symmetry breaking.  For the first
time, it became possible to investigate strong and weak
interaction processes using high-order perturbation theory.
A remarkable property of these theories ---
\index{Asymptotic freedom}%
asymptotic freedom --- also made it possible in principle to describe
interactions of elementary particles up to center-of-mass
energies  ${\rm E}\sim {\rm M}_{\rm P}\sim 10^{19}$~GeV,
that is, up to the
\index{Planck energy}%
Planck energy, where quantum gravity effects become important.

Here we will recount only the main stages in the
development of gauge theories, rather than discussing their
properties in detail.  In the 1960s,
\index{Glashow--Weinberg--Salam theory}%
Glashow, Weinberg, and
Salam proposed a unified theory of the weak and
electromagnetic interactions [\cite{1}], and real progress
was made in this area in 1971--1973 after the theories were
shown to be renormalizable [\cite{2}].  It was proved in 1973 that
many such theories, with quantum chromodynamics in
particular serving as a description of strong interactions,
possess the property of
\index{Asymptotic freedom}%
asymptotic freedom (a decrease in
the coupling constant with increasing energy [\cite{3}]).  The
first unified gauge theories of strong, weak, and
electromagnetic interactions with a simple symmetry group, the so-called
\index{Grand unified theories}%
grand unified theories [4], were proposed in
1974.  The first theories to unify all of the fundamental
interactions, including gravitation, were proposed in 1976
within the context of
\index{Supergravity theory}%
supergravity theory.  This was followed by the development of
\index{Kaluza--Klein theories}%
Kaluza--Klein theories, which
maintain that our four-dimensional space-time results from
the spontaneous compactification of a higher-dimensional
space [\cite{6}].  Finally, our most recent hopes for a unified
theory of all interactions have been invested in
\index{Superstring theory}%
superstring theory [\cite{7}]. Modern theories of elementary
particles are covered in a number of excellent reviews and
monographs (see [\cite{8}--\cite{17}], for example).

The rapid development of elementary particle theory has not
only led to great advances in our understanding of particle
interactions at superhigh energies, but also (as a
consequence) to significant progress in the theory of
\index{Superdense matter}%
superdense matter.  Only fifteen years ago, in fact, the
term {\it superdense matter} meant matter with a
\index{Density}%
density somewhat higher than nuclear values,
$\rho\sim10^{14}$--$10^{15}$~$\mbox{g}\cdot\mbox{cm}^{-3}$ and
it was virtually
impossible to conceive of how one might describe matter
with $\rho\gg 10^{15}$~$\mbox{g}\cdot\mbox{cm}^{-3}$.
The main problems involved strong-interaction
theory, whose typical coupling constants at
$\rho\ga10^{15}$~$\mbox{g}\cdot\mbox{cm}^{-3}$ were large,
making standard perturbation-theory predictions of the
properties of such matter unreliable.  Because of
\index{Asymptotic freedom}%
asymptotic freedom in quantum chromodynamics, however, the
corresponding coupling constants decrease with increasing
temperature (and density).  This enables one to describe
the behavior of matter at temperatures approaching
${\rm T}\sim{\rm M}_{\rm P}\sim10^{19}$~GeV, which corresponds
to a density
$\rho_{\rm P}\sim{\rm M}^4_{\rm P}\sim10^{94}$~$\mbox{g}\cdot\mbox{cm}^{-3}$.
Present-day elementary particle theories thus make it
possible, in principle, to describe the properties of
matter more than 80 orders of magnitude denser than nuclear
matter!

The study of the properties of superdense matter described
by unified gauge theories began in 1972 with the work of
Kirzhnits [\cite{18}], who showed that the classical scalar
field $\varphi$ responsible for symmetry breaking should
disappear at a high enough temperature T.  This means that a
\index{Phase transitions|(}%
phase transition (or a series of phase transitions)
occurs at a sufficiently high temperature ${\rm T}> {\rm T}_c$,
after which symmetry is restored between various
types of interactions.  When this happens, elementary
particle properties and the laws governing their
interaction change significantly.

This conclusion was confirmed in many subsequent
publications [\cite{19}--\cite{24}].  It was found that similar
\index{Phase transitions|)}%
phase transitions could also occur when the density
of cold matter was raised [\cite{25}--\cite{29}], and in
the presence of external fields and currents [\cite{22},
\cite{23}, \cite{30}, \cite{33}].  For brevity, and to
conform with current terminology, we will hereafter refer
to such processes as phase transitions in gauge theories.

Such phase transitions typically take place at exceedingly
high temperatures and densities.  The critical temperature
for a phase transition in the
\index{Glashow--Weinberg--Salam theory}%
Glashow--Weinberg--Salam theory of weak and
electromagnetic interactions [\cite{1}],
for example, is of the order of
$10^{2}\;\mbox{GeV}\sim 10^{15}\;\mbox{K}$.
The \index{Temperature}%
temperature  at which symmetry is
restored between the strong and electroweak interactions in
\index{Grand unified theories}%
grand unified theories is even higher,
${\rm T}_c\sim10^{15}\;\mbox{GeV}\sim 10^{28}\;\mbox{K}$.  For
comparison, the highest temperature attained in a supernova
explosion is about $10^{11}$~K.  It is therefore impossible
to study such phase transitions in a laboratory. However,
the appropriate extreme conditions could exist at the
earliest stages of the evolution of the universe.

According to the standard version of the
\index{Hot universe theory}%
hot universe theory, the universe could have expanded from a state in
which its temperature was at least ${\rm T}\sim10^{19}$~GeV
[\cite{34}, \cite{35}], cooling all the while.  This means
that in its earliest stages, the symmetry between the
strong, weak, and electromagnetic interactions should have
been intact.  In cooling, the universe would have gone
through a number of phase transitions, breaking the
symmetry between the different interactions
[\cite{18}--\cite{24}].

This result comprised the first evidence for the importance
of unified theories of elementary particles and the theory
of superdense matter for the development of the theory of
the evolution of the universe.  Cosmologists became
particularly interested in recent theories of elementary
particles after it was found that
\index{Grand unified theories}%
grand unified theories provide a natural
framework within which the observed baryon
\index{Universe!baryon asymmetry of}%
\index{Baryon asymmetry}%
\index{Asymmetry, baryon}%
asymmetry of the universe (that is, the lack of
antimatter in the observable part of the universe) might
arise [\cite{36}--\cite{38}].  Cosmology has likewise
turned out to be an important source of information for
elementary particle theory.  The recent rapid development
of the latter has resulted in a somewhat unusual situation
in that branch of theoretical physics.  The reason is that
typical elementary particle energies required for a direct
test of grand unified theories are of the order of
$10^{15}$~GeV, and direct tests of supergravity,
Kaluza--Klein theories, and superstring theory require
energies of the order of $10^{19}$~GeV.  On the other hand,
currently planned
\index{Accelerators}%
accelerators will only produce particle
beams with energies of about $10^4$~GeV.  Experts estimate
that the largest
\index{Accelerators}%
accelerator that could be built on earth
(which has a radius of about $6000$~km) would enable us to
study particle interactions at energies of the order of
$10^7$~GeV, which is typically the highest (center-of-mass)
energy encountered in cosmic ray experiments.  Yet this is
twelve orders of magnitude lower than the Planck energy
\index{Planck energy}%
${\rm E}_{\rm P}\sim {\rm M}_{\rm P}\sim10^{19}$~GeV.

The difficulties involved in studying interactions at
superhigh energies can be highlighted by noting that
$10^{15}$~GeV is the kinetic energy of a small car, and
$10^{19}$~GeV is the kinetic energy of a medium-sized
airplane.  Estimates indicate that accelerating particles
to energies of the order of  $10^{15}$~GeV using
present-day technology would require an
\index{Accelerators}%
accelerator approximately one light-year long.

It would be wrong to think, though, that the elementary
particle theories currently being developed are totally
without experimental foundation --- witness the experiments
on a huge scale which are under way to detect the
\index{Proton decay}%
\index{Decay of proton}%
decay of the proton, as predicted by grand unified theories.
It is also possible that
\index{Accelerators}%
accelerators will enable us to detect
some of the lighter particles (with mass
$m\sim10^2$--$10^{3}$~GeV) predicted by certain versions of
supergravity and superstring theories.  Obtaining
information solely in this way, however, would be similar
to trying to discover a unified theory of weak and
electromagnetic interactions using only radio telescopes,
detecting radio waves with an energy ${\rm E}_\gamma$ no
greater than $10^{-5}$~eV (note that
$\displaystyle\frac{{\rm E}_{\rm P}}{{\rm E}_{\rm W}}
\sim\frac{{\rm E}_{\rm W}}{{\rm E}_\gamma}$,
where ${\rm E}_{\rm W}\sim10^2$~GeV is the
characteristic energy in the unified theory of weak and
electromagnetic interactions).

The only laboratory in which particles with energies of
$10^{15}$--$10^{19}$ GeV could ever exist and interact with
one another is our own universe in the earliest stages of
its evolution.

At the beginning of the 1970s, Zeldovich wrote that the
universe is the poor man's accelerator:  experiments don't
need to be funded, and all we have to do is collect the
experimental data and interpret them properly [\cite{39}].
More recently, it has become quite clear that the universe
is the only accelerator that could ever produce particles
at energies high enough to test unified theories of all
fundamental interactions directly, and in that sense it is
not just the poor man's accelerator but the richest man's
as well.  These days, most new elementary particle theories
must first take a
\index{``Cosmological validity''}%
``cosmological validity'' test --- and only a very few pass.

It might seem at first glance that it would be difficult to
glean any reasonably definitive or reliable information
from an experiment performed more than ten billion years
ago, but recent studies indicate just the opposite.  It has
been found, for instance, that phase transitions, which
should occur in a hot universe in accordance with the grand
unified theories, should produce an abundance of magnetic
\index{Monopoles}%
monopoles, the density of which ought to exceed the
observed density of matter at the present time,
$\rho\sim10^{-29}$~$\mbox{g}\cdot\mbox{cm}^{-3}$, by
approximately fifteen orders of magnitude [\cite{40}].  At
first, it seemed that uncertainties inherent in both the
hot universe theory and the grand unified theories, being
very large, would provide an easy way out of the primordial
monopole problem.  But many attempts to resolve this
problem within the context of the standard hot universe
theory have not led to final success.  A similar situation
has arisen in dealing with theories involving spontaneous
breaking of a discrete symmetry (spontaneous CP-invariance
breaking, for example).  In such models, phase transitions
ought to give rise to supermassive domain walls, whose
existence would sharply conflict with the astrophysical
data [\cite{41}--\cite{43}].  Going to more complicated
theories such as ${\rm N} = 1$
\index{Supergravity theory!${\rm N} = 1$}%
supergravity has engendered
new problems rather than resolving the old ones.  Thus it
has turned out in most theories based on ${\rm N} = 1$
supergravity that the decay of gravitinos ($\mbox{spin}=3/2$
superpartners of the graviton) which existed in the early
stages of the universe leads to results differing from the
observational data by about ten orders of magnitude
[\cite{44}, \cite{45}].  These theories also predict the
existence of so-called scalar
\index{Polonyi fields}%
Polonyi fields [\cite{15}, \cite{46}].
The energy density that would have been
accumulated in these fields by now differs from the
cosmological data by fifteen orders of magnitude
[\cite{47}, \cite{48}].  A number of
\index{Axion theories}%
axion theories [\cite{49}] share this difficulty, particularly in the
simplest models based on superstring theory [\cite{50}].
Most Kaluza--Klein theories based on supergravity in an
11-dimensional space lead to
\index{Energies!vacuum}%
vacuum energies of order
$-{\rm M}_{\rm P}^4\sim-10^{94}$~$\mbox{g}\cdot\mbox{cm}^{-3}$
[\cite{16}], which differs from the cosmological data by
approximately 125 orders of magnitude\ldots

This list could be continued, but as it stands it suffices
to illustrate why elementary particle theorists now find
\index{Cosmology}%
cosmology so interesting and important.  An even more
general reason is that no real unification of all
interactions including gravitation is possible without an
analysis of the most important manifestation of that
unification, namely the existence of the universe itself.
This is illustrated especially clearly by Kaluza--Klein and
superstring theories, where one must simultaneously
investigate the properties of the space-time formed by
compactification of ``extra'' dimensions, and the
phenomenology of the elementary particles.

It has not yet been possible to overcome some of the problems
listed above.  This places important constraints on elementary
particle theories currently under development.  It is all the
more surprising, then, that many of these problems, together with
a number of others that predate the hot universe theory, have
been resolved in the context of one fairly simple scenario for
the development of the universe --- the so-called
\index{Inflationary universe scenario}%
inflationary universe scenario [\cite{51}--\cite{57}].
According to this scenario, the
universe, at some very early stage of its evolution, was in an
unstable vacuum-like state and expanded exponentially (the stage
of inflation).  The vacuum-like state then decayed, the universe
heated up, and its subsequent evolution can be described by the
usual hot universe theory.

Since its conception, the inflationary universe scenario
has
progressed from something akin to science fiction to a
well-established theory of the evolution of the universe
accepted by most cosmologists.  Of course this doesn't mean
that we have now finally achieved total enlightenment as to
the physical processes operative in the early universe.
The incompleteness of the current picture is reflected by
the very word {\it scenario}, which is not normally found
in the working vocabulary of a theoretical physicist.  In
its present form, this scenario only vaguely resembles the
simple models from which it sprang.  Many details of the
inflationary universe scenario are changing, tracking
rapidly changing (as noted above) elementary particle
theories.  Nevertheless, the basic aspects of this scenario
are now well-developed, and it should be possible to
provide a preliminary account of its progress.

Most of the present book is given over to discussion of
inflationary cosmology.  This is preceded by an outline of the
general theory of spontaneous symmetry breaking and a discussion of
phase transitions in superdense matter, as described by present-day
theories of elementary particles.  The choice of material has been
dictated by both the author's interests and his desire to make the
contents useful both to quantum field theorists and
astrophysicists.  We have therefore tried to concentrate on those
problems that yield an understanding of the basic aspects of the
theory, referring the reader to the original papers for further
details.

In order to make this book as widely accessible as possible, the
main exposition has been preceded by a long introductory chapter,
written at a relatively elementary level.  Our hope is that by
using this chapter as a guide to the book, and the book itself as a
guide to the original literature, the reader will gradually be able
to attain a fairly complete and accurate understanding of the
present status of this branch of science.  In this regard, he might
also be assisted by an acquaintance with the books {\it Cosmology
of the Early Universe}, by A. D. Dolgov, Ya. B. Zeldovich, and M.
V. Sazhin; {\it How the Universe Exploded}, by I. D. Novikov;
{\it A  Brief History of Time:  From the Big Bang to Black Holes}, by S.
W. Hawking; and {\it An Introduction to Cosmology and Particle
Physics}, by R. Dominguez-Tenreiro and M. Quiros.  A good
collection of early papers on inflationary cosmology and galaxy
formation can also be found in the book {\it Inflationary
Cosmology}, edited by L. Abbott and S.-Y.  Pi.  We apologize in
advance to those authors whose work in the field of inflationary
cosmology we have not been able to treat adequately.  Much of the
material in this book is based on the ideas and work of S. Coleman,
J. Ellis, A. Guth, S. W. Hawking, D. A. Kirzhnits, L. A. Kofman, M.
A. Markov, V. F.  Mukhanov, D. Nanopoulos, I. D. Novikov, I. L.
Rozental', A. D. Sakharov, A. A.  Starobinsky, P. Steinhardt, M.
Turner, and many other scientists whose contribution to modern
cosmology could not possibly be fully reflected in a single
monograph, no matter how detailed.

I would like to dedicate this book to the memory of Yakov
Borisovich Zeldovich, who should by rights be considered the
founder of the Soviet school of cosmology.
\cleardoublepage


\setcounter{page}{1}
\pagenumbering{arabic}
\chapter{\label{c1}Overview of Unified Theories of Elementary Particles
and the Inflationary  Universe Scenario}
\section{\label{s1.1}The scalar field and spontaneous symmetry breaking}

Scalar fields $\varphi$ play a fundamental role in unified
\index{Scalar fields}%
theories of the weak, strong, and electromagnetic interactions.
Mathematically, the theory of these fields is simpler than that of the
\index{Spinor fields}%
spinor fields $\psi$ describing electrons or quarks, for
instance, and it is simpler than the theory of the
\index{Vector fields}%
vector fields ${\rm A}_\mu$ which describes photons, gluons, and so on.
The most interesting and important properties of these fields for
both elementary particle theory and cosmology, however, were
grasped only fairly recently.

\begin{figure}[t]\label{f1}
\centering \leavevmode\epsfysize=6cm
\epsfbox{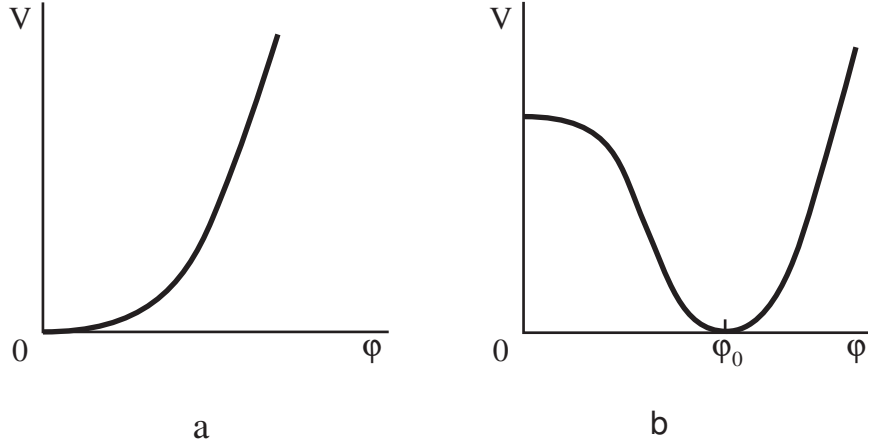}
\caption{Effective potential
\index{Effective potential}%
${\rm V}(\varphi)$ in the simplest
theories of the scalar field $\varphi$.
{\it a}) ${\rm V}(\varphi)$ in the theory (\ref{1.1.1}), and
{\it b}) in the theory (\ref{1.1.5}).}
\end{figure}

Let us recall the basic properties of such fields.  Consider
first the simplest theory of a one-component real scalar field
$\varphi$ with the Lagrangian\footnote{In this book we employ
units such that $\hbar=c=1$, the system commonly used in
elementary particle theory. In order to transform expressions to
\index{Conventional units}%
\index{Units conventional}%
conventional units, corresponding terms must be multiplied by
appropriate powers of $\hbar$ or $c$ to give the correct
dimensionality (note that $\hbar=6.6\cdot10^{-22}\;\mbox{MeV}
\cdot \mbox{sec}\approx10^{-27}\;\mbox{erg}\cdot\mbox{sec}$,
$c\approx3\cdot10^{10}\;\mbox{cm}\cdot\mbox{sec}^{-1}$). Thus,
for example Eq. (\ref{1.1.1}) would acquire the form
$${\rm L}=\frac{1}{2}\,(\partial_\mu\varphi)^2-
\frac{m^2\,c^2}{2\hbar^2}\,\varphi^2-\frac{\lambda}{4}\,\varphi^4\ .$$}
\be
\label{1.1.1}
{\rm L}=\frac{1}{2}\,(\partial_\mu\varphi)^2-\frac{m^2}{2}\,\varphi^2-
\frac{\lambda}{4}\,\varphi^4\ .
\ee
In this equation, $m$ is the mass of the scalar field, and
$\lambda$ is its coupling constant.  For simplicity, we assume
throughout that $\lambda \ll 1$.  When $\varphi$ is small and we
can neglect the last term in (\ref{1.1.1}), the field satisfies the
\index{Klein--Gordon equation}%
 Klein--Gordon equation
\be
\label{1.1.2}
(\dla+m^2)\,\varphi=\ddot\varphi-\Delta\varphi+m^2\,\varphi=0\ ,
\ee
where a dot denotes differentiation with respect to time.  The
general solution of this equation is expressible as a
superposition of plane waves, corresponding to the propagation of
particles of mass $m$ and momentum $k$ [\cite{58}]:
\ba
\label{1.1.3}
\varphi(x)&=&(2\pi)^{-3/2}\,\int d^4k\:
\delta(k^2-m^2)[e^{i\,k\,x}\,\varphi^+(k)
+e^{-i\,k\,x}\,\varphi^-(k)]\nonumber \\
&=&(2\pi)^{-3/2}\,\int \frac{d^3k}{\sqrt{2k_0}}[e^{i\,k\,x}\,a^+({\bf k})
        +e^{-i\,k\,x}\,a^-({\bf k})]\ ,
\ea
where $\displaystyle a^\pm({\bf
k})=\frac{1}{\sqrt{2k_0}}\,\varphi^\pm(k)$,
$k_0=\sqrt{{\bf k}^2+m^2}$, $k\,x=k_0\,t-{\bf k}\cdot{\bf x}$.  According to
(\ref{1.1.3}), the field $\varphi(x)$ will oscillate about the
point $\varphi=0$ density for the field $\varphi$ (the so-called
\index{Effective potential}%
effective potential)
\be
\label{1.1.4}
{\rm V}(\varphi)=\frac{1}{2}\,(\nabla\varphi)^2+
    \frac{m^2}{2}\,\varphi^2+\frac{\lambda}{4}\,\varphi^4
\ee
occurs at $\varphi=0$ (see Fig.~\ref{f1}a).




Fundamental advances in the unification of the weak, strong, and
electromagnetic interactions were finally achieved when simple
theories based on Lagrangians like (\ref{1.1.1}) with $m^2> 0$
gave way to what were at first glance somewhat strange-looking
theories with
\index{Mass!negative}%
\index{Negative mass}%
negative mass squared:
\be
\label{1.1.5}
{\rm L}=\frac{1}{2}\,(\partial_\mu\varphi)^2+\frac{\mu^2}{2}\,\varphi^2
-\frac{\lambda}{4}\,\varphi^4\ .
\ee

Instead of oscillations about $\varphi = 0$, the solution
corresponding to (\ref{1.1.3}) gives modes that grow
exponentially near $\varphi = 0$ when ${\bf k}^2 < m^2$:
\be
\label{1.1.6}
\delta\varphi({\bf k})\sim\exp\left(\pm\sqrt{\mu^2-{\bf k}^2}\,t\right)\cdot
\exp(\pm i\,{\bf k}\,{\bf x})\ .
\ee
What this means is that the minimum of the
\index{Effective potential}%
effective potential
\be
\label{1.1.7}
{\rm V}(\varphi)=\frac{1}{2}\,(\nabla\varphi)^2-\frac{\mu^2}{2}\,\varphi^2+
\frac{\lambda}{4}\,\varphi^4
\ee
will now occur not at $\varphi = 0$, but at
$\varphi_c={\pm}\mu/\sqrt{\lambda}$ (see Fig.~\ref{f1}b).\footnote{
${\rm V}(\varphi)$ usually attains a minimum for homogeneous fields
$\varphi$, so gradient terms in the expression for ${\rm
V}(\varphi)$ are often omitted.}
Thus, even if the field
$\varphi$ is zero initially, it soon undergoes a transition
(after a time of order  $\mu^{-1}$) to a stable state with the
\index{Classical fields}%
classical field $\varphi_c={\pm}\mu/\sqrt{\lambda}$, a phenomenon
known as
\index{Spontaneous symmetry breaking}%
spontaneous symmetry breaking.

After spontaneous symmetry breaking, excitations of the field
$\varphi$ near $\varphi_c={\pm}\mu/\sqrt{\lambda}$ can also be
described by a solution like (\ref{1.1.3}).  In order to do so,
we make the change of variables
\be
\label{1.1.8}
\varphi\rightarrow\varphi+\varphi_0\ .
\ee
The Lagrangian (\ref{1.1.5}) thereupon takes the form
\ba
\label{1.1.9}
{\rm L}(\varphi+\varphi_0)&=&\frac{1}{2}\,(\partial_\mu(\varphi+\varphi_0))^2
+\frac{\mu^2}{2}\,(\varphi+\varphi_0)^2-
\frac{\lambda}{4}\,(\varphi+\varphi_0)^4\nonumber \\
&=&\frac{1}{2}\,(\partial_\mu\varphi)^2-
\frac{3\,\lambda\,\varphi_0^2-\mu^2}{2}\,\varphi^2
-\lambda\,\varphi_0\,\varphi^3-\frac{\lambda}{4}\,\varphi^4\nonumber \\
&+&\frac{\mu^2}{2}\,\varphi_0^2-\frac{\lambda}{4}\,\varphi_0^4
-\varphi\,(\lambda\varphi_0^2-\mu^2)\,\varphi_0\ . \ea We see from
(\ref{1.1.9}) that when $\varphi_0\neq0$, the effective
\index{Mass!squared}%
mass squared of the field $\varphi$ is not equal to $-\mu^2$, but rather
\be
\label{1.1.10}
m^2=3\,\lambda\,\varphi_0^2-\mu^2\ ,
\ee
and when $\varphi_0={\pm}\mu/\sqrt{\lambda}$, at the minimum of
the potential ${\rm V}(\varphi)$ given by (\ref{1.1.7}), we have
\be
\label{1.1.11}
m^2=2\,\lambda\,\varphi_0^2=2\,\mu^2>0\ ;
\ee
in other words, the mass squared of the field $\varphi$ has the
correct sign.  Reverting to the original variables, we can write
the solution for $\varphi$ in the form
\be
\label{1.1.12}
\varphi(x)=\varphi_0+(2\pi)^{-3/2}\,
\int\frac{d^3k}{\sqrt{2k_0}}\:[e^{i\,k\,x}\,a^+({\bf k})
        +e^{-i\,k\,x}\,a^-({\bf k})]\ .
\ee
The integral in (\ref{1.1.12}) corresponds to particles (quanta)
of the field $\varphi$ with mass given by (\ref{1.1.11}),
propagating against the background of the constant classical
\index{Constant classical fields}%
\index{Classical fields!constant}%
field $\varphi_0$.

The presence of the constant classical field $\varphi_0$ over all
space will not give rise to any preferred reference frame
associated with that field:  the Lagrangian (\ref{1.1.9}) is
covariant, irrespective of the magnitude of $\varphi_0$.
Essentially, the appearance of a uniform field $\varphi_0$ over
all space simply represents a restructuring of the vacuum state.
In that sense, the space filled by the field $\varphi_0$ remains
``empty.''  Why then is it necessary to spoil the good theory
(\ref{1.1.1})?

The main point here is that the advent of the field $\varphi_0$
changes the masses of those particles with which it interacts.
We have already seen this in considering the example of the sign
``correction'' for the mass squared of the field $\varphi$ in the
theory (\ref{1.1.5}).  Similarly,
\index{Scalar fields}%
scalar fields can change the mass of both
\index{Fermions}%
fermions and vector particles.

Let us examine the two simplest models. The first is the
simplified $\sigma$-model, which is sometimes used for a
phenomenological description of strong interactions at high
energy [\cite{26}].  The Lagrangian for this model is a sum of
the Lagrangian (\ref{1.1.5}) and the Lagrangian for the massless
fermions $\psi$, which interact with $\varphi$ with a coupling
constant $h$:
\be
\label{1.1.13}
{\rm L}=\frac{1}{2}\,(\partial_\mu\varphi)^2+\frac{\mu^2}{2}\,\varphi^2
-\frac{\lambda}{4}\,\varphi^4+\bar\psi\,
(i\,\partial_\mu\gamma_\mu-h\,\varphi)\,\psi\ .
\ee
After symmetry breaking, the fermions will clearly acquire a mass
\be
\label{1.1.14}
m_\psi=h\,|\varphi_0|=h\,\frac{\mu}{\sqrt{\lambda}}\ .
\ee

The second is the so-called
\index{Higgs model}%
Higgs model [\cite{59}], which
describes an Abelian vector field ${\rm A}_\mu$ (the analog of
the electromagnetic field) that interacts with the complex scalar
field $\chi=(\chi_1+i\,\chi_2)/\sqrt{2}$.  The Lagrangian for
this theory is given by
\ba
\label{1.1.15}
{\rm L}&=&-\frac{1}{4}\,(\partial_\mu{\rm A}_\nu-\partial_\nu{\rm A}_\mu)^2
+(\partial_\mu+i\,e\,{\rm A}_\mu)\,\chi^*\,
(\partial_\mu-i\,e\,{\rm A}_\mu)\,\chi\nonumber \\
&+&\mu^2\,\chi^*\,\chi-\lambda\,(\chi^*\,\chi)^2\ .
\ea
As in (\ref{1.1.7}), when $\mu^2<0$ the scalar field $\chi$
acquires a classical component.  This effect is described most
easily by making the change of variables
\ba
\label{1.1.16}
\chi(x)&\rightarrow&\frac{1}{\sqrt{2}}\left(\varphi(x)+
\varphi_0\right)\, \exp\frac{i\,\zeta(x)}{\varphi_0}\ ,\nonumber \\
{\rm A}_\mu(x)&\rightarrow&{\rm A}_\mu(x)+
\frac{1}{e\,\varphi_0}\,\partial_\mu\zeta(x)\ ,
\ea
whereupon the Lagrangian (\ref{1.1.15}) becomes
\ba
\label{1.1.17}
{\rm L}&=&-\frac{1}{4}\,\,(\partial_\mu{\rm A}_\nu-\partial_\nu{\rm A}_\mu)^2
+\frac{e^2}{2}(\varphi+\varphi_0)^2\,
{\rm A}_\mu^2+\frac{1}{2}\,(\partial_\mu\varphi)^2\nonumber \\
&-&\frac{3\,\lambda\,\varphi_0^2-\mu^2}{2}\,\varphi^2
-\lambda\,\varphi_0\,\varphi^3-\frac{\lambda}{4}\,\varphi^4
+\frac{\mu^2}{2}\,\varphi_0^2-\frac{\lambda}{4}\,\varphi_0^4\nonumber \\
&-&\varphi(\lambda\,\varphi_0^2-\mu^2)\,\varphi_0\ .
\ea
Notice that the auxiliary field $\zeta(x)$ has been entirely
canceled out of (\ref{1.1.17}), which describes a theory of
vector particles of mass  $m_{\rm A}=e\,\varphi_0$ that interact
with a scalar field having the
\index{Effective potential}%
effective potential (\ref{1.1.7}).
As before, when $\mu^2>0$, symmetry breaking occurs, the field
$\varphi_0=\mu/\sqrt{\lambda}$ appears, and the vector particles
of ${\rm A}_\mu$ acquire a mass
$m_{\rm A}=e\,\mu/\sqrt{\lambda}$. This scheme for making vector mesons
\index{Vector mesons}%
massive is called the
\index{Higgs mechanism}%
Higgs mechanism, and the fields $\chi$,
$\varphi$ are known as\index{Higgs fields} Higgs fields.
The appearance of the classical field  $\varphi_0$ breaks the symmetry of
(\ref{1.1.15}) under ${\rm U}(1)$ gauge transformations:
\ba
\label{1.1.18}
{\rm A}_\mu&\rightarrow&{\rm A}_\mu+\frac{1}{e}\,\partial_\mu\zeta(x)\
\nonumber \\
\chi&\rightarrow&\chi\,\exp\,[i\,\zeta(x)]\ .
\ea

The basic idea underlying unified theories of the weak, strong,
and electromagnetic interactions is that prior to symmetry
breaking, all
\index{Vector mesons}%
vector mesons (which mediate these interactions)
are massless, and there are no fundamental differences among the
interactions.  As a result of the symmetry breaking, however, some of the
\index{Bosons!vector}%
vector bosons do acquire mass, and their
corresponding interactions become short-range, thereby destroying
the symmetry between the various interactions.  For example,
prior to the appearance of the constant scalar Higgs field H, the
Glashow--Weinberg--Salam model [\cite{1}] has
$\mbox{SU}(2)\times{\rm U}(1)$ symmetry, and electroweak
interactions are mediated by massless vector bosons.
\index{Bosons!vector}%
After the appearance of the constant scalar field H, some of the vector
bosons
\index{Bosons!W}%
(${\rm W}_\mu^\pm$  and ${\rm Z}_\mu^0$) acquire
masses of order $e{\rm H} \sim 100$ GeV, and the corresponding interactions
become short-range (weak interactions), whereas the
electromagnetic field ${\rm A}_\mu$ remains massless.

The Glashow--Weinberg--Salam model was proposed in the 1960's
[\cite{1}], but the real explosion of interest in such theories
did not come until 1971--1973, when it was shown that gauge
theories with spontaneous symmetry breaking are renormalizable,
which means that there is a regular method for dealing with the
ultraviolet divergences, as in ordinary quantum electrodynamics
[\cite{2}].
\index{Renormalizability, proof of}%
The proof of renormalizability for unified field
theories is rather complicated, but the basic physical idea
behind it is quite simple.  Before the appearance of the scalar
field $\varphi_0$, the unified theories are renormalizable, just
like ordinary quantum electrodynamics.  Naturally, the appearance
of a classical
\index{Classical fields}%
\index{Scalar fields}%
scalar field $\varphi_0$ (like the presence of the
ordinary classical electric and magnetic fields) should not
affect the high-energy properties of the theory;  specifically,
it should not destroy the original renormalizability of the
theory.  The creation of unified gauge theories with spontaneous
symmetry breaking and the proof that they are renormalizable
carried elementary particle theory in the early 1970's to a
qualitatively new level of development.

The number of scalar field types occurring in unified theories
can be quite large.  For example, there are two Higgs fields in
the simplest theory with
\index{Symmetry!SU(5)}%
\index{SU(5) symmetry}%
$\mbox{SU}(5)$ symmetry [\cite{4}].  One
of these, the field $\Phi$, is represented by a traceless $5\times 5$
matrix.  Symmetry breaking in this theory results from
the appearance of the classical field
\be
\label{1.1.19}
\Phi_0=\sqrt{\frac{2}{15}}\,\varphi_0\,\left(
\begin{array}{ccccc}
1&&&&0\\
 &1\\
 &&1\\
&&&-3/2\\
0&&&&-3/2
\end{array}\right)\ ,
\ee
where the value of the field $\varphi_0$ is extremely large ---
$\varphi_0\sim10^{15}$  GeV.  All vector particles in this theory
are massless prior to symmetry breaking, and there is no
fundamental difference between the weak, strong, and
electromagnetic interactions.  Leptons can then easily be
transformed into quarks, and vice versa.  After the appearance of
the field (\ref{1.1.19}), some of the vector
\index{Bosons!X}%
mesons (the X and Y mesons responsible for transforming quarks into leptons)
acquire enormous mass:
$m_{\rm X,Y}=(5/3)^{1/2}\,g\,\varphi_0/2\sim10^{15}$ GeV,
where $g^2\sim0.3$ is the SU(5) gauge coupling constant.  The
transformation of quarks into leptons thereupon becomes strongly
inhibited, and the proton becomes almost stable.  The original
SU(5) symmetry breaks down into
$\mbox{SU}(3) \times \mbox{SU}(2) \times {\rm U}(1)$;
that is, the strong interactions (SU(3)) are
separated from the electroweak ($\mbox{SU}(2) \times {\rm U}(1)$).
Yet another classical scalar field ${\rm H}\sim10^2$
GeV then makes its appearance, breaking the symmetry between the
weak and electromagnetic interactions, as in the
\index{Glashow--Weinberg--Salam theory}%
Glashow--Weinberg--Salam theory [\cite{4}, \cite{12}].

The Higgs effect and the general properties of theories with
\index{Higgs effect}%
spontaneous symmetry breaking are discussed in more detail in
Chapter \ref{c2}.  The elementary theory of spontaneous symmetry
breaking is discussed in Section \ref{s2.1}.
In Section \ref{s2.2}, we further
study this phenomenon, with quantum corrections to the
\index{Effective potential}%
effective potential ${\rm V}(\varphi)$ taken into consideration.  As will
be shown in Section \ref{s2.2}, quantum corrections can in some cases
significantly modify the general form of the potential
(\ref{1.1.7}).  Especially interesting and unexpected properties
of that potential will become apparent when we study it in the
$1/{\rm N}$ approximation.

\section{\label{s1.2}Phase transitions in
\index{Gauge theories!phase transitions in|(}%
\index{Phase transitions!in gauge theories|(}%
gauge theories}

The idea of spontaneous symmetry breaking, which  proved to be so
useful in building unified gauge theories, has an extensive
history in solid-state theory and quantum statistics, where it
has been used to describe such phenomena as ferromagnetism,
superfluidity,
\index{Superconductivity}%
superconductivity, and so forth.

Consider, for example, the expression for the energy of a
superconductor in the phenomenological
\index{Ginzburg--Landau theory}%
Ginzburg--Landau theory [\cite{60}] of superconductivity:
\be
\label{1.2.1}
{\rm E}={\rm E}_0+\frac{{\rm H}^2}{2}+\frac{1}{2m}\,
|(\nabla-2\,i\,e\,{\rm A})\,\Psi|^2-\alpha\,|\Psi|^2+\beta\,|\Psi|^4\ .
\ee
Here ${\rm E}_0$ is the energy of the normal metal without a
magnetic field H, $\Psi$ is the field describing the
Cooper-pair
\index{Bose condensate!cooper-pair}%
\index{Cooper-pair Bose condensate}%
Bose condensate, and $\alpha$ and $\beta$ are positive parameters.

Bearing in mind, then, that the potential energy of a field
enters into the Lagrangian with a negative sign, it is not hard
to show that the Higgs model (\ref{1.1.15}) is simply a
relativistic generalization of the
\index{Ginzburg--Landau theory}%
Ginzburg--Landau theory of
\index{Superconductivity}%
superconductivity (\ref{1.2.1}), and the classical field
$\varphi$ in the Higgs model is the analog of the Cooper-pair
Bose condensate.\footnote{Where this does not lead to confusion,
we will simply denote the classical scalar field by $\varphi$, rather
then $\varphi_0$. In certain other cases, we will also denote the
initial value of the classical scalar field $\varphi$ by $\varphi_0$.
We hope that the meaning of $\varphi$ and $\varphi_0$ in each particular
case will be clear from the context.}

The analogy between unified theories with spontaneous symmetry
breaking and theories of
\index{Superconductivity}%
superconductivity has been found to be
extremely useful in studying the properties of superdense matter
described by unified theories.  Specifically, it is well known
that when the temperature is raised, the Cooper-pair condensate
shrinks to zero and superconductivity disappears.  It turns out
that the uniform scalar field $\varphi$ should also disappear
when the temperature of matter is raised;  in other words, at
superhigh temperatures, the symmetry between the weak, strong,
and electromagnetic interactions ought to be restored
[\cite{18}--\cite{24}].

\begin{figure}[t]\label{f2}
\centering \leavevmode\epsfysize=6cm \epsfbox{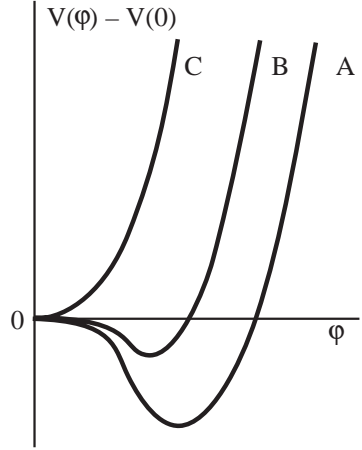}
\caption{Effective
\index{Effective potential}%
potential ${\rm V}(\varphi, {\rm T})$ in
the theory (\ref{1.1.5}) at finite temperature.
A) ${\rm T} = 0$;  B) $0<{\rm T}<{\rm T}_c$; C) ${\rm T}>{\rm T}_c$.
As the temperature rises, the field $\varphi$ varies smoothly,
corresponding to a
\index{Phase transitions!second-order}%
\index{Second-order phase transitions}%
second-order phase transition.}
\end{figure}

A theory of phase transitions involving the disappearance of the
classical field $\varphi$ is discussed in detail in
Ref.~\cite{24}. In gross outline, the basic idea is that the
equilibrium value of the field $\varphi$ at fixed temperature
${\rm T} \neq 0$ is governed not by the location of the minimum
of the potential energy density ${\rm V}(\varphi)$, but by the
location of the minimum of the free energy density ${\rm
F}(\varphi, {\rm T}) \equiv {\rm V}(\varphi, {\rm T})$, which
equals ${\rm V}(\varphi)$ at ${\rm T} = 0$.  It is well-known
that the temperature-dependent contribution to the free energy F
from ultrarelativistic scalar particles of mass $m$ at
temperature ${\rm T}\gg m$ is given [\cite{61}] by
\be
\label{1.2.2}
\Delta{\rm F}=\Delta{\rm V}(\varphi,{\rm T})=
-\frac{\pi^2}{90}\,{\rm T}^4+\frac{m^2}{24}\,{\rm T}^2\,
\left(1+{\rm O}\left(\frac{m}{{\rm T}}\right)\right)\ .
\ee
If we then recall that
$$
m^2(\varphi)=\frac{d^2{\rm V}}{d\varphi^2}=3\,\lambda\,\varphi^2-\mu^2
$$
in the model (\ref{1.1.5}) (see Eq. (\ref{1.1.10})), the complete
expression for ${\rm V}(\varphi,{\rm T})$ can be written in the
form \be \label{1.2.3} {\rm V}(\varphi,{\rm
T})=-\frac{\mu^2}{2}\,\varphi^2+
\frac{\lambda\,\varphi^4}{4}+\frac{\lambda\,{\rm
T}^2}{8}\,\varphi^2 +\ldots\ , \ee where we have omitted terms
that do not depend on $\varphi$.  The behavior of ${\rm
V}(\varphi,{\rm T})$ is shown in Fig.~\ref{f2} for a number of
different temperatures.

It is clear from (\ref{1.2.3}) that as T rises, the equilibrium
value of $\varphi$ at the minimum of ${\rm V}(\varphi, {\rm T})$
decreases, and above some critical temperature
\be
\label{1.2.4}
{\rm T}_c=\frac{2\,\mu}{\sqrt{\lambda}}\ ,
\ee
the only remaining minimum is the one at $\varphi = 0$, i.e.,
symmetry is restored (see Fig.~\ref{f2}).  Equation (\ref{1.2.3})
then implies that the field $\varphi$ decreases continuously to
zero with rising temperature;  the restoration of symmetry in the
theory (\ref{1.1.5}) is a second-order phase transition.

Note that in the case at hand, when $\lambda \ll 1$, ${\rm T}_c
\gg m$ over the entire range of values of $\varphi$ that is of
interest ($\varphi\la\varphi_c$), so that a high-temperature
expansion of ${\rm V}(\varphi, {\rm T})$ in powers of $m/{\rm T}$
in (\ref{1.2.2}) is perfectly justified.  However, it is by no
means true that phase transitions take place only at ${\rm T}\gg
m$ in all theories.  It often happens that at the instant of a
phase transition, the potential ${\rm V}(\varphi, {\rm T})$ has
two local minima, one giving a stable state and the other an
unstable state of the system (Fig.~1.3).  We then have a
\index{First-order phase transitions}%
first-order phase transition, due to the formation and subsequent
expansion of bubbles of a stable phase within an unstable one, as
in boiling water.  Investigation of the first-order phase
transitions in gauge theories [\cite{62}] indicates that such
transitions are sometimes considerably delayed, so that the
transition takes place (with rising temperature) from a strongly
superheated state, or (with falling temperature) from a strongly
supercooled one.  Such processes are explosive, which can lead to
many important and interesting effects in an expanding universe. The
\index{Bubble production}%
formation of bubbles of a new phase
is typically a barrier tunneling process;  the theory of this process
at a finite temperature was given in [\cite{62}].

\begin{figure}[t]
\centering \leavevmode\epsfysize=6cm \epsfbox{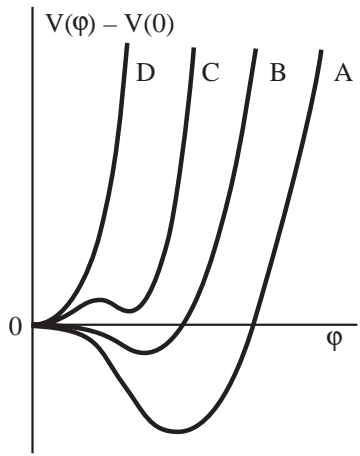}
\caption{Behavior of the
\index{Effective potential}%
effective potential ${\rm V}(\varphi, {\rm T})$ in theories in
which
\index{First-order phase transitions}%
phase transitions are first-order. Between ${\rm T}_{c_1}$ and
${\rm T}_{c_2}$, the effective potential has two minima;  at ${\rm
T}={\rm T}_c$, these minima have the same depth. A) ${\rm T} = 0$;
B) ${\rm T}_{c_1} < {\rm T} < {\rm T}_c$; C) ${\rm T}_c < {\rm T}
< {\rm T}_{c_2}$; D) ${\rm T} > {\rm T}_{c_2}$.}
\end{figure}\label{f3a}

It is well known that
\index{Superconductivity}%
superconductivity can be destroyed not only
by heating, but also by external fields {\bf H} and currents
{\bf j};  analogous effects exist in unified gauge theories
[\cite{22}, \cite{23}].  On the other hand, the value of the
field $\varphi$, being a scalar, should depend not just on the
currents {\bf j}, but on the square of current $j^2=\rho^2-{\bf j}^2$,
where $\rho$ is the charge density.  Therefore, while
increasing the current {\bf j} usually leads to the restoration
of symmetry in gauge theories, increasing the charge density
$\rho$ usually results in the enhancement of symmetry breaking
[\cite{27}].  This effect and others that may exist in superdense
cold matter are discussed
\index{Gauge theories!phase transitions in|)}%
in Refs. \cite{27}--\cite{29}.
\index{Phase transitions!in gauge theories|)}%

\section{\label{s1.3}Hot universe theory
\index{Hot universe theory|(}%
}

There have been two important stages in the development of
twentieth-century cosmology.  The first began in the 1920's, when
Friedmann used the general theory of relativity to create a
theory of a homogeneous and isotropic expanding universe with
metric [\cite{63}--\cite{65}]
\be
\label{1.3.1}
ds^2=dt^2-a^2(t)\,
\left[\frac{dr^2}{1-k\,r^2}+r^2\,(d\theta+\sin^2\theta\,d\varphi^2)\right]\ ,
\ee
where $k = +1$, $-1$, or 0 for a closed, open, or flat Friedmann
\index{Friedmann universe}%
\index{Universe!Friedmann}%
universe, and $a(t)$ is the ``radius'' of the universe, or more
precisely, its
\index{Universe!scale factor of}%
\index{Scale factor of universe}%
scale factor (the total size of the universe may
be infinite).  The term
\index{Universe!flat}%
\index{Flat universe}%
{\it flat universe} refers to the fact that when
$k = 0$, the metric (\ref{1.3.1}) can be put in the form
\be
\label{1.3.2}
ds^2=dt^2-a^2(t)\,(dx^2+dy^2+dz^2)\ .
\ee
At any given moment, the spatial part of the metric describes an
ordinary three-dimensional Euclidean (flat) space, and when
$a(t)$ is constant (or slowly varying, as in our universe at
present), the flat-universe metric describes Minkowski space.

For $k = \pm 1$, the geometrical interpretation of the
three-dimensional space part of (\ref{1.3.1}) is somewhat more
complicated [\cite{65}].  The analog of a closed world at any
given time $t$ is a sphere ${\rm S}^3$ embedded in some auxiliary
four-dimensional space $(x, y, z, \tau)$.  Coordinates on this
sphere are related by \be \label{1.3.3} x^2+y^2+z^2+\tau^2=a^2(t)\
. \ee The metric on the surface can be written in the form \be
\label{1.3.4} dl^2=a^2(t)\,
\left[\frac{dr^2}{1-r^2}+r^2\,(d\theta^2+\sin^2\theta\,d\varphi^2)\right]\
, \ee where $r$, $\theta$, and $\varphi$ are spherical coordinates
on the surface of the sphere ${\rm S}^3$.

\begin{figure}[t]\label{f4}
\centering \leavevmode\epsfysize=5cm \epsfbox{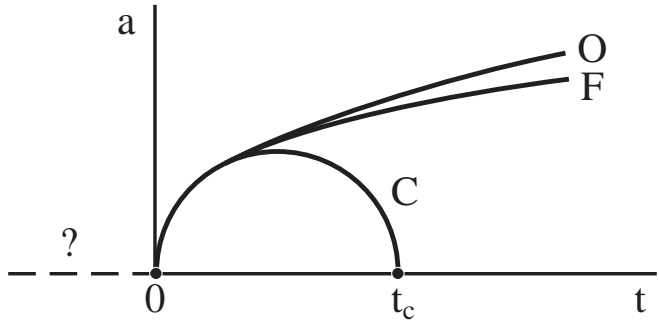}
\caption{Evolution of the scale factor $a(t)$ for three different
versions of the Friedmann hot universe theory: open (O), flat (F),
and closed (C).}
\end{figure}

The analog of an
\index{Open universe}%
\index{Universe!open}%
open universe at fixed $t$ is the surface of the hyperboloid
\be
\label{1.3.5} x^2+y^2+z^2-\tau^2=a^2(t)\ . \ee
The evolution of
the scale factor $a(t)$ is given by the Einstein equations \ba
\label{1.3.6}
\ddot a&=&-\frac{4\,\pi}{3}\,{\rm G}\,(\rho+3\,p)\,a\ ,\\
\label{1.3.7} {\rm H}^2+\frac{k}{a^2}&\equiv&\left(\frac{\dot
a}{a}\right)^2+\frac{k}{a^2} =\frac{8\,\pi}{3}\,{\rm G}\,\rho\ .
\ea Here $\rho$ is the energy density of matter in the universe,
and $p$ is its pressure. The gravitational constant ${\rm G}={\rm
M}_{\rm P}^{-2}$, where  ${\rm M}_{\rm P}=1.2\cdot10^{19}$ GeV is
the Planck mass,\footnote{The reader should be warned that in the
recent literature the authors often use a different definition of
the Planck mass, which is smaller than the one used in our book by
a factor of $\sqrt {8\pi}$.} and $\displaystyle {\rm H}=\frac{\dot
a}{a}$ is the
\index{Hubble ``constant''}%
Hubble ``constant'', which in general is a function of time.
Equations (\ref{1.3.6}) and (\ref{1.3.7}) imply an energy
conservation law, which can be written in the form
\be
\label{1.3.8}
\dot \rho\,a^3+3\,(\rho+p)\,a^2\,\dot a=0\ .
\ee

To find out how this universe will evolve in time, one also needs to
know the so-called
\index{Equation of state}%
equation of state, which relates the energy density of
matter to its pressure.  One may assume, for instance, that the equation of
state for matter in the universe takes the form $p = \alpha\,\rho$. From the
energy conservation law, one then deduces that
\be
\label{1.3.9}
\rho\sim a^{-3(1+\alpha)}\ .
\ee
In particular, for nonrelativistic cold matter with $p = 0$,
\be
\label{1.3.10}
\rho\sim a^{-3}\ ,
\ee
and for a hot ultrarelativistic gas of noninteracting particles with
$\displaystyle p=\frac{\rho}{3}$,
\be
\label{1.3.11}
\rho\sim a^{-4}\ .
\ee
In either case (and in general for any medium with
$\displaystyle p>-\frac{\rho}{3}$), when $a$ is small, the
quantity $\displaystyle \frac{8\,\pi}{3}\,{\rm G}\,\rho$ is much
greater than $\displaystyle \frac{k}{a^2}$.  We then find from
(\ref{1.3.7}) that for small $a$, the expansion of the universe
goes as
\be
\label{1.3.12}
a\sim t^{\frac{2}{3(1+a)}}\ .
\ee

In particular, for nonrelativistic cold matter
\be
\label{1.3.13}
a\sim t^{2/3}\ ,
\ee
and for the ultrarelativistic gas
\be
\label{1.3.14}
a\sim t^{1/2}\ .
\ee
Thus, regardless of the model used ($k = \pm 1$, $0$), the scale
factor vanishes at some time $t = 0$, and the matter density at
that time becomes infinite.  It can also be shown that at that
time, the
\index{Curvature tensor}%
curvature tensor ${\rm R}_{\mu\nu\alpha\beta}$ goes to
infinity as well.  That is why the point $t = 0$ is known as the
point of the initial cosmological singularity
\index{Big Bang}%
(Big Bang).

An open or flat universe will continue to expand forever.  In a
\index{Closed universe}%
\index{Universe!closed}%
closed universe with $\displaystyle p>-\frac{\rho}{3}$, on the
other hand, there will be some point in the expansion when the
term $\displaystyle \frac{1}{a^2}$ in (\ref{1.3.7}) becomes equal
to $\displaystyle \frac{8\,\pi}{3}\,{\rm G}\,\rho$.  Thereafter,
the scale constant $a$ decreases, and it vanishes at some time $t_c$
\index{Big Crunch}%
(Big Crunch).  It is straightforward to
show [\cite{65}] that the lifetime of a closed universe filled
with a total mass M of cold nonrelativistic matter is
\be
\label{1.3.15}
t_c=\frac{4\,{\rm M}}{3}\,{\rm G}=\frac{4\,{\rm M}}{3\,{\rm M}_{\rm P}^2}
\sim\frac{{\rm M}}{{\rm M}_{\rm P}}\cdot10^{-43}\;\mbox{sec}\ .
\ee

The lifetime of a closed universe filled with a hot
ultrarelativistic gas of particles of a single species may be
conveniently expressed in terms of the total
\index{Universe!total entropy of}%
\index{Entropy, total!of universe}%
entropy of the universe, ${\rm S}=2\,\pi^2\,a^3\,s$,
where $s$ is the entropy
density.  If the total entropy of the universe does not change
(adiabatic expansion), as is often assumed, then
\be
\label{1.3.16}
t_c=\left(\frac{32}{45\,\pi^2}\right)^{1/6}\,
\frac{{\rm S}^{2/3}}{{\rm M}_{\rm P}}\sim
{\rm S}^{2/3}\cdot10^{-43}\;\mbox{sec}\ .
\ee
These estimates will turn out to be useful in discussing the
difficulties encountered by the standard theory of expansion of
the hot universe.

Up to the mid-1960's, it was still not clear whether the early
universe had been hot or cold.  The critical juncture marking the
beginning of the second stage in the development of modern
cosmology was Penzias and Wilson's 1964--65 discovery of the 2.7 K microwave
\index{Microwave background radiation}%
\index{Background radiation, microwave}%
background radiation arriving from the farthest
reaches of the universe.  The existence of the microwave
background had been predicted by the hot universe theory [\cite{66},
\cite{67}], which gained immediate and widespread acceptance after the
discovery.

According to that theory, the universe, in the very early stages
of its evolution, was filled with an ultrarelativistic gas of
photons, electrons, positrons, quarks, antiquarks, etc.  At that
epoch, the excess of baryons over antibaryons was but a small
fraction (at most $10^{-9}$) of the total number of particles.
As a result of the decrease of the effective coupling constants
for weak, strong, and electromagnetic interactions with
increasing density, effects related to interactions among those
particles affected the equation of state of the superdense matter
only slightly, and the quantities $s$, $\rho$, and $p$ were given
[\cite{61}] by
\ba
\label{1.3.17}
\rho&=&3\,p=\frac{\pi^2}{30}\,{\rm N}({\rm T})\,{\rm T}^4\ ,\\
\label{1.3.18}
s&=&\frac{2\,\pi^2}{45}\,{\rm N}({\rm T})\,{\rm T}^3\ ,
\ea
where the effective number of particle species $\rm N(T)$ is
$\displaystyle {\rm N}_{\rm B}({\rm T})+\frac{7}{8}{\rm N}_{\rm F}({\rm T})$,
and ${\rm N}_{\rm B}$ and ${\rm N}_{\rm F}$ are the
number of boson and fermion species\footnote{To be more precise,
${\rm N}_{\rm B}$ and ${\rm N}_{\rm F}$ are the number of boson
and fermion degrees of freedom. For example, ${\rm N}_{\rm B}=2$
for photons, ${\rm N}_{\rm F}=2$ for neutrinos, ${\rm N}_{\rm
F}=4$ for electrons, etc.}
with masses $m \ll {\rm T}$.

In realistic elementary particle theories, ${\rm N}({\rm T})$ increases with
increasing T, but it typically does so relatively slowly, varying over the
range $10^2$ to $10^4$.  If the universe expanded adiabatically, with
$s\,a^3\approx\mbox{const}$, then (\ref{1.3.18}) implies that during
the expansion, the quantity $a{\rm T}$ also
remained approximately constant.  In other words, the
\index{Temperature!of universe}%
\index{Universe!temperature of}%
temperature of the universe dropped off as
\be
\label{1.3.19}
{\rm T}(t)\sim a^{-1}(t)\ .
\ee

The background radiation detected by Penzias and Wilson is a
result of the cooling of the hot photon gas during the expansion
of the universe.  The exact equation for the time-dependence of
the temperature in the early universe can be derived from
(\ref{1.3.7}) and (\ref{1.3.17}):
\be
\label{1.3.20}
t=\frac{1}{4\,\pi}\,\sqrt{\frac{45}{\pi\,{\rm N}({\rm T})}}\,
\frac{{\rm M}_{\rm P}}{{\rm T}^2}\ .
\ee
In the later stages of the evolution of the universe, particles
and antiparticles annihilate each other, the
\index{Photon-gas energy density}%
photon-gas energy density falls off relatively rapidly (compare
(\ref{1.3.10}) and (\ref{1.3.11})), and the main contribution to the
matter density starts to come from the small excess of baryons over
antibaryons, as well as from other fields and particles which now comprise
the so-called
\index{Hidden mass}%
\index{Mass!hidden}%
hidden mass in the universe.

The most detailed and accurate description of the hot universe
theory can be found in the fundamental monograph by Zeldovich and
Novikov [\cite{34}] (see also [\cite{35}]).

Several different avenues were pursued in the 1970's in
developing this theory.  Two of these will be most important in
the subsequent discussion:  the development of the hot universe
theory with regard to the theory of phase transitions in
superdense matter [\cite{18}--\cite{24}], and the theory of
formation of the baryon asymmetry of the universe
[\cite{36}--\cite{38}].

Specifically, as just stated in the preceding paragraph, symmetry
should be restored in\index{Grand unified theories} grand unified
theories at superhigh
temperatures.  As applied to the simplest\index{SU(5) theory}
SU(5) model, for
instance, this means that at a temperature ${\rm T}\ga10^{15}$
GeV, there was essentially no difference between the weak,
strong, and electromagnetic interactions, and quarks could easily
transform into leptons;  that is, there was no such thing as
baryon number conservation.  At $t_1 \sim 10^{-35}$ sec after the
\index{Big Bang}Big Bang, when the temperature had dropped to
${\rm T} \sim {\rm T}_{c_1}\sim 10^{14}$--$10^{15}$ GeV,
the universe underwent the first symmetry-breaking phase transition,
with SU(5) perhaps being broken into
$\mbox{SU}(3) \times \mbox{SU}(2) \times {\rm U}(1)$.
After this transition, strong interactions were
separated from electroweak and leptons from quarks, and
superheavy-meson decay processes ultimately leading to the baryon
asymmetry of the universe were initiated.  Then, at
$t_2 \sim 10^{-10}$ sec, when the temperature had dropped to
${\rm T}_{c_2} \sim 10^2$ GeV, there was a second phase transition,
which\index{Electromagnetic and weak interactions, symmetry
broken between} broke the symmetry between the weak and
electromagnetic interactions,
$\mbox{SU}(3) \times \mbox{SU}(2)\times {\rm U}(1)
\rightarrow \mbox{SU}(3) \times {\rm U}(1)$.  As the temperature
dropped still further to ${\rm T}_{c_3} \sim 10^2$ MeV, there was
yet another phase transition (or perhaps two distinct ones), with
the formation of baryons and mesons from quarks and the breaking
of chiral invariance in strong interaction theory.  Physical
processes taking place at later stages in the evolution of the
universe were much less dependent on the specific features of
unified gauge theories (a description of these processes can be
found in the books cited above [\cite{34}, \cite{35}]).

Most of what we have to say in this book will deal with events
that transpired approximately $10^{10}$ years ago, in the time up
to about $10^{-10}$ seconds after the \index{Big Bang}Big Bang.
This will make it possible to examine the global structure of the
universe, to derive a more adequate understanding of the present state
of the universe and its future, and finally, even to modify considerably
the very notion of the Big Bang.\index{Hot universe theory|)}

\section{\label{s1.4}Some\index{Friedmann models|(}
properties of the Friedmann models}

In order to provide some orientation for the problems of modern
cosmology, it is necessary to present at least a rough idea of
typical values of the quantities appearing in the equations, the
relationships among these quantities, and their physical meaning.

We start with the Einstein equation (\ref{1.3.7}), which we will find to be
particularly important in what follows.  What can one say about the Hubble
parameter $\displaystyle {\rm H}=\frac{\dot a}{a}$, the density
$\rho$, and the quantity $k$?

At the earliest stages of the evolution of the universe (not long
after the singularity), H and $\rho$ might have been arbitrarily
large.  It is usually assumed, though, that at densities
$\rho\ga\m^4\sim10^{94}$ g/cm$^3$, quantum gravity effects are so
significant that quantum fluctuations of the metric exceed the
classical value of $g_{\mu\nu}$, and classical space-time does
not provide an adequate description of the universe [\cite{34}].
We therefore restrict further discussion to phenomena for which
$\rho\la\m^4$, ${\rm T}\la\m\sim10^{19}$ GeV, ${\rm H} <\m$, and
so on.  This restriction can easily be made more precise by
noting that quantum corrections to the Einstein equations in a
hot universe are already significant for
$\displaystyle{\rm T}\sim\frac{\m}{\sqrt{{\rm N}}}\sim10^{17}$--$10^{18}$
GeV and
$\displaystyle\rho\sim\frac{\m^4}{{\rm N}}\sim10^{90}$--$10^{92}$ g/cm$^3$.
It is also worth noting that in an expanding universe,
thermodynamic equilibrium cannot be established immediately, but
only when the temperature T is sufficiently low.  Thus in SU(5)
models, for example, the typical time for equilibrium to be
established is only comparable to the age $t$ of the universe
from (\ref{1.3.20}) when ${\rm T}\la {\rm T}^*\sim10^{16}$ GeV
(ignoring hypothetical graviton processes that might lead to
equilibrium even before the Planck time has elapsed, with
$\rho\gg\m^4$).

The behavior of the nonequilibrium universe at densities of the
order of the Planck density\index{Planck density}\index{Density!Planck}
is an important problem to which we
shall return again and again.  Notice, however, that
${\rm T}^*\sim10^{16}$ GeV exceeds the typical critical temperature for
a phase transition in grand unified theories, ${\rm T}_c\la10^{15}$ GeV.

At the present time, the values of H and $\rho$ are not
well-determined. For example, \be \label{1.4.1} {\rm
H}=100\,h\,\frac{\mbox{km}}{\mbox{sec}\cdot\mbox{Mpc}}\sim
h\cdot(3\cdot10^{17})^{-1}\;\mbox{sec}^{-1} \sim
h\cdot10^{-10}\;\mbox{yr}^{-1}\ , \ee where the factor $h = 0.7\pm
0.1$ (1 \index{Mpc (megaparsec)}\index{Megaparsec
(Mpc)}megaparsec (Mpc) equals $3.09\cdot 10^{24}$ cm or $3.26\cdot
10^6$ light years). For a flat universe, H and $\rho$ are uniquely
related by Eq. (\ref{1.3.7});  the corresponding value
$\rho=\rho_c({\rm H})$ is known as the\index{Critical
density}\index{Density!critical} critical density, since the
universe must be closed (for given H) at higher density, and open
at lower: \be \label{1.4.2} \rho_c=\frac{3\,{\rm
H}^2}{8\,\pi\,{\rm G}}= \frac{3\,{\rm H}^2\,\m^2}{8\,\pi}\ , \ee
and at present, the critical density of the universe is \be
\label{1.4.3} \rho_c\approx2\cdot 10^{-29}\,h^2\;\mbox{g/cm$^3$}\
. \ee The ratio of the actual density of the universe to the
critical density is given by the quantity $\Omega$, \be
\label{1.4.4} \Omega=\frac{\rho}{\rho_c}\ . \ee Contributions to
the density $\rho$ come both from luminous baryon matter, with
$\rho_{\rm LB}\sim10^{-2}\,\rho_c$, and from dark (hidden,
missing) matter, which should have a density at least an order of
magnitude higher.  The observational data imply that\footnote{The
estimate of $h$ and $\Omega$ are changed from their values given
in the original edition of the book with an account taken of the
recent observational data. The age of the universe will be somewhat bigger than the one given in (1.4.8) (about 13.7 billion years) for the presently accepted cosmological model where 70 percent of matter corresponds to dark energy with  $p \approx -\rho$.} \be \label{1.4.5} \Omega = 1.01\pm
0.02 . \ee

The present-day universe is thus not too far from being flat
(while according to the inflationary universe scenario, $\Omega = 1$
to high accuracy; see below).  Furthermore, as we remarked
previously, the early universe not far from being spatially flat
because of the relatively small value of $\displaystyle \frac{k}{a^2}$
compared to $\displaystyle \frac{8\,\pi\,{\rm G}}{3}\,\rho$ in
(\ref{1.3.7}).  From here on, therefore, we
confine our estimates to those for a flat universe ($k = 0$).

Equations (\ref{1.3.13}) and (\ref{1.3.14}) imply that the age of
\index{Age!of universe}\index{Universe!age of}a universe
filled with ultrarelativistic gas is related to the
quantity $\displaystyle {\rm H}=\frac{\dot a}{a}$ by
\be
\label{1.4.6}
t=\frac{1}{2\,{\rm H}}\ ,
\ee
and for a universe with the equation of state $p = 0$,
\be
\label{1.4.7}
t=\frac{2}{3\,{\rm H}}\ .
\ee
If, as is often supposed, the major contribution to the missing
mass comes from nonrelativistic matter, the age of the universe
will presently be given by Eq.  (\ref{1.4.7}):
\be
\label{1.4.8}
t\sim\frac{2}{3\,h}\cdot 10^{10}\;\mbox{yr}\  .
\ee

${\rm H}(t)$ not only determines the age, but the distance to the
horizon as well, that is, the radius of the observable part of
the universe.

To be more precise, one must distinguish between two horizons ---
the\index{Particle horizon}
particle horizon and the\index{Event horizon} event horizon [\cite{35}].

The particle horizon delimits the causally connected part of the
universe that an observer can see in principle {\it at $a$ given
time $t$}. Since light propagates on the light cone $ds^2=0$, we
find from (\ref{1.3.1}) that the rate at which the radius $r$ of a
wavefront changes is \be \label{1.4.9}
\frac{dr}{dt}=\frac{\sqrt{1-k\,r^2}}{a(t)}\ , \ee and the physical
distance traveled by light in time $t$ is \be \label{1.4.10} {\rm
R}_p(t)=a(t)\,\int^{r(t)}_0\frac{dr}{\sqrt{1-k\,r^2}}=
a(t)\,\int^t_0\frac{dt'}{a(t')}\ . \ee In particular, for
$a(t)\sim t^{3/2}$ (\ref{1.3.13}), \be \label{1.4.11} {\rm
R}_p=3\,t=2\,[{\rm H}(t)]^{-1}\ . \ee

The quantity ${\rm R}_p$ gives the size of the observable part of
the universe at time $t$.  From (\ref{1.4.1}) and (\ref{1.4.11}),
we obtain the present-day value of ${\rm R}_p$ (i.e., the distance
to the particle horizon) for the cold dark matter dominated
universe \index{Particle horizon} \be \label{1.4.12} {\rm R}_p\sim
2 \,h^{-1}\cdot 10^{28}\;\mbox{cm}\ . \ee

In a certain conceptual sense, the event horizon is the complement
of the particle horizon:  it delimits that part of the universe
from which we can ever (up to some time $t_{max}$) receive
information about events taking place {\it now} (at time $t$): \be
\label{1.4.13} {\rm
R}_e(t)=a(t)\,\int^{t_{max}}_t\frac{dt'}{a(t')}\ . \ee For a flat
universe with $a(t)\sim t^{2/3}$, there is no event horizon: ${\rm
R}_e(t)  \rightarrow \infty$ as $t_{max}\rightarrow \infty$. In
what follows, we will be particularly interested in the case
$a(t)\sim e^{{\rm H}t}$, where ${\rm H}=\mbox{const}$. This
corresponds to the Sitter metric, and gives\index{de Sitter
metric} \be \label{1.4.14} {\rm R}_e(t)={\rm H}^{-1}\ . \ee The
thrust of this result is that an observer in an exponentially
expanding universe sees only those events that take place at a
distance no farther away than ${\rm H}^{-1}$.  This is completely
analogous to the situation for a black hole\index{Black holes},
from whose surface no information can escape.  The difference is
that an observer in\index{de Sitter space}\index{Space!de Sitter}
de Sitter space (in an exponentially expanding universe) will find
himself effectively {\it surrounded} by a ``black hole'' located
at a distance ${\rm H}^{-1}$.

In closing, let us note one more rather perplexing circumstance.
Consider two points separated by a distance R at time $t$ in a
flat Friedmann universe.  If the spatial coordinates of these
points remain unchanged (and in that sense, they remain
stationary), the distance between them will nevertheless
increase, due to the general expansion of the universe, at a rate
\be
\label{1.4.15}
\frac{d{\rm R}}{dt}=\frac{\dot a}{a}\,{\rm R}={\rm HR}\ .
\ee
What this means, then, is that two points more than a distance
${\rm H}^{-1}$ apart will move away from one another faster than
the speed of light $c = 1$.  But there is no paradox here, since
what we are concerned with now is the rate at which two objects
subject to the general cosmological expansion separate from each
other, and not with a signal propagation velocity at all, which
is related to the local variation of particle spatial
coordinates.  On the other hand, it is just this effect that
provides the foundation for the existence of an event horizon in
de Sitter space.\index{Friedmann models|)}

\section{\label{s1.5}Problems
of\index{Hot universe theory!problems of|(}
the standard scenario}

Following the discovery of the microwave background radiation,
the hot universe theory immediately gained widespread acceptance.
Workers in the field have indeed pointed out certain difficulties
which, over the course of many years, have nevertheless come to
be looked upon as only temporary.  In order to make the changes
now taking place in cosmology more comprehensible, we list here
some of the problems of the standard hot universe theory.

\subsection{The singularity problem\index{Singularity problem|(}}

Equations (\ref{1.3.9}) and (\ref{1.3.12}) imply that for all
``reasonable'' equations of state, the density of matter in the
universe goes to infinity as $t\rightarrow 0$, and the
corresponding solutions cannot be formally continued to the domain
$t < 0$.

One of the most distressing questions facing cosmologists is
whether anything existed {\it before} $t = 0$;  if not, then
where did the universe come from?  The birth and death of the
universe, like the birth and death of a human being, is one of
the most worrisome problems facing not just cosmologists, but all
of contemporary science.

At first, there seemed to be some hope that even if the problem
could not be solved, it might at least be possible to circumvent
it by considering a more general model of the universe than the
Friedmann model --- perhaps an inhomogeneous, anisotropic
universe filled with matter having some exotic equation of state.
Studies of the general structure of space-time near a singularity
[\cite{68}] and several important theorems on singularities in
the general theory of relativity [\cite{69}, \cite{70}] proven by
topological methods, however, demonstrated that it was highly
unlikely that this problem could be solved within the framework
of classical gravitation theory.\index{Singularity problem|)}

\subsection{The flatness of space
\index{Flatness of space|(}\index{Space!flatness of|(}}

This problem admits of several equivalent or almost equivalent
formulations, differing somewhat in the approach taken.\vspace{1pc}

a. THE EUCLIDICITY PROBLEM.  We all learned in grade school that
our world is described by Euclidean geometry, in which the angles
of a triangle sum to $180^\circ$ and parallel lines never meet
(or they ``meet at infinity'').  In college, we were told that it
was\index{Geometry, Riemann }\index{Riemann  geometry}
Riemann  geometry that described the world, and that
parallel lines {\it could} meet or diverge at infinity.  But
nobody ever explained why what we learned in school was also true
(or almost true) --- that is, why the world is Euclidean to such
an incredible degree of accuracy.  This is even more surprising
when one realizes that there is but one natural scale length in
general relativity, the\index{Planck length}
Planck length  $l_{\rm P}\sim\m^{-1}\sim10^{-33}$ cm.

One might expect that the world would be close to Euclidean
except perhaps at distances of the order of $l_{\rm P}$ or less
(that is, less than the characteristic radius of curvature of
space).  In fact, the opposite is true:  on small scales
$l\la l_{\rm P}$, quantum fluctuations of the metric make it impossible
in general to describe space in classical terms (this leads to
the concept of space-time foam [\cite{71}]).  At the same time,
for reasons unknown, space is almost perfectly Euclidean on large
scales, up to $l\sim10^{28}$ cm --- 60 orders of magnitude
greater than the\index{Planck length} Planck length.\vspace{1pc}

b.  THE FLATNESS PROBLEM.  The seriousness of the preceding problem
is\index{Flatness problem}
most easily appreciated in the context of the Friedmann model
(\ref{1.3.1}).  We have from Eq. (\ref{1.3.7}) that
\be
\label{1.5.1}
|\Omega-1|=\frac{|\rho(t)-\rho_c|}{\rho_c}=[\dot a(t)]^{-2}\ ,
\ee
where $\rho$ is the energy density in the universe, and $\rho_c$
is the critical density for a flat universe with the same value
of the Hubble parameter ${\rm H}(t)$.

As already mentioned in Section \ref{s1.4}, the present-day value of
$\Omega$ is known only roughly, $0.1\la\Omega\la2$, or in other
words our universe could presently show a fairly sizable
departure from flatness.  On the other hand, $(\dot a)^{-2}\sim t$
in the early stages of evolution of a hot universe (see
(\ref{1.3.14})), so the quantity
$\displaystyle |\Omega-1|=\left|\frac{\rho}{\rho_c}-1\right|$
was extremely small.  One can show that in order for $\Omega$ to
lie in the range $0.1\la\Omega\la2$ now, the early universe must have had
$\displaystyle |\Omega-1|\la10^{-59}\frac{\m^2}{{\rm T}^2}$, so
that at ${\rm T}\sim \m$,
\be
\label{1.5.2}
|\Omega-1|=\left|\frac{\rho}{\rho_c}-1\right|\la 10^{-59}\ .
\ee
This means that if the density of the universe were initially (at
the\index{Planck time}
Planck time $t_{\rm P}\sim\m^{-1}$) greater than $\rho_c$,
say by $10^{-55}\,\rho_c$, it would be closed, and the limiting
value $t_c$ would be so small that the universe would have
collapsed long ago.  If on the other hand the density at the
Planck time were $10^{-55}\,\rho_c$ less than $\rho_c$, the
present energy density in the universe would be vanishingly low,
and the life could not exist.  The question of why the energy
density $\rho$ in the early universe was so fantastically close
to the critical density (Eq.  (\ref{1.5.2})) is usually known as
the flatness problem.\vspace{1pc}

c.  THE TOTAL ENTROPY AND TOTAL MASS PROBLEM.\index{Mass!total|(}%
\index{Total mass|(}
The question here%
\index{Entropy, total!and total mass problem|(}%
\index{Mass!total entropy and total, problem|(}
is why the total entropy S and total mass M of matter in the
observable part of the universe, with ${\rm R}_p\sim10^{28}$ cm,
is so large.  The total entropy S is of order
$({\rm R}_p{\rm T}_\gamma)^3\sim10^{87}$, where
${\rm T}_\gamma \sim 2.7$ K is the temperature of the primordial
background radiation.  The total mass is given by
${\rm M}\sim {\rm R}_p^3\,\rho_c\sim10^{55}\;{\rm g}\sim10^{49}$ tons.

If the universe were open and its density at the Planck time had
been subcritical, say, by $10^{-55}\,\rho_c$, it would then be
easy to show that the total mass and entropy of the observable
part of the universe would presently be many orders of magnitude
lower.

The corresponding problem becomes particularly difficult for a
closed universe.  We see from (\ref{1.3.15}) and (\ref{1.3.16})
that the total lifetime $t_c$ of a closed universe is of order
$\m^{-1}\sim10^{-43}$  sec, and this will be a long timespan
($\sim10^{10}$ yr) only when the total mass and energy of the
entire universe are extremely large.  But why is the total
entropy of the universe so large, and why should the mass of the
universe be tens of orders of magnitude greater than the Planck
mass $\m$, the only parameter with the dimension of mass in the
general theory of relativity?  This question can be formulated in
a paradoxically simple and apparently na\"\i ve way:  Why are
there so many different things in the
universe?\vspace{1pc}\index{Entropy, total!and total mass problem|)}%
\index{Mass!total|)}\index{Total mass|)}%
\index{Mass!total entropy and total, problem|)}

d.  THE PROBLEM OF THE SIZE OF THE UNIVERSE.
\index{Size of the universe, problem of}%
\index{Universe!problem of size}Another problem
associated with the flatness of the universe is that according to
the hot universe theory, the total size $l$ of the part of the
universe currently accessible to observation is proportional to
$a(t)$;  that is, it is inversely proportional to the temperature
T (since the quantity $a\,{\rm T}$ is practically constant in an
adiabatically expanding hot universe --- see Section \ref{s1.3}).  This
means that at ${\rm T}\sim\m\sim10^{19}\;\mbox{GeV}\sim10^{32}\;\mbox{K}$,
the region from which the observable part of the universe (with a size of
$10^{28}$ cm) formed was of the order of $10^{-4}$ cm in size, or
29 orders of magnitude greater than the Planck length
$l_{\rm P}\sim\m^{-1}\sim10^{-33}$ cm.  Why, when the universe was at the
Planck density, was it 29 orders of magnitude bigger than the
Planck length?  Where do such large numbers come from?

We discuss the flatness problem here in such detail not only
because an understanding of the various aspects of this problem
turns out to be important for an understanding of the
difficulties inherent in the standard hot universe theory, but
also in order to be able to understand later which versions of
the inflationary universe scenario to be discussed in this book
can resolve this problem.\index{Flatness of space|)}%
\index{Space!flatness of|)}

\subsection{The problem of the large-scale homogeneity and
isotropy\index{Homogeneity, large-scale, problem of}%
\index{Isotropy problem}%
\index{Large-scale homogeneity problem}
of the universe}

In Section \ref{s1.3}, we assumed that the universe was initially
absolutely homogeneous and isotropic.  In actuality, or course,
it is not completely homogeneous and isotropic even now, at least
on a relatively small scale, and this means that there is no
reason to believe that it was homogeneous {\it ab initio}.  The
most natural assumption would be that the initial conditions at
points sufficiently far from one another were chaotic and
uncorrelated [\cite{72}].  As was shown by Collins and Hawking
[\cite{73}] under certain assumptions, however, the class of
initial conditions for which the universe tends asymptotically
(at large $t$) to a Friedmann universe (\ref{1.3.1}) is one of
measure zero among all possible initial conditions.  This is the
crux of the problem of the homogeneity and isotropy of the
universe.  The subtleties of this problem are discussed in more
detail in the book by Zeldovich and Novikov [\cite{34}].

\subsection{The horizon problem\index{Horizon problem|(}}

The severity of the isotropy problem is somewhat ameliorated by
the fact that effects connected with the presence of matter and
elementary particle production in an expanding universe can make
the universe locally isotropic [\cite{34}, \cite{74}].  Clearly,
though, such effects cannot lead to global isotropy, if only
because causally\index{Disjoint regions} disjoint regions
separated by a distance greater
than the particle horizon (which in the simplest cases is given
by ${\rm R}_p\sim t$, where $t$ is the age of the universe)
cannot influence each other.  In the meantime, studies of the
microwave background have shown that at $t\sim 10^5$ yr, the
universe was quite accurately homogeneous and isotropic on scales
orders of magnitude greater than $t$, with temperatures T in
different regions differing by less than ${\rm O}(10^{-4}){\rm T}$.
Inasmuch as the observable part of the universe presently
consists of about $10^6$ regions that were causally unconnected
at $t\sim 10^5$ yr, the probability of the temperature T in these
regions being fortuitously correlated to the indicated accuracy
is at most $10^{-24}$--$10^{-30}$.  It is exceedingly difficult to
come up with a convincing explanation of this fact within the
scope of the standard scenario.  The corresponding problem is
known as the horizon problem or the \index{Causality problem}
causality problem [\cite{48},
\cite{56}].

There is one more aspect of the horizon problem which will be
important for our purposes.  As we mentioned in the earlier
discussion of the flatness problem, at the Planck time
$t_{\rm P}\sim\m^{-1}\sim10^{-43}$ sec, when the size (the radius of the
particle horizon) of each causally connected region of the
universe was $l_{\rm P}\sim10^{-33}$ cm, the size of the overall
region from which the observable part of the universe formed was
of order $10^{-4}$ cm.  The latter thus consisted of
$(10^{29})^3\sim10^{87}$ causally unconnected regions.  Why then
should the expansion of the universe (or its emergence from the
space-time foam with the Planck density $\rho\sim\m^4$) have
begun simultaneously (or nearly so) in such a huge number of
causally unconnected regions?  The probability of this occurring
at random is close to $\exp(-10^{90})$.\index{Horizon problem|)}

\subsection{The galaxy formation problem\index{Galaxy formation problem}}

The universe is of course not perfectly homogeneous.  It contains
such important inhomogeneities as stars, galaxies, clusters of
galaxies, etc.  In explaining the origin of galaxies, it has been
necessary to assume the existence of initial inhomogeneities
[\cite{75}] whose spectrum is usually taken to be almost
scale-invariant [\cite{76}].  For a long time, the origin of such
density inhomogeneities remained completely obscure.

\subsection{The baryon asymmetry problem\index{Baryon asymmetry problem|(}}

The essence of this problem is to understand why the universe is
made almost entirely of matter, with almost no antimatter, and
why on the other hand baryons are many orders of magnitude
scarcer than photons, with
$\displaystyle \frac{n_{\rm B}}{n_\gamma}\sim10^{-9}$.

Over the course of time, these problems have taken on an almost
metaphysical flavor.  The first is self-referential, since it can
be restated by asking ``What was there before there was anything
at all?'' or ``What was at the time at which there was no
space-time at all?''  The others could always be avoided by
saying that by sheer good luck, the initial conditions in the
universe were such as to give it precisely the form it finally
has now, and that it is meaningless to discuss initial
conditions.  Another possible answer is based on the so-called
\index{Anthropic Principle|(}Anthropic Principle, and seems
almost purely metaphysical:  we
live in a homogeneous, isotropic universe containing an excess of
matter over antimatter simply because in an inhomogeneous,
anisotropic universe with equal amounts of matter and antimatter,
life would be impossible and these questions could not even be
asked [\cite{77}].

Despite its cleverness, this answer is not entirely satisfying,
since it explains neither the small ratio
$\displaystyle \frac{n_{\rm B}}{n_\gamma}\sim10^{-9}$, nor the high
degree of homogeneity and isotropy in the universe, nor the observed
spectrum of galaxies.  The \index{Anthropic Principle|)}Anthropic
Principle is also incapable
of explaining why all properties of the universe are
approximately uniform over its entire observable part
($l\sim10^{28}$ cm) --- it would be perfectly possible for life
to arise if favorable conditions existed, for example, in a
region the size of the solar system, $l\sim 10^{14}$ cm.
Furthermore, Anthropic Principle rests on an implicit assumption
that either universes are constantly created, one after another,
or there exist many different universes, and that life arises in
those universes which are most hospitable.  It is not clear,
however, in what sense one can speak of different universes if
ours is in fact unique.  We shall return to this question later
and provide a basis for a version of the Anthropic Principle in
the context of inflationary cosmology [\cite{57}, \cite{78},
\cite{79}].

The first breach in the cold-blooded attitude of most physicists
toward the foregoing ``metaphysical'' problems appeared after
Sakharov discovered [\cite{36}] that the baryon asymmetry problem
could be solved in theories in which baryon number is not
conserved by taking account of nonequilibrium processes with C
and CP-violation in the very early universe.  Such processes can
occur in all grand unified theories [\cite{36}--\cite{38}].  The
discovery of a way to generate the observed baryon asymmetry of
the universe was considered to be one of the greatest successes
of the hot universe cosmology.  Unfortunately, this success was
followed by a whole series of disappointments.
\index{Baryon asymmetry problem|)}

\subsection{The domain wall problem\index{Domain wall problem|(}}

As we have seen, symmetry is restored in the theory (\ref{1.1.5})
when ${\rm T}>2\,\mu/\sqrt{\lambda}$.  As the temperature drops
in an expanding universe, the symmetry is broken.  But this
symmetry breaking occurs independently in all causally
unconnected regions of the universe, and therefore in each of the
enormous number of such regions comprising the universe at the
time of the symmetry-breaking phase transition, both the field
$\varphi=+\mu/\sqrt{\lambda}$ and the field
$\varphi=-\mu/\sqrt{\lambda}$ can arise.  Domains filled by the
field $\varphi=+\mu/\sqrt{\lambda}$ are separated from those with
the field $\varphi-\mu/\sqrt{\lambda}$ by domain walls.  The
energy density of these walls turns out to be so high that the
existence of just one in the observable part of the universe
would lead to unacceptable cosmological consequences [\cite{41}].
This implies that a theory with spontaneous breaking of a
discrete symmetry is inconsistent with the cosmological data.
Initially, the principal theories fitting this description were
those with spontaneously broken CP invariance [\cite{80}].  It
was subsequently found that domain walls also occur in the
simplest version of the SU(5) theory, which has the discrete
invariance $\Phi \rightarrow-\Phi$ [\cite{42}], and in most axion
theories [\cite{43}].  Many of these theories are very appealing,
and it would be nice if we could find a way to save at least some
of them.\index{Domain wall problem|)}

\subsection{The primordial monopole problem
\index{Monopoles!primordial!problem}\index{Primordial monopole problem}}

Other structures besides domain walls can be produced following
symmetry-breaking phase transitions.  For example, in the Higgs
model with broken U(1) symmetry and certain others, strings of
the Abrikosov superconducting vortex tube type can occur
[\cite{81}].  But the most important effect is the creation of
superheavy t'Hooft--Polyakov magnetic monopoles [\cite{82},
\cite{83}], which should be copiously produced in practically all
of the grand unified theories [\cite{84}] when phase transitions
take place at ${\rm T}_{c_1}\sim10^{14}$--$10^{15}$ GeV.  It was
shown by Zeldovich and Khlopov [\cite{40}] that
\index{Monopole annihilation}monopole
annihilation proceeds very slowly, and that the monopole density
at present should be comparable to the baryon density.  This
would of course have catastrophic consequences, as the mass of
each monopole is perhaps $10^{16}$ times that of the proton,
giving an energy density in the universe about 15 orders of
magnitude higher than the critical density $\rho_c\sim10^{29}$
g/cm$^3$.  At that density, the universe would have collapsed
long ago.  The primordial monopole problem is one of the sharpest
encountered thus far by elementary particle theory and cosmology,
since it relates to practically all unified theories of weak,
strong, and electromagnetic interactions.

\subsection{The primordial gravitino
problem\index{Primordial gravitino problem}%
\index{Gravitino!primordial, problem|(}}

One of the most interesting directions taken by modern elementary
particle physics is the study of\index{Supersymmetry} supersymmetry,
the symmetry
between fermions and bosons [\cite{85}].  Here we will not list
all the advantages of supersymmetric theories, referring the
reader instead to the literature [\cite{13}, \cite{14}].  We
merely point out that phenomenological supersymmetric theories,
and\index{Supergravity theory!${\rm N} = 1$} ${\rm N} = 1$
supergravity in particular, may provide a way
to solve the mass hierarchy problem of unified field theories
[\cite{15}];  that is, they may explain why there exist such
drastically differing mass scales $\m\gg {\rm M}_{\rm X}\sim10^{15}$ GeV
and ${\rm M}_{\rm X}\gg m_{\rm W}\sim 10^2$ GeV.

One of the most interesting attempts to resolve the mass
hierarchy problem for ${\rm N} = 1$ supergravity is based on the
suggestion that the\index{Gravitino} gravitino (the spin-$3/2$
superpartner of the
graviton) has mass $m_{3/2}\sim m_{\rm W}\sim 10^2$ GeV
[\cite{15}].  It has been shown [\cite{86}], however, that
gravitinos with this mass should be copiously produced as a
result of high-energy particle collisions in the early universe,
and that gravitinos decay rather slowly.

Most of these gravitinos would only have decayed by the later
stages of evolution of the universe, after helium and other light
elements had been synthesized, which would have led to many
consequences that are inconsistent with the observations
[\cite{44}, \cite{45}].  The question is then whether we can
somehow rescue the universe from the consequences of gravitino
decay; if not, must we abandon the attempt to solve the hierarchy
problem?

Some particular models [\cite{87}] with superlight or superheavy
gravitinos manage to avoid these difficulties.  Nevertheless, it
would be quite valuable if we could somehow avoid the stringent
constraints imposed on the parameters of ${\rm N} = 1$
supergravity by the hot universe
\index{Gravitino!primordial, problem|)}theory.

\subsection{The problem of Polonyi fields\index{Polonyi fields|(}}

The gravitino problem is not the only one that arises in
phenomenological theories based on ${\rm N} = 1$ supergravity
(and superstring theory).  The so-called scalar Polonyi fields
$\chi$ are one of the major ingredients of these theories
[\cite{46}, \cite{15}].  They are relatively low-mass fields that
interact weakly with other fields.  At the earliest stages of the
evolution of the universe they would have been far from the
minimum of their corresponding effective potential ${\rm V}(\chi)$.
Later on, they would start to oscillate about the
minimum of ${\rm V}(\chi)$, and as the universe expanded, the
Polonyi field energy density $\rho_\chi$ would decrease in the
same manner as the energy density of nonrelativistic matter
($\rho_\chi\sim a^{-3}$), or in other words much more slowly than
the energy density of hot plasma.  Estimates indicate that for
the most likely situations, the energy density presently stored
in these fields should exceed the critical density by about 15
orders of magnitude [\cite{47}, \cite{48}].  Somewhat more
refined models give theoretical predictions of the density
$\rho_\chi$ that no longer conflict with the observational data
by a factor of $10^{15}$, but only by a factor of $10^6$
[\cite{48}], which of course is also highly
undesirable.\index{Polonyi fields|)}

\subsection{The vacuum energy problem\index{Energies!vacuum}
\index{Vacuum energy problem|(}}

As we have already mentioned, the advent of a constant
homogeneous scalar field $\varphi$ over all space simply
represents a restructuring of the vacuum,\index{Vacuum} and in some sense,
space filled with a constant scalar field $\varphi$ remains
``empty'' --- the constant scalar field does not carry a
preferred reference frame with it, it does not disturb the motion
of objects passing through the space that it fills, and so forth.
But when the scalar field appears, there is a change in the
vacuum energy density, which is described by the quantity
${\rm V}(\varphi)$. If there were no gravitational effects, this
change in the energy density of the vacuum would go completely
unnoticed. In general relativity, however, it affects the
properties of space-time.  ${\rm V}(\varphi)$ enters into the
Einstein equation in the following way:
\be
\label{1.5.3}
{\rm R}_{\mu\nu}-\frac{1}{2}\,g_{\mu\nu}\,{\rm R}=
8\,\pi\,{\rm G}\,{\rm T}_{\mu\nu}=
8\,\pi\,{\rm G}\,(\tilde{\rm T}_{\mu\nu}+g_{\mu\nu}\,{\rm V}(\varphi))\ ,
\ee
where ${\rm T}_{\mu\nu}$ is the
total energy-momentum tensor,
$\tilde{\rm T}_{\mu\nu}$ is
the\index{Energy-momentum tensor of matter}%
\index{Matter!energy-momentum tensor of}
energy-momentum tensor of
substantive matter (elementary particles), and
$g_{\mu\nu}\,{\rm V}(\varphi)$ is the energy-momentum
tensor of the vacuum (the
constant scalar field $\varphi$).  By comparing the usual
energy-momentum tensor of matter
\be
\label{1.5.4}
\tilde {\rm T}_\mu{}^\nu=\left(
\begin{array}{cccc}
\rho\\
&-p\\
&&-p\\
&&&-p
\end{array}\right)
\ee
with $g_\mu{}^\nu\,{\rm V}(\varphi)$, one can see that the
``pressure'' exerted by the vacuum and its energy density have
opposite signs, $p = -\rho = -{\rm V}(\varphi)$.

The cosmological data imply that the present-day vacuum energy
density $\rho_{\rm vac}$ is not much greater in absolute value
than the critical density $\rho_c\sim10^{-29}$ g/cm$^3$: \be
\label{1.5.5} |\rho_{\rm vac}|=|{\rm V}(\varphi_0)|\la
10^{-29}\;\mbox{g/cm$^3$}\ . \ee

This value of ${\rm V}(\varphi)$ was attained as a result of a
series of symmetry-breaking phase transitions.  In the SU(5)
theory, after the first phase transition
$\mbox{SU}(5)\rightarrow\mbox{SU}(3)\times\mbox{SU}(2)\times{\rm U}(1)$,
the vacuum energy (the value of ${\rm V}(\varphi)$) decreased by
approximately $10^{80}$ g/cm$^3$.  After the
$\mbox{SU}(3)\times\mbox{SU}(2)\times{\rm U}(1)\rightarrow
\mbox{SU}(3) \times {\rm U}(1)$
transition, it was reduced by about another $10^{25}$
g/cm$^3$.  Finally, after the phase transition that formed the
baryons from quarks, the vacuum energy again decreased, this time
by approximately $10^{14}$ g/cm$^3$, and surprisingly enough
after all of these enormous drops, it turned out to equal zero to
an accuracy of $\pm10^{-29}$ g/cm$^3$!  It seems unlikely that
the complete (or almost complete) cancellation of the vacuum
energy should occur merely by chance, without some deep physical
reason.  The vacuum energy problem in theories with spontaneous
symmetry breaking [\cite{88}] is presently deemed to be one of
the most important problems facing elementary particle theories.

The vacuum energy density multiplied by $8\,\pi\,{\rm G}$ is
usually called the \index{Cosmological constant}cosmological
constant $\Lambda$ [\cite{89}];
in the present case, $\Lambda =8\,\pi\,{\rm G}\,{\rm V}(\varphi)$
[\cite{88}]. The vacuum energy problem is therefore also often
called the cosmological constant problem.

Note that by no means do all theories ensure, even in principle,
that the vacuum energy at the present epoch will be small.  This
is one of the most difficult problems encountered in
Kaluza--Klein theories based on ${\rm N}=1$ supergravity in
11-dimensional space [\cite{16}].  According to these theories,
the vacuum energy would now be of order $-\m^4\sim-10^{94}$
g/cm$^{-3}$.  On the other hand, indications that the vacuum
energy problem may be solvable in superstring theories
[\cite{17}] have stimulated a great deal of interest in the
latter.\index{Vacuum energy problem|)}

\subsection{The problem of the uniqueness of the universe}
\index{Universe!problem of uniqueness|(}%
\index{Uniqueness of universe, problem of|(}%

The essence of this problem was most clearly enunciated by
Einstein, who said that ``we wish to know not just the structure
of Nature (and how natural phenomena are played out), but insofar
as we can, we wish to attain a daring and perhaps utopian goal
--- to learn why Nature is just the way it is, and not
otherwise'' [\cite{90}].  As recently as a few years ago, it
would have seemed rather meaningless to ask why our space-time is
four-dimensional, why there are weak, strong, and electromagnetic
interactions and no others, why the fine-structure constant
$\displaystyle \alpha=\frac{e^2}{4\,\pi}$ equals $1/137$, and so
on. Of late, however, our attitude toward such questions has
changed, since unified theories of elementary particles
frequently provide us with many different solutions of the
relevant equations that in principle could describe our universe.

In theories with spontaneous symmetry breaking, for example, the
effective potential will often have several local minima --- in
the theory (\ref{1.1.5}), for instance, there are two, at
$\varphi=\pm\mu/\sqrt{\lambda}$.  In the minimal supersymmetric
SU(5) grand unification theory, there are three local minima of
the effective potential for the field $\Phi$ that have nearly the
same depth [\cite{91}].  The degree of degeneracy of the
effective potential in supersymmetric theories (the number of
different types of vacuum states having the same energy) becomes
even greater when one takes into account other Higgs fields H
which enter into the theory [\cite{92}].

The question then arises as to how and why we come to be in a
minimum in which the broken symmetry is $\mbox{SU}(3) \times {\rm U}(1)$
(this question becomes particularly complicated if we
recall that the early high-temperature universe was at an
SU(5)-symmetric minimum $\Phi = {\rm H} = 0$ [\cite{93}], and
there is no apparent reason for the entire universe to jump to
the $\mbox{SU}(3) \times {\rm U}(1)$ minimum upon cooling).

It is assumed in the Kaluza--Klein and superstring theories that
we live in a space with $d > 4$ dimensions, but that $d - 4$ of
these dimensions have
been\index{Compactified dimensions}\index{Dimensions!compactified}
compactified --- the radius of
curvature of space in the corresponding directions is of order
$\m^{-1}$.  That is why we cannot move in those directions, and
space is apparently four-dimensional.

Presently, the most popular theories of that kind have $d = 10$
[\cite{17}], but others with $d = 26$ [\cite{94}] and $d = 506$
[\cite{95}, \cite{96}] have also been considered.  One of the
most fundamental questions that comes up in this regard is why
precisely $d - 4$ dimensions
were\index{Compactified dimensions}\index{Dimensions!compactified}
compactified, and not $d - 5$
or $d - 3$.  Furthermore, there are usually a great many ways to
compactify $d-4$ dimensions, and each results in its own peculiar
laws of elementary particle physics in four-dimensional space.  A
frequently asked question is then why Nature chose just that
particular vacuum state which leads to the strong, weak, and
electromagnetic interactions with the coupling constants that we
measure experimentally.  As the dimension $d$ of the parent space
rises, this problem becomes more and more acute.  Thus, it has
variously been estimated that in $d =10$ superstring theory,
there are perhaps $10^{1500}$  ways of\index{Compactified dimensions}
compactifying the
ten-dimensional space into four dimensions (some of which may
lead to unstable compactification), and there are many more ways
to do this in space with $d >10$.  The question of why the world
that surrounds us is structured just so, and not otherwise, has
therefore lately turned into one of the most fundamental problems
of modern physics.

We could continue this list of problems facing cosmologists and
elementary particle theorists, of course, but here we are only
interested in those that bear some relation to our basic theme.

The vacuum energy problem has yet to be solved definitively.
There are many interesting attempts to do so, some of which are
based on quantum cosmology and on the inflationary universe
scenario.  A solution to the baryon asymmetry problem was
proposed by Sakharov long before the advent of the inflationary
universe scenario [\cite{36}], but the latter also introduces
much that is new [\cite{97}--\cite{99}].  As for the other ten
problems, they can all be solved either partially or completely
within the framework of inflationary cosmology, and we now turn
to a description of that theory.\index{Hot universe theory!problems of|)}%
\index{Uniqueness of universe, problem of|)}%
\index{Universe!problem of uniqueness|)}

\section[development of the inflationary universe scenario]%
{\label{s1.6}A sketch of the development
of\index{Inflationary universe scenario!development of|(}
the inflationary
universe scenario}

The main idea underlying all existing versions of the
inflationary universe scenario is that in the very earliest
stages of its evolution, the universe could be in an unstable
vacuum-like state having high energy density.  As we have already
noted in the preceding section, the vacuum pressure\index{Vacuum pressure}
and energy
density\index{Density!vacuum energy}\index{Energy density, vacuum}%
\index{Vacuum energy density}
are related by Eq. (\ref{1.5.4}), $p=-\rho$. This means,
according to (\ref{1.3.8}), that the vacuum energy density does
not change as the universe expands (a ``void'' remains a
``void'', even if it has weight).  But (\ref{1.3.7}) then implies
that at large times $t$, the universe in an unstable vacuum state
$\rho > 0$ should expand exponentially, with
\be
\label{1.6.1}
a(t)={\rm H}^{-1}\,\cosh{\rm H}\,t
\ee
for $k = +1$ (a closed Friedmann universe),
\be
\label{1.6.2}
a(t)={\rm H}^{-1}\,e^{{\rm H}t}
\ee
for $k = 0$ (a flat universe), and
\be
\label{1.6.3}
a(t)={\rm H}^{-1}\,\sinh{\rm H}\,t
\ee
for $k = -1$ (an open universe).
Here $\displaystyle {\rm H}=\sqrt{\frac{8\,\pi}{3}\,{\rm G}\,\rho}=
\sqrt{\frac{8\,\pi\,\rho}{3\,\m^2}}$.  More generally,
during expansion the magnitude of H in the inflationary universe
scenario changes, but very slowly,
\be
\label{1.6.4}
\dot {\rm H}\ll{\rm H}^2\ .
\ee
Over a characteristic time $\Delta t={\rm H}^{-1}$ there is
little change in the magnitude of H, so that one may speak of
a\index{Quasiexponential expansion of universe}%
\index{Universe!quasiexponential expansion of}
quasiexponential expansion of the universe,
\be
\label{1.6.5}
a(t)=a_0\,\exp\left[\int^t_0{\rm H}(t)\:dt\right]\sim a_0\,e^{{\rm H}t}\ ,
\ee
or of a quasi-de Sitter stage in its expansion;  just this regime of
quasiexponential expansion is known as\index{Inflation} inflation.

Inflation comes to an end when H begins to decrease rapidly.  The
energy stored in the vacuum-like state is then transformed into
thermal energy, and the universe becomes extremely hot.  From
that point onward, its evolution is described by the standard hot
universe theory, with the important refinement that the initial
conditions for the expansion stage of the hot universe are
determined by processes which occurred at the inflationary stage,
and are practically unaffected by the structure of the universe
prior to inflation.  As we shall demonstrate below,   just this
refinement enables us to solve many of the problems of the hot
universe theory discussed in the preceding section.

The space (\ref{1.6.1})--(\ref{1.6.3}) was first described in the
1917 papers of\index{de Sitter space} de Sitter [\cite{100}], well
before the appearance
of Friedmann's theory of the expanding universe.  However, de
Sitter's solution was obtained in a form differing from
(\ref{1.6.1})--(\ref{1.6.3}), and for a long time its physical
meaning was somewhat obscure.  Before the advent of the
inflationary universe scenario, de Sitter space was employed
principally as a convenient staging area for developing the
methods of general relativity and quantum field theory in curved
space.

The possibility that the universe might expand exponentially
during the early stages of its evolution, and be filled with
superdense matter with the equation of state $p=-\rho$, was first
suggested by Gliner [\cite{51}]; see also
[\cite{101}--\cite{103}].  When they appeared, however, these
papers did not arouse much interest, as they dealt mainly with
superdense baryonic matter, which, as we now believe, has an
equation of state close to $\displaystyle p=\frac{\rho}{3}$,
according to asymptotically free theories of weak, strong, and
electromagnetic interactions.

It was subsequently realized that the constant (or almost
constant) scalar field $\varphi$ appearing in unified theories of
elementary particles could play the role of a vacuum state with
energy density ${\rm V}(\varphi)$ [\cite{88}].  The magnitude of
the field $\varphi$ in an expanding universe depends on the
temperature, and at times of phase transitions that change
$\varphi$, the energy stored in the field is transformed into
thermal energy [\cite{21}--\cite{24}].  If, as sometimes happens,
the phase transition takes place from a highly supercooled
metastable vacuum state, the total entropy of the universe can
increase considerably afterwards [\cite{23}, \cite{24},
\cite{104}], and in particular, a cold Friedmann universe can
become hot.  The corresponding model of the universe was
developed by Chibisov and the present author (in this regard, see
[\cite{24}, \cite{105}]).

In 1979--80, a very interesting model of the evolution of the
universe was proposed by Starobinsky\index{Starobinsky model}
[\cite{52}].  His model was
based on the observation of Dowker and Critchley [\cite{106}]
that the de Sitter metric is a solution of the Einstein equations
with quantum corrections.  Starobinsky noted that this solution
is unstable, and after the initial vacuum-like state decays (its
energy density is related to the curvature of space R), de Sitter
space transforms into a hot Friedmann universe [\cite{52}].

Starobinsky model proved to be an important step on the road
towards the inflationary universe scenario.  However, the
principal advantages of the inflationary stage had not yet been
recognized at that time.  The main objective pursued in
[\cite{52}] was to solve the problem of the initial
\index{Cosmological singularity!initial}cosmological
singularity.  The goal was not reached at that time, and the
question of initial conditions for the model remained unclear.
In that model, furthermore, the density inhomogeneities that
appeared after decay of de Sitter space turned out to be too
large [\cite{107}].  All of these considerations required that
the foundations of the model be significantly altered
[\cite{108}--\cite{110}].  In its modified form, the Starobinsky
model has become one of the most actively developed versions of
the inflationary universe scenario (or, to be more precise, the
chaotic inflation scenario;  see below).

The necessity of considering models of the universe with a stage
of quasiexponential expansion was fully recognized only after the
work of Guth [\cite{53}], who suggested using
the exponential
expansion\index{Inflation} (inflation) of the universe
in a supercooled vacuum
state $\varphi = 0$ to solve three of the problems discussed in
Section \ref{s1.5}, namely the flatness problem, the horizon
problem, and the primordial monopole problem (a similar
possibility for solving the flatness problem was independently
suggested by Lapchinsky, Rubakov, and Veryaskin [\cite{111}]).
The scenario suggested by\index{Guth scenario} Guth was based
on three fundamental propositions:

1. The universe initially expands in a state with superhigh
temperature and restored symmetry, $\varphi_{({\rm T})}=0$.

2. One considers theories in which the potential ${\rm
V}(\varphi)$ retains a local minimum at $\varphi = 0$ even at a
low temperature T.  As a result, the evolving universe remains in
the supercooled metastable state $\varphi = 0$ for a long time.
Its temperature in this state falls off, the energy-momentum
tensor gradually becomes equal to ${\rm
T}_{\mu\nu}=g_{\mu\nu}{\rm V}(0)$, and the universe expands
exponentially (inflates) for a long time.

3. Inflation continues until the end of a phase transition to a
stable state $\varphi_0\neq0$. This phase transition proceeds by
\index{Bubble production}forming bubbles containing the field
$\varphi=\varphi_0$. The
universe heats up due to bubble-wall collisions, and its
subsequent evolution is described by the hot universe theory.

The exponential expansion of the universe in stage (2) is
introduced to make the term $\displaystyle \frac{k}{a^2}$ in the
Einstein equation (\ref{1.3.7}) vanishingly small as compared
with $\displaystyle \frac{8\,\pi\,{\rm G}}{3}\,\rho$, i.e., in
order to make the universe flatter and flatter.  This same
process is invoked to ensure that the observable part of the
universe, some $10^{28}$ cm in size, came about as the result of
inflation of a very small region of space that was initially
causally connected.  In this scenario, monopoles are created at
places where the walls of several exponentially large bubbles
collide, and they therefore have exponentially low density.

The main idea behind the Guth scenario is very simple and
extremely attractive.  As noted by Guth himself [\cite{53}],
however, collisions of the walls of very large bubbles should
lead to an unacceptable destruction of homogeneity and isotropy
in the universe after inflation.  Attempts to improve this
situation were unsuccessful [\cite{112}, \cite{113}] until
cosmologists managed to surmount a certain psychological barrier
and renounce all three of the aforementioned assumptions of the
Guth scenario, while retaining the idea that the universe might
have undergone inflation during the early stages of its
evolution.

The invention of\index{Inflationary universe scenario!new}
the so-called new inflationary universe scenario
[\cite{54}, \cite{55}] marked the departure from assumptions (2)
and (3). This scenario is based on the fact that inflation can
occur not only in a supercooled state $\varphi = 0$, but also
during the process of growth of the field $\varphi$ if this field
increases to its equilibrium value $\varphi_0$ slowly enough, so
that the time $t$ for $\varphi$ to reach the minimum of
${\rm V}(\varphi)$ is much longer than ${\rm H}^{-1}$.  This condition
can be realized if the effective potential of the field $\varphi$
has a sufficiently flat part near $\varphi = 0$.  If inflation
during the stage when $\varphi$ is rolling downhill is large
enough, the walls of bubbles of the field $\varphi$ (if they are
formed) will, after inflation, be separated from one another by
much more than $10^{28}$ cm, and will not engender any
inhomogeneities in the observable part of the universe.  In this
scenario, the universe is heated after inflation not because of
collisions between bubble walls, but because of the creation of
elementary particles by the classical field $\varphi$, which
executes damped oscillations about the minimum of ${\rm V}(\varphi)$.

The new inflationary universe scenario is free of the major
shortcomings of the old scenario.  In the context of this
scenario, it is possible to propose solutions not just to the
flatness, horizon, and primordial monopole problems, but also to
the homogeneity and isotropy problems, as well as many of the
others referred to in Section \ref{s1.5}. It has been found, in
particular, that at the time of inflation in this scenario,
density inhomogeneities are produced with a spectrum that is
virtually independent of the logarithm of the wavelength (a
so-called flat, scale-free, or\index{Harrison--Zeldovich spectrum}
Harrison--Zeldovich spectrum
[\cite{214}, \cite{76}]).  This marked an important step on the
road to solving the problem of the origin of the large-scale
structure of the universe.

The successes of the new inflationary universe scenario were so
impressive that even now, many scientists who speak of the
inflationary universe scenario mean this new scenario [\cite{54},
\cite{55}].  In our opinion, however, this scenario is still far
from perfect;  there are at least three problems that stand in
the way of its successful implementation:

1.  The new scenario requires a realistic theory of elementary
particles in which the effective potential satisfies many
constraints that are rather unnatural.  For example, the potential
${\rm V}(\varphi)$ must be very close to flat (${\rm
V}(\varphi)\approx\mbox{const}$) for values of the field close to
$\varphi=0$.  If for instance, the behavior of ${\rm V}(\varphi)$
at small $\varphi$ is close to $\displaystyle {\rm
V}(0)-\frac{\lambda}{4}\,\varphi^4$, then in order for density
inhomogeneities generated at the time of inflation to have the
required amplitude \be \label{1.6.6}
\frac{\delta\rho}{\rho}\sim\mbox{$10^{-4}$--$10^{-5}$}\ , \ee the
constant $\lambda$ must be extremely small [\cite{114}], \be
\label{1.6.7} \lambda\sim\mbox{$10^{-12}$--$10^{-14}$}\ . \ee On
the other hand, the curvature of the effective potential ${\rm
V}(\varphi)$ near its minimum at $\varphi=\varphi_0$ must be great
enough to make the field $\varphi$ oscillate at high frequency
after inflation, thereby heating the universe to a rather high
temperature T.  It has turned out to be rather difficult to
suggest a natural yet realistic theory of elementary particles
that satisfies all the necessary requirements.

2.  The second problem is related to the fact that the weakly
interacting field $\varphi$ (see (\ref{1.6.7})) is most likely
not to be in a state of thermodynamic equilibrium with the other
fields present in the early universe.  But even if it were in
equilibrium, if $\lambda$ is small, high-temperature corrections
to ${\rm V}(\varphi)$ of order $\lambda\,{\rm T}^2\,\varphi^2$
cannot alter the initial value of the field $\varphi$ and make it
zero in the time between the birth of the universe and the
assumed start of inflation [\cite{115}, \cite{116}].

3.  Yet another problem relates to the fact that in both the old
and new scenarios, inflation will only begin when the temperature
of the universe has dropped sufficiently far, ${\rm T}^4\la{\rm V}(0)$.
However, the condition (\ref{1.6.6}) implies not only
the constraint (\ref{1.6.7}) on $\lambda$, but also (in most
models) a constraint on the value of ${\rm V}(\varphi)$ in the
last stages of inflation, which in the new inflationary universe
scenario is practically equal to ${\rm V}(0)$ [\cite{116},
\cite{117}]:
\be
\label{1.6.8}
{\rm V}(0)\la10^{-13}\,\m^4\ .
\ee
This means that inflation will start when ${\rm
T}^2\la10^{-7}\,\m^2$, i.e., at a time $t$ following the
beginning of expansion of the universe that exceeds the Planck
time $t_{\rm P}\sim\m^{-1}$ (\ref{1.3.20}) by 6 orders of
magnitude.  But for a hot, closed universe to live that long, its
total entropy must at the very outset be greater than ${\rm
S}\sim10^9$ (\ref{1.3.16}).  Thus, the flatness problem for a
closed universe has not been solved [\cite{116}], either in the
context of the Guth scenario or the new inflationary universe
scenario.  One could look upon this result as an argument in
favor of the universe being either open or flat.  We think,
however, that this is not a problem of the theory of a closed
universe;  rather, it is just one more shortcoming of the new
inflationary universe scenario.

Fortunately, there is another version of the inflationary
universe scenario, the so-called chaotic inflation scenario
[\cite{56}, \cite{57}], which does not share these problems.
Rather than being based on the theory of high-temperature phase
transitions, it is simply concerned with the evolution of a
universe filled with a chaotically (or almost chaotically --- see
below) distributed scalar field $\varphi$.  In what follows, we
will discuss this scenario and the considerable changes that have
taken place in recent years in our ideas about the early stages
of the evolution of the universe, and about its large-scale
structure.\index{Inflationary universe scenario!development of|)}

\section{\label{s1.7}The chaotic inflation scenario
\index{Chaotic inflation scenario|(}}

We will now illustrate the basic idea of the chaotic inflation
scenario with an example drawn from the simplest theory of the
scalar field $\varphi$ minimally coupled to gravity, with the
Lagrangian
\be
\label{1.7.1}
{\rm L}=\frac{1}{2}\,\partial_\mu\varphi\,\partial^\mu\varphi-
{\rm V}(\varphi)\ .
\ee
We shall also assume that when $\varphi\ga\m$, the potential
${\rm V}(\varphi)$ rises more slowly than (approximately)
$\displaystyle \exp\left(\frac{6\,\varphi}{\m}\right)$.  In
particular, this requirement is satisfied by any potential that
follows a power law for $\varphi\ga\m$:
\be
\label{1.7.2}
{\rm V}(\varphi)=\frac{\lambda\,\varphi^n}{n\,\m^{n-4}}\ ,
\ee
$n > 0$, $0 < \lambda \ll 1$.

In order to study the evolution of a universe filled with a
scalar field $\varphi$, we must somehow set the initial values of
the field and its derivatives at different points in space, and
also specify the topology of the space and its metric in a manner
consistent with the initial conditions for $\varphi$.  We might
assume, for example, that from the very beginning, the field
$\varphi$ over all space is in the equilibrium state
$\varphi=\varphi_0$ corresponding to a minimum of ${\rm V}(\varphi)$.
But this would be even more unconvincing than
assuming that the whole universe is perfectly uniform and
isotropic from the very beginning.  Actually, regardless of
whether the universe was originally hot or its dynamical behavior
was determined solely by the classical field $\varphi$, at a time
$t\sim t_{\rm P}\sim\m^{-1}$ after the singularity (or after the
quantum birth of the universe --- see below) the energy density
$\rho$ (and consequently the value of ${\rm V}(\varphi)$) was
determined only to accuracy ${\rm O}(\m^4)$ by virtue of the
Heisenberg uncertainty principle.  Assuming that the field
$\varphi$ initially taken value $\varphi=\varphi_0$  is therefore
no more plausible than assuming it taken any other value with
\ba
\label{1.7.3}
\partial_0\varphi\,\partial^0\varphi&\la&\m^4\ ,\\
\label{1.7.4}
\partial_i\varphi\,\partial^i\varphi&\la&\m^4\ ,\quad i=1,2,3\ ,\\
\label{1.7.5}
{\rm V}(\varphi)&\la&\m^4\ ,\\
\label{1.7.6}
{\rm R}^2&\la&\m^4\ .
\ea
The last of these inequalities is taken to mean that invariants
constructed from the\index{Curvature tensor}
curvature tensor ${\rm R}_{\mu\nu\alpha\beta}$
are less than corresponding powers of the Planck mass
(${\rm R}_{\mu\nu\alpha\beta}\,{\rm R}^{\mu\nu\alpha\beta}\la\m^4$,
${\rm R}_\mu{}^\nu\,{\rm R}_\nu{}^\alpha\,{\rm R}_\alpha{}^\mu\la\m^6$,
etc.).  It is usually assumed that the first instant at which the
foregoing conditions hold is the instant after which the region of the
universe under consideration can be described as a classical
space-time (in nonstandard versions of gravitation theory, the
corresponding conditions may generally differ from
(\ref{1.7.3})--(\ref{1.7.6})).  It is precisely this instant
after which one can speak of specifying the initial distribution
of a classical scalar field $\varphi$ in a region of classical
space-time.

Since there is absolutely no {\it a priori} reason to expect that
$\partial_\mu\varphi\,\partial^\mu\varphi\ll\m^4$, ${\rm R}^2\ll\m^4$,
or ${\rm V}(\varphi)\ll\m^4$, it seems reasonable to suppose that
the most natural initial conditions at the moment when the
classical description of the universe first becomes feasible are
\ba
\label{1.7.7}
\partial_0\varphi\,\partial^0\varphi&\sim&\m^4\ ,\\
\label{1.7.8}
\partial_i\varphi\,\partial^i\varphi&\sim&\m^4\ ,\quad i=1,2,3\ ,\\
\label{1.7.9}
{\rm V}(\varphi)&\sim&\m^4\ ,\\
\label{1.7.10}
{\rm R}^2&\sim&\m^4\ .
\ea
We shall return to a discussion of initial conditions in the
early universe more than once in the main body of this book, but
for the moment, we will attempt to understand the consequences of
the assumption made above [\cite{56}, \cite{118}].

Investigation of the expansion of the universe with initial
conditions (\ref{1.7.7})--(\ref{1.7.10}) is still an extremely
complicated problem, but there is a simplifying circumstance that
carries one a long way toward a solution.  Specifically, we are
most interested in studying the possibility that regions of the
universe will form that look like part of an exponentially
expanding Friedmann universe.  As we have already noted in
Section \ref{s1.4}, the latter is a\index{de Sitter space}
de Sitter space, with only a small
part of that space, of radius ${\rm H}^{-1}$, being accessible to
a stationary observer.  This observer sees himself as surrounded
by a black hole situated at a distance ${\rm H}^{-1}$,
corresponding to the event horizon of the de Sitter space.  It is
well known that nothing entering a black hole can\index{Black holes}
reemerge, nor can anything that has been captured affect physical processes
outside the black hole.  This assertion (with certain
qualifications that need not concern us here) is known as the
theorem that ``a black hole has no hair'' [\cite{119}].  There is
an analogous theorem for de Sitter space as well:  all particles
and other inhomogeneities within a sphere of radius ${\rm H}^{-1}$
will have left that sphere (crossed the event horizon)
by a time of order ${\rm H}^{-1}$, and will have no effect on
events taking place within the horizon (de Sitter space ``has no
hair'' [\cite{120}, \cite{121}]).  As a result, the local
geometrical properties of an expanding universe with
energy-momentum tensor
${\rm T}_{\mu\nu}\approx g_{\mu\nu}\,{\rm V}(\varphi)$
approach those of de Sitter space at an
exponentially high rate;  that is, the universe becomes
homogeneous and isotropic, and the total size of the homogeneous
and isotropic region rises exponentially [\cite{120}--\cite{122}].

In order for such behavior to be feasible, the size of the domain
within which the expansion takes place must exceed $2\,{\rm H}^{-1}$.
When ${\rm V}(\varphi)\sim\m^4$, the horizon is as
close as it can be, with ${\rm H}^{-1}\sim\m^{-1}$; that is, we
are dealing with the smallest domains that can still be described
in terms of classical space-time.  Moreover, it is necessary that
expansion be approximately exponential in order for the event
horizon ${\rm H}^{-1}(t)$ to recede slowly enough, and for
inhomogeneities at the time of expansion to escape beyond the
horizon, without engendering any back influence on the expansion
taking place within the horizon.  This condition will be
satisfied if $\dot {\rm H}\ll{\rm H}^2$, and this is just the
situation during the stage of inflation.

Thus, to assess the possibility of inflationary regions arising
in a universe with initial conditions
(\ref{1.7.7})--(\ref{1.7.10}), it is sufficient to consider
whether inflationary behavior could arise at the Planck epoch in
an isolated domain of the universe, with the minimum size $l$
that could still be treated in terms of classical space-time,
$l\sim{\rm H}^{-1}(\varphi)\sim\m^{-1}$.

The significance of (\ref{1.7.9}) is that the typical initial
value $\varphi_0$  of the field $\varphi$ in the early universe
is exceedingly large.  For example, in a theory with
$\displaystyle {\rm V}(\varphi)=\frac{\lambda}{4}\,\varphi^4$
and $\lambda \ll 1$,
\be
\label{1.7.11}
\varphi_0(x)\sim\lambda^{-1/4}\,\m\gg\m\ .
\ee
According to (\ref{1.7.4}) and (\ref{1.7.11}), in any region
whose size is of the order of the event horizon
${\rm H}^{-1}(\varphi)\sim\m^{-1}$, the field $\varphi_0(x)$
changes by a relatively insignificant amount,
$\Delta\varphi\sim\m\ll\varphi_0$.  In each such domain, as we
have said, the evolution of the field proceeds independently of
what is happening in the rest of the universe.

Let us consider such a region of the universe having initial size
${\rm O}(\m^{-1})$, in which
$\partial_\mu\varphi\,\partial^\mu\varphi$ and the squares of the
components of the\index{Curvature tensor}
curvature tensor ${\rm R}_{\mu\nu\alpha\beta}$, which are
responsible for the inhomogeneity and anisotropy of the
universe,\footnote{Note that the quantities
$\partial_\mu\partial^\mu\varphi$ and ${\rm R}^2$ cannot exceed
${\rm V}(\varphi)$ in one small {\it part} of the region
considered and be less than ${\rm V}(\varphi)$ in another, since
it is not possible to
subdivide\index{Classical space}\index{Space!classical}
{\it classical space} into parts
less than $\m^{-1}$ in size and consider the\index{Classical fields}
{\it classical
field} $\varphi$ separately in each of these parts, due to the
large quantum fluctuations of the metric at this scale.}
are several times smaller than ${\rm V}(\varphi)\sim\m^4$.
Since all these quantities are typically of the same order of
magnitude according to (\ref{1.7.7})--(\ref{1.7.10}), the
probability that regions of the specified type do exist should
not be much less than unity.  The subsequent evolution of such
regions turns out to be extremely interesting.

In fact, the relatively low degree of anisotropy and
inhomogeneity of space in such regions enables one to treat each
of them as being a locally Friedmann space, with a metric of the
type (\ref{1.3.1}), governed by Eq.  (\ref{1.3.7}):
\be
\label{1.7.12}
{\rm H}^2+\frac{k}{a^2}\equiv\left(\frac{\dot a}{a}\right)^2+\frac{k}{a^2}=
\frac{8\,\pi}{3\,\m^2}\,\left(\frac{\dot \varphi^2}{2}+
\frac{(\nabla\varphi)^2}{2}+{\rm V}(\varphi)\right)\ .
\ee
At the same time, the field $\varphi$ satisfies the equation
\be
\label{1.7.13}
\dla\varphi=\ddot\varphi+3\,\frac{\dot a}{a}\,\dot\varphi
-\frac{1}{a^2}\,\Delta\varphi=-\frac{d{\rm V}}{d\varphi}\ ,
\ee
where $\dla$ is the\index{d'Alembertian operator, covariant}
\index{Covariant d'Alembertian operator}covariant
d'Alembertian operator, and
$\Delta$ is the\index{Laplacian} Laplacian
in three-dimensional space with the time-independent metric
\be
\label{1.7.14}
dl^2=\frac{dr^2}{1-k\,r^2}+r^2\,(d\theta^2+\sin^2\theta\,d\varphi^2)\ .
\ee
For a sufficiently uniform and slowly varying field $\varphi$
$\displaystyle \left(\dot\varphi^2,(\nabla\varphi)^2\ll{\rm V};
\ddot\varphi\ll\frac{d{\rm V}}{d\varphi}\right)$,
Eqs. (\ref{1.7.12}) and (\ref{1.7.13}) reduce to
\ba
\label{1.7.15}
{\rm H}^2+\frac{k}{a^2}&\equiv&\left(\frac{\dot a}{a}\right)^2+\frac{k}{a^2}
=\frac{8\,\pi}{3\,\m^2}\,{\rm V}(\varphi)\ ,\\
\label{1.7.16} 3\,{\rm H}\,\dot\varphi&=&-\frac{d{\rm
V}}{d\varphi}\ . \ea It is not hard to show that if the universe
is expanding ($\dot a>0$) and, as we have said, the initial value
for $\varphi$ satisfies (\ref{1.7.11}), then the solution of the
system of equations (\ref{1.7.15}) and (\ref{1.7.16}) rapidly
proceeds to its asymptotic limit of quasiexponential
expansion\index{Inflation} (inflation), whereupon the term
$\displaystyle \frac{k}{a^2}$ in (\ref{1.7.15}) can be neglected.
Such behavior is understandable, inasmuch as Eq. (\ref{1.7.15})
tells us that when $a^2$  is large, $\displaystyle {\rm
H}^2=\frac{8\,\pi\,{\rm V}(\varphi)}{3\,\m^2}$. It then follows
from (\ref{1.7.16}) that \be \label{1.7.17}
\frac{1}{2}\,\dot\varphi^2=\frac{\m^2}{48\,\pi\,{\rm V}}\,
\left(\frac{d{\rm V}}{d\varphi}\right)^2\ . \ee
Hence, for ${\rm
V}(\varphi)\sim\varphi^n$, we have that \be \label{1.7.18}
\frac{1}{2}\,\dot\varphi^2
=\frac{n^2\,\m^2}{48\,\pi\,\varphi^2}\,{\rm V}(\varphi)\ , \ee
i.e., that  $\displaystyle \frac{1}{2}\,\dot\varphi^2\ll{\rm
V}(\varphi)$ when \be \label{1.7.19}
\varphi\gg\frac{n}{4\,\sqrt{3\,\pi}}\,\m\ . \ee

This means that for large $\varphi$, the energy-momentum tensor
${\rm T}_{\mu\nu}$ of the field $\varphi$ is determined almost
entirely by the quantity $g_{\mu\nu}\,{\rm V}(\varphi)$, or in
other words, $p \approx-\rho$, and the universe expands
quasiexponentially. Because of the fact that when $\varphi\gg\m$
the rates at which the field $\varphi$ and potential ${\rm V}(\varphi)$
vary are much less than the rate of expansion of the universe
$\displaystyle \left(\frac{\dot\varphi}{\varphi}\ll{\rm H},
\dot {\rm H}\ll {\rm H}^2\right)$, over time intervals
$\displaystyle \Delta t\la\frac{{\rm H}}{\dot{\rm H}}\gg{\rm H}^{-1}$
the universe looks approximately like de Sitter space with the expansion law
\be
\label{1.7.20}
a(t)\sim e^{{\rm H}t}
\ee
where the quantity
\be
\label{1.7.21}
{\rm H}(\varphi(t))=\sqrt{\frac{8\,\pi\,{\rm V}(\varphi)}{3\,\m^2}}
\ee
decreases slowly with time [\cite{56}].

\begin{figure}[t]\label{f5}
\centering \leavevmode\epsfysize=7cm \epsfbox{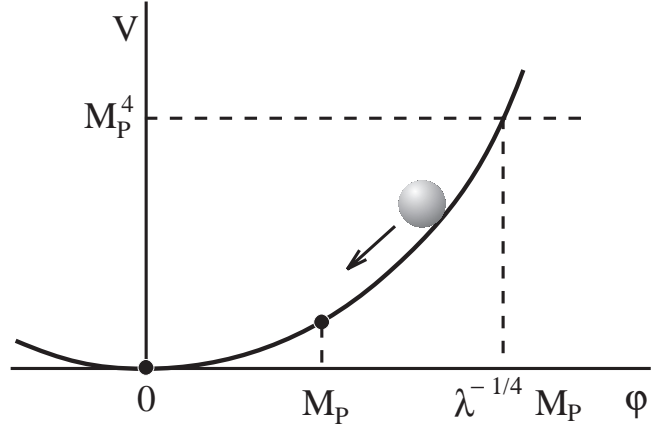}
\caption{Evolution of a homogeneous classical scalar field
$\varphi$ in a theory with $\displaystyle {\rm
V}(\varphi)=\frac{\lambda}{4}\,\varphi^4$, neglecting quantum
fluctuations of the field. When $\varphi>\lambda^{-1/4}\,\m$, the
energy density of the field $\varphi$ is greater than the Planck
density, and the evolution of the universe cannot be described
classically.  When $\displaystyle
\frac{\m}{3}\la\varphi\la\lambda^{-1/4}\,\m$, the field $\varphi$
slowly decreases, and the universe then expands quasiexponentially
(inflates).  When $\displaystyle \varphi\la\frac{\m}{3}$, the
field $\varphi$ oscillates rapidly about the minimum of ${\rm
V}(\varphi)$, and transfers its energy to the particles produced
thereby (reheating of the universe).}
\end{figure}
Under these conditions, the behavior of the field $\varphi(t)$
(see Fig. 1.5) is \be \label{1.7.22}
\varphi(t)=\varphi_0\,\exp\left(-\sqrt{\frac{\lambda}{6\,\pi}}\,\m\,t\right)
\ee for a theory with $\displaystyle
\vf=\frac{\lambda}{4}\,\varphi^4$, and \be \label{1.7.23}
\varphi(t)^{2-\frac{n}{2}}=\varphi_0^{2-\frac{n}{2}}+
t\,\left(2-\frac{n}{2}\right)\,\sqrt{\frac{n\,\lambda}{24\,\pi}}\,
\m^{3-\frac{n}{2}} \ee for $\vf\sim\varphi^n$ (\ref{1.7.2}), with
$n \neq 4$. In particular, for a theory with $\displaystyle
\vf=\frac{m^2\varphi^2}{2}$ (i.e., for $n = 2$ and
$\lambda\,\m^2=m^2$), \be \label{1.7.24}
\varphi(t)=\varphi_0-\frac{m\,\m}{2\,\sqrt{3\,\pi}}\,t\ . \ee
Meanwhile,\index{Universe!scale factor of}\index{Scale factor of
universe} the behavior of the scale factor of the universe is
given by the general equation \be \label{1.7.25}
a(t)=a_0\,\exp\frac{4\,\pi}{n\,\m^2}\,(\varphi_0^2-\varphi^2(t))\
, \ee which yields Eq. (\ref{1.7.20}) for sufficiently small $t$.
Making use of the estimate (\ref{1.7.18}), one can easily see that
this regime (the inflation regime) ends when $\displaystyle
\varphi\la\frac{n}{12}\,\m$. If $\varphi_0\gg\m$, then
(\ref{1.7.24}) implies that the
overall\index{Inflation factor of universe}%
\index{Universe!inflation factor of}
inflation factor P for the universe at that time is
\be
\label{1.7.26}
P\approx\exp\left(\frac{4\,\pi}{n\,\m^2}\varphi_0^2\right)\ .
\ee

According to (\ref{1.7.26}), then, the degree of inflation is
small for small initial values of the field $\varphi$, and it
grows exponentially with increasing $\varphi_0$.  This means that
most of the physical volume of the universe comes into being not
by virtue of the expansion of regions which initially, and
randomly, contained a small field $\varphi$ (or a markedly
inhomogeneous and rapidly varying field $\varphi$ that failed to
lead to exponential expansion of the universe), but as a result
of the inflation of regions of a size exceeding the radius of the
event horizon ${\rm H}^{-1}(\varphi)$ which were initially filled
with a sufficiently homogeneous, slowly varying, extremely large
field $\varphi = \varphi_0$.  The only fundamental constraint on
the magnitude of the homogeneous, slowly varying field $\varphi$
is $\vf\la\m^4$ (\ref{1.7.5}).  As we have already mentioned, the
probability that domains of size
$\Delta l\ga{\rm H}^{-1}(\varphi)\sim\m^{-1}$ exist in the early
universe with
$\dot\varphi^2$, $(\nabla\varphi)^2\la\vf\sim\m^{-1}$ should not
be significantly suppressed.  In conjunction with (\ref{1.7.26}),
this leads one to believe that most of the physical volume of the
present-day universe came into being precisely as a result of the
exponential expansion of regions of the aforementioned type.

\begin{figure}[t]\label{f6}
\centering \leavevmode\epsfysize=7cm \epsfbox{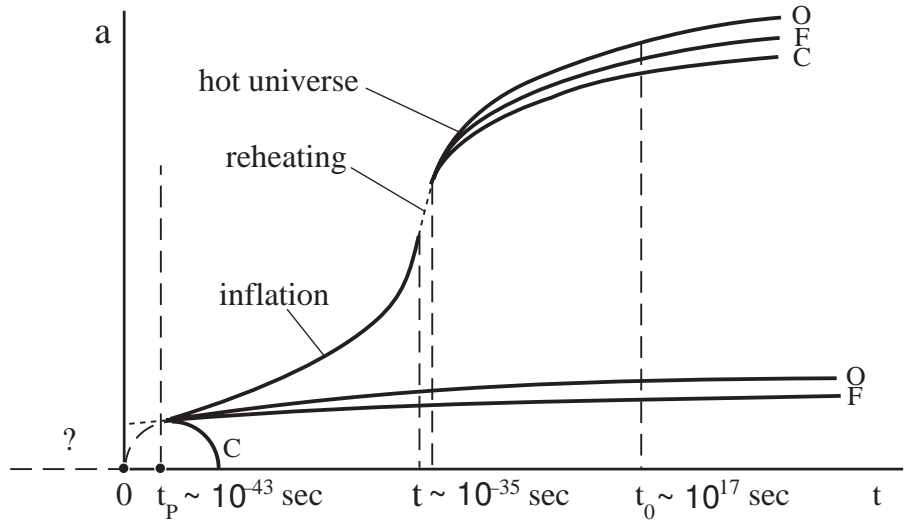} \caption{The
lighter set of curves depicts the behavior of the size of the hot
universe (or more precisely, its scale factor) for three Friedmann
models:  open (O), flat (F), and closed (C).  The heavy curves
show the evolution of an inflationary region of the universe.
Because of quantum gravitational fluctuations, the classical
description of the expansion of the universe cannot be valid prior
to $t\sim t_{\rm P}=\m^{-1}\sim10^{-43}$ sec after the Big Bang at
$t = 0$ (or after the start of inflation in the given region).  In
the simplest models, inflation continues for approximately
$10^{-35}$ sec.  During that time, the inflationary region of the
universe grows by a factor of from $10^{10^7}$ to $10^{10^{14}}$.
Reheating takes place afterwards, and the subsequent evolution of
the region is described by the hot universe theory.}
\end{figure}
If in the initial state, as we are assuming, \be \label{1.7.27}
{\rm
V}(\varphi_0)\sim\frac{\lambda\,\varphi_0^n}{n\,\m^{n-4}}\sim\m^4\
, \ee the inflation factor of the corresponding region is \be
\label{1.7.28} {\rm P}\sim\exp\left[\frac{4\,\pi}{n}
\left(\frac{\lambda}{n}\,\right)^{-\frac{2}{n}}\right]\ . \ee In
particular, for a $\displaystyle \frac{\lambda}{4}\,\varphi^4$
theory \be \label{1.7.29} {\rm
P}\sim\exp\left(\frac{2\pi}{\sqrt{\lambda}}\right)\ , \ee while
for an $\displaystyle \frac{m^2\,\varphi^2}{2}$ theory, \be
\label{1.7.30} {\rm P}\sim\exp\frac{4\pi\,\m^2}{m^2}\ . \ee

After the field $\varphi$ decreases in magnitude to a value of
order $\m$ (\ref{1.7.18}), the quantity H, which plays the role of
a coefficient of friction in Eq. (\ref{1.7.13}), is no longer
large enough to prevent the field $\varphi$ from rapidly rolling
down to the minimum of the effective potential.  The field
$\varphi$ starts its oscillations near the minimum of ${\rm
V}(\varphi)$, and its energy is transferred to the particles that
are created as a result of these oscillations.  The particles thus
created collide with one another, and approach a state of
thermodynamic equilibrium --- in other words, the universe heats
up [\cite{53}, \cite{123}, \cite{124}] (see Fig. 1.6).

If this reheating of the universe occurs rapidly enough (during
the time $\Delta t\la{\rm H}^{-1}(\varphi\sim\m)$), virtually all
of the energy from the oscillating field will be transformed into
thermal energy, and the temperature of the universe after
reheating will be given by
\be
\label{1.7.31}
\frac{\pi^2\,{\rm N}({\rm T}_{\rm R})}{30}\,{\rm T}_{\rm R}^4\sim
{\rm V}\left(\varphi\sim\frac{n}{12}\,\m\right)\ .
\ee
For example, with ${\rm N}({\rm T})\sim10^{3}$ for the
$\displaystyle \vf=\frac{\lambda}{4}\,\varphi^4$  theory,
${\rm T}_{\rm R}=c\,\lambda^{1/4}\,\m$, where $c={\rm O}(10^{-1})$.  In
many realistic versions of the inflationary scenario, however,
the temperature of the universe after reheating is found to be
many orders of magnitude lower than
$\displaystyle {\rm V}^{1/4}\left(\varphi\sim\frac{n}{12}\,\m\right)$
because of the inefficiency of the reheating process that results from
the weak interaction of the field $\varphi$ with itself and with other
fields (see below).

One circumstance that is especially important is that both the
value and the behavior of the field $\varphi$ near
$\varphi\sim\m$ are essentially independent of its initial value
$\varphi_0$ when $\varphi_0 \gg \m$; that is, the initial
temperature of the universe after reheating depends neither on
the initial conditions during the inflationary stage nor its
duration, etc.  The only parameter that changes during inflation
is the scale factor, which grows exponentially in accordance with
(\ref{1.7.28})--(\ref{1.7.30}).  This is precisely the
circumstance that enables us to solve the majority of the
problems recounted in Section \ref{s1.5}.

\begin{figure}[t]\label{f7}
\centering \leavevmode\epsfysize=10cm \epsfbox{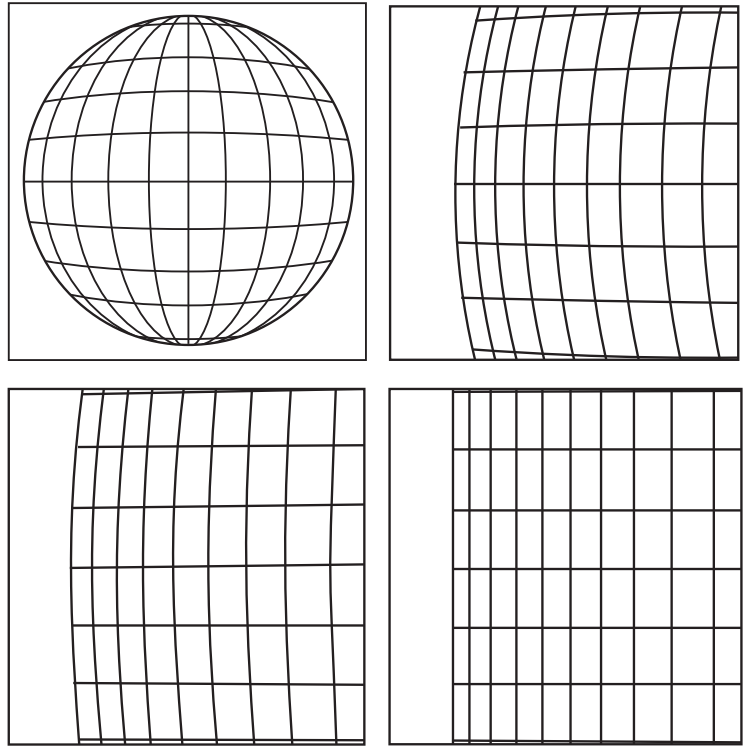}
\caption{When an object increases enormously in size, its surface
geometry becomes almost Euclidean.  This effect is fundamental to
the solution of\index{Flatness problem! solution of} the flatness,
homogeneity, and isotropy problems in the observable part of the
universe, by virtue of the exponentially rapid inflation of the
latter.}
\end{figure}

First of all, let us discuss the problems of the flatness,
homogeneity, and isotropy of space.  Note that during the
quasiexponential expansion of the universe, the right-hand side
of Eq. (\ref{1.7.12}) decreases very slowly, while the term
$\displaystyle \frac{k}{a^2}$ on the left-hand side falls off
exponentially.  Thus, the local difference between the
three-dimensional geometry of the universe and the geometry of
flat space also falls off exponentially, although the global
topological properties of the universe remain unchanged.  To
solve the\index{Flatness problem! solution of} flatness
problem, it is necessary that during inflation
a region of initial size $\Delta l\sim\m^{-1}\sim10^{-33}$ cm
grow by a factor of roughly $10^{30}$ (see Section \ref{s1.5}).  This
condition is amply satisfied in most specific realizations of the
chaotic inflation scenario (see below), and in contrast to the
situation in the new inflationary universe scenario, inflation
can begin in the present scenario at energy densities as high as
one might wish, and arbitrarily soon after the universe starts to
expand, i.e., prior to the moment when a closed universe starts
to recollapse.  After a closed universe has passed through its
inflationary stage, its size (and therefore its lifetime) becomes
exponentially large.  The flatness problem in the chaotic
inflation scenario is thereby solved, even if the universe is
closed.

The solution of the flatness problem in this scenario has a
simple, graphic interpretation:  when a sphere inflates, its
topology is unaltered, but its geometry becomes flatter (Fig. \ref{f7}).
The analogy is not perfect, but it is reasonably
useful and instructive.  It is clear, for instance, that if the
Himalayas were drastically stretched horizontally while their
height remained fixed, we would find a plain in place of the
mountains.  The same thing happens during inflation of the
universe.  Thus, for example, rapid inflation inhibits
time-dependent changes in the amplitude of the field $\varphi$
(the term $3\,{\rm H}\,\dot\varphi$ in (\ref{1.7.13}) plays the
role of viscous damping), i.e., the distribution of the field
$\varphi$ in coordinates $r$, $\theta$, $\varphi$  is ``frozen
in.'' At the same time, the overall scale of the universe $a(t)$
grows exponentially, so that the distribution of the classical
field $\varphi$ {\it per unit physical volume} approaches spatial
uniformity at an exponential rate,
$\partial_i\varphi\,\partial^i\varphi\rightarrow0$.  At the same
time, the energy-momentum tensor rapidly approaches
$g_{\mu\nu}\,\vf$ (to within small corrections
$\sim\dot\varphi^2$)\index{Curvature tensor}
curvature tensor acquires the form
\ba
\label{1.7.32}
{\rm R}_{\mu\nu\alpha\beta}={\rm H}^2\,(g_{\mu\nu}\,g_{\alpha\beta}-
g_{\mu\beta}\,g_{\nu\alpha})\ ,\\[4pt]
\label{1.7.33}
{\rm R}_{\mu\nu}=3\,{\rm H}^2\,g_{\mu\nu}\ ,\\[4pt]
\label{1.7.34}
{\rm R}=12\,{\rm H}^2=\frac{32\,\pi}{\m^2}\,\vf\ ,
\ea
and the difference between the properties of this domain of the
universe and those of the homogeneous, isotropic Friedmann
universe (\ref{1.3.1}) becomes exponentially small (in complete
accord with the ``no hair'' theorem for de Sitter space).  After
inflation, this homogeneous and isotropic domain becomes
exponentially large.  This explains the homogeneity and isotropy
of the observable part of the universe [\cite{54}--\cite{56},
\cite{120}--\cite{122}].

The stretching of the scales of all inhomogeneities leads to an
exponential decrease in the density of monopoles, domain walls,
gravitinos, and other entities produced before or during
inflation.  If ${\rm T}_{\rm R}$, the temperature of the universe
after reheating, is not high enough to produce monopoles, domain
walls, and gravitinos again, the corresponding problems
disappear.

{\looseness=1
Simultaneously with the smoothing of the original inhomogeneities
and ejection of monopoles and domain walls beyond the limits of
the observable universe, inflation itself gives rise to specific
large-scale inhomogeneities [\cite{107}, \cite{114}, \cite{125}].
The theory of this phenomenon is quite complicated;  it will be
considered in Section \ref{s7.5}.  Physically, the reason for the
appearance of large-scale inhomogeneities in an inflationary
universe is related to the restructuring of the vacuum state
resulting from the exponential expansion of the universe.  It is
well known that the expansion of the universe often leads to the
\index{Elementary particles!production of}production
of elementary particles [\cite{74}].  It turns out
that the usual particles are produced at a very low rate during
inflation,  but inflation converts short-wavelength quantum
fluctuations $\delta\varphi$ of the field $\varphi$ into
long-wavelength fluctuations.  In an inflationary universe,
short-wavelength fluctuations of the field $\varphi$ are no
different from short-wavelength fluctuations in the Minkowski
space (\ref{1.1.13}) (a field with momentum $k \gg {\rm H}$ does
not ``feel'' the curvature of space).  After the wavelength of a
fluctuation $\delta\varphi$ exceeds the horizon ${\rm H}^{-1}$ in
size, however, its amplitude is ``frozen in'' (due to the damping
term $3\,{\rm H}\dot\varphi$ in (\ref{1.7.13})); that is, the
field $\delta\varphi$ stops oscillating, but the wavelength of
the field $\delta\varphi$ keeps growing exponentially.  Looked at
from the standpoint of conventional scalar field quantization in
a Minkowski space, the appearance of such scalar field
configurations may be interpreted not as the production of
particles of the field $\varphi$ (\ref{1.1.13}), but as the
creation of an inhomogeneous (quasi)classical field
$d\varphi(x)$, where the degree to which it can be considered
semiclassical rises exponentially as the universe expands.  One
could say that in a certain sense an inflationary universe works
like a laser, continuously generating waves of the classical
field $\varphi$ with wavelength $l\sim k^{-1}\sim{\rm H}^{-1}$.
There is an important difference, however, in that the wavelength
of the inhomogeneous classical field $\delta\varphi$ that is
produced then grows exponentially with time.  Small-scale
inhomogeneities of the field $\varphi$ that arise are therefore
stretched to exponentially large sizes (with their amplitudes
changing very slowly), while new small-scale inhomogeneities
$\delta\varphi(x)$ are generated in their place.

}

The typical time scale in an inflationary universe is of course
$\Delta t={\rm H}^{-1}$.  The mean amplitude of the field
$\delta\varphi(x)$ with wavelength $l\sim k^{-1}\sim {\rm H}^{-1}$
generated over this period is [\cite{126}--\cite{128}] \be
\label{1.7.35} |\delta\varphi(x)|\sim\frac{{\rm
H}(\varphi)}{2\,\pi}\ . \ee Since ${\rm H}(\varphi)$ varies very
slowly during inflation, the amplitude of perturbations of the
field $\varphi$ that are formed over $\Delta t={\rm H}^{-1}$ a
time  will have only weak time dependence.  Bearing in mind, then,
that the wavelength $l\sim k^{-1}$ of fluctuations
$\delta\varphi(x)$ depends exponentially on the inflation time
$t$, it can be shown that the spectrum of inhomogeneities of the
field $\varphi$ formed during inflation and the spectrum of
density inhomogeneities $\delta\rho$ proportional to
$\delta\varphi$ are almost independent of wavelength $l$ (momentum
$k$) on a logarithmic scale. As we have already
mentioned,\index{Inhomogeneity spectra} inhomogeneity spectra of
this type were proposed long ago by cosmologists studying galaxy
formation [\cite{76}, \cite{214}].  The theory of galaxy formation
requires, however, that the relative amplitude of
density\index{Density perturbation} fluctuations with such a
spectrum be fairly low, \be \label{1.7.36}
\frac{\delta\rho(k)}{\rho}\sim\mbox{$10^{-4}$--$10^{-5}$}\ . \ee
At the same time, estimates of the quantity $\displaystyle
\frac{\delta\rho}{\rho}$  in the $\displaystyle
\vf\sim\frac{\lambda}{4}\,\varphi^4$ theory yield [\cite{114},
\cite{116}] \be \label{1.7.37}
\frac{\delta\rho}{\rho}\sim10^2\,\sqrt{\lambda}\ , \ee whereupon
we find that the constant $\lambda$ should be extremely small, \be
\label{1.7.38} \lambda\sim\mbox{$10^{-13}$--$10^{-14}$}\ , \ee
exactly as in the new inflationary universe scenario.  With this
value of $\lambda$, the typical inflation factor for the universe
is of order \be \label{1.7.39} {\rm
P}\sim\exp\frac{\pi}{\sqrt{\lambda}}\sim10^{10^5}\ . \ee During
inflation, a region of initial size $\Delta l\sim l_{\rm
P}\sim\m^{-1}\sim10^{-33}$ cm will grow to \be \label{1.7.40} {\rm
L}\sim\m^{-1}\exp\frac{\pi}{\sqrt{\lambda}}\sim10^{10^5}\;\mbox{cm}\
, \ee which is many orders of magnitude larger than the observable
part of the universe, ${\rm R}_p\sim10^{28}$ cm.  According to
(\ref{1.7.22}), the total duration of inflation will be \be
\label{1.7.41}
\tau\sim\frac{1}{4}\,\sqrt{\frac{6\,\pi}{\lambda}}\,
\m^{-1}\,\ln\frac{1}{\lambda}\sim10^8\,\m^{-1}
\sim10^{-35}\;\mbox{sec}\ . \ee

The estimates (\ref{1.7.39}) and (\ref{1.7.40}) make it clear how
the\index{Horizon problem!resolved} horizon problem
is resolved in the chaotic inflation
scenario:  expansion began practically simultaneously in
different regions of the observable part of the universe with a
size $l\la10^{28}$ cm, since they all came into being as a result
of inflation of a region of the universe no bigger than
$10^{-33}$ cm, which started simultaneously to within
$\Delta t\sim t_{\rm P}\sim10^{-43}$ sec.  The exponential expansion of
the universe makes it causally connected at scales many orders of
magnitude greater than the horizon size in a hot universe,
$R_{\rm P}\sim c\,t$.

These results may seem absolutely incredible, especially when one
realizes that the entire observable part of the universe, which
according to the hot universe theory has now been expanding for
about $10^{10}$ years, is incomparably smaller than a single
inflationary domain which started with the smallest possible
initial size, $\Delta l\sim l_{\rm P}\sim\m^{-1}\sim10^{-33}$  cm
(\ref{1.7.40}), and expanded in a matter of $10^{-37}$ sec.  Here
we must again emphasize that such a rapid increase in the size of
the universe is not at variance with the conventional limitation
on the speed at which a signal can propagate, ${\rm v} \le c = 1$
(see Section \ref{s1.4}).  On the other hand, it must also be understood
that the actual numerical estimates (\ref{1.7.39}) and
(\ref{1.7.40}) depend heavily on the model used.  For example, if
$\displaystyle\frac{\delta\rho}{\rho}\sim10^{-5}$ in the theory
with $\displaystyle \vf=\frac{m^2\,\varphi^2}{2}$, the
characteristic inflation factor P becomes $10^{10^{14}}$ instead
of $10^{10^7}$;  the inflation factor is much smaller in some
other models.  For our purposes, it will only be important that
after inflation, the typical size of regions of the universe that
we consider become many orders of magnitude larger than the
observable part of the universe.  Accordingly, the quantity
$\displaystyle \frac{k}{a^2}$ in (\ref{1.3.7}) will then be many
orders of magnitude less than
$\displaystyle \frac{8\,\pi}{3}\,{\rm G}\,\rho$; i.e., the universe after
inflation becomes (locally) indistinguishable from a flat
universe.  This implies that the\index{Universe!density}%
\index{Density!of universe}
density of the universe at the
present time must be very close to the critical value,
\be
\label{1.7.42}
\Omega=\frac{\rho}{\rho_c}=1\ ,
\ee
to within $\displaystyle \frac{\delta\rho}{\rho_c}\sim 10^{-3}$--$10^{-4}$,
which is related to local density\index{Density!of universe}%
\index{Universe!density}
inhomogeneities in the observable part of the universe.  This is
one of the most important predictions of the inflationary
universe scenario, and in principle it can be verified using
astronomical observations.

Let us now turn to the problem
of\index{Inflation!reheating of the universe after}%
\index{Reheating of the universe after inflation}%
\index{Universe!reheating of, after inflation}
reheating of the universe after
inflation.  For $\lambda\sim 10^{-14}$ in the
$\displaystyle \frac{\lambda}{4}\,\varphi^4$ theory, the temperature
of the universe after reheating, according to (\ref{1.7.31}), cannot
typically exceed
\be
\label{1.7.43}
{\rm T}_{\rm R}\sim10^{-1}\,\lambda^{1/4}\,\m
\sim3\cdot 10^{14}\;\mbox{GeV}\ .
\ee
As a rule, ${\rm T}_{\rm R}$ actually turns out to be even lower.
In the first place, in this theory, the oscillation frequency of
the field $\varphi$ near the minimum of ${\rm V}(\varphi)$ is
$\sqrt{\lambda}\,\m\sim10^{12}$ GeV at most, and in some theories
it is impossible to reheat the universe to higher temperatures.
Furthermore, the weakness with which $\varphi$ interacts with
other fields retards reheating.  As a result, the oscillation
amplitude of the field $\varphi$ decreases as the universe
expands, due to the term $3\,{\rm H}\,\dot\varphi$ in the
equation of motion for $\varphi$, and the temperature of the
universe after inflation turns out to be much lower in certain
theories than the value in (\ref{1.7.43}).

\begin{figure}[t]\label{f8}
\centering \leavevmode\epsfysize=13cm \epsfbox{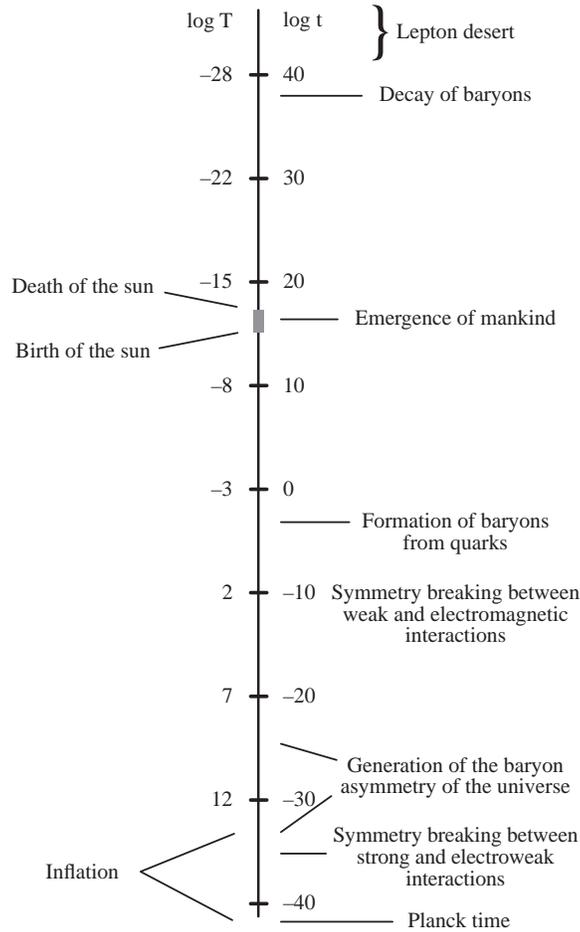} \caption{The
main stages in the evolution of an inflationary region of the
universe.  The time $t$ is measured in seconds from the start of
inflation, and the temperature T is measured in GeV
($1\;\mbox{GeV} \approx10^{13}$ K).  The typical lifetime of a
region of the universe, between the start of inflation and the
region's collapse (if it exceeds the critical density), is many
orders of magnitude greater than the proton decay lifetime in the
simplest grand unified theories.}
\end{figure}
Generally speaking, this can lead to some difficulties in
treating the baryon asymmetry problem.  Actually, during
inflation, any initial baryon asymmetry in the universe dies out
exponentially, and for such asymmetry to arise after inflation
becomes not just aesthetically attractive, as in the usual hot
universe theory, but necessary.  Moreover, the mechanism for
producing the baryon asymmetry that was proposed in
[\cite{36}--\cite{38}] and worked out in the context of grand
unified theories is only effective if the temperature T is high
enough that superheavy particles appear in the hot plasma, with
their subsequent decay producing an excess of baryons over
antibaryons.  Usually, for this to happen, the temperature of the
universe must be higher than $10^{15}$ GeV, and this is seldom
achievable in the inflationary universe scenario.  Fortunately,
however, baryon production can also proceed at much lower
temperatures after inflation, due to nonequilibrium processes
that take place during reheating [\cite{123}].  Moreover, a
number of models have recently been suggested which allow for the
onset of baryon asymmetry even if the temperature of the universe
after inflation never exceeds $10^2$ GeV [\cite{97}--\cite{99},
\cite{129}].  Thus, the inflationary universe scenario can
successfully incorporate all basic results of the hot universe
theory, and the resulting theory proves to be free of the main
difficulties of the standard hot Big Bang cosmology.

The main stages in the evolution of an inflationary domain of the
universe are shown in Fig. 8.1.  The initial and final stages
in the development of each individual inflationary domain depend
on the global structure of the inflationary universe, which we
will discuss in Section \ref{s1.8}.\index{Chaotic inflation scenario|)}

\section{\label{s1.8}The self-reproducing universe%
\index{Self-reproducing inflationary universe|(}%
\index{Universe!self-reproducing|(}}

The attentive reader probably already has noticed that in
discussing the problems resolved with the aid of the inflationary
universe scenario, we have silently skirted the most important
one --- the problem of the\index{Cosmological singularity}
cosmological singularity.  We have
also said nothing about the global structure of the inflationary
universe, having limited ourselves to statements to the effect
that its local properties are very similar to those of the
observable world.  The study of the global structure of the
universe and the problem of the cosmological singularity within
the scope of the inflationary universe scenario conceals a number
of surprises.  Prior to the advent of this scenario, there was
absolutely no reason to suppose that our universe was
markedly\index{Inhomogeneous universe}\index{Universe!inhomogeneous}
inhomogeneous on a large scale. On the contrary, the astronomical
data attested to the fact that on large scales, up to the very size
of the entire\index{Universe!observable part of} observable part of
the universe ${\rm R}_p\sim10^{28}$ cm,
inhomogeneities $\displaystyle \frac{\delta\rho}{\rho}$ on the
average were at most $10^{-3}$.  To understand the evolution of
the universe, it was in large measure thought to be sufficient to
investigate homogeneous (or slightly inhomogeneous) cosmological
models like the Friedmann model (or anisotropic Bianchi models)
[\cite{65}].

Meanwhile, the results of the preceding section make it clear
that the observable part of the universe is most likely just a
minuscule part of the universe as a whole, and it is an
impermissible extrapolation to draw any conclusions about the
homogeneity of the latter based on observations of such a tiny
component.  On the contrary, an investigation of the global
geometry of the inflationary universe shows that the universe,
which is locally Friedmann, should be completely inhomogeneous on
the largest scales, and its global geometry and dynamical
behavior as a whole have nothing in common with the geometry and
dynamics of a Friedmann universe [\cite{57}, \cite{78},
\cite{132}, \cite{133}].

\begin{figure}[t]\label{f9}
\centering \leavevmode\epsfysize=6cm \epsfbox{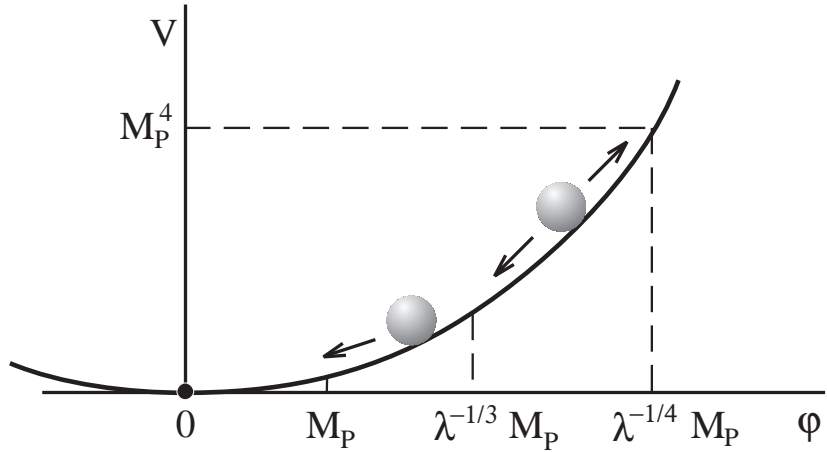}
\caption{\index{Scalar fields!evolution of}Evolution of the scalar
field $\varphi$ in the simplest field theory with the potential
$\displaystyle \vf=\frac{\lambda}{4}\,\varphi^4$, with quantum
fluctuations of the field $\varphi$ taken into account. When
$\varphi\ga\lambda^{-1/4}\,\m$ ($\vf\ga\m^4$), quantum
gravity\index{Quantum gravity} fluctuations of the metric are
large, and no classical description of space is possible in the
simplest theories.  For $\displaystyle
\frac{\m}{3}\la\varphi\la\lambda^{-1/4}\,\m$, the field $\varphi$
evolves relatively slowly, and the universe expands
quasiexponentially.  When
$\lambda^{-1/6}\,\m\la\varphi\ll\lambda^{-1/4}\,\m$, the amplitude
of $\varphi$ fluctuates markedly, leading to the endless birth of
ever newer regions of the universe.  For $\displaystyle
\frac{\m}{3}\la\varphi\ll\lambda^{-1/6}\,\m$, fluctuations of the
field are of relatively low amplitude.  The field $\varphi$ rolls
downhill, and fluctuations engender the density inhomogeneities
required for the formation of galaxies. When $\displaystyle
\varphi\la\frac{\m}{3}$, the field starts to oscillate rapidly
about the point $\varphi = 0$, particle pairs are produced, and
all of the energy of the oscillating field is converted into
heat.}
\end{figure}

In order to obtain a simple derivation of this important and
somewhat surprising result, let us consider more carefully the
behavior of the scalar field $\varphi$ for the minimal model
(\ref{1.7.1}) with $\displaystyle
\vf=\frac{\lambda}{4}\,\varphi^4$ in the chaotic inflation
scenario, taking into account long-wave fluctuations of $\varphi$
that arise during inflation [\cite{57}].  We have from
(\ref{1.7.21}) and (\ref{1.7.22}) that in a typical time \be
\label{1.8.1} \Delta t={\rm H}^{-1}(\varphi)
=\sqrt{\frac{3}{2\,\pi\,\lambda}}\,\frac{\m}{\varphi^2}\ , \ee the
classical homogeneous field $\varphi$ decreases by \be
\label{1.8.2} \Delta\varphi=\frac{\m^2}{2\,\pi\,\varphi}\ . \ee
During this same period of time, according to (\ref{1.7.36}),
inhomogeneities of the field $\varphi$ are generated having
wavelength $l\ga{\rm H}^{-1}$ and mean amplitude \be \label{1.8.3}
|\delta\varphi(x)|\approx\frac{{\rm H}(\varphi)}{2\,\pi}
=\sqrt{\frac{\lambda}{6\,\pi}}\,\frac{\varphi^2}{\m}\ . \ee It is
not hard to show that when $\varphi \ll \varphi^*$, where \be
\label{1.8.4} \varphi^*=\lambda^{-1/6}\,{\rm M}\ , \ee quantum
fluctuations of $\varphi$ have a negligible influence on its
evolution, $|\delta\varphi(x)|\ll\Delta\varphi$.  It is precisely
at the later stages of inflation, when the field $\varphi$ becomes
less than $\varphi^*=\lambda^{-1/6}\,\m$ that small
inhomogeneities $\delta\varphi$ in the field $\varphi$ and small
density inhomogeneities $\delta\rho$ are produced, leading
to\index{Galaxy formation} the formation of galaxies.  On the
other hand, when $\varphi\gg \varphi^*$, only the mean field
$\varphi$ is governed by Eq. (\ref{1.7.22}), and the role played
by fluctuations becomes extremely significant (see Fig. 1.9).

Consider a region of an inflationary universe of size
$\Delta l\sim{\rm H}^{-1}(\varphi)$ that contains the field
$\varphi \gg \varphi^*$.  According to the ``no hair'' theorem
for de Sitter space, inflation in this region of space proceeds
independently of what happens in other regions.  In such a region,
the field can be assumed to be largely homogeneous, as initial
inhomogeneities of the field $\varphi$ are reduced by inflation,
and nascent inhomogeneities (\ref{1.8.3}) that make their
appearance during inflation have wavelengths $l>{\rm H}^{-1}$.
In the typical time $\Delta t={\rm H}^{-1}$, the region in
question will have grown by a factor of $e$, and its volume will
have grown by a factor of $e^3\approx20$, so that it could be
subdivided approximately into $e^3$ regions of size
${\rm O}({\rm H}^{-1})$, each once again containing an almost
homogeneous field $\varphi$, which differs from the original field
$\varphi$ by $\delta\varphi(x)-\Delta\varphi \approx\delta\varphi(x)$.
This means, however, that in something like $\displaystyle \frac{e^3}{3}$
regions of size ${\rm O}({\rm H}^{-1})$, instead of decreasing,
the field $\varphi$ increases by a quantity of order
$\displaystyle |\delta\varphi(x)|\sim\frac{{\rm H}}{2\,\pi}\gg\Delta\varphi$
(see Fig. 1.10).  This process repeats during the next time
interval $\Delta t={\rm H}^{-1}$, and so on.  It is not hard to show
that the total volume of the universe occupied by
the\index{Scalar fields!continually growing}
{\it continually growing} field $\varphi$ increases approximately as

\begin{figure}[t]\label{f10}
\vskip -5cm \centering \leavevmode\epsfysize=12cm \epsfbox{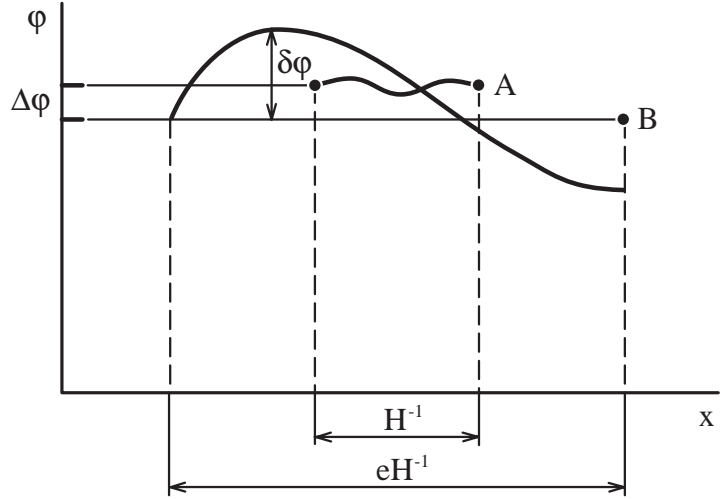}
\caption{\index{Scalar fields!evolution of}Evolution of a field
$\varphi\gg\varphi^*=\lambda^{-1/6}\,\m$ in an inflationary region
of the universe of initial size $\Delta l={\rm H}^{-1}(\varphi)$.
Initially (A), the field $\varphi$ in this domain is relatively
homogeneous, since inhomogeneities $\delta\varphi(x)$ with a
wavelength $l\sim {\rm H}^{-1}(\varphi)$ that result from
inflation are of order $\displaystyle \delta\varphi\sim\frac{{\rm
H}}{2\,\pi}\ll\varphi$. After a time $\Delta t={\rm
H}^{-1}$,\index{Scalar fields!continually growing} the size of the
region has grown (B) by a factor of $e$.  When
$\varphi\gg\varphi^*$, the average decrement $\Delta\varphi$ of
the field $\varphi$ in this region is much less than
$\displaystyle |\delta\varphi|\sim\frac{{\rm H}}{2\,\pi}$.  This
means that in almost half the region under consideration, the
field $\varphi$ grows instead of shrinking. Thus, in a time
$\Delta t={\rm H}^{-1}$, the volume occupied by {\it increasing}
$\varphi$ values\index{Scalar fields!continually growing} grows by
a factor of approximately $\displaystyle \frac{e^3}{2}\approx10$.
\vspace{-3pt}
}
\end{figure}
$$
\exp[(3-\ln2)\,{\rm H}\,t]
\ga\exp\left(3\,\sqrt{\lambda}\,\frac{\varphi^2}{\m}\,t\right)\ ,
$$
while the total volume occupied by the {\it non-decreasing} field
$\varphi$\index{Scalar fields!continually growing} grows
almost as fast as $\exp[3\,{\rm H}(\varphi)\,t]$.

This means that regions of space containing a field $\varphi$
continually spawn brand new regions with even higher field
values, and as $\varphi$ grows the birth and expansion process in
the new regions takes place at an ever increasing rate.

To better understand the physical meaning of this phenomenon, it
is useful to examine those regions which, while rare, are still
constantly appearing, where the field $\varphi$ {\it increases
continuously}, i.e., it is typically augmented by
$\displaystyle \delta\varphi\sim\frac{{\rm H}(\varphi)}{2\,\pi}$ in each
successive time interval $\Delta t={\rm H}^{-1}(\varphi)$.  The
rate of\index{Scalar fields!continually growing} growth
of the field in such regions is given by
\be
\label{1.8.5}
\frac{d\varphi}{dt}=\frac{{\rm H}^2(\varphi)}{2\,\pi}
=\frac{4\,\vf}{3\,\m^2}=\frac{\lambda\,\varphi^4}{3\,\m^2}\ ,
\ee
whereupon
\be
\label{1.8.6}
\varphi^{-3}(t)=\varphi_0^{-3}-\frac{\lambda\,t}{\m^2}\ .
\ee
This means that in a time
\be
\label{1.8.7}
\tau=\frac{\m^2}{\lambda\,\varphi_0^3}
\ee
the field $\varphi$ should become infinite.  Actually, of course,
one can only say that the field in such regions approaches the
limiting value $\varphi$ for which $\vf\sim\m^4$ (i.e.,
$\varphi\sim\lambda^{-1/4}\,\m$).  At higher densities, it
becomes impossible to treat such regions of space classically.
Furthermore, formal consideration of inflationary regions with
$\vf\gg\m^4$  indicates that most of their field energy is
concentrated not at ${\rm V}(\varphi)$, but at a value related to
the inhomogeneities $\delta\varphi(x)$ and proportional to
$\displaystyle {\rm H}^4\sim\frac{{\rm V}^2}{\m^4}$.  Therefore,
in the overwhelming majority of regions of the universe with
$\vf\gg\m^4$, inflation is most likely cut short, and in any case
we cannot describe it in terms of classical space-time.

To summarize, then, many inflationary regions with $\vf\sim\m^4$
are created over a time
$\displaystyle \tau\sim\frac{\m^2}{\lambda\,\varphi_0^3}$
in a part of the universe originally filled with a field
$\varphi_0 \gg \varphi^*$.  Some fraction of these regions will ultimately
expand to become regions with $\vf\gg\m^4$, that is, a
\index{Space-time foam}space-time
foam, which we are presently not in a position to describe.  It
will be important for us, however, that the volume of the
universe filled by the extremely large and non-decreasing field
$\varphi$, such that $\vf\sim\m^4$, continue
to\index{Scalar fields!continually growing} grow at the
highest possible rate, as $\exp(c\,\m\,t)$, $c = {\rm O}(1)$. The
net result is that in time $t\gg\tau$ (\ref{1.8.7}) (in a
synchronous coordinate system; see Section \ref{s10.3}), most of the
physical volume of an original inflationary region of the
universe with $\varphi=\varphi_0\gg\varphi^*$ should contain an
exceedingly large field $\varphi$, with $\vf\sim\m^4$.

This does not at all mean that the whole universe must be in a
state with the Planck density.  Fluctuations of the field
$\varphi$ constantly lead to the formation of regions not just
with $\varphi  \gg \varphi^*$, but with $\varphi \ll \varphi^*$ as
well.  Just such regions form
gigantic\index{Universe!homogeneous regions of}
homogeneous regions of the
universe like our own.  After inflation, the typical size of each
such region exceeds
\be
\label{1.8.8}
l\sim \m^{-1}\,\exp\left[\frac{\pi(\varphi^*)^2}{\m^2}\right]
\sim\m^{-1}\,\exp(\pi\,\lambda^{-1/3})
\sim 10^{6\cdot 10^4}\;\mbox{cm}\ .
\ee
when $\lambda\sim10^{-14}$.
This is much less than
$l\sim\m^{-1}\,\exp(\pi\,\lambda^{-1/2})\sim10^{10^7}$ cm
(\ref{1.7.40}), which we obtained by neglecting quantum
fluctuations, but it is still hundreds of orders of magnitude
larger than the observable part of the universe.

A more detailed justification of the foregoing results
[\cite{132}, \cite{133}] has been obtained within the context of
a stochastic approach to the inflationary universe theory
[\cite{134}, \cite{135}] (see Sections \ref{s10.2}--\ref{s10.4}).  We now
examine two of the major consequences of these results.

\subsection{The self-reproducing universe and the singularity problem}

As we have already pointed out in the preceding section, the most
natural initial value of the field $\varphi$ in an inflationary
region of the universe is
$\varphi\sim\lambda^{-1/4}\,\m\gg\varphi^*\sim\lambda^{-1/6}\,\m$.
Such a region endlessly produces ever newer regions of the
inflationary universe containing the field $\varphi  \gg \varphi^*$.
As a whole, therefore, this entire universe will
never collapse, even if it starts out as a closed Friedmann
universe (see Fig. 1.11). In other words, contrary to
conventional expectations, even in a closed (compact) universe
there will never be a global singular spacelike\index{Hypersurface}
hypersurface ---
the universe as a whole will never just vanish into nothingness.
Similarly, there is no sufficient reason for assuming that such a
hypersurface ever existed in the past --- that at some instant of
time $t = 0$, the universe as a whole suddenly appeared out of
nowhere.
\begin{figure}[t]\label{f11}
\vskip -0.5cm \centering \leavevmode\epsfysize=9.5cm
\epsfbox{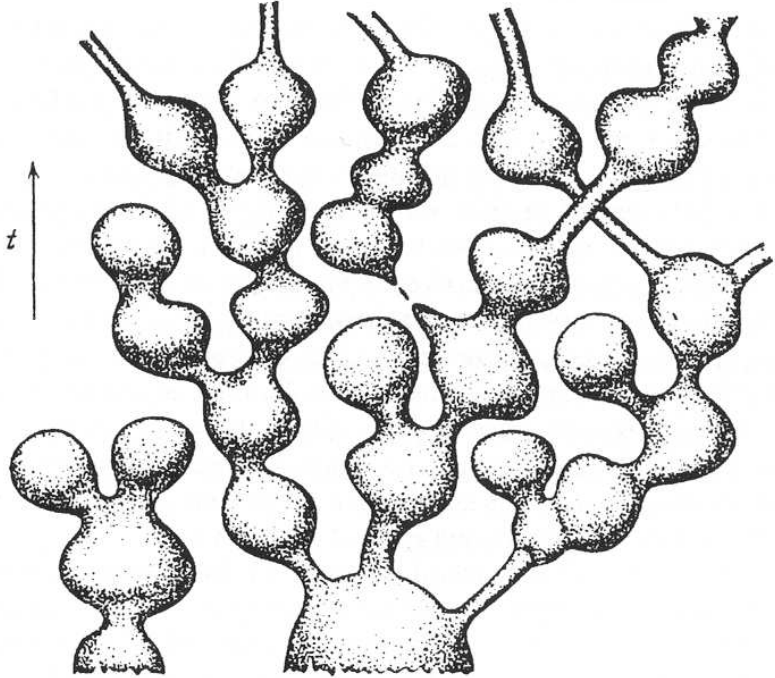} \caption{An attempt to convey some
impression of the global structure of the inflationary universe.
One region of the inflationary universe gives rise to a multitude
of new \index{Inflationary regions}inflationary regions; in
different regions, the properties of space-time and elementary
particle interactions  may be utterly different.  In this
scenario, the evolution of the universe as a whole has no end, and
may have no beginning.}
\end{figure}

This of course does not mean that there are no singularities in an
inflationary universe.  On the contrary, a considerable part of
the physical volume of the universe is constantly in a state that
is close to singular, with energy density approaching the Planck
density $\vf\sim\m^4$.  What is important, however, is that
different regions of the universe pass through a singular state
at different times, so there is no unique end of time, after
which space and time disappear.  It is also quite possible that
there was no unique beginning
of\index{Time in universe}\index{Universe!time in} time
in the universe.

It is worth noting that the standard assertion about the
occurrence of a general cosmological singularity (i.e., a global
singular spacelike hypersurface in the universe, or what is the
same thing, a unique beginning or end of time for the universe as
a whole) is {\it not} a direct consequence of existing
topological theorems on singularities in general relativity
[\cite{69}, \cite{70}], or of the behavior of general solutions
of the Einstein equations near a singularity [\cite{68}].  This
assertion is primarily based on analyzes of homogeneous
cosmological models like the Friedmann or Bianchi models.
Certain authors have emphasized that there might in fact not be a
unique beginning and end of time in the universe as a whole if
our own universe is only locally a Friedmann space but is
globally inhomogeneous (a
so-called\index{Quasihomogeneous universe}%
\index{Universe!quasihomogeneous}
quasihomogeneous universe;
see [\cite{34}, \cite{136}]).  However, in the absence of any
experimental basis for hypothesizing significant large-scale
inhomogeneity of the universe, this approach to settling the
problem of an overall cosmological singularity has not elicited
much interest.

The present attitude toward this problem has changed
considerably.  Actually, the only explanation that we are aware
of for the homogeneity of the observable part of the universe is
the one provided by the inflationary universe scenario.  But as
we have just shown, this same scenario implies that on the
largest scales the universe must be absolutely inhomogeneous,
with density excursions ranging from $\rho\la10^{-29}$ g/cm$^{3}$
(as in the observable part of the universe) to
$\rho\sim\m^4\sim10^{94}$ g/cm$^3$.  Hence, there is presently no
compelling basis for maintaining that there is a unique beginning
or end of the universe as a whole.  (For a more detailed
discussion of this question, see also Section \ref{s10.4}.)

It is not impossible in principle that the universe as a whole
might have been born ``out of nothing,'' or that it might have
appeared from a unique initial singularity.  Such a suggestion
could be fairly reasonable if in the process a compact (closed,
for example) universe of size $l={\rm O}(\m^{-1})$ were produced.
But for a noncompact universe, this hypothesis is not only hard
to interpret, it is completely implausible, since there would
seem to be absolutely no likelihood that all causally
disconnected regions of an infinite universe could spring
{\it simultaneously} from a singularity (see the discussion of the
horizon problem in Section \ref{s1.5}).  Fortunately, this
hypothesis turns out to be unnecessary for the scenario being
developed, and in that sense it seems possible to avoid one of
the main conceptual difficulties associated with the problem of
the cosmological singularity [\cite{57}].

\subsection{The problem of the uniqueness of the universe, and the
\protect\\ Anthropic Principle\index{Anthropic Principle|(}}

The ceaseless creation of new regions of the inflationary universe
takes place at the stage when
$\lambda^{-1/6}\,\m\la\varphi\la\lambda^{-1/4}\,\m$; then
$\lambda^{-1/3}\,\m^4\la\vf\la\m^4$ ($10^{-5}\,\m^4\la\vf\la\m^4$
for $\lambda\sim10^{-14}$). In other words, it is not necessary to
appeal for a description of this process to speculative phenomena
that take place above the Planck density.

On the other hand, it is important that much of the physical
volume of the universe must at all times be close to the Planck
density, and expanding exponentially with a Hubble constant H of
order $\m$.  In realistic elementary particle theories, apart
from the scalar field $\varphi$ responsible for inflation, there
are many other types of scalar fields $\Phi$, H, etc., with
masses $m \ll \m$.  Inflation leads to the generation of
long-wave fluctuations not just in the field $\varphi$, but in
all scalar fields with $m \ll {\rm H}\sim\m$.  As a result, the
universe becomes filled with fields $\varphi$, $\Phi$, and so
forth, which vary slowly in space and take on all allowable
values for which ${\rm V}(\varphi,\Phi,\ldots)\la\m^4$.  In those
regions where inflation has ended, the scalar fields ``roll
down'' to the nearest minimum of the effective potential ${\rm
V}(\varphi,\Phi,\ldots)$, and the universe breaks up into
exponentially large
domains\index{Mini-universes}\index{Universe!mini-universes}
(mini-universes) filled with the
fields $\varphi$, $\Phi$, etc., which in the different domains
take on values corresponding to all the local minima of
${\rm V}(\varphi,\Phi,\ldots)$.  In Kaluza--Klein and superstring
theories,\index{Quantum fluctuations}
quantum fluctuations can result in a local change in the type
of compactification on a scale ${\rm O}({\rm H}^{-1})\sim{\rm O}(\m^{-1})$.
If the region continues to inflate after this change, then by
virtue of the ``no hair'' theorem for de Sitter space, the
properties of the universe outside this region (its size and the
type of compactification) cannot exert any influence on the
region, and after inflation an exponentially large mini-universe
with altered compactification will have been created
[\cite{333}].

{\looseness=1%
The result is that the universe breaks up into mini-universes in
which all possible types of (metastable) vacuum states and all
possible types of compactification that support inflationary
behavior are realized.  We live in a region of the universe in
which there are weak, strong, and electromagnetic interactions,
and in which space-time is four-dimensional.  We cannot rule out
the possibility, however, that this is so not because our region
is the only one or the best one, but because such regions exist,
are exponentially plentiful (or more likely, infinitely
plentiful), and life of our type would be impossible in any other
kind of region [\cite{57}, \cite{78}].

}

This discussion is based on the Anthropic Principle, whose
validity the author himself previously cast into doubt (in
Section \ref{s1.5}).  But now the situation is different --- it
is not at all necessary for someone to sit down and create one
universe after another until he finally succeeds.  Once the
universe has come into being (or if it has existed eternally),
that universe itself will create exponentially large regions
(mini-universes), each having different elementary-particle and
space-time properties.  If good conditions for the appearance of
life in a solar-system environment are then to ensue, it turns
out that the same conditions necessarily appear on a scale much
larger than the entire observable part of the universe.  In fact,
for galaxies to arise in the simple model we are considering, one
must have $\lambda\sim10^{-14}$, and as we have seen, this leads
to a characteristic size $l\ga10^{6\cdot 10^4}$ cm for the
uniform region.  Thus, within the context of the approach being
developed, it is possible to remove the main objections to the
cosmological application of the Anthropic Principle (or to be
more precise, of the \index{Anthropic Principle!Weak}%
\index{Weak Anthropic Principle}
Weak Anthropic Principle;  see Sections
(\ref{s10.5} and \ref{s10.7}, where stronger versions of the
Anthropic Principle are also discussed).

This result may have important methodological implications.
Attempts to construct a theory in which the observed state of the
universe and the observed laws of interaction between elementary
particles are the only ones possible and are realized throughout
the entire universe become unnecessary.  Instead, we are faced
with the problem of constructing theories that can produce large
regions of the universe that resemble our own.  The question of
the most reasonable initial conditions near a singularity and the
probability of an inflationary universe being created is
supplanted by the question of what values the physical fields
might take, what the properties of space are in most of the
inflationary universe, and what the most likely way is to form a
region of the universe of size $R_p\sim10^{28}$ cm with
observable properties and observers resembling our own.

This new statement of the problem greatly enhances our ability to
construct realistic models of the inflationary universe and
realistic elementary particle theories.%
\index{Anthropic Principle|)}%
\index{Self-reproducing inflationary universe|)}%
\index{Universe!self-reproducing|)}

\section{\label{s1.9}Summary}

In this introductory chapter, we have discussed some basic
features of inflationary cosmology.  One should take into
account, however, that many details of the inflationary universe
scenario look different in the context of different theories of
elementary particles.  For example, it is not necessary to assume
that the field $\varphi$ which drives inflation is an elementary
scalar field.  In certain theories, the role played by this field
can be assumed by the curvature scalar R, a fermion condensate
$\langle \bar\psi\,\psi\rangle $ or vector meson condensate
$\langle {\rm G}_{\mu\nu}^a\,{\rm G}_{\mu\nu}^a\rangle $, or even
the logarithm of the radius of a compactified space.  More
detailed discussions of phase transitions in the unified theories
of weak, strong, and electromagnetic interactions, of various
versions of the inflationary universe scenario, and of different
aspects of quantum inflationary cosmology are to be found in
subsequent sections of this book.


\chapter{\label{c2}Scalar Field, Effective Potential,
and Spontaneous Symmetry Breaking}
\section{\label{s2.1}Classical and quantum scalar
\index{Classical scalar fields|(}%
\index{Scalar fields!classical|(}fields}

As we have seen, classical (or semiclassical) scalar fields play
an essential role in present-day cosmological models (and also in
modern elementary particle theories). We will often deal with
homogeneous and inhomogeneous classical fields, and there is
sometimes a question as to which fields can be considered
classical,\index{Classical fields} and in what sense.

Let us recall, first of all, that in accordance with the standard
approach to quantization of the scalar field $\varphi(x)$, the
functions $a^+({\bf k})$ and $a^-({\bf k})$ in (\ref{1.1.3}) can be put
into correspondence with the creation and annihilation operators
$a^+_k$ and $a^-_k$ for particles with momentum $k$.  The
commutation relations take the form [\cite{58}]
\be
\label{2.1.1}
\frac{1}{2k_0}\,[\varphi_k^-,\varphi_q^+]\equiv
[a_k^-,a_q^+]=\delta({\bf k}-{\bf q})\ ,
\ee
where the operator $a_k^-$ acting on the vacuum gives zero:
\be
\label{2.1.2}
a^-_k\,|0\rangle=0\ ;\quad \langle0|\,a^+_k=0\ ; \quad
\langle 0|\varphi(x)|0\rangle\ .
\ee
The operator $a_k^+$ creates a particle with momentum $k$,
\be
\label{2.1.3}
a^+_k\,|\psi\rangle =|\psi,{\bf k}\rangle\ ,
\ee
while the operator $a_k^-$ annihilates it.
\be
\label{2.1.4}
a_k^-\,|\psi,{\bf k}\rangle =|\psi\rangle\ .
\ee

Now consider the Green's function for the scalar field $\varphi$ [\cite{58}],
\be \label{2.1.5} {\rm G}(x)=\langle 0|{\rm
T}[\varphi(x)\,\varphi(0)]|0\rangle =
\frac{i}{(2\,\pi)^4}\int\frac{e^{-ikx}}{m^2-k^2-i\,\varepsilon}\:d^4k\
. \ee Here T is the time-ordering operator, and $\varepsilon$
shows how to perform integration near the singularity at $k^2=m^2$
(from here on, we omit both symbols). Evaluation of this
expression indicates that when $t=0$ and $x\ga m^{-1}$, ${\rm
G}(x)$ falls off exponentially with increasing $x$; that is, the
correlation between $\varphi(x)$ and $\varphi(0)$ becomes
exponentially small.  When $m=0$, ${\rm G}(x)$ has a power-law
dependence on $x$.

It is also useful to calculate ${\rm G}(0)$, which after
\index{Wick rotation}%
transforming to Euclidean space (by a Wick rotation
$k_0\rightarrow-i\,k_4$) may be written in the form
\be
\label{2.1.6}
{\rm G}(0)=\langle 0|\varphi^2|0\rangle =\frac{1}{(2\,\pi)^4}
\int\frac{d^4k}{k^2+m^2}=
\frac{1}{(2\,\pi)^3}\int\frac{d^3k}{2\,\sqrt{{\bf k}^2+m^2}}\ .
\ee
If averaging is carried out, for example, over a state containing
particles rather than over the conventional vacuum in Minkowski
space, we can represent the quantity
$\langle 0|\varphi^2|0\rangle \equiv \langle \varphi^2\rangle $ in the
form
\ba
\label{2.1.7}
\langle \varphi^2\rangle &=&\frac{1}{(2\,\pi)^3}
\int\frac{d^3k}{2\sqrt{{\bf k}^2+m^2}}\:
(1+2\langle a^+_k\,a^-_k\rangle )\nonumber \\
&=&\frac{1}{(2\,\pi)^3}
\int\frac{d^3k}{\sqrt{{\bf k}^2+m^2}}\:\left(\frac{1}{2}+n_k\right)\ .
\ea
Here $n_k$ is the number density of particles with momentum $k$.
For instance, for a Bose gas at nonzero temperature T, one has
\be
\label{2.1.8}
n_k=\frac{1}{
\exp\left(\frac{\displaystyle\sqrt{k^2+m^2}}{\displaystyle {\rm T}}
\right)-1}\ .
\ee
Another important example is a\index{Bose condensate}
Bose condensate $\varphi_0$ of
noninteracting particles of the field $\varphi$, with mass $m$
and vanishing momentum ${\bf k}$, for which
\be
\label{2.1.9}
n_k=(2\,\pi)^3\,\varphi_0^2\,m\,\delta({\bf k})\ ,
\ee
or a coherent wave of particles with momentum ${\bf p}$:
\be
\label{2.1.10}
n_k=(2\,\pi)^3\,\varphi_p^2\,\sqrt{{\bf p}^2+m^2}\,\delta({\bf k}-{\bf p})\ .
\ee
In both cases, $n_k$ tends to infinity at some value of ${\bf k}$.
The fact that the $a_k^\pm$ operators of (\ref{2.1.1}) do not commute
can then be ignored, as $n_k\gg 1$ in (\ref{2.1.7}). Therefore, the
\index{Condensate}condensate
$\varphi_0$ and the coherent wave $\varphi_p$ can be
called {\it classical} scalar fields. In performing calculations,
it is convenient to separate the field $\varphi$ into a classical
field (condensate) $\varphi_0$ $(\varphi_p)$ and field
excitations (scalar particles), with quantum effects being
associated only with the latter. This is formally equivalent to
the appearance of a nonvanishing vacuum average of the original
field $\varphi$, $\langle0| \varphi|0\rangle =\varphi_0$, and
reversion to the standard formalism (\ref{2.1.2}) requires that we
subtract the classical part $\varphi_0$ from the field $\varphi$;
see (\ref{1.1.12}).

The foregoing instances are not the most general. If the
condensate results from dynamic effects (minimization of a
relativistically invariant effective potential), the properties
of its constituent particles will be altered, and the condensate
itself (in contrast to (\ref{2.1.9}) and (\ref{2.1.10})) can turn out to be
relativistically invariant. This is precisely what happens in
\index{Glashow--Weinberg--Salam theory}%
theories incorporating the Glashow--Weinberg--Salam model, where
$\langle \varphi^2\rangle $ can be put in the form
\be
\label{2.1.11}
\langle \varphi^2\rangle =\frac{1}{(2\,\pi)^3}\,
\int\frac{d^3k}{2\,\sqrt{{\bf k}^2+m^2}}
+\frac{1}{(2\,\pi)^3}\,
\int\frac{d^3k}{\sqrt{{\bf k}^2}}\:n_k
\ee
with $k=\sqrt{{\bf k}^2}$, and
\be
\label{2.1.12}
n_k=(2\,\pi)^3\,\varphi^2_0\,k\,\delta({\bf k})\ .
\ee
The gist of this representation is that the constant classical
scalar field $\varphi_0$ (\ref{1.1.12}) is Lorentz invariant, and
it can therefore only form a condensate if the particles
comprising it have zero momentum and zero energy --- in other
words, zero mass (compare (\ref{2.1.11}) and (\ref{2.1.7})).

It is not obligatory that the constant classical field be
interpreted as a condensate, but this proves to be a very useful,
fruitful approach to the analysis of phase transitions in gauge
theories. There, the relativistically invariant form of the
condensate (\ref{2.1.11}), (\ref{2.1.12}) leads to a number of effects which
are lacking from solid-state theory with a condensate (\ref{2.1.9}). We
shall return to this problem in the next chapter.

Note that $n_k \gg 1$ when $\sqrt{{\bf k}^2+m^2}\ll\rm T$ for an
ultrarelativistic Bose gas\index{Bose gas}
(\ref{2.1.8}). We can therefore tentatively
divide the field $\varphi$ into a quantum part corresponding to
$\sqrt{{\bf k}^2+m^2}\ga\rm T$, and a (quasi)classical part with
$\sqrt{{\bf k}^2+m^2}\ll\rm T$. This sort of partitioning is not
very useful, however, since it is excitations with
$\sqrt{{\bf k}^2+m^2}\sim\rm T$ that make the main contribution to
most thermodynamic functions.

Much more interesting effects arise in the inflationary
universe,\linebreak[10000]
where the main contribution to $\langle \varphi^2\rangle $, to
density inhomo\-ge\-neities, and to a number of other quantities
comes precisely from long-wavelength modes with $k\ll \rm H$, for
which $n_k \gg 1$. The interpretation of these modes as
inhomogeneous classical fields $\delta \varphi $ significantly
facilitates one's understanding of a great many of the
fundamental features of the inflationary universe scenario.
Corresponding effects were discussed in Section \ref{s1.8}, and
we shall continue the discussion in Chapters \ref{c7} and
\ref{c10}.

Let us formulate a few more criteria that could help one decide
whether the field $\varphi $ is (quasi)classical. One has already
been discussed, namely the presence of modes with $n_k \gg 1$.
Another is the behavior of the correlation function of ${\rm G}(x)$
at large $x$. At large $x$, this function usually (when
there are no classical fields) falls off either exponentially or
according to a power law (as $x^{-2})$. When there really is a
condensate (\ref{2.1.9}), (\ref{2.1.11}) or a coherent wave
(\ref{2.1.10}), the
correlation function no longer decreases at large $x$ (since the
condensate is everywhere the same, i.e., the values at different
points are correlated). The onset of long-range order is thus
another criterion for the existence of a classical field in a
medium, one that has long been successfully applied in the theory
of phase transitions. As will be shown in Chapter \ref{c7}, the
corresponding correlation function in the inflationary universe
theory falls off only at exponentially large distances
$x \sim {\rm H}^{-1}\,\exp ({\rm H}\,t)$, ${\rm H}\,t \gg 1$,
which enables us
to speak of a classical field $\delta \varphi (x)$ being produced
during inflation.

Somewhat surprisingly, the classical field $\varphi $ cannot be overly
non-\linebreak[10000]
uniform (unless it is a coherent wave with a single well-defined
momentum (\ref{2.1.10})). Suppose, in fact, that in some region
of space $\nabla\varphi\sim k\,\varphi\gg m\,\varphi$. In order
for this field to be distinguishable from the quantum fluctuation
background, the field $\varphi $ must be greater than the
contribution to the rms value $\sqrt{\langle \varphi^2\rangle }$
coming from quantum fluctuations with momentum $\sim k \gg m$.
Making use of (\ref{2.1.6}), we obtain
\be
\label{2.1.13}
\varphi^{2} \ga {\rm C}\,k^{2}\ ,
\ee
where $\rm C = O(1)$, or
\be
\label{2.1.14}
(\nabla\varphi)^2\la \varphi^4\ .
\ee
In particular, this means that the initial value of the {\it classical}
scalar field $\varphi $ cannot be arbitrary; inhomogeneities in
the classical scalar field cannot exceed a certain limit.

Even more important constraints can be obtained by taking quantum
gravitation into consideration. At energy densities of the order
of the Planck density, fluctuations of the metric become so large
that one can no longer speak of classical space-time with a
classical metric
\index{Classical metric}\index{Metric!classical}$g_{\mu \nu }$
(in the same sense as one would
speak of the classical field $\varphi$). This means that it is
impossible to treat fields $\varphi $ as being classical unless
\ba
\partial_\mu\varphi\,\partial^\mu\varphi&\la& {\rm M_P^4}\ , \quad
\mu=0,1,2,3\ ,\\
\label{2.1.16}
{\rm V}(\varphi)&\la& {\rm M_P^4}\ .
\ea
We made essential use of these constraints in discussing initial
conditions in the inflationary universe in Section
\ref{s1.7}.\index{Classical scalar fields|)}%
\index{Scalar fields!classical|)}

\section[Quantum corrections to the effective potential]%
{\label{s2.2}Quantum corrections to the effective potential
 V($\varphi$)}
\index{Effective potential!quantum corrections to|(}%
\index{Quantum corrections to effective potential|(}%
\index{Effective potential|(}%

In Section \ref{s1.1}, we investigated the theory of symmetry
breaking in the simplest quantum field theoretical models,
neglecting quantum corrections to the effective potential of the
scalar field $\varphi $. Nevertheless, in some cases quantum
corrections to ${\rm V}(\varphi)$ are substantial.

\begin{figure}[t]\label{f12}
\centering \leavevmode\epsfysize=5.5cm \epsfbox{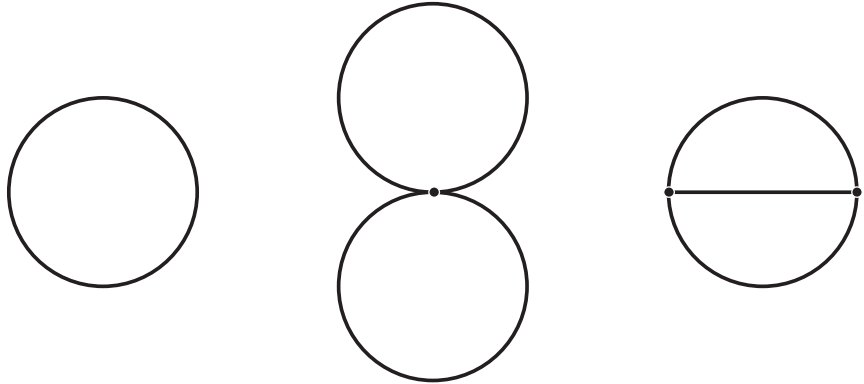}
\caption{One- and two-loop diagrams for ${\rm V}(\varphi )$ in the
theory of the scalar field (1.1.5).}
\end{figure}

According to [\cite{137}, \cite{138}], quantum corrections to the
classical expression for the effective potential are given by a
set of all one-particle irreducible vacuum diagrams (diagrams
that do not dissociate into two when a single line is cut) in a
theory with the Lagrangian ${\rm L}(\varphi  + \varphi_{0})$
without the terms linear in $\varphi $.  Corresponding diagrams
with one, two, or more loops for the theory (\ref{1.1.5}) have
been drawn in Fig.~\ref{f12}. In the present case, expansion in
the number of loops corresponds to expansion in the small
coupling constant $\lambda $. In the one-loop approximation
\index{One-loop approximation}%
(taking only the first diagram of Fig.~\ref{f12} into account),
\be
\label{2.2.1}
{\rm V}(\varphi)=-\frac{\mu^2}{2}\,\varphi^2+\frac{\lambda}{4}\,\varphi^4+
\frac{1}{2\,(2\,\pi)^4}\,\int d^4k\:\ln\left[k^2+m^2(\varphi)\right]\ .
\ee
Here $k^{2} = k_4^2+{\bf k}^2$ (i.e., we have carried out a Wick
\index{Wick rotation}%
rotation $k_{0} \rightarrow -i\,k_{4}$ and integrated over
Euclidean momentum space), and the effective mass squared of the
field $\varphi$ is
\be
\label{2.2.2}
m^{2}(\varphi ) = 3\,\lambda\,\varphi^{2} - \mu^{2}\ .
\ee
As before, we have omitted the subscript 0 from the classical
field $\varphi $ in Eqs. (\ref{2.2.1}) and (\ref{2.2.2}).
The integral in (\ref{2.2.1})
diverges at large $k$. To supplement the definition given by
(\ref{2.2.1}), it is necessary to renormalize the wave function, mass,
coupling constant, and vacuum energy [\cite{2}, \cite{8}, \cite{9}].
To do so, we may
add to ${\rm L}(\varphi  + \varphi_{0})$ of (\ref{1.1.5}) the counterterms
${\rm C}_1\,\partial_\mu(\varphi  + \varphi_{0})\,
\partial^\mu(\varphi  + \varphi_{0})$,
${\rm C}_2\,(\varphi  + \varphi_{0})^2$,
${\rm C}_3\,(\varphi  + \varphi_{0})^4$ and ${\rm C}_{4}$.

The meaning of (\ref{2.2.1}) becomes particularly clear after
integrating over $k_{4}$. The result (up to an infinite constant that
is eliminated by renormalization of the vacuum energy, i.e., by
the addition of ${\rm C}_{4}$ to ${\rm L}(\varphi  + \varphi_{0})$), is
\be \label{2.2.3} {\rm
V}(\varphi)=-\frac{\mu^2}{2}\,\varphi^2+\frac{\lambda}{4}\,\varphi^4+
\frac{1}{2(2\,\pi)^3}\,\int d^3k\:\sqrt{k^2+m^2(\varphi)}\ . \ee
\index{One-loop approximation}%
Thus, in the one-loop approximation, the effective potential
${\rm V}(\varphi )$ is given by the sum of the classical
expression for the potential energy of the field $\varphi $ and
a $\varphi $-dependent vacuum energy shift due to quantum
fluctuations of the field $\varphi $. To determine the quantities
${\rm C}_{i}$, normalization conditions must be imposed on the
potential, and these, for example, can be chosen to be [\cite{139}]
\ba
\label{2.2.4}
\frac{d{\rm V}}{ d\varphi}\biggl|_{\varphi=\mu/\sqrt{\lambda}}
&=&0\ ,\nonumber \\
\frac{d^2{\rm V}}{d\varphi^2}\biggl|_{\varphi=\mu/\sqrt{\lambda}}
&=&2\,\mu^2\ .
\ea
These normalization conditions are chosen to ensure that the
location of the minimum of ${\rm V}(\varphi )$ for
$\varphi  = \mu/\sqrt{\lambda}$ and the curvature of
${\rm V}(\varphi )$ at
the minimum (which is the same to lowest order in $\lambda $ as
the mass squared of the scalar field $\varphi )$ remain the same
as in the classical theory. Other types of normalization
conditions also exist; for example, the Coleman--Weinberg
conditions\index{Coleman--Weinberg conditions} [\cite{137}] are
\ba
\label{2.2.5}
\frac{d^2{\rm V}}{d\varphi^2}\biggl|_{\varphi=0}&=&m^2\ ,\nonumber \\
\frac{d^4{\rm V}}{d\varphi^4}\biggl|_{\varphi={\rm M}}&=&\lambda\ ,
\ea
where ${\rm M}$ is some normalization point. All physical results
obtained via the normalization conditions (\ref{2.2.4}) and
(\ref{2.2.5}) are equivalent, after one establishes the
appropriate correspondence between the parameters $\mu$, $m$, M,
and $\lambda $ in the renormalized expressions for ${\rm
V}(\varphi )$ in the two cases. The conditions (\ref{2.2.4}) are
usually more convenient for practical purposes in work with
theories that have spontaneous symmetry breaking, although
(\ref{2.2.5}) is sometimes the most suitable approach in certain
instances involving the study of the fundamental features of the
theory, since the first condition determines the mass squared of
the scalar field prior to symmetry breaking. Since we are most
interested in the present section in the properties of
${\rm V}(\varphi )$ for certain values of
$\displaystyle m^{2}(\varphi ) =\frac{d^2{\rm V}}{d\varphi^2} $
at the minimum of ${\rm V}(\varphi )$,
we will use the conditions (\ref{2.2.4}).  The
effective potential $V(\varphi )$ then takes the form [\cite{23}]
\ba
\label{2.2.6}
{\rm V}(\varphi)&=&-\frac{\mu^2}{2}\,\varphi^2+\frac{\lambda}{4}\,\varphi^4+
\frac{(3\,\lambda\,\varphi^2-\mu^2)^2}{64\,\pi^2}\,
\ln\left(\frac{3\,\lambda\,\varphi^2-\mu^2}{2\,\mu^2}\right)\nonumber \\
&+&\frac{21\,\lambda\,\mu^2}{64\,\pi^2}\,\varphi^2-
\frac{27\,\lambda^2}{128\,\pi^2}\,\varphi^4\ .
\ea

\begin{figure}[t]\label{f13}
\centering \leavevmode\epsfysize=3.1cm \epsfbox{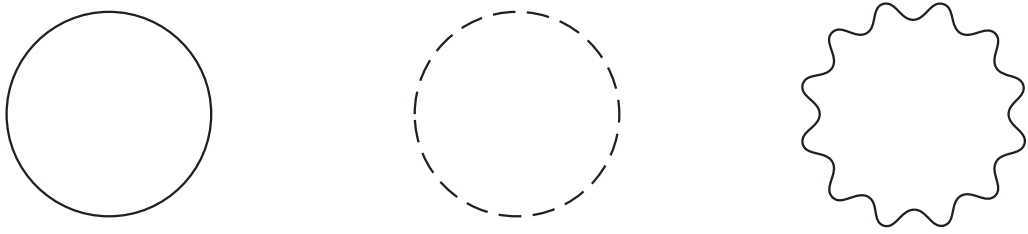}
\caption{Diagrams for ${\rm V}(\varphi )$ in the Higgs
\index{Higgs model}%
model. The solid, dashed, and wavy lines correspond to the
$\chi_{1}, \chi_{2}$, and ${\rm A}_{\mu }$ fields, respectively.}
\end{figure}
Clearly, for $\lambda\ll 1$, quantum corrections only become
important for asymptotically large $\varphi$  (when
$\lambda\,\ln(\varphi/\mu)\gg 1$), where it becomes necessary to
take account of all higher-order corrections. When $\lambda>0$,
it becomes extremely difficult to sum all higher-order
corrections to the expression for ${\rm V}(\varphi )$ at large
$\varphi $. This problem can only be solved for a special class
of $\lambda\,\varphi^{4}$ theories discussed in the next section.

We can make much more progress in clarifying the role of quantum
corrections in theories with several different coupling
\index{Higgs model}%
constants. As an example, let us consider the Higgs model
(\ref{1.1.15}) in the transverse gauge $\partial_\mu{\rm A}_\mu$.
\index{One-loop approximation}%
In the one-loop approximation, the effective potential in that
case is given by the diagrams in Fig.~\ref{f13}.

For $e^{2} \ll \lambda$, the contribution of vector particles can
be neglected, and the situation is analogous to the one described
above. When $e^{2}\gg \lambda$, we can ignore the contribution of
scalar particles. In that event, the expression for ${\rm V}(\varphi )$
takes the form [\cite{139}]
\ba
\label{2.2.7}
{\rm V}(\varphi)&=&-\frac{\mu^2\,\varphi^2}{2}\,
\left(1-\frac{3\,e^4}{16\,\pi^2\,\lambda}\right)+
\frac{\lambda\,\varphi^4}{4}\,
\left(1-\frac{9\,e^4}{32\,\pi^2\,\lambda}\right)\nonumber \\
&+&\frac{3\,e^4\,\varphi^4}{64\,\pi^2}\,
\ln\left(\frac{\lambda\,\varphi^2}{\mu^2}\right)\ .
\ea
Clearly, then, when $\displaystyle \lambda<\frac{3\,e^4}{16\,\pi^2}$,
the effective potential acquires an additional minimum at
$\varphi  = 0$, and when
$\displaystyle \lambda<\frac{3\,e^4}{32\,\pi^2}$, this
minimum becomes even deeper than the usual minimum at $\varphi  =
\varphi_{0} = \frac{\mu}{\sqrt{\lambda}}$; see Fig.~2.3.
\begin{figure}[t]\label{f14}
\centering \leavevmode\epsfysize=8cm \epsfbox{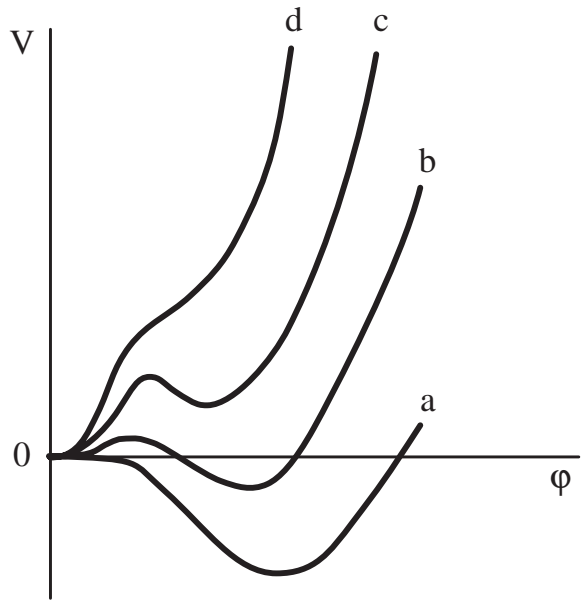}
\caption{Effective potential in the Higgs model. a)
\index{Higgs model}%
$\displaystyle \lambda>\frac{3\,e^4}{16\,\pi^2}$; b)
$\displaystyle
\frac{3\,e^4}{16\,\pi^2}>\lambda>\frac{3\,e^4}{32\,\pi^2}$; c)
$\displaystyle \frac{3\,e^4}{32\,\pi^2}>\lambda>0$; d) $\lambda  =
0$.}
\end{figure}

Hence, when $\displaystyle \lambda<\frac{3\,e^4}{16\,\pi^2}$,
symmetry breaking in the Higgs model becomes energetically
\index{Higgs model!symmetry breaking in}%
\index{Symmetry breaking!in Higgs model}%
unfavorable. This effect is due not to large logarithmic factors
like $\displaystyle \lambda\,\ln\frac{\varphi}{\mu}\ga 1$, but to
special relations between $\lambda $ and $e^{2}$ ($\lambda  \sim e^{4}$),
whereby the classical terms in the expression for the
effective potential (\ref{2.2.7}) become of the same order as the
quantum corrections to order $e^{2}$. Higher-order corrections to
(\ref{2.2.7}) are proportional to $\lambda^{2}$ and $e^{6}$, and
do not lead to any substantial modification of the form of
${\rm V}(\varphi)$ given by (\ref{2.2.7}), over the range
$\varphi\la\mu/\sqrt{\lambda}$ in which we are most interested.

We remark here that
$\displaystyle m^2_{\rm A}=e^2\,\varphi_0^2=\frac{e^2\,\mu^2}{\lambda}$,
$m_\varphi^2=2\,\lambda\,\varphi_0^2$
up to higher-order corrections in $e^{2}$. This
\index{Higgs model}%
means that symmetry breaking is only favorable in the Higgs model
if
\be
\label{2.2.8}
m_\varphi^2>\frac{3\,e^4}{16\,\pi^2}\,m_{\rm A}^2\ .
\ee
The significance of this result for the Glashow--Weinberg--Salam
\index{Glashow--Weinberg--Salam theory}%
\index{Higgs boson}%
model is that the mass of the Higgs boson in that theory (more
precisely, in the standard version with one kind of Higgs boson
and no superheavy fermions, and with
$\sin^{2}\theta_{\rm W} \sim 0.23$)
should be more than approximately 7~GeV [\cite{139}, \cite{140}],
\be
\label{2.2.9}
m_\varphi\ga 7\;{\rm GeV}\ .
\ee
From Eq. (\ref{2.2.7}), we also obtain bounds on the coupling constant
\index{Higgs boson}%
between Higgs bosons,
$\displaystyle \lambda\, (\varphi  = \varphi _{0}) = \frac{1}{6}\,
\frac{d^4{\rm V}}{d\varphi^4}\biggl|_{\varphi=\varphi_0}$ [\cite{139}].
In fact, $\lambda>0$, and
\be
\label{2.2.10}
\lambda(\varphi_0)=\lambda+\frac{e^4}{2\,\pi^2}\ ,
\ee
and this means that $V(\varphi )$ has a minimum at $\varphi_0  \ne 0$ if
\be
\label{2.2.11}
\lambda(\varphi_0)>\frac{11\,e^4}{16\,\pi^2}\ .
\ee
and the minimum at $\varphi  = \varphi_{0}$ is deeper than the
one at $\varphi  = 0$ if
\be
\label{2.2.12}
\lambda(\varphi_0)>\frac{19\,e^4}{32\,\pi^2}\ ,
\ee
A bound like (\ref{2.2.12}) in the Weinberg--Salam model yields
\be
\label{2.2.13}
\lambda(\varphi_0)\ga 3\cdot10^{-3}\ .
\ee

With cosmological considerations taken into account, the
cor\-res\-pon\-ding bound can be improved somewhat. As we have already
said in the Introduction, symmetry was restored in the early
universe at ${\rm T}\ga 10^2$~GeV in Glashow--Weinberg--Salam
\index{Glashow--Weinberg--Salam theory}%
theory, and  the only minimum of ${\rm V}(\varphi ,{\rm T})$ was
the one at $\varphi  = 0$.  A minimum appears at $\varphi  \ne 0$
only as the universe cools, and if the effective potential then
continues to have a minimum at $\varphi  = 0$, it is not clear
{\it a priori} whether the field $\varphi $ will be able to jump
out of the local minimum at $\varphi  = 0$ to a global minimum at
$\varphi  = \varphi_{0} \sim 250$~GeV, nor is it clear what the
properties of the universe would be after such a phase
transition. By making use of high-temperature tunneling theory
[\cite{62}], it has been shown that this transition has an
exceedingly low probability of occurrence in the
\index{Glashow--Weinberg--Salam theory}%
Glashow--Weinberg--Salam model. The phase transition can
therefore only take place if the minimum of ${\rm V}(\varphi )$
at $\varphi  = 0$ is very shallow,
$\displaystyle \frac{d^2{\rm V}}{d\varphi^2}\biggl|_{\varphi=0}\ll \mu^2$.
This then leads to a somewhat more rigorous bound on the mass of the
\index{Higgs boson}%
Higgs boson [\cite{141}--\cite{144}],
\be
\label{2.2.14}
m_\varphi\ga 10\;{\rm GeV}\ .
\ee

One particular case which is especially interesting from the
standpoint of cosmology (as well as from the standpoint of
elementary particle theory) is that in which
$\displaystyle \frac{d^2{\rm V}}{d\varphi^2}\biggl|_{\varphi=0}=0$.
This is known as the\index{Coleman--Weinberg theory}
Coleman--Weinberg theory [\cite{137}]. The effective
\index{Higgs model}%
potential in this theory, which is based on the Higgs model
(\ref{1.1.15}), takes the form
\be
\label{2.2.15}
{\rm V}(\varphi)=\frac{25\,e^4}{128\,\pi^2}\,
\left(\varphi^4\,\ln\frac{\varphi}{\varphi_0}-\frac{\varphi^4}{4}+
\frac{\varphi_0^4}{4}\right)\ .
\ee
We have added the term
$\displaystyle \frac{25\,e^4}{512\,\pi^2}\,\varphi_0^4$
here in order to ensure that ${\rm V}(\varphi_{0}) = 0$.  In the SU(5)
model, the corresponding effective potential takes the form
\be
\label{2.2.16}
{\rm V}(\varphi)=\frac{25\,g^4}{128\,\pi^2}\,
\left(\ln\frac{\varphi}{\varphi_0}-\frac{1}{4}\right)
+\frac{9}{32\,\pi^2}\,{\rm M}^4_{\rm X}\ ,
\ee
where $g^{2}$ is the SU(5) gauge coupling constant,
${\rm M}_{\rm X}$ is the mass of the X boson, and $\varphi $ is defined by
Eq.~(\ref{1.1.19}).  Equation (\ref{2.2.16}) lay at the
foundation of the first version of the new inflationary universe
scenario, so we will have a number of occasions to return to it.

\begin{figure}[t]\label{f15}
\centering \leavevmode\epsfysize=6cm \epsfbox{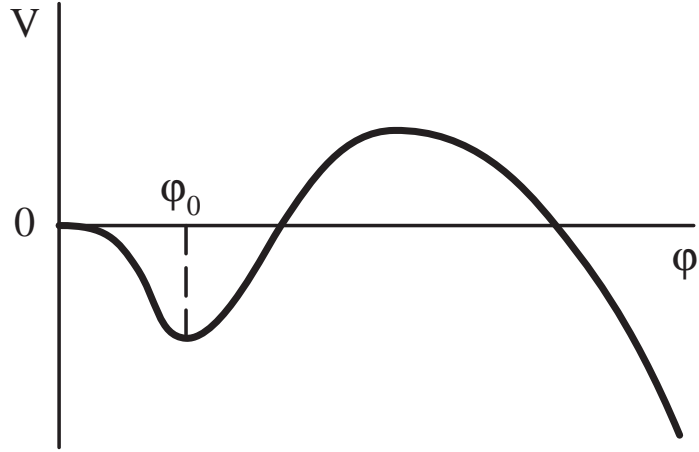}
\caption{Effective potential in the theory (1.1.13) with
$m_\psi\gg m_\varphi$.}
\end{figure}
Whereas quantum fluctuations of vector fields stimulate the
dynamical restoration of symmetry, quantum fluctuations of
fermions enhance symmetry breaking. We now consider the
simplified $\sigma $-model (\ref{1.1.13}) as an example. At large $\varphi$,
the effective potential in this theory is given by [\cite{145}]
\be \label{2.2.17} {\rm
V}(\varphi)=-\frac{\mu^2}{2}\,\varphi^2+\frac{\lambda}{4}\,\varphi^4+
\frac{9\,\lambda^2-4\,h^4}{64\,\pi^2}\,\varphi^4\,
\ln\left(\frac{\lambda\,\varphi^2}{\mu^2}\right)\ . \ee Fermions
evidently make a negative contribution at large $\varphi $, and
when $3\,\lambda < 2\,h^{2}$, the effective potential ${\rm
V}(\varphi )$ is unbounded from below (Fig. 2.4).

Of course when $\varphi\rightarrow\infty$, the one-loop
approximation is longer applicable. However, if $\lambda  \ll h^{2}$,
there is a range of values of the field $\varphi $
($\displaystyle\varphi^2\sim\frac{\mu^2}{\lambda}\,\exp\frac{\lambda}{h^4}$)
for which
$\displaystyle {\rm V}(\varphi)<
{\rm V}\left(\frac{\mu}{\sqrt{\lambda}}\right)$, and the one-loop
approximation still gives reliable results. Thus, in the $\sigma$-model
with $\lambda\ll h^2$, or what is the same thing, with
$m_\varphi\ll m_\psi$, the state
$\displaystyle \varphi=\frac{\mu}{\sqrt{\lambda}}$ is unstable, and strong
dynamical symmetry breaking takes place.

\begin{figure}[t]\label{f16}
\centering \leavevmode\epsfysize=7cm \epsfbox{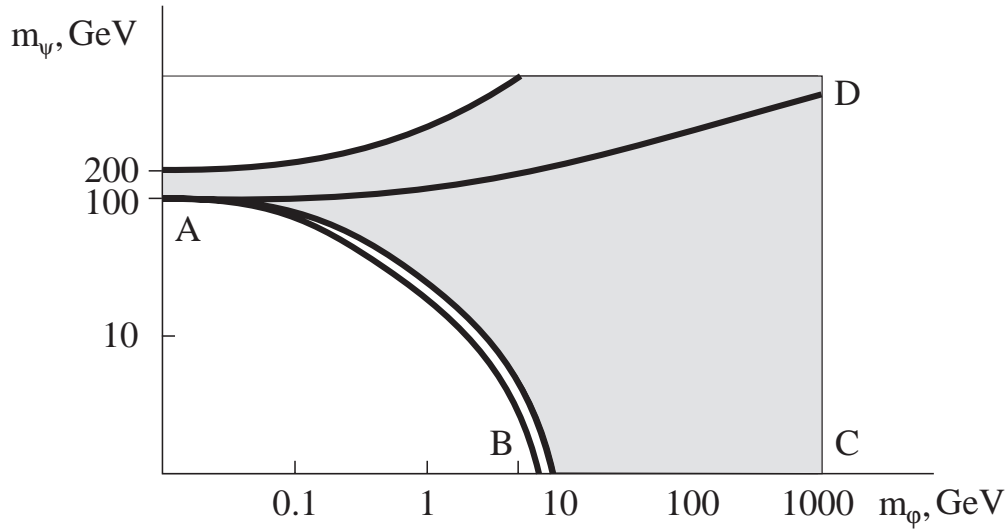} \caption{The
hatched region corresponds to allowable mass
\index{Higgs boson}%
values for the Higgs boson, $m_\varphi$, and heavy fermions,
$m_\psi$ (or more precisely, $\displaystyle
\sum_i(m_{\psi_i}^4)^{1/4}$) when one takes account of both
cosmological considerations and quantum corrections to the
effective potential in the
\index{Glashow--Weinberg--Salam theory}%
Glashow--Weinberg--Salam model. The area bounded by the curve ABCD
is the region of absolute phase stability with spontaneous
symmetry breaking, $\varphi=\displaystyle
\frac{\mu}{\sqrt{\lambda}}$.}
\end{figure}

We can readily generalize this result to a wider class of
\index{Glashow--Weinberg--Salam theory}%
theories, including the Glashow--Weinberg--Salam theory, which
leads to a set of constraints on the mass of the Higgs meson and
the fermion masses in this theory [\cite{139}--\cite{151}];
see Fig.~2.5. We shall take advantage of cosmological
considerations in Chapter \ref{c6} to strengthen these
\index{Effective potential!quantum corrections to|)}%
\index{Quantum corrections to effective potential|)}%
constraints.

\section[The $1/{\rm N}$
expansion and the effective potential]%
{\label{s2.3}The
\ $1/{\rm N}$ expansion and the effective potential
in the\protect\\
\index{O(N) symmetric theory|(}%
$\lambda \varphi^{4}/\bf N$
theory}

As a rule, it is not possible to study the behavior of the
effective potential in standard perturbation theory as
$\varphi\rightarrow\infty$, but theories that are asymptotically
free in all coupling constants constitute an important exception.
For example, it can be shown that in a massless $\lambda\,\varphi^{4}$
theory with negative $\lambda$, ${\rm V}(\varphi )$
decreases without bound as $\varphi\rightarrow\infty$ both in the
classical approximation and when quantum corrections are taken
into consideration [\cite{137}, \cite{152}]. It is difficult to
use standard perturbation theory in $\lambda $ to investigate the
behavior of ${\rm V}(\varphi )$ as $\varphi\rightarrow\infty$ in
the $\lambda\,\varphi^{4}$ theory with $\lambda  > 0$.  There does
exist a class of theories, however, in which one can make
substantial progress toward understanding the properties of
${\rm V}(\varphi )$ for both small and large $\varphi $, bringing with
it a number of surprising results.

Consider the O(N) symmetric theory of the scalar field
$\Phi=\{\Phi_1,$\linebreak[1000]
$\ldots,\Phi_{\rm N}\}$, with the Lagrangian
\be
\label{2.3.1}
\rm L=\frac{1}{2}\,(\partial_\mu\Phi)^2-\frac{\mu^2}{2}\,\Phi^2-
\frac{\lambda}{\rm 4!N}\,\left(\Phi^2\right)^2\ ,
\ee
where $\displaystyle \Phi^2=\sum_i\Phi^2_i$. The field $\Phi $
may have a classical part $\Phi_0=\sqrt{\rm N}\,\{\varphi,0,\ldots,0\}$.
Let us also introduce the composite field

\be
\label{2.3.2}
\hat\chi=\mu^2+\frac{\lambda}{6\,\rm N}\,\Phi^2
\ee
with a classical part $\chi $, and let us add to (\ref{2.3.1}) the term
\be
\label{2.3.3}
\Delta{\rm L}=\frac{3\,{\rm N}}{2\,\lambda}
\left(\hat\chi-\mu^2-\frac{\lambda}{6\,{\rm N}}\,\Phi^2\right)^2\ ,
\ee
so that
\be \label{2.3.4} \rm L'=L+\Delta
L=L=\frac{1}{2}\,(\partial_\mu\Phi)^2-
\frac{3\,N}{\lambda}\,\mu^2\,\hat\chi
+\frac{3\,N}{\lambda}\,{\hat\chi}^2-\frac{1}{2}\,\hat\chi\,\Phi^2\
. \ee The theory described by (\ref{2.3.4}) is equivalent to the
theory (\ref{2.3.1}), since the Lagrange equation for the field
$\hat\chi$ in the theory (\ref{2.3.4}) is exactly (\ref{2.3.2}),
while the Lagrange equation for the field $\Phi $ in the theory
(\ref{2.3.4}), taking (\ref{2.3.2}) into account, gives the
Lagrange equation for the field $\Phi $ in the theory
(\ref{2.3.1}) [\cite{153}]. In the one-loop approximation, the
effective potential ${\rm V}(\varphi ,\chi )\equiv {\rm N}\,{\bf
V}(\varphi, \chi)$ which corresponds to the theory (\ref{2.3.4})
is given by [\cite{154}]
\ba \label{2.3.5} {\bf V}(\varphi,\chi)&=&-\frac{3}{2}\,
\left(\frac{1}{\lambda}+\frac{1}{96\,\pi^2}\right)\,
\chi\,(\chi-2\,\mu^2)+\frac{1}{2}\,\chi\,\varphi^2\nonumber \\
&+&\frac{\chi^2}{128\,\pi^2}\,\left(2\,\ln\frac{\chi}{\rm M^2}-1\right)\ ,
\ea
where M is the normalization parameter, and
\be \label{2.3.6}
(\chi-\mu^2)\,\left(\frac{1}{\lambda}+\frac{1}{96\,\pi^2}\right)=
\frac{\varphi^2}{6}+\frac{\chi}{96\,\pi^2}\,\ln\frac{\chi}{\rm
M^2}\ . \ee The effective potential ${\rm V}(\varphi)\equiv {\rm
N}\,{\bf V}(\varphi)$ in the original theory (\ref{2.3.1}) is
equal to ${\rm V}(\varphi ,\chi (\varphi ))$.  It is important to
note that all of the higher-order corrections to Eqs.
(\ref{2.3.5}) and (\ref{2.3.6}) contain higher powers of $1/{\rm
N}$, and vanish in the limit as $\rm N\rightarrow\infty$.  In that
sense, Eqs. (\ref{2.3.5}) and (\ref{2.3.6}) are {\it exact} in the
limit $\rm N\rightarrow\infty$.

We now impose the following normalization conditions on $\mu^{2}$
and $\lambda $ in (\ref{2.3.5}) and (\ref{2.3.6}):
\ba
\label{2.3.7}
{\rm Re}\frac{d^2{\bf V}}{d\varphi^2}\biggl|_{\varphi=0}&=&\mu^2\ ,\\
\label{2.3.8} {\rm Re}\frac{d^4{\bf
V}}{d\varphi^4}\biggl|_{\varphi=0}&=&\lambda\ . \ea This then
tells us that after renormalization, the parameter ${\rm M}^{2}$
in (\ref{2.3.5}) should be put equal to $\mu^{2}$.

The signs of $\mu^{2}$ and $\lambda $ in (\ref{2.3.7}) and (\ref{2.3.8}) are
arbitrary. For simplicity, we will consider the case in which
$\mu^2>0$, $\lambda>0$.  The field $\chi $ is found to be a
double-valued function of $\varphi $ when $\varphi<\bar\varphi$,
where
\be \label{2.3.9}
1-\frac{\lambda}{96\,\pi^2}\,\ln\frac{\chi(\bar\varphi)}{\mu^2}=0\
. \ee As a result, for $\varphi<\bar\varphi$, the effective
potential ${\bf V}(\varphi)$ turns out to be a double-valued
function of $\varphi$  (with branches ${\bf V}^{\rm I}(\varphi)$
and ${\bf V}^{\rm II}(\varphi)$, ${\bf V}^{\rm I}>{\bf V}^{\rm
II}$; see Fig.~2.6) [\cite{154}]. The normalization
conditions (\ref{2.3.7}) and (\ref{2.3.8}) hold on the upper
branch of ${\bf V}(\varphi)$.

\begin{figure}[t]\label{f17}
\centering \leavevmode\epsfysize=5cm \epsfbox{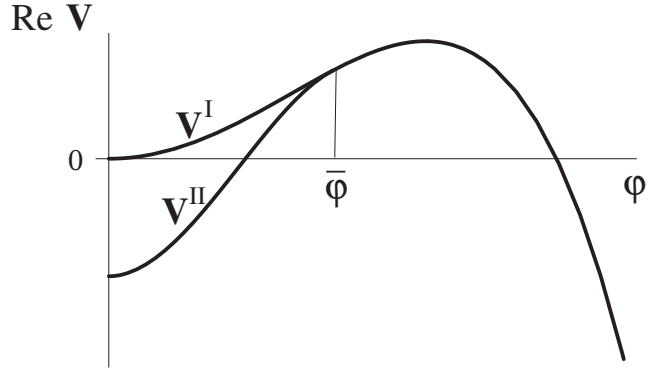}
\caption{Effective potential in the theory (\ref{2.3.1}) with
$\mu^{2} > 0$.}
\end{figure}
On the branch ${\bf V}^{\rm II}$, the field $\chi $ is extremely
large ($\displaystyle
\frac{\lambda}{96\,\pi^2}\,\ln\frac{\chi}{\mu^2}>1$), and one may
well ask whether Eqs.~(\ref{2.3.5}) and (\ref{2.3.6}) are actually
valid for such large $\chi $ and for any large but finite N. The
answer to this question is affirmative, since on the branch ${\bf
V}^{\rm II}$, $\chi $ is large in magnitude but finite, and is
independent of N. For any arbitrarily large $\chi$, there should
therefore exist an N such that corrections $\sim {\rm O}(1/{\rm
N})$ to Eqs.~(\ref{2.3.5}) and (\ref{2.3.6}) for this $\chi $ are
small [\cite{155}].

When $\varphi  = 0$, as was shown in [\cite{153}] to lowest order
in $1/{\rm N}$, the Green's function ${\rm G}_{\chi \chi }(k^{2})$
of the field $\chi $ on the upper branch ${\bf V}^{\rm I}$ has a
\index{Tachyon}%
tachyon pole at $k^{2} = -\mu^{2}\,e^{1/\lambda}$. Using the same
arguments as above, it can be shown that higher-order corrections
in $1/{\rm N}$ to ${\rm G}_{\chi \chi }(k^{2})$ can change the type of
singularity at $k^{2} < 0$, but they cannot alter the fact that
${\rm G}_{\chi \chi }(k^{2})$ changes sign at $k^{2} < 0$.  Such
behavior of ${\rm G}_{\chi \chi }(k^{2})$ is incompatible with the
\index{K\"all\'en--Lehmann theorem}%
K\"all\'en--Lehmann theorem, and indicates that the theory is
unstable against production of the classical field $\chi $, the
reason simply being that on the branch ${\bf V}^{\rm I}$, the
point $\varphi  = 0$ is not a minimum but a saddle point of the
potential ${\bf V}(\varphi,\chi)$, and a transition takes place to
the minimum at $\varphi  = 0$ on the branch ${\bf V}^{\rm II}$.

However, even this point is not an absolute minimum of ${\bf
V}(\varphi)$. In fact, according to (\ref{2.3.5}) and
(\ref{2.3.6}),

\be \label{2.3.10} {\bf V}(\varphi)=-4\pi^2\,
\frac{\varphi^4}{\displaystyle\ln\frac{\varphi^2}{\mu^2}}
\left(1+\frac{i\,\pi}{\displaystyle\ln\frac{\varphi^2}{\mu^2}}\right)
\ee as $\varphi\rightarrow\infty$. This means that the potential
${\bf V}(\varphi)$ is not bounded from below, and the theory
(\ref{2.3.1}) is unstable against production of arbitrarily large
fields $\varphi$  [\cite{155}].

A number of objections can be raised to this conclusion, the
principal one being the following. Equation (\ref{2.3.10}) holds
when ${\rm N} = \infty$, but for any finite N there might exist a
field $\varphi  = \varphi_{\rm N}$ so large that when $\varphi  >
\varphi_{\rm N}$ the expression (\ref{2.3.10}) becomes unreliable;
an absolute minimum of ${\bf V}(\varphi)$ might thus exist for
$\varphi  > \varphi_{\rm N}$.

\begin{figure}[t]\label{f18}
\centering \leavevmode\epsfysize=5cm \epsfbox{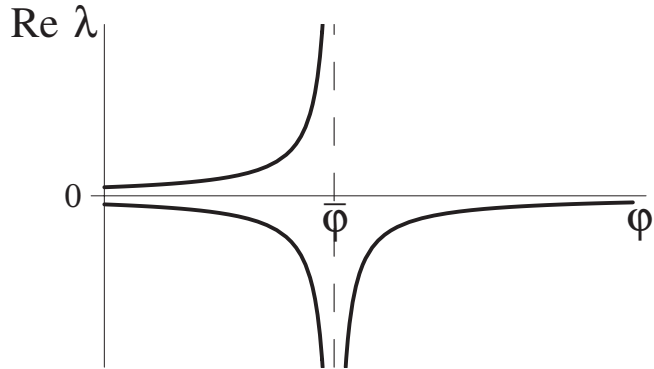}
\caption{Effective coupling constant $\lambda (\varphi )$ in the
theory (\ref{2.3.1}).}
\end{figure}
One response to this objection can be found by combining the
$1/{\rm N}$ expansion and the renormalization group equation
[\cite{155}]. In order to do so, we first note that the magnitude
of the effective coupling constant $\displaystyle
\lambda(\varphi)=\frac{d^4{\rm V}}{d\varphi^4}$ which can be
calculated using (\ref{2.3.5}) and (\ref{2.3.6}), behaves as
shown in Fig.~2.7. This then leads to several consequences:

a) for large enough N, a $\displaystyle \frac{\lambda}{\rm N}\,\varphi^4$
theory with $\lambda  > 0$ is equivalent to a theory with $\lambda<0$,
representing merely another branch of the same theory;

b) contrary to the usual expectations, a
$\displaystyle\frac{\lambda}{\rm N}\,\varphi^4$ theory with
$\lambda>0$ is unstable, while a theory with $\lambda<0$ is
metastable for small $\varphi $;

c) for large enough $\varphi$, ${\rm Re}\, \lambda$ becomes
negative, and tends to zero with increasing $\varphi $.

The last of these is the decisive point. We may choose such a
large value $\varphi  = \varphi_{1}$ that $\lambda $ is in fact
small and negative, and such a large value of ${\rm
N}(\varphi_{1})$ that the higher-order corrections in powers of
$1/{\rm N}$ to the value of $\lambda (\varphi )$ at
$\varphi\sim\varphi_1$ are also small. We can then make use of the
renormalization group equation to continue the quantity $\lambda
(\varphi )$ from $\varphi  = \varphi_{1}$ to
$\varphi\rightarrow\infty$, since the $\displaystyle
\frac{\lambda}{\rm N}\,\varphi^4$ theory is asymptotically free
when $\lambda  < 0$. We must then integrate $\lambda (\varphi )$
with respect to $\varphi $ and obtain the value of ${\bf
V}(\varphi)$.  These calculations result in a value for ${\bf
V}(\varphi)$ identical to that obtained from (\ref{2.3.10}), and
thereby confirm that the effective potential in this theory is
actually unbounded from below for large $\varphi$  [\cite{155}].

This conclusion turns out to be valid regardless of the sign of
$\mu^{2}$ and $\lambda $ at $\varphi  = 0$. Interestingly enough,
spontaneous symmetry breaking, which ought to occur in the theory
(\ref{2.3.1}) when $\mu^2<0$, actually takes place only on the
upper (unstable) branch of ${\bf V}(\varphi)$; on the lower
(metastable) branch, the effective mass squared of the field
$\varphi $ is always positive, and symmetry breaking does not
occur [\cite{154}].

These results are fairly surprising, and in many respects they
are quite instructive. Quantum corrections are found to lead to
instability even in theories where this might be least expected,
such as (\ref{2.3.1}) with $\mu^{2} > 0$ and $\lambda  > 0$. It
turns out that for large N, there is no spontaneous symmetry
breaking in this theory when $\mu^{2} < 0$; it has also been
found that in the theory (\ref{2.3.1}), $\lambda < 0$ and $\lambda>0$
actually represent two branches of a single theory.  These
branches coalesce at exponentially large values of $\varphi $,
and in the limit of very large $\varphi $, the effective constant
$\lambda (\varphi )$ becomes negative and tends to zero from
below. The latter result, however, is not so very surprising, as
that is just how the effective constant $\lambda $ ought to
behave for large fields and large momenta, according to a study
based on the renormalization group equation (see [\cite{58}], for
example). This sort of pathological behavior of the effective
coupling constant $\lambda $ also lay at the basis of the
\index{Zero-charge problem}%
so-called zero-charge problem [\cite{156}, \cite{157}]. For a long
time, the corresponding results were viewed as being rather
unreliable, and it was considered plausible that in many realistic
situations the zero-charge problem actually does not appear; for
example, see [\cite{158}]. On the other hand, the principal
objections to the reliability of the results presented in
[\cite{156}, \cite{157}] do not seem to apply to derivations based
on the $1/{\rm N}$ expansion [\cite{155}, \cite{159}]. More
recently, the existence of the zero-charge problem in the theory
$\lambda\, \varphi^{4}$ has been sufficiently well proven, both
analytically [\cite{160}] and numerically [\cite{161}]
(``triviality'' of the theory $\lambda\,\varphi^{4})$.

The foregoing analysis aids in an understanding of the essence of
this problem using the theory (\ref{2.3.1}) as an example:
according to our results, for large N, no theory of the form
(\ref{2.3.1}) possesses both a stable vacuum and a nonvanishing
coupling constant $\lambda $.

One question that then emerges is whether this result has any
bearing on realistic elementary particle theories with spontaneous
symmetry breaking. First of all, then, let us analyze just how
serious the shortcomings of the theory (\ref{2.3.1}) actually are.
At first glance, the presence of a pole at $k^{2} = -\mu
^{2}\,e^{1/\lambda}$ on the upper branch of ${\bf V}(\varphi)$
does not seem so terrible, since it is usually taken to mean that
the low-energy physics does not ``feel'' the structure of the
theory at superhigh momenta and masses. This is actually so at
large $k^{2} > 0$. But the example of the theory (\ref{1.1.5})
with symmetry breaking demonstrates that the
\index{Tachyon}%
presence of a tachyon pole at $k^{2} = -\mu^{2} < 0$ leads to more
rapid development of an instability than would a large tachyon
mass; see (\ref{1.1.6}). The upper branch of the potential ${\bf
V}(\varphi)$ therefore actually corresponds to an unstable vacuum
state (an analogous instability also occurs in a multicomponent
formulation of quantum electrodynamics at sufficiently large N
[\cite{159}, \cite{162}]). On the other hand, when $\lambda \ll
1$, the life-time of the universe at the point $\varphi  = 0$ on
the lower branch turns out to be exponentially large, so the
putative instability of the vacuum in this theory in no way
implies that it cannot correctly describe our universe. One
possible problem here is that at a temperature ${\rm T}\ga \mu\,
e^{1/\lambda}$, the local minimum at $\varphi  = 0$ on the branch
${\bf V}^{\rm II}$ also vanishes [\cite{155}, \cite{163}], but in
the inflationary universe theory the temperature can never reach
such high values.

Proceeding to a discussion of more realistic theories, it must be
\index{Tachyon}%
pointed our that when $\lambda  \ll 1$, the tachyon pole on the
upper branch of ${\bf V}(\varphi)$ is situated at $|k^2|\gg\rm
M^2_P$, and at the point $\bar \varphi$ where ${\bf V}^{\rm I}$
and ${\bf V}^{\rm II}$ merge, the effective potential ${\rm
V}(\varphi )$ exceed the Planck energy density $\rm M_P^4$.  In
that event, as will be shown in the following section, all of the
major qualitative and quantitative results obtained neglecting
quantum gravitation become unreliable. Furthermore, quantum
corrections to ${\rm V}(\varphi )$ associated with the presence of
other matter fields can become important at lower momenta and
densities. These corrections will not change the form of ${\rm
V}(\varphi)$ at small $\varphi $, but they can completely
eliminate the instability that arises with large fields and
momenta. Exactly the same thing happens with the instability in
the zero-charge problem when one makes the transition to
asymptotically free theories [\cite{3}, \cite{152}].

The basic practical conclusion to be drawn from the last two
sections is that for the most reasonable relationship between the
coupling constants ($\lambda\sim e^2\sim h^2\ll 1$), quantum
corrections to ${\rm V}(\varphi )$ in theories of the weak,
strong, and electromagnetic interactions become important only
for exponentially strong fields, so that the classical expression
for ${\rm V}(\varphi )$ is often a perfectly good approximation.
Quantum corrections can often lead to instability of the vacuum
when the fields or momenta are exponentially large, but this
difficulty can in principle be avoided by a small modification of
the theory without altering the shape of the effective potential
at small $\varphi $.
\index{O(N) symmetric theory|)}%

\section[effective potential and quantum gravity]
{\label{s2.4}The
\index{Effective potential!quantum gravitational effects and|(}%
\index{Quantum gravity!effective potential and|(}%
effective potential and quantum gravitational effects}

In our discussion of the inflationary universe scenario in
Chapter \ref{c1}, we often turned our attention to fields
$\varphi  \gg\rm M_{P}$.  There is some question as to whether
quantum gravitational effects might substantially modify
${\rm V}(\varphi )$ under such conditions, ultimately
invalidating the chaotic inflation scenario. Such suspicions have
been voiced by a number of authors (see [164], for example), and
we must therefore dwell specifically upon this question.

Gravitational corrections $\Delta{\rm V}(\varphi )$ to the
potential ${\rm V}(\varphi )$ are of a two\-fold nature. On the one
hand, they are associated with the gravitational interaction
between vacuum fluctuations, as in the Feynman diagrams shown in
Fig.~2.8. The entire set of such diagrams can be summed,
with the final result being [\cite{165}]
\begin{figure}[t]\label{f19}
\centering \leavevmode\epsfysize=5cm \epsfbox{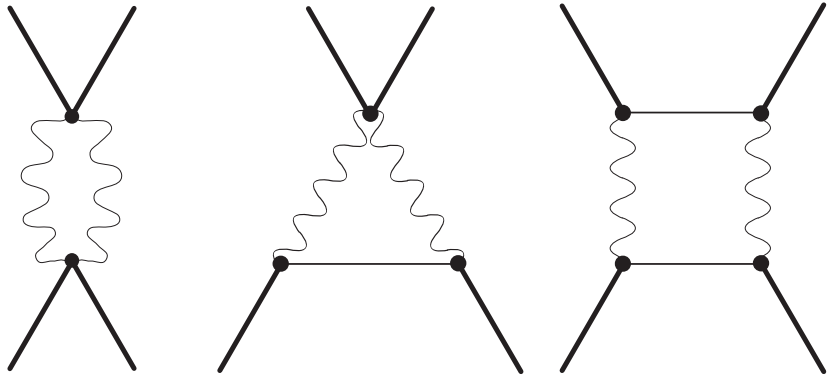} \caption{
Typical diagrams for ${\rm V}(\varphi )$ with gravitational
effects taken into account. The heavy lines correspond to the
external classical field $\varphi $, the lighter lines to scalar
particles of $\varphi $, and the wavy lines to gravitons.}
\end{figure}
\be
\label{2.4.1}
\Delta{\rm V}(\varphi)={\rm C}_1\,\frac{d^2{\rm V}}{d\varphi^2}\cdot
\frac{\rm V}{\rm M_P^2}\,\ln\frac{\Lambda^2}{\rm M_P^2} +
{\rm C}_2\,\frac{\rm V^2(\varphi)}{\rm M_P^4}\,
\ln\frac{\Lambda^2}{\rm M_P^2}\ .
\ee
The ${\rm C}_{i}$ here are numerical coefficient of order unity,
and $\Lambda $ is the ultraviolet cutoff. These corrections
clearly diverge as $\Lambda\rightarrow\infty$, and generally
speaking, they fail to converge simply to a renormalized version
of the original potential ${\rm V}(\varphi )$. This is
manifestation of the well known difficulty associated with the
nonrenormalizability of quantum gravitation. One usually assumes,
however, that at momenta of order $\m$, there should exist a
natural cutoff due either to the nontrivial structure of the
gravitational vacuum, or to the fact that when $|k^2|\ga\rm M_{\rm P}^2$,
gravitation becomes part of a more general theory
with no divergences. If, in accordance with this assumption,
$\Lambda^{2}$ does not exceed $\rm M_{\rm P}^2$ by many orders of
magnitude, then
\be
\label{2.4.2}
\Delta{\rm V}={\rm \tilde C}_1\,\frac{d^2{\rm V}}{d\varphi^2}\cdot
\frac{\rm V}{\m^2}+{\rm \tilde C}_2\,\frac{\rm V^2}{\m^4}\ ,
\ee
where  ${\rm \tilde C}_i= {\rm O}(1)$.

Note that, contrary to the often expressed belief, these
corrections {\it do not} contain any terms of the type $O(1)
{\phi^n \over {\rm M_P^n}}$, which would make the theory
ill-defined at $\phi
> M_p$. The main reason why these terms are not generated by
quantum gravity effects is that the field $\phi$ by itself does
not have any physical meaning. It enters the theory only via its
effective potential $V$ and mass squared $\frac{d^2{\rm
V}}{d\varphi^2}$, which is why the quantum gravity corrections
have the structure shown in Eq. (\ref {2.4.2}).

It can readily be shown that when
\ba
\label{2.4.3}
m^2_\varphi=\frac{d^2{\rm V}}{d\varphi^2}&\ll&\m^2\ ,\\
\label{2.4.4}
{\rm V}(\varphi)&\ll&\m^4\ ,
\ea
gravitational corrections to ${\rm V}(\varphi )$ are negligible.
In particular, for the theory $\lambda\, \varphi^{4}$,
(\ref{2.4.4}) is a much stronger condition than (\ref{2.4.3}); it
holds when
\be
\label{2.4.5}
\varphi\ll\varphi_{\rm P}=\lambda^{-1/4}\,\m\ .
\ee
For $\lambda  \sim 10^{-14}$, we obtain a very weak bound on
$\varphi $ from (\ref{2.4.5}):
\be
\label{2.4.6}
\varphi\ll 3000\,\m\ .
\ee
Thus, in a classical space-time in which (\ref{2.4.5}) holds (see
Section \ref{s1.7}), the indicated quantum gravitational corrections to
${\rm V}(\varphi )$ are negligible.

The other type of correction to ${\rm V}(\varphi )$ relates to
the change in the spectrum of vacuum fluctuations in an external
gravitational field.  However, inasmuch as the magnitude of the
field itself is proportional to ${\rm V}(\varphi )$, the
corresponding corrections (for ${\rm V}(\varphi ) \ll \m^4)$
are usually negligible. The most important exception to this is
the contribution to ${\rm V}(\varphi )$ from long-wavelength
fluctuations of the scalar field $\varphi $ that are generated at
the time of inflation. But as we already noted in Section
\ref{s1.8}, taking this effect into account does not lead to any
problems with the realization of the chaotic inflation scenario;
in fact, on the contrary, it engenders a self-sustaining
inflationary regime over most of the physical volume of the
universe. We shall return to this question in Chapter \ref{c10}.
\index{Effective potential|)}%
\index{Effective potential!quantum gravitational effects and|)}%
\index{Quantum gravity!effective potential and|)}%


\chapter{\label{c3}Restoration of Symmetry at High Temperature}
\index{Symmetry!at high temperature|(}%
\index{Temperature!high, symmetry at|(}%

\section[Phase transitions in the simplest models]%
{\label{s3.1}Phase transitions in the simplest models with spontaneous
\index{Phase transitions!in gauge theories|(}%
\index{Spontaneous symmetry breaking!phase transitions with|(}%
symmetry breaking}

Having discussed the basic features of spontaneous symmetry
breaking in quantum field theory, we can now turn to a
consideration of symmetry behavior in systems of particles in
thermodynamic equilibrium which interact in accord with unified
theories of the weak, strong, and electromagnetic interactions.
We will primarily examine systems of scalar particles $\varphi$
with the Lagrangian (\ref{1.1.5}).  Such particles carry no
conserved charge, nor is their number a conserved quantity.  The
chemical potential therefore vanishes for such particles, and
their density in momentum space is
\be
\label{3.1.1}
n_k=\frac{1}{\displaystyle \exp\left(\frac{k_0}{{\rm T}}\right)-1}\ ,
\ee
where $k_0=\sqrt{{\bf k}^2+m^2}$ is the energy of a particle with
momentum {\bf k} and mass $m$.  All particles disappear at
${\rm T} = 0$ ($n_k\rightarrow0$), and
we revert to the situation described in the previous chapter.

At finite temperature, all physically interesting quantities
(thermodynamic potentials, Green's functions, etc.) in this
\index{Gibbs averages}%
system are given not by vacuum averages, but by Gibbs averages
\be
\label{3.1.2}
\langle \ldots\rangle =
\frac{\displaystyle\tr
\left[\exp\left(-\frac{{\rm H}}{{\rm T}}\right)\ldots\right] }%
{\displaystyle\tr
\left[\exp\left(-\frac{{\rm H}}{{\rm T}}\right)\right]}
\ee
where H is the system Hamiltonian.  In particular, the symmetry
breaking parameter (the ``classical'' scalar field $\varphi$) in
this system is given by $\varphi({\rm T})=\langle \varphi\rangle $,
rather than by $\langle 0|\varphi|0\rangle $.

In order to investigate the behavior of $\varphi({\rm T})$ at
$T\neq0$, let us consider the Lagrange equation for the field
$\varphi$ in the theory (\ref{1.1.5}),
\be
\label{3.1.3}
(\dla+\mu^2-\lambda\,\varphi^2)\,\varphi=0\ ,
\ee
and let us take the Gibbs average of this equation, giving
\be
\label{3.1.4}
\dla\varphi({\rm T})-[\lambda\,\varphi^2({\rm T})-\mu^2]\,\varphi({\rm T})
-3\,\lambda\,\varphi({\rm T})\,\langle \varphi^2\rangle
-\lambda\,\langle \varphi^3\rangle =0\ .
\ee
Here, as in the analysis of spontaneous symmetry breaking at
${\rm T}=0$, we have separated out the analog of the classical
field $\varphi$ by carrying out the shift
$\varphi\rightarrow\varphi+\varphi({\rm T})$, such that
\be
\label{3.1.5}
\langle \varphi\rangle =0\ .
\ee
To lowest order in $\lambda$, $\langle \varphi^3\rangle $  is
equal to zero, whereas
\ba
\label{3.1.6}
\langle \varphi^2\rangle &=&\frac{1}{(2\,\pi)^3}\,\int
\frac{d^3k}{2\,\sqrt{k^2+m^2}}\:
(1+2\,\langle a_k^+\,a_k^-\rangle)\nonumber \\
&=&\frac{1}{(2\,\pi)^3}\,\int
\frac{d^3k}{\sqrt{k^2+m^2}}\:\left(\frac{1}{2}+n_k\right)\ .
\ea
The first term in (\ref{3.1.6}) vanishes after renormalizing the
mass of the field $\varphi$ in the field theory (at ${\rm T}=0$).
As a result,
\ba
\label{3.1.7}
\langle \varphi^2\rangle &=&{\rm F}({\rm T},m(\varphi))\nonumber \\
&=&\frac{1}{2\,\pi^2}\,\int^\infty_0
\frac{k^2\:dk}{\displaystyle \sqrt{k^2+m^2(\varphi)}\,
\left(\exp\frac{\sqrt{k^2+m^2(\varphi)}}{{\rm T}}-1\right)}\ .\nonumber \\
\ea Clearly, all interesting effects in this theory
($\lambda\ll1$) take place at ${\rm T}\gg m$, where we can neglect
$m$ in (\ref{3.1.7}).  Then \be \label{3.1.8} \langle
\varphi^2\rangle ={\rm F}({\rm T},0)=\frac{{\rm T}^2}{12}\ , \ee
and Eq. (\ref{3.1.4}) becomes \be \label{3.1.9} \dla\varphi({\rm
T})-\left[\lambda\,\varphi^2({\rm T})-\mu^2
+\frac{\lambda}{4}\,{\rm T}^2\right]\,\varphi({\rm T})=0\ . \ee
From (\ref{3.1.9}), we obtain for the constant field $\varphi({\rm
T})$ \be \label{3.1.10} \varphi({\rm
T})\,\left[\lambda\,\varphi^2({\rm T})-\mu^2
+\frac{\lambda}{4}\,{\rm T}^2\right]=0\ . \ee At sufficiently low
temperature, this equation has two solutions, \ba \label{3.1.11}
1)\quad \varphi({\rm T})&=&0\ ;\nonumber \\
2)\quad \varphi({\rm T})&=&
\sqrt{\frac{\mu^2}{\lambda}-\frac{{\rm T}^2}{4}}\ .
\ea
The second of these vanishes above a critical temperature
\be
\label{3.1.12}
{\rm T}_c=\frac{2\,\mu}{\sqrt{\lambda}}=2\,\varphi_0\ .
\ee

To derive the excitation spectrum at ${\rm T}\neq0$, we must
carry out the shift $\varphi\rightarrow\varphi+\delta\varphi$ in
(\ref{3.1.9}).  When $\varphi({\rm T})=0$, the corresponding
equation takes the form
\be
\label{3.1.13}
\dla\delta\varphi-\left(\mu^2+\frac{\lambda}{4}\,{\rm T}^2\right)\,
\delta\varphi=0\ ,
\ee
which corresponds to a mass
\be
\label{3.1.14}
m^2=-\mu^2+\frac{\lambda}{4}\,{\rm T}^2
\ee
for the scalar field at $\varphi=0$.  This quantity is negative
when ${\rm T}<{\rm T}_c$, and it becomes positive when ${\rm T}>{\rm T}_c$.
For the solution of the second of Eqs. (\ref{3.1.11}),
$\displaystyle \varphi({\rm T})=
\sqrt{\frac{\mu^2}{\lambda}-\frac{{\rm T}^2}{4}}$, it takes the value
\be
\label{3.1.15}
m^2=3\,\lambda\,\varphi^2({\rm T})-\mu^2+\frac{\lambda}{4}\,{\rm T}^2
=2\,\lambda\,\varphi^2({\rm T})\ .
\ee
This solution is therefore stable for ${\rm T}<{\rm T}_c$, and it
vanishes for ${\rm T}>{\rm T}_c$ at the instant when the solution
$\varphi=0$ becomes stable.  This then means that a phase
transition with restoration of symmetry takes place at a
temperature ${\rm T}={\rm T}_c$  [\cite{18}--\cite{24}].

We illustrate the foregoing results in Fig. 3.1.  The quantity
clearly decreases smoothly with increasing temperature,
\index{Phase transitions!second-order}%
corresponding to a second-order phase transition.
\begin{figure}[t]\label{f20}
\centering \leavevmode\epsfysize=5cm \epsfbox{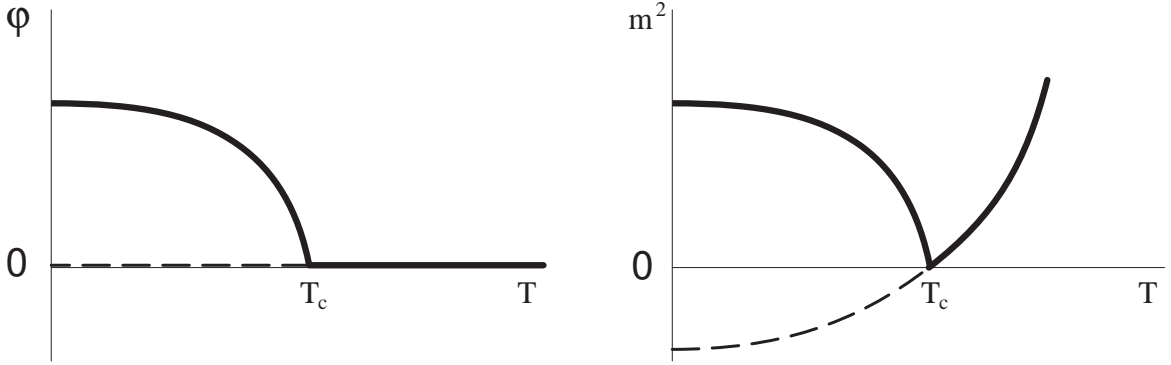} \caption{The
quantities $\varphi({\rm T})$ and $m^2({\rm T})$ in the theory
(1.1.5).  The dashed lines correspond to the unstable phase
$\varphi = 0$ at ${\rm T}<{\rm T}_c$.}
\end{figure}

These results can also be obtained in a different way, based on a
finite-temperature generalization of the concept of the effective
potential ${\rm V}(\varphi)$.  We will not dwell on this problem,
noting simply that at its extrema, the effective potential
${\rm V}(\varphi,{\rm T})$ coincides with the free energy
${\rm F}(\varphi,{\rm T})$.  To calculate ${\rm V}(\varphi,{\rm T})$,
it suffices to recall that at ${\rm T}\neq0$, quantum statistics
is equivalent to Euclidean quantum field theory in a space which
is periodic, with period $1/{\rm T}$ along the ``imaginary time''
axis [\cite{166}, \cite{20}].  To go from ${\rm V}(\varphi,0)$ to
${\rm V}(\varphi,{\rm T})$ one should replace all boson momenta
$k_4$ in the Euclidean integrals by $2\,\pi\,n\,{\rm T}$  for
bosons and $(2\,n+1)\,\pi\,{\rm T}$ for fermions, and sum over
$n$ instead of integrating over $k_4$:
$\displaystyle \int dk_4\rightarrow2\,\pi\,{\rm T}\sum^\infty_{n=-\infty}$.
For example, at ${\rm T}\neq0$, Eq. (\ref{2.2.1}) for
${\rm V}(\varphi)$ in the theory (\ref{1.1.5}) transforms into
\ba
\label{3.1.16}
{\rm V}(\varphi,{\rm T})&=&-\frac{\mu^2}{2}\,\varphi^2+
\frac{\lambda}{4}\,\varphi^4\nonumber \\
&+&\frac{{\rm T}}{2\,(2\,\pi)^3}\sum^\infty_{n=-\infty}
\int d^3k\:\ln[(2\,\pi\,n\,{\rm T})^2+k^2+m^2(\varphi)]\ ,\nonumber \\
\ea
where $m^2(\varphi)=3\,\lambda\,\varphi^2-\mu^2$.  This
expression can be renormalized using the same counterterms as for
${\rm T}=0$.  Equation (\ref{1.2.3}) gives the result of
calculating ${\rm V}(\varphi,{\rm T})$  for ${\rm T}\gg m$.  It
is straightforward to show that the equation $\displaystyle
\frac{d{\rm V}}{d\varphi}=0$, which determines the equilibrium
values of $\varphi({\rm T})$, is the same as (\ref{3.1.10}), and
that the quantity $\displaystyle \frac{d^2{\rm V}}{d\varphi^2}$,
which determines the mass squared of the field $\varphi$,
coincides with (\ref{3.1.14}) and (\ref{3.1.15}) (for equilibrium
$\varphi({\rm T})$).  The description of the phase transition in
terms of the behavior of ${\rm V}(\varphi,{\rm T})$ is given in
Section \ref{s1.2}.

The methods developed above can readily be generalized to more
complicated models.  In the Higgs model (\ref{1.1.15}), for
example, in the transverse gauge $\partial_\mu{\rm A}_\mu=0$, we
have
\be
\label{3.1.17}
\left\langle \frac{\delta{\rm L}}{\delta\varphi}\right\rangle =
\varphi({\rm T})\,[\mu^2-\lambda\,\varphi^2({\rm T})
-3\,\lambda\,\langle \chi^2_1\rangle -\lambda\,\langle \chi_2^2\rangle
+e^2\,\langle {\rm A}_\mu^2\rangle ]=0
\ee
instead of Eq. (\ref{3.1.4}).
To start with, let us assume that $\lambda\sim e^2$.  Then the
phase transition takes place at ${\rm T}\gg m_\chi$, $m_{\rm A}$,
as in the theory (\ref{1.1.5}); hence
\be
\label{3.1.18}
\langle \chi_1^2\rangle =\langle \chi_2^2\rangle
=-\frac{1}{3}\,\langle {\rm A}_\mu^2\rangle =\frac{{\rm T}^2}{12}\ ,
\ee
and Eq. (\ref{3.1.17}) becomes
\be
\label{3.1.19}
\varphi\,\left(\lambda\,\varphi^2-\mu^2+
\frac{4\,\lambda+3\,e^2}{12}{\rm T}^2\right)=0\ .
\ee
This then implies that the phase transition takes place in the
\index{Higgs model!phase transition in}%
\index{Phase transitions!in Higgs model}%
Higgs model at a critical temperature
\be
\label{3.1.20}
{\rm T}^2_{c_1}=\frac{12\,\mu^2}{4\,\lambda+3\,e^2}\ .
\ee
According to (\ref{3.1.19}), $\varphi({\rm T})$ is a continuous
\index{Phase transitions!second-order}%
function of T, that is, this is a second-order phase transition
[\cite{18}--\cite{20}].

If we consider the case $\lambda\la e^4$, however, we find that
$\displaystyle m_{\rm A}({\rm T}_{c_1})
\approx\frac{e\,\mu}{\lambda}\ga{\rm T}_{c_1}$, i.e., we no longer
have ${\rm T}\gg m_{\rm A}$,
and the contribution of vector particles to (\ref{3.1.19}) at
${\rm T}\sim{\rm T}_{c_1}$  is strongly
suppressed.  We can then no longer neglect $m_{\rm A}$
compared with T when calculating
$\langle {\rm A}_\mu^2\rangle =-{\rm F}({\rm T},m_{\rm A})$,
and all of the equations are significantly altered.  The simplest way to
understand this is to note that when $m < {\rm T}$,
the quantity ${\rm F}({\rm T},m_{\rm A})$ can be represented by
a power series in $\displaystyle \frac{m}{{\rm T}}$:
\be
\label{3.1.21}
{\rm F}({\rm T},m)=\frac{{\rm T}^2}{12}\,\left[1-
\frac{3}{\pi}\,\frac{m}{{\rm T}}
+{\rm O}\left(\frac{m^2}{{\rm T}^2}\right)\right]\ .
\ee
Bearing in mind, then, that in the lowest order of perturbation theory
$m_{\rm A}=e\,\varphi$, Eq. (\ref{3.1.19}) can be rewritten as
\be
\label{3.1.22}
\varphi\,\left(\lambda\,\varphi^2-\mu^2+
\frac{4\,\lambda+3\,e^2}{12}\,{\rm T}^2-
\frac{3\,e^3}{4\,\pi}\,{\rm T}\,\varphi\right)=0\ .
\ee
\begin{figure}[t]\label{f21}
\centering \leavevmode\epsfysize=6cm \epsfbox{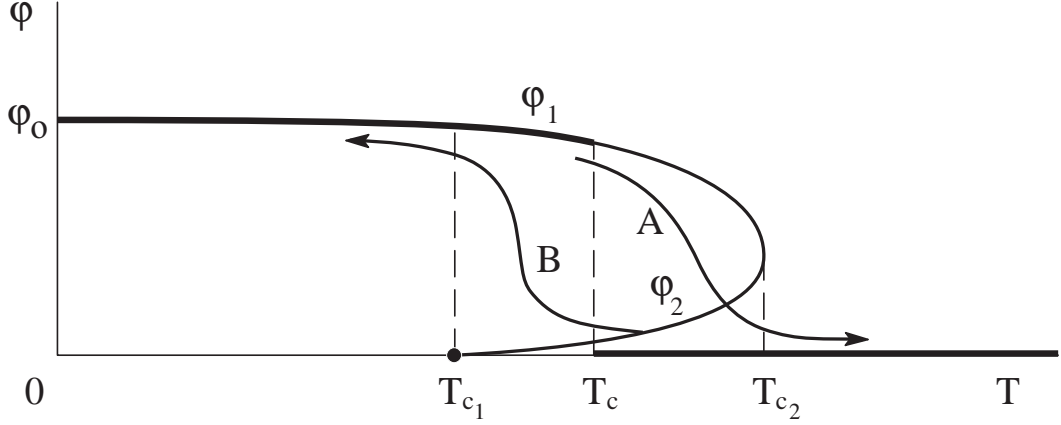} \caption{The
function $\varphi({\rm T})$ in the Higgs model with $\displaystyle
\frac{3\,e^4}{16\,\pi^2}<\lambda\la e^4$. The heavy curve
corresponds to the stable state of the system.  Arrows indicate
the behavior of $\varphi$ with increasing (A) and decreasing (B)
temperature.}
\end{figure}
In contrast to (\ref{3.1.19}), this equation has three solutions
rather than two in a certain temperature range
${\rm T}_{c_1}<{\rm T}<{\rm T}_{c_2}$, corresponding to three different
extrema of ${\rm V}(\varphi,{\rm T})$; see Fig. 3.2.  The
solution $\varphi=0$ is metastable when ${\rm T}>{\rm T}_{c_1}$.
Upon heating, a phase transition from the phase
\index{Higgs model!phase transition in}%
\index{Phase transitions!in Higgs model}%
$\varphi=\varphi_1$ to the phase $\varphi=0$
begins at a temperature ${\rm T}_c$, where
\be
\label{3.1.23}
{\rm V}(\varphi_1({\rm T}_c),{\rm T}_c)={\rm V}(0,{\rm T}_c)\ .
\ee
In the Higgs model with $\lambda\la e^4$,
\be
\label{3.1.24}
{\rm T}_c=\left(\frac{15\,\lambda}{2\,\pi^2}\right)^{1/4}\,\mu\ ;
\ee
see [\cite{23}].  Clearly, the phase transition in the present
case is a discontinuous one --- a first-order phase transition
\index{Phase transitions!first-order}%
(see Fig. 3.2).

\begin{figure}[t]\label{f22}
\centering \leavevmode\epsfysize=6cm \epsfbox{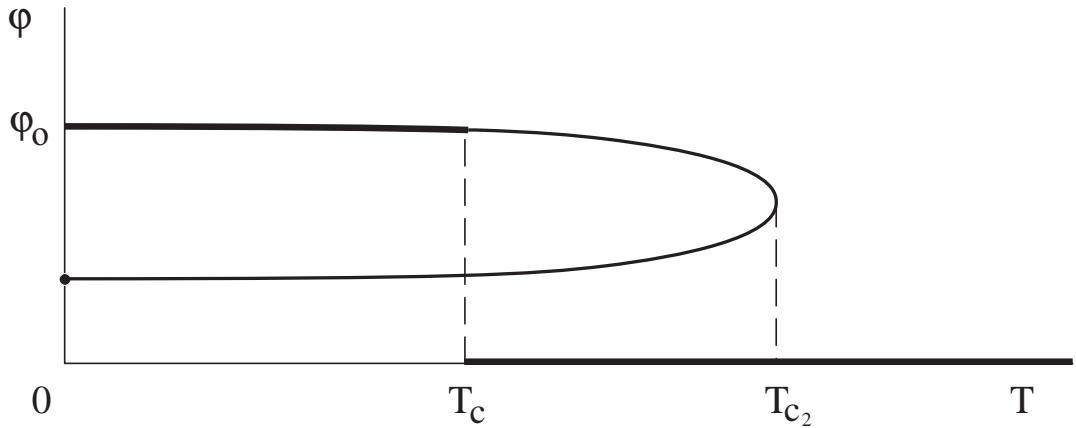} \caption{The
function $\varphi({\rm T})$ in the Higgs model with $\displaystyle
\lambda<\frac{3\,e^4}{16\,\pi^2}$.}
\end{figure}
Recall now that when $\displaystyle \lambda\la\frac{3\,e^4}{16\,\pi^2}$,
quantum corrections to ${\rm V}(\varphi,{\rm T})$ lead to the
existence of a local minimum of ${\rm V}(\varphi)$ even at
${\rm T}=0$ (see Fig. 3.3), and when
$\displaystyle \lambda<\frac{3\,e^4}{32\,\pi^2}$, this minimum
becomes deeper than the usual one at
$\displaystyle \varphi=\frac{\mu}{\sqrt{\lambda}}$; see
Section \ref{s2.2}. Thus, as
$\displaystyle \lambda\rightarrow\frac{3\,e^4}{32\,\pi^2}$ the critical
temperature ${\rm T}_c\rightarrow0$.  This does not mean,
however, that a phase transition in such a theory becomes easy to
produce in a laboratory.  The point here is that a first-order
\index{Phase transitions!first-order}%
phase transition occurs by virtue of the sub-barrier creation and
subsequent growth of bubbles of the new phase.  Bubble formation
is often strongly suppressed, so it may take an extremely long
time for the phase transition to occur.  When the system is
heated, the phase transition therefore really takes place from a
superheated phase $\varphi_1$ at some temperature higher than
${\rm T}_c$.  Likewise, when the system is cooled, a first-order
\index{Phase transitions!first-order}%
phase transition takes place from a supercooled phase at
${\rm T}<{\rm T}_c$.  We will consider the theory of bubble production
in Chapter \ref{c5}, and the cosmological consequences of
first-order phase transitions will be discussed in Chapters
\ref{c6} and \ref{c7}.
\index{Spontaneous symmetry breaking!phase transitions with|)}%

\section[Phase transitions in realistic theories]%
{\label{s3.2}Phase transitions in realistic theories of the
\index{Phase transitions!in weak, strong, and electromagnetic interactions|(}%
weak, strong, and electromagnetic interactions}

As we have shown in the preceding section, when the coupling
constants $\lambda$ and $e^2$ are related in the most natural
\index{Higgs model!phase transition in}%
\index{Phase transitions!in Higgs model}%
\index{Phase transitions!second-order}%
way, the phase transition in the Higgs model is second-order, but
when $\lambda\sim e^4$, it becomes first-order.  One can readily
show that the same is true of a phase transition in the
Weinberg--Salam theory.

Thus, for example, when $\lambda\sim e^2$, the counterpart of Eq.
(\ref{3.1.19}) in the Weinberg--Salam theory becomes [\cite{24}]
\be
\label{3.2.1}
\varphi\left(\lambda\,\varphi^2-\mu^2+
\left[\lambda+
\frac{e^2\,(1+2\,\cos^2\theta_{\rm W})}{\sin^22\,\theta_{\rm W}}\right]\,
\frac{{\rm T}^2}{2}\right)=0\ ,
\ee
where $\theta_{\rm W}$ is the
\index{Weinberg angle}%
Weinberg angle, $\sin^2\theta_{\rm W}\sim0.23$.
Equation (\ref{3.2.1}) then gives
\be
\label{3.2.2}
{\rm T}^2_c=\frac{2\,\mu^2}{
\displaystyle \lambda+\frac{e^2\,
(1+2\,\cos^2\theta_{\rm W})}{\sin^22\,\theta_{\rm W}}}=
\frac{2\,\varphi_0^2}{
\displaystyle 1+\frac{e^2\,(1+2\cos^2\theta_{\rm W})}{\lambda\,
\sin^22\,\theta_{\rm W}}}\ ,
\ee
where $\varphi_0\approx250$ GeV.  Putting $\lambda\sim e^2\sim0.1$,
we obtain
\be
\label{3.2.3}
{\rm T}_c\sim100\;\mbox{GeV}\ ,
\ee
which is more than twice the mass of the ${\rm W}^\pm$ and Z
particles and the mass of the Higgs boson for $\lambda\sim e^2$,
${\rm T}=0$.  In the case at hand, an analysis similar to the one
carried out in Section \ref{s3.1} indicates that to high
\index{Phase transitions!second-order}%
accuracy, the phase transition can be called a second-order
transition:  the jump in the field $\varphi$ at the moment of the
phase transition turns out to be more than an order of magnitude
less than $\varphi_0$.
The phase transition in
the Weinberg-Salam model could be first order only if the Higgs mass were
sufficiently small.\footnote{This analysis should be modified taking into account the interaction of the scalar field with the top quarks and using realistic values of the Higgs mass. For a more recent discussion of this issue see [\cite{Dine:1992wr}].}

On the other hand, the transitions that take place in grand
unified theories at ${\rm T}\ga10^{14}$ GeV, as a rule, prove to
be first-order transitions with a considerable jump in the field
$\varphi$ at the critical point [\cite{104}].  There are two
reasons why this is so.  First, at ${\rm T}\sim10^{14}$ GeV, the
effective gauge constant $g^2\sim 0.3$; that is, it is three
times the value of $e^2$ at ${\rm T}\sim10^2$ GeV.  Second, there
are a great many particles in grand unified theories that
contribute to temperature corrections to the effective potential.
The net result is that the critical temperature ${\rm T}_{c_1}$
of the phase transition turns out to be approximately of the same
order of magnitude as the particle masses at that temperature.
As we showed in Section \ref{s3.1}, this is precisely the
circumstance that leads to a first-order phase transition.

As an example, consider a theory with SU(5) symmetry [\cite{91}].  The
\index{SU(5) symmetry}%
\index{Symmetry!SU(5)}%
effective potential in the simplest version of this theory is
\be
\label{3.2.4}
{\rm V}(\Phi)=-\frac{\mu^2}{2}\,\tr\Phi^2+
\frac{a}{4}\,(\tr\Phi^2)^2+\frac{b}{2}\,\tr\Phi^4\ ,
\ee
where $\Phi$ is a traceless $5 \times 5$ matrix.  If one takes
$b>0$, $\lambda>0$, where $\displaystyle \lambda=a+\frac{7}{15}\,b$,
the symmetric state $\Phi=0$ is unstable with respect to the
appearance of the scalar field (\ref{1.1.19}),
\be
\label{3.2.5}
\Phi=\sqrt{\frac{2}{15}}\,\varphi\cdot
\mbox{diag}\,\left(1,1,1,-\frac{3}{2},-\frac{3}{2}\right)\ ,
\ee
\index{Symmetry!SU(5)}%
\index{SU(5) symmetry}%
which breaks the SU(5) symmetry to
$\mbox{SU}(3) \times \mbox{SU}(2) \times {\rm U}(1)$.
At ${\rm T}=0$, the minimum of ${\rm V}(\varphi)$ corresponds to
$\displaystyle \varphi_0=\frac{\mu}{\sqrt{\lambda}}$.
Temperature corrections to ${\rm V}(\varphi)$ come from 24
different kinds of Higgs bosons and 12 X and Y vector bosons.  As
a result, the counterpart of Eq. (\ref{3.2.1}) for $\varphi({\rm T})$
\index{Symmetry!SU(5)}%
\index{SU(5) symmetry}%
in the SU(5) theory is of the form [\cite{104}]
\be
\label{3.2.6}
\varphi\,\left(\mu^2-\beta\,{\rm T}^2-\lambda\,\varphi^2
-\frac{{\rm T}\,\varphi}{30\,\pi}\,{\rm Q}(g^2,\lambda,b)\right)=0\ ,
\ee
where
\ba
\label{3.2.7}
\beta&=&\frac{75\,g^2+130\,a+94\,b}{60},\\
\label{3.2.8}
{\rm Q}&=&7\,\lambda\,\sqrt{10\,b}+\frac{16}{3}\,b\,\sqrt{10\,b}
+3\,\sqrt{15}\,\lambda^{3/2}\nonumber \\
&+&2\,\sqrt{15}\,\lambda\,g
+\frac{75}{4}\,\sqrt{2}\,g^3\ .
\ea
As we have already noted, a phase transition with symmetry
breaking upon cooling takes place somewhere in the temperature
range between ${\rm T}_{c_1}$ and ${\rm T}_c$, where ${\rm T}_{c_1}$
is given by
\be
\label{3.2.9}
{\rm T}_{c_1}=\frac{\mu}{\sqrt{\beta}}\ .
\ee
To estimate the size of the jump at the phase transition point, let us
determine the quantity $\varphi_1({\rm T}_{c_1})$, Fig. 3.2.
At ${\rm T}={\rm T}_{c_1}$, the first two terms in (\ref{3.2.6}) cancel,
and we find that
\be
\label{3.2.10}
\varphi_1({\rm T}_{c_1})=
\frac{{\rm Q}\,\varphi_0}{30\,\pi\,\sqrt{\beta\,\lambda}}\ .
\ee
For the most reasonable values of the parameters $a\sim b\sim g^2=0.3$,
Eqs. (\ref{3.2.7})--(\ref{3.2.10}) imply that
\be
\label{3.2.11}
\varphi_1({\rm T}_{c_1})\sim0.75\,\varphi_0\ ,
\ee
i.e., the jump in the field at the time of the phase transition
is very large (of the same order of magnitude as $\varphi_0$).

{\looseness=1
In the discussion above, we studied only one ``channel'' of the phase
transition, in which the transition goes directly from the SU(5) phase
to the $\mbox{SU}(3) \times \mbox{SU}(2) \times {\rm U}(1)$ phase.
In actuality, the phase transition usually entails
the formation of an $\mbox{SU}(4) \times {\rm U}(1)$ intermediate
phase, and other phases as well [\cite{67}, \cite{42}].  Each of
the intermediate phase transitions is also a first-order
transition.  The kinetics of the phase transition in the minimal
SU(5) theory will be discussed in Chapter \ref{c6}.
\index{Phase transitions!in gauge theories|)}%
\index{Phase transitions!in weak, strong, and electromagnetic interactions|)}%

}

\section[Higher-order perturbation theory]%
{\label{s3.3}Higher-order perturbation theory and the infrared\protect\\
\index{Gauge fields, perturbation theory in|(}%
\index{Perturbation theory in gauge fields|(}%
problem in the thermodynamics of gauge fields}

Our analysis of the high-temperature restoration of symmetry in
the theory (\ref{1.1.5}) (Section \ref{s3.1}) was based on the
use of the lowest-order perturbation theory in $\lambda$.  One
may then ask how reliable the results obtained in this manner
really are.

This is not a completely trivial question.  For example, apart from
small terms $\sim\lambda^n\,{\rm T}^4$, $\lambda^n\,{\rm T}^2\,m^2$,
high-order corrections in $\lambda$ to the expression
for ${\rm V}(\varphi,{\rm T})$ at ${\rm T}\neq0$
could contain terms proportional to $m^{-n}$.
Such terms become large when $m$ is small.

In order to analyze this question more thoroughly, let us examine
the N-th order diagrams in $\lambda$ for ${\rm V}(\varphi,{\rm T})$
in the theory (\ref{1.1.5}), for $\varphi=0$.  The
contribution of these diagrams to ${\rm V}(0,{\rm T})$ can be
written out as an expression of the form
\ba
\label{3.3.1}
{\rm V}_{\rm N}(0,{\rm T})&\sim&(2\,\pi\,{\rm T})^{{\rm N}+1}\,
\lambda^{\rm N}\,\int d^3p_1\ldots d^3p_{\rm N+1}\:\nonumber \\
&\times&\sum^\infty_{n_i=-\infty}\prod_{k=1}^{2{\rm N}}
[(2\,\pi\,r_k\,{\rm T})^2+{\bf q}_k^2+m^2({\rm T})]^{-1}\ ,
\ea
where ${\bf q}_k$
is a homogeneous linear combination of the ${\bf p}_i$,
and $r_k$  is a corresponding
combination of the $n_i$, $i=1,\ldots,{\rm N}+1$,
$k=1,\ldots,2{\rm N}$.  When $m\rightarrow0$,
the leading term in the sum over $n_i$ is the one for
which all $n_i=0$ ($r_k=0$),
since the factors containing the terms $(2\,\pi\,r_k\,{\rm T})^2$
are nonsingular as $m\rightarrow0$, ${\bf q}_k\rightarrow0$.
This leading term is given by
\ba
\label{3.3.2}
\Delta{\rm V}_{\rm N}(0,{\rm T})&\sim&(2\,\pi\,{\rm T})^{{\rm N}+1}\,
\lambda^{\rm N}\,\int d^3p_1\ldots d^3p_{\rm N+1}\:
\prod_{k=1}^{2{\rm N}}[{\bf q}_k^2+m^2({\rm T})]^{-1}\nonumber \\
&\sim&\lambda^3\,{\rm T}^4\,
\left(\frac{\lambda\,{\rm T}}{m({\rm T})}\right)^{{\rm N}-3}\ .
\ea
It can be seen that dangerous terms
$\displaystyle \sim\left(\frac{\lambda\,{\rm T}}{m}\right)^{{\rm N}-3}$
appear in the expression for ${\rm V}(0,{\rm T})$ starting
with perturbations of order ${\rm N} = 4$,
and these make it impossible to obtain
reliable results from perturbation theory with $m < \lambda\,{\rm T}$.
Fortunately,
however, (\ref{3.1.14}) can be used to show that $m \gg \lambda\,{\rm T}$
everywhere outside a
small region near the critical temperature ${\rm T}_c$, within which
\be
\label{3.3.3}
|{\rm T}-{\rm T}_c|\la\lambda\,{\rm T}_c\ .
\ee
Everywhere outside this region, the results obtained in the preceding two
sections are reliable.

Matters are much more difficult when it comes to dealing with
phase transitions in the non-Abelian gauge theories that describe
\index{Yang--Mills fields}%
the interaction of Yang--Mills fields ${\rm A}_\mu^a$ with one
another and with scalar fields $\varphi$ with a coupling constant
$g^2$.  At ${\rm T}\neq 0$ in such theories, higher-order
perturbation terms $\sim g^{2{\rm N}}$ grow with increasing N (as
${\rm N}\rightarrow\infty$) for $m_{\rm A}\la g^2\,{\rm T}$.
Since in the classical approximation, $m_{\rm A}$ goes to zero
($m_{\rm A}\sim g\,\varphi({\rm T})$) at all temperatures above
the critical value ${\rm T}={\rm T}_c$, we are left with the
question of whether high-temperature corrections lead to
sufficiently high mass $m_{\rm A}({\rm T})\neq0$, with a
corresponding cutoff of infrared-divergent powers of the form
$\displaystyle \left(\frac{g^2\,{\rm T}}{m_{\rm A}}\right)^{\rm N}$.

The authors of [\cite{168}, \cite{169}] have shown that
high-temperature effects give rise to a pole of the Green's
\index{Yang--Mills fields}%
function of the Yang--Mills fields ${\rm G}_{\mu\nu}^{ab}(k)$ at
$k_0\sim g\,{\rm T}$, ${\bf k}=0$.  This might be
interpreted as the appearance of an infrared cutoff at a mass
$m_{\rm A}\sim g\,{\rm T}$, making the terms
$\displaystyle \left(\frac{g^2\,{\rm T}}{m_{\rm A}}\right)^{\rm N}$
small; actually, however, that would not be correct.  The
foregoing analysis tells us that the leading infrared divergences
as $m_{\rm A}\rightarrow0$ are associated not with the behavior
of the Green's functions at ${\bf k}=0$, $k_0\neq0$, but with
their behavior when $k_0=0$, ${\bf k}\rightarrow0$ ($k_0=0$
corresponds to $n_i=0$ in (\ref{3.3.1})).  In this limit, the
behavior of ${\rm G}_{\mu\nu}^{ab}(k)$ is most easily studied in
the\index{Coulomb gauge}
Coulomb gauge, for which [\cite{166}, \cite{24}]
\ba
\label{3.3.4}
{\rm G}_{00}^{ab}&=&\delta^{ab}\,[k^2+\pi_{00}(k)]^{-1}\ ,\\
\label{3.3.5}
{\rm G}_{i0}^{ab}&=&{\rm G}_{0j}^{ab}=0\ , \\
\label{3.3.6}
{\rm G}_{ij}^{ab}&=&\delta^{ab}\,
\left(\delta_{ij}-\frac{k_i\,k_j}{k^2}\right)\,{\rm G}(k)\ ,
\ea
where $k=|{\bf k}|$, $a$ and $b$ are isotopic spin indices,
$\pi_{00}(0)\sim g^2\,{\rm T}^2$ to lowest order $g^2$, and $i$,
$j=1$, 2, 3.

Thus, there really {\it is} an infrared cutoff at $m_0\sim g\,{\rm T}$
in ${\rm G}_{00}^{ab}$, corresponding to the usual
Debye screening of the electromagnetic field in hot plasma
[\cite{166}].  A well known result in quantum electrodynamics,
however, is that a static magnetic field in plasma cannot be
screened, and there is consequently no infrared cutoff in ${\rm G}_{ij}$
($k=0$, ${\bf k}\rightarrow\infty$) to any order of
perturbation theory [\cite{166}].  In the Yang--Mills gas, there
will likewise be no infrared cutoff at $k_0=0$,
${\bf k}\rightarrow0$  with momentum $k\sim g\,{\rm T}$.  There may in
principle be one, however, at momentum $k\sim g^2\,{\rm T}$,
inasmuch as massless Yang--Mills particles (in contrast to
photons) interact with each other directly, and the same infrared
divergences appear in the thermodynamics of a Yang--Mills gas as
\index{Phase transitions!second-order}%
in scalar field theory at the point of a second-order phase
transition.  The difference is that the mass of a scalar field at
a phase transition point vanishes ``by definition'' (the
curvature of ${\rm V}(\varphi)$ changes sign at the phase
transition point), while in the thermodynamics of a Yang--Mills
gas, the presence or absence of an infrared cutoff does not
follow from any general considerations, which imply only that the
expected scale of the infrared cutoff is $k\sim g^2\,{\rm T}$.
One can reach the same conclusion by analyzing the most strongly
infrared-divergent part of the theory [\cite{170}], as well as
the specific diagrams that could contribute to such a cutoff
[\cite{24}, \cite{171}, \cite{172}].  Unfortunately, when
$k\sim g^2\,{\rm T}$, all high-order corrections to the diagrams for the
\index{Yang--Mills fields}%
polarization operator of the Yang--Mills field are of comparable
size, so the infrared behavior of the Green's functions of the
Yang--Mills field at $k\la g^2\,{\rm T}$ thus far remains an open
problem.  Meanwhile, the degree of confidence that we have in our
understanding of many of the fundamental features of gauge-theory
thermodynamics depends on the solution to this problem.  Let us
consider the three main possibilities, which illustrate the
significance of this problem.

1)  {\it There is no infrared cutoff at $k\sim g^2\,{\rm T}$ in
the thermodynamics of a Yang--Mills gas.} In that case,
higher-order perturbations will be larger than lower-order ones
for all thermodynamic quantities, making it impossible to use
perturbation theory to study the thermodynamic properties of
gauge theories at ${\rm T}>{\rm T}_c$.  The only thing we might
possibly be able to verify with reasonable assurance is that the
energy density should be proportional to ${\rm T}^4$ at superhigh
temperature (from dimensional considerations).  This would only
suffice for the crudest approach to the theory of the evolution
of a hot universe at ${\rm T}>{\rm T}_c$.

2) {\it There is a tachyon pole or a sign change at some momentum
$k\sim g^2\,{\rm T}$ in the Green's function ${\rm G}_{ij}^{ab}(k)$.}
This implies instability with respect to creation of classical
Yang--Mills fields.  The second case is particularly interesting
--- the instability could result in the spontaneous
crystallization of the Yang--Mills gas at superhigh temperature,
which might lead to nontrivial cosmological consequences.

3) {\it In the best possible case (from the standpoint of
perturbation theory), the theory contains a cutoff by virtue of
the fact that ${\rm G}^{-1/2}(0)$ turns out to be a positive
quantity $m({\rm T})$ of order $g^2\,{\rm T}$.} In that event,
one can reliably calculate several of the lowest-order
perturbation terms in $g^2$ for the thermodynamic potential of
the Yang--Mills gas (up to $\sim g^6\,{\rm T}^4$) [\cite{171},
\cite{172}].  In principle, the appearance of such a cutoff can
lead to monopole confinement in a hot Yang--Mills plasma
[\cite{173}];  see Chapter \ref{c6}.

Thus, the thermodynamic properties of hot dense matter described
by gauge theories are still far from being well-understood, and
one must not lose sight of the inherent difficulties and
uncertainties.  Nevertheless, many results have been reasonably
reliably established.  As applied to the theory of phase
transitions studied in this chapter, the infrared problem in the
thermodynamics of a Yang--Mills gas does not modify the results
obtained for ${\rm T}<{\rm T}_c$.  It can also be shown that
$\varphi({\rm T})$ should be much smaller than $\varphi_0$ when
${\rm T}>{\rm T}_c$, and that at large T it cannot exceed
${\rm O}(g\,{\rm T})$. (If one assumes that
$\varphi({\rm T})\gg g\,{\rm T}$, then the Yang--Mills fields
acquire a mass $m_{\rm A}\gg g^2\,{\rm T}$, perturbation theory
becomes reliable, and the latter predicts that $\varphi({\rm T})= 0$
at ${\rm T}>{\rm T}_c$.) One should keep these uncertainties
in mind in discussing such complicated problems as production and
evolution of monopoles in grand unified theories.  However, for
most of the effects to be discussed below, these uncertainties
will not be important, and we will usually presume that
$\varphi({\rm T})$ at a sufficiently high temperature
${\rm T}>{\rm T}_c$, in accordance with the results obtained in
the preceding sections.

We must now state one last (but very important) reservation.  We
have assumed throughout that the field $\varphi$ has sufficient
time to roll down to a minimum of ${\rm V}(\varphi,{\rm T})$.
This natural assumption is valid if the field $\varphi$ is not
too large initially --- violations of this ``rule'' are just what
lead to the chaotic inflation scenario, which we discussed in
Chapter \ref{c1} and will examine further in Chapter \ref{c7}.
\index{Gauge fields, perturbation theory in|)}%
\index{Perturbation theory in gauge fields|)}%
\index{Symmetry!at high temperature|)}%
\index{Temperature!high, symmetry at|)}%


\chapter{\label{c4}Phase Transitions in Cold Superdense Matter}
\index{Phase transitions!in cold superdense matter|(}%
\index{Superdense matter!cold, phase transitions in|(}%

\section[Restoration of symmetry]{\label{s4.1}Restoration of symmetry
\index{Neutral currents!restoration of symmetry in theories with no|(}%
\index{Symmetry!restoration of, in theories with no neutral currents|(}%
in theories with no neutral\protect\\ currents}

In Chapters \ref{c2} and \ref{c3}, we studied phase transitions
in hot superdense matter, where the increase in density resulted
from an increase in temperature.  But it is also possible to
examine phase transitions in {\it cold} superdense matter, where
the density is increased by increasing the density of conserved
charge or the number of particles at zero temperature T.  In the
first papers in which this problem was studied it was claimed
that raising the density of cold matter would also result in the
restoration of symmetry [\cite{25}, \cite{26}].  The basic idea
\index{Fermions}%
in those papers was that the energy of fermions interacting with
a scalar field is proportional to
$g\,\varphi\,\langle \bar\psi\,\psi\rangle $.  When the fermion density
$j_0=\langle \bar\psi\,\gamma_0\,\psi\rangle $ increases, so does
$\langle \bar\psi\,\psi\rangle $, and states with $\varphi\neq0$
become energetically unfavorable.

As an example, consider the theory (\ref{1.1.13}) with the Lagrangian
\be
\label{4.1.1}
{\rm L}=\frac{1}{2}\,(\partial_\mu\varphi)^2+\frac{\mu^2}{2}\,\varphi^2
-\frac{\lambda}{4}\,\varphi^4+\bar\psi\,
(i\,\partial_\mu\gamma_\mu-h\,\varphi)\,\psi\ .
\ee
It is possible for fermions with $j_0\neq0$ to exist if they have
\index{Fermions}%
a chemical potential $\alpha$, which (for $\alpha\gg m_\psi=h\,\varphi$)
is related to $j_0$ [\cite{61}] by
\be
\label{4.1.2}
j_0=\frac{\alpha^3}{3\,\pi^2}\ .
\ee
To calculate the corrections to ${\rm V}(\varphi)$ which appear
due to the existence of the current $j_0$ (\ref{4.1.2}), we must
add $i\,\alpha$ to the component $k_4$ of the momentum of the
fermions when the one-loop contribution of fermions to ${\rm V}(\varphi)$
is computed [\cite{166}].  As a result, the equation
for the equilibrium value of the field $\varphi$ in the theory
(\ref{4.1.1}) takes the form [\cite{25}, \cite{26}]
\be
\label{4.1.3}
\frac{d{\rm V}}{d\varphi}=0
=\varphi\,\left[\lambda\,\varphi^2-\mu^2+\frac{h^2}{2}\,
\left(\frac{g\,j_0^2}{\pi^2}\right)^{1/3}\right]\ .
\ee
When $j_0=0$, we obtain
$\displaystyle \varphi=\pm\frac{\mu}{\sqrt{\lambda}}$,
\index{Fermions}%
as before.  But it is clear that the presence of fermions with
$j_0\neq0$ changes the effective value of $\mu^2$, and for
$j_0>j_c$, where
\be
\label{4.1.4}
j_c=\frac{2\,\pi\,\sqrt{2}}{3}\,\left(\frac{\mu}{h}\right)^3\ ,
\ee
symmetry is restored in the theory (\ref{4.1.1}).
\index{Neutral currents!restoration of symmetry in theories with no|)}%
\index{Symmetry!restoration of, in theories with no neutral currents|)}%

\section[Enhancement of symmetry breaking]%
{\label{4.2}Enhancement of symmetry breaking and the \protect\\
\index{Neutral currents!condensation of vector mesons in theories with|(}%
\index{Vector mesons!condensation of|(}%
\index{Symmetry breaking!enhancement of, in theories with neutral currents|(}%
condensation of vector mesons in theories with\protect\\
neutral currents}

Effects leading to the restoration of symmetry in the theory
(\ref{4.1.1}) appear only due to quantum corrections to
\index{Fermion current}%
${\rm V}(\varphi)$ for $\alpha\neq0$.  This is because the fermion
current $j_\mu=\langle \bar\psi\,\gamma_\mu\,\psi\rangle $ in the
model (\ref{4.1.1}) does not interact directly with any physical
fields.  At the same time [\cite{27}, \cite{24}], for realistic
theories with neutral currents, in which the fermion current
$j_\mu$ interacts with the neutral massive vector field
${\rm Z}_\mu$, an increase in the fermion density $j_0$ leads to an
enhancement of symmetry breaking, while the effects considered in
[\cite{25}, \cite{26}] are but minor quantum corrections relative
to the effects examined in [\cite{27}, \cite{24}].  Subsequent
study of this problem has shown that the effects appearing in
cold superdense matter do not simply amount to enhanced symmetry
breaking.  At high enough density, a condensate of charged vector
\index{Condensate!vector field}%
fields appears, and a redistribution of charge takes place among
bosons and fermions [\cite{28}, \cite{29}].

As an example, let us consider the effects that take place in the
\index{Glashow--Weinberg--Salam theory}%
Glashow--Weinberg--Salam theory with $\lambda\gg e^4$ in the
presence of a nonvanishing neutrino density $\displaystyle
n_\nu=\frac{1}{2}\,\langle \bar\nu_e\,\gamma_0\,(1-\gamma_5)\,\nu_e\rangle $.  The conserved
fermion density in this theory is
\be
\label{4.2.1}
l=\bar e\,\gamma_0\,e+\frac{1}{2}\,\bar\nu_e\,\gamma_0\,
(1-\gamma_5)\,\nu_e\ .
\ee
Clearly, given a lepton charge density $\langle l\rangle $, the
\index{Lepton charge density}%
\index{Fermion distribution}%
most energetically favorable fermion distribution is
$$
n_{e_{\rm R}}=\frac{1}{2}\,\langle \bar e\,\gamma_0\,(1+\gamma_5)\,e\rangle
=n_{e_{\rm R}}\,\frac{1}{2}\,\langle\bar e\,\gamma_0\,(1-\gamma_5)\,e\rangle
=n_\nu\ .
$$
This would imply the appearance of a large charge density of
electrons, however, which is only possible if some sort of
charge-canceling subsystem comes into being at the same time.
\index{Glashow--Weinberg--Salam theory}%
In the Glashow--Weinberg--Salam model, this subsystem may appear
in the form of a condensate of W\index{W bosons}\index{Bosons!W}
bosons.  Recall that in this
theory, there are three fields ${\rm A}_\mu^\alpha$, $\alpha=1$,
2, 3, and a field ${\rm B}_\mu$, from which --- after symmetry
breaking --- the electromagnetic field
\be
\label{4.2.2}
{\rm A}_\mu={\rm B}_\mu\,\cos\theta_{\rm W}+
{\rm A}^3_\mu\,\sin\theta_{\rm W}\ ,
\ee
the massive neutral field
\be
\label{4.2.3}
{\rm Z}_\mu={\rm B}_\mu\,\sin\theta_{\rm W}
-{\rm A}^3_\mu\,\cos\theta_{\rm W}\ ,
\ee
and the charged field
\be
\label{4.2.4}
{\rm W}^\pm_{\mu}=\frac{1}{\sqrt{2}}\,({\rm A}_\mu\mp{\rm A}_\mu^2)
\ee
are formed.  To be able to describe effects associated with
nonvanishing lepton density, we must append to the Lagrangian of
the theory a term $\alpha\,l$, where $\alpha$ is the chemical potential
corresponding to the lepton charge density.  The vector field
\index{Condensate!vector field}condensate
that arises at sufficiently high fermion density is of
the form
\ba
\label{4.2.5}
{\rm W}_1^\pm&=&{\rm C}\ ,\\
\label{4.2.6}
{\rm W}_0^\pm&=&{\rm W}_2^\pm={\rm W}_3^\pm={\rm A}_i^3= 0\ ,\\
\label{4.2.7}
{\rm A}_0^3&=&\pm\frac{\varphi}{2}\ ,
\ea
where C and $\varphi$ are determined by the equations
\ba
\label{4.2.8}
\left\langle \frac{\delta{\rm L}}{\delta{\rm A}_0^3}\right\rangle &=&
\frac{2\,e^2}{\sin\theta_{\rm W}}\,{\rm C}^2\,{\rm A}^3_0
+e\,(n_{e_{\rm L}}+n_{e_{\rm R}})= 0\ ,\\
\label{4.2.9}
\left\langle \frac{\delta{\rm L}}{\delta\varphi}\right\rangle &=&
\varphi\,\left(\mu^2-\lambda\,\varphi^2+
\frac{e^2\,{\rm Z}^2_0}{\sin^22\,\theta_{\rm W}}+
\frac{e^2\,{\rm C}^2}{2\,\sin^2\theta_{\rm W}}\right)=0\ ,
\ \mbox{\hspace*{10pt}}\\
\label{4.2.10}
\left\langle \frac{\delta{\rm L}}{\delta{\rm Z}_0}\right\rangle &=&
\frac{e^2\,\varphi^2\,{\rm Z}_0}{2\,\sin2\,\theta_{\rm W}}+
e\,(2n_{e_{\rm R}}+n_{e_{\rm L}}+2\,n_\nu)=0\ ,
\ea
$n_\nu$, $n_{e_{\rm R}}$, and $n_{e_{\rm L}}$ are given by
\ba
\label{4.2.11}
n_\nu&=&\frac{1}{6\,\pi^2}\,\left(\alpha
+\frac{e\,{\rm Z}_0}{\sin2\,\theta_{\rm W}}\right)^3\ ,\\
\label{4.2.12}
n_{e_{\rm R}}&=&\frac{1}{6\,\pi^2}\,(\alpha+e\,{\rm Z}_0\,\tan\theta_{\rm W}+
e\,{\rm A}_0)^3\ ,\\
\label{4.2.13}
n_{e{\rm L}}&=&\frac{1}{6\,\pi^2}\,(\alpha-e\,{\rm Z}_0\,\cot\theta_{\rm W}+
e\,{\rm A}_0)^3\ .
\ea
The solution ${\rm W}_1^\pm={\rm C}\neq0$ can only appear at high
enough lepton density, $n_{\rm L}=n_\nu+n_{e_{\rm R}}+n_{e_{\rm L}}$.
To determine the critical value $n_{\rm L}=n_{\rm L}^c$,
one should take into account that (as can be verified {\it
a posteriori}) at $n_{\rm L}\sim n_{\rm L}^c$ the field ${\rm Z}_0$
is of higher order in $e^2$ than are ${\rm A}_0^3$ or C.
In determining $n_{\rm L}^c$, we can therefore put ${\rm Z}_0=0$
in (\ref{4.2.8})--(\ref{4.2.13}), making subsequent analysis
quite simple.

Specifically, only the trivial solution ${\rm W}_1^\pm=0$ exists
at low  densities, and we deduce from
(\ref{4.2.8})--(\ref{4.2.13}) that $\displaystyle
{\rm A}_0^3={\rm A}_0\,\sin\theta_{\rm W} =
-\frac{\alpha}{e}\,\sin\theta_{\rm W}$, $n_{e_{\rm L}}=n_{e_{\rm R}}=0$.
Starting with $\displaystyle {\rm A}_0^3
=\frac{\alpha}{e}\,\sin\theta_{\rm W}
=\frac{\varphi_0}{2}$, the condensate solution
${\rm W}_1^\pm={\rm C}\neq0$ appears.  At that point
\be
\label{4.2.14}
n_\nu=n_{\rm L}^c=\frac{1}{6\,\pi^2}\,
\left(\frac{e\,\varphi_0}{2\,\sin\theta_{\rm W}}\right)^3
=\frac{M_{\rm W}^3}{6\,\pi^2}\ ,
\ee
where $M_{\rm W}$ is the mass of the
\index{W bosons}\index{Bosons!W}W boson.  With a further
increase in the fermion density, this solution becomes
energetically favored over the solution ${\rm C}=0$.  As the
reader can verify, this is because the energy required to create
the classical field ${\rm W}^\pm$ is small compared with the
energy gain achieved by redistributing the lepton charge among
the neutrinos and electrons, thereby reducing the Fermi energy of
the leptons.

Our final result is that both C and ${\rm Z}_0$ increase in
magnitude with increasing fermion density.  When the latter is
high enough, charged \index{W bosons}\index{Bosons!W}${\rm W}^\pm$
boson condensates appear. This then leads to an asymptotic
equalization of the partial densities of right- and left-handed
leptons (baryons) of various kinds in superdense matter:
$n_{\nu_e}=n_{e_{\rm R}}=n_{e_{\rm L}}$, $n_{\nu_\mu}=n_{\mu_{\rm
R}}=n_{\mu_{\rm L}}$, etc.  On the other hand, according to
(\ref{4.2.9}), the growth of both C and ${\rm Z}_0$ will lead to
growth of the field $\varphi$,  i.e., to the enhancement of
symmetry breaking between the weak and electromagnetic
interactions.

One should note that for a particular chemical composition of
cold superdense matter ($\displaystyle n_{\rm
B}=\frac{4}{3}\,n_{\rm L}$, where $n_{\rm B}$ and $n_{\rm L}$ are
the baryon and lepton densities, respectively), the fermion
matter proves to be neutral both with respect to the field ${\rm A}_0$
and the field ${\rm Z}_0$.  In that case, no W condensate
is formed in the superdense matter, and at high enough density,
the field $\varphi$ will tend to vanish [\cite{29}]. Interesting
nonperturbative effects may then come to the fore [\cite{174}].
For the time being, we have no idea what might cause this special
regime to be realized during expansion of the universe.  Strictly
speaking, this reservation also pertains to the more general case
$\displaystyle n_{\rm B}\neq\frac{4}{3}\,n_{\rm L}$ considered
above, the point being that at the present time
$n_{\rm L}\sim n_{\rm B}\ll n_\gamma$.
The\index{Density!neutrino}\index{Neutrino density}
neutrino density is not known
accurately, but in grand unified theories with nonconservation of
baryon charge, it is most reasonable to expect that at present
$n_{\rm L}\sim n_{\rm B}\ll n_\gamma$.  The dominant effects, at
least at very early stages in the evolution of the universe, are
then those that are related not to the chemical potential
$\alpha$ of cold fermionic matter, but to the temperature
${\rm T}\gg\alpha$.  It is possible in principle that effects
considered in this chapter may be important in the study of
certain intermediate stages in the evolution of the universe,
after which there was an abrupt increase in the specific entropy
$\displaystyle \frac{n_\gamma}{n_{\rm B}}$, as induced by
processes considered in [\cite{97}, \cite{98}, \cite{129}], for
example.  A combined study of both high-temperature effects and
effects related to nonvanishing lepton- and baryon-charge
density, as well as an investigation of concomitant
nonperturbative effects, can be found in a number of recent
papers on this topic;  for example, see
[\cite{130}, \cite{175}--\cite{178}].
\index{Neutral currents!condensation of vector mesons in theories with|)}%
\index{Vector mesons!condensation of|)}%
\index{Phase transitions!in cold superdense matter|)}%
\index{Superdense matter!cold, phase transitions in|)}%
\index{Symmetry breaking!enhancement of, in theories with neutral currents|)}%


\chapter{\label{c5}Tunneling Theory and the Decay of a Metastable Phase in a
\index{First-order phase transition!tunneling theory in|(}%
\index{Tunneling theory in first-order phase transition|(}%
First-Order Phase Transition}

\section[General theory: bubbles of a new phase]%
{\label{s5.1}General theory of the formation of bubbles of
a new phase\index{Bubble production!general theory of|(}}

One important and somewhat surprising property of field theories
with spontaneous symmetry breaking is that the lifetime of the
universe in an energetically unfavorable metastable vacuum state
can be exceptionally long.  This phenomenon forms the basis of
the first versions of the inflationary universe scenario,
according to which inflation takes place from a supercooled
metastable vacuum state (``false vacuum'') $\varphi=0$
\index{False vacuum}%
\index{Vacuum!false}%
[\cite{53}--\cite{55}].  This same phenomenon can lead to a
partitioning of the universe into enormous, exponentially
long-lived regions in different metastable vacuum states, each
corresponding to a different local minimum of the effective
potential.

For definiteness, we shall discuss the decay of the vacuum state
with $\varphi=0$ in the theory with the Lagrangian
\be
\label{5.1.1}
{\rm L}(\varphi)=\frac{1}{2}\,(\partial_\mu\varphi)^2-\vf\ ,
\ee
where the effective potential ${\rm V}(\varphi)$ has a local
minimum at $\varphi=0$ and a global minimum at
$\varphi=\varphi_0$.  Decay of the vacuum state with $\varphi=0$
proceeds via tunneling, with the formation of bubbles of the
field $\varphi$.  A theory of bubble production at zero
temperature was suggested in [\cite{179}], and was substantially
developed in [\cite{180}, \cite{181}], where the Euclidean
approach to the theory of the decay of a metastable vacuum state
was proposed.

We know from elementary quantum mechanics that the tunneling of a
particle through a one-dimensional potential barrier ${\rm V}(x)$
can be treated as motion with imaginary energy, or to  put it
differently, as motion in imaginary time, i.e., in Euclidean
space.  To generalize this approach to the case of tunneling
through the barrier $V(\varphi)$, one should consider the wave
\index{Wave functional}%
{\it functional} $\Psi(\varphi(x,t))$ in place of the particle
wave function $\psi(x,t)$, and investigate its evolution in
Euclidean space.  This generalization was proposed in
[\cite{180}, \cite{181}].  The Euclidean approach to tunneling
theory is simple and elegant, and it enables one to progress
rather far in calculating the decay probability of the false
vacuum.  We shall therefore refer below (without proof) to the
basic results obtained in [\cite{180}, \cite{181}], and to their
generalization to nonzero temperature [\cite{62}].  In subsequent
sections, these methods will be applied to the study of tunneling
in several specific theories.

As in conventional quantum mechanics, determination of the
\index{Tunneling probability}%
tunneling probability requires first of all that we solve the
classical equation of motion for the field $\varphi$ in Euclidean
space,
\be
\label{5.1.2}
\dla\varphi=\frac{d^2\varphi}{dt^2}+\Delta\varphi
=\frac{d{\rm V}}{d\varphi}\ ,
\ee
with boundary condition $\varphi\rightarrow0$ as
$x^2+t^2\rightarrow\infty$.  If we then normalize ${\rm V}(\varphi)$
so that ${\rm V}(0) = 0$ (that is, we redefine ${\rm V}(\varphi)$
by $\vf\rightarrow\vf-{\rm V}(0)$), the tunneling probability per
unit time and per unit volume will be given by
\be
\label{5.1.3}
\Gamma={\rm A}\,e^{-{\rm S}_4(\varphi)}\ ,
\ee
where ${\rm S}_4(\varphi)$
is the Euclidean action corresponding to the solution of Eq. (\ref{5.1.2})
\be
\label{5.1.4}
{\rm S}_4(\varphi)=\int d^4x\:
\left[\frac{1}{2}\,\left(\frac{d\varphi}{dt}\right)^2+
\frac{1}{2}\,(\nabla\varphi)^2+\vf\right]\ ,
\ee
and the factor A preceding the exponential is given by
\be
\label{5.1.5}
{\rm A}=\left(\frac{{\rm S}_4}{2\,\pi}\right)^2\,
\left(\frac{\det'[-\dla+{\rm V}''(\varphi)]}%
{\det[-\dla+{\rm V}''(0)]}\right)^{-1/2}\ .
\ee
Here  $\displaystyle {\rm V}''(\varphi)=\frac{d^2{\rm V}}{d\varphi^2}$,
and the notation ``$\det'$'' means that in calculating the
functional determinant of the operator $[-\dla+{\rm V}''(\varphi)]$,
its vanishing eigenvalues, corresponding to the so-called zero
modes of the operator, are to be omitted.  This operator has four
zero modes, corresponding to the possibility of translating the
solution $\varphi(x)$ along any of the four axes in Euclidean
space.  Contributions of
$\displaystyle \left(\frac{{\rm S}_4}{2\,\pi}\right)^{1/2}$
from each of the zero modes result in the factor
$\displaystyle \left(\frac{{\rm S}_4}{2\,\pi}\right)^2$ in (\ref{5.1.5}).

The derivation of Eqs. (\ref{5.1.3}) and (\ref{5.1.5}) may be
found in [\cite{181}], and is based on a calculation of the
imaginary part of the magnitude of the potential ${\rm V}(\varphi)$
at $\varphi=0$.  To a large extent, the equations
here are analogous to the corresponding expressions in the theory
\index{Instantons!Yang--Mills}%
\index{Yang--Mills instantons}%
of Yang--Mills instantons [\cite{182}].  In essence, the
solutions of Eq. (\ref{5.1.2}) with the indicated boundary
\index{Scalar instantons}%
\index{Instantons!scalar}%
conditions are scalar instantons in the theory (\ref{5.1.1}).  We
now wish to make several remarks before moving on to a
generalization of these results to ${\rm T}\neq0$.

First of all, notice that in order to obtain a complete answer,
it is necessary to sum over the contributions to $\Gamma$ from
all possible solutions of Eq. (\ref{5.1.2}).  Fortunately,
however, it is usually sufficient to limit consideration to the
simplest O(4)-symmetric solution $\varphi({\bf x}^2+t^2)$,
inasmuch as these are usually the very solutions that minimize
the action ${\rm S}_4$.  In that event, Eq.  (\ref{5.1.2}) takes
on an even simpler form,
\be
\label{5.1.6}
\frac{d^2\varphi}{dr^2}+\frac{3}{r}\,\frac{d\varphi}{dr}={\rm V}'(\varphi)\ ,
\ee
where $r=\sqrt{{\bf x}^2+t^2}$, with boundary conditions
$\varphi\rightarrow0$ as $r\rightarrow\infty$  and
$\displaystyle \frac{d\varphi}{dr}=0$  at $r=0$.

The high degree of symmetry inherent in the solution
$\varphi({\bf x}^2+t^2)$ helps us to obtain a graphic description
of the structure and evolution of a bubble of the field $\varphi$
after it is created.  To do so, we analytically continue the
solution to conventional time, $t\rightarrow i\,t$, or in other
words $\varphi({\bf x}^2+t^2)\rightarrow\varphi({\bf x}^2-t^2)$.
Since the solution $\varphi({\bf x}^2-t^2)$ depends only on the
invariant quantity ${\bf x}^2-t^2$, the corresponding bubble will
look the same in any reference frame, and the expansion speed of
a region filled with the field $\varphi$ (the speed of the bubble
``walls'' will asymptotically approach the speed of light.  The
creation and growth of bubbles is an interesting mathematical
problem [\cite{180}], but one that we shall not discuss here, as
our main objective is to study those situations in which the
probability of bubble creation is negligible.

Unfortunately, Eq. (\ref{5.1.6}) can seldom be solved
analytically, so that both the solution and the associated value
of the Euclidean action ${\rm S}_4(\varphi)$ must often be
computed numerically.  In this sort of situation, determinants
can only be calculated in certain special cases.  It turns out,
however, that in most practical problems just a rough estimate of
the pre-exponential factor A will suffice.  We can come up with
such an estimate by noting that the factor A has dimensionality
$m^4$, and its value is determined by three different quantities
with dimensionality $m$, namely $\varphi(0)$, $\sqrt{{\rm V}''(\varphi)}$,
and $r^{-1}$, where $r$ is the typical size of a bubble.  In the
theories that interest us most, all of these quantities lie
within an order of magnitude of one another, so for a rough
estimate one may assume that
\be
\label{5.1.7}
\frac{\det'[-\dla+{\rm V}''(\varphi)]}%
{\det[-\dla+{\rm V}''(0)]}={\rm O}(r^{-4},\varphi^4(0),({\rm
V}'')^2)\ , \ee where we denote by $r$ and ${\rm V}''(\varphi)$
typical mean values of these parameters for the solution
$\varphi(r)$ of Eq. (\ref{5.1.6}).
\begin{figure}[t]\label{f23}
\vskip-2cm \centering \leavevmode\epsfysize=12cm \epsfbox{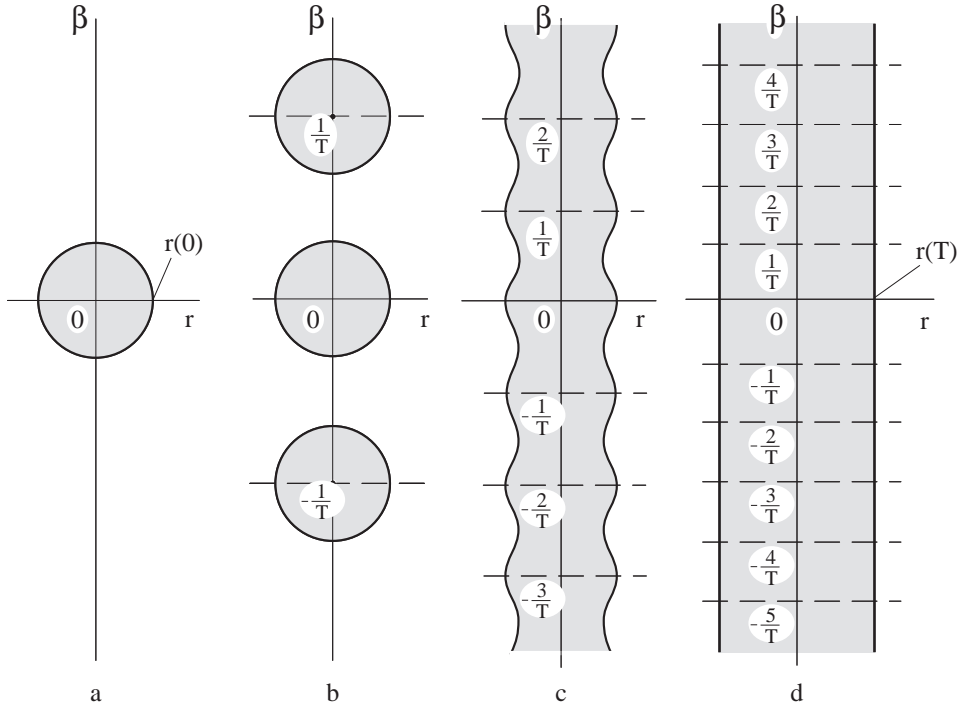}
\caption{The form taken by the solution to Eq. (\ref{5.1.2}) at
various temperatures:  a) ${\rm T}=0$; b) ${\rm T}\ll r^{-1}(0)$;
c) ${\rm T}\sim r^{-1}(0)$; d) ${\rm T}\gg r^{-1}(0)$.  The shaded
regions contain the classical field $\varphi\neq0$.  For
simplicity, we have drawn bubbles for those cases in which the
thickness of their walls is much less than their
radii.\vspace{10pt}
}
\end{figure}
Next, let us proceed to the case in which ${\rm T}\neq0$
[\cite{62}].  In order to generalize the preceding results to
this instance, it suffices to recall that the quantum statistics
of bosons (fermions) at ${\rm T}\neq0$ are formally equivalent to
quantum field theory in a Euclidean space with a periodicity
(antiperiodicity) of $1/{\rm T}$ in the ``time'' $\beta$ (see
[\cite{166}], for example).  When one considers processes at
fixed temperature, the quantity ${\rm V}(\varphi,{\rm T})$ plays
the role of the potential energy.  The imaginary part of this
function in an unstable vacuum can be calculated in just the same
way as is done in [\cite{181}] for the case ${\rm T}=0$.  The
only essential difference is that instead of finding an
O(4)-symmetric solution of Eq. (\ref{5.1.2}), one must find an
O(3)-symmetric (in the spatial coordinates) solution periodic in
the ``time'' $\beta$, with period $1/{\rm T}$.  As ${\rm T}\rightarrow0$,
the solution of Eq. (\ref{5.1.2}) that minimizes the action
${\rm S}_4(\varphi)$ consists of an O(4)-symmetric bubble with some
typical radius $r(0)$ (Fig. 5.1 a). As ${\rm T}\rightarrow r^{-1}(0)$,
the solution becomes a series of such bubbles separated from one
another by a distance $1/{\rm T}$ in the ``time'' direction (Fig.
5.1 b). When ${\rm T}\sim r^{-1}(0)$, the bubbles start to
overlap (Fig. 5.1 c).  Finally, when ${\rm T}\gg r^{-1}(0)$
(which is just the case that is of most importance and interest
to us), the solution becomes a cylinder whose spatial cross
section is an O(3)-symmetric bubble of some new radius $r({\rm T})$
(Fig. 5.1 d).

When we calculate ${\rm S}_4(\varphi)$ in the latter case, the
integration over $\beta$ reduces simply to a multiplication by
$1/{\rm T}$ --- that is, $\displaystyle {\rm
S}_4(\varphi)=\frac{1}{{\rm T}}\,{\rm S}_3(\varphi)$, where
${\rm S}_3(\varphi)$ is the three-dimensional action corresponding to
the O(3)-symmetric bubble,

\be
\label{5.1.8}
{\rm S}_3(\varphi)=\int d^3x\:
\left[\frac{1}{2}\,(\nabla\varphi)^2+{\rm V}(\varphi,{\rm T})\right]\ .
\ee
To calculate ${\rm S}_3(\varphi)$, we must solve the equation
\be
\label{5.1.9}
\frac{d^2\varphi}{dr^2}+\frac{2}{r}\,\frac{d\varphi}{dr}=
\frac{d{\rm V}(\varphi,{\rm T})}{d\varphi}={\rm V}'(\varphi,{\rm T})
\ee
with boundary conditions $\varphi\rightarrow0$ as
$r\rightarrow\infty$ and $\displaystyle \frac{d\varphi}{dr}=0$ as
$r\rightarrow0$.  In the high-temperature limit (${\rm T}\gg r^{-1}(0)$),
the complete expression for the tunneling
probability per unit time and per unit volume is obtained in a
manner completely analogous to that employed in [\cite{181}] to
derive Eqs. (\ref{5.1.4}) and (\ref{5.1.5}); the result is

\be \label{5.1.11} \Gamma({\rm T})\sim{\rm T}^4 ~
\exp\left(-\frac{{\rm S}_3(\varphi,{\rm T})}{{\rm T}}\right)\ .
\ee

It can be seen from Eq. (\ref{5.1.11}) that the main problem to be
solved in determining the probability of bubble creation is to
find ${\rm S}_3(\varphi, {\rm T})$ (or ${\rm S}_4$ at ${\rm
T}=0$).  Furthermore, if we are to obtain reasonable estimates of
the determinants, and wish to be able to study the expansion of
the bubbles that are formed, we must know the form taken by the
function $\varphi(r)$ and the typical size of a bubble.  As we
pointed out earlier, the corresponding results are obtained, as a
rule, by computer solution of the equations, which seriously
complicates the investigation of phase transitions in realistic
theories.  It is therefore of particular interest to study those
instances in which the problem can be solved analytically, and we
treat one such case in the next section. From here on, we shall
consider not only the case ${\rm T}\gg r^{-1}(0)$, but the case
${\rm T}=0$ as well, as the latter gives us information on the
probability of bubble creation in the limit of a strongly
supercooled metastable phase, where ${\rm T}\ll
r^{-1}(0)$\index{Bubble production!general theory of|)}.

\section{\label{s5.2}The
\index{Thin-wall approximation|(}%
thin-wall approximation}

In tunneling theory, there are two limiting cases in which the
problem simplifies considerably.  One of these is associated with
the situation where ${\rm V}(\varphi)$ at a minimum with
$\varphi=\varphi_0(\tau)\neq0$ is much larger in absolute value
than the height of the potential barrier in ${\rm V}(\varphi)$
between $\varphi=0$ and $\varphi=\varphi_0$; that case will be
taken up in the next chapter.  Here we examine the other limit,
in which $|{\rm V}(\varphi_0)|=\varepsilon$ is much lower than
the barrier height (see Fig. 5.2).

\begin{figure}[t]\label{f24}
 \centering \leavevmode\epsfysize=5cm \epsfbox{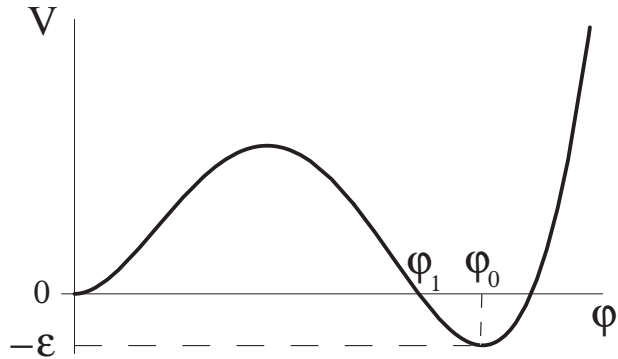}
\caption{Effective potential ${\rm V}(\varphi)$ in the case of
slight supercooling of the phase $\varphi = 0$ (i.e., the quantity
$\varepsilon={\rm V}(0,{\rm T})-{\rm V}(\varphi_0,{\rm T})$ is
small).}
\end{figure}
It is readily seen that as  decreases, the volume energy involved
in bubble creation ($\sim\varepsilon\, r^3$) becomes large compared
to the surface energy ($\sim r^2$) for large $r$ only if the
bubble is big enough.  When the size of a bubble greatly exceeds
the bubble wall thickness (the wall being that region where
derivatives $\displaystyle \frac{d\varphi}{dr}$  are large), one
can neglect the second term in (\ref{5.1.6}) and (\ref{5.1.9})
compared with the first. In other words, these equations
effectively reduce to one that describes tunneling in
one-dimensional space-time:
\be
\label{5.2.1}
\frac{d^2\varphi}{dr^2}={\rm V}'(\varphi,{\rm T})\ .
\ee
In the limit as $\varepsilon\rightarrow0$, the solution of this
equation takes the form
\be
\label{5.2.2}
r=\int^{\varphi_0}_\varphi\frac{d\varphi}{\sqrt{2\,\vf}}\ ,
\ee
where the functional form of $\varphi(r)$ has been sketched in
Fig. 5.3.

\begin{figure}[t]\label{f25}
\centering \leavevmode\epsfysize=6cm \epsfbox{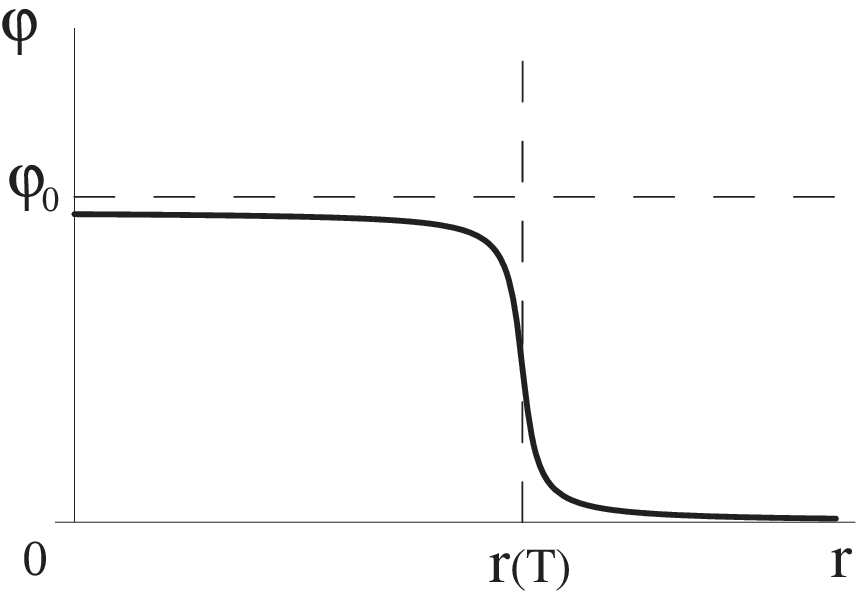}
\caption{Typical form of the solution of Eqs. (\ref{5.1.6}) and
(\ref{5.1.9}) when $\varepsilon \rightarrow0$.}
\end{figure}
Let us first consider tunneling in quantum field theory (${\rm T}=0$).
In an O(4)-symmetric bubble (\ref{5.2.2}), the action
${\rm S}_4$ is given by
\ba
\label{5.2.3}
{\rm S}_4&=&2\,\pi^2\,\int^\infty_0r^3\,dr\:
\left[\frac{1}{2}\,\left(\frac{d\varphi}{dr}\right)^2+{\rm V}\right]
\nonumber \\
&=&-\frac{\varepsilon}{2}\,\pi^2\,r^4+2\,\pi^2\,r^3\,{\rm S}_1\ ,
\ea
where ${\rm S}_1$ is the surface energy of the bubble wall
(surface tension), and is equal to the action in the
corresponding one-dimensional problem (\ref{5.2.1}):
\ba
\label{5.2.4}
{\rm S}_1&=&\int^\infty_0dr\:
\left[\frac{1}{2}\,\left(\frac{d\varphi}{dr}\right)^2+{\rm V}\right]
\nonumber \\
&=&\int^{\varphi_0}_0d\varphi\:\sqrt{2\,\vf}\ ,
\ea
the integral in (\ref{5.2.4}) being calculated in the limit as
$\varepsilon\rightarrow0$.  We find the\index{Bubble radius}
bubble radius $r(0)$ by minimizing (\ref{5.2.3}):
\be
\label{5.2.5}
r(0)=\frac{3\,{\rm S}_1}{\varepsilon}\ ,
\ee
whereupon
\be
\label{5.2.6}
{\rm S}_4=\frac{27\,\pi^2\,{\rm S}_1^4}{2\,\varepsilon^3}\ .
\ee
Notice that to order of magnitude, the bubble wall thickness is
simply $({\rm V}''(0))^{-1/2}$.  Taking (\ref{5.2.5}) into
account, therefore, the condition for the present approximation
(the so-called thin-wall approximation) to be valid is
\be
\label{5.2.7}
\frac{3\,{\rm S}_1}{\varepsilon}\gg({\rm V}''(0))^{-1/2}\ .
\ee
The foregoing results were derived by Coleman [\cite{180}].  We
can now readily generalize these results to the high-temperature
case, ${\rm T}\gg r^{-1}(0)$.  To do so, we merely point out that
\ba
\label{5.2.8}
{\rm S}_3&=&4\,\pi\,\int^\infty_0r^2\,dr\:
\left[\frac{1}{2}\,\left(\frac{d\varphi}{dr}\right)^2+
{\rm V}(\varphi,{\rm T})\right] \nonumber \\
&=&-\frac{4\,\pi}{3}\,r^3\,\varepsilon+4\,\pi\,r^2\,{\rm S}_1({\rm T})\ ,
\ea
so that
\be
\label{5.2.9}
r({\rm T})=\frac{2\,{\rm S}_1}{\varepsilon}
\ee
and
\be
\label{5.2.10}
{\rm S}_3=\frac{16\,\pi\,{\rm S}_1^3}{3\,\varepsilon^2}\ .
\ee
The expression thus obtained for the probability of bubble formation,
\be
\label{5.2.11}
\Gamma\sim
\exp\left(-\frac{16\,\pi\,{\rm S}_1^3}{3\,\varepsilon^2\,{\rm T}}\right)\ ,
\ee
is consistent with the well-known expression found in textbooks
[\cite{61}].  The only difference (but an important one) is that
we have the closed expression (\ref{5.2.4}) for the surface
tension ${\rm S}_1$, where instead of ${\rm V}(\varphi)$ one
should use ${\rm V}(\varphi,{\rm T})$.  In many cases of
interest, the function ${\rm V}(\varphi, {\rm T})$ plotted in
Fig. 5.2 can be approximated by the expression
\be
\label{5.2.12}
\vf=\frac{{\rm M}}{2}\,\varphi^2-\frac{\delta}{3}\,\varphi^3
+\frac{\lambda}{4}\,\varphi^4\ .
\ee

Let us investigate bubble formation in this theory in more
detail, since for the potential (\ref{5.2.12}) one can evaluate
the integral in (\ref{5.2.4}) exactly, and it thereby becomes
possible to obtain analytic expressions for ${\rm S}_1$, ${\rm S}_3$,
${\rm S}_4$ and $r({\rm T})$.

In fact, it can readily be demonstrated that for values of the
parameters M, $\delta$,  and $\lambda$ such that the minima at
$\varphi = 0$ and $\varphi=\varphi_0$ are of equal depth
($\varepsilon\rightarrow0$), Eq. (\ref{5.2.12}) becomes
\be
\label{5.2.13}
\vf=\frac{\lambda}{4}\,\varphi^2\,(\varphi-\varphi_0)^2\ ,
\ee
and in that event $\varphi_0$ is
\be
\label{5.2.14}
\varphi_0=\frac{2\,\delta}{\lambda}\ ,
\ee
while M, $\delta$,  and $\lambda$ are related by
\be
\label{5.2.15}
2\,\delta^2=9\,{\rm M}^2\,\lambda\ .
\ee
From (\ref{5.2.8}) and (\ref{5.2.13})--(\ref{5.2.15}), it follows that
\be
\label{5.2.16}
{\rm S}_1=\sqrt{\frac{\lambda}{2}}\,\frac{\varphi_0^3}{6}=
2^{3/2}\,3^{-4}\,\delta^3\,\lambda^{-5/2}\ ,
\ee
whereupon for ${\rm T}=0$ one obtains
\be
\label{5.2.17}
{\rm S}_4=
\frac{\pi^2\,2^5\,\delta^{12}}{3^{13}\,\lambda^{10}\,\varepsilon^3}\ ,
\qquad
r(0)=\frac{2^{3/2}\,\delta^3}{3^3\,\lambda^{5/2}\,\varepsilon}\ ,
\ee
while for ${\rm T}\gg r^{-1}(0)$,
\be
\label{5.2.18}
{\rm S}_3
=\frac{2^{17/2}\,\pi\,\delta^9}{3^{13}\,\lambda^{15/2}\,\varepsilon^2}\ ,
\qquad
r({\rm T})=\frac{2^{5/2}\,\delta^3}{3^4\,\lambda^{5/2}\,\varepsilon}\ .
\ee

We now turn to the specific case of phase transitions in gauge
theories at high temperature.  Here a typical expression for
${\rm V}(\varphi, {\rm T})$ is
\be
\label{5.2.19}
{\rm V}(\varphi,{\rm T})
=\frac{\beta\,({\rm T}^2-{\rm T}_{c_1}^2)}{2}\,\varphi^2
-\frac{\alpha}{3}\,{\rm T}\,\varphi^3+\frac{\lambda}{4}\,\varphi^4\ ,
\ee
where ${\rm T}_{c_1}$ is the temperature above which the
symmetric phase $\varphi = 0$ is metastable, and $\beta$ and
$\alpha$ are numerical coefficients (compare (\ref{3.1.21}),
(\ref{3.1.22})).  The temperature ${\rm T}_c$ at which the values
of ${\rm V}(\varphi, {\rm T})$ for the phases with $\varphi = 0$
and $\varphi=\varphi_0({\rm T})$ are equal is given by
\be
\label{5.2.20}
{\rm T}_c^2={\rm T}^2_{c_1}\,
\left(1-\frac{2\,\alpha^2}{9\,\beta\,\lambda}\right)^{-1}\ .
\ee
One can readily determine the quantity e as a function of the
departure of the temperature T from its equilibrium value:
\be
\label{5.2.21}
\varepsilon
=\frac{4\,{\rm T}_c\,{\rm T}^2_{c_1}\,\alpha^2\,\beta}{9\,\lambda^2}\,
\Delta{\rm T}\ ,
\ee
where $\Delta{\rm T}={\rm T}_c-{\rm T}$.  It is then
straightforward, using Eqs. (\ref{5.2.14})--(\ref{5.2.20}), to
derive expressions for the quantities of interest, namely ${\rm S}_3$
and $r({\rm T})$.  These may be written out for the most
frequently encountered situation, with
$\displaystyle x=\frac{{\rm T}_c-{\rm T}}{{\rm T}_c}\ll1$:
\ba
\label{5.2.22}
{\rm S}_4&=&\frac{{\rm S}_3}{{\rm T}}
=\frac{2^{9/2}\,\pi\,\alpha^5}{3^9\,\beta^2\,\lambda^{7/2}}\,
\frac{1}{x^2}\ ,\\
\label{5.2.23}
r&=&\sqrt{\frac{2}{\lambda}}\,\frac{\alpha}{9\,\beta\,{\rm T}_c}\,
\frac{1}{x}\ .
\ea

Thus, the thin-wall approximation makes it possible to progress
rather far towards an understanding of the formation of bubbles
of a new phase.  Unfortunately, however, this method can only be
applied to relatively slow phase transitions, or to be more
precise, those for which
\be
\label{5.2.24}
{\rm S}_4=\frac{{\rm S}_3}{{\rm T}}\ga10\,\alpha\,\lambda^{-3/2}\ .
\ee
This restriction is not satisfied in many cases of interest,
forcing us to seek ways of proceeding beyond the scope of the
thin-wall approximation.
\index{Thin-wall approximation|)}%

\section{\label{s5.3}Beyond the thin-wall approximation}

We have already remarked that there is one more instance in which
the theory of bubble creation may be considerably simplified.
Specifically, if the minimum of ${\rm V}(\varphi)$ at the point
$\varphi_0$ is deep enough, the maximum value of the field
$\varphi(r)$ corresponding to the solution of Eqs. (\ref{5.1.6})
and (\ref{5.1.9}) becomes of order $\varphi_1$, where
${\rm V}(\varphi_1)={\rm V}(0)$, $\varphi_1\ll\varphi_0$.  In solving
(\ref{5.1.6}) and (\ref{5.1.9}), one can then neglect the details
of the behavior of ${\rm V}(\varphi)$ for $\varphi\gg\varphi_1$,
and when $\varphi\la\varphi_1$, it is often possible to
approximate the potential ${\rm V}(\varphi)$ with one of two
basic types of functions:
\ba
\label{5.3.1}
{\rm V}^1(\varphi)&=&
\frac{{\rm M}^2}{2}\,\varphi^2-\frac{\lambda}{4}\,\varphi^4\ ,\\
\label{5.3.2}
{\rm V}^2(\varphi)&=&
\frac{{\rm M}^2}{2}\,\varphi^2-\frac{\delta}{3}\,\varphi^3\ .
\ea

At zero temperature and with ${\rm M}=0$, Eq. (\ref{5.1.6}) for
the theory (\ref{5.3.1}) can be solved exactly [\cite{182}],
\be
\label{5.3.3}
\varphi=\sqrt{\frac{8}{\lambda}}\,\frac{\rho}{r^2+\rho^2}\ ,
\ee
where $\rho$ is an arbitrary parameter with dimensionality of
length (the arbitrariness in the choice of $\rho$ is a
consequence of the absence of any mass parameter in the theory
(\ref{5.3.1}) for ${\rm M}=0$).  For all $\rho$, the action
corresponding to the solutions of (\ref{5.3.3}) is
\be
\label{5.3.4}
{\rm S}_4=\frac{8\,\pi^2}{3\,\lambda}\ .
\ee

To find the total probability of bubble formation, one must
integrate (with a certain weight) the contributions from
\index{Instantons}%
\index{Instantons!Yang--Mills}%
\index{Yang--Mills instantons}%
solutions (instantons) for all values of $\rho$, as in the theory
of Yang--Mills instantons [\cite{183}].

At ${\rm T}=0$ and arbitrary ${\rm M}\neq0$, Eq. (\ref{5.1.6}) in
the theory (\ref{5.3.1}) has no exact instanton solutions of the
type we have studied [\cite{184}], for the same reason that there
\index{Instantons}%
are no instantons in the theory of massive Yang--Mills fields.
On the other hand, for $\rho\ll{\rm M}^{-1}$, the solution of
(\ref{5.3.3}) is essentially insensitive to the presence of a
mass M in the theory (\ref{5.3.1}).  Therefore, for ${\rm T}=0$
and ${\rm M}\neq0$, the theory (\ref{5.3.1}) admits of ``almost
exact solutions'' of Eq. (\ref{5.1.6}) that are identical to
(\ref{5.3.3}) when $\rho\ll{\rm M}^{-1}$.  This means that an
entire class of trajectories (\ref{5.3.3}) exists in Euclidean
space that describes formation of a bubble of the field
$\varphi\neq0$.  To high accuracy, the action corresponding to
each of these trajectories in the theory (\ref{5.3.1}) with
${\rm M}\neq0$ is the same as (\ref{5.3.4}) when $\rho\ll{\rm M}^{-1}$,
and it tends to the minimum (\ref{5.3.4}) as $\rho\rightarrow0$.
As a result, tunneling does exist at ${\rm T}=0$ in the theory
(\ref{5.3.1}), and to describe it, one must integrate over $\rho$
the contributions to $\Gamma$ from all ``solutions''
(\ref{5.3.3}) with $\rho\ll{\rm M}^{-1}$, as is done in the
\index{Instantons}%
theory of instantons when the Yang--Mills field acquires mass
[\cite{183}].  In this somewhat inexact sense, we will speak of
solutions of Eq. (\ref{5.1.6}) in the theory (\ref{5.3.1}) with
${\rm M}\neq0$ and with the action (\ref{5.3.4}) (a similar
situation is investigated in [\cite{185}]).

\begin{figure}[t]\label{f26}
\centering \leavevmode\epsfysize=10cm \epsfbox{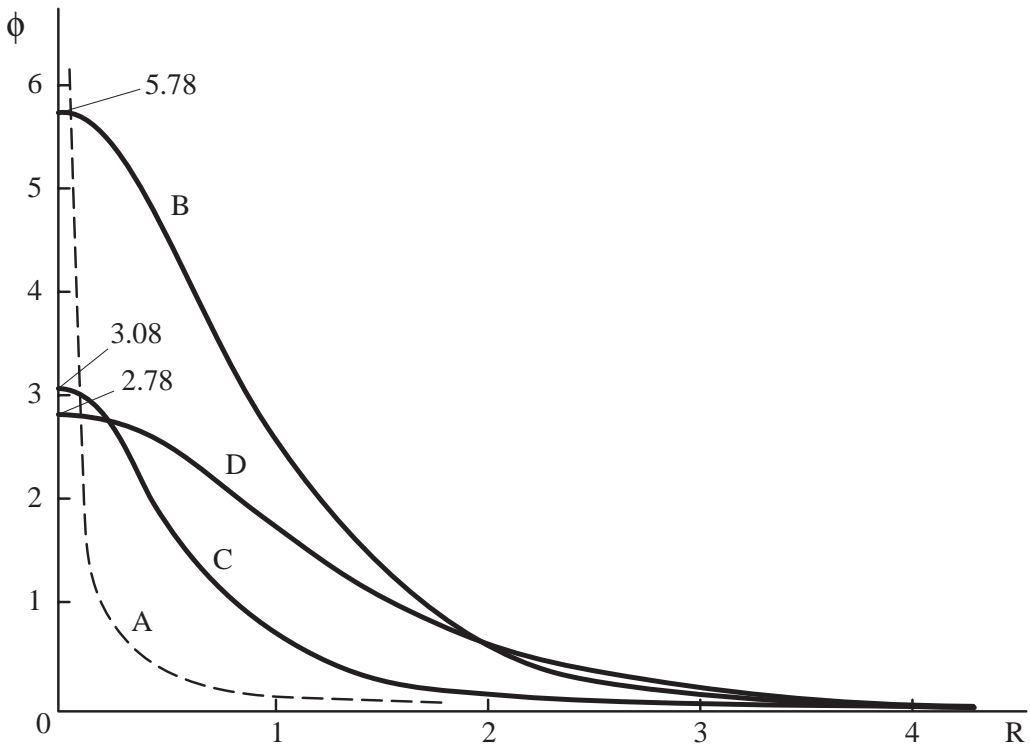} \caption{The
form of bubbles $\varphi(r)$ in the theories (\ref{5.3.1}) and
(\ref{5.3.2}) at ${\rm T} = 0$ and ${\rm T}\gg r^{-1}(0)$.  The
behavior of $\varphi$ as a function of $r$ has been plotted in
this figure in terms of the dimensionless variables ${\rm R} =
r\,{\rm M}$ and $\Phi=\varphi/\varphi_1$, where $\varphi_1$ is
defined by ${\rm V}(\varphi_1,{\rm T})={\rm V}(0,{\rm T})$. Curves
A and B are O(4)-symmetric bubbles in the theories (\ref{5.3.1})
and (\ref{5.3.2}) respectively;  curves C and D are O(3)-symmetric
bubbles in those same theories.}
\end{figure}
In all other cases under consideration (at high temperature in
the theory (\ref{5.3.1}), and at both high and low temperature in
the theory (\ref{5.3.2})), exact solutions do exist.  The form
taken by the solutions is shown in Fig.  5.4.  At
${\rm T}=0$, the action ${\rm S}_4$ corresponding to the solution
$\varphi(r)$ in the theory (\ref{5.3.2}) is
\be
\label{5.3.5}
{\rm S}_4(\varphi)\approx205\,\frac{{\rm M}^2}{\delta^2}\ .
\ee
In the high-temperature limit, the action
$\displaystyle {\rm S}_4(\varphi)=\frac{{\rm S}_3}{{\rm T}}$ for
the solutions corresponding to the theories (\ref{5.3.1}) and
(\ref{5.3.2}) is
\be
\label{5.3.6}
{\rm S}_4(\varphi)\approx\frac{19\,{\rm M}}{\lambda\,{\rm T}}
\ee
and
\be
\label{5.3.7}
{\rm S}_4(\varphi)\approx44\,\frac{{\rm M}^3}{\delta^2\,{\rm T}}
\ee
respectively.

Note that the results obtained above do not just refer to the
limiting cases ${\rm T}=0$ and ${\rm T}\gg{\rm M}$.  An analysis
of this problem shows that Eqs.  (\ref{5.3.4}) and (\ref{5.3.5})
continue to hold down to temperatures ${\rm T} \le 0.7\, {\rm M}$
(${\rm T} \le 0.2\, {\rm M}$), and at higher temperatures one can
make use of the results (\ref{5.3.6}) and (\ref{5.3.7})
[\cite{62}].

To conclude this chapter, let us consider briefly the most
typical case, in which the potentials ${\rm V}^1$ and ${\rm V}^2$
are of the form
\ba
\label{5.3.8}
{\rm V}^1(\varphi,{\rm T})&=&
\frac{\beta\,({\rm T}^2-{\rm T}^2_{c_1})}{2}\,\varphi^2
-\frac{\lambda}{4}\,\varphi^4\ ,\\
\label{5.3.9}
{\rm V}^2(\varphi,{\rm T})&=&
\frac{\beta\,({\rm T}^2-{\rm T}^2_{c_1})}{2}\,\varphi^2
-\frac{\alpha}{3}\,{\rm T}\,\varphi^3\ .
\ea
From the previous results, it follows that at high enough
temperature in the theory (\ref{5.3.8}),
\be
\label{5.3.10}
{\rm S}_4=
\frac{19\,\sqrt{\beta\,({\rm T}^2-{\rm T}^2_{c_1})}}{\lambda\,{\rm T}}\ ,
\ee
while in the theory (\ref{5.3.9}),
\be
\label{5.3.11}
{\rm S}_4=
\frac{44\,[\beta\,({\rm T}^2-{\rm T}^2_{c_1})]^{3/2}}{\alpha^2\,{\rm T}^3}\ .
\ee

In many realistic situations, the effective potential near a
phase transition point is well-approximated by one of the types
considered in Sections \ref{s5.2} and \ref{s5.3}. The results
obtained above may therefore often be directly applied to studies
of the kinetics of the first-order phase transitions in realistic
theories.  We shall use these results to analyze a number of
specific effects in Chapters \ref{c6} and \ref{c7}.

At this point, we would like to add two remarks in connection
with the foregoing results.  We see from Eqs.
(\ref{5.3.4})--(\ref{5.3.7}) that for certain values of the
parameters that enter into these equations, the probability that
a metastable phase will decay can be exceedingly low.  For
example, when $\lambda\sim10^{-2}$, tunneling in the theory
(\ref{5.3.1}) is suppressed by a factor
\be
\label{5.3.12}
{\rm P}\sim \exp\left(-\frac{8\,\pi^2}{3\,\lambda}\right)
\sim\exp(-10^3)\ .
\ee
This explains why in realistic theories, metastable vacuum states
can turn out to be almost indistinguishable from the stable one.
In particular, one has essentially no reason to think that the
vacuum state in which we now reside is the one corresponding to
the absolute minimum of energy.  One might try, in principle, to
carry out an experiment to test the stability of our vacuum by
attempting to create a nucleus of a new phase (e.g., via heavy
ion collisions), but both the technological feasibility and,
understandably, the advisability of such an experiment are highly
dubious.\footnote{One might argue, of course, that such an experiment
would be quite enlightening, regardless of the results. If the experiment
were to confirm the stability of our vacuum state, it would make us
all very proud. On the other hand, if a bubble of a more energetically
advantageous vacuum state were produced, the observable part of the
universe would gradually be transformed into a better vacuum state,
and no observes would remain to be dissatisfied with the experimental
results.}

The second remark bears on the range of applicability of the
foregoing results.  These were obtained by neglecting effects
associated with the expansion of the universe, an approximation
that is perfectly adequate if the curvature ${\rm V}''(\varphi)$
of the effective potential is much greater than the curvature
tensor ${\rm R}_{\mu\nu\alpha\beta}$.  But in the inflationary
universe scenario, ${\rm V}''(\varphi)\ll{\rm R}=12\,{\rm H}^2$
during inflation.  Tunneling during inflation must therefore be
studied as a separate issue.  We shall return to this question in
Chapter \ref{c7}.
\index{First-order phase transition!tunneling theory in|)}%
\index{Tunneling theory in first-order phase transition|)}%


\chapter{\label{c6}Phase Transitions in a Hot Universe}
\index{Phase transitions!in hot universe|(}%
\index{Hot universe!phase transitions in|(}%

\section[Phase transitions with symmetry breaking]%
{\label{s6.1}Phase transitions with symmetry breaking between the
\index{Phase transitions!with symmetry breaking|(}%
\index{Symmetry breaking!phase transitions with|(}%
weak, strong, and electromagnetic interactions}

We have already pointed out in Chapter \ref{c1} that according to
\index{Universe!expansion of}%
the standard hot universe theory, the expansion of the universe
started from a state of enormously high density, at a temperature
T much higher than the critical temperature of a phase transition
\index{Grand unified theories!phase transition in}%
\index{Phase transitions!in grand unified theories}%
with symmetry restoration between the strong and electroweak
interactions in grand unified theories.  Therefore, the symmetry
between these interactions should have been restored in the very
early stages of the evolution of the universe.

As the temperature decreases to ${\rm T}\sim{\rm
T}_{c_1}\sim10^{14}$--$10^{15}$ GeV (see Eq. (\ref{3.2.9})), a phase
transition (or several) takes place, generating a classical
scalar field $\Phi\sim10^{15}$ GeV, which breaks the symmetry
between the strong and electroweak interactions.  When the
temperature drops to ${\rm T}_{c_2}\sim200$ GeV, the symmetry between the weak and electromagnetic
interactions breaks.  Finally, at ${\rm T}\sim10^2$ MeV, there
should be a phase transition (or two separate transitions) which
breaks the chiral invariance of the theory of strong interactions
and leads to the coalescence of quarks into hadrons
(confinement).

Here we must voice some reservations.  The
\index{Glashow--Weinberg--Salam theory}%
Glashow--Weinberg--Salam theory of electroweak interactions has
withstood experimental tests quite well, but the situation with
grand unified theories is not nearly so satisfactory.  Prior to
the 1980's, there seemed to be little doubt of the existence of
grand unification at energies ${\rm E}\sim10^{15}$ GeV, with the
\index{Unified theory}%
most likely candidate for the role of a unified theory being
minimal SU(5).  Subsequently, unified theories became more and
more complicated, starting with ${\rm N}=1$ supergravity, then
the Kaluza--Klein theory, and finally superstring theory.  As the
theories have changed, so has our picture of the evolution of the
universe at high temperatures.  But all versions of this picture
have at least one thing in common:  without an inflationary
stage, they all lead to consequences in direct conflict with
existing cosmological data.  In the present section, in order to
expose the sources of these problems and point out some
possibilities for overcoming them, we will study the kinetics of
phase transitions in minimal SU(5) theory.

In that theory, the potential in the field $\Phi$ responsible for
symmetry breaking between the strong and electroweak interactions
takes the form (see Section \ref{s3.2})
\be
\label{6.1.1}
{\rm V}(\Phi)=-\frac{\mu^2}{2}\,\tr\Phi^2
+\frac{a}{4}\,(\tr\Phi^2)^2+\frac{b}{2}\,\tr\Phi^4\ .
\ee
At ${\rm T}\gg\mu$, the main modification to ${\rm V}(\Phi)$
consists of a change of sign of the effective parameter $\mu^2$,
\be
\label{6.1.2}
\mu^2({\rm T})=\mu^2-\beta\,{\rm T}^2\ ,
\ee
see (\ref{3.2.6}).  This leads to the restoration of symmetry at
high temperatures.  According to (\ref{3.2.6}), however, at
${\rm T}\la\mu$ the modification of the effective potential does not
reduce to a change in $\mu^2$; the effective potential
${\rm V}(\Phi, {\rm T})$ can acquire additional local minima that
correspond not just to $\mbox{SU}(3) \times \mbox{SU}(2) \times {\rm U}(1)$
symmetry breaking (see Chapter \ref{c1}), but also to symmetry
breaking described by the groups $\mbox{SU}(4)\times{\rm U}(1)$,
$\mbox{SU}(3)\times(\mbox{SU}(1))^2$, or
$(\mbox{SU}(2))^2\times(\mbox{SU}(1))^2$ [\cite{167}].  This,
plus the fact that phase transitions in grand unified theories
are first-order transitions, greatly complicates investigation of
the kinetics of the transition from the SU(5) phase to the
$\mbox{SU}(3) \times \mbox{SU}(2) \times {\rm U}(1)$ phase.  Here
we present the main results from this investigation [\cite{187}].

First of all, recall that according to [\cite{167}], the
effective potential ${\rm V}(\varphi,{\rm T})$ of the minimal
SU(5) theory takes the form
\be
\label{6.1.3}
{\rm V}(\varphi,{\rm T})=-\frac{{\rm N}\,\pi\,{\rm T}^4}{90}
-\frac{\mu^2({\rm T})}{2}\,\varphi^2
-\alpha_i\,{\rm T}\,\varphi^3+\gamma_i\,\varphi
\ee
for each of the four types of symmetry breaking mentioned above,
where $\varphi^2=\tr\Phi^2$, and $\alpha_i$ and $\gamma_i$, $i = 1$,
2, 3, 4 are certain constants calculated in [\cite{167}].
This effective potential is the same as the potential
(\ref{5.2.12}), so that all of the results we obtained in the
thin-wall approximation concerning tunneling from a state
$\varphi\neq0$, with formation of bubbles of a field
$\varphi\neq0$, also apply to the theory (\ref{6.1.3}).  On the
other hand, in those cases where the thin-wall approximation does
not work, the field $\varphi$ within a bubble is small, the last
term in (\ref{6.1.3}) can be discarded, and the potential is the
same as (\ref{5.3.2}), for which we also studied tunneling in
Chapter \ref{c5}.

Our plan of attack is thus as follows.  We must understand how
the quantity ${\rm V}(\varphi,{\rm T})$ in (\ref{6.1.3}) depends
on time in an expanding universe, calculate the rate of
production of each of the four types of bubbles enumerated above,
determine the moment at which the bubbles thus formed occupy the
whole universe, explore what happens to bubbles formed in earlier
stages, and find the typical volume occupied by regions that are
filled with the various phases at the end of the whole process.

Since we have already developed the theory of bubble formation,
the solution of the foregoing problem should not be particularly
difficult.  Nevertheless, it does turn out to be a fairly tedious
task, since the numerical calculations must be rerun for every
new choice of parameters $a$ and $b$ in (\ref{6.1.1}).  Below we
present and discuss the main results that we have obtained for
the most natural case, $a\sim b\sim 0.1$ [\cite{187}].

For this case, the phase transition takes place from a
supercooled state in which the temperature of the universe
approaches ${\rm T}_{c_1}$; starting with this temperature, the
symmetric phase $\varphi$ becomes absolutely unstable.  The jump
in the field $\varphi$ at the phase transition point is then
large (of order $\varphi_0$).  In that sense, the phase
transition is a ``strong'' first-order transition.

The phase transition proceeds with the simultaneous production of
all four types of phases listed above, the overwhelming majority
of the bubbles containing the $\mbox{SU}(4) \times {\rm U}(1)$
phase, and not the energetically more favorable
$\mbox{SU}(3) \times \mbox{SU}(2) \times {\rm U}(1)$ phase, which
initially occupies only a few percent of the whole volume.
$\mbox{SU}(3) \times \mbox{SU}(2) \times {\rm U}(1)$ bubbles
eventually start expanding within the
$\mbox{SU}(4) \times {\rm U}(1)$ phase, ``devouring'' both it and
the bubbles of the other two phases.  At such time as the
$\mbox{SU}(3) \times \mbox{SU}(2) \times {\rm U}(1)$ bubbles
coalesce, they have a typical size of
\be
\label{6.1.4}
r\sim {\rm T}^{-1}_{c_1}\ .
\ee

Prior to the formation of a homogeneous
$\mbox{SU}(3) \times \mbox{SU}(2) \times {\rm U}(1)$ phase, the
kinetics of processes during the intermediate phase is very
complex, depending on the values of $a$, $b$, and $g^2$.  The
duration of this intermediate stage, as well as that of the stage
preceding the end of the phase transition, can only be
significant in theories with certain specific relations between
the coupling constants.

Despite the large jump in the field $\varphi$ at the phase
transition point, the amount of energy liberated in the phase
transition process is relatively minor, as a rule, so that given
the most reasonable values of the coupling constants, a
symmetry-breaking transition from a supercooled SU(5)-symmetric
phase will not result in a discontinuous rise in temperature, nor
will it produce a marked increase in the total entropy of the
expanding universe.

As the temperature drops further to ${\rm T}_{c_2}\sim10^2$ GeV,
the phase transition
$\mbox{SU}(3) \times \mbox{SU}(2) \times {\rm U}(1) \rightarrow
\mbox{SU}(3) \times {\rm U}(1)$ takes place, and with it the
\index{Symmetry!between weak and electromagnetic interactions}%
\index{Weak and electromagnetic interactions, symmetry broken between}%
symmetry between the weak and electromagnetic interactions is
broken.  At the time of this transition, the temperature is many
orders of magnitude lower than the mass of the superheavy bosons
with ${\rm M}_{\rm X}\sim10^{14}$ GeV that appear after the first
phase transition.  Lighter particles in this theory are described
\index{Glashow--Weinberg--Salam theory}%
by the Glashow--Weinberg--Salam theory, so the phase transition
at ${\rm T}_{c_2}\sim10^2$ GeV proceeds exactly as in the latter;
see Chapter \ref{c3}.

Generally speaking, the foregoing pattern of phase transitions is
only relevant to the simplest grand unified theories with the
most natural relation between coupling constants.  In more
complicated theories, phase transitions may occur with many more
steps;  for example, see [\cite{42}, \cite{167}].  A somewhat
unusual picture also emerges for certain special relations among
the parameters of a theory, for which imply the effective
potential of scalar fields contains a local minimum or a
relatively flat region at small $\varphi$.

By way of example, let us consider the Glashow--Weinberg--Salam
\index{Glashow--Weinberg--Salam theory}%
model with
\ba
\label{6.1.5}
\lambda(\varphi_0)=\frac{1}{6}\,
\frac{d^4{\rm V}}{d\varphi^4}\biggr|_{\varphi=\varphi_0}\,
\!\!\!\!&<&\!\!\!\!
\frac{11\,e^4}{16\,\pi^2}\cdot
\frac{2\,\cos^4\theta_{\rm W}+1}{\sin^22\,\theta_{\rm W}}
\approx3\cdot 10^{-3}\ ,\\
\label{6.1.6}
m_\varphi^2(\varphi_0)=
\frac{d^2{\rm V}}{d\varphi^2}\biggr|_{\varphi=\varphi_0}
\!\!\!\!&<&\!\!\!\!
\frac{e^4\,\varphi_0^2}{16\,\pi^2}\cdot
\frac{2\,\cos^4\theta_{\rm W}+1}{\sin^22\,\theta_{\rm W}}\nonumber \\
\!\!\!\!&\approx&\!\!\!\!(10\;\mbox{GeV})^2\ ,
\ea
where $\sin^2\theta_{\rm W}\approx0.23$, $\varphi_0\approx250$
GeV.  For these values of $\lambda(\varphi_0)$ and $m_\varphi^2$,
the effective potential ${\rm V}(\varphi)$ has a local minimum at
$\varphi=0$ even at zero temperature [\cite{139}--\cite{141}];
see Section \ref{s2.2}.

In that case, symmetry was restored in the early universe as
usual, with $\varphi=0$.  As the universe cooled, a minimum of
${\rm V}(\varphi)$ then appeared at $\varphi\sim\varphi_0$,
becoming deeper than the one at $\varphi=0$ shortly thereafter.
Nevertheless, the universe remained in the state $\varphi=0$
until such time as bubbles of a new phase with $\varphi\neq0$
formed and filled the entire universe.  The formation of bubbles
\index{Glashow--Weinberg--Salam theory}%
of a new phase in the Glashow--Weinberg--Salam theory was studied
in [\cite{141}, \cite{142}].  It turns out that if $m_\varphi$ is
even one percent less than the limiting value $m_\varphi\sim10$
GeV (\ref{6.1.6}), the probability of bubble formation with
$\varphi\neq0$ becomes exceedingly small.

The reason for this is not far to seek if we hark back to the
results of the previous chapter.  Consider the limiting case
\be
\label{6.1.7}
m_\varphi^2=\frac{e^4\,\varphi_0^2}{16\,\pi^2}\cdot
\frac{2\,\cos^4\theta_{\rm W}+1}{\sin^22\,\theta_{\rm W}}\ .
\ee
The curvature of ${\rm V}(\varphi)$ at $\varphi=0$, ${\rm T}=0$
then tends to zero (the\index{Coleman--Weinberg theory}
Coleman--Weinberg model [\cite{137}];
see Section \ref{s2.2}).  At ${\rm T}\neq0$, the mass of the
scalar field in the vicinity of $\varphi=0$ is, according to
(\ref{3.2.1}),
\be
\label{6.1.8}
m_\varphi\sim\frac{e\,{\rm T}}{\sin2\,\theta_{\rm W}}\,
\sqrt{1+2\,\cos^2\theta_{\rm W}}
\ee
(recall that in the present case $\lambda\sim e^4\ll e^2$).  In
this model, at small $\varphi$, the potential ${\rm V}(\varphi)$
is approximately
\be
\label{6.1.9}
\vf={\rm V}(0)+\frac{3\,e^4\,\varphi^4}{32\,\pi^2}\,
\left(\frac{2\,\cos^4\theta_{\rm W}+1}{\sin^22\,\theta_{\rm W}}\right)\,
\ln\frac{\varphi}{\varphi_0}+\frac{m_\varphi^2\,\varphi^2}{2}\ .
\ee
Now $\displaystyle \ln\frac{\varphi}{\varphi_0}$
is a fairly slowly varying function of $\varphi$, so to determine the
probability P of tunneling out of the local minimum at $\varphi=0$,
we can make use of Eq. (\ref{5.3.6}) [\cite{144}]:
\ba
\label{6.1.10}
{\rm P}&\sim&
\exp\left(-\frac{19\,m_\varphi({\rm T})}{\lambda\,{\rm T}}\right)
\sim\exp\left(-\frac{19\,\sin2\,\theta_{\rm W}}{\displaystyle
\frac{3\,e^3}{8\,\pi^2}
\sqrt{1+\cos^4\theta_{\rm W}}\,
\ln\frac{\varphi}{\varphi_0}}\right)\nonumber \\
&\sim&\exp\left(-\frac{15000}{\displaystyle
\ln\frac{\varphi}{\varphi_0}}\right)\ .
\ea
The typical value of the field $\varphi$ appearing in
(\ref{6.1.10}) corresponds to a local maximum of ${\rm V}(\varphi)$
in (\ref{6.1.9}) located at $\varphi\sim 10\,{\rm T}$;  that is,
\be
\label{6.1.11}
{\rm P}\sim\exp\left(-\frac{15000}{\displaystyle
\ln\frac{{\rm T}}{\varphi_0}}\right)\ .
\ee
Hence, we find that in the theory under consideration, the phase
transition in which bubbles of the field $\varphi$ are formed can
only take place if the temperature of the universe is
exponentially low.  A similar phenomenon in the
\index{Coleman--Weinberg theory}%
\index{Glashow--Weinberg--Salam theory}%
Coleman--Weinberg SU(5) theory lays at the
basis of the new inflationary universe
scenario (see Chapter \ref{c8}).  But in the Glashow--Weinberg--Salam
theory with
$$
\frac{d^2{\rm V}}{d\varphi^2}\biggr|_{\varphi=0}=0\ ,
$$
supercooling is actually not so strong
as might be construed from\linebreak[10000]
(\ref{6.1.11}):  the phase transition occurs at ${\rm T}\sim10^2$
MeV on account of effects associated with strong interactions
[\cite{144}].  When it takes place, the specific entropy of the
universe $\displaystyle \frac{n_\gamma}{n_{\rm B}}$ should rise
approximately $10^5$--$10^6$-fold [\cite{144}], due to liberation
of the energy stored in the metastable vacuum $\varphi=0$.  Even
if the effective potential ${\rm V}(\varphi)$  has only a very
shallow minimum at $\varphi=0$, the increase in the specific
entropy of the universe may become unacceptably large
[\cite{143}, \cite{144}].  Furthermore, the lifetime of the
\index{Lifetime of universe}%
\index{Universe!lifetime of}%
universe in a metastable vacuum state with
${\rm V}''(0)\ga(10^2\;\mbox{MeV})^2$ will be greater than the
age of the observable part of the universe, $t\sim10^{10}$ yr
[\cite{141}, \cite{142}].  Bubbles formed as a result of such a
phase transition would make the universe strongly anisotropic and
inhomogeneous.  The universe would be homogeneous only inside
each of the bubbles, which would be devoid of matter of any kind.
This leads to the strong constraint (\ref{2.2.14}) on the mass of
\index{Glashow--Weinberg--Salam theory}%
the Higgs boson in the Glashow--Weinberg--Salam theory without
\index{Fermions!superheavy}%
\index{Superheavy fermions}%
superheavy fermions\footnote{To avoid misunderstanding, we
should emphasize that these bounds refer only to the simplest version
of the Glashow--Weinberg--Salam theory, with a single type of scalar
field $\varphi$}:
\be
\label{6.1.12}
m_\varphi\ga 10\;\mbox{GeV}\ .
\ee

As we showed in Chapter \ref{c2}, the absolute minimum of
${\rm V}(\varphi)$ in a theory with superheavy fermions may turn
out not to be at
$\displaystyle \varphi=\varphi_0=\frac{\mu}{\sqrt{\lambda}}$, but at
$\varphi\gg\varphi_0$, which constrains the allowable fermion
masses in the theory [\cite{146}--\cite{151}].  When cosmological
effects are taken into account, the corresponding bounds are
softened somewhat, since the universe will not always succeed in
going from a state $\varphi=\varphi_0$ to an energetically more
favorable one [\cite{188}].  The complete set of bounds on the
masses of fermions and the Higgs boson, including the
cosmological constraints, is shown in Fig. 2.5 (Chapter
\ref{c2}, page \pageref{f16}).  Notice, however, that in studying
tunneling, the authors of [\cite{141}--\cite{151}, \cite{188}]
did not discuss the possibility of tunneling induced by
collisions of cosmic rays with matter.  If such processes could
substantially increase the probability of decay of a metastable
vacuum [\cite{189}] (see, however, [\cite{360}]), then the region
above the curve AD in Fig. 2.5 would turn out to be
forbidden, and the most stringent constraint on $m_\varphi$ would
be set by Eq. (\ref{2.2.9}).  This problem requires more detailed
investigation.
\index{Phase transitions!with symmetry breaking|)}%
\index{Symmetry breaking!phase transitions with|)}%

\section{\label{s6.2}Domain walls, strings, and monopoles}

In the preceding section, we pointed out that a phase transition
with SU(5) symmetry breaking takes place with the formation of
bubbles containing several different phases, and only
subsequently does all space fill with matter in a single
energetically most favorable phase.  For this to happen, at least
two conditions must be satisfied:  only one energetically most
favorable phase may exist, and the typical size $r$ of the
bubbles must not exceed $t$, where $t$ is the time at which the
entire universe should have made the transition to a single
phase.  In the hot universe theory (in contrast to the
inflationary universe theory), bubbles typically do not grow to
be very large --- $r\sim m^{-1}$  or $r\sim{\rm T}^{-1}$ --- so the second requirement
is usually met.  But there are a great many theories in which the
effective potential has several minima of the same (or almost the
same) depth.  The simplest example is the theory (\ref{1.1.5}),
which has minima at $\displaystyle \varphi=\frac{\mu}{\sqrt{\lambda}}$
and $\displaystyle \varphi=-\frac{\mu}{\sqrt{\lambda}}$ of equal
depth.  When a phase transition occurs at some time $t=t_c$
during the expansion of the universe, symmetry breaking takes
place independently in different causally disconnected regions of
size ${\rm O}(t_c)$.  As a result, the universe is partitioned
into approximately equal numbers of regions filled with the
fields $\displaystyle \varphi=\frac{\mu}{\sqrt{\lambda}}$ and
$\displaystyle \varphi=-\frac{\mu}{\sqrt{\lambda}}$.  These
regions are separated from one another by domain walls of
\index{Domain walls}%
thickness ${\rm O}(\mu^{-1})$, with the field changing from
$\displaystyle \varphi=\frac{\mu}{\sqrt{\lambda}}$ to
$\displaystyle \varphi=-\frac{\mu}{\sqrt{\lambda}}$ from one side
of the wall to the other.

Actually, as a rule, the regions in which symmetry breaking takes
place independently initially have sizes of order ${\rm T}_c^{-1}$;
that is, they have dimensions much smaller than the horizon
$\displaystyle t\sim10^{-2}\,\frac{\m}{{\rm T}_c^2}$ at the time
when the phase transition starts.  One example of this is the
formation of  regions with different phases at the time of the
phase transition in the SU(5) model;  see (\ref{6.1.4}).

Regions that are filled with different phases at the same energy
density also tend to ``eat'' each other, as the presence of
\index{Domain walls}%
domain walls is energetically unfavorable.  But this mutual
consumption proceeds independently in regions separated by
distances of order $t$, where $t$ is the age of the universe.  As
we have already noted in Section \ref{s1.5}, at time $t\sim10^5$
yr, the presently observable part of the universe consisted of
approximately $10^6$ causally disconnected regions, or, in other
\index{Domain walls}%
words, of $10^6$ domains separated by superheavy domain walls.
Since in the last $\sim10^5$ years the observable part of the
universe has been transparent to photons, the existence of such
domains would lead to considerable anisotropy in the primordial
background radiation.  The astronomical observations, however,
indicate that the background radiation is isotropic to within
$\displaystyle \frac{\Delta{\rm T}}{{\rm T}}\sim3\cdot 10^{-5}$.
This is the essence of the domain wall problem in the hot
universe theory [\cite{41}].  These results would force us to
renounce theories with discrete symmetry breaking, such as the
theory (\ref{1.1.5}), theories with spontaneously broken CP
invariance, the minimal SU(5) theory, in which the potential
${\rm V}(\Phi)$ (\ref{6.1.1}) is invariant under reflection
$\Phi \rightarrow-\Phi$, etc.  Most theories of the axion field
$\theta$ encounter similar difficulties:  the potential
${\rm V}(\theta)$ in many versions of the axion theory has several
minima of the same depth [\cite{49}].  In some theories, this
difficulty can be overcome (for example, by adding a term
$c\,\tr\Phi^3$ to ${\rm V}(\Phi)$ (\ref{6.1.1})), but usually the
problems are insurmountable without changing the theory
fundamentally (or reverting to the inflationary universe
scenario).

Besides domain walls, phase transitions can give rise to other
\index{Domain walls}%
nontrivial entities as well.  Consider, for example, a model of a
complex scalar field $\chi$ with the Lagrangian
\be
\label{6.2.1}
{\rm L}=\partial_\mu\chi^*\,\partial_\mu\chi+m^2\,\chi^*\,\chi
-\lambda\,(\chi^*\,\chi)^2\ .
\ee
This is the Higgs model of (\ref{1.1.15}) prior to the inclusion
\index{Higgs model!symmetry breaking in}%
\index{Symmetry breaking!in Higgs model}%
of the vector fields ${\rm A}_\mu$.  In order to study symmetry
breaking in this theory, it is convenient to change variables:
\be
\label{6.2.2}
\chi(x)\rightarrow\frac{1}{\sqrt{2}}\,\varphi(x)\,
\exp\left(\frac{i\,\zeta(x)}{\varphi_0}\right)\ .
\ee
The effective potential ${\rm V}(\chi, \chi^*)$ has a minimum at
$\displaystyle \varphi(x)=\varphi_0=\frac{\mu}{\sqrt{\lambda}}$
irrespective of the value of the constant part of the phase
$\zeta_0$.  ${\rm V}(\chi, \chi^*)$ is thus shaped like the
bottom of a basin, with a maximum in the middle (at $\chi(x)=0$),
and rather than being characterized simply by the scalar
$\varphi_0$, symmetry breaking is characterized by the vector
$\displaystyle \varphi(x)\,\exp\left(\frac{i\,\zeta(x)}{\varphi_0}\right)$
in the $(\chi,\chi^*)$ isotopic space.

The existence of fields with different phases $\zeta(x)$ in
different regions of space is energetically unfavorable.  But
just as in the case of domain walls, the value of the phase ---
\index{Domain walls}%
that is, the direction of the vector
$\displaystyle \varphi(x)\,\exp\left(\frac{i\,\zeta(x)}{\varphi_0}\right)$
--- cannot be correlated over distances greater than the size of
the horizon, $\sim t$.  Moreover, immediately after the phase
transition, the direction of this vector at different points $x$
cannot be correlated over distances much greater than ${\rm
O}({\rm T}_c^{-1})$.

\begin{figure}[t]\label{f27}
\centering \leavevmode\epsfysize=5cm \epsfbox{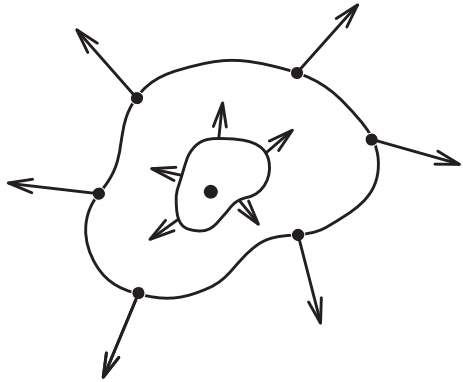}
\caption{Distribution of the field
$\chi=\varphi(x)\,e^{i\theta(x)}$ in isotopic space over a path
surrounding a string $\varphi(x) = 0$.}
\end{figure}
Let us consider some two-dimensional surface in our
three-dimen\-si\-onal space and study the possible configurations of
the field $\varphi$ there.  Among these configurations, there is
one such that upon traversing some closed contour in $x$-space,
the vector
$\displaystyle \varphi(x)\,\exp\left(\frac{i\,\zeta(x)}{\varphi_0}\right)$
executes a complete rotation in $(\chi,\chi^*)$ isotopic space
(i.e., the function $\displaystyle \frac{\zeta(x)}{\varphi_0}$
changes by $2\pi$); see Fig. 6.1.  The appearance of such
an initial field distribution for $\varphi$ as a result of a
phase transition is in no way forbidden.  Now let us gradually
constrict this contour, remaining all the while in a region with
$\varphi(x)\neq0$.  Since the field $\chi(x)$ is continuous and
differentiable, the vector $\chi(x)$ should also execute a
complete rotation in traveling along the shrinking contour.  If
we could shrink the contour to a point at which $\varphi(x)\neq0$
in this manner, the field $\chi(x)$ would no longer be
differentiable there;  that is, the equations of motion would not
hold at that point.  This implies that somewhere within the
original contour there must be a point at which $\varphi(x)=0$.
Let us suppose for the sake of simplicity that there is just one.
Now change the section of space under consideration,
appropriately moving our contour in space so that as before it
does not pass through any region with $\varphi(x)=0$.  By
continuity, then, in circling the contour, the vector $\chi(x)$
will also rotate by $2\,\pi$.

Thus, there will be a point within each such contour at which
$\varphi(x)=0$.  This implies that somewhere in space there
exists a curve --- either closed or infinite --- upon which
$\varphi(x)=0$.  The existence of such a curve is energetically
unfavorable, since $\varphi\ll\varphi_0$ nearby and the gradient
of the field $\varphi$ is also nonzero.  However, topological
considerations indicate that such a curve, once having been
produced during a phase transition, cannot break;  only if it is
closed can it shrink to a point and disappear.  The curve
$\varphi(x)=0$ owes its topological stability to the fact that as
one goes around this curve, the vector $\chi(x)$ executes either
no full rotations, or one, two, or three, but there is no
continuous transformation between the corresponding distributions
of the field $\chi$ (in traveling along the closed contour, and
returning to the same point $x$, the vector $\chi(x)$  cannot
make 0.99 full rotations in $(\chi,\chi^*)$ space).  Such curves,
together with their surrounding regions of inhomogeneous field $\chi(x)$,
are called strings.

Similar configurations of the field $\chi$ can also arise in the
Higgs model itself.  In that case, however, everywhere except on
the curve $\varphi(x)=0$ one can carry out a gauge transformation
like (\ref{1.1.16}) and transform away the field $\zeta(x)$.
However, this leads to the appearance of a field
${\rm A}_\mu(x)\neq0$ within the string, which contains a quantum of
magnetic flux ${\bf H}=\nabla\times{\bf A}$.  Such strings are
entirely analogous to
\index{Abrikosov filaments}%
\index{Filaments, Abrikosov}%
Abrikosov filaments in the theory of
superconductivity [\cite{190}].  Just as before, it is impossible
to break such a string, in the present case by virtue of the
conservation of magnetic flux.  In order to distinguish such
strings from those devoid of gauge fields, the latter are
\index{Global strings}%
\index{Strings!global}%
sometimes called global strings (their existence being related to
global symmetry breaking).

Inasmuch as the directions of the isotopic vectors $\chi(x)$ are
practically uncorrelated in every region of size ${\rm O}({\rm T}^{-1}_c)$
immediately after the phase transition, strings
initially look like Brownian trajectories with ``straight''
segments whose characteristic length is ${\rm O}({\rm T}_c^{-1})$.
Gradually straightening out, these strings then accelerate as a
result of their tension, and start to move at close to the speed
of light.  The end result is that small closed strings (with
sizes less than ${\rm O}(t)$) start to collapse, intersect,
radiate their energy in the form of gravitational waves, and
finally disappear.  Very long strings, with sizes of the order of
the distance to the horizon $\sim t$, became almost straight. If,
\index{Strings!intersecting}%
\index{Intersecting strings}%
as appears possible, intersecting strings have a non-negligible
probability of coalescing, thereby forming small closed strings,
the number of long, straight strings remaining within the horizon
ought to decrease to a value of order unity.

Let $\alpha$ be the energy density of a string of unit length.
In theories with coupling constants of order unity,
$\alpha\sim\varphi_0^2$.  The mass of a string inside the horizon
is of order $\delta{\rm M}\sim\alpha\,t\sim\varphi_0^2\,t$, while
according to (\ref{1.3.20}) the total mass of matter inside the
horizon is ${\rm M}\sim10\,\m^2\,t$.  This means that due to the
evolution of strings, the universe will eventually contain
density inhomogeneities [\cite{192}, \cite{81}]
\index{Inhomogeneities!density}%
\index{Density!perturbations}%
\be
\label{6.2.3}
\frac{\delta\rho}{\rho}\sim\frac{\delta{\rm M}}{{\rm M}}\sim
10\,\frac{\alpha}{\m^2}\sim10\,\frac{\varphi_0^2}{\m^2}\ .
\ee
For $\alpha\sim10^{-6}\m^2$, $\varphi_0\sim10^{16}$ GeV, we
obtain $\displaystyle \frac{\delta\rho}{\rho}\sim10^{-5}$, as
required for galaxy formation.

In deriving this estimate, we have assumed that small closed
strings rapidly (in a time of order $t$) radiate their energy and
vanish.  Actually, this will only happen if the value of $\alpha$
is large enough.  More refined estimates [\cite{193}] lead to
values of $\alpha$ similar to those obtained above,
$$
\alpha\sim2\cdot 10^{-6}\m^2\ .
$$
Notice that the typical mass scale and the value of $\varphi_0$
that appear here are close to those associated with symmetry
breaking in grand unified theories.  Such strings can exist in
some grand unified theories.  Unfortunately, it is far from
simple to arrange for such heavy strings to be created after
inflation, since the temperature of the universe after inflation
is typically much lower than $\varphi_0\sim10^{16}$ GeV, and the
phase transition which leads to heavy string formation typically
does not occur.  Some possibilities for the formation of heavy
strings in the inflationary universe scenario will be discussed
in the next chapter.

Now let us look at another important class of topologically
stable objects which might be formed at the time of phase
transitions.  To this end, we analyze symmetry breaking in the
O(3)-symmetric model of the scalar field $\varphi^a$, $a=1$, 2,
3:
\be
\label{6.2.4}
{\rm L}=\frac{1}{2}\,(\partial_\mu\varphi^a)^2
+\frac{\mu^2}{2}\,(\varphi^a)^2
-\frac{\lambda}{4}\,[(\varphi^a)^2]^2\ .
\ee
Symmetry breaking occurs in this model as a result of the
appearance of the scalar field $\varphi^a$, with absolute value
$\varphi_0$ equal to $\displaystyle \frac{\mu}{\sqrt{\lambda}}$,
but with arbitrary direction in isotopic space
$(\varphi^1,\varphi^2,\varphi^3)$.  At the time of the phase
transition, domains can appear such that the vector $\varphi^a$
can look either ``out of'' or ``into'' each domain (in isotopic
space) at all points on its surface.  One example is the
so-called ``hedgehog'' distribution shown in Fig. 6.2,
\be
\label{6.2.5}
\varphi^a({\bf x})=\varphi_0\,f(r)\,\frac{x^a}{r}\ ,
\ee
where $\displaystyle \varphi_0=\frac{\mu}{\sqrt{\lambda}}$,
$r=\sqrt{{\bf x}^2}$, and $f(r)$ is some function that tends to
$\pm1$ for $r\gg\mu^{-1}$, and tends to zero as $r\rightarrow0$
(the latter condition derives from the continuity of to the
function $\varphi^a({\bf x})$).  Such a distribution is a solution
of equations of motion in the theory (\ref{6.2.4}) (for a
specific choice of function $f(r)$ with the indicated
properties), and this solution turns out to be topologically
stable for the same reason as do the global strings considered
above.
\begin{figure}[t]\label{f28}
\centering \leavevmode\epsfysize=6cm \epsfbox{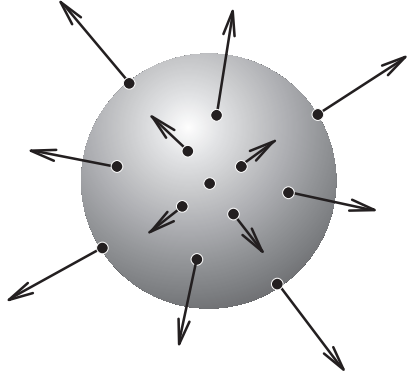}
\caption{Distribution of the field $\varphi^a$ (\ref{6.2.5}) about
the center of a hedgehog (global monopole).}
\end{figure}

At large $r$, the main contribution to the hedgehog energy comes
\index{Hedgehog energy}%
\index{Energies!hedgehog}%
from gradient terms arising from the change in direction of the
unit vector $\displaystyle \frac{x^a}{r}$ at different points,
\be
\label{6.2.6}
\rho\approx\frac{1}{2}\,(\partial_i\varphi)^2
=\frac{3}{2}\,\frac{\varphi_0^2}{r^2}\ ,
\ee
whereupon that part of the hedgehog energy contained within a
\index{Hedgehog energy}%
\index{Energies!hedgehog}%
sphere of radius $r$ centered at $x=0$ is
\be
\label{6.2.7}
{\rm E}(r)=6\,\pi\,\varphi_0^2\,r\ .
\ee

In infinite space, the total hedgehog energy thus goes to
infinity (as $r$).  That is why the hedgehog solution
(\ref{6.2.5}), discovered more than ten years ago in the same
paper as monopoles [\cite{83}], failed until fairly recently to
elicit much interest in and of itself.

When phase transitions take place in an expanding universe,
however, hedgehogs can most certainly be created.  The theory of
hedgehog formation is similar to the theory of string creation,
and in fact the first estimates of the number of monopoles
created during a phase transition [\cite{40}] were based
implicitly on an analysis of hedgehog production.  An
investigation of this problem shows that  rather than being
created singly, hedgehogs are typically created in
hedgehog-antihedgehog pairs (corresponding to $f(r)=\pm1$ for
$r\gg m^{-1}$ in (\ref{6.2.5})).  At large distances, such a pair
exerts a mutually compensatory influence on the field $\varphi$,
\index{Hedgehog energy}%
\index{Energies!hedgehog}%
and instead of the infinite energy of the individual hedgehogs,
we obtain the energy of the pair, which is proportional to their
mutual separation $r$ (\ref{6.2.7}).  This is the simplest
example of a realization of the idea of confinement that we know
of.

Subsequent evolution of a hedgehog-antihedgehog molecule depends
strongly on hedgehog interactions with matter.  In a hot
universe, such a molecule is initially not very large,
$r\la10^2\,{\rm T}_c^{-1}$.  Actually, though, according to the
results of the previous section, domains filled with the
homogeneous field $\varphi$ have characteristic sizes of order
$10\,{\rm T}_c^{-1}$ (see (\ref{6.1.4})).  Simple combinatoric
arguments indicate that in a region containing $10^2$--$10^3$
such domains with uncorrelated values of $\varphi^a$, one will
surely find at least one hedgehog, thereby yielding the preceding
estimate.

If the fields $\varphi^a$ interact weakly with matter, hedgehogs
and antihedgehogs quickly approach one another, start executing
oscillatory motion, radiate Goldstone bosons and gravitational
waves, approach still closer, and finally annihilate, radiating
their energy in the same way as do closed (global) strings.  But
if hedgehog motion is strongly damped by matter, the annihilation
process can take much longer.  We shall return to the discussion
of possible cosmological effects associated with hedgehogs when
we consider the production of density inhomogeneities in the
inflationary universe scenario.

If we supplement the theory (\ref{6.2.4}) with O(3)-symmetric
Yang--Mills fields with a coupling constant $e$, the resultant
theory will also have a solution of the equations of motion like
(\ref{6.2.5}) for the field $\varphi^a$, but classical
Yang--Mills fields will show up as well.  By a gauge
transformation of the fields $\varphi^a$ and ${\rm A}_\mu^a$, we
can ``comb out'' the hedgehog, i.e., send the fields $\varphi^a$
off in one direction (for example, $\varphi^a\sim
x^3\,\delta^a_3$) everywhere except along some infinitely thin
filament emanating from the point $x=0$.  Far from the point
$x=0$, then, the vector fields ${\rm A}_\mu^{1,2}$ acquire a mass
$m_{\rm A}=e\,\varphi_0$, while the vector field ${\rm A}_\mu^3$
remains massless.  The most important feature of the resulting
configuration of the fields $\varphi^a$ and ${\rm A}_\mu^a$ is
then the presence of a magnetic field ${\bf H}=\nabla\times{\bf A}^3$
which falls off far from the center,
\be
\label{6.2.8}
{\bf H}=\frac{1}{e}\,\frac{{\bf x}}{r^3}\ .
\ee
Hence, this theory gives rise to particles analogous to the Dirac
mono\-pole ('t~Hooft--Polyakov monopoles) with magnetic charge
\be
\label{6.2.9}
g=\frac{4\,\pi}{e}\ ,
\ee
and these particles have an extremely high mass,
\be
\label{6.2.10}
{\rm M}=c\,\left(\frac{\lambda}{e^2}\right)\,\frac{4\,\pi\,m_{\rm A}}{e^2}
=\frac{c\,m_{\rm A}}{\alpha}\ ,
\ee
where $\displaystyle \alpha=\frac{e^2}{4\,\pi}$, and
$\displaystyle c\,\left(\frac{\lambda}{e^2}\right)$ is a quantity
approximately equal to unity:  ($c(0)=1$, $c(0.5)=1.42$,
$c(10)=1.44$).

In contrast to hedgehogs (\ref{6.2.5}), 't~Hooft--Polyakov
monopoles ought to exist in all grand unified theories, in which
\index{Monopoles!'t~Hooft--Polyakov}%
\index{'t~Hooft--Polyakov monopoles}%
the weak, strong, and electromagnetic interactions prior to
symmetry breaking are described by a single theory with a simple
symmetry group (SU(5), O(10), ${\rm E}_6$, \ldots).  Just as for
hedgehogs, monopoles are produced during phase transitions,
separated from each other by distance of order $10^2\,{\rm T}_c^{-1}$.
Their initial density $n_{\rm M}$ at the phase transition epoch
was thereby $10^{-6}$ times the photon density $n_\gamma$.
Zeldovich and Khlopov study of the monopole-antimonopole
annihilation rate [\cite{40}] has shown that annihilation
proceeds very slowly, so that at present we should find
$\displaystyle \frac{n_{\rm M}}{n_\gamma}\sim10^{-9}$--$10^{-10}$, i.e.,
$n_{\rm M}\approx n_{\rm B}$, where $n_{\rm B}$ is the baryon
(proton and neutron)
\index{Baryon density}%
\index{Density!baryon}%
density. The present density $\rho_{\rm B}$
\index{Density!baryon matter}%
\index{Baryon matter density}%
\index{Matter!baryon density}%
of baryon matter in the universe differs from the critical
density by no more than one or two orders of magnitude,
$\rho_{\rm B}\sim10^{-29}$ g/cm$^3$.  In grand unified theories,
according to (\ref{6.2.10}), monopoles should have a mass of
$10^2\,{\rm M}_{\rm X}\sim10^{16}$--$10^{17}$ GeV;  that is,
$10^{16}$--$10^{17}$ times the mass of the proton.  But that
would mean, if we believe the estimate $n_{\rm M}\approx n_{\rm B}$,
that the density of matter in the universe exceeds the critical
value by 16 orders of magnitude.  Such a universe would already
have collapsed long ago!

Even more stringent limits are placed on the allowable
present-day density by the existence of the galactic magnetic
field [\cite{194}], and by theoretical estimates of pulsar
luminosity [\cite{195}] due to monopole catalysis of proton decay
[\cite{196}].  These constraints lead one to conclude that at
present, most likely
$\displaystyle \frac{n_{\rm M}}{n_{\rm B}}\la10^{-25}$--$10^{-30}$.
Such an enormous disparity between the observational constraints
on the\index{Density!of monopoles}\index{Monopoles!density of}
density of monopoles in the universe and the theoretical
predictions have led us to the brink of a crisis:  modern
elementary particle theory is in direct conflict with the hot
universe theory.  If we are to get rid of this contradiction, we
have three options:

a) renounce grand unified theories;

b) find conditions under which monopole annihilation proceeds
much more efficiently;

c) renounce the standard hot universe theory.

At the end of the 1970's, the first choice literally amounted to
blasphemy.  Later on, after the advent of more complicated
theories based on supergravity and superstring theory, the
general attitude toward grand unified theories began to change.
But for the most part, rather than helping to solve the
primordial monopole problem, the new theories engender fresh
conflicts with the hot universe theory that are just as serious;
see Section \ref{s1.5}.

The second possibility has so far not been carried through to
completion.  The basic conclusions of the theory of monopole
annihilation proposed in [\cite{40}] have since been confirmed by
many authors.  On the other hand, it has been argued [\cite{173}]
that nonperturbative effects in a high-temperature Yang--Mills
gas can lead to monopole confinement, accelerating the
annihilation process considerably.

\begin{figure}[t]\label{f29}
\centering \leavevmode\epsfysize=3cm \epsfbox{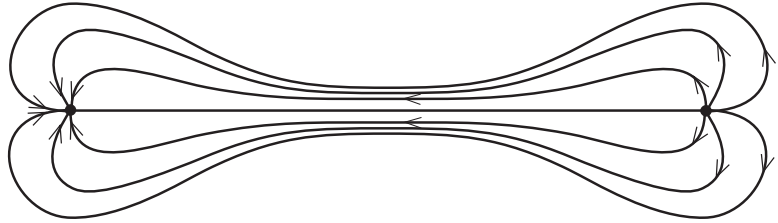}
\caption{Magnetic field configuration for a monopole-antimonopole
pair embedded in a superconductor.}
\end{figure}
The basic idea here is that far from a monopole, its field is
effectively Abelian, ${\bf H}^a=\nabla\times{\bf A}^a\cdot \delta^a_3$.
Such a field satisfies Gauss' theorem identically, $\nabla\cdot{\bf H}=0$,
so its flux is conserved.  However, if the Yang--Mills fields in
a hot plasma acquire an effective magnetic mass $m_{\rm A}\sim e^2\,{\rm T}$
(see Section \ref{s3.3}), then the monopole magnetic field will
be able to penetrate the medium only out to a distance $m_{\rm A}^{-1}$.
The only way to make this condition compatible with the magnetic
field version of Gauss' theorem is to invoke filaments of
thickness $\Delta l\sim m_{\rm A}^{-1}$ emanating from the
monopoles and incorporating their entire magnetic field.  But
this is exactly how the magnetic field of a monopole embedded in
a superconductor behaves (and for the same reason):  Abrikosov
filaments
\index{Abrikosov filaments}%
\index{Filaments, Abrikosov}%
(strings) of the magnetic field come into being between
monopoles and
\index{Monopoles}%
\index{Antimonopoles}%
antimonopoles [\cite{190}]; see Fig. 6.3.
Since the energy of each such string is proportional to its
length, monopoles in a superconductor ought to be found in a
confinement phase [\cite{197}].  If the analogous phenomenon
comes into play in the hot Yang--Mills gas, then the monopoles
there should be bound to antimonopoles by strings of thickness
$\Delta l\sim (e^2\,{\rm T})^{-1}$.  Monopole-antimonopole pairs
will therefore annihilate much more rapidly than when they are
bound solely by conventional attractive Coulomb forces.

Unfortunately, we still have too imperfect an understanding of
the thermodynamics of the Yang--Mills gas to be able to confirm
or refute the existence of monopole confinement in a hot plasma.
Nonperturbative analysis of this problem using Monte Carlo
lattice simulations [\cite{198}, \cite{199}] is not particularly
informative, since the use of the lattice gives rise to
fictitious light monopoles whose mass is of the order of the
reciprocal lattice spacing $a^{-1}$.  These fictitious monopoles
act to screen out the mutual interaction of 't~Hooft--Polyakov
monopoles during the Monte Carlo simulations;  they are difficult
\index{Monopoles!'t~Hooft--Polyakov}%
\index{'t~Hooft--Polyakov monopoles}%
to get rid of with presently available computing capabilities.

Besides the
\index{Monopole confinement mechanism}%
monopole confinement mechanism discussed above, there
is another that is even simpler [\cite{200}].  Specifically, it
is well known that in addition to not being able to penetrate a
superconductor, a magnetic field cannot penetrate the bulk of a
perfect conductor either (if the field was not present in the
conductor from the very start), the reason being that induced
currents cancel the external magnetic field.  The conductivity of
\index{Yang--Mills plasma}%
the Yang--Mills plasma is extremely high, and that is why when
\index{Monopoles}%
monopoles appear during a phase transition, their magnetic field
does not make its appearance in the medium right away.  The
entire magnetic field must first be concentrated into some string
joining a monopole and antimonopole, as in Fig. 6.3 (due to
conservation of total magnetic flux, induced currents cannot
cancel the entire magnetic flux, which passes along the
filament).  The string will gradually thicken, and the field will
adopt its usual Coulomb configuration.  If, however, the rate at
which the thickness of the string grows is small compared with
the rate at which the monopoles separate from one another due to
the expansion of the universe, the field distribution will remain
one-dimensional for a long time;  in other words, a confinement
regime will prevail once again.  Our estimates show that such a
regime is actually possible in grand unified theories.

A preliminary analysis of the annihilation of monopoles in a
\index{Monopoles}%
confinement phase indicate that the monopole density at the
present epoch may be 10--20 orders of magnitude lower than was
first thought.  A complete solution of this problem is
exceedingly difficult, however, and it is not clear whether
monopole confinement will provide a way to reconcile theoretical
estimates of their density with the most stringent experimental
limits, namely those based on the existence of galactic magnetic
fields and the observed lack of strong X-ray emission from
pulsars.

The theory of the interaction of monopoles with matter may yet
\index{Monopoles}%
harbor even more surprises.  But even if a way were found to
solve the primordial monopole problem within the framework of the
standard hot universe theory, it would be hard to overstate the
value of the contribution made by analysis of this problem to the
development of contemporary cosmology.  It is precisely the
numerous attempts to resolve this problem that have led to
wide-ranging discussions of the internal inconsistencies of the
hot universe theory, and to a recognition of the need to
reexamine its foundations.  These attempts served as an impetus
for the development of the inflationary universe scenario, and
for the appearance of new concepts relating both to the initial
stages of the evolution of the observable part of the universe
and the global structure of the universe as a whole.  We now turn
to a description of these concepts.
\index{Phase transitions!in hot universe|)}%
\index{Hot universe!phase transitions in|)}%


\chapter{\label{c7}General Principles of Inflationary Cosmology}
\vspace{2pc}
\index{Inflationary universe!general principles of|(}%

\section{\label{s7.1}Introduction}

In Chapter \ref{c1}, we discussed the general structure of the
inflationary universe scenario.  Recent developments have gone in
three main directions:

a) studies of the basic features of the scenario and the
revelation of its potential capacity for a more accurate
description of the observable part of the universe.  These
studies deal basically with problems related to the production of
density inhomogeneities at the time of inflation, the reheating
of the universe, and the generation of the post-inflation baryon
asymmetry, along with those predictions of the scenario that
might be tested by analysis of the available observational
evidence;

b) the construction of realistic versions of the inflationary
universe scenario based on modern elementary particle theories;

c) studies of the global properties of space and time within the
framework of quantum cosmology, making use of the inflationary
universe scenario.

The first of these avenues of research will be discussed in the
present chapter.  The second will form the subject of Chapters
\ref{c8} and \ref{c9}, and the third, Chapter \ref{c10}.
\al_break 

\section{\label{s7.2}The inflationary universe and de Sitter
\index{de Sitter space!inflationary universe and|(}%
\index{Inflationary universe!de Sitter space and|(}%
space}

As we have already noted in Chapter \ref{c1}, the main feature of
the inflationary stage of evolution of the universe is the slow
variation (compared with the rate of expansion of the universe)
of the energy density $\rho$.  In the limiting case
$\rho=\mbox{const}$, the Einstein equation (\ref{1.3.7}) for a
homogeneous universe has the de Sitter space
(\ref{1.6.1})--(\ref{1.6.3}) as its solution.

It is easy to see that when ${\rm H}\,t\gg1$, the distinction
between an open, closed, and flat de Sitter space tends to
vanish.  Much less obvious is the fact that all three of the
solutions (\ref{1.6.1})--(\ref{1.6.3}) actually describe the very
same de Sitter space.

To facilitate an intuitive interpretation of a curved
\index{Four-dimensional space}%
\index{Space!four-dimensional}%
four-dimensional space, it is often convenient to imagine it to
be a curved four-dimensional hypersurface embedded in a
higher-dimensional space.  De Sitter space is most easily
\index{Hyperboloid}%
represented as a hyperboloid
\be
\label{7.2.1}
z_0^2-z_1^2-z_2^2-z_3^2-z_4^2=-{\rm H}^{-2}
\ee
in the five-dimensional Minkowski space $(z_0,z_1,\ldots,z_4)$.
In order to represent de Sitter space as a flat Friedmann
universe (\ref{1.3.2}), (\ref{1.6.2}), it suffices to consider a
coordinate system $t$, $x_i$ on the hyperboloid (\ref{7.2.1})
defined by the relations
\ba
\label{7.2.2}
z_0&=&{\rm H}^{-1}\,\sinh{\rm H}\,t+
\frac{1}{2}\,{\rm H}\,e^{{\rm H}\,t}\,{\bf x}^2\ ,\nonumber \\
z_4&=&{\rm H}^{-1}\,\cosh{\rm H}\,t-
\frac{1}{2}\,{\rm H}\,e^{{\rm H}\,t}\,{\bf x}^2\ ,\nonumber \\
z_i&=&e^{{\rm H}\,t}\,x_i\ ,\qquad i=1,2,3\ .
\ea
\begin{figure}[t]\label{f30}
\centering \leavevmode\epsfysize=9cm \epsfbox{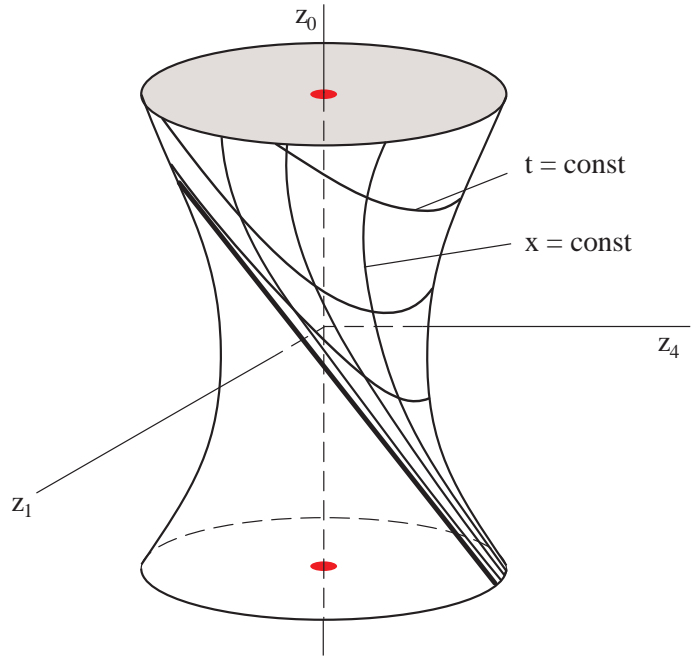} \caption{De
Sitter space represented as a hyperboloid in
\index{Hyperboloid}%
\index{Space-time!five-dimensional}%
five-dimensional space-time (with two dimensions omitted).  In
the coordinates (\ref{7.2.2}), three-dimensional space at
$t=\mbox{const}$ is flat, expanding exponentially with increasing
$t$ --- see (\ref{7.2.3}).  The coordinates (\ref{7.2.2}) span
only half the hyperboloid.}
\end{figure}
This coordinate system spans the half of the hyperboloid with
$z_0+z_4>0$ (see Fig. 7.1), and
its
metric takes the form
\be
\label{7.2.3}
ds^2=dt^2-e^{2{\rm H}\,t}\,d{\bf x}^2\ .
\ee

De Sitter space looks like a closed Friedmann universe in the
coordinate system $(t,\chi,\theta,\varphi)$ defined by
\ba
\label{7.2.4}
z_0&=&{\rm H}^{-1}\,\sinh{\rm H}\,t\nonumber \\
z_1&=&{\rm H}^{-1}\,\cosh{\rm H}\,t\,\cos\chi\ ,\nonumber \\
z_2&=&{\rm H}^{-1}\,\cosh{\rm H}\,t\,\sin\chi\,\cos\theta\ ,\nonumber \\
z_3&=&{\rm H}^{-1}\,\cosh{\rm H}\,t\,\sin\chi\,\sin\theta\,\cos\varphi\ ,
                                        \nonumber \\
z_4&=&{\rm H}^{-1}\,\cosh{\rm H}\,t\,\sin\chi\,\sin\theta\,\sin\varphi\ .
\ea
The metric then becomes
\be
\label{7.2.5}
ds^2=dt^2-{\rm H}^{-2}\cosh^2{\rm H}\,t\,
[d\chi^2+\sin^2\chi\,(d\theta^2+\sin^2\theta\,d\varphi^2)]\ .
\ee
It is important to note that in contrast to the flat-universe
metric (\ref{7.2.3}) and the metric for an open de Sitter space
(which we will not write out here), the
\index{Closed-universe metric}%
\index{Metric!closed-universe}%
\index{Hyperboloid}%
closed-universe metric (\ref{7.2.5}) describes the entire
hyperboloid.  In the terminology of general relativity, one can
say that the closed de Sitter space, as distinct from the flat or
open one, is geodesically complete (see Fig. 7.2).

To gain some understanding of this situation, it is useful here
to draw an analogy with what happens near a
\index{Black holes}%
black hole. In particular, the Schwarzschild metric does not provide a
\index{Metric!Schwarzschild}%
\index{Schwarzschild metric}%
description of events near the gravitational radius $r_g$ of the
\index{Black holes}%
black hole, but there do exist coordinate systems that enable one
to describe what occurs within the black hole.  In the present
instance, the analog of the Schwarzschild metric is the metric
for a flat (or open) de Sitter space.  An even more complete
analog is given by the static coordinates $(r,t,\theta,\varphi)$:
\ba
\label{7.2.6}
z_0&=&\sqrt{{\rm H}^{-2}-r^2}\,\sinh{\rm H}\,t\ ,\nonumber \\
z_1&=&\sqrt{{\rm H}^{-2}-r^2}\,\cosh{\rm H}\,t\ ,\nonumber \\
z_2&=&r\,\sin\theta\,\cos\varphi\ ,\nonumber \\
z_3&=&r\,\sin\theta\,\sin\varphi\ ,\nonumber \\
z_4&=&r\,\cos\theta\ ,\qquad 0\le r\le{\rm H}^{-1}\ .
\ea
\begin{figure}[t]\label{f31}
\centering \leavevmode\epsfysize=9cm \epsfbox{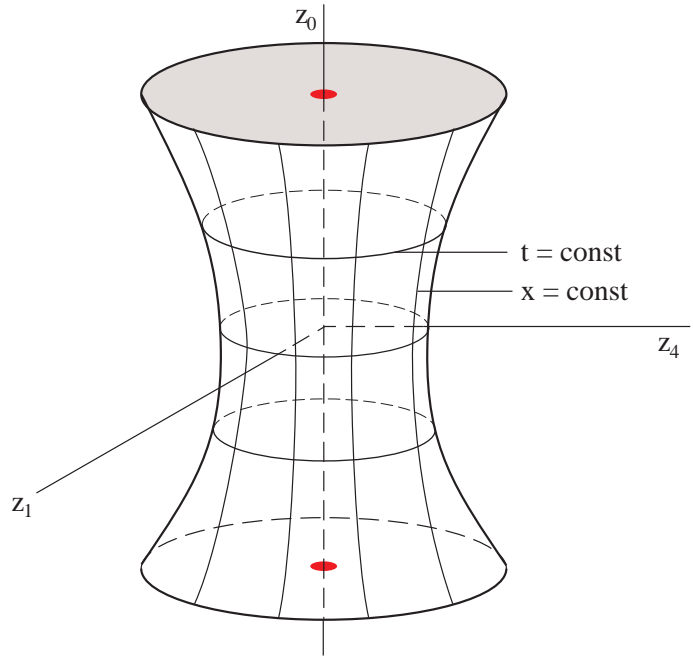} \caption{De
Sitter space, represented as a closed
\index{Hyperboloid}%
Friedmann universe with coordinates (\ref{7.2.4}), (\ref{7.2.5}).
These coordinates span the entire hyperboloid.}
\end{figure}
These coordinates span that part of the de Sitter space with
$z_0+z_1>0$, and the metric takes the form
\be
\label{7.2.7}
ds^2=(1-r^2\,{\rm H}^2)\,dt^2-(1-r^2\,{\rm H}^2)^{-1}\,dr^2
-r^2\,(d\theta^2+\sin^2\theta\,d\varphi^2)\ ,
\ee
resembling the form of the Schwarzschild metric
\be
\label{7.2.8}
ds^2=(1-r_g\,r^{-1})\,dt^2-(1-r_g\,r^{-1})^{-1}\,dr^2
-r^2\,(d\theta^2+\sin^2\theta\,d\varphi^2)\ ,
\ee
where $\displaystyle r_g=\frac{2\,{\rm M}}{\m^2}$, and M is the
mass of the
\index{Black holes}%
black hole. Equations (\ref{7.2.7}) and
(\ref{7.2.8}) demonstrate that de Sitter space in static
coordinates comprises a region of radius ${\rm H}^{-1}$ that
looks as if it were {\it surrounded} by a black hole.  This
result was provided with a physical interpretation in Chapter
\ref{c1} (see Eq. (\ref{1.4.14})) by introducing the concept of
\index{Event horizon}%
the event horizon.  The analogy between the properties of de
Sitter space and those of a
\index{Black holes}%
black hole is a very important one
for an understanding of many of the features of the inflationary
universe scenario, and it therefore merits further discussion.

It is well known that any perturbations of the black hole
\index{Perturbations!of black holes}%
(\ref{7.2.8}) are rapidly damped out, and the only observable
characteristic that remains is its mass (and its electric charge
and angular momentum if it is rotating).  No information about
physical processes occurring inside a black hole leaves its
surface (that is, the horizon located at $r=r_g$). This set of
statements (along with some qualifications and additions), is
often known in the literature as the theorem that {\it a black
hole has no hair};  for example, see [\cite{119}].

The generalization of this ``theorem'' to de Sitter space
[\cite{120}, \cite{121}] reads that any perturbation of the
latter will be ``forgotten'' at an exponentially high rate;  that
is, after a time $t\gg{\rm H}^{-1}$, the universe will become
locally indistinguishable from a completely homogeneous and
isotropic de Sitter space.  On the other hand, because of the
existence of an event horizon, all physical processes in a given
region of de Sitter space are independent of anything that
happens at a distance greater than ${\rm H}^{-1}$ from that
region.

The physical meaning of the first part of the theorem is
especially transparent in the coordinate system (\ref{7.2.3}) (or
(\ref{7.2.5}) when $t\gg{\rm H}^{-1}$):  any perturbation of de
Sitter space that is entrained by the general cosmological
expansion will be exponentially stretched.  Accordingly, spatial
gradients of the metric, which characterize the local
inhomogeneity and anisotropy of the universe, are exponentially
damped.  This general statement, which has been verified for a
wide class of specific models [\cite{122}], forms the basis for
the solution of the homogeneity and isotropy problems in the
inflationary universe [\cite{54}--\cite{56}].

The second part of the theorem means that if the initial size of
an inflationary region exceeds the distance to the horizon
($r>{\rm H}^{-1}$), then no events outside that region can hinder
its inflation, since no information about those events can ever
reach it.  The indifference of inflationary regions to what goes
on about them might be characterized as a sort of relatively
\index{Inflationary regions!growth of}%
harmless egoism:  the growth of inflationary regions takes place
basically by virtue of their inherent resources, rather than
those of neighboring regions of the universe.  This kind of
process (chaotic inflation) eventually leads to a universe with
very complex structure on enormous scales, but within any
inflationary region, the universe looks locally uniform to high
accuracy.  This circumstance plays an important role in any
discussion of the initial conditions required for the onset of
inflationary behavior (see Sections \ref{s1.7} and \ref{s9.1}) or
investigation of the global structure of the universe (Sections
\ref{s1.8} and \ref{s10.2}).

We shall return in the next section to the analogy between
physical processes inside a
\index{Black holes}%
black hole and those in the
inflationary universe, but here we should like to make one more
remark apropos of de Sitter space and its relation to the
inflationary universe theory.

Many classic textbooks on general relativity theory treat de
Sitter space as nothing but the static space (\ref{7.2.7}).  As
we have already pointed out, however, the space described by the
metric (\ref{7.2.7}) is geodesically incomplete; that is, there
exist geodesics that carry one out of the space (\ref{7.2.7}).
In much the same way that an observer falling into a
\index{Black holes}%
black hole does not notice anything exceptional as he makes the final plunge
through the Schwarzschild sphere  $r=r_g$, so an observer in de
Sitter space who is located at some initial point $r=r_0<{\rm H}^{-1}$
emerges from the region described by the coordinates
(\ref{7.2.7}) after a definite proper time interval (as measured
by his own clocks).  (While this is going on, a stationary
observer located at $r=\infty$ in the metric (\ref{7.2.8}) or at
$t=0$ in the metric (\ref{7.2.7}) will never expect to see his
friend disappear beyond the horizon, but he will receive less and
less information.)  At the same time, the geodesically complete
space (\ref{7.2.5}) is non-static.

In the absence of observers, matter, or even test particles, this
lack of stationarity  is a ``thing unto itself,'' since the
invariant characteristics of de Sitter space itself that are
associated with the curvature tensor are time-independent.  Thus,
for example, the scalar curvature of de Sitter space is
\be
\label{7.2.9}
{\rm R}=12\,{\rm H}^2=\mbox{const}\ .
\ee
Therefore, if the inflationary universe were simply an empty de
Sitter space, it would be difficult to speak of its expansion.
It would always be possible to find a coordinate system in which
de Sitter space looked, for example, as if it were contracting,
or as if it had a size $\sim{\rm H}^{-1}$ (Eqs. (\ref{7.2.5}),
(\ref{7.2.7})).  But in the inflationary universe, the de Sitter
invariance is either spontaneously broken (due to the decay of
the initial de Sitter vacuum), or is broken on account of an
initial disparity between the actual universe and de Sitter space.
In particular, the energy-momentum tensor ${\rm T}_{\mu\nu}$ in the
chaotic inflation scenario, even though it is close to $\vf\,g_{\mu\nu}$,
is never exactly equal to the latter, and in the last stages of
inflation, the relative magnitude of the field kinetic energy
$\displaystyle \frac{1}{2}\,\dot\varphi^2$ becomes large compared
to ${\rm V}(\varphi)$, and the difference between
${\rm T}_{\mu\nu}$ and $\vf\,g_{\mu\nu}$ becomes significant.  The
distinction between static de Sitter space and the inflationary
universe becomes especially clear at the quantum level, when one
analyzes density inhomogeneities
$\displaystyle\frac{\delta\rho}{\rho}$ that arise at the time of
inflation.  As we will show in Section \ref{s7.5}, by the end of
inflation, these inhomogeneities grow to $\displaystyle
\frac{\delta\rho}{\rho}\sim\frac{{\rm H}^2}{\dot\varphi}$.  Thus,
if the field $\varphi$ were constant and the inflating universe
were indistinguishable from de Sitter space, then after inflation
ended our universe would be highly inhomogeneous.  In other
words, a correct treatment of the inflationary universe requires
that we not only take its similarities to de Sitter space into
account, but its differences as well, especially in the latest
stages of inflation, when the structure of the observable part of
the universe was
formed.\index{de Sitter space!inflationary universe and|)}%
\index{Inflationary universe!de Sitter space and|)}

\section{\label{s7.3}Quantum fluctuations
\index{Inflationary universe!quantum fluctuations in|(}%
\index{Quantum fluctuations in!inflationary universe|(}%
in the inflationary universe}

The analogy between a
\index{Black holes}%
black hole and de Sitter space is also
useful in studying quantum effects in the inflationary universe.
It is well known, for example, that black holes evaporate,
\index{Hawking temperature}%
\index{Temperature!Hawking}%
emitting radiation at the Hawking temperature
$\displaystyle{\rm T}_{\rm H}= \frac{\m^2}{8\,\pi\,{\rm M}}=
\frac{1}{4\,\pi\,r_g}$, where M is the mass of the black hole
[\cite{119}].  A similar phenomenon exists in de Sitter space,
where an observer will feel as if he is in a thermal bath at a
temperature $\displaystyle {\rm T}_{\rm H}=\frac{{\rm H}}{2\,\pi}$.
Formally, we can see this making the substitution $t\rightarrow
i\,\tau$ in Eq. (\ref{7.2.5}) in order to make the transition to
the Euclidean formulation of quantum field theory in de Sitter
space.  The metric then becomes that of a four-sphere ${\rm S}^4$,
\be
\label{7.3.1}
-ds^2=d\tau^2+{\rm H}^{-2}\,\cos^2{\rm H}\,\tau\,
[d\chi^2+\sin^2\chi\,(d\theta^2+\sin^2\theta\,d\varphi^2)]\ .
\ee
Bose fields on the sphere are periodic in $\tau$ with period
$\displaystyle \frac{2\,\pi}{{\rm H}}$, which is equivalent to
considering quantum statistics at a temperature
$\displaystyle {\rm T}_{\rm H}=\frac{{\rm H}}{2\,\pi}$  [\cite{201}].
Physically, the appearance of a temperature ${\rm T}_{\rm H}$ in
de Sitter space (as is also the case for a\index{Black holes} black hole)
is related
to the necessity of averaging over states beyond the event
horizon [\cite{119}, \cite{120}].  However, the ``temperature''
of de Sitter space is highly unusual, in that the Euclidean
sphere ${\rm S}^4$ is periodic in {\it all four directions}, so
the vacuum fluctuation spectrum turns out to be quite unlike the
usual spectrum of thermal fluctuations.

Averages like $\langle \varphi(x)\,\varphi(y)\rangle $ and
$\langle \varphi(x)^2\rangle $ will play a particularly important
role in our investigation.  In Minkowski space at a finite
temperature T

\be
\label{7.3.2}
\langle \varphi(x)^2\rangle =\frac{{\rm T}}{(2\,\pi)^3}\,
\sum^\infty_{n=-\infty}
\frac{d^3k}{(2\,\pi\,n\,{\rm T})^2+{\bf k}^2+m^2}\ ,
\ee
which reduces to Eq. (\ref{3.1.7}) for $\langle \varphi^2\rangle $
after summing over $n$.  In  ${\rm S}^4$-space, {\it all}
integrations are replaced by summations over $n_i$, $i=1$, 2, 3,
4,  and the temperature is replaced by the quantity
$\displaystyle \frac{{\rm H}}{2\,\pi}$.  A term with $n_i=0$ is
especially important in summing over $n_i$, since it makes the
leading contribution to $\langle \varphi^2\rangle $ as
$m^2\rightarrow0$.  It is readily shown that this contribution
will be proportional to $\displaystyle \frac{{\rm H}^2}{m^2}$;
for $m^2\ll{\rm H}^2$, the corresponding calculation gives
\be
\label{7.3.3}
\langle \varphi^2\rangle =\frac{3\,{\rm H}^4}{8\,\pi^2\,m^2}
\ee
(a result first obtained by a different method [\cite{202},
\cite{126}--\cite{128}]).  The pathological behavior of
$\langle \varphi^2\rangle $ as $m^2\rightarrow0$ is noteworthy.
Formally, it occurs because now instead of one summation, we have
four, and the corresponding infrared divergences of scalar field
theory in de Sitter space are found to be three orders of magnitude
stronger than in quantum statistics.\footnote{Note that the vector or
spinor field theory, sums over $n_i$ contain no terms that are singular
in the limit $m\rightarrow0$}
It will be very important to understand the physical basis of
such a strange result.

To this end, one quantizes the massless scalar field $\varphi$ in
de Sitter space in the coordinates (\ref{7.2.3}) in much the same
way as in Minkowski space [\cite{202}, \cite{126}--\cite{128}].
The scalar field operator $\varphi(x)$ can be represented in the
form
\be
\label{7.3.4}
\varphi({\bf x},t)=(2\,\pi)^{-3/2}\,
\int d^3p\:[a_p^+\,\psi_p(t)\,e^{i\,{\bf p\,x}}
+a_p^-\,\psi^*_p(t)\,e^{-i\,{\bf p\,x}}]\ ,
\ee
where according to (\ref{1.7.13}), $\psi_p(t)$ satisfies the equation
\be
\label{7.3.5}
\ddot\psi_p(t)+3\,{\rm H}\dot\psi_p(t)
+{\bf p}^2\,e^{-2\,{\rm H}\,t}\,\psi_p(t)=0\ .
\ee
In Minkowski space, the function
$\displaystyle \frac{1}{\sqrt{2\,p}}\,e^{-ipt}$ takes on the role of
$\psi_p(t)$, where $p=\sqrt{{\bf p}^2}$; see (\ref{1.1.3}).  In
de Sitter space (\ref{7.2.3}), the general solution of
(\ref{7.3.5}) takes the form
\be
\label{7.3.6}
\psi_p(t)=\frac{\sqrt{\pi}}{2}\,{\rm H}\,\eta^{3/2}\,
[{\rm C}_1(p)\,{\rm H}^{(1)}_{3/2}(p\,\eta)
+{\rm C}_2(p)\,{\rm H}^{(2)}_{3/2}(p\,\eta)]\ ,
\ee
where $\eta=-{\rm H}^{-1}\,e^{-{\rm H}\,t}$ is the conformal time,
and the ${\rm H}_{3/2}^{(i)}$ are Hankel functions:
\be
\label{7.3.7}
{\rm H}_{3/2}^{(2)}(x)=[{\rm H}_{3/2}^{(1)}(x)]^*=
-\sqrt{\frac{2}{\pi\,x}}\,e^{-i\,x}\,\left(1+\frac{1}{i\,x}\right)\ .
\ee
Quantization in de Sitter space and Minkowski space should be
identical in the high-frequency limit, i.e., ${\rm C}_1(p)\rightarrow0$,
${\rm C}_2(p)\rightarrow-1$ as $p\rightarrow\infty$.  In particular,
this condition is satisfied\footnote{It is important that if the
inflationary stage is long enough, all physical results are
independent of the specific choice of functions ${\rm C}_1(p)$
and ${\rm C}_2(p)$ if ${\rm C}_1(p)\rightarrow0$,
${\rm C}_2(p)\rightarrow-1$ as $p\rightarrow\infty$.}
for ${\rm C}_1\equiv0$, ${\rm C}_2\equiv-1$.  In that case,
\be
\label{7.3.8}
\psi_p(t)=\frac{i\,{\rm H}}{p\,\sqrt{2\,p}}\,
\left(1+\frac{p}{i\,{\rm H}}e^{-{\rm H}\,t}\right)\,
\exp\left(\frac{i\,p}{{\rm H}}\,e^{-{\rm H}\,t}\right)\ .
\ee
Notice that at sufficiently large $t$ (when $p\,e^{-{\rm H}\,t}<{\rm H}$),
$\psi_p(t)$ ceases to oscillate, and becomes
equal to $\displaystyle \frac{i\,{\rm H}}{p\,\sqrt{2\,p}}$.

The quantity $\langle \varphi^2\rangle $
may be simply expressed in terms of $\psi_p$:
\be
\label{7.3.9}
\langle \varphi^2\rangle =\frac{1}{(2\,\pi)^3}\,
\int|\psi_p|^2\:d^3p=
\frac{1}{(2\,\pi)^3}\,\int
\left(\frac{e^{-2{\rm H}\,t}}{2\,p}+\frac{{\rm H}^2}{2\,p^3}\right)\:d^3p\ .
\ee
The physical meaning of this result becomes clear when one
transforms from the conformal momentum $p$, which is
time-independent, to the conventional physical momentum
$k=p\,e^{-{\rm H}\,t}$, which decreases as the universe expands:
\be
\label{7.3.10}
\langle \varphi^2\rangle =\frac{1}{(2\,\pi)^3}\,\int
\frac{d^3k}{k}\:\left(\frac{1}{2}+\frac{{\rm H}^2}{2\,k^2}\right)\ .
\ee
The first term is the usual contribution from vacuum fluctuations
in Minkowski space (for ${\rm H}=0$;  see (\ref{2.1.6}),
(\ref{2.1.7})).  This contribution can be eliminated by
renormalization, as in the theory of phase transitions (see
(\ref{3.1.6})).  The second term, however, is directly related to
inflation.  Looked at from the standpoint of quantization in
Minkowski space, this term arises because of the fact that de
Sitter space, apart from the usual quantum fluctuations that are
present when ${\rm H}=0$, also contains $\varphi$-particles with
\index{Occupation numbers}%
occupation numbers
\be
\label{7.3.11}
n_k=\frac{{\rm H}^2}{2\,k^2}\ .
\ee
It can be seen from (\ref{7.3.10}) that the contribution to
$\langle \varphi^2\rangle $  from long-wave fluctuations of the
$\varphi$ field diverges, and that is why the value of
$\langle \varphi^2\rangle $  in Eq. (\ref{7.3.3}) becomes
infinite as $m^2\rightarrow0$.

However, the value of $\langle \varphi^2\rangle $ for a massless
field $\varphi$ is infinite only in de Sitter space that exists
forever, and not in the inflationary universe, which expands
exponentially (or quasiexponentially) starting at some time $t=0$
(for example, when the density of the universe becomes smaller
than the Planck density).  Indeed, the spectrum of vacuum
fluctuations (\ref{7.3.10}) differs from the fluctuation spectrum
in Minkowski space only when $k\la{\rm H}$.  If the fluctuation
spectrum before inflation has a cutoff at $k\la k_0\sim{\rm T}$
resulting from high-temperature effects [\cite{127}], or at
$k\la k_0\sim{\rm H}$ due to an inflationary region of the universe
having an initial size ${\rm O}({\rm H}^{-1})$, then the spectrum
will change at the time of inflation, due to exponential growth
in the wavelength of vacuum fluctuations.  The spectrum
(\ref{7.3.10}) will gradually be established, but only at momenta
$k\ga k_0\,e^{-{\rm H}\,t}$.  There will then be a cutoff in the
integral (\ref{7.3.9}).  Restricting our attention to
contributions made by long-wave fluctuations with $k\la{\rm H}$,
which are the only ones that will subsequently be important for
us, and assuming that $k_0={\rm O}({\rm H})$, we obtain
\ba
\label{7.3.12}
\langle \varphi^2\rangle &\approx& \frac{{\rm H}^2}{2\,(2\,\pi)^3}\,
\int^{\rm H}_{He^{-{\rm H}\,t}}\frac{d^3k}{k}
=\frac{{\rm H}^2}{4\,\pi^2}\,\int^0_{-{\rm H}\,t}
d\ln\frac{k}{{\rm H}}\nonumber \\
&\equiv&\frac{{\rm H}^2}{4\,\pi^2}\int^{{\rm H}\,t}_0d\ln\frac{p}{{\rm H}}
=\frac{{\rm H}^3}{4\,\pi^2}\,t\ .
\ea
As $t\rightarrow\infty$, $\langle \varphi^2\rangle$ tends to
infinity in accordance with (\ref{7.3.3}).  A similar result is
obtained for a massive scalar field $\varphi$.  In that case,
long-wave fluctuations with $m^2\ll{\rm H}^2$ behave as
\be
\label{7.3.13}
\langle \varphi^2\rangle =\frac{3\,{\rm H}^4}{8\,\pi^2\,m^2}\,
\left[1-\exp\left(-\frac{2\,m^2}{3\,{\rm H}}\,t\right)\right]\ .
\ee

When $\displaystyle t\la\frac{3\,{\rm H}}{m^2}$,
$\langle \varphi^2\rangle $ grows linearly, just as in the case
of the massless field (\ref{7.3.12}), and it then tends to its
asymptotic value (\ref{7.3.3}).

Let us now try to provide an intuitive physical interpretation of
these results.  First, note that the main contribution to
$\langle \varphi^2\rangle $ (\ref{7.3.12}) comes from integrating
over exponentially small $k$ (with $k\sim{\rm H}\,\exp(-{\rm H}\,t)$).
The corresponding occupation numbers $n_k$ (\ref{7.3.11}) are
then exponentially large.  For large $l=|{\bf x}-{\bf y}|\,e^{{\rm H}\,t}$,
the correlation function $\langle \varphi(x)\,\varphi(y)\rangle $
for the massless field $\varphi$ is [\cite{203}]
\be
\label{7.3.14}
\langle \varphi({\bf x},t)\,\varphi({\bf y},t)\rangle \approx
\langle \varphi^2({\bf x},t)\rangle \,
\left(1-\frac{1}{{\rm H}\,t}\,\ln{\rm H}\,l\right)\ .
\ee
This means that the magnitudes of the fields $\varphi(x)$ and
$\varphi(y)$ will be highly correlated out to exponentially large
separations $l\sim{\rm H}^{-1}\,\exp({\rm H}\,t)$.  By all these
criteria, long-wave quantum fluctuations of the field $\varphi$
with $k\ll{\rm H}^{-1}$ behave like a weakly inhomogeneous
(quasi)classical field $\varphi$ generated during the
inflationary stage;  see the discussion of this point in Section
\ref{s2.1}.

\begin{figure}[t]\label{f32}
\centering \leavevmode\epsfysize=6cm \epsfbox{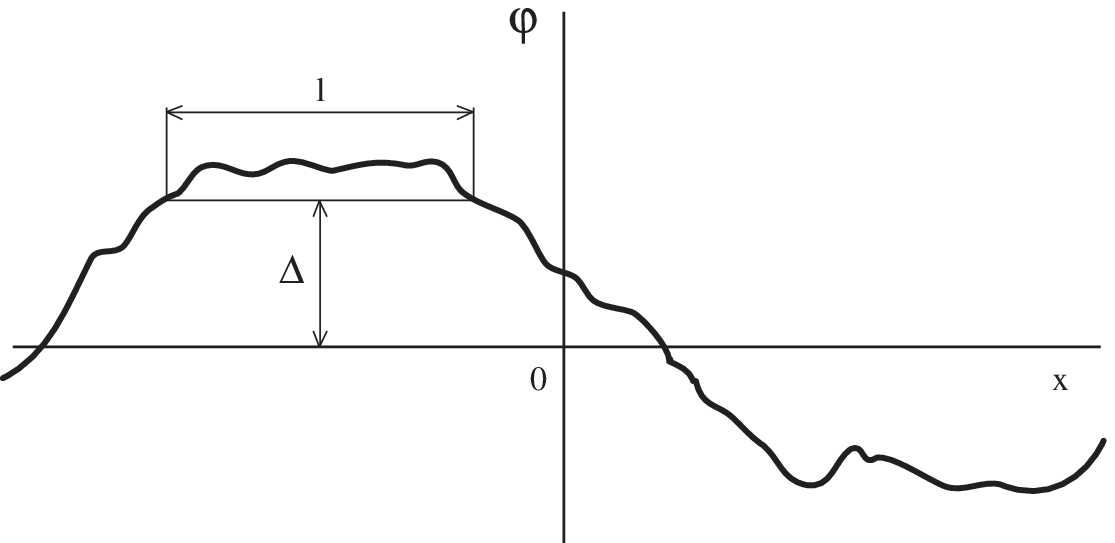}
\caption{Distribution of the semiclassical field $\varphi$
generated at the time of inflation.  For a massless field,
dispersion $\Delta$ is equal to $\displaystyle \frac{{\rm
H}}{2\,\pi}\,\sqrt{{\rm H}\,t}$, and a typical correlation length
$l$ is equal to ${\rm H}^{-1}\,\exp({\rm H}\,t)$.  For a massive
field with $m\ll{\rm H}$, an equilibrium distribution is
established in a time $\displaystyle \Delta t\la\frac{{\rm
H}}{m^2}$, having $\displaystyle \Delta\sim\frac{{\rm H}^2}{m}$
and $\displaystyle l\sim{\rm H}^{-1}\,\exp\left(\frac{3\,{\rm
H}^2}{2\,m^2}\right)$.} \vspace{-5pt}
\end{figure}
Analogous results also hold for a massive field with $m^2\ll{\rm H}^2$.
There, the principal contribution to $\langle \varphi^2\rangle $
comes from modes with
$\displaystyle k\sim{\rm H}\exp\left(-\frac{3\,{\rm H}^2}{2\,m^2}\right)$,
and the correlation length is of order\linebreak[10000]
$\displaystyle {\rm H}^{-1}\,\exp\left(\frac{3\,{\rm H}^2}{2\,m^2}\right)$;
see Fig. 7.3.

An important remark is in order here.  In constructing theories
of particle creation in an expanding universe, elementary
particle theorists have had to come to grips with the fact that
distinguishing real particles from vacuum fluctuations in the
general theory of relativity is a rather ambiguous problem
[\cite{74}].  What we have encountered here in our example is a
similar situation.  Specifically, from the standpoint of
quantization in the coordinate system (\ref{7.2.3}), long-wave
fluctuations with ${\rm H}\,e^{-{\rm H}\,t}\la k\la{\rm H}$
correspond to momenta ${\rm H}\la p\la{\rm H}\,e^{{\rm H}\,t}$.
The corresponding occupation numbers in $p$-space show no
exponential rise with time whatsoever.  The correlation between
$\varphi(x)$ and $\varphi(y)$ at large $|{\bf x}-{\bf y}|$ is
negligible.  Thus, from the standpoint of quantization in de
Sitter space (\ref{7.2.3}), we are dealing with quantum
fluctuations.  But from the standpoint of occupation numbers at
{\it physical} momenta $k=p\,\exp(-{\rm H}\,t)$ and the
correlation at large {\it physical} separations $l=|{\bf x}-{\bf
y}|\,e^{{\rm H}\,t}$, we are dealing with a semiclassical weakly
inhomogeneous field $\varphi$.

The difference in question is quite evident when we compare the
functions $\psi_p(t)$ (\ref{7.3.8}) and
$\displaystyle \psi_k(t)=\psi_p\exp\left(\frac{3}{2}\,{\rm H}\,t\right)$,
the square of which also gives the spectrum (\ref{7.3.10}) in
terms of the physical momentum $k$:
\be
\label{7.3.15}
\psi_k(t)=-\frac{1}{\sqrt{2}}\,e^{-\frac{i\,k}{{\rm H}}}\,
\left(1+\frac{{\rm H}}{i\,k}\right)\ .
\ee
When $k\gg {\rm H}$, we are dealing with a field that oscillates
at constant amplitude $\displaystyle \frac{1}{\sqrt{2}}$.  But in
the course of time, when the magnitude of $k\sim p\,e^{-{\rm H}\,t}$
($p = \mbox{const}$) falls below H, oscillations will
cease, and the amplitude of the field distribution for
$\psi_k(t)$, which has had its phase frozen in, begins to grow
exponentially,
\be
\label{7.3.16}
\psi_k(t)=\frac{i\,{\rm H}}{\sqrt{2}\,k}
\sim\frac{i\,{\rm H}}{\sqrt{2}\,p}\,e^{{\rm H}\,t}\ ,
\ee
which is just the reason for the appearance of exponentially
large occupation numbers.

We have already encountered this phenomenon in discussing the
problem of Bose condensation and symmetry breaking in field
theory.  This is exactly the way in which the exponential
instability involving the creation of the classical Higgs field
develops;  see Eq. (\ref{1.1.6}).  The difference here is that
for symmetry breaking in Minkowski space, it is the mode with
vanishing momentum $k$ that grows most rapidly.  In the
inflationary universe, the momentum $k$ at any of the modes falls
off exponentially.  This leads to an almost identical growth of
modes with different initial momenta $k$, as a result of which
the classical field $\varphi$ becomes inhomogeneous, although
this inhomogeneity becomes significant only at exponentially
large distances $l\sim{\rm H}^{-1}\,\exp({\rm H}\,t)$;  see
(\ref{7.3.14}).  Another important difference between the
phenomenon at hand and spontaneous symmetry breaking in Minkowski
space is that the production of a classical field $\varphi$ in de
Sitter space is an induced phenomenon.  The growth of long-wave
perturbations of the field $\varphi$ occurs even when it is
energetically unfavorable, as for instance when $m^2>0$ (but only
when $m^2\ll{\rm H}^2$).

The process for generating a classical scalar field $\varphi(x)$
in the inflationary universe can be interpreted to be the result
of the Brownian motion of the field $\varphi$ induced by the
conversion of quantum fluctuations of that field into a
semiclassical field $\varphi(x)$.  For any given mode with fixed
$p$, this conversion occurs whenever the physical momentum
$k\sim p\,e^{-{\rm H}\,t}$ becomes comparable to H.  A
``freezing'' of the amplitude of the field $\psi_p(t)$ then
occurs;  see (\ref{7.3.8}).  Due to a phase mismatch
$e^{i\,{\bf p}\,{\bf x}}$, waves with different momenta contribute to the
classical field $\varphi(x)$ with different signs, and this also
shows up in Eq. (\ref{7.3.9}), which characterizes the variance
in the random distribution of the field that arises at the time
of inflation.  As in the standard diffusion problem for a
particle undergoing Brownian motion, the mean squared particle
distance from the origin is directly proportional to the duration
of the process (\ref{7.3.12}).

At any given point, the diffusion of the field $\varphi$ can
conveniently be described by the probability distribution
${\rm P}_c(\varphi,t)$ to find the field $\varphi$ at that point
and at a given instant of time $t$.  The subscript $c$ here
serves to indicate the fact that this distribution, as can
readily be shown, also corresponds to the fraction of the
original \b{c}oordinate volume $d^3x$ (\ref{7.2.3}) filled by the
field $\varphi$ at time $t$.  The evolution of the distribution
of the massless field $\varphi$ in the inflationary universe can
be found by solving the diffusion equation [\cite{204},
\cite{134}, \cite{135}]:
\be
\label{7.3.17}
\frac{\partial{\rm P}_c(\varphi,t)}{\partial t}={\rm D}\,
\frac{\partial^2{\rm P}_c(\varphi,t)}{\partial\varphi^2}\ .
\ee
\index{Diffusion coefficient}%
To determine the diffusion coefficient D in (\ref{7.3.17}), we
take advantage of the fact that
$$
\langle \varphi^2\rangle \equiv
\int\varphi^2\,{\rm P}_c(\varphi,t)\:d\varphi
=\frac{{\rm H}^3}{4\,\pi^2}\,t\ .
$$
Differentiating this relation with respect to $t$
and using (\ref{7.3.17}), we obtain
$$
{\rm D}=\frac{{\rm H}^3}{8\,\pi^2}\ .
$$
It is readily shown that the solution of (\ref{7.3.17}) with
initial condition ${\rm P}_c(\varphi,0)=\delta(\varphi)$  is a
Gaussian distribution
\be
\label{7.3.18}
{\rm P}_c(\varphi,t)=\sqrt{\frac{2\,\pi}{{\rm H}^3\,t}}\,
\exp\left(-\frac{2\,\pi^2\,\varphi^2}{{\rm H}^3\,t}\right)\ ,
\ee
with dispersion squared $\displaystyle \Delta^2=
\langle \varphi^2\rangle =\frac{{\rm H}^3\,t}{4\,\pi^2}$ (\ref{7.3.12}).

When we consider the production of a massive classical scalar
field with mass $|m^2|\ll{\rm H}^2$, the diffusion coefficient D,
\index{Diffusion coefficient}%
which is related to the rate at which quantum fluctuations with
$k>{\rm H}$ are transferred to the range $k<{\rm H}$, remains
unchanged, since the contribution to $\langle \varphi^2\rangle $
from modes with $k\sim{\rm H}$ does not depend on $m$ for
$|m^2|\ll{\rm H}^2$.  For the same reason, $\langle \varphi^2\rangle $
of (\ref{7.3.13}) grows in just the same way
as for a massless field, as given by (\ref{7.3.12}).  But
subsequently, the long-wave classical field $\varphi$, which
appears during the first stages of the process, begins to
decrease as a result of the slow roll down toward the point
$\varphi=0$, in accordance with the classical equation of motion
\be
\label{7.3.19}
\ddot\varphi+3\,{\rm H}\,\dot\varphi=-\frac{d{\rm V}}{d\varphi}
=-m^2\,\varphi\ .
\ee
This finally leads to stabilization of the quantity
$\langle \varphi^2\rangle $ at its limiting value
$\displaystyle \frac{3\,{\rm H}^4}{8\,\pi^2\,m^2}$
(\ref{7.3.13}).  To describe this process, we must write the
diffusion equation in a more general form [\cite{205}]:
\be
\label{7.3.20}
\frac{\partial{\rm P}_c}{\partial t}
={\rm D}\,\frac{\partial^2{\rm P}_c}{\partial\varphi^2}
+b\,\frac{\partial}{\partial\varphi}
\left({\rm P}_c\,\frac{d{\rm V}}{d\varphi}\right)\ ,
\ee
where as before
$\displaystyle {\rm D}=\frac{{\rm H}^3}{8\,\pi^2}$ and $b$
is the mobility coefficient, defined by the equation
$\displaystyle \dot\varphi=-b\,\frac{d{\rm V}}{d\varphi}$.
Using (\ref{7.3.19}) for the slowly varying field $\varphi$
($\ddot\varphi\ll3\,{\rm H}\,\dot\varphi$), we obtain
\be
\label{7.3.21}
\frac{\partial{\rm P}_c}{\partial t}=
\frac{{\rm H}^3}{8\,\pi^2}\,\frac{\partial^2{\rm P}_c}{\partial\varphi^2}
+\frac{1}{3\,{\rm H}}\,\frac{\partial}{\partial\varphi}
\left({\rm P}_c\,\frac{d{\rm V}}{d\varphi}\right)\ .
\ee
This equation was first derived by Starobinsky [\cite{134}] by
another, more rigorous method;  a more detailed derivation can be
found in [\cite{186}, \cite{135}, \cite{132}, \cite{206}].
Solution of this equation for the case
$\displaystyle {\rm V}(\varphi)=\frac{m^2}{2}\,\varphi^2+{\rm V}(0)$
actually leads to the distribution ${\rm P}_c(\varphi,t)$ with
dispersion determined by (\ref{7.3.13}).  Solutions that are
valid for a more general class of potentials ${\rm V}(\varphi)$
will be discussed in the next section in connection with the
problem of tunneling in the inflationary universe.

To conclude, we note that in deriving Eq. (\ref{7.3.21}), it has
been assumed that H is independent of the field $\varphi$.  More
generally, Eq. (\ref{7.3.21}) can be written in the form
\be
\label{7.3.22}
\frac{\partial{\rm P}_c}{\partial t}=
\frac{\partial^2}{\partial\varphi^2}
\left(\frac{{\rm H}^3\,{\rm P}_c}{8\,\pi^2}\right)+
\frac{\partial}{\partial\varphi}\left(\frac{{\rm P}_c}{3\,{\rm H}}\,
\frac{d{\rm V}}{d\varphi}\right)\ .
\ee
Strictly speaking, this equation also holds only when variations
in the field $\varphi$ are small enough that the back reaction of
inhomogeneities of the field on the metric is not too large.
Nevertheless, with the help of this equation, one can obtain
important information on the global structure of the universe;
see Chapter \ref{c10}.
\index{Inflationary universe!quantum fluctuations in|)}%
\index{Quantum fluctuations in!inflationary universe|)}%

\section{\label{s7.4}Tunneling in
\index{Inflationary universe!tunneling in|(}%
\index{Tunneling in!inflationary universe|(}%
the inflationary universe}

The first versions of the inflationary universe scenario were
based on the theory of decay of a supercooled vacuum state
$\varphi=0$ due to tunneling with creation of bubbles of the
field $\varphi$ at the time of inflation [\cite{53}--\cite{55}].
The theory of such processes in Minkowski space, which is
discussed in Chapter \ref{c4}, turns out to be inapplicable to
the most interesting situations, where the curvature of the
effective potential near its local minimum is small compared to
${\rm H}^2$.  Coleman and De Luccia [\cite{207}] have developed a
Euclidean theory of tunneling in de Sitter space, but the general
applicability of this theory to the study of tunneling during
inflation was confirmed only very recently [\cite{208}].  One of
the main problems was that according to [\cite{207}] both the
scalar field $\varphi$ inside a bubble and the metric
$g_{\mu\nu}(x)$ experience a quantum jump. However, in certain
situations, there is a barrier only in the direction of change of
the field $\varphi$.  The analog of this problem is that of the
motion of a particle in the $(x, y)$-plane in a potential ${\rm V}(x, y)$
having the form of a barrier only in the $x$-direction.
A particle encountering the barrier in this situation tunnels
through in the $x$-direction, but it may continue to move
undisturbed along its classical trajectory in the $y$-direction.
To  investigate tunneling under these circumstances, in general
one cannot simply transform to imaginary time (imaginary energy);
instead, one must undertake an honest solution of the
Schr\"{o}dinger equation for the wave function $\Psi(x,y)$,
allowing for the fact that some of the components of the particle
momentum may have an imaginary part.

\begin{figure}[t]\label{f33}
\hskip 3cm \centering \leavevmode\epsfysize=6cm \epsfbox{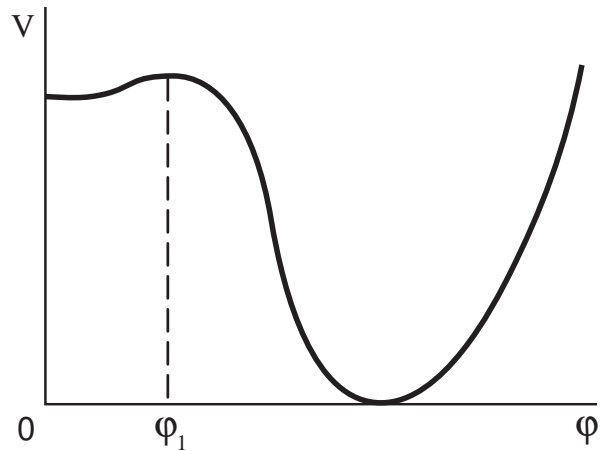}
\caption{The potential ${\rm V}(\varphi)$ used by Hawking and Moss
to study tunneling.}
\end{figure}
Another problem was an ambiguity of interpretation of the results
of Euclidean approach to tunneling.  Let us consider e.g. a
theory with the effective potential with a small local minimum a
$\varphi=0$, such that $\displaystyle m^2 =
\frac{d^2{\rm V}}{d\varphi^2}\biggr|_{\varphi=0}\ll{\rm H}^2$;
tunneling in such a theory was studied by Hawking and Moss
[\cite{121}].  Their expression for the probability of tunneling
from the point $\varphi=0$ through a barrier with a maximum at
the point $\varphi_1$ (Fig. 7.4) looks like
\be
\label{7.4.1}
{\rm P}\sim{\rm A}\,\exp\left[-\frac{3\,\m^4}{8}\,
\left(\frac{1}{{\rm V}(0)}-\frac{1}{{\rm V}(\varphi_1)}\right)\right]\ ,
\ee
where A is some multiplicative factor with dimensionality $m^4$.
Hawking and Moss assumed in deriving this equation that by virtue
of the ``no hair'' theorem for de Sitter space (see Section
\ref{s7.2}), tunneling in an exponentially expanding de Sitter
space (\ref{7.2.3}) should occur in just the same way it occurs
in a closed space (\ref{7.2.5}).  Tunneling in the latter case is
most likely at the throat of the hyperboloid (i.e., at $t=0$,
$a={\rm H}^{-1}$), while according to [\cite{207}], a description
of that tunneling requires that we calculate the appropriate
action in the Euclidean version of the space (\ref{7.2.5}), that
is, on a sphere ${\rm S}^4$ of radius ${\rm H}^{-1}(\varphi)$.
Since our concern is with tunneling in which ${\rm H}(\varphi)$
increases (i.e.  $a$ decreases), which is classically forbidden,
the preceding argument against a Euclidean approach to this case
does not apply.  A calculation of the action S on the sphere
leads to the quantity
\be
\label{7.4.2}
{\rm S}_{\rm E}(\varphi)=-\frac{3\,\m^4}{8\,\vf}\ .
\ee
Adhering to the ideology developed in the work of Coleman and De
Luccia, Hawking and Moss asserted that the probability of
tunneling is proportional to
$\exp({\rm S}_{\rm E}(0)-{\rm S}_{\rm E}(\varphi))$.  This also
leads to Eq. (\ref{7.4.1}), but the contribution to the action
from the bubble walls was not taken into account --- in other
words, they treated purely homogeneous tunneling suddenly taking
place over all space [\cite{121}].  This result was later
``confirmed'' by numerous authors;  however, simultaneous
tunneling throughout an entire exponentially large universe seems
quite unlikely.

In order to study this problem in more detail, a Hamiltonian
approach to the theory of tunneling at the time of inflation was
developed, and has succeeded in showing that the probability of
homogeneous tunneling over an entire inflationary universe is
actually vanishingly small [\cite{186}].  Hawking and Moss
themselves later remarked that their result should be interpreted
not as the probability of homogeneous tunneling throughout the
entire universe, but as the probability of tunneling which looks
homogeneous only on some scale $l\ga {\rm H}^{-1}$  [\cite{209}].
They argued that bubble walls and other inhomogeneities should
have no effect on tunneling, due to the ``no hair'' theorem for
de Sitter space (see Section \ref{s7.2}).

The validity of that argument, and in fact the overall
applicability of the Euclidean approach to this problem, was open
to doubt.  Only much later was it learned that when $m^2\ll{\rm H}^2$,
the contribution to the Euclidean action from gradients of
the field $\varphi$ is small [\cite{186}] (in contrast to the
situation in a Minkowski space, where this contribution is of the
same order as that of the potential energy of the field
$\varphi$), and that tunneling in this instance is effectively
one-dimensional (basically occurring because of a change in the
scalar field).  Equation (\ref{7.4.1}) was thereby partially
justified.  But a genuine understanding of the physical essence
of this phenomenon was not achieved until an approach to the
theory of tunneling based on the diffusion equation
(\ref{7.3.21}) was developed [\cite{134}, \cite{135}].

The fundamental idea is that for tunneling to occur, it suffices to
have a bubble with a field exceeding $\varphi_1$ and radius
$$
r>{\rm H}^{-1}(\varphi)=
\sqrt{\frac{3\,\m^2}{8\,\pi\,{\rm V}(\varphi_1)}}\ .
$$
Further evolution of the field $\varphi$ inside this bubble does
not depend on what goes on outside it;  in other words, the field
will start to roll down to the absolute minimum of ${\rm V}(\varphi)$
with $\varphi>\varphi_1$.  It only remains to evaluate the
probability that a region of this type will form --- but that is
exactly the problem we studied in the previous section!

As we have already stated, the distribution ${\rm P}_c(\varphi,t)$
actually characterizes that fraction of the
original coordinate volume $d^3x$ (\ref{7.2.3}) which at time $t$
contains the field $\varphi$; the latter is homogeneous on a
scale $l\ga{\rm H}^{-1}$.  The problem of tunneling at the time
of inflation thereby reduces to the solution of the diffusion
equation (\ref{7.3.21}) with initial condition
${\rm P}_{c}(\varphi,0)=\delta(\varphi)$.

At this point, we must distinguish between two possible regimes.

1) In the initial stage of this process, the dispersion
$\sqrt{\langle \varphi^2\rangle }$ grows as
$\displaystyle \frac{{\rm H}}{2\,\pi}\,\sqrt{{\rm H}\,t}$
(\ref{7.3.12}).  If at that stage it becomes larger in magnitude
than $\varphi_1$, which characterizes the position of the local
maximum of ${\rm V}(\varphi)$, the process will proceed as if
there were no barrier at all [\cite{127}].  In that event,
diffusion will end when the field $\varphi$ encounters a steep
slope in ${\rm V}(\varphi)$, where the rate of diffusive growth
of the field drops below the rate of classical rolling.  Under
typical conditions, the diffusion stage lasts for a time
\be
\label{7.4.3}
t\sim\frac{4\,\pi^2\,\varphi_1^2}{{\rm H}^3}\ ,
\ee
and the typical shape of the regions within which the field
$\varphi$ exceeds some given value (say $\varphi_1$) is far from
that of a spherical bubble.

2) If the dispersion stops growing when
$\sqrt{\langle \varphi^2\rangle }\ll\varphi_1$, the distribution
${\rm P}_c(\varphi,t)$ will become quasistationary.  It should then be
possible to find it by putting
$\displaystyle \frac{\partial{\rm P}_c(\varphi,t)}{\partial t}=0$
in Eq.  (\ref{7.3.21}), or in the more general equation (\ref{7.3.22}).
To clarify the physical meaning of solutions of these equations, it
is convenient to rewrite Eq.  (\ref{7.3.22}) in the form
\ba
\label{7.4.4}
\frac{\partial{\rm P}_c}{\partial t}&=&
-\frac{\partial j_c}{\partial\varphi}\ , \\
\label{7.4.5}
-j_c&=&\frac{1}{3}\,\sqrt{\frac{3\,\m^2}{8\,\pi\,{\rm V}}}
\left[\frac{8\,{\rm V}^2}{3\,\m^4}\,\frac{\partial{\rm P}_c}{\partial\varphi}
+{\rm P}_c\,\frac{d{\rm V}}{d\varphi}\,
\left(1+\frac{4\,{\rm V}}{\m^4}\right)\right]\ .
\ea
Here we have introduced the\index{Probability current}
probability current $j_c(\varphi,t)$
in $(\varphi, t)$-space [\cite{205}], so that Eq. (\ref{7.4.4})
takes on the standard form of the continuity equation for the
probability\index{Density!probability}\index{Probability density}
density ${\rm P}_c(\varphi,t)$.  Consideration of the
standard condition $\displaystyle \frac{\partial{\rm P}_c}{\partial t}=0$
amounts to examination of the case in which
the probability current is constant for all $\varphi$ from
$-\infty$ to $\infty$.  As a rule, there are no reasonable
initial conditions under which an unattenuated, nonvanishing
diffusion current $j_c=\mbox{const}\neq0$ arises between
$\varphi=-\infty$ and $\varphi=+\infty$ (see [\cite{135}],
however).  Furthermore, the diffusion process itself is usually
feasible only within certain limited ranges of variation of the
field $\varphi$ (namely, where
$\displaystyle \frac{d^2{\rm V}}{d\varphi^2}\ll{\rm H}^2$ and $\vf\ll\m^4$).
Outside these zones, the first (diffusion) term in (\ref{7.4.5})
does not appear, and if the potential ${\rm V}(\varphi)$ is an
even function of $\varphi$, Eq. (\ref{7.4.5}) implies that ${\rm P}_c$
must be an odd function of $\varphi$, which is impossible,
inasmuch as ${\rm P}_c(\varphi,t)\ge0$.  For all of these
reasons, we will consider only the case $j_c=0$ (in this regard,
see also Chapter \ref{c10}).

When $j_c=0$, and also when $\vf\ll\m^4$, Eq. (\ref{7.4.5})
reverts to a very  simple form,
\be
\label{7.4.6}
\frac{\partial\ln{\rm P}_c}{\partial\varphi}=
-\frac{3\,\m}{8\,{\rm V}^2(\varphi)}\,\frac{d{\rm V}}{d\varphi}\ ,
\ee
whereupon
\be
\label{7.4.7}
{\rm P}_c={\rm N}\,\exp\left(\frac{3\,\m^4}{8\,\vf}\right)\ ,
\ee
where N is a normalizing factor such that $\displaystyle
\int{\rm P}_c\:d\varphi=1$.  In the present instance, where the rms
deviation of the field is much less than the width of the
potential well ($\sqrt{\langle \varphi^2\rangle }\ll\varphi_1$),
the function $\displaystyle \exp\left(\frac{3\,\m}{8\,\vf}\right)$
has a clear-cut maximum at $\varphi = 0$, and therefore, to
within an unimportant sub-exponential factor,
\be
\label{7.4.8}
{\rm P}_c(\varphi)=\exp\left(-\frac{3\,\m^4}{8}\,
\left(\frac{1}{{\rm V}(0)}-\frac{1}{{\rm V}(\varphi)}\right)\right)\ .
\ee
According to (\ref{7.4.8}), the probability that the field at a
given point (or more precisely, in a neighborhood of size
$l\ga{\rm H}^{-1}$ at a given point) will equal $\varphi_1$ is
given precisely by the exponential term in the Hawking--Moss
\index{Hawking--Moss equation}%
equation (\ref{7.4.1}).  This is not just a coincidence, since
the mean diffusion time from $\varphi=0$ to $\varphi =\varphi_1$,
that is, the mean time that tunneling goes on at a given point,
is in fact proportional to ${\rm P}_c(\varphi_1)$.  The
corresponding result in the theory of Brownian motion is well
known [\cite{210}];  for the present case, it was derived in
[\cite{134}, \cite{135}].  Its physical meaning is most easily
understood if we consider motion along a Brownian trajectory at
(approximately) constant speed (as happens here when
${\rm H}(\varphi) \approx \mbox{const}$).  The value of
${\rm P}_c(\varphi)$ indicates the number density of points on this
trajectory at which the value of the field is equal to $\varphi$.
This means that the mean time $\tau$ to move from the point
$\varphi = 0$ to the point $\varphi=\varphi_1$ along a Brownian
trajectory is proportional to $[{\rm P}_c(\varphi)]^{-1}$, and
consequently the tunneling (diffusion) probability P per unit
time $\tau$ is proportional to ${\rm P}_c(\varphi)$.

Strictly speaking, the tunneling process is not stationary, but
if the time required for relaxation to the quasistationary state
is much less than the time for tunneling, then Eq. (\ref{7.4.8})
will provide a good representation of the distribution ${\rm P}_c(\varphi)$
This condition is satisfied if
\be
\label{7.4.9}
\frac{3\,\m^4}{8}\,
\left(\frac{1}{{\rm V}(0)}-\frac{1}{{\rm V}(\varphi_1)}\right)\gg1\ .
\ee
One can readily show that this is equivalent to requiring that
$\sqrt{\langle \varphi^2\rangle }\ll\varphi_1$.  In the present
context, the probability of forming large nonspherical regions of
a field $\varphi>\varphi_1$ that are bigger than ${\rm H}^{-1}(\varphi_1)$
is much lower than that of forming spherical bubbles of the field
$\varphi$.

As an example, consider the theory with effective potential \be
\label{7.4.10} \vf={\rm
V}(0)+\frac{m^2}{2}\,\varphi^2-\frac{\lambda}{4}\,\varphi^4\ . \ee
For this theory, $\displaystyle
\varphi_1=\frac{m}{\sqrt{\lambda}}$, and Eq. (\ref{7.4.1}) for
${\rm V}(\varphi_1)-{\rm V}(0)\ll{\rm V}(0)$ becomes \be
\label{7.4.11} {\rm
P}_c\sim\exp\left[-\frac{3\,\m^4\,m^4}{32\,\lambda\,{\rm
V}^2(0)}\right]
=\exp\left[-\frac{2\pi^2}{3\,\lambda}\,\left(\frac{m}{{\rm
H}}\right)^4\right]\ , \ee while (\ref{7.4.9}), together with the
condition that $m^2\ll{\rm H}^2$, may be cast in the form \be
\label{7.4.12} \sqrt{\lambda}<\frac{m^2}{{\rm H}^2}\ll1\ . \ee

A more detailed study of the solutions of Eq. (\ref{7.3.22})
makes it possible to obtain expressions for the mean duration of
tunneling which hold both for $\sqrt{\langle \varphi^2\rangle }\gg\varphi_1$
and $\sqrt{\langle \varphi^2\rangle }\ll\varphi_1$  [\cite{135}].
Most important for us here has been the elucidation of the
general properties of phase transitions at the time of inflation,
a subject discussed in more detail elsewhere [\cite{186}].  One
of the most surprising features of such phase transitions is the
possibility of diffusion from one local minimum of ${\rm V}(\varphi)$
to another with an {\it increase} in energy density
[\cite{211}].  This effect and related ones are extremely
important for an understanding of the global structure of the
universe.  We shall return to this question in Chapter \ref{c10}.

Thus, with a stochastic approach, one can justify the
\index{Hawking--Moss equation}%
Hawking--Moss equation (\ref{7.4.1}) [\cite{121}] and confirm
their interpretation of this equation [\cite{209}].  On the other
hand, this same approach has provided a means of appreciating the
limits of applicability of Eq. (\ref{7.4.1}).  The ``derivation''
of this result given in [\cite{121}] imposed no constraints on
the form of the potential ${\rm V}(\varphi)$, and it was not
clear why tunneling should occur through the nearest maximum of
${\rm V}(\varphi)$, rather than directly to its next minimum.
The answer to this last question is obvious within the context of
our present approach, and it is also clear that Eq. (\ref{7.4.1})
itself is only valid if the curvature of ${\rm V}(\varphi)$ is
much less than ${\rm H}^2$ over the whole domain of variation of
$\varphi$ from 0 to $\varphi_1$.

Another important observation to be made in studying the theory
of tunneling in the inflationary universe relates to the
properties of the walls of bubbles of a new phase.  In Minkowski
space, the total energy of a bubble of a new phase that is
created from vacuum is exactly zero.  As the bubble grows, so
does its negative energy, which is proportional to its volume
$\displaystyle \sim\frac{4}{3}\,\pi\,r^3\,\varepsilon$, and which
is related to the energy gain $\varepsilon$ realized in the
transition to a new phase.  At the same time (and the same rate),
the positive\index{Bubble-wall energy}\index{Energies!bubble-wall}
bubble-wall energy grows as
$4\,\pi\,r^2\,\sigma(t)$, where $\sigma$ is the surface energy
density of the bubble.  These two terms cancel, which is only
possible because the surface energy is also proportional to $r$,
the reason being that the speed of a wall approaches the speed of
light, while its thickness decreases.  Thus, even if the
thin-wall approximation were inappropriate in a description of
the bubble creation process, it could be usable in a description
of its subsequent evolution [\cite{212}, \cite{213}].  Formally,
this occurs because any bubble of the field $\varphi$ created
from vacuum by O(4)-symmetric tunneling can be described by some
function of the form
\be
\label{7.4.13}
\varphi=\varphi({\bf r}^2-t^2)\ ;
\ee
see [\cite{180}].  If this bubble has a characteristic initial size of
$r_0$ at $t=0$, then the field at large $t$ will reach a value
$\varphi(0)$ at a distance
\be
\label{7.4.14}
\Delta r=\frac{r_0^2}{2\,r}\approx\frac{r_0^2}{2\,t}
\ee
from the bubble boundary (i.e., from the sphere on which
$\varphi({\bf r}^2-t^2)=\varphi(r_0^2)\approx0$).  With time, then,
the wall thickness quickly decreases.

In the inflationary universe, everything is completely different.
The total energy of the field $\varphi$ within a bubble does not
vanish, and is not conserved as the universe expands.  This is
attributable to the same gravitational forces that drive the
exponential growth of the total energy of the scalar field at the
time of inflation (${\rm E}\sim\vf\,a^3(t)\sim\vf\,e^{3\,{\rm H}\,t}$).
Tunneling results from the formation of perturbations
$\delta\varphi(x)$ with wavelengths $l\ga{\rm H}^{-1}$.  All
gradients of these perturbations are very small, and do not
affect their evolution.  This is also the reason why in the final
\index{Hawking--Moss equation}%
analysis the Hawking--Moss equation, neglecting the contribution
of boundary terms in the Euclidean action, is found to be
correct.  In the study of bubbles engendered by the foregoing
mechanism, therefore, the thin-wall approximation is often
inapplicable at any stage of bubble evolution.  But if the
regions that are formed contain matter in many different phase
states, then the domain walls that appear between these regions
in the late stages of inflation, or after inflation has
terminated, can actually become thin.  The powerful methods
developed in [\cite{212}, \cite{213}] can be utilized to
investigate the structure of the universe in the vicinity of such
regions.
\index{Inflationary universe!tunneling in|)}%
\index{Tunneling in!inflationary universe|)}%

\section[Quantum fluctuations and perturbations]%
{\label{s7.5}Quantum fluctuations and the generation of adiabatic
\index{Adiabatic density perturbations|(}%
\index{Perturbations!adiabatic density|(}%
\index{Density!perturbations|(}%
\index{Quantum fluctuations!and generation
of adiabatic density perturbations|(}%
density perturbations}

We now continue our study of perturbations of a scalar field with
exponentially large wavelength that come into being during an
inflationary stage.  From the standpoint of quantization in the
coordinate system (\ref{7.2.3}), the wavelengths of these
fluctuations do not grow ($p=\mbox{const}$ in Eq. (\ref{7.3.4})),
and they differ but little from conventional vacuum fluctuations.
In particular, one can calculate the corrections to the
energy-momentum tensor $g_{\mu\nu}\,{\rm V}(\varphi)$ that are
associated with these fluctuations;  in the stationary state
($\varphi=\mbox{const}$), these are [\cite{202}, \cite{203}]
\be
\label{7.5.1}
\Delta{\rm T}_{\mu\nu}=
\frac{1}{4}\,\langle \varphi^2\rangle \,m^2\,g_{\mu\nu}
=\frac{3\,{\rm H}^4}{32\,\pi^2}\,g_{\mu\nu}
\ee
regardless of the mass of the field $\varphi$ (for $m^2\ll{\rm H}^2$).
These corrections have a relativistically invariant form (despite
the presence of the Hawking ``temperature''
$\displaystyle {\rm T}_{\rm H}=\frac{{\rm H}}{2\,\pi}$ in de
Sitter space).

But as we have already pointed out, from the point of view of a
stationary observer armed with measuring rods that do not stretch
during inflation of the universe, fluctuations of a scalar field
that have wavelengths greater than the distance to the horizon
($k^{-1}\ga{\rm H}^{-1}$) look like a classical field
$\delta\varphi$ that is weakly inhomogeneous on scales $l\ga{\rm H}^{-1}$.
These fluctuations give rise to density
inhomogeneities on an exponentially large scale.  During
inflation, the magnitude of these inhomogeneities is
\be
\label{7.5.2}
\delta\rho\approx{\rm V}'\,\delta\varphi\ ,
\ee
where $\displaystyle {\rm V}'=\frac{d{\rm V}}{d\varphi}$.  In the
last stages of inflation, an ever increasing fraction of the
field energy is tied up in the kinetic energy of the field
$\varphi$, rather than in ${\rm V}(\varphi)$.  This energy then
transforms into heat, and energy density inhomogeneities
$\delta\rho$ lead to temperature inhomogeneities $\delta {\rm T}$.
The original density inhomogeneities (\ref{7.5.39}) are
thereby transformed into hot-plasma density inhomogeneities, and
then into cold-matter density inhomogeneities.  The corresponding
density inhomogeneities result in so-called {\it adiabatic}
\index{Isothermal perturbations}%
\index{Perturbations!isothermal}%
\index{Metric perturbations}%
\index{Perturbations!metric}%
perturbations of the metric, in contrast to the {\it isothermal}
perturbations associated with inhomogeneities of the metric that
arise at constant temperature.

The appearance of long-wave density (metric) perturbations is
necessary for the subsequent formation of the large scale
structure of the universe (galaxies, clusters of galaxies, voids,
and so on).  Another possible mechanism for generating density
perturbations is related to the theory of cosmic strings produced
during phase transitions in a hot universe.  But it is very
difficult to get along without an inflationary stage, and
therefore the prospect of obtaining inhomogeneities of the type
required simply by virtue of quantum effects at the time of
inflation, without invoking any additional mechanisms, seems
especially interesting.  The fact that the contribution to
$\langle \varphi^2\rangle $  (\ref{7.3.12}) due to integration
over the fixed interval $\displaystyle \Delta\ln\frac{k}{{\rm H}}$
is independent of $k$ leads to a flat spectrum $\delta\rho(k)$
(\ref{7.5.2}) which does not depend on $k$ (as the momentum varies
on a logarithmic scale).  This is exactly the sort of spectrum
suggested earlier by Harrison and Zeldovich [\cite{76}] (see also
[\cite{214}]) as the initial perturbation spectrum required for
the subsequent formation of galaxies.  If we normalize this
spectrum so that $\delta\rho(k)$ denotes the contribution to
$\delta\rho$ from all perturbations per unit interval in
$\displaystyle \ln\frac{k}{{\rm H}}$, then the desired spectral
amplitude should be \be \label{7.5.3}
\frac{\delta\rho(k)}{\rho}\sim\mbox{$10^{-4}$--$10^{-5}$} \ee on a
galactic scale ($l_g\sim10^{22}$ cm at the present epoch;
$l_g\sim10^{-5}$  cm at the instant that inflation ended). Notice,
however, that rather than referring to perturbations $\delta\rho$
at the inflationary stage (\ref{7.5.2}), condition (\ref{7.5.3})
relates to their progeny at a later stage, after reheating of the
universe, when its equation of state becomes $\displaystyle
p=\frac{\rho}{3}$  (or $p=0$, when cold nonrelativistic matter
predominates).  The question of how these perturbations actually
relate to the initial perturbations (\ref{7.5.2}) is a very
complicated one.  Some important steps in the development of a
theory of adiabatic density perturbations formed during the
exponential expansion stage of the universe are to be found in
[\cite{101}, \cite{215}--\cite{217}].  The corresponding problem
for the inflationary universe scenario was first solved by
Mukhanov and Chibisov [\cite{107}] in their investigation of the
Starobinsky model [\cite{52}].  The quantity $\displaystyle
\frac{\delta\rho}{\rho}$ for the new inflationary universe
scenario was calculated by four other groups practically
simultaneously [\cite{114}].  The results obtained by all of these
authors, using different approaches, agreed to within a numerical
factor ${\rm C}\sim {\rm O}(1)$: \be \label{7.5.4}
\frac{\delta\rho(k)}{\rho}={\rm C}\, \frac{{\rm
H}(\varphi)\,\delta\varphi(k)}{\dot\varphi} \biggr|_{k\sim{\rm
H}}\ . \ee What this expression means is that in order to
calculate $\displaystyle \frac{\delta\rho(k)}{\rho}$ on a
logarithmic scale in $k$, one should calculate the value of the
function $\displaystyle \frac{{\rm
H}[\varphi(t)]}{\dot\varphi(t)}$ at a time when the corresponding
wavelength $k^{-1}$ is of the order of the distance to the horizon
${\rm H}^{-1}$ (that is, when the field $\delta\varphi(k)$ becomes
semiclassical).  For $\delta\varphi(k)$ here we can take the rms
value defined by (see (\ref{7.3.12})) \be \label{7.5.5}
[\delta\varphi(k)]^2=\frac{{\rm H}^2(\varphi)}{4\,\pi^2}\,
\int^{\ln\frac{k}{{\rm H}}+1}_{\ln\frac{k}{{\rm
H}}}d\ln\frac{k}{{\rm H}} =\frac{{\rm H}^2(\varphi)}{4\,\pi^2}\ ,
\ee or in other words \be \label{7.5.6}
|\delta\varphi(k)|=\frac{{\rm H}(\varphi)}{2\,\pi}\ . \ee In the
final analysis, these same results are found to be correct for the
chaotic inflation scenario as well [\cite{218}].

It would be hard to overestimate the significance of the first
papers on the density perturbations produced during inflation
[\cite{114}].  However, the validity of some of the assumptions
made in these papers is not obvious.  Furthermore, it was not
entirely clear what the connection was between the density
perturbations occurring during the inflationary stage
(\ref{7.5.2}) and Eq.  (\ref{7.5.4}), and in fact the value of
the parameter C in the latter equation differed somewhat in the
different papers.  This situation engendered an extensive
literature on the problem;  for a review, see [\cite{219}].  In
our opinion, a final clarification of the situation was
particularly facilitated by Mukhanov [\cite{218}].  In what
follows, we will describe the basic ingredients of his work, and
use his results to obtain an equation for
$\displaystyle \frac{\delta\rho}{\rho}$  that will be valid for a
large class of inflationary models.

Consider a region of the inflationary universe of initial size
$\Delta l\ga{\rm H}^{-1}$, containing a sufficiently homogeneous
field $\varphi$ ($\partial_i\varphi\,\partial^i\varphi\ll\vf$).
All initial inhomogeneities of this field die out exponentially,
and the total field in this region can therefore be put in the
form
\be
\label{7.5.7}
\varphi({\bf x},t)\rightarrow\varphi(t)+\delta\varphi({\bf x},t)\ ,
\ee
where inhomogeneities $\delta\varphi({\bf x},t)$ appear because
of long-wave fluctuations that are generated with $k\la{\rm H}$.
The leading contribution to $\delta\varphi({\bf x},t)$ comes from
fluctuations with exponentially long wavelengths.  Therefore, the
main contribution to inhomogeneities in the mean energy-momentum
tensor ${\rm T}_{\mu}^\nu$  on the large scales that we are
interested in come from terms like
$\partial_0\varphi\cdot \partial_0[\delta\varphi({\bf x},t)]$ or
$\displaystyle \delta\varphi({\bf x},t)\cdot \frac{d{\rm V}}{d\varphi}$,
rather than from spatial gradients
$(\partial_i[\delta\varphi({\bf x},t)])^2$ (the second of these
is in fact the leading contribution at the time of inflation; see
(\ref{7.5.2})).  This implies that $\delta{\rm T}_\mu^\nu$ is
diagonal to first order in $\delta\varphi$.  For perturbations of
this type, the corresponding perturbations of the metric in a
flat universe can be represented by [\cite{220}]
\be
\label{7.5.8}
ds^2=(1+2\,\Phi)\,dt^2-(1-2\,\Phi)\,a^2(t)\,d{\bf x}^2\ .
\ee
The function $\Phi({\bf x},t)$ plays a role similar to that of
the Newtonian potential used to describe weak gravitational
fields (compare the metric (\ref{7.5.8}) with the Schwarzschild
metric (\ref{7.2.8})).  The coordinate system (\ref{7.5.8}) is
more convenient for the investigation of density perturbations
than the more frequently used synchronous system [\cite{65}],
since after a synchronous system is chosen through the condition
$\delta g_{i0}=0$, one still has the freedom to change the
coordinate system;  this leads to the existence of two
nonphysical perturbation modes, which makes the calculations and
their interpretation rather complicated and ambiguous.  However,
these modes do not contribute to $\Phi({\bf x},t)$.  Density
inhomogeneities of the type we are considering, with wavelengths
$k^{-1}>{\rm H}^{-1}$, are related to the function $\Phi({\bf x},t)$
in a very simple way,
\be
\label{7.5.9}
\frac{\delta\rho}{\rho}=-2\,\Phi\ .
\ee
A more detailed discussion of the use of the relativistic potential
$\Phi({\bf x},t)$ may be
found in [\cite{220}--\cite{222}, \cite{133}].  Linearizing the
Einstein equations and the equation for the field
$\varphi({\bf x},t)$ in terms of $\delta\varphi$ and $\Phi$,
one can obtain a system of
differential equations for $\delta\varphi$ and $\Phi$:
\be
\label{7.5.10}
\ddot\Phi+
\left(\frac{\dot a}{a}-2\,\frac{\ddot\varphi}{\dot\varphi}\right)\,\dot\Phi
-\frac{1}{a^2}\,\Delta\Phi+2\,\left(\frac{\ddot a}{a}
-\left(\frac{\dot a}{a}\right)^2-
\frac{\dot a}{a}\,\frac{\ddot\varphi}{\dot\varphi}\right)\,\Phi=0\ ,
\ee
\be
\label{7.5.11}
\frac{1}{a}\,(a\,\Phi)\dot{}_{,\beta}=
\frac{4\,\pi}{\m^2}\,(\dot\varphi\,\delta\varphi)_{,\beta}\ ,
\ee
\be
\label{7.5.12}
\delta\ddot\varphi+3\,\frac{\dot a}{a}\,\delta\dot\varphi
-\frac{1}{a^2}\,\Delta\delta\varphi
+\frac{d^2{\rm V}}{d\varphi^2}\,\delta\varphi
+2\frac{d{\rm V}}{d\varphi}\,\Phi-4\,\dot\varphi\,\dot\Phi=0\ .
\ee
Here $\varphi(t)$ and $a(t)$ are the solutions of the unperturbed
equations (see Section \ref{s1.7}), and a dot signifies
differentiation with respect to time.  Using one of the
consequences of the Einstein equations for $a(t)$,
\be
\label{7.5.13}
\left(\frac{\dot a}{a}\right)\dot{\vphantom{\frac{\dot a}{a}}}
=-\frac{4\,\pi}{\m^2}\,\dot\varphi^2\ ,
\ee
Eq. (\ref{7.5.10}) can be cast in the form
\be
\label{7.5.14}
u''-\Delta u-
\frac{(a'/a^2\,\varphi')''}{a'/a^2\,\varphi'}\,u=0\ ,
\ee
where $\displaystyle u=\frac{a}{\varphi'}\,\Phi$, and primes in
the (and only in this) equation stand for differentiation with
respect to the conformal time $\displaystyle
\eta=\int\frac{dt}{a(t)}$.  In the long-wave limit ($k\ll{\rm H}$,
$\displaystyle k^2\ll\frac{d^2{\rm V}}{\delta\varphi^2}$),
the solution of Eq. (\ref{7.5.14}) can be put in the form
\be
\label{7.5.15}
\Phi={\rm C}\,\left(1-\frac{\dot a}{a^2}\,\int^t_0a\:dt\right)\ ,
\ee
where C is some constant, and ${\rm H}\,tt\gg1$.
Thereupon, making use of (\ref{7.5.11}), we obtain
\be
\label{7.5.16}
\delta\varphi={\rm C}\,\dot\varphi\cdot \frac{1}{a}
\int^t_0a\:dt\ .
\ee

The desired result follows from (\ref{7.5.15}) and
(\ref{7.5.16}), namely the relationship between long-wave
fluctuations of the field $\varphi$, perturbations of the metric
$\Phi$, and density inhomogeneities [\cite{218}]:
\be
\label{7.5.17}
\frac{\delta\rho}{\rho}=-2\,\Phi
=-2\,\left[\frac{a}{\displaystyle \int^t_0a\:dt}\,
\frac{\delta\varphi}{\dot\varphi}\right]\,
\left(1-\frac{\dot a}{a^2}\,\int^t_0a\:dt\right)\ .
\ee
The quantity in square brackets is the constant C of
(\ref{7.5.15}), the value of which can be computed at any stage
of inflation.  This is most conveniently done when the wavelength
of a perturbation $\delta\varphi(k)$ is equal to the distance to
the horizon, $k\sim{\rm H}$.  The amplitude at that time can be
estimated with the help of Eq.  (\ref{7.5.6}).

We now make use of the preceding results to compare Eqs.
(\ref{7.5.2}) and (\ref{7.5.4}), and we apply these results to
the calculation of $\displaystyle \frac{\delta\rho}{\rho}$ in the
simple models.  One can easily verify that during the stage of
inflation, when $\dot{\rm H}\ll{\rm H}^2$, $\ddot{\rm H}\ll{\rm H}^3$,
and ${\rm H}\,t\gg1$,
\be
\label{7.5.18}
\frac{a}{\displaystyle \int^t_0a\:dt}={\rm H}(t)\,
\left(1+\frac{\dot {\rm H}}{{\rm H}^2}\,
\left[1+{\rm O}\left(
\frac{\dot {\rm H}}{{\rm H}^2},\frac{\ddot{\rm H}}{{\rm H}^3}\right)
\right]\right)\ .
\ee
The expression in square brackets in (\ref{7.5.17}) is thus equal to
\be
\label{7.5.19}
{\rm C}={\rm H}(\varphi(t))\,\frac{\delta\varphi}{\dot\varphi}\ .
\ee
On the other hand, Eq. (\ref{7.5.18}) implies that during the
inflationary stage, the parenthesized expression in
(\ref{7.5.17}) equals $\displaystyle \frac{\dot{\rm H}}{{\rm H}^2}\ll1$.
In that event, it is readily verified using
(\ref{7.5.17}) and (\ref{7.5.19}) that density inhomogeneities
during inflation are given by
\be
\label{7.5.20}
\frac{\delta\rho}{\rho}\approx\frac{\delta{\rm V}}{{\rm V}}=
\frac{{\rm V}'}{{\rm V}}\,\delta\varphi\ ,
\ee
as one would find from (\ref{7.5.2}), and not by (\ref{7.5.4}).
The difference between (\ref{7.5.20}) and (\ref{7.5.4}) lies in a
small factor that is
$\displaystyle {\rm O}\left(\frac{\dot{\rm H}}{{\rm H}^2}\right)\ll1$.

However, if the universe is hot ($a\sim t^{1/2}$) or cold
($a\sim t^{2/3}$), Eqs. (\ref{7.5.17}) and (\ref{7.5.19}) lead to Eq.
(\ref{7.5.4}), with ${\rm C}=-4/3$ and ${\rm C}=-6/5$ for these
two respective cases [\cite{218}, \cite{220}].  If for
$\delta\varphi(k)$ we then take $\displaystyle \frac{{\rm H}}{2\,\pi}$
as in (\ref{7.5.6}) in order to find the rms value
of $\displaystyle \frac{\delta\rho}{\rho}$ per unit interval in
$\displaystyle \ln\frac{k}{{\rm H}}$, we obtain
\be
\label{7.5.21}
\frac{\delta\rho(k)}{\rho}
={\rm C}\,\frac{[{\rm H}(\varphi)]^2}{2\,\pi\,\dot\varphi}
\biggr|_{k\sim{\rm H}}\ .
\ee
Expressing $\dot\varphi$ and ${\rm H}(\varphi)$ in terms of
${\rm V}(\varphi)$ during inflation, we find that at the stage
when cold matter predominates (when galaxy formation presumably
started),
\be
\label{7.5.22}
\frac{\delta\rho(k)}{\rho}=\frac{48}{5}\,\sqrt{\frac{2\,\pi}{3}}\,
\frac{{\rm V}^{3/2}}{\m^3\,\displaystyle \frac{d{\rm V}}{\delta\varphi}}
\biggr|_{k\sim{\rm H}(\varphi)}
\ee
(an irrelevant minus sign has been omitted from (\ref{7.5.22})).
As an example of the use of this equation, one can obtain the
amplitude of density inhomogeneities in the theory
$\displaystyle \frac{\lambda}{4}\,\varphi^4$ for the chaotic
inflation scenario,
\be
\label{7.5.23}
\frac{\delta\rho(k)}{\rho}=\frac{6}{5}\,\sqrt{\frac{2\,\pi\,\lambda}{3}}\,
\left(\frac{\varphi}{\m}\right)^3\biggr|_{k\sim{\rm H}(\varphi)}\ .
\ee
If we are to compare (\ref{7.5.23}) with the value of
$\displaystyle \frac{\delta\rho(k)}{\rho}$  on a galactic scale
($l_g\sim10^{22}$ cm) or on the scale of the horizon ($l_{\rm H}\sim10^{28}$
cm), we must follow the behavior of a wave with
momentum $k$ during and after inflation.  According to
(\ref{1.7.25}), a wave emitted at some value of $\varphi$ will,
over the course of inflation, increase in wavelength by a factor
$\displaystyle \exp\left(\pi\,\frac{\varphi^2}{\m^2}\right)$.
After reheating to a temperature ${\rm T}_{\rm R}$ and cooling to
a temperature ${\rm T}_\gamma$, where ${\rm T}_\gamma\sim 3$ K is
the present-day temperature of the microwave background
radiation, the universe will again typically expand by another
factor of $\displaystyle \frac{{\rm T}_{\rm R}}{{\rm T}_\gamma}$.
Presuming that reheating takes place immediately after the end of
inflation (with $\displaystyle \varphi\sim\frac{\m}{3}$),
${\rm T}_{\rm R}$ will be of order $\displaystyle \left[{\rm V}\,
\left(\frac{\m}{3}\right)\right]^{1/4}\sim\frac{\lambda^{1/4}}{10}\,\m$.
(The final results will be a very weak (logarithmic) function of
the duration of reheating and the magnitude of ${\rm T}_{\rm R}$.)
Thus, the present wavelength of a perturbation produced at
the moment when the scalar field had some value $\varphi$ is of
order
\ba
\label{7.5.24}
l(\varphi)&\sim&
{\rm H}^{-1}(\varphi)\,\frac{{\rm T}_{\rm R}}{{\rm T}_\gamma}\,
\exp\left(\frac{\pi\,\varphi^2}{\m^2}\right)\nonumber \\
&\sim&\m^{-1}\,\left(\frac{\m}{\varphi}\right)\,
\frac{\m}{\lambda^{1/4}\,{\rm T}_\gamma}\,
\exp\left(\frac{\pi\,\varphi^2}{\m^2}\right)\ .
\ea
Bearing in mind that 1 GeV corresponds approximately to $10^{13}$ K,
$\m\sim10^{-33}$ cm, and $\varphi\sim5\,\m$ at the time
of interest, we obtain (for $\lambda\sim10^{-14}$;  see below)
\be
\label{7.5.25}
l(\varphi) \sim \exp(\pi\varphi^2/{\m^2}),
\ee
whereupon
\be
\label{7.5.26}
\varphi^2\approx\frac{\m^2}{\pi}\,\ln l\ ,
\ee
where $l$ here is measured in centimeters.

Equation (\ref{7.5.26}) tells us that density perturbations on a
scale $l_{\rm H}\sim10^{28}$ cm are produced at
$\varphi=\varphi_{\rm H}$, where
\be
\label{7.5.27}
\varphi_{\rm H}\approx4.5\,\m\approx5.5\cdot 10^{19}\;\mbox{GeV}\ ,
\ee
and those on a galactic scale $l_g\sim10^{22}$ cm
come into existence at $\varphi=\varphi_g$, where
\be
\label{7.5.28}
\varphi_g\approx4\,\m\approx5\cdot 10^{19}\;\mbox{GeV}\ .
\ee
Equations (\ref{7.5.23}) and (\ref{7.5.26}) yield a general
equation for $\displaystyle \frac{\delta\rho}{\rho}$ in the
theory $\displaystyle \frac{\lambda}{4}\,\varphi^4$:
\be
\label{7.5.29}
\frac{\delta\rho}{\rho}\sim\frac{2\,\sqrt{6}}{5\,\pi}\,\sqrt{\lambda}\,
\ln^{3/2}l\;(\mbox{cm})
\ee
with the amplitude of inhomogeneities on the scale of the horizon
being
\be
\label{7.5.30}
\frac{\delta\rho}{\rho}\sim150\,\sqrt{\lambda}\ ,
\ee
while for those on a galactic scale,
\be
\label{7.5.31}
\frac{\delta\rho}{\rho}\sim110\,\sqrt{\lambda}\ .
\ee
The spectrum of $\displaystyle \frac{\delta\rho}{\rho}$ is
evidently almost flat, increasing slightly (logarithmically) at
long wavelengths.

We now discuss in somewhat more detail what the magnitude of the
constant $\lambda$ should be in order for the predicted
inhomogeneities to be consistent with the observational data and
the theory of galaxy formation.

Apparently, the most exacting constraints imposed by the
cosmological data are not on
$\displaystyle \frac{\delta\rho}{\rho}$ itself, but on the quantity A that
determines the anisotropy of the microwave background
$\displaystyle \frac{\Delta{\rm T}}{{\rm T}}$ produced by
\index{Metric perturbations}%
\index{Perturbations!metric}%
adiabatic perturbations of the metric [\cite{223}--\cite{227}],
\be \label{7.5.32} \left(\frac{\Delta{\rm T}}{{\rm T}}\right)_{l}=
\frac{{\rm A}}{\sqrt{l\,(l+1)}}\,\frac{{\cal
K}_l}{10\,\sqrt{\pi}}\ , \ee where $l$ is the order of the
harmonic in the multipole expansion of $\displaystyle \frac{\Delta
{\rm T}}{{\rm T}}$ ($l\ge2$ in (\ref{7.5.32})).  The relationship
between A in (\ref{7.5.32}) and metric perturbations is
[\cite{220}--\cite{222}]
\be \label{7.5.33}
\frac{\delta\rho(k)}{\rho}=-2\,\Phi(k)=-\frac{\sqrt{2}}{\pi}\,
\alpha\,{\rm A}(k)\ . \ee
The numerical factors $\alpha$ and
${\cal K}_l$ in (\ref{7.5.32}) and (\ref{7.5.33}) depend on the
specific assumptions made about the nature of the missing mass in
the universe.  The magnitude of ${\cal K}_l$ is usually of the
order of one.  As for $\alpha$, that quantity is $2/3$ for a hot
universe, and $3/5$ for a cold one.  In either case, \be
\label{7.5.34} {\rm A}(k)\approx16\,\pi\, \sqrt{\frac{\pi}{3}}\,
\frac{{\rm V}^{3/2}}{\m^3\,\displaystyle \frac{d{\rm
V}}{d\varphi}} \Biggr|_{k\sim{\rm H}(\varphi)}\ . \ee In
particular, for the $\displaystyle \frac{\lambda}{4}\,\varphi^4$
theory, \be \label{7.5.35} {\rm
A}=\frac{2\,\sqrt{\lambda}}{3}\,\ln^{3/2}l\;(\mbox{cm})
\sim1.2\,\sqrt{\lambda}\,\ln^{3/2}l\;(\mbox{cm})\sim 6\cdot
10^2\,\sqrt{\lambda} \ee on the scale of the horizon.  From the
observational constraints on $\displaystyle \frac{\Delta {\rm
T}}{{\rm T}}$, it follows that A lies somewhere in the range \be
\label{7.5.36} 5\cdot 10^{-5}\la{\rm A}\la5\cdot 10^{-4}\ , \ee
depending on the physical nature of the dark matter in the
universe [\cite{227}].  Condition (\ref{7.5.2}) is thus a
consequence of (\ref{7.5.33}) and (\ref{7.5.36}) as well.  To
determine the constraints on $\lambda$, it is most convenient to
use (\ref{7.5.35}) and (\ref{7.5.36}) directly: \be \label{7.5.37}
0.5\cdot 10^{-14}\la\lambda\la0.5\cdot 10^{-12}\ . \ee From here
on, we assume for definiteness that \be \label{7.5.38}
\lambda\sim10^{-14}\ , \ee which is closer to estimates in the
context of the theory of galaxy formation in a universe filled
with cold dark matter.  As the theory of large scale structure in
the universe is developed and the observational limits on
$\displaystyle \frac{\Delta{\rm T}}{{\rm T}}$ are refined
[\cite{228}], this estimate will improve.

Let us now consider another important example, the theory of a massive
scalar field with $\displaystyle \vf=\frac{m^2}{2}\,\varphi^2$.
For a cold Friedmann universe in this case,
\be
\label{7.5.39}
\frac{\delta\rho(k)}{\rho}=\frac{24}{5}\,\sqrt{\frac{\pi}{3}}\,
\frac{m}{\m}\,\left(\frac{\varphi}{\m}\right)^2\biggr|_{k\sim{\rm H}}\ .
\ee
Likewise, for either a hot or cold universe,
\be
\label{7.5.40}
{\rm A}(k)=4\,\sqrt{2}\,\pi\,
\sqrt{\frac{\pi}{3}}\,\frac{m}{\m}\,\left(\frac{\varphi}{\m}\right)^2
\biggr|_{k\sim{\rm H}}\ .
\ee
In this theory, both $\varphi_{\rm H}$ and $\varphi_g$ are a
factor of $\sqrt{2}$ less than in the
$\displaystyle \frac{\lambda}{4}\,\varphi^4$ theory.  The analog
of Eq.  (\ref{7.5.29}) for the present theory is
\be
\label{7.5.41}
\frac{\delta\rho}{\rho}\sim0.8\,\frac{m}{\m}\,\ln l\;[\mbox{cm}]\ ,
\ee
and on the scale of the horizon, Eq. (\ref{7.5.35}) for A becomes
\be
\label{7.5.42}
{\rm A}\sim200\,\frac{m}{\m}\ ,
\ee
whence
\be
\label{7.5.43}
3\cdot 10^{12}\;\mbox{GeV}\approx
2.5\cdot 10^{-7}\,\m\la m\la2.5\cdot 10^{-6}\m\approx3\cdot 10^{13}\;
\mbox{GeV}\ .
\ee

Next, consider the more general theory with potential
\be
\label{7.5.44}
\vf=\frac{\lambda\,\varphi^4}{n}\,\left(\frac{\varphi}{\m}\right)^{n-4}\ .
\ee
For such a theory,
\be
\label{7.5.45}
{\rm A}=16\,\pi\,\sqrt{\frac{\pi}{3}}\,
\left(\frac{\vf}{\m^4}\right)^{1/2}\,\frac{\varphi}{n\,\m}
\ee
and the field $\varphi_{\rm H}$ is
\be
\label{7.5.46}
\varphi_{\rm H}\sim2\,\sqrt{n}\,\m\ .
\ee
Perturbations on the scale of the horizon are thus characterized by
\be
\label{7.5.47}
{\rm A}\sim32\,\pi\,\sqrt{\frac{\pi}{3\,n}}\,
\left(\frac{\vf}{\m^4}\right)^{1/2}\ .
\ee
Specifically, when ${\rm A}\sim10^{-4}$, we find from
(\ref{7.5.46}) that in the last stages of inflation, when the
structure of the observable part of the universe had been formed,
the value of the effective potential was of order
\be
\label{7.5.48}
{\rm V}(\varphi_{\rm H})\sim10^{-12}\,n\,\m
\sim n\cdot 10^{82}\;{\rm g}\cdot \mbox{cm}^{-3}\ .
\ee
The rate of expansion of the universe was then
\be
\label{7.5.49}
{\rm H}(\varphi_{\rm H})\sim3\cdot 10^{-6}\,\sqrt{n}\,\m
\sim3.5\,\sqrt{n}\cdot 10^{13}\;\mbox{GeV}\ ;
\ee
that is, the universe increased in size by a factor of $e$ in a time
\be
\label{7.5.50}
t\sim{\rm H}^{-1}\sim n^{-1/2}\cdot 10^{-37}\;\mbox{sec}\ .
\ee
In such a theory, the constant $\lambda$ should be (for
${\rm A}\sim5\cdot 10^{-5}$)
\be
\label{7.5.51}
\lambda\sim2.5\cdot 10^{-13}\,n^2\,(4\,n)^{-n/2}\ .
\ee

These results give a general impression of the orders of
magnitude which might be encountered in realistic versions of the
inflationary universe scenario.  The estimate of
${\rm V}(\varphi_{\rm H})$ deserves special attention:  a similar
estimate can also be obtained from an analysis of the theory of
gravitational wave production at the time of inflation
[\cite{117}].  In the new inflationary universe scenario, an
analogous result implies that {\it at all stages of inflation},
${\rm V}(\varphi)$ should be at least ten to twelve orders of
magnitude less than $\m^4$ [\cite{107}, \cite{229}--\cite{231}].
Within the framework of the chaotic inflation scenario, a similar
statement is incorrect.  The value of A, which is $10^{-4}$ when
$\varphi\sim\varphi_{\rm H}$, tends to increase in accordance
with (\ref{7.5.45}) at large $\varphi$, and the observational
data impose no upper limits whatever on ${\rm V}(\varphi)$.  On
the other hand, we can derive a rather general constraint on the
magnitude of ${\rm V}(\varphi)$ in the last stages of inflation
from (\ref{7.5.34}).  In fact, at the end of inflation, the rate
at which the potential energy ${\rm V}(\varphi)$ decreases
becomes large --- the energy density ${\rm V}(\varphi)$ is
reduced by a quantity that is ${\rm O}({\rm V}(\varphi))$ within
a typical time $\Delta t={\rm H}^{-1}$.  In other words, the
criterion $\dot{\rm H}\ll{\rm H}^2$ is no longer satisfied.  One
can readily show that this means that at the end of inflation,
$\displaystyle {\rm V}'\sim\frac{{\rm V}}{\m}\,\sqrt{8\,\pi}$.
In that case, (\ref{7.5.34}) tells us that the quantity A, which
is related to fluctuations of the field $\varphi$ that are
generated during the very last stage of inflation, is given to
order of magnitude by
\be
\label{7.5.52}
{\rm A}\sim 10\,\sqrt{\frac{\vf}{\m^4}}\ .
\ee
With ${\rm A}\la 10^{-4}$, we then find that at the end of inflation,
\be
\label{7.5.53}
{\rm V}\la10^{-10}\,\m^4\ .
\ee
This constraint applies both to the new inflationary universe
scenario and the chaotic inflation scenario.

The formalism that we have employed in this chapter rests on an
assumption of the relative smallness of
$\displaystyle \frac{\delta\rho}{\rho}$.  During the inflationary
stage, as a rule, this condition is met.  For example, in the
$\displaystyle \frac{\lambda}{4}\,\varphi^4$ theory,
\be
\label{7.5.54}
\frac{\delta\rho}{\rho}\sim\frac{{\rm V}'\,\delta\varphi}{{\rm V}}\sim
\frac{4\,\delta\varphi}{\varphi}\sim\frac{2\,{\rm H}(\varphi)}{\pi\,\varphi}
\sim\frac{\sqrt{\lambda}\,\varphi}{\m}\ll1
\ee
for $\vf\la\m^4$, $\lambda\ll1$.

On relatively small scales ($l\sim{\rm H}^{-1}$), gradient terms
$\partial_i(\delta\varphi)\,\partial^i(\delta\varphi)\sim{\rm H}^4$
make a sizable contribution to $\displaystyle \frac{\delta\rho}{\rho}$.
We have not considered these terms, since in the last analysis we
were interested in perturbations with exponentially long
wavelengths.  This contribution is also much less than ${\rm V}(\varphi)$
when $\vf\ll\m^4$.

However, density perturbations produced at large $\varphi$ become large
after inflation.  In particular, according to (\ref{7.5.23}),
$\displaystyle \frac{\delta\rho}{\rho}\sim1$  in the
$\displaystyle \frac{\lambda}{4}\,\varphi^4$ theory for
perturbations produced when $\varphi=\varphi^*$, where
\be
\label{7.5.55}
\varphi^*\sim\lambda^{-1/6}\,\m\ .
\ee
By (\ref{7.5.25}), this means that after inflation ends, the
universe only looks like a homogeneous Friedmann space on a scale
\be
\label{7.5.56}
l_*\la\exp(\pi\,\lambda^{-1/3})\;\mbox{cm}\sim10^{6\cdot 10^4}\;\mbox{cm}\ .
\ee
for $\lambda\sim10^{-14}$.
This is many orders of magnitude larger than the observable part
of the universe, with $l_{\rm H}\sim10^{28}$ cm, so for a
present-day observer such inhomogeneities would lie beyond his
radius of visibility.  From the standpoint of the global
structure of the universe, however, nonuniformity on scales $l\gg l_*$
is exceedingly important, as we have discussed in Chapter
\ref{c1}.  We shall return to this question in Chapter \ref{c10}.

We make one more remark in closing.  We have been accustomed to
calling the quantity $l_{\rm H}\sim3\,t\sim10^{28}$ cm the
distance to the horizon, as in the usual Friedmann model (see
(\ref{1.4.11})).  But strictly speaking, the distance to the
actual particle horizon in the inflationary universe is
exponentially large.  Denoting this distance by ${\rm R}_{\rm H}$
(so as to distinguish it from $l_{\rm H}\sim3\,t$), we may use
(\ref{1.4.10}) and (\ref{1.7.28}) for the
$\displaystyle \frac{\lambda}{4}\,\varphi^4$ theory to obtain
\be
\label{7.5.57}
{\rm R}_{\rm H}\sim\m^{-1}\,\exp\frac{\pi}{\sqrt{\lambda}}
\sim10^{10^7}\;\mbox{cm}
\ee
(see also (\ref{1.7.39})).  Nevertheless, this quantity can only
tentatively be called the horizon.  The photons which presently
enable us to view the universe only permit us to see back to
$t\ga10^5$ years after the end of inflation in our part of the
universe, the reason being that the hot plasma that filled the
universe at $t\la10^5$ years was opaque to photons.  Thus, the
size of that part of the universe accessible to electromagnetic
observations is in fact $l_{\rm H}$ to high accuracy.  A similar
argument holds for neutrino astrophysics as well.  We can proceed
a bit further, studying metric perturbations [\cite{136}].
According to the standard hot universe theory, gravitational
waves provide the opportunity to obtain information about any
\index{Gravitational waves}%
\index{Waves, gravitational}%
process in the universe that takes place at less than the Planck
density, since the universe is transparent to gravitational waves
when ${\rm T}\la\m$.  This is not true in the inflationary
universe scenario, however.

Let us consider a gravitational wave with wavelength $l\la l_{\rm H}$
(since these are the only waves we can study experimentally).
At the stage of inflation, when the scalar field $\varphi$ was
equal to $\varphi_{\rm H}$, the wavelength of this gravitational
wave would have been of order $l\sim{\rm H}^{-1}\sim10^5\,\m^{-1}$
(\ref{7.5.48}), whereas at $\varphi\ga1.05\,\varphi_{\rm H}$, its
wavelength would have been less than $\m^{-1}$.  Quantum
fluctuations of the metric on this length scale are so large that
no measurements of gravitational waves inside the present horizon
(at $l\le l_{\rm H}$) could give us any information about the
structure of the universe with $\varphi\ga1.05\,\varphi_{\rm H}$.
In that sense, the range $\varphi\ga1.05\,\varphi_{\rm H}$,
corresponding to scales
$\displaystyle l\ga l_{\rm H}\cdot \frac{\m}{{\rm H}}\sim10^5\,l_{\rm H}$,
is ``opaque'' even to gravitational waves.  Thus, by analyzing
perturbations of the metric, we can in principle study phenomena
beyond the visibility horizon (at $l>l_{\rm H}$), but here we
cannot progress beyond a factor of
$\displaystyle \frac{\m}{{\rm H}(\varphi_{\rm H})}\sim10^5$.  The
energy density at the corresponding epoch (with
$\varphi\sim\varphi_{\rm H}$) was seven orders of magnitude less
than the Planck density (\ref{7.5.48}).  What this means is that
we cannot obtain information about the initial stages of
inflation (with $\vf\sim\m^4$) --- that is, the present state of
the observable part of the universe is essentially independent of
the choice of initial conditions in the inflationary
universe.
\index{Perturbations!adiabatic density|)}%
\index{Adiabatic density perturbations|)}%
\index{Quantum fluctuations!and generation
of adiabatic density perturbations|)}%

\section[Are flat-spectrum adiabatic perturbations sufficient?]%
{\label{s7.6}Are scale-free adiabatic perturbations sufficient\protect\\
\index{Universe!large scale structure of|(}%
to produce the observed large scale structure\protect\\
of the universe?}

The creation of the theory of adiabatic perturbations in
inflationary cosmology has been an unqualified success.
Beginning in 1982, when the theory was constructed in broad
outline, theoretical investigations of the formation of large
scale structure in the inflationary universe have, as a rule,
been based on two assumptions:

1) to high accuracy, the parameter $\displaystyle
\Omega=\frac{\rho}{\rho_0}$ is presently equal to unity (the
universe is almost flat);

2) initial density perturbations leading to galaxy formation were
adiabatic perturbations with a flat (or almost flat) spectrum,
$\displaystyle \frac{\delta\rho}{\rho}\sim10^{-5}$.

The possibility of describing all of the existing data on large
scale structure of the universe on the basis of these simple
assumptions is quite attractive, but we should recall at this
point the analogy between the universe and a giant accelerator.
Experience has taught us that the correct description of a large
body of diverse experimental data is seldom provided by the
simplest possible theory.  For example, the simplest description
of the weak and electromagnetic interactions would be given by
\index{Georgi--Glashow model}%
the Georgi--Glashow model [\cite{232}], which is based on the
symmetry group O(3).  But the experimental discovery of neutral
currents forced us to turn to the far more complicated
\index{Glashow--Weinberg--Salam theory}%
Glashow--Weinberg--Salam model [\cite{1}], based on the symmetry
group $\mbox{SU}(2) \times {\rm U}(1)$.  The latter contains
about 20 different parameters whose values are not grounded in
any aesthetic considerations at all.  For instance, almost all
coupling constants in this theory are ${\rm O}(10^{-1})$, while
the coupling constant for interaction between the electron and
the scalar (Higgs) field is $2\cdot 10^{-6}$.  The reason for the
appearance of such a small coupling constant (just like the
reason for the appearance of the constant $\lambda\sim10^{-14}$
in the simplest versions of the inflationary universe scenario)
is as yet unclear.

It seems unlikely that cosmology will turn out to be a much
simpler science than elementary particle theory.  After all, the
number of different types of large-scale objects in the universe
(quasars, galaxies, clusters of galaxies, filaments and voids,
etc.) is very large.  The sizes of these objects form a hierarchy
of scales that is absent from the flat spectrum of the initial
perturbations.  In principle, some of these scales may be related
to the properties of the dark matter comprising most of the mass
of the universe; see, for example, [\cite{224}, \cite{235},
\cite{236}].  Nevertheless, it is not at all obvious how to
consistently describe the formation of a large number of diverse
large-scale objects, starting with the simple assumptions (1) and
(2).  The requisite theory encounters a number of difficulties
[\cite{235}] which, while not insurmountable, have nonetheless
stimulated a search for alternative versions of the theory of
formation of large scale structure;  see, for example,
[\cite{236}].

Another potential problem that a theory based on assumptions (1)
and (2) may encounter is related to measurements of the
anisotropy $\displaystyle \frac{\Delta{\rm T}(\theta)}{{\rm T}}$
of the
\index{Anisotropy of microwave background radiation}%
\index{Microwave background radiation!anisotropy of}%
\index{Background radiation, microwave!anisotropy of}%
microwave background radiation, where $\theta$ is the
angular scale of observation.  So far, only a dipole anisotropy
$\displaystyle \frac{\Delta{\rm T}}{{\rm T}}$ associated with the
earth's motion through the microwave background has been
\index{Quadrupole anisotropy}%
detected, and neither a quadrupole anisotropy nor a small-angle
anisotropy in $\displaystyle \frac{\Delta{\rm T}}{{\rm T}}$ has
been found at a level
$\displaystyle \frac{\Delta{\rm T}}{{\rm T}}\ga2\cdot 10^{-5}$
[\cite{228}].  Meanwhile, flat-spectrum adiabatic density
perturbations should lead to
$\displaystyle \frac{\Delta{\rm T}}{{\rm T}}\ga{\rm C}\cdot 10^{-5}$
[\cite{223}--\cite{227}], where the function ${\rm
C}(\theta)={\rm O}(1)$ depends on the angle $\theta$ and on the
properties of dark matter.  The function ${\rm C}(\theta)$ is
especially large at large angles $\theta$.  The comparison
between experimental constraints on
$\displaystyle \frac{\Delta{\rm T}}{{\rm T}}$
and theoretical predictions of
\index{Quadrupole anisotropy}%
quadrupole anisotropy is therefore a particularly important
question, with a bearing on perturbations $\displaystyle
\frac{\delta\rho}{\rho}$ on a scale $l\sim l_{\rm H}\sim10^{28}$
cm.  The complexity of the situation is exacerbated by the fact
that the inflationary universe scenario gives an adiabatic
perturbation spectrum that is not perfectly flat.  In most models,
$\displaystyle \frac{\delta\rho}{\rho}$ grows with increasing $l$.
For example, in a theory with $\displaystyle
\vf\sim\frac{\lambda}{4}\,\varphi^4$, as we progress from the
galactic scale $l_g$ to the size of the horizon $l_{\rm H}$, the
quantity $\displaystyle \frac{\delta\rho}{\rho}$ increases by a
factor of about 1.4 (see (\ref{7.5.29}) and (\ref{7.5.30})),
resulting in a concomitant enhancement of the
\index{Quadrupole anisotropy}%
quadrupole anisotropy $\displaystyle \frac{\Delta{\rm T}}{{\rm T}}$.
An assessment of the predictions of anisotropy
$\displaystyle \frac{\Delta{\rm T}}{{\rm T}}$ in the simplest
versions of the inflationary universe scenario has already made
it possible to discard the simplest models of baryonic dark
matter, and they cast some doubt upon the validity of models in
which the missing mass is concentrated in massive neutrinos
[\cite{225}, \cite{226}].  However, in the cold dark matter
models, in which the dark matter consists of\index{Axion fields}
axion fields
[\cite{233}, \cite{234}], Polonyi fields [\cite{46}, \cite{15}],
or any weakly interacting nonrelativistic particles, the
theoretical estimates of $\displaystyle \frac{\Delta{\rm T}}{{\rm T}}$
are perfectly consistent with current observational limits
[\cite{225}, \cite{226}].

Thus, it remains possible that a theory of the formation of the
large scale structure of the universe can be completely assembled
within the framework of the very simple assumptions (1) and (2)
(that is, a flat universe with a flat spectrum of adiabatic
perturbations).  However, as the inhabitants of new housing
projects know only too well, the simplest project is almost never
the most successful.  It would therefore be well to understand
whether we can somehow modify assumptions (1) and (2) while
remaining within the framework of the inflationary universe
scenario.  Specifically, we would like to single out five basic
questions.

1) Is it possible to get away from the condition $\Omega=1$?

2) Is it possible to obtain non-adiabatic perturbations after inflation?

3) Is it possible to obtain perturbations with a spectrum that
decreases at $l\sim l_{\rm H}$, so as to reduce the
\index{Quadrupole anisotropy}%
quadrupole anisotropy of
$\displaystyle \frac{\Delta{\rm T}}{{\rm T}}$?

4) Is it possible to obtain perturbations with a spectrum having
one or perhaps several maxima, which would help to explain the
origin of the hierarchy of scales (galaxies, clusters, \ldots)?

5) Is it possible to produce the large scale structure of the
universe through nonperturbative effects associated with
inflation?

For the time being, the answer to the first question is negative:
we know of no way to obtain $\Omega\neq1$ in a natural manner
within the context of inflationary cosmology.  Even if we could,
it would most likely be only for some special choice of the
potential ${\rm V}(\varphi)$ and after painstaking adjustment of
the parameters, for which there is as yet no particular
justification.

Building a model in which the spectrum of adiabatic perturbations
falls off monotonically at long wavelengths is possible in
principle, but rather difficult.  The only  reasonable theory of
\index{Shafi--Wetterich model}%
this type that we are aware of is the Shafi--Wetterich model,
based on a study of inflation in the Kaluza--Klein theory
[\cite{237}].  A peculiar feature of this model is that inflation
and the evolution of a scalar field $\varphi$ (the role of which
is played by the logarithm of the compactification radius) are
described by two different effective potentials, ${\rm
V}(\varphi)$ and ${\rm W}(\varphi)$.  Unfortunately, it is
difficult to realize the initial conditions required for
inflation in this model --- see Chapter \ref{c9}.  Another
suggestion that has been made is to study the spectra produced by
double inflation, driven first by one scalar field $\varphi$,
then by another $\Phi$ [\cite{238}].  For the most natural
initial conditions, however, the last stages of inflation are
governed by the field with the flattest potential (the smallest
parameters $m^2$ and $\lambda$).  As a rule, therefore, rather
than leading to cutoff, two-stage inflation will lead to a more
abrupt rise in $\displaystyle \frac{\delta\rho}{\rho}$  at the
long wavelengths generated during the stage when the ``heavier''
field $\varphi$ is dominant.

Nevertheless, all of the questions posed above (except the first)
can be answered in the affirmative.  There is a rather broad
class of models which, in addition to adiabatic perturbations,
can also produce isothermal perturbations [\cite{239},
\cite{240}], with spectra that fall off at long wavelengths
[\cite{239}, \cite{241}].  Particularly interesting effects are
associated with phase transitions, which can occur at the later
stages of inflation (when the universe still has another factor
of $e^{50}$--$e^{60}$ left to expand).  In particular, such phase
transitions can result in density perturbations having a spectrum
with one or several maxima [\cite{242}], and to the appearance of
exponentially large strings, domain walls, bubbles, and other
objects that can play a significant role in the formation of the
large scale structure of the universe [\cite{125}, \cite{243}].
We shall discuss some of the possibilities mentioned above in the
next two sections.
\index{Universe!large scale structure of|)}%

\section[Nonflat isothermal and adiabatic perturbations]%
{\label{s7.7}Isothermal perturbations
\index{Isothermal perturbations|(}%
\index{Perturbations!isothermal|(}%
and adiabatic perturbations\protect\\
with a nonflat spectrum}

The theory of the formation of density perturbations discussed in
Section \ref{s7.5} was based upon a study of the simplest models,
describing only a single scalar field $\varphi$ responsible for
the dynamics of inflation.  In realistic elementary particle
\index{Inflation}%
theories, there exist many scalar fields $\Phi_i$ of various
kinds.  To understand how inflation comes into play and what
sorts of density inhomogeneities arise in such theories, let us
consider first the simplest model, describing two noninteracting
fields $\varphi$ and $\Phi$ [\cite{239}]:
\be
\label{7.7.1}
{\rm L}=\frac{1}{2}\,(\partial_\mu\varphi)^2
+\frac{1}{2}\,(\partial_\mu\Phi)^2-\frac{m_\varphi^2}{2}\,\varphi^2
-\frac{m_\Phi^2}{2}\,\Phi^2-\frac{\lambda_\varphi}{4}\,\varphi^4
-\frac{\lambda_\Phi}{4}\,\Phi^4\ .
\ee
We assume for simplicity that
$\lambda_\varphi\ll\lambda_\Phi\ll1$, and
$m_\varphi^2,m_\Phi^2\ll\lambda_\varphi\,\m^2$.  Then for large
$\varphi$ and $\Phi$, terms quadratic in the fields can be
neglected.  The only constraint on the initial amplitudes of the
fields $\varphi$ and $\Phi$ is
\be
\label{7.7.2}
\vf+{\rm V}(\Phi)\approx
\frac{\lambda_\varphi}{4}\,\varphi^4+\frac{\lambda_\Phi}{4}\,\Phi^4
\la\m^4\ .
\ee
This means that the most natural initial values of the fields
$\varphi$ and $\Phi$ are $\varphi\sim\lambda_\varphi^{-1/4}\,\m$,
$\Phi\sim\lambda_\Phi^{-1/4}\,\m$; that is, initially,
$\vf\sim{\rm V}(\Phi)\sim\m^4$, $\varphi\gg\Phi\gg\m$.  Since the
curvature of the potential ${\rm V}(\Phi)$ is much greater than
that of ${\rm V}(\varphi)$, it is clear that under the most
natural initial conditions, the field $\Phi$ and its energy ${\rm V}(\Phi)$
fall off much more rapidly than the field $\varphi$ and
its energy ${\rm V}(\varphi)$.  The total energy density
therefore quickly becomes equal to ${\rm V}(\varphi)$; i.e., the
Hubble parameter ${\rm H}(\varphi,\Phi)$ becomes
\be
\label{7.7.3}
{\rm H}(\varphi,\Phi)\approx{\rm H}(\varphi)=
\sqrt{\frac{2\,\pi\,\lambda_\varphi}{3}}\,\frac{\varphi^2}{\m}\ .
\ee
Thus, within a short time, inflation will be governed solely by
the field $\varphi$ having the potential ${\rm V}(\varphi)$ with
the least curvature (the smallest coupling constant
$\lambda_\varphi$).  For this reason, the field $\varphi$ can be
called the inflaton.  It evolves just as if the field $\Phi$ did
not exist (see (\ref{1.7.21})):
\be
\label{7.7.4}
\varphi(t)=
\varphi_0\,\exp\left(-\sqrt{\frac{\lambda_\varphi}{6\,\pi}}\,\m\,t\right)\ .
\ee
In that case, the equation
\be
\label{7.7.5}
\ddot\Phi+3\,{\rm H}\,\dot\varphi=-\lambda_\Phi\,\Phi^3
\ee
implies that during the inflationary stage
\be
\label{7.7.6}
\Phi(t)=\sqrt{\frac{\lambda_\varphi}{\lambda_\Phi}}\,\varphi(t)\ ,
\ee
and therefore
\be
\label{7.7.7}
m^2(\varphi)=m^2(\Phi)=
\frac{3\,\m\sqrt{3\,\lambda_\varphi}}{\sqrt{2\,\pi}}\,{\rm H}(\varphi)\ ,
\ee
where (for $m_\Phi^2\ll\lambda_\varphi\,\m^2$)
\ba
\label{7.7.8}
m^2(\varphi)&=&\frac{d^2{\rm V}}{\delta\varphi^2}=
3\,\lambda_\varphi\,\varphi^2\ ,\nonumber \\
m^2(\Phi)&=&\frac{d^2{\rm V}}{d\Phi^2}=3\,\lambda_\Phi\,\Phi^2\ .
\ea
In the last stage of inflation, $\varphi\sim\m$ and
$\displaystyle \Phi\sim\sqrt{\frac{\lambda_\varphi}{\lambda_\Phi}}\,\m$.
The perturbations of the $\varphi$ and
$\Phi$ fields have equal amplitudes (see (\ref{7.5.6})):
\be
\label{7.7.9}
\delta\varphi=\delta\Phi=\frac{{\rm H}}{2\,\pi}=
\sqrt{\frac{\lambda_\varphi}{6\,\pi}}\,\frac{\varphi^2}{\m}
\sim\sqrt{\lambda_\varphi}\,\m\ .
\ee
At that time, however, the contribution $\delta\rho_\Phi$ that
the field $\Phi$ makes to density inhomogeneities $\delta\rho$ is
much less than $\delta\rho_\Phi$, the corresponding contribution
from $\varphi$:
\be
\label{7.7.10}
\delta\rho_\Phi=\frac{d{\rm V}}{d\Phi}\,\delta\Phi=
\sqrt{\frac{\lambda_\varphi}{\lambda_\Phi}}\,
\lambda_\varphi\,\varphi^3\,\delta\varphi=
\sqrt{\frac{\lambda_\varphi}{\lambda_\Phi}}\,\delta\rho_\Phi
\ll\delta\rho_\varphi\ .
\ee

It is therefore precisely the fluctuations $\delta\rho_\varphi$
of the inflaton field that govern the amplitude of adiabatic
density perturbations.  At the stage of inflation with
$\varphi\sim\m$,
\be
\label{7.7.11}
\frac{\delta\rho_\varphi}{\rho_\varphi}\approx\frac{\delta\rho}{\rho}
=\frac{1}{{\rm V}}\,\frac{d{\rm V}}{\delta\varphi}\,\delta\varphi
\sim4\,\frac{\delta\varphi}{\varphi}\sim\sqrt{\lambda_\varphi}\ ,
\ee
where $\rho=\rho_\varphi+\rho_\Phi\approx\rho_\varphi$.  Meanwhile,
\be
\label{7.7.12}
\frac{\delta\rho_\Phi}{\rho_\Phi}\sim\frac{4\,\delta\Phi}{\Phi}\sim
\sqrt{\lambda_\Phi}\ .
\ee
After inflation, the inhomogeneities (\ref{7.7.11}) produce the
adiabatic perturbations (\ref{7.5.29}), which will in fact remain
the dominant density perturbations {\it if} upon further
expansion of the universe, the quantity $\rho_\Phi$ falls off in
the same way as $\rho_\varphi$.  However, in some cases this
condition is not satisfied, since the evolution of $\rho_\Phi$
and $\rho_\varphi$ depends on the interaction of these fields
with other fields, and on the shape of ${\rm V}(\varphi)$ and
${\rm V}(\Phi)$.  Let us assume, for example,  that the field
$\Phi$ interacts very weakly with other fields.  Such weakly
interacting scalar fields do exist in many realistic theories
(axions, Polonyi fields, etc.).
\index{Axion fields}%
If the field $\varphi$ interacts
strongly with other fields, its energy is quickly transformed
into heat, $\rho_\varphi\rightarrow{\rm T}_{\rm R}^4$, and falls
off with the expansion of the universe as ${\rm T}^4\sim a^{-4}$.
At the same time, the field $\Phi$, without decaying, oscillates
in the neighborhood of the point $\Phi=0$ with frequency
$\kappa_0=m_\Phi$.  As it does so, its energy falls off in the
same way as the energy of nonrelativistic particles,
$\rho_\Phi\sim a^{-3}$ (see Section \ref{s7.9}) i.e., much more
slowly that the energy of the decay products from the field
$\varphi$.  In the later stages of evolution of the universe, the
energy of the field $\Phi$ can therefore become greater than the
energy of decay products of the inflaton field,
$\rho=\rho_\varphi+\rho_\Phi\approx\rho_\Phi$.

This is just the effect that underlies the possibility, discussed
in [\cite{49}], that the
\index{Axion fields}%
axion field $\theta$ may be responsible
for the missing mass of the universe at the present epoch.

Prior to the stage at which the field $\Phi$ is dominant, both
the mean density $\rho_\Phi$ and the quantity
$\rho_\Phi+\delta\rho_\Phi$ fall off in the same way,
$\rho_\Phi\sim\rho_\Phi+\delta\rho_\Phi\sim a^{-3}$;  the
quantity $\displaystyle
\frac{\delta\rho_\Phi}{\rho_\Phi}\sim\sqrt{\lambda_\Phi}$
therefore remains constant.  From the start, inhomogeneities
$\delta\rho_\Phi$ are in no way associated with the temperature
inhomogeneities $\delta{\rm T}$ of decay products of the field
$\varphi$, and in that sense they are isothermal.  They might
\index{Inhomogeneities!isoinflaton}%
also be called {\it isoinflaton} inhomogeneities, as they are
independent of fluctuations of the inflaton field $\varphi$.
Consequently, due to the increasing fraction of the overall
matter density $\rho$ accounted for by $\rho_\Phi$, isothermal
perturbations $\delta\rho_\Phi$, with
$\sqrt{\lambda_\Phi}\ga10^2\,\sqrt{\lambda_\varphi}$, begin to
dominate, generating adiabatic perturbations
\be
\label{7.7.13}
\frac{\delta\rho}{\rho}\approx\frac{\delta\rho_\Phi}{\rho_\Phi}\sim
\sqrt{\lambda_\Phi}
\ee
in the process.  Note that in (\ref{7.7.13}), there is no
enhancement factor ${\rm O}(10^2)$ associated with the transition
from the inflationary stage to the expansion with $a\sim t^{1/2}$
or $a\sim t^{2/3}$.

Thus, even in the simplest theory of two noninteracting fields,
the process by which density perturbations are generated can
unfold in a fairly complicated manner:  in addition to adiabatic
density perturbations, isothermal perturbations can also come
about, and for $\lambda_\varphi\ll10^{-14}$,
$\lambda_\Phi\ga10^{-10}$, the latter can dominate.

Even more interesting possibilities are revealed when we allow for
interactions between the fields $\varphi$ and $\Phi$. Consider,
for example, a theory with the effective potential \be
\label{7.7.14} {\rm
V}(\varphi,\Phi)=\frac{m_\varphi^2}{3}\,\varphi^2
+\frac{\lambda_\varphi}{4}\,\varphi^4
-\frac{m^2_\Phi}{2}\,\Phi^2+\frac{\lambda_\Phi}{4}\,\Phi^4
+\frac{\nu}{2}\,\varphi^2\,\Phi^2+{\rm V}(0)\ . \ee Let us suppose
that $0<\lambda_\varphi\ll\nu\ll\lambda_\Phi$ and
$\lambda_\varphi\,\lambda_\Phi>\nu^2$, and assume also that
$m_\varphi^2\ll\lambda_\varphi\,\m^2$  and $m_\Phi^2\ll{\rm
C}\,\nu\,\m^2$, where ${\rm C} = {\rm O}(1)$. Just as in the
theory (\ref{7.7.1}), the most natural initial values for
$\varphi$ and $\Phi$ satisfy the conditions $\varphi\gg\m$,
$\varphi\gg\Phi$.  With $\varphi\gg\m$, the minimum of ${\rm
V}(\varphi, \Phi)$ is located at $\Phi=0$, and the effective mass
of the field $\Phi$ at $\Phi=0$ is \be \label{7.7.15}
m^2_\Phi(\varphi,0)= \frac{\partial^2{\rm
V}}{\partial\Phi^2}\biggr|_{\Phi=0}
=\nu\,\varphi^2-m^2_\Phi\approx\nu\,\varphi^2\ . \ee This is much
greater than the mass of the field $\varphi$, \be \label{7.7.16}
m^2_\varphi(\varphi,0)=m^2_\varphi+3\,\lambda_\varphi\,\varphi^2
\approx3\,\lambda_\varphi\,\varphi^2\ll\nu\,\varphi^2\ . \ee The
field $\Phi$ is therefore rapidly dumped into the minimum of ${\rm
V}(\varphi, \Phi)$, and as in the theory (\ref{7.7.1}), inflation
becomes driven by the field $\varphi$.

In the last stage of inflation, when the field $\varphi$ becomes
less than
\be
\label{7.7.17}
\varphi_c=\sqrt{{\rm C}}\,\m\ ,
\ee
the minimum of ${\rm V}(\varphi, \Phi)$ is located at
\be
\label{7.7.18}
\Phi^2=\frac{m_\Phi^2-\nu\,\varphi^2}{\lambda_\Phi}
=\nu\,\frac{{\rm C}\m^2-\varphi^2}{\lambda_\Phi}\ ,
\ee
and the effective mass of the field $\Phi$ is then
\be
\label{7.7.19}
m^2_\Phi(\varphi,\Phi)=2\,\nu\,({\rm C}\m^2-\varphi^2)\ .
\ee

Notice that both when $\varphi\gg\varphi_c$ and
$\varphi\ll\varphi_c$, the effective mass of the field $\Phi$ is
\index{Hubble ``constant''}%
much greater than the Hubble constant
$\displaystyle{\rm H}\sim\frac{\sqrt{\lambda_\varphi}\,\varphi^2}{\m}$.
Long-wave fluctuations $\delta\Phi$ of the field $\Phi$ are
therefore generated only in some neighborhood of the phase
transition point at $\varphi\sim\varphi_c$.  In studying the
density perturbations produced in this model, it turns out to be
important that the amplitude of fluctuations $\delta\Phi$
displays a variety of temporal behaviors, depending on what
precisely the value of the field $\varphi$ was when these
fluctuations arose.  Numerical calculations [\cite{242}] taking
this fact into account have shown that for certain relationships
among the parameters of the theory (\ref{7.7.14}), the spectrum
of adiabatic perturbations produced during inflation may have a
reasonably narrow maximum that is slightly shifted with respect
to $\displaystyle l\sim\exp\left(\frac{\pi\,\varphi_c^2}{\m^2}.\right)$.

It should be pointed out here that many different types of scalar
fields figure into realistic elementary particle theories.  It
would therefore be hard to doubt that phase transitions should
actually take place at the time of inflation, and in fact most
likely not one, but many.  The only question is whether these
phase transitions take place fairly late, when the field
$\varphi$ is changing between $\varphi_{\rm H}$ (\ref{7.5.27})
and $\varphi_g$ (\ref{7.5.28}).  This is a condition that is
satisfied, given an appropriate choice of parameters in the
theory.  The actual parameter values chosen (like the parameters
used in building the theories of the weak and electromagnetic
interactions) should be based on experimental data, rather than
on some {\it a priori} judgment about their naturalness (since
according to that criterion one could reject the
\index{Glashow--Weinberg--Salam theory}%
Glashow--Weinberg--Salam model --- see the preceding section).
In the case at hand, these data consist of observations of the
large scale structure of the universe and the anisotropy of the
cosmic microwave radiation background.  In our opinion, the
possibility of studying the phase structure of unified theories
of elementary particles and determining the parameters of these
theories through astrophysical observations is extremely
interesting.

To conclude this section, let us briefly deal with the production
of isothermal perturbations in axion models.  To this end, we
examine the theory of a complex scalar field $\Phi$ which
interacts with an inflaton field $\varphi$:
\ba
\label{7.7.20}
{\rm V}(\varphi,\Phi)&=&\frac{m_\varphi^2}{2}\,\varphi^2
+\frac{\lambda_\varphi}{4}\,\varphi^4-m_\Phi^2\,\Phi^*\,\Phi\nonumber \\
&+&\lambda_\Phi\,(\Phi^*\,\Phi)^2+\frac{\nu}{2}\,\varphi^2\,\Phi^*\,\Phi
+{\rm V}(0)\ .
\ea
After the spontaneous symmetry breaking which occurs at
$\displaystyle \varphi<\varphi_c=\frac{m_\Phi}{\sqrt{\nu}}$, the
field $\Phi$ can be represented in the form
\be
\label{7.7.21}
\Phi(x)=\Phi_0\,\exp\left(\frac{i\,\theta(x)}{\sqrt{2}\,\Phi_0}\right)\ ,
\ee
where $\displaystyle \Phi_0=\frac{m_\Phi}{\sqrt{\lambda_\Phi}}$
for $\varphi\ll\varphi_c$.  The field $\theta(x)$ is a massless
\index{Goldstone field}%
Goldstone scalar field [\cite{244}] with vanishing effective
potential, ${\rm V}(\theta)=0$.

In contrast to the familiar Goldstone field described above, the
\index{Axion fields}axion field is not massless.
Because of nonperturbative
corrections to ${\rm V}(\varphi, \Phi)$ associated with strong
interactions, the effective potential ${\rm V}(\theta)$ becomes
[\cite{233}, \cite{234}]
\be
\label{7.7.22}
{\rm V}(\theta)={\rm C}\,m^4_\pi\,
\left(1-\cos\frac{{\rm N}\,\theta}{\sqrt{2}\,\Phi_0}\right)\ .
\ee
Here ${\rm C} \sim {\rm O}(1)$, and N is an integer that depends
on the detailed structure of the theory;  for simplicity, we
henceforth consider only the case ${\rm N}=1$.  We thus see from
(\ref{7.7.22}) that axions can now have a small mass
$\displaystyle
m_\theta\sim\frac{m^2_\pi}{\Phi_0}\sim10^{-2}\;\mbox{GeV}^2/\Phi_0$.

From the standpoint of elementary particle theory, the main reason for
considering the axion field $\theta$ is that the field value that minimizes
${\rm V}(\theta)$
automatically leads to cancellation of the effects of the strong CP violation
that are associated with the nontrivial vacuum structure in the theory of
strong interactions [\cite{233}, \cite{234}].  Cosmologists became interested in this field
for another reason.  It turns out that at temperatures ${\rm T}\gg10^2$ MeV, the
nonperturbative effects resulting in nonvanishing ${\rm V}(\theta)$
are strongly inhibited.
Therefore, the field $\theta$
is equally likely to take any initial value in the
range $-\sqrt{2}\,\pi\,\Phi_0\le\theta\le\sqrt{2}\,\pi\,\Phi_0$.
As the temperature of the universe drops to
${\rm T}\la10^2$ MeV, the effective
potential ${\rm V}(\theta)$
takes the form (\ref{7.7.22}), so that in the mean, the field $\theta$
acquires an energy density of order
$m_\pi^4\sim10^{-4}$ GeV$^4$.
This field interacts with other fields
extremely weakly, and its mass is extraordinarily small
($m_\theta\sim10^{-5}$ eV for the realistic
value $\Phi_0\sim10^{12}$ GeV;  see below).
It therefore mainly loses its energy not through
radiation, but through damping of its oscillations
near $\theta=0$ as the universe
expands (by virtue of the term $3\,{\rm H}\,\dot\theta$
in the equation for the field $\theta$).  As we have already
said, the energy density of any noninteracting massive field,
which oscillates near the minimum of its effective potential,
falls off in the same way as the energy density of a gas of
nonrelativistic particles, $\rho_\theta\sim a^{-3}$, i.e., more
slowly than that of a relativistic gas.  As a result, the
relative contribution of the axion field to the total energy
density increases.

The present-day value of the ratio $\displaystyle
\frac{\rho_\theta}{\rho}$ depends on the value of $\Phi_0$.  For
$\Phi_0\sim10^{12}$ GeV, most of the total energy density of the
universe should presently be concentrated in an almost
homogeneous, oscillating axion field, which would then account
for the missing mass.  As has been claimed in [\cite{49}], a
value of $\Phi_0\gg10^{12}$ GeV would be difficult to reconcile
with the available cosmological data (see Section \ref{s10.5},
however).  When $\Phi_0\ll10^{12}$ GeV, the relative contribution
of axions to the energy density of the universe falls off as
$(\Phi_0/10^{12}\;\mbox{GeV})^2$.

If the phase transition with symmetry breaking and the creation
\index{Pseudo-Goldstone field}%
of the pseudo-Goldstone field $\theta$ takes place during the
inflationary stage, then inflation leads to fluctuations of the
field $\theta$; as before,
$\displaystyle \delta\theta=\frac{{\rm H}}{2\,\pi}$ per unit
interval $\Delta\ln k$.  When ${\rm T}<10^2$ MeV, density
inhomogeneities $\displaystyle
\frac{\delta\rho_\theta}{\rho_\theta}\sim
\frac{\delta{\rm V}(\theta)}{{\rm V}(\theta)}$  appear, which are
associated with these fluctuations.  But because of the
periodicity of the potential ${\rm V}(\theta)$, these
inhomogeneities are related in a much more complicated way to the
magnitude of the fluctuations in the field $\theta$.  Let us
assume, for example, that after a phase transition, inflation
continues long enough that the rms value $\displaystyle
\sqrt{\langle \theta^2\rangle }=\frac{{\rm H}}{2\,\pi}\,\sqrt{{\rm H}\,t}$
becomes much greater than $\Phi_0$.  The classical field $\theta$
will then take on any value in the range
$-\sqrt{2}\,\pi\,\Phi_0\le\theta\le\sqrt{2}\,\pi\,\Phi_0$ with
uniform probability.  When temperature drops down to ${\rm T}<10^2$ MeV,
the surfaces at which the field $\theta$ is equal
to $\sqrt{2}\,\Phi_0\,(2\,n+1)\,\pi$ form domain walls
corresponding to the maximum of energy density (\ref{7.7.22})
[\cite{361}].  Therefore, in addition to small isothermal density
perturbations proportional to perturbations $\delta\theta$, in
the axion cosmology one may get also exponentially large (or even
infinitely large) domain walls.  In order to avoid disastrous
cosmological consequences of existence of such walls one should
consider such theories where the Hubble parameter H at the end of
inflation becomes much smaller than $\Phi_0$.  In such a case the
probability of formation of domain  walls becomes exponentially
suppressed and no such walls appear in the observable part of the
universe.  It would be especially interesting to investigate an
intermediate case, where H is smaller than $\Phi_0$, but not too
much smaller.  In such a case the axionic domain walls will form
small bubbles displaced exponentially far away from each other.
Is it possible that explosions of such bubbles may be responsible
for formation of the large-scale structure of the universe?  We
will return to a discussion of similar possibilities in the next
section.

Another interesting possibility which may appear in the axion
cosmology is related to the time-dependence of the radius of the
axion potential $\Phi_0$.  Indeed, isothermal perturbations
generated in the axion cosmology are proportional to perturbations
of the angle, $\displaystyle\frac{\delta\rho_\theta}{\rho_\theta}
\sim\frac{\delta\theta}{\sqrt{2}\,\Phi_0}$.  During inflation, the
value of $\Phi_0$ in the model (\ref{7.7.20}) rapidly changes.
This leads to the modification of the spectrum of density
perturbations, which acquires a strong maximum on the scale
corresponding to the moment of the phase transition (at which
$\displaystyle \varphi=\varphi_c=\frac{m_\Phi}{\sqrt{{\rm V}}}$,
$\Phi_0=0$). Again, one should worry about overproducing density
perturbations and axionic domain walls on this scale [\cite{362}].
Our main purpose in discussing all these possibilities was just to
demonstrate that by a simple extension of minimal inflationary
models containing only one scalar field one can easily obtain
perturbations with a non-flat spectrum, either adiabatic or
isothermal.  One should not use such possibilities without a
demonstrated need, but it is better to know that if the experts in
the theory of galaxy formation will tell us that they are unhappy
with adiabatic perturbation with  a flat  spectrum, inflation will
have something else to offer.
\index{Density!perturbations|)}%
\index{Isothermal perturbations|)}%
\index{Perturbations!isothermal|)}%

\section[Nonperturbative effects:  strings, etc.]%
{\label{s7.8}Nonperturbative effects:  strings, hedgehogs, walls,\protect\\
bubbles, \ldots}

In Section \ref{s7.5} we studied mechanisms for generating
small density perturbations in the inflationary universe.
However, as we have seen in the previous section,  phase
transitions at the time of inflation can lead not just to small
density perturbations;  they can also produce nontrivial
structures of exponentially large size.  Herewith, we present
some examples.

1.  {\it Strings.}  The theory of the formation of density
\index{Strings}%
inhomogeneities during the evolution of cosmic strings [\cite{81}]
was long considered to be the only real alternative to
inflationary theory for the formation of flat-spectrum adiabatic
perturbations.  It is now quite clear that there exists a wide
range of other possibilities --- see Section \ref{s7.7} and the
discussion below.  Furthermore, without inflation, string theory
provides no help in solving the problems of standard Friedmann
cosmology, and the formation of superheavy strings through
high-temperature phase transitions following inflation is
complicated by the fact that in most models, the universe after
inflation is not hot enough.  But it is perfectly possible to
produce strings during phase transitions in the inflationary stage
[\cite{125}, \cite{246}, \cite{247}].  About the simplest model in
which one could treat such a process would be a theory describing
the interaction of an inflaton $\varphi$ with a complex scalar
field $\Phi$ having effective potential (\ref{7.7.20}).  In the
early stages of inflation, when $\displaystyle
\varphi^2>\frac{m_\Phi^2}{\nu}$, symmetry is restored in the
theory (\ref{7.7.20}).  As the field $\varphi$ falls off toward
$\displaystyle \varphi=\varphi_c=\frac{m_\Phi}{\sqrt{\nu}}$, the
symmetry-breaking phase transition leading to the production of
strings takes place, as happens in the case of a phase transition
with a decrease in temperature;  see Section \ref{s6.2}.  The
difference here is that during inflation, the typical size of the
strings produced increases by a factor of $\displaystyle
\exp\left(\frac{\pi\,\varphi_c^2}{\m^2}\right)=
\exp\left(\frac{\pi\,m_\Phi^2}{\nu\,\m^2}\right)$. If this factor
is not too large, then most of the results obtained in the theory
of formation of density inhomogeneities due to strings [\cite{81}]
remain valid.

\index{Hedgehogs|(}%
2.  {\it Hedgehogs.} Phase transitions at the time of inflation
also lead to the production of hedgehog-antihedgehog pairs (see
Section \ref{s6.2}).  A typical separation $r_0$ between a
hedgehog and antihedgehog is of order ${\rm H}^{-1}(\varphi_c)$,
but as a result of inflation, this separation increases
exponentially.  The pair energy is proportional to $r$.  Hedgehog
annihilation begins when the size of the horizon $\sim t$ growth
to a size comparable with the distance between the hedgehog and
antihedgehog.  This gives rise to density inhomogeneities
$\displaystyle \frac{\delta\rho}{\rho}$ of order
$\displaystyle \frac{\Phi_0^2}{\m^2}$, for the same reason as in
the theory of strings (\ref{6.2.3}).  In the present case,
however, the spectrum of density inhomogeneities will have a
pronounced maximum at a wavelength of the order of the typical
distance between hedgehogs,
$\displaystyle\sim \exp\left(\frac{\pi\,\varphi_c^2}{\m^2}\right)$.
\index{Hedgehogs|)}%

3.  {\it Monopoles.}  Monopoles can be produced as a byproduct of
\index{Monopoles}%
phase transitions at the time of inflation.  The density of such
monopoles will be reduced by factors like
$\displaystyle \exp\left(-\frac{3\,\pi\,\varphi_c^2}{\m^2}\right)$,
but for sufficiently small $\varphi_c$, attempts to detect them
experimentally may have some chance of succeeding.

4.  {\it Monopoles connected by strings.}  Such objects also crop
up in certain theories.  Just as for hedgehogs, such monopoles
\index{Monopoles!connected by strings}%
\index{Strings!monopoles connected by}%
appear in a confinement phase, and in the hot universe theory,
where the typical distance between monopoles is of order ${\rm
T}_c^{-1}$, they are rapidly annihilated [\cite{81}].  In the
inflationary universe scenario, they can lead to approximately the
same consequences as hedgehogs.

5.  {\it Domain walls bounded by strings.}  The axion theory
\index{Domain walls!bounded by strings|(}%
\index{Strings!domain walls bounded by|(}%
discussed in the preceding section qualifies as one of a number of
theories in which strings are produced following symmetry
breaking.  With currently accepted model parameter values, axion
strings, in and of themselves, are too light to induce large
enough density inhomogeneities.  But a more careful analysis shows
that every axion string is the boundary of a domain wall
[\cite{43}, \cite{81}].  This is related to the fact that in
proceeding around a string, with the quantity $\displaystyle
\frac{\theta(x)}{\sqrt{2}\,\Phi_0}$ changing by $2\,\pi$, we
necessarily pass through a maximum of ${\rm V}(\theta)$
(\ref{7.7.22}).  Energetically, the most favorable configuration
of the field $\theta(x)$ is that in which the field $\theta$ does
not change as one proceeds around the string;  this corresponds to
having a minimum of ${\rm V}(\theta)$ everywhere except at a wall
whose thickness is of order $m_\theta^{-1}$, and having the
quantity $\displaystyle \frac{\theta(x)}{\sqrt{2}\,\Phi_0}$ change
by $2\,\pi$ when the latter is traversed.  The surface energy of
the wall is of order $m_\pi^2\,\Phi_0$.

Analysis of the evolution of a system of strings that acts like a
wireframe supporting a soap film composed of domain walls shows
that the initial field configuration resembles a single
infinitely curved surface containing a large number of holes.
Moreover, there also exist isolated surfaces of finite size, but
they contribute negligibly to the total energy of the universe
[\cite{81}].  Portions of these surfaces eventually begin to
intersect and tear one another over a surface region of small
extent, and resemble frothy ``pancakes,'' which subsequently
oscillate and radiate their energy away in the form of
gravitational waves.  If these surfaces form as a result of phase
transitions in a hot universe, the pancakes turn out to be
extremely small, and they quickly disappear.  But surfaces
created at the time of inflation produce pancakes that are
exponentially large [\cite{81}, \cite{125}].  The possible role
of such objects in the formation of large scale structure in the
universe requires further investigation.
\index{Domain walls!bounded by strings|)}%
\index{Strings!domain walls bounded by|)}%

6.  {\it Bubbles.}
\index{Bubbles}%
In studying the cosmological consequences of the phase transitions
occurring at the time of inflation, we have implicitly assumed
that they were soft transitions, with no tunneling through
barriers, as in the second-order phase transition considered in
Section \ref{s7.7}.  Meanwhile, the phase transitions can also be
first-order --- see Section \ref{s7.4}.  Bubbles of the field
$\Phi$ could then be produced.

During inflation, there is much less energy in the fields $\Phi$
than in the inflaton field $\varphi$.  For that reason, the
appearance of such bubbles has practically no effect on the rate
of expansion of the universe, and after inflation, the sizes of
bubbles of the field $\Phi$ wind up being exponentially large;
more specifically, all bubbles turn out to have a typical size of
order $\displaystyle \exp\left(\frac{\pi\,\varphi_c^2}{\m^2}\right)$ cm.
If the rate of bubble production is high, then the resulting
distribution of the field $\Phi$ will resemble a foam (cells),
with maximum energy density on the walls of adjoining bubbles and
with voids within them.  If the bubble creation rate is low, then
mutually separated regions will arise within which the matter
density is lower than that outside.  In the later stages of the
evolution of the universe, when the energy of the field $\Phi$
may become dominant, the corresponding density contrast may turn
out to be quite high [\cite{125}, \cite{240}, \cite{243}].

\index{Domains}%
7.  {\it Domains.}  Especially interesting effects can transpire
when the universe acquires a domain structure at the time of
inflation.  As the simplest example of this we consider the
possible kinetics of the SU(5)-breaking phase transition at that
epoch.  As we already noted in the preceding chapter, as the
temperature T drops, phase transitions in the SU(5) theory entail
the formation of bubbles which can contain a field $\Phi$ that
corresponds to any one of four different types of symmetry
breaking:  $\mbox{SU}(3) \times \mbox{SU}(2) \times {\rm U}(1)$,
$\mbox{SU}(4) \times {\rm U}(1)$, $\mbox{SU}(3) \times {\rm U}(1)
\times {\rm U}(1)$, or $\mbox{SU}(2) \times \mbox{SU}(2) \times
{\rm U}(1) \times {\rm U}(1)$.  An analogous phase transition can
also take place during inflation, but in the latter event,
inflation ensures that bubbles of the different phases becomes
exponentially large.  This results in the formation of large
domains filled with matter in different phases --- that is, with
slightly different  density.  In the standard SU(5) model, only
the phase $\mbox{SU}(3) \times \mbox{SU}(2) \times {\rm U}(1)$ is
stable after inflation, so ultimately the entire universe is
transformed to this phase, and the domain walls that separate the
different phases disappear.  However, the corresponding density
inhomogeneities that appeared during the epoch when domains were
still present somehow remain imprinted on the subsequent density
distribution of matter in the universe.
\index{Domains}%

If the probability of bubble formation is significantly different
for bubbles containing matter in different phases, the universe
will eventually consist of islands of reduced or enhanced density
superimposed on a relatively uniform background.  In principle,
we could associate such islands with galaxies, clusters of
galaxies, or even the insular structure of the universe proposed
in [\cite{248}].

If on the other hand the universe simultaneously spawns
comparable numbers of bubbles in different phases, the resulting
density distribution takes on a sponge-like structure.
Specifically, there will be cells containing phases of different
density, but a significant fraction of the cells of a given phase
will be connected to one another, so that one could pass from one
part of the universe to another through cells that are all of the
same type (percolation).  Concepts involving a sponge-like
universe have lately become rather popular.

Recent results [\cite{249}] indicating that the universe
effectively consists of contiguous bubbles 50--100 Mpc
($1.5\cdot 10^{26}$--$3\cdot 10^{26}$ cm) in size containing few
luminous entities, so that galaxies are basically concentrated at
bubble walls, have been especially noteworthy.  Particularly
interesting in that regard is the fact that such structures may
appear as a natural consequence of phase transitions at the time
of inflation [\cite{125}, \cite{244}].

In the context of the present model, the advent of regions of the
universe containing most of the luminous (baryon) matter is not
at all necessarily connected with density enhancements above the
mean.  Firstly, post-inflation baryon production (see the next
section) proceeds entirely differently in the different phases
($\mbox{SU}(3) \times \mbox{SU}(2) \times {\rm U}(1)$ or
$\mbox{SU}(4) \times {\rm U}(1)$).  It could turn out, in
principle, that baryons are only produced in those regions filled
with a phase whose density lies below the mean, and these would
then be just the regions in which we would see galaxies.
Secondly, if galaxy formation is associated with isothermal
perturbations of the field $\Phi$, then one should take into
account that the amplitude of such perturbations will also depend
on the phase inside each of the domains.  Isothermal
perturbations can therefore only be large enough for subsequent
galaxy formation in those domains filled with some particular
phase, and these are precisely the regions in which galaxies,
clusters of galaxies, and so forth should form.  Thus, depending
on the specific elementary particle theory chosen, galaxies will
preferentially form in regions of either enhanced or reduced
density, and either in the space outside bubbles (for example, at
bubble walls) or within them.

If certain phases remain metastable after inflation, then as a
rule their characteristic decay time will turn out to be much
greater than the age of the observable part of the universe,
$t\sim10^{10}$ yr.  In that event, the universe should be
partitioned right now into domains that contain matter in a
variety of phase states.  This is just the situation in
supersymmetric SU(5) models, where the minima corresponding to
SU(5), $\mbox{SU}(3) \times \mbox{SU}(2) \times {\rm U}(1)$, and
$\mbox{SU}(4) \times {\rm U}(1)$ symmetries are of almost
identical depth and are separated from one another by high
potential barriers [\cite{91}--\cite{93}].  During inflation, the
universe is partitioned into exponentially large domains, each of
which contains one of the foregoing phases, and we happen to live
in one such domain corresponding to the
$\mbox{SU}(3) \times \mbox{SU}(2) \times {\rm U}(1)$ phase
[\cite{211}].  If inflation were to go on long enough after the
phase transition (that is, if the phase transition had taken
place with $\varphi_c\ga5\,\m$ in the $\displaystyle
\frac{\lambda}{4}\,\varphi^4$ theory), then there would be not a
single domain wall in the observable part of the universe.  In
the opposite case, domains would be less than $10^{28}$ cm in
size.  If regions containing different phases have the same
probability of forming (as in a theory that is symmetric with
respect to the interchange $\varphi\rightarrow-\varphi$), then
for $\varphi_c\la5\,\m$, we encounter the domain wall problem
discussed in Section \ref{s6.2}.  However, the probability of
producing bubbles containing matter in different phases depends
on the height of the walls separating different local minima of
${\rm V}(\Phi)$, and in general differs significantly among the
phases.  The universe is therefore mostly filled with just one of
its possible phases, and the other phases are present in the form
of widely-spaced, exponentially large, isolated domains.  Those
domains containing energetically unfavorable phases later
collapse.  As we have already remarked in Section \ref{s7.4},
those regions whose probability of formation is fairly low should
be close to spherical, and the collapse of such regions proceeds
in an almost perfectly spherically symmetric manner.  The entire
gain in potential energy due to compression of a bubble of a
metastable phase would then be converted into kinetic energy of
its collapsing wall.  If the wall is comprised of a scalar field
$\Phi$ that interacts strongly enough both with itself and with
other fields, the bubble wall energy will be transformed after
collapse into the energy of those elementary particles created at
the instant of wall collapse.  The particles thus produced fly
off in all directions, forming a spherical shell.  We thereby
have yet another mechanism capable of producing a universe with
bubble-like structure.  The process will be more complicated if
the original bubble is significantly nonspherical, and the cloud
of newly-created particles will also no longer be spherically
symmetric.

This model resembles the model of Ostriker and Cowie [\cite{250}]
for the explosive formation of the large scale structure of the
universe.  However, the physical mechanism discussed above differs
considerably from that suggested in Ref. \cite{250}.

The investigation of nonperturbative mechanisms for the formation
of the large scale structure of the universe has just begun, but
it is already apparent from the foregoing discussion how many new
possibilities the study of the cosmological consequences of phase
transitions during the inflationary stage carries with it.  The
overall conclusion is that inflation can result in the appearance
of various exponentially large objects.  The latter may be of
interest not just as the structural material out of which
galaxies could subsequently be built, but, for example, as a
possible source of intense radio emission [\cite{251}].  They
could also turn into supermassive black holes, and finally they
might turn out to be responsible for anomalous exoergic processes
in the universe.  This abundance of new possibilities does not
mean that anything goes, but it does substantially expand the
horizons of those seeking the correct theory of formation of the
large scale structure of the universe.

\section{\label{s7.9}Reheating
\index{Inflation!reheating of universe after|(}%
\index{Universe!reheating of, after inflation|(}%
\index{Reheating of universe after inflation|(}%
of the universe after inflation}

The production of inhomogeneities in the inflationary universe
has eli\-cited a great deal of interest of late, as this process is
directly reflected in the structure of the observable part of the
universe.  Of no less value is the study of the process whereby
the universe is reheated and its baryon asymmetry generated,
since this process is a mandatory connecting link between the
inflationary universe in its vacuum-like state and the hot
Friedmann universe.  In the present section, we study the
reheating process through the example of the simplest theory of a
massive scalar field $\varphi$ that interacts with a scalar field
$\chi$ and a spinor field $\psi$, with the Lagrangian
\ba
\label{7.9.1}
{\rm L}&=&\frac{1}{2}\,(\partial_\mu\varphi)^2
-\frac{m^2_\varphi}{2}\,\varphi^2+\frac{1}{2}\,(\partial_\mu\chi)^2
-\frac{m_\chi^2}{2}\,\chi^2
+\bar\psi\,(i\,\gamma_\mu\,\partial_\mu-m_\psi)\,\psi\nonumber \\
&+&\nu\,\sigma\,\varphi\,\chi^2-h\,\bar\psi\,\psi\,\varphi
-\Delta{\rm V}(\varphi,\chi)\ .
\ea
Here $\nu$ and $h$ are small coupling constants, and $\sigma$ is
a parameter with the dimensionality of mass.  In realistic
theories, the constant part of the field $\varphi$, for example,
can play the role of $\sigma$.  What we mean by $\Delta{\rm
V}(\varphi,\chi)$ is that part of ${\rm V}(\varphi,\chi)$ that is
of higher order in $\varphi^2$ and $\chi^2$.  We shall assume
(allowing for $\Delta{\rm V}(\varphi,\chi)$) that in the last
stages of inflation, the role of the inflaton field is taken on
by the field $\varphi$, and then go on to investigate the process
by which the energy of this field is converted into particles
$\chi$ and $\psi$.  We suppose for simplicity that $m_\varphi\gg m_\chi$,
$m_\psi$, and that at the epoch of interest,
$\nu\,\sigma\,\varphi\ll m_\chi^2$, $h\,\varphi\ll m_\psi$.

If we ignore effects associated with particle creation, the field
$\varphi$ after inflation will oscillate near the point
$\varphi=0$ at a frequency $k_0=m_\varphi$.  The oscillation
amplitude will fall off as $[a(t)]^{-3/2}$, and the energy of the
field $\varphi$ will decrease in the same way as the density of
nonrelativistic $\varphi$ particles of mass $m_\varphi$:
$\displaystyle
\rho_\varphi=\vf=\frac{m_\varphi^2}{2}\,\varphi^2\sim a^{-3}$,
where $\varphi$ is the amplitude of oscillations of the field
[\cite{252}].  The physical meaning of this is that a homogeneous
scalar field $\varphi$, oscillating at frequency $m_\varphi$, can
be represented as a coherent wave of $\varphi$-particles with
vanishing momenta and particle density $\displaystyle
n_\varphi=\frac{\rho_\varphi}{m_\varphi}=\frac{m_\varphi}{2}\,\varphi^2$.
If the total number of particles $\sim n_\varphi\,a^3$
is conserved (no pair production), the amplitude of the field $\varphi$
will fall
off as $a^{-3/2}$.  The equation of state of matter at that time is
$p=0$;  i.e., $a(t)\sim t^{3/2}$, $\displaystyle {\rm H}=\frac{2}{3\,t}$,
$\varphi\sim a^{-3/2}\sim t^{-1}$.

In order to describe the particle production process with its
concomitant decrease in the amplitude of the field $\varphi$, let
us consider the quantum corrections to the equation of motion for
the homogeneous field $\varphi$, oscillating at a frequency
$k_0=m_\varphi\gg{\rm H}(t)$:
\be
\label{7.9.2}
\ddot\varphi+3\,{\rm H}(t)\,\dot\varphi+
[m_\varphi^2+\Pi(k_0)]\,\varphi=0\ .
\ee
\index{Polarization operator}%
Here $\Pi(k_0)$ is the polarization operator for the field
$\varphi$ at a\linebreak[10000]
four-momentum $k=(k_0,0,0,0)$, $k_0=m_\varphi$.

The real part of $\Pi(k_0)$ gives only a small correction to
$m_\varphi^2$, but when $k_0>2\,m_\chi$ (or $k_0>2\,m_\psi$),
$\Pi(k_0)$ acquires an imaginary part $\Pi(k_0)$.  For
$m_\varphi^2\gg{\rm H}^2$ and $m^2_\varphi\gg\mbox{Im}\,\Pi$, and
neglecting the time-dependence of H, we obtain a solution of Eq.
(\ref{7.9.2}) that describes the damped oscillations of the field
$\varphi$ near the point $\varphi=0$:
\be
\label{7.9.3}
\varphi=\varphi_0\,\exp(i\,m_\varphi\,t)\cdot
\exp\left[-\frac{1}{2}\,\left(3\,{\rm H}+
\frac{\mbox{Im}\,\Pi(m_\varphi)}{m_\varphi}\right)\,t\right]\ .
\ee
From the unitarity relations [\cite{10}, \cite{124}], it follows that
\be
\label{7.9.4}
\mbox{Im}\,\Pi(m_\varphi)=m_\varphi\,\Gamma_{tot}\ ,
\ee
where $\Gamma_{tot}$ is the total decay probability for a
$\varphi$-particle.  Hence, when $\Gamma_{tot}\gg3\,{\rm H}$, the
energy density of the field $\varphi$ decreases exponentially in
a time less than the typical expansion time of the universe
$\Delta t\sim{\rm H}^{-1}$:
\be
\label{7.9.5}
\rho_\varphi=\frac{m^2\,\varphi^2}{2}\sim\rho_0\,e^{-\Gamma_{tot}\,t}\ .
\ee
This is exactly the result one would expect on the basis of the
interpretation of the oscillating field $\varphi$ as a coherent
wave consisting of (decaying) $\varphi$-particles.

The probability of decay of a $\varphi$-particle into a pair of
$\chi$-particles or $\psi$-particles is known --- see, for example,
[\cite{10}, \cite{122}, \cite{123}].
For $m_\varphi\gg m_\chi$, $m_\psi$,
\ba
\label{7.9.6}
\Gamma(\varphi\rightarrow\chi\,\chi)&=&
\frac{\nu^2\,\sigma}{8\,\pi\,m_\varphi}\ ,\\
\label{7.9.7}
\Gamma(\varphi\rightarrow\bar\psi\,\psi)&=&
\frac{h^2\,m_\varphi}{8\,\pi}\ .
\ea
If the constants $\nu\,\sigma$ and $h^2$ are small, then initially
$\displaystyle \Gamma_{tot}=\Gamma(\varphi\rightarrow\chi\,\chi)
+\Gamma(\varphi\rightarrow\bar\psi\,\psi)<3\,{\rm H}(t)=\frac{2}{t}$.
Basically, in that event, the energy density of the field
$\varphi$ decreases from the outset simply due to the expansion
of the universe, $\displaystyle \frac{m^2\,\varphi^2}{2}\sim t^{-2}$.
The fraction of the total energy converted into energy of the
particles that are produced remains small right up to the time
$t^*$ at which $3\,{\rm H}(t^*)$ becomes less than $\Gamma_{tot}$.
Particles produced prior to this time can also be thermalized, in
principle, and their temperature in certain situations can be
even higher than the final temperature ${\rm T}_{\rm R}$ [\cite{253}].
But the contribution of newly-created particles to the overall
matter density becomes significant only starting at the time
$t^*$, after which practically all the energy of the field
$\varphi$ is transformed into the energy of newly-created $\chi$-
and $\psi$-particles within a time $\Delta t\sim t^*\la{\rm H}^{-1}$.
The condition $3\,{\rm H}(t^*)\sim\Gamma_{tot}$ tells us that the
energy density of these particles at the time $t^*$ is of order
\be
\label{7.9.8}
\rho^*\sim\frac{\Gamma^2_{tot}\,\m^2}{24}\ .
\ee
If the $\chi$- and $\psi$-particles interact strongly enough with
each other, or if they can rapidly decay into other species, then
thermodynamic equilibrium quickly sets in, and matter acquires a
temperature ${\rm T}_{\rm R}$, where according to (\ref{1.3.17})
and (\ref{7.9.8})
\be
\label{7.9.9}
\rho^*\sim\frac{\pi\,{\rm N}({\rm T}_{\rm R})}{30}\,{\rm T}_{\rm R}^4
\sim\frac{\Gamma^2_{tot}\,\m^2}{24}
\ee
Here $\rm N(T_{\rm R})$ is the effective number of degrees of
freedom at $\rm T=T_{\rm R}$, with
${\rm N}({\rm T}_{\rm R})\sim10^2$--$10^3$, so that
\be
\label{7.9.10}
{\rm T}_{\rm R}\sim10^{-1}\,\sqrt{\Gamma_{tot}\,\m}\ .
\ee
Note that as we said earlier, ${\rm T}_{\rm R}$ does not depend
on the initial value of the field $\varphi$, and is determined
solely by the parameters of the elementary particle theory.

Let us now estimate ${\rm T}_{\rm R}$ numerically.  In order for
adiabatic inhomogeneities
$\displaystyle \frac{\delta\rho}{\rho}\sim10^{-5}$ to appear in
the present theory, it is necessary that $m_\varphi$ be of order
$10^{-6}\,\m\sim10^{13}$ GeV.  One can readily verify that
quantum corrections investigated in Chapter \ref{c2} do not
significantly alter the form of ${\rm V}(\varphi)$ at
$\varphi\la\m$ only if
$\displaystyle h^2\la\frac{8\,m_\varphi}{\m}\sim10^{-5}$ and
$\nu\,\sigma\la5\,m_\varphi\sim10^{14}$ GeV.  Under these
conditions,
\ba
\label{7.9.11}
\Gamma(\varphi\rightarrow\chi\,\chi)&\la&m_\varphi\sim10^{-6}\,\m\ ,\\
\label{7.9.12}
\Gamma(\varphi\rightarrow\bar\psi\,\psi)&\la&
\frac{m^2_\varphi}{\m}\sim 10^{-12}\,\m\ .
\ea

For completeness, we note that in theories like the Starobinsky
model or supergravity, the magnitude of $\Gamma$ for
$\varphi$-field decays induced by gravitational effects is
usually [\cite{135}, \cite{286}]
\be
\label{7.9.13}
\Gamma_g\sim\frac{m_\varphi^3}{\m^2}\sim10^{-18}\,\m\ .
\ee
Thus, if direct decay of the field $\varphi$ into scalar
$\chi$-particles is possible, then one might expect that in
general this will be the leading process [\cite{123}].  We see
from (\ref{7.9.11}) that the rate at which the $\varphi$-field
decays into $\chi$-particles can be of the same order of
magnitude as the rate at which the $\varphi$-field oscillates;
it can therefore divest itself of most of its energy in several
cycles of oscillation (or even simply in the time it takes to
roll down from $\varphi\sim\m$ to $\varphi=0$ [\cite{254}]).
Since ${\rm H}(\varphi)\sim m_\varphi$ at the end of inflation,
the universe has almost no time to expand during reheating, and
almost all the energy stored in the field $\varphi$ can be
converted into energy for the production of $\chi$-particles.
This same result follows from (\ref{7.9.8}):
\be
\label{7.9.14}
\rho^*=\frac{m_\varphi^2}{2}\,\varphi^2\la\frac{m_\varphi^2\,\m}{24}\ ,
\ee
whence $\varphi(t^*)\la\m$ and
\be
\label{7.9.15}
{\rm T}_{\rm R}\la10^{-1}\,\sqrt{m_\varphi\,\m}\sim10^{15}\;
\mbox{GeV}\ .
\ee
Reheating to ${\rm T}_{\rm R}\sim10^{15}$ GeV occurs only for a
special choice of parameters.  Moreover, in some models the
universe cannot be heated up to a temperature much higher than
$m_\varphi$;  see Section \ref{s7.10}.  Nevertheless, one must
keep in mind the possibility of such an efficient reheating,
which can take place immediately after inflation ends, despite
the weakness of the interaction between the $\varphi$- and
$\chi$-fields.  A similar possibility can come into play if the
potential ${\rm V}(\varphi)$ takes on a more complicated form ---
for example, if the curvature of ${\rm V}(\varphi)$ near its
minimum is much greater than at $\varphi\sim\m$ [\cite{255}].

If the field $\varphi$ can only decay into fermions, then we find
from (\ref{7.9.12}) and (\ref{7.9.10}) that in the simplest
models, the temperature of the universe after reheating will be
at least three orders of magnitude lower,
\be
\label{7.9.16}
{\rm T}_{\rm R}\la10^{-1}\,m_\varphi\sim10^{12}\;\mbox{GeV}\ ,
\ee
and if gravitational effects are dominant, then
\be
\label{7.9.17}
{\rm T}_{\rm R}\la10^{-1}\,m_\varphi\,\sqrt{\frac{m_\varphi}{\m}}
\sim 10^9\;\mbox{GeV}\ .
\ee

The foregoing estimates have been made using the simplest model
and assuming that the oscillating field is small.  But if the
field $\varphi$ is large ($\nu\,\sigma\,\varphi>m_\chi^2$ or
$h\,\varphi>m_\psi$), it is not enough to calculate just the
polarization operator;  one must then either calculate the
imaginary part of the effective action ${\rm S}(\varphi)$ in the
external field $\varphi(t)$ [\cite{122}, \cite{256}], or employ
methods based on the Bogoliubov transformation [\cite{74}].

We shall not discuss this point in detail here, since for the
theory (\ref{7.9.1}), an investigation of the case in which
$\nu\,\sigma\,\varphi>m^2_\chi$  and $h\,\varphi>m_\psi$ results
only in a change in the numerical coefficients in (\ref{7.9.6})
and (\ref{7.9.7}).  More important changes arise in the theories
with Lagrangians that lack ternary interactions like
$\varphi\,\chi^2$  and $\varphi\,\bar\psi\,\psi$, having only
vertices like $\varphi^4$, $\varphi^2\,\chi^2$ or
$\varphi^2\,{\rm A}_\mu^2$, with a field $\varphi$ that has no
classical part $\varphi_0$.

Thus, for example, in the
$\displaystyle \frac{\lambda}{4}\,\varphi^4$ theory of the
massless field, evaluation of the imaginary part of the effective
Lagrangian ${\rm L}(\varphi)$ leads to an expression for the
probability of pair production [\cite{122}]:
\be
\label{7.9.18}
{\rm P}\approx2\,\mbox{Im}\,{\rm L}(\varphi)\sim
\lambda^2\,\varphi^4\cdot {\rm O}(10^{-3})\ .
\ee
An analogous expression holds for a $\lambda\,\varphi^2\,\chi^2$
theory.  The energy density of particles created in a time
$\Delta t\sim{\rm H}^{-1}$ is
\be
\label{7.9.19}
\Delta\rho\sim 10^{-3}\,\lambda^2\,\varphi^4\cdot \sqrt{\lambda}\,\varphi
\cdot {\rm H}^{-1}
\sim10^{-3}\,\lambda^2\,\varphi^3\,\m\ ,
\ee
where the effective mass of the $\varphi$- and $\chi$-fields is
${\rm O}(\sqrt{\lambda}\,\varphi)$.  This quantity becomes
comparable to the total energy density
$\displaystyle \rho(\varphi)\sim\frac{\lambda}{4}\,\varphi^4$ when
\be
\label{7.9.20}
\varphi\la10^{-2}\,\lambda\,\m\ ,
\ee
that is, when
\be
\label{7.9.21}
\rho(\varphi)\sim10^{-8}\,\lambda^5\,\m^4\ ,
\ee
whereupon
\be
\label{7.9.22}
{\rm T}_{\rm R}\la10^{-3}\,\lambda^{5/4}\,\m\ .
\ee
For $\lambda\sim10^{-14}$,
\be
\label{7.9.23}
{\rm T}_{\rm R}\la3\cdot 10^{-21}\,\m\sim3\cdot 10^{-2}\;\mbox{GeV}\ .
\ee
If the field $\varphi$ in the theory $\displaystyle
\frac{\lambda}{4}\,\varphi^4$ has a nonvanishing mass
$m_\varphi$, the reheating of the universe becomes ineffectual at
$\displaystyle \varphi\la\frac{m_\varphi}{\sqrt{\lambda}}$, since
at small $\varphi$, the value of $\Delta\rho$ from (\ref{7.9.19})
always turns out to be less than
$\displaystyle \rho(\varphi)\sim\frac{m_\varphi^2}{2}\,\varphi^2$.
In such a situation, the energy of the field $\varphi$ basically
falls due to the expansion of the universe, $\rho(\varphi)\sim a^{-3}$,
rather than due to the field decay.  This implies that
after expansion of the universe to its present state, even a
strongly interacting oscillating classical field $\varphi$
($10^{-14}\ll\lambda\la1$) can turn out to be largely undecayed
into elementary particles, and can make a sizable contribution to
the density of dark matter in the universe.  This topic is
treated in more detail in [\cite{254}].
\index{Inflation!reheating of universe after|)}%
\index{Reheating of universe after inflation|)}%
\index{Universe!reheating of, after inflation|)}%

\section[origin of the baryon asymmetry of the universe]%
{\label{s7.10}The origin of the baryon asymmetry
\index{Asymmetry baryon|(}%
\index{Baryon asymmetry|(}%
\index{Universe!baryon asymmetry of|(}%
of the universe}

As we have already noted, the elaboration of feasible mechanisms
for generating an excess of baryons over antibaryons in the
inflationary universe [\cite{36}--\cite{38}] was one of the most
important stages in the development of modern cosmology.  The
baryon asymmetry problem served to demonstrate quite clearly that
questions which to many had seemed meaningless, or at best
metaphysical (``Why is the universe structured as it is, and not
otherwise?''), could actually have a physical answer.  Without a
solution of the\index{Baryogenesis} baryogenesis
problem, the inflationary universe
scenario would be impossible, since the density of baryons that
exist at the earliest stages of evolution of the universe becomes
exponentially small after inflation.  The generation of a baryon
asymmetry in the universe is therefore just as indispensable an
element of the inflationary universe scenario as the reheating of
the universe discussed in the previous section.

As the first treatment of the origin of baryons in the universe
made clear [\cite{36}], an asymmetry between the number of
baryons and 
antibaryons arises when three conditions are
satisfied:

1.  The processes involved violate baryon charge conservation.

2.  These processes also violate C and CP invariance.

3.  Baryon production processes take place in a nonequilibrium
thermodynamic state.  One example would be the decay of particles with mass
$\rm M\gg T$.

The need for the first condition is obvious.  The second is
needed in order for the decay of particles and antiparticles to
produce different numbers of baryons and antibaryons.  The third
condition is primarily needed to prevent inverse processes which
might destroy baryon asymmetry.

Genuine interest in the possibility of generating the baryon
asymmetry of the universe was kindled by the advent of grand
unified theories, in which baryons could freely transform into
leptons prior to symmetry breaking between the strong and
electroweak interactions.  After the symmetry breaking,
\index{Bosons!superheavy}%
\index{Superheavy bosons}%
superheavy scalar and vector particles ($\Phi$, H, X, and Y)
decay into baryons and leptons.  If the decay of these particles
takes place in a state far removed from thermodynamic
equilibrium, so that the inverse processes of baryon and lepton
reversion to superheavy particles are inhibited, and if C and CP
invariance is violated, then the decays will produce slightly
different numbers of baryons and antibaryons.  This difference,
after annihilation of baryons and antibaryons,  is exactly what
produces the baryonic matter that we see in our universe.  The
small number $\displaystyle \frac{n_{\rm B}}{n_\gamma}\sim10^{-9}$
comes about as a product of the gauge coupling constant in grand
unified theories, the constant related to the strength of CP
violation, and the relative abundance of the particles that
produce baryon asymmetry after their decay [\cite{38}].

We shall not dwell here on a detailed description of this
mechanism for \index{Baryogenesis}baryogenesis,
referring the reader instead to the
excellent reviews in [\cite{105}, \cite{257}, \cite{258}].  What
is important for us is that theories leading to the desired
result $\displaystyle \frac{n_{\rm B}}{n_\gamma}\sim10^{-9}$
actually do exist.  A similar mechanism can also operate as a
part of the inflationary universe scenario, where it is even more
effective, since the universe is reheated after inflation in what
is essentially a nonequilibrium process, and during this process,
superheavy particles can be produced with masses much greater
than the temperature of the universe after reheating ${\rm T}_{\rm R}$
[\cite{122}].  However, in the minimal SU(5) theory with a single
family of Higgs bosons H and the most natural relationship
between coupling constants, $\displaystyle \frac{n_{\rm
B}}{n_\gamma}$   turns out to be many orders of magnitude less
than $10^{-9}$.  In order to get
$\displaystyle \frac{n_{\rm B}}{n_\gamma}\sim10^{-9}$, it is
necessary either to introduce two additional families of Higgs
bosons, or to consider the possibility of a complex sequence of
phase transitions as the universe cools in the SU(5) theory
[\cite{259}].  Furthermore, it is far from easy to obtain the
fairly large number of
\index{Bosons!superheavy}%
\index{Superheavy bozons}%
superheavy bosons needed to implement this
mechanism at the time of post-inflation reheating. This is
especially difficult in supergravity-type theories, where the
temperature to which the universe is reheated is usually
$10^{12}$ GeV at most.  Finally, one more potential problem
became apparent fairly recently.  It was found that
nonperturbative effects lead to efficient annihilation of baryons
and leptons at a temperature that is higher than or of the same
order as the phase transition temperature ${\rm T}_c\sim200$ GeV
\index{Glashow--Weinberg--Salam theory}%
in the Glashow--Weinberg--Salam model [\cite{129}].  This means
that if equal numbers of baryons and leptons are generated in the
early stages of evolution of the universe, so that $\rm B-L=0$,
where B and L are the baryon and lepton charges respectively (and
this is exactly the situation in the simplest models of
\index{Baryogenesis}baryogenesis [\cite{38}]), then the entire
baryon asymmetry of the universe arising at ${\rm T}>10^2$ GeV
subsequently vanishes. If this is so, then either it is necessary
to have theories that begin with an asymmetry $\rm B-L\neq0$,
which makes these theories even more complicated, or mechanisms
for baryogenesis which could operate efficiently even at a
temperature ${\rm T}\la10^2$ GeV must be worked out.  Several
possible mechanisms of this kind have been proposed in recent
years. Below we describe one of them, the details of which are
probably the closest of any to the inflationary universe scenario.

The basic idea behind that scheme was proposed in a paper by
Affleck and Dine [\cite{97}];  their mechanism was implemented in
the context of the inflationary universe scenario [\cite{98}].
Later, it was demonstrated that this mechanism could work in
models based on superstring theory [\cite{260}].  Referring the
reader to the original literature for details, we discuss here
the basic outlines of this new mechanism for
\index{Baryogenesis}baryogenesis.

As an example, consider a supersymmetric SU(5) grand unification
\index{Grand unified theories!supersymmetric SU(5)}%
\index{SU(5) theory!supersymmetric grand unification theory}%
\index{Supersymmetric SU(5) grand unification theory}%
\index{Slepton field}%
\index{Squarks field}%
theory.  In this theory there exist squarks and sleptons, which
are scalar fields, the superpartners of quarks and leptons.  An
analysis of the shape of the effective squark and slepton
potential shows that it has valleys --- flat directions --- in
which the effective potential approaches zero [\cite{97}].  We
will refer to the corresponding linear combinations of squark and
slepton fields in the flat directions as the scalar field
$\varphi$.  After supersymmetry breaking in the model in question,
the value of the effective potential $\vf$ in the valleys rises
slightly, and the field $\varphi$ acquires an effective mass
$m\sim10^2$ GeV.  Excitations of this field consist of
electrically neutral unstable particles having baryon and lepton
charge $\rm B=L=\pm1$.  The baryon charge of each such particle is
not conserved by their interactions, but the difference $\rm B-L$
is.  These particles interact among themselves via the same gauge
coupling constant $g$ as do the quarks.  The coupling constant of
the baryon-nonconserving interactions of the $\varphi$-particles
is $\displaystyle \lambda={\rm O}\left(\frac{m^2}{{\rm M}_{\rm
X}^2}\right)$, where ${\rm M}_{\rm X}$ is the X-boson mass in the
SU(5) theory.  For large values of the classical field $\varphi$,
many of the particles that interact with it acquire a very high
mass that is ${\rm O}(g\,\varphi)$.  However, there are also light
particles such as quarks, leptons, W mesons, and so forth, that
interact only indirectly with the field $\varphi$ (via radiative
corrections), with an effective coupling constant $\displaystyle
\tilde \lambda\sim\left(\frac{\alpha_s}{\pi}\right)^2\,
\frac{m^2}{\varphi^2}$, where $\displaystyle
\alpha_s=\frac{g^2}{4\,\pi}$. In a rigorous treatment, it would be
necessary to consider the dynamics of the two fields $v$ and $a$,
corresponding to different combinations of squark-slepton fields
in the valley of the effective potential [\cite{97}].  A thorough
study of a system of such fields in the SU(5) theory would be
fairly complicated, but fortunately, in the most important
instances, it can be reduced to the study of one simple model that
describes the complex scalar field $\displaystyle
\varphi=\frac{1}{\sqrt{2}}\,(\varphi_1+i\,\varphi_2)$, with the
somewhat unusual potential [\cite{97}, \cite{98}] \be
\label{7.10.1} \vf=
m^2\,\varphi^*\,\varphi+\frac{i}{2}\,\lambda\,[\varphi^4-(\varphi^*)^4]\
, \ee The quantity $\displaystyle j_\mu=-i\,\varphi^*\,
\mathop{\partial_\mu}^\leftrightarrow\varphi =\frac{1}{2}\,
(\varphi_1\,\partial_\mu\varphi_2-\varphi_2\,\partial_\mu\varphi_1)$
corresponds to the baryon current of scalar particles in the SU(5)
model, while $j_0$ is the baryon charge density $n_{\rm B}$ of the
field $\varphi$.  The equations of motion of the fields
$\varphi_1$ and $\varphi_2$ are \ba \label{7.10.2}
\ddot\varphi_1+3\,{\rm H}\,\dot\varphi_1&=& -\frac{\partial{\rm
V}}{\partial\varphi_1}
=-m^2\,\varphi_1+3\,\lambda\,\varphi_1^2\,\varphi_2-\lambda\,\varphi_2^3\ ,\\
\label{7.10.3}
\ddot\varphi_2+3\,{\rm H}\,\dot\varphi_2&=&
-\frac{\partial{\rm V}}{\partial\varphi_2}
=-m^2\,\varphi_2-3\,\lambda\,\varphi_2^2\,\varphi_1+\lambda\,\varphi_1^3\ .
\ea
During inflation, when H is a very large quantity, the fields
$\varphi_i$ evolve slowly, so that as usual the terms
$\ddot\varphi_i$ in (\ref{7.10.2}) and (\ref{7.10.3}) can be
neglected.  This then leads to an expression for the density
$n_{\rm B}$ at the time of inflation:
\ba
\label{7.10.4}
n_{\rm B}\equiv j_0&=&\frac{1}{3\,{\rm H}}
\left(\varphi_1\,\frac{\partial{\rm V}}{\partial\varphi_2}
-\varphi_2\,\frac{\partial{\rm V}}{\partial\varphi_1}\right)\nonumber \\
&=&\frac{\lambda}{3\,{\rm H}}
(\varphi_1^4-6\,\varphi_1^2\,\varphi_2^2+\varphi_2^4)\ .
\ea
If, for example, we take the initial conditions to be
$\displaystyle \varphi_2\ga\frac{1}{4}\,\varphi_1>0$,
$\lambda\,\varphi_i^2\ll m^2$, then we find from
(\ref{7.10.2})--(\ref{7.10.4}) that during inflation the field
$\varphi_1$ evolves very slowly, and it remains much smaller than
$\varphi_2$, so that during the inflationary stage  is $n_{\rm B}$
approximately constant in magnitude and equal to its initial value,
$\displaystyle n_{\rm B}\approx\frac{\lambda}{3\,{\rm H}}\,\varphi_2^4$.

To clarify the physical meaning of this result, let us write down
the equation for the partially conserved current in our model in
the following form:
\be
\label{7.10.5}
a^{-3}\,\frac{d(n_{\rm B}\,a^3)}{dt}\equiv\dot n_{\rm B}+3\,n_{\rm B}\,{\rm H}
=i\,\left(\varphi^*\,\frac{\partial{\rm V}}{\partial\varphi}-
\varphi\,\frac{\partial{\rm V}}{\partial\varphi^*}\right)\ ,
\ee
where $a(t)$ is the scale factor.  If there were no term
$\sim i\,\lambda\,(\varphi^4-(\varphi^*)^4)$  in (\ref{7.10.1})
leading to nonconservation of baryon charge, the total baryon
charge of the universe ${\rm B}\sim n_{\rm B}\,a^3$ would be
constant, and the baryon charge density $n_{\rm B}$  would become
exponentially small at the time of inflation.  In our example,
however, the right-hand side of (\ref{7.10.5}) does not vanish,
and serves as a source of baryon charge.  Bearing in mind then
that all fields vary very slowly during inflation, so that
$\dot n_{\rm B}\ll3\,n_{\rm B}\,{\rm H}$, (\ref{7.10.5}) again
implies (\ref{7.10.4}), as obtained previously.

In other words, due to the presence of the last term in
(\ref{7.10.1}), the baryon charge density varies very slowly
during inflation, as do the fields $\varphi_i$ (see
(\ref{7.10.4})), while the total baryon charge of the part of the
universe under consideration grows exponentially.  The baryon
charge density and its sign depend on the initial values of the
fields $\varphi_i$, and will be different in different parts of
the universe.

When the rate of expansion of the universe becomes low, the field
begins to oscillate in the vicinity of the minimum of ${\rm V}(\varphi)$
at $\varphi=0$.  While this is going on, the gradual
decrease in the amplitude of oscillation reduces terms
$\sim\lambda\,\varphi^4$ in (\ref{7.10.5}) which are responsible
for the nonconservation of baryon charge; the total baryon charge
of the field $\varphi$ will then be conserved, and its density
will fall off as $a^{-3}(t)$.  Notice that at that point the
energy density of the scalar field
$\displaystyle \rho\sim\frac{m^2\,\varphi^2}{2}$ will also fall
off as $a^{-3}(t)$.  This coincidence has a very simple meaning.
As we have already discussed, a homogeneous field $\varphi$
oscillating with frequency $m$ can be represented as a coherent
wave consisting of particles of the field $\varphi$ with particle
density $\displaystyle  n_\varphi=\frac{\rho}{m}=\frac{m}{2}\,\varphi^2$,
where $\varphi$ is the amplitude of the oscillating field. Some
of these particles have baryon charge $\rm B=+1$, and some have
$\rm B=-1$.  The baryon charge density $n_{\rm B}$ is thus
proportional to $n_\varphi$, so that the ratio
$\displaystyle \frac{n_{\rm B}}{n_\varphi}$ is time-independent
and cannot be bigger than unity in absolute value:
\be
\label{7.10.6}
\frac{|n_{\rm B}|}{n_\varphi}=\mbox{const}\le1\ .
\ee
This ratio
$\displaystyle \frac{n_{\rm B}}{n_\varphi}$ is determined by the
initial conditions.  The oscillatory regime sets in at ${\rm H}\sim m$,
so (\ref{7.10.4}) implies that at that stage
\be
\label{7.10.7}
\frac{n_{\rm B}}{n_\varphi}\approx
\frac{\lambda\,\tilde \varphi_2^4}{3\,m}\ ,
\ee
where $\tilde \varphi_2$ is the value of the field $\varphi_2$ at
the onset of the oscillation stage.  In the realistic SU(5)
model, Eq. (\ref{7.10.7}) also contains a factor $\cos
2\,\theta$, where $\theta$ is the angle between the $v$ and $a$
fields in the complex plane.  The $\varphi$-particles are
\index{Temperature!of universe}%
\index{Universe!temperature of}%
unstable and decay into leptons and quarks.  The temperature of
the universe rises at that point, but it cannot rise much higher
than $m$, since at high temperature the quarks have an effective
mass $m_q\sim g\,{\rm T}\sim {\rm T}$, so that the field
$\varphi$ cannot decay when ${\rm T}\gg m$.  The field $\varphi$
therefore oscillates and decays gradually, rather than suddenly,
and in the process it warms the universe up to a constant
temperature ${\rm T}\sim m\sim 10^2$ GeV.  By the end of this
stage, the entire baryon charge of the scalar field has been
transformed into the baryon charge of the quarks, and for every
quark or antiquark produced through the decay of a
$\varphi$-particle, there is approximately one photon of energy
${\rm E}\sim{\rm T}\sim m$.  This means that the density of
photons $n_\gamma$ produced by the decaying field $\varphi$ is of
the same order of magnitude as $n_\varphi$.  The baryon asymmetry
of the universe thus engendered is
\be
\label{7.10.8}
\frac{n_{\rm B}}{n_\varphi}\sim\frac{n_{\rm B}}{n_\gamma}\sim
\cos2\,\theta\cdot \frac{\lambda\tilde \varphi_2^2}{m^2}\sim
\cos2\,\theta\cdot \frac{\tilde \varphi_2^2}{{\rm M}_{\rm X}^2}\ .
\ee
Note that this equation is only valid when $\lambda\,\tilde
\varphi_2^2\ll m$, that is, when $\tilde \varphi_2\ll{\rm M}_{\rm X}$,
so only from that point onward can the violation of baryon charge
conservation be neglected, with the quantity
$\displaystyle \frac{n_{\rm B}}{n_\varphi}$  becoming constant.
As expected from (\ref{7.10.6}), the baryon asymmetry of the
universe as given by (\ref{7.10.8}) then turns out to be less
than unity.  However, from (\ref{7.10.8}), it follows that the
mechanism of \index{Baryogenesis}baryogenesis
discussed above may even be too
efficient.  For example, for $\tilde\varphi_2\sim {\rm M}_{\rm X}$,
Eq. (\ref{7.10.8}) yields
$\displaystyle \frac{n_{\rm B}}{n_\gamma}={\rm O}(1)$.  We must
therefore try to understand what $\tilde \varphi_2$ should be
equal to, and how to reduce $\displaystyle \frac{n_{\rm B}}{n_\varphi}$
to the desirable value $\displaystyle \frac{n_{\rm B}}{n_\varphi}\sim10^{-9}$.

Research into this question has shown that just like money,
baryon asymmetry is hard to come by but easy to get rid of
[\cite{98}].  One mechanism for reducing the baryon asymmetry is
the previously cited nonperturbative scheme [\cite{129}].  If,
\index{Temperature!of universe}%
\index{Universe!temperature of}%
for example, the temperature of the universe following decay of
the field $\varphi$ exceeds approximately 200 GeV, then virtually
the entire baryon asymmetry that has been produced will
disappear, with the exception of a small part resulting from
processes that violate $\rm B-L$ invariance.  This residual can
in fact account for the observed asymmetry
$\displaystyle \frac{n_{\rm B}}{n_\varphi}\sim10^{-9}$.  Another
possibility is that the temperature is less than 200 GeV when
decay of the $\varphi$-field ends; that is, the baryons do not
burn up, but the initial value of the field $\varphi$ is fairly
small.  This could happen, for instance, if the fields
$\varphi_i$ were to vanish due to high-temperature effects or
interaction with the fields responsible for inflation.  The role
played by these fields would then be taken up by their long-wave
quantum fluctuations with an amplitude proportional to
$\displaystyle \frac{{\rm H}}{2\,\pi}\,\sqrt{{\rm H}\,t}$  (see
(\ref{7.3.12})), which could be several orders of magnitude less
than ${\rm M}_{\rm X}$.

Finally, the\index{Anthropic Principle} Anthropic Principle
provides one more plausible explanation of why $\displaystyle
\frac{n_{\rm B}}{n_\gamma}$ is so small in the observable part of
the universe.  The fields $\varphi_i$ and the quantity $\cos
2\,\theta$ take on all possible values in different regions of the
universe.  In most such regions, $\varphi$ can be extremely large,
and $|\cos2\,\theta|\sim 1$.  But these are regions with
$\displaystyle \frac{n_{\rm B}}{n_\gamma}\gg10^{-9}$, and life of
our type it is impossible.  The reason is that for a given
amplitude of perturbations $\displaystyle
\frac{\delta\rho}{\rho}$, elevating the baryon density by just two
to three orders of magnitude results in the formation of galaxies
having extremely high matter density and a completely different
complement of stars.  It is therefore not inconsistent to think
that there are relatively few regions of the universe with small
initial values of $\varphi$ and $\cos 2\,\theta$ --- but these are
just the regions with the highest likelihood of supporting life of
our type.  We will discuss this problem in a more detailed way in
Chapter \ref{c10}.

In addition to the mechanism discussed above, several more that
could operate at temperatures ${\rm T}\la10^2$ GeV have recently
been proposed [\cite{130}, \cite{131}, \cite{178},
\cite{261}--\cite{263}].  It is still difficult to say which of
these are realistic.  One important point is that many ways have
been found to explain the baryon asymmetry of the universe, and
superhigh temperatures
${\rm T}\sim {\rm M}_{\rm X}\sim10^{14}$--$10^{15}$ GeV, which
occur only after extremely efficient reheating, are not at all
mandatory.  In principle, baryon asymmetry could even occur if
the temperature of the universe never exceeded 100 GeV!  This
then substantially facilitates the construction of realistic
models of the inflationary universe.  On the other hand, the
realization that it is possible to construct a consistent theory
of the evolution of the universe in which the temperature may
never exceed ${\rm T}\sim 10^2\;\mbox{GeV}\sim10^{-17}\,\m$
leads us yet again to ponder the extent to which our notions have
changed over the past few years, and to wonder what surprises
might await us in the
future.
\index{Inflationary universe!general principles of|)}%
\index{Asymmetry baryon|)}%
\index{Baryon asymmetry|)}%
\index{Universe!baryon asymmetry of|)}%


\chapter{\label{c8}The New Inflationary Universe Scenario}
\vspace{-1pc}
\index{Inflationary universe scenario!new|(}%

\section[Introduction. The old scenario]%
{\label{s8.1}Introduction.
\index{Inflationary universe scenario!old|(}%
The old inflationary universe scenario}

In the previous chapter, we described the building blocks needed
for a complete theory of the inflationary universe.  It is now
time to demonstrate how all parts of the theory that we have
described thus far may be combined into a single scenario,
implemented in the context of some of recently developed theories
of elementary particles.  As we have already pointed out,
however, there are presently two significantly different
fundamental versions of inflation theory, namely the new
inflationary universe scenario [\cite{54}, \cite{55}], and the
chaotic inflation scenario [\cite{56}, \cite{57}].  Although we
lean toward the latter, in view of its greater naturalness and
simplicity, it is still too soon to render a final decision.
Moreover, many of the results obtained in the course of
constructing the new inflationary universe scenario will prove
useful, even if the scenario itself is to be abandoned.  We
therefore begin our exposition with a description of the various
versions of the new inflationary universe scenario, and in the
next chapter we turn to a description of the chaotic inflation
scenario.  Our description of the former would be incomplete,
however, if we did not say a few words about the old inflationary
\index{Guth scenario}%
universe scenario proposed in the important paper by Guth
[\cite{53}].

As stated in Chapter \ref{c1}, the old scenario was based on the
study of phase transitions from a strongly supercooled unstable
phase $\varphi=0$  in grand unified theories.  The theory of such
phase transitions had been worked out long before Guth's effort
(see Chapter \ref{c5}), but nobody had attempted to use that
theory to resolve such cosmological problems as the flatness of
the universe or the horizon problem.

Guth drew attention to the fact that upon strong supercooling,
\index{Guth scenario}%
the energy density of relativistic particles, being proportional
to ${\rm T}^4$, becomes negligible in comparison with the vacuum
energy ${\rm V}(\varphi)$ in the vacuum state $\varphi=0$.  This
then means that in the limit of extreme supercooling, the energy
density $\rho$ of an expanding (and cooling) universe tends to
${\rm V}(0)$ and ceases to depend on time.  At large $t$, then,
according to (\ref{1.3.7}), the universe expands exponentially,
\be
\label{8.1.1}
a(t)\sim e^{{\rm H}\,t}\ ,
\ee
where the Hubble constant at that time is
\be
\label{8.1.2}
{\rm H}=\sqrt{\frac{8\,\pi\,{\rm V}(0)}{3\,\m}}\ .
\ee
If all of the energy is rapidly transformed into heat at the time
of a phase transition to the absolute minimum of ${\rm V}(\varphi)$,
the universe will be reheated to a temperature
${\rm T}_{\rm R}\sim[{\rm V}(0)]^{1/4}$ after the transition, regardless of
how long the previous expansion went on (this circumstance was
exploited earlier by Chibisov and the present author to construct
a model of the universe which could initially be cold, but would
ultimately be reheated by a strongly exoergic phase transition;
this model has been reviewed in Refs. [\cite{24}, \cite{105}]).

Since the temperature ${\rm T}_{\rm R}$ to which the universe is reheated
after the phase transition does not depend on the duration of the
exponential expansion stage in the supercooled state, the only
quantity that depends on the length of that stage is the scale
factor $a(t)$, which grows exponentially at that time.  But as we
have already remarked, the universe becomes flatter and flatter
during exponential expansion (inflation).  This is an especially
clear-cut effect when one considers why the total entropy of the
universe is so high, ${\rm S}\ga10^{87}$ (as noted in Chapter
\ref{c1}, this problem is closely related to the flatness
problem).

Prior to the phase transition, the total entropy of the universe
could be fairly low.  But afterwards, it increases markedly, with
$$
{\rm S}\ga a^3\,{\rm T}_{\rm R}^3\sim a^3\,[{\rm V}(0)]^{3/4}\ ,
$$
where $a^3$ can be exponentially large.  For example, let the
exponential expansion begin in a closed universe at a time when
its radius is $a_0=c_1\,\m^{-1}$, and the vacuum energy is
${\rm V}(0)=c_2\,\m^4$, where $c_1$ and $c_2$ are certain constants.
In realistic theories, $c_1$ lies between 1 and $10^{10}$, and
$c_2$ is of order $10^{-10}$; we will soon see that the quantity
of interest depends very weakly on $c_1$ and $c_2$.  Following
exponential expansion lasting for a period , the total entropy of
the universe becomes
\be
\label{8.1.3}
{\rm S}\sim
a_0^3\,e^{3\,{\rm H}\,\Delta t}\,{\rm T}_{\rm R}^3\sim c_1^3\,c_2^{3/4}\,
e^{3\,{\rm H}\,\Delta t}\ ,
\ee
whereupon S exceeds $10^{87}$ if
\be
\label{8.1.4}
\Delta t\ga  {\rm H}^{-1}\,(67-\ln c_1\,c_2^{1/4})\ .
\ee
Under typical conditions, the absolute value of
$\ln c_1\,c_2^{1/4}$ will be 10 at most.  The implication is that in
order to solve the flatness problem, it is necessary that the
universe be in a supercooled state $\varphi=0$ for a period
\be
\label{8.1.5}
\Delta t\ga70\,{\rm H}^{-1}=70\,\m\,
\sqrt{\frac{3}{8\,\pi\,{\rm V}(0)}}\ .
\ee

It must be noted here that if $\Delta t$ is much greater than
$70\,{\rm H}^{-1}$ (as will happen in any realistic version of
the inflationary universe scenario), then after inflation and
reheating, the universe will be almost perfectly flat, with
$\displaystyle \Omega=\frac{\rho}{\rho_c}=1$.  Allowing for
moderate local variations of $\rho$ on the scale of the observable
part of the universe, this is one of the most important
observational predictions of the inflationary universe scenario.

It can readily be shown that the condition $(a\,{\rm T})^3\ga10^{87}$
\index{``Radius'' of universe}%
\index{Universe!``radius'' of}%
means that the ``radius'' of the universe $a\sim c_1\m^{-1}$
after expansion up through the present epoch will exceed the size
of the observable part of the universe, $l\sim10^{28}$ cm (see
the preceding chapter).  But what this means is that in a time
only slightly greater (by ${\rm H}^{-1}\,\ln c_1$) than $70\,{\rm H}^{-1}$,
any region of space of size $\Delta l\sim\m^{-1}$ would
have inflated so much that by the present epoch it would be
larger than the observable part of the universe.

If we then bear in mind that we are considering processes taking
place in the post-Planckian epoch ($\rho<\m^4$, ${\rm T}<\m$,
$t>\m^{-1}$), it becomes clear that a region $\Delta
l\sim\m^{-1}$ in size at the onset of exponential expansion must
necessarily be causally connected.  Thus, in this scenario, the
entire observable part of the universe results from the inflation
of a single causally connected region, and the horizon problem is
thereby solved.

The primordial monopole problem could in principle also be solved
\index{Monopoles!primordial}%
\index{Primordial monopole problem}%
within the framework of the proposed scenario.  Primordial
monopoles are produced only at the points of collisions of
several different bubbles of the field $\varphi$ that are formed
during the phase transition.  If the phase transition is
significantly delayed by supercooling, then the bubbles will
become quite large by the time they begin to fill the entire
universe, and the density of the monopoles produced in the
process will be extremely low.

Unfortunately, however, as noted by Guth himself, the scenario
\index{Guth scenario}%
that he had proposed led to a number of undesirable consequences
with regard to the properties of the universe after the phase
transition.  Specifically, within the bubbles of the new phase,
the field $\varphi$ rapidly approached the equilibrium field
$\varphi_0$ corresponding to the absolute minimum of ${\rm V}(\varphi)$,
and all the energy of the unstable vacuum with
$\varphi=0$ within the bubble was transformed into kinetic energy
of the walls, which moved away from the center of the bubble at
close to the speed of light.  Reheating of the universe after the
phase transition would have to result from collisions of the
bubble walls, but due to the large size of the bubbles in this
scenario, the universe after collisions between bubble walls
would become highly inhomogeneous and anisotropic, a result
flatly inconsistent with the observational data.

Despite all the problems encountered by the first version of the
inflationary universe scenario, it engendered a great deal of
\index{Guth scenario}%
interest, and in the year following the publication of Guth's
work this scenario was diligently studied and discussed by many
workers in the field.  These investigations culminated in the
papers by Hawking, Moss, and Stewart [\cite{112}] and Guth and
Weinberg [\cite{113}], where it was stated that the defects
inherent in this scenario could not be eliminated.  Fortunately,
the new inflationary universe scenario had been suggested by then
[\cite{54}, \cite{55}];  it was not only free of some of the
\index{Guth scenario}%
shortcomings of the Guth scenario, but also held out the
possibility of solving a number of other cosmological problems
enumerated in Section \ref{s1.5}.
\index{Inflationary universe scenario!old|)}%

\section[Coleman--Weinberg SU(5) theory]%
{\label{s8.2}The
\index{Coleman--Weinberg theory|(}%
\index{SU(5) theory!Coleman--Weinberg|(}%
\index{Inflationary universe scenario!new!initial simplified version|(}%
Coleman--Weinberg SU(5) theory and the new\protect\\
inflationary universe scenario (initial simplified version)}

The first version of the new inflationary universe scenario was
based on the study of the phase transition with the symmetry breaking
$\mbox{SU}(5) \rightarrow \mbox{SU}(3)\times \mbox{SU}(2) \times {\rm U}(1)$
in the SU(5)-symmetric Coleman--Weinberg theory (\ref{2.2.16}).
The theory of this phase transition is very complicated.  We
therefore start by giving somewhat of a simplified description of
this phase transition, so as to elucidate the general idea behind
the new scenario.

First of all, we examine how the effective potential in this theory
behaves with respect to the symmetry breaking $\mbox{SU}(5) \rightarrow
\mbox{SU}(3) \times \mbox{SU}(2) \times {\rm U}(1)$
(\ref{2.2.16}) at a finite temperature.

As we said in Chapter \ref{c3}, symmetry is restored in gauge
theories, as a rule, at high enough temperatures.  It can be
shown in the present case that when ${\rm T}\gg{\rm M}_{\rm X}$, the
function ${\rm V}(\varphi,{\rm T})$ in the Coleman--Weinberg theory
becomes
\be
\label{8.2.1}
{\rm V}(\varphi,{\rm T})=\frac{5}{8}\,g^2\,{\rm T}^2\,\varphi^2+
\frac{25\,g^4\,\varphi^4}{128\,\pi^2}\,
\left(\ln\frac{\varphi}{\varphi_0}-\frac{1}{4}\right)+
\frac{9\,{\rm M}_{\rm X}^4}{32\,\pi^2}+c\,{\rm T}^4\ .
\ee
where $c$ is some constant of order 10.  An analysis of this
expression shows that at high enough temperature T, the only
minimum of ${\rm V}(\varphi,{\rm T})$ is the one at $\varphi=0$;  that
is, symmetry is restored.  When ${\rm T}\ll{\rm M}_{\rm X}\sim10^{14}$
GeV, all high-temperature corrections to ${\rm V}(\varphi)$ at
$\varphi\sim\varphi_0$ vanish.  However, the masses of all
particles in the Coleman--Weinberg theory tend to zero as
$\varphi\rightarrow0$, so in the neighborhood of the point
$\varphi=0$, Eq. (\ref{8.2.1}) for ${\rm V}(\varphi,{\rm T})$ holds
for ${\rm T}\ll10^{14}$ GeV as well.  This means that the point
$\varphi=0$ remains a local minimum
of the potential ${\rm V}(\varphi,{\rm T})$
at any temperature T, regardless of the fact that
the minimum at $\varphi\approx\varphi_0$ is much deeper when
${\rm T}\ll{\rm M}_{\rm X}$ (Fig. 8.1).

\begin{figure}[t]\label{f34}
\centering \leavevmode\epsfysize=5.5cm \epsfbox{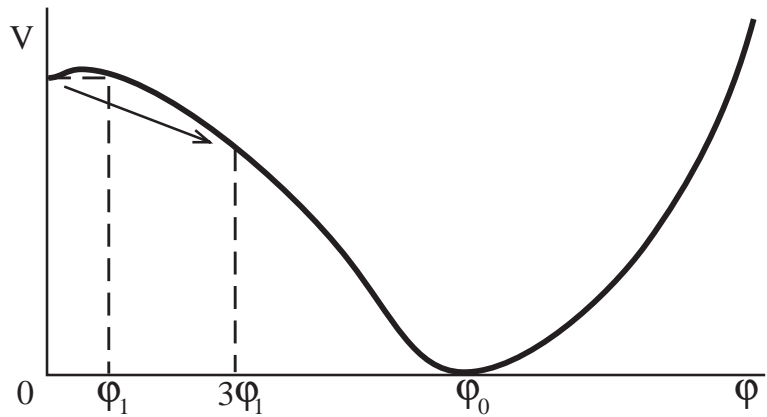}
\caption{Effective potential in the Coleman--Weinberg theory at
finite temperature.  Tunneling proceeds via formation of bubbles
of the field $\varphi\la3\,\varphi_1$, where ${\rm V}(\varphi_1,
{\rm T}) = {\rm V}(0, {\rm T})$.}
\end{figure}
In an expanding universe, a phase transition from the local
minimum at $\varphi=\varphi_0$ to a global minimum at
$\varphi=\varphi_0$ takes place when the typical time required
for the multiple production of bubbles with $\varphi\neq0$
becomes less than the age of the universe $t$.  Study of this
question has led many researchers to conclude that the phase
transition in the Coleman--Weinberg theory is a long, drawn-out
affair that takes place only when the temperature T of the
universe has fallen to approximately ${\rm T}_c\sim10^6$ GeV (this is
not an entirely correct statement, but for simplicity we shall
temporarily assume that it is, and return to this point in
Section \ref{s8.3}).  It is clear, however, that at such a low
temperature, the barrier separating the minimum at $\varphi=0$
from the minimum at $\varphi=\varphi_0$ will be located at
$\varphi\ll\varphi_0$ (see Fig. 8.1), and the bubble
formation process will be governed solely by the shape of
$V(\varphi, {\rm T})$ near $\varphi=0$, rather than by the value
of $\varphi_0$.  As a result, the field $\varphi$ within the
bubbles of the new phase formed in this way is at first very
small,
\be
\label{8.2.2}
\varphi\la3\,\varphi_1\approx
\frac{12\,\pi\,{\rm T}_c}{\displaystyle
g\,\sqrt{5\,\ln\frac{{\rm M}_{\rm X}}{{\rm T}_c}}}
\ll\varphi_0\ ,
\ee
where the field $\varphi_1$ is determined by the condition
${\rm V}(0,{\rm T})={\rm V}(\varphi_1,{\rm T})$  (see Fig. 8.1).
With this value of the field, the curvature of the effective
potential is relatively small.
\be
\label{8.2.3}
|m^2|=\left|\frac{d^2{\rm V}}{d\varphi^2}\right|\la
75\,g^2\,{\rm T}_c^2\sim25\,{\rm T}_c^2\ .
\ee
The field $\varphi$ within the bubble will clearly grow to its
equilibrium value $\varphi\sim\varphi_0$ in a time
$\Delta t\ga|m^{-1}|\sim0.2\,{\rm T}_c^{-1}$.  For most of this time,
the field $\varphi$ will remain much smaller than $\varphi_0$.
This means that over a period of order $0.2\,{\rm T}_c^{-1}$, the
vacuum energy of ${\rm V}(\varphi,{\rm T})$ will remain almost exactly
equal to ${\rm V}(0)$, and consequently the part of the universe
inside the bubble will continue to expand exponentially, just as
at the beginning of the phase transition.  Here we have the
fundamental difference between the new inflationary universe
\index{Guth scenario}%
scenario and the scenario of Guth, in which it is assumed that
exponential expansion ceases at the moment that bubbles are
formed.

When $\varphi\ll\varphi_0$ and ${\rm M}_{\rm X}\sim5\cdot 10^{14}$ GeV,
the Hubble constant H (is given by)
\be
\label{8.2.4}
{\rm H}=\sqrt{\frac{8\,\pi}{3\,\m^2}\,{\rm V}(0)}=
\frac{{\rm M}_{\rm X}^2}{2\,\m}\,\sqrt{\frac{3}{\pi}}\approx
10^{10}\;\mbox{GeV}\ .
\ee
In a time $\Delta t\sim0.2\,{\rm T}_c^{-1}$,
the universe expands by a factor $e^{{\rm H}\,\Delta t}$,
where
\be
\label{8.2.5}
e^{{\rm H}\,\Delta t}\sim e^{0.2\,{\rm H}\,{\rm T}_c^{-1}}
\sim e^{2000}\sim10^{800}\ .
\ee
To order of magnitude, the typical\index{Bubble sizes}
size of a bubble at the
instant it is formed is ${\rm T}_c^{-1}\sim10^{-20}$ cm.  After
expansion, this size becomes $\sim10^{800}$ cm, which is
enormously greater than the size of the observable part of the
universe, $l\sim10^{-28}$ cm.  Thus, within the scope of this
scenario, the entire observable part of the universe should lie
{\it within a single bubble}.  We therefore see no
inhomogeneities that might arise from bubble wall collisions.

As in the Guth scenario, exponential  expansion by a factor of
\index{Guth scenario}%
more than $e^{70}$ (\ref{8.2.5}) enables one to resolve the
horizon and flatness problems.  But more than that, it makes it
possible to explain the large-scale homogeneity and isotropy of
the universe (see Chapter \ref{c7}).

Since bubble sizes exceed the dimensions of the observable part of the
universe, and since monopoles and domain walls are only produced near bubble
walls, there should be not a single monopole or domain wall in the observable
part of the universe, which removes the corresponding problems discussed in
Section \ref{s1.5}.

{\looseness=2
Note that the curvature of the effective potential (\ref{8.2.1})
grows rapid\-ly with increasing field $\varphi$.  The slow-growth
stage of the field $\varphi$, which is accompanied by exponential
expansion of the universe, is therefore replaced by a stage with
extremely rapid attenuation of the field $\varphi$ to its
equilibrium value $\varphi=\varphi_0$, where it oscillates about
the minimum of the effective potential.  In the model in
question, the oscillation frequency is equal to the mass of the
Higgs field $\varphi$ when $\varphi=\varphi_0$,
$m=\sqrt{{\rm V}''(\varphi_0)}\sim10^{14}$ GeV.  The typical period of
oscillation $\sim m^{-1}$ is evidently many orders of magnitude
less than the characteristic expansion time of the universe ${\rm H}^{-1}$.
In studying oscillations of the field $\varphi$ near
the point $\varphi_0$, one can therefore neglect the expansion of
the universe.  This means that at the stage we are considering,
all of the potential energy ${\rm V}(0)$ is transformed into the
energy of the oscillating scalar field.  The oscillating
classical field $\varphi$ produces Higgs bosons and vector
bosons, which quickly decay.  In the end, all of the energy of
the oscillating field $\varphi$ is transformed into the energy of
relativistic particles, and the universe is reheated to a
temperature [\cite{123}, \cite{124}]
$$
{\rm T}_{\rm R}\sim[{\rm V}(0)]^{1/4}\sim10^{14}\;\mbox{GeV}\ .
$$
The mechanism for reheating the universe in the new scenario is
\index{Guth scenario}%
thus quite different from the corresponding mechanism in the Guth
scenario.

}

{\looseness=1
The baryon asymmetry of the universe is produced when scalar and
vector mesons decay during the reheating of the universe
[\cite{36}--\cite{38}].  Because of the fact that processes
taking place at that time are far from equilibrium, however, the
baryon asymmetry is produced much more efficiently in this model
than in the standard hot universe theory [\cite{123}].

}

{\looseness=1
We see, then, that the fundamental idea behind the new
inflationary universe scenario is quite simple:  it requires that
symmetry breaking due to growth of the field $\varphi$ proceed
fairly slowly at first, giving the universe a chance to inflate
by a large factor, and that in the later stages of the process,
the rate of growth and oscillation frequency of the field
$\varphi$ near the minimum of ${\rm V}(\varphi)$ be large enough
to ensure that the universe is reheated efficiently after the
phase transition.  This idea has been used both in a refined
version of the new scenario, which we discuss next, and in all
subsequent variants of the inflationary universe
scenario.
\index{Coleman--Weinberg theory|)}%
\index{Inflationary universe scenario!new!initial simplified version|)}%
\index{SU(5) theory!Coleman--Weinberg|)}%

}

\section[Refinement of the new scenario]%
{\label{s8.3}Refinement of the
\index{Inflationary universe scenario!new!refinement of|(}%
new inflationary universe scenario}

The description of the new inflationary universe scenario in the
previous section was oversimplified, its main drawback being our
neglect of the effects of exponential expansion of the universe
on the kinetics of a phase transition.  When ${\rm T}\gg{\rm
H}\sim10^{10}$ GeV, such a simplification is completely
admissible, but according to the discussion in Section
\ref{s8.2}, the phase transition can only begin when ${\rm T}_c\ll{\rm H}$.
In that event, high-temperature effects exert practically
no influence on the kinetics of the phase transition.  Indeed,
the typical time over which bubbles might be formed at a
temperature ${\rm T}_c$ must certainly be greater than
$$
m^{-1}(\varphi=0,{\rm T}={\rm T}_c)\sim(g\,{\rm T}_c)^{-1}\gg{\rm H}^{-1}\ .
$$
But in that much time, the universe expands by a factor of
approximately $e^{{\rm H}/g{\rm T}_c}$, and the temperature falls from
${\rm T}={\rm T}_c$ practically to zero.  Thus, the role of
high-temperature effects is just to place the field $\varphi$ at
the point $\varphi=0$, and one can then neglect all
high-temperature effects in describing the formation of bubbles
of the field $\varphi$ and the process by which $\varphi$ rolls
down to $\varphi_0$.  It is necessary, however, to take account
of effects related to the rapid expansion of the universe, since
at that time ${\rm H}\gg{\rm T}$.

The resulting refinement of the scenario takes place in several
steps.

1) In studying the evolution of the field $\varphi$ in an
inflationary universe, one must make allowance for the fact that
the equation of motion of the field is modified, and takes the
form
\be
\label{8.3.1}
\ddot\varphi+3\,{\rm H}\,\dot\varphi
-\frac{1}{a^2}\,\nabla^2\varphi=-\frac{d{\rm V}}{d\varphi}\ .
\ee
If the effective potential is not too steep, the $\ddot\varphi$
term in (\ref{8.3.1}) can be discarded, so that the homogeneous
field $\varphi$ satisfies the equation
\be
\label{8.3.2}
\dot\varphi=-\frac{1}{3\,{\rm H}}\,\frac{d{\rm V}}{d\varphi}\ .
\ee
In particular, (\ref{8.3.2}) implies that with ${\rm H}=\mbox{const}\gg m$
in a theory with
$\displaystyle {\rm V}={\rm V}(0)+\frac{m^2}{2}\,\varphi^2$,
\be
\label{8.3.3}
\varphi\sim\varphi_0\,\exp\left(-\frac{m^2}{3\,{\rm H}}\,t\right)\ ,
\ee
and in a theory with
$\displaystyle {\rm V}={\rm V}(0)-\frac{m^2}{2}\,\varphi^2$,
\be
\label{8.3.4}
\varphi\sim\varphi_0\,\exp\left(+\frac{m^2}{3\,{\rm H}}\,t\right)\ .
\ee
This means, in particular, that the curvature of the
effective\index{Curvature of effective potential}%
\index{Effective potential!curvature of}
potential at $\varphi=0$ need not necessarily be zero.  To solve
the flatness and horizon problems, it is sufficient that the
field $\varphi$ (as well as the quantity ${\rm V}(\varphi)$) vary
slowly over a time span $\Delta t\ga70\,{\rm H}^{-1}$.  In
conjunction with (\ref{8.2.4}), this condition leads to the
constraint
\be
\label{8.3.5}
|m^2|\la\frac{{\rm H}^2}{20}\ .
\ee

It will also be useful to investigate the evolution of a classical
field in the theory described by
\be
\label{8.3.6}
\vf={\rm V}(0)-\frac{\lambda}{4}
\,\varphi^4\ .
\ee
In that case, it follows from (\ref{8.3.2}) that
\be
\label{8.3.7}
\frac{1}{\varphi_0^2}-\frac{1}{\varphi^2}=\frac{2\,\lambda}{3\,{\rm H}}\,
(t-t_0)\ ,
\ee
where $\varphi_0$ is the initial value of the field $\varphi$.
This means that the field becomes infinitely large in a finite time
\be
\label{8.3.8}
t-t_0=\frac{3\,{\rm H}}{2\,\lambda\,\varphi_0^2}\ .
\ee
If $\lambda\,\varphi_0^2\ll{\rm H}^2$, then $t-t_0\gg{\rm H}^{-1}$,
and $\varphi$ will spend most of this time span in its
slow downhill roll.  It is only at the end of the interval
(\ref{8.3.8}) that the field quickly rolls downward, with
$\varphi \rightarrow \infty$, in a time $\Delta t\sim{\rm H}^{-1}$.
For $\lambda\,\varphi_0^2\ll{\rm H}^2$, therefore, the
\index{Inflationary stage, duration of}%
duration of the inflationary stage in the theory (\ref{8.3.6}) as
the field $\varphi$ rolls down from $\varphi=\varphi_0$ is
$\displaystyle \frac{3\,{\rm H}}{2\,\lambda\,\varphi_0^2}$
(\ref{8.3.8}) (to within $\Delta t\sim{\rm H}^{-1}$).  This
result will shortly prove useful.

2) Corrections to the expression (\ref{8.2.1}) for ${\rm V}(\varphi)$
arise in  de Sitter space.  If we limit attention,
as before, to the contribution to ${\rm V}(\varphi)$ from heavy
vector particles (see Chapter \ref{c2}), then for small $\varphi$
($e\,\varphi\ll{\rm H}$), ${\rm V}(\varphi)$ takes the form
[\cite{264}, \cite{265}]
\be
\label{8.3.9}
{\rm V}(\varphi,{\rm R})=\frac{\mu_1^2}{2}\,{\rm R}
+\frac{e^2\,{\rm R}}{64\,\pi^2}\,\varphi^2\,\ln\frac{{\rm R}}{\mu_2^2}
+\frac{3\,e^4\,\varphi^4}{64\,\pi^2}\,\ln\frac{{\rm R}}{\mu_3^2}
+{\rm V}(0,{\rm R})\ ,
\ee
where R is the\index{Curvature scalar}
curvature scalar (${\rm R}=12\,{\rm H}^2$), and
the $\mu_i$ are some normalization factors with dimensions of
mass, whose magnitude is determined by the normalization
conditions imposed on ${\rm V}(\varphi, {\rm R})$.  When
$\vf\ll\m^4$, the corresponding corrections to the effective
potential ${\rm V}(\varphi)$ itself are extremely small, although
they can induce significant corrections to the quantity
$\displaystyle m^2=\frac{d^2{\rm V}}{d\varphi^2}\biggr|_{\varphi=0}$
that are of order $e^2\,{\rm H}^2$, and these can prevent
(\ref{8.3.5}) from being satisfied.  Fortunately, there does
exist a choice of normalization conditions (i.e., a redefinition
of the Coleman--Weinberg theory in curved space) for which this
does not happen, and for which $m^2$ remains equal to zero.  We
shall not pursue this problem any further here, referring the
reader to Ref. [\cite{265}] for a discussion of the
renormalization of ${\rm V}(\varphi, {\rm R})$ for the
Coleman--Weinberg theory in de Sitter space.

3) The most important refinement of the scenario has to do with
the first stage of growth of the field $\varphi$.  As stated
earlier, some time $\tau\sim{\rm O}({\rm H}^{-1})$ after the
temperature of the universe has dropped to ${\rm T}\sim{\rm H}$, the
temperature and effective mass of the field $\varphi$ at the
point $\varphi=0$ become exponentially small.  At that time, the
effective potential ${\rm V}(\varphi)$ (\ref{8.2.1}) in the
neighborhood of interest around $\varphi=0$ (with
$\displaystyle {\rm H}\la\varphi\la\frac{{\rm H}}{\sqrt{\lambda}}$)
can be approximated by (\ref{8.3.6}), where
\be
\label{8.3.10}
\lambda\approx\frac{25\,g^4}{32\,\pi^2}\,
\left(\ln\frac{{\rm H}}{\varphi_0}-\frac{1}{4}\right),\qquad
{\rm V}(0)=\frac{9\,{\rm M}_{\rm X}^4}{32\,\pi^2}\ .
\ee
According to Eq. (\ref{8.3.8}), the classical motion of the field
$\varphi$, starting out from the point $\varphi_0=0$, would go on
for an infinitely long time.  As we noted in Section \ref{s7.3},
however, quantum fluctuations of the field $\varphi$ in the
inflationary universe engender long-wave fluctuations in the
field, and on a scale $l\sim{\rm H}^{-1}$, these look like a
homogeneous classical field.  Making use of (\ref{7.3.12}), the
rms value of this field (averaged over many independent regions
of size $l\ga{\rm H}^{-1}$), is
\be
\label{8.3.11}
\varphi\sim\frac{{\rm H}}{2\,\pi}\,\sqrt{{\rm H}\,(t-t_0)}\ .
\ee
In the case at hand, $t_0$ is the time at which the effective
mass squared of the field $\varphi$ at $\varphi=0$ becomes much
less than ${\rm H}^2$.

Long-wave fluctuations of the field $\varphi$ can play the role
of the initial nonzero field $\varphi$ in Eq. (\ref{8.3.7}).
Here, however, we must voice an important reservation.  In
different regions of the universe, the fluctuating field
$\varphi$ will take on different values;  in particular, there
will always be regions in which $\varphi$ does not decrease at
\index{Self-reproducing inflationary universe}%
all, giving rise to a self-reproducing inflationary universe
[\cite{266}, \cite{267}, \cite{204}] analogous to that of the
chaotic inflation scenario [\cite{57}, \cite{132}, \cite{133}]
(see Section \ref{s1.8}). Further on, we shall discuss the
average behavior of the fluctuating field $\varphi$
(\ref{8.3.11}).

During the first stage of the process, fluctuating (diffusive)
growth of the field $\varphi$ takes place more rapidly than the
classical rolling\index{Classical rolling}:
\be
\label{8.3.12}
\dot\varphi\sim\frac{{\rm H}^2}{4\,\pi\,\sqrt{{\rm H}\,(t-t_0)}}\gg
\frac{\lambda\,\varphi^3}{3\,{\rm H}}\sim
\frac{\lambda\,{\rm H}^2\,[{\rm H}\,(t-t_0)]^{3/2}}{6\,\pi\,\sqrt{2\,\pi}}
\ .
\ee
This stage lasts for a time
\be
\label{8.3.13}
\Delta t=t-t_0\sim\frac{\sqrt{2}}{{\rm H}\,\sqrt{\lambda}}\ ,
\ee
during which the mean field $\varphi$ (\ref{8.3.11}) rises to
\be
\label{8.3.14}
\varphi_0\sim\frac{{\rm H}}{2\,\pi}\,\left(\frac{2}{\lambda}\right)^{1/4}\ .
\ee
To a good approximation, subsequent evolution of the field
$\varphi$ may be described by Eq. (\ref{8.3.7}), where we must
substitute $t_0+\Delta t$ for $t_0$.  The overall duration of the
rolling of the field $\varphi$ from $\varphi=\varphi_0$ to
$\varphi=\infty$ is
\be
\label{8.3.15}
t-(t_0+\Delta t)=\frac{3\,{\rm H}}{2\,\lambda\,\varphi_0^2}
=\frac{3\,\sqrt{2}\,\pi}{\sqrt{\lambda}\,{\rm H}}\ ,
\ee
and the total duration of inflation is given by
\be
\label{8.3.16}
t-t_0\sim\frac{4\,\sqrt{2}\,\pi}{\sqrt{\lambda}\,{\rm H}}\ .
\ee
During that time, the size of the universe grows by approximately
a factor of
\be
\label{8.3.17}
\exp({\rm H}\,(t-t_0))\sim
\exp\left(\frac{4\,\sqrt{2}\,\pi}{\sqrt{\lambda}}\right)\ .
\ee
The condition ${\rm H}\,(t-t_0)\ga70$ leads to the constraint
[\cite{265}, \cite{128}, \cite{134}, \cite{135}]
\be
\label{8.3.18}
\lambda\la\frac{1}{20}\ ,
\ee
which can also be satisfied, in principle, in the SU(5)
Coleman--Weinberg theory.

With minor modifications, the foregoing discussion also applies
to the case in which $m^2\equiv {\rm V}''(0)<0$, $|m^2|\ll{\rm H}^2$,
as well as to the case in which the effective potential
has a shallow local minimum at $\varphi=0$ --- that is, when
$0<m^2\ll{\rm H}^2$.

In the first of these two instances, the process whereby the
field rolls down from the point $\varphi=0$ is analogous to the
previous situation. In the second, diffusion of the field
$\varphi$ looks like tunneling, the theory of which was discussed
in an Section \ref{s7.4}.

Clearly, the details of the behavior of the scalar field
$\varphi$ as it undergoes a phase transition from the point
$\varphi=0$ to a minimum of ${\rm V}(\varphi)$ at
$\varphi=\varphi_0$ differ from the description given in the
preceding section.  Nevertheless, most of the qualitative
conclusions having to do with the existence of an inflationary
regime in the Coleman--Weinberg theory remain valid.

Unfortunately, however, the original version of the new
inflationary universe scenario, based on the theory
(\ref{8.2.1}), is not entirely realistic, the point being that
fluctuations of the scalar field $\varphi$ that are generated
during the inflationary stage give rise to large density
inhomogeneities by the time inflation has ended.  Specifically,
according to (\ref{7.5.22}), after the inflation, reheating, and
subsequent cooling of the universe, density inhomogeneities
\be
\label{8.3.19}
\frac{\delta\rho(\varphi)}{\rho}=\frac{48}{5}\,\sqrt{\frac{2\,\pi}{3}}\,
\frac{[\vf]^{3/2}}{\m^3\,{\rm V}'(\varphi)}
\ee
will be produced.  In this expression, $\varphi$ is the value of
the field at the time when the corresponding fluctuations
$\delta\varphi$ had a wavelength $l\sim k^{-1}\sim{\rm H}^{-1}$.
In the new inflationary universe scenario, $\vf\approx{\rm V}(0)$
at the time of inflation.  Let us estimate the present-day
wavelength of a perturbation whose wavelength was previously
$l\sim[{\rm H}(\varphi)]^{-1}$.  Equation (\ref{8.3.8}) tells us
that after the field becomes equal to $\varphi$, the universe
still has an inflation factor of
$\displaystyle \exp\left(\frac{3\,{\rm H}^2}{2\,\lambda\,\varphi^2}\right)$
to go.  Estimates analogous to those made in the previous chapter
then show that after inflation and the subsequent stage of hot
universe expansion, the wavelength $l\sim[{\rm H}(\varphi)]^{-1}$
typically increases to
\be
\label{8.3.20}
l\sim\exp\left(\frac{3\,{\rm H}^2}{2\,\lambda\,\varphi^2}\right)\;
\mbox{cm}\ .
\ee
From (\ref{8.3.19}) and (\ref{8.3.20}), we obtain\vspace{-2pt}
\be
\label{8.3.21}
\frac{\delta\rho}{\rho}\sim\frac{9}{5\,\pi}\,
\frac{{\rm H}^3}{\lambda\,\varphi^3}\sim
\frac{2\,\sqrt{6}}{5\,\pi}\,\sqrt{\lambda}\,\ln^{3/2}l\;[\mbox{cm}]\ ,
\vspace{-2pt}
\ee
just as in the chaotic inflation scenario (\ref{7.5.29}).
At the galactic scale $l_g\sim10^{22}$ cm,\vspace{-2pt}
\be
\label{8.3.22}
\frac{\delta\rho}{\rho}\sim110\,\sqrt{\lambda}\ .
\vspace{-2pt}
\ee
This means that $\displaystyle \frac{\delta\rho}{\rho}\sim10^{-5}$
when\vspace{-2pt}
\be
\label{8.3.23}
\lambda\sim10^{-14}\ ,
\vspace{-2pt}
\ee
again as in the chaotic inflation scenario;  see (\ref{7.5.38}).
In the original version of the new inflationary universe
scenario, the condition (\ref{8.3.23}) was not satisfied.  This
made it necessary to seek other more realistic models in which
the new inflationary universe scenario might be realizable, and
we now turn to a discussion of the models proposed.\vspace{-1pc}
\index{Inflationary universe scenario!new!refinement of|)}%

\section[Primordial inflation]%
{\label{s8.4}
Primordial inflation in
N  = 1 supergravity}
\index{Primordial inflation in ${\rm N} = 1$ supergravity|(}%
\index{Supergravity theory!${\rm N} = 1$!primordial inflation in|(}%

The main reason why the new inflationary universe scenario has
not been fully implemented in the Coleman--Weinberg SU(5) theory
is that the scalar field interacts with vector particles, and as
a result it acquires an effective coupling constant $\lambda\sim
g^4\gg10^{-14}$.  The conclusion to be drawn, then, is that the
\index{Inflation field}%
(inflaton) field $\varphi$ responsible for the inflation of the
universe must interact both with itself and with other fields
extremely weakly.  In particular, it must not interact with
vector fields,  or in other words it ought to be a singlet under
gauge transformations in grand unified theories.

A long list of requirements has been formulated which must be
satisfied in order for a theory to provide a feasible setting for
the new inflationary universe scenario [\cite{268}].
Specifically, the effective potential at small $\varphi$ must be
extremely flat (as can be seen from (\ref{8.3.5}) and
(\ref{8.3.23})), and near its minimum at $\varphi=\varphi_0$, it
must be steep enough to ensure efficient reheating of the
universe.  Next, after the main requirements for a theory have
been formulated, the search for a realistic elementary particle
theory of the desired type begins.  Since the step following the
construction of grand unified theories was the development of
phenomenological theories based on ${\rm N} = 1$ supergravity,
there has been a great deal of work attempting to describe
inflation within the scope of these theories (for example, see
[\cite{269}]--[\cite{271}]).

{\looseness=-2
In ${\rm N} = 1$ supergravity, the inflaton field $\varphi$
responsible for inflation of the universe is represented by the
scalar component $z$ of an additional singlet of the chiral
superfield $\Sigma$.  In the theories considered, the Lagrangian
for this field can be put into the form [\cite{272}]\vspace{-2pt}
\ba
\label{8.4.1}
{\rm L}&=&
{\rm G}_{zz^*}\,\partial_\mu z\,\partial^\mu z^*-{\rm V}(z,z^*)\ ,\\
\label{8.4.2}
{\rm V}(z,z^*)&=&e^{\rm G}\,({\rm G}_{z}\,{\rm G}_{zz^*}^{-1}\,
{\rm G}_{z^*}-3)\ ,\vspace{-2pt}
\ea
where G is an arbitrary real-valued function of $z$ and $z^*$,
${\rm G}_z$ is its derivative with respect to $z$, and ${\rm
G}_{zz^*}$ is its derivative with respect to $z$ and $z^*$.  In
the minimal versions of the theory, one imposes the constraint
${\rm G}_{zz^*}= 1/2$ on G so that the kinetic term in
(\ref{8.4.1}) takes on the standard (minimal) form
$\partial_\mu z\,\partial^\mu z^*$ (up to a factor $1/2$), while
the form chosen for G itself is\vspace{-2pt}
\be
\label{8.4.3}
{\rm G}(z,z^*)=\frac{z\,z^*}{2}+\ln|g(z)|^2\ ,
\vspace{-2pt}
\ee
where $g(z)$ is an arbitrary function of the field $z$, called
\index{Superpotential}%
the superpotential; all dimensional terms in (\ref{8.4.3}) are
expressed in units of $\displaystyle \frac{\m}{\sqrt{8\,\pi}}$.
The effective potential is then given by\vspace{-2pt}
\be
\label{8.4.4}
{\rm V}(z,z^*)=e^{z\,z^*/2}\,
\left(2\,\left|\frac{dg}{dz}+\frac{z^2}{2}\,g\right|^2-3\,|g^2|\right)\ .
\vspace{-2pt}
\ee
The function $g$ is subject to two constraints, namely ${\rm V}(z_0)=0$
and $g(z_0)\ll1$, where $z_0$ is the point at which
${\rm V}(z, z^*)$ has its minimum.  The first condition means
that the vacuum energy vanishes at the minimum of ${\rm V}(z, z^*)$,
and the second is required in order that the mass of the
gravitino $m_{3/2}$, which is proportional to $g(z_0)$, be much
lower than the other masses that occur in the theory.  This
requirement is necessary for the solution of the mass hierarchy
problem in the context of ${\rm N} = 1$ supergravity [\cite{15}].

}

{\looseness=-1
The superpotential $g(z)$ can be expressed as a product
$\mu^3\,f(z)$, where m is some parameter with dimensions of mass.
The potential ${\rm V}(z, z^*)$, and thus the effective coupling
constants for the $z$ and $z^*$ fields, are consequently
proportional to $\mu^6$.  The choice $\mu\sim10^{-2}$--$10^{-3}$,
which seems to be a fairly natural one, therefore results in the
appearance of extremely small effective coupling constants
$\lambda\sim10^{-12}$--$10^{-18}$, which is just what is needed
to obtain the desired amplitude
$\displaystyle \frac{\delta\rho}{\rho}\sim10^{-4}$--$10^{-5}$ if
inflation takes place in the theory (\ref{8.4.4}).  This variant
of the new inflationary universe scenario was called the
\index{Primordial inflation scenario}%
primordial inflation scenario by its authors [\cite{270}], since
it was expected to be played out on an energy scale far exceeding
that of the grand unified theories.  In fact, it has turned out
that the corresponding energy scales are virtually identical.

}
{\looseness=-1
The development of the primordial inflation scenario was party to
\index{Primordial inflation scenario}%
many interesting ideas and considerable ingenuity.
Unfortunately, however, no realistic versions of this scenario
(or indeed of any other versions of the new inflationary universe
scenario) have yet been suggested.  The principal reason for this
is that particles of the field $z$, interacting very weakly
(either gravitationally or through a coupling constant
$\mu^6\sim10^{-14}$) with one another and with other fields, were
not in a state of thermodynamic equilibrium in the early
universe.  Furthermore, even if they had been, the corresponding
corrections of the type $\lambda\,z\,z^*\,{\rm T}^2$ to ${\rm V}(z, z^*)$
are so small that they are incapable of changing the
initial value of the field $z$; that is, in most models of this
kind, they cannot raise the field $z$ to a maximum of the
potential ${\rm V}(z, z^*)$, as required for the onset of
inflation in this scenario [\cite{115}, \cite{116}] (this point
will be treated in more detail in Section \ref{s8.5}).
Meanwhile, as we shall show in Chapter \ref{c9}, the chaotic
inflation scenario can be implemented in ${\rm N}=1$ supergravity
[\cite{273}, \cite{274}].
\index{Primordial inflation in ${\rm N} = 1$ supergravity|)}%
\index{Supergravity theory!${\rm N} = 1$!primordial inflation in|)}%

}
\section{\label{s8.5}The
\index{Shafi--Vilenkin model|(}%
Shafi--Vilenkin model}

It was Shafi and Vilenkin who came closest to a consistent
implementation of the new inflationary universe scenario
[\cite{275}] (see also [\cite{276}]).  They returned to a
consideration of the SU(5)-symmetric theory of Coleman and
Weinberg, with symmetry breaking due to the Coleman1 Weinberg
mechanism occurring not in the field $\Phi$, which interacts with
vector bosons through a gauge coupling constant $g$, but in a
new, specially introduced field $\chi$, an SU(5) singlet, which
interacts very weakly with the superheavy $\Phi$ and
${\rm H}_5^{\vphantom{+}}$
Higgs fields.  The effective potential in this model is
\ba
\label{8.5.1}
{\rm V}&=&\frac{1}{4}\,a\,\tr(\Phi^2)^2+\frac{1}{2}\,b\,\tr\Phi^4
-\alpha\,({\rm H}_5^+\,{\rm H}_5^{\vphantom{+}})\,\tr\Phi^2+
\frac{\gamma}{4}\,({\rm H}_5^+\,{\rm H}_5^{\vphantom{+}})^2\nonumber \\
&-&\beta\,{\rm H}_5^+\,\Phi^2\,{\rm H}_5^{\vphantom{+}}+
\frac{\lambda_1}{4}\,\chi^4
-\frac{\lambda_2}{2}\,\chi^2\,\tr\Phi^2
+\frac{\lambda_3}{2}\,\chi^2\,{\rm H}_5^+\,{\rm H}_5^{\vphantom{+}}
\nonumber \\
&+&{\rm A}\,\chi^4\,\left(\ln\frac{\chi^2}{\chi_0^2}+{\rm C}\right)+
{\rm V}(0)\ .
\ea
where $a$, $b$, $\alpha$, and $\gamma$ are all proportional to
$g^2$;  C is some normalization constant; $0<\lambda_i\ll g^2$,
$\lambda_1\ll\lambda_2^2,\lambda_3^2$, and the magnitude of A is
determined by radiative corrections associated with the
interaction of the field $\chi$ with the fields $\Phi$,
${\rm H}_5^{\vphantom{+}}$,
and (indirectly) with the X and Y vector mesons.

In the present case, it is not an entirely trivial matter to
calculate A, and the procedure requires some explanation.
Spontaneous SU(5) symmetry breaking takes place when the nonzero
classical field $\chi$ emerges, thanks to the term
$\displaystyle -\frac{1}{2}\,\lambda_2\,\chi^2\tr\Phi^2$ in
(\ref{8.5.1}).  The symmetry breaks down to $\mbox{SU}(3) \times
\mbox{SU}(2) \times {\rm U}(1)$ by virtue of the emergence of the
field
$$
\Phi=\sqrt{\frac{2}{15}}\,\varphi\cdot
\mbox{diag}\left(1,1,1,-\frac{3}{2},-\frac{3}{2}\right)
$$
(see (\ref{1.1.19})), where
\be
\label{8.5.2}
\varphi^2=\frac{2\,\lambda_2}{\lambda_c}\,\chi^2
\ee
and $\displaystyle \lambda_c=a+\frac{7}{15}\,b$.  The time needed
for the field $\varphi$ to grow to the value (\ref{8.5.2}) is
then $\tau\sim(\sqrt{\lambda_2}\,\chi)^{-1}$, which is much less
than the typical time scale of variations in $\chi$ at the time
of inflation (see below).  The field $\varphi$ thus continuously
follows the behavior of the field $\chi$.  Consequently, not only
does a change in $\chi$ alter the masses of those particles with
which this field interacts directly (such as $\Phi$ and
${\rm H}_5^{\vphantom{+}}$),
but also those that interact with the field   $\varphi$
--- in particular, the X and Y vector mesons.  Here the behavior
of the ${\rm H}_5^{\vphantom{+}}$
boson masses is especially interesting.  The
first two components of ${\rm H}_5^{\vphantom{+}}$
\index{Higgs model field doublet}%
play the role of the Higgs
field doublet in $\mbox{SU}(2)\times{\rm U}(1)$ symmetry
breaking.  These should be quite light, with
$m_2\sim10^2\;\mbox{GeV}\ll m_3$, ${\rm M}_{\rm X}$, ${\rm
M}_{\rm Y}$,$\ldots\,$.  To lowest order, one may therefore put
$m_2=0$ at the minimum of ${\rm V}(\varphi, \chi)$.

The general expression for the doublet and triplet masses of the H
fields follows from (\ref{8.5.1}):
\ba
\label{8.5.3}
m_2^2&=&\lambda_3\,\chi^2-(\alpha+0.3\,\beta)\,\varphi^2\ ,\\
\label{8.5.4}
m_3^2&=&m_2^2+\frac{\beta}{6}\,\varphi^2\ .
\ea
Making use of (\ref{8.5.2}), one obtains
\be
\label{8.5.5}
\lambda_3=\frac{2\,\lambda_2}{\lambda_c}\,(\alpha+0.3\,\beta)\ .
\ee
This implies that not just at the minimum of ${\rm V}(\varphi,
\chi)$, but anywhere along a trajectory along which the field
$\chi$ varies,
\be
\label{8.5.6}
m_2^2=0,\qquad m_3^2=\frac{\beta}{6}\,\varphi^2\ .
\ee
The constant $\lambda_3$ is thus not independent, and the value
of $m_3^2$ is proportional to $\displaystyle \frac{\beta}{6}\,\varphi^2$,
rather than $\lambda_3\,\chi^2$.  A calculation of the radiative
corrections to ${\rm V}(\varphi, \chi)$ in the vicinity of a
trajectory down which the field $\chi$ is rolling, with
$\lambda_i$, $\beta\ll g^2$, finally gives [\cite{277}]
\be
\label{8.5.7}
{\rm A}=\frac{\lambda_2^2}{16\,\pi^2}\,
\left(1+\frac{25\,g^4}{16\,\lambda_c^2}+
\frac{14\,b^2}{9\,\lambda_c^2}\right)\ .
\ee
(This expression differs slightly from the one given in Ref.
[\cite{275}].) If one takes for simplicity $a\sim b\sim g^2$,
then Eq. (\ref{8.5.7}) yields
\be
\label{8.5.8}
{\rm A}\sim1.5\cdot 10^{-2}\,\lambda_2^2\ .
\ee

The effective potential ${\rm V}(\varphi, \chi)$
in the theory (\ref{8.5.1}) looks like
\be
\label{8.5.9}
{\rm V}=\frac{\lambda_c}{16}\,\varphi^4
-\frac{\lambda_2}{4}\,\varphi^2\,\chi^2+\frac{\lambda_1}{4}\,\chi^4+
{\rm A}\,\chi^4\,\left(\ln\frac{\chi}{{\rm M}}+{\rm C}\right)+{\rm V}(0)\ ,
\ee
where M and C are some normalization parameters.  To determine M,
C, and ${\rm V}(0)$, one should use Eq. (\ref{8.5.2}):
\be
\label{8.5.10}
{\rm V}=-\frac{\lambda_2^2}{4\,\lambda_c}\,\chi^4+
{\rm A}\,\chi^4\,\left(\ln\frac{\chi}{{\rm M}}+{\rm C}\right)+{\rm V}(0)\ .
\ee
With an appropriate choice of the normalization constant C, the
effective potential (\ref{8.5.10}) can be put in the standard
form
\be
\label{8.5.11}
{\rm V}(\chi)={\rm A}\,\chi^4\,
\left(\ln\frac{\chi}{\chi_0}-\frac{1}{4}\right)
+\frac{{\rm A}\,\chi_0^4}{4}\ ,
\ee
where $\chi_0$ gives the position of the minimum of ${\rm
V}(\chi)$.  The corresponding minimum in $\varphi$ is located at
$\displaystyle
\varphi_0=\sqrt{\frac{2\,\lambda_2}{\lambda_c}}\,\chi_0$ (see
(\ref{8.5.2})), and the mass of the X boson
\index{X bosons}\index{Bosons!X}is equal to
$$
{\rm M}_{\rm X}=\sqrt{\frac{5}{3}}\,\frac{g\,\varphi_0}{2}
\sim10^{14}\;\mbox{GeV}\ .
$$
Hence,
$$
\chi_0\sim\frac{{\rm M}_{\rm X}}{g}\,
\sqrt{\frac{6\,\lambda_c}{5\,\lambda_2}}\ ,
$$
and
$$
{\rm V}(0)=\frac{{\rm A}}{4}\,\chi_0^4\approx{\rm M}_{\rm X}^4\ .
$$

The high-temperature correction to the effective potential
(\ref{8.5.11}) is given by
\be
\label{8.5.12}
\Delta{\rm V}(\chi,{\rm T})=
\left(\frac{5}{12}\,\lambda_3-\lambda_2\right)\,{\rm T}^2\,\chi^2\ ,
\ee
which for $\displaystyle \lambda_3>\frac{12}{5}\,\lambda_2$ could
lead to the restoration of symmetry, $\chi \rightarrow 0$ (see
the next section, however).  Upon cooling, the process of
inflation would begin, which would be very similar to the process
described in Section \ref{s8.3}.

To determine the numerical value of the parameter A, one should first
determine the value of
$\displaystyle \ln\left(\frac{\chi_0}{\chi}\right)$
at which the observable structure of the universe is actually
formed --- this occurs when a time $t\sim60\,{\rm H}^{-1}$
remains prior to the end of inflation.  According to
(\ref{8.3.8}), the magnitude of the field $\chi$ at that point is
given by
\be
\label{8.5.13}
\chi^2\sim\frac{{\rm H}^2}{40\,\lambda(\chi)}
\ee
where for
$\displaystyle \ln\left(\frac{\chi}{\chi_0}\right)\gg1$,
the effective coupling constant $\lambda(\chi)$  and the Hubble
constant H are
\ba
\label{8.5.14}
\lambda(\chi)&\approx&4\,{\rm A}\,\ln\left(\frac{\chi_0}{\chi}\right)\sim
10^{-14}\ ,\nonumber \\
{\rm H}&=&\sqrt{\frac{8\,\pi\,{\rm V}(0)}{3\,\m^2}}
\sim3\,\frac{{\rm M}_{\rm X}^2}{\m}\sim3\cdot 10^9\;\mbox{GeV}\ ,
\ea
(see (\ref{8.3.23})), whereupon
\be
\label{8.5.15}
\chi\sim5\cdot 10^{15}\;\mbox{GeV}\ .
\ee
Inserting these values,
$\displaystyle \ln\left(\frac{\chi_0}{\chi}\right)$ is found to
be of order 3 (see below).  From (\ref{8.5.8}) and
(\ref{8.5.14}), it follows that
\ba
\label{8.5.16}
\lambda_2&\sim&3\cdot 10^{-6}\ ,\\
\label{8.5.17}
\chi_0&\sim&\frac{{\rm M}_{\rm X}}{g}\,
\sqrt{\frac{6\,\lambda_c}{5\,\lambda_2}}
\sim10^{17}\;\mbox{GeV}
\ea
According to (\ref{8.5.11}), the value of $\lambda_3$ should be
greater than $\displaystyle \frac{12\,\lambda_2}{5}$.  But
$\lambda_3$ cannot be {\it much} greater than $\lambda_2$, since
it can be shown that if it were, $\varphi$ would not vanish at
high temperature [\cite{275}].  In accordance with [\cite{275}],
we shall therefore assume that $\lambda_3\sim3\cdot 10^{-6}$,
like $\lambda_2$.

We have from (\ref{8.3.17}) that in the present model, a typical
\index{Inflation factor of universe}%
\index{Universe!inflation factor of}%
inflation factor is of order
\be
\label{8.5.18}
\exp\left(\frac{4\,\sqrt{2}\,\pi}{\sqrt{\lambda(\chi)}}\right)\sim
10^{10^8}\ ,
\ee
which is more than adequate.

Unfortunately, both reheating and the baryon asymmetry production
in the post-inflation universe are rather inefficient in this
model.  After inflation, the field $\chi$ oscillates about the
minimum of ${\rm V}(\chi)$ at $\chi=\chi_0$ at a very low
frequency,
\be
\label{8.5.19}
m_\chi=22\,\sqrt{{\rm A}}\,\chi_0\sim10^{11}\;\mbox{GeV}\ .
\ee
The principal decay mode of the field $\chi$ is
$\chi\,\chi\rightarrow{\rm H}^+_3\,{\rm H}_3^{\vphantom{+}}$,
where ${\rm H}_3^{\vphantom{+}}$
\index{Higgs boson! heavy, triplet of}%
is the triplet of heavy Higgs bosons.  Subsequent decay of the
${\rm H}_3^{\vphantom{+}}$
bosons gives rise to the baryon asymmetry of the
universe.  The corresponding part of the effective Lagrangian
responsible for decay of the field $\chi$ is of the form
$\displaystyle
\frac{\beta\,\lambda_2}{6\,\lambda_c}\,\chi^2\,
{\rm H}^+_3\,{\rm H}_3^{\vphantom{+}}$.
But such a process is only possible if
$m_3<m_\chi\sim10^{11}\;\mbox{GeV}$, and if ${\rm H}_3^{\vphantom{+}}$
had such
\index{Proton lifetime}%
a mass, the proton lifetime would be unacceptably short, making
the entire scheme unrealistic.

Let us digress from this problem for a moment, since in any case
the SU(5) model in question is in need of modification --- it
gives a high probability of\index{Proton decay}\index{Decay of proton}
proton decay even when $m_3\gg
m_\chi$.  In order for the decay $\chi\,\chi\rightarrow{\rm
H}_3^+\,{\rm H}_3^{\vphantom{+}}$
to occur, let us take $m_\chi\sim m_{{\rm H}_3^+}$,
that is, $\beta\sim10^{-6}$.  In that event,
\be
\label{8.5.20}
\Gamma(\chi\,\chi\rightarrow{\rm H}_3^+\,{\rm H}_3^{\vphantom{+}})\sim
\frac{(10^{-11}\,\chi)^2}{m_\chi}\cdot {\rm O}(10^{-2})
\sim10^{-2}\;\mbox{GeV}\ ,
\ee
and so, according to (\ref{7.9.10}),
\be
\label{8.5.21}
{\rm T}_{\rm R}\sim10^{-1}\,\sqrt{\Gamma\,\m}\sim3\cdot 10^7\;\mbox{GeV}\ .
\ee
Baryon asymmetry formation is a possible process in this model,
since ${\rm H}_3^{\vphantom{+}}$  bosons are created and destroyed, but each
decay generates\linebreak[10000]
$\displaystyle {\rm O}\left(\frac{m_3}{{\rm T}_{\rm R}}\right)\sim
3\cdot 10^3$ photons of energy ${\rm E}\sim{\rm T}$.  This tends to
make the occurrence of a baryon asymmetry less likely by a factor
of $3\cdot 10^3$.  To circumvent this problem, one should either
invoke alternative baryon production mechanisms (see Chapter
\ref{c7}), or modify the Shafi--Vilenkin model.  We shall return
to this question in the next chapter;  for the time being, let us
try to analyze the main results obtained above, and evaluate the
prospects for further development of the new inflationary
universe scenario.
\index{Shafi--Vilenkin model|)}%

\section[The new scenario: problems and prospects]%
{\label{s8.6}The new inflationary universe scenario:
\index{Inflationary universe scenario!new!problems and prospects|(}%
problems and prospects}

As we have already seen, the basic problems with the new
inflationary universe scenario have to do with the need to obtain
small density inhomogeneities in the observable part of the
universe after inflation.  This question deserves more detailed
discussion.

1.  As we mentioned in Section \ref{s7.5}, the condition
${\rm A}\la10^{-4}$ imposes an overall constraint on ${\rm V}(\varphi)$
in the new inflationary universe scenario,
\be
\label{8.6.1}
\vf\la10^{-10}\,\m^4\ .
\ee
This means that in any version of this scenario, including the
primordial inflation scenario, the process of inflation can only
\index{Inflation!process begun}%
begin at a time
$$
t\ga{\rm H}^{-1}\sim\sqrt{\frac{3\,\m^2}{8\,\pi\,{\rm V}}}\sim
10^{-36}\;\mbox{sec}\ ,
$$
or in other words, six orders of magnitude later than the Planck
time $t_{\rm P}\sim\m^{-1}\sim10^{-43}$ sec.  Bearing in mind,
then, that a typical total lifetime for a hot, closed universe is
of order $t\sim t_{\rm P}$ (see Chapter \ref{c1}), it is clearly
almost always the case that a closed universe simply fails to
survive until the beginning of inflation; i.e., the flatness
problem cannot be solved for a closed universe.  The new
inflationary universe scenario can therefore only be realized in a
topologically nontrivial or a noncompact (infinite) universe, and
only in those parts of the latter which don't collapse and are
sufficiently large ($l\ga10^5\,\m^{-1}$) at the moment when the
matter density therein becomes less than $\vf\sim10^{-10}\,\m^4$.

2.  Let us now direct our attention to the fact that the new
inflationary universe scenario can only be realized in theories
in which the potential energy ${\rm V}(\varphi)$ takes on a
highly specific form where, as we have seen, the coupling
constants are strongly interrelated.  Considerable ingenuity is
required to construct such theories, with the result that the
original simplicity underlying the idea of inflation of the
universe is gradually lost in the profusion of conditions and
reservations required for its implementation.

3.  The basic difficulty of the new inflationary universe
scenario relates to the question of how the field $\varphi$
reaches the maximum of the effective potential ${\rm V}(\varphi)$
at $\varphi=0$.  This problem turned out to be an especially
serious one as soon as it was realized that the field $\varphi$
ought to interact extremely weakly with other fields.

To get to the heart of the problem, let us examine some region of
a hot universe in which the field $\varphi$ has the initial value
$\varphi\sim\varphi_0$.  Suppose that high-temperature
corrections lead to a correction to ${\rm V}(\varphi)$ of the
form
\be
\label{8.6.2}
\Delta {\rm V}\sim\frac{\alpha^2}{2}\,\varphi^2\,{\rm T}^2\ .
\ee
The age of the hot universe equals
\index{Age!of hot universe}%
\index{Hot universe!age of}%
$\displaystyle \frac{{\rm H}^{-1}}{2}$  (\ref{1.4.6}):
\be
\label{8.6.3}
t=\frac{{\rm H}^{-1}}{2}=\frac{\m}{2}\,\sqrt{\frac{3}{8\,\pi\,\rho}}
<\frac{\m}{2}\,\sqrt{\frac{3}{8\,\pi\,\Delta{\rm V}}}
\sim\frac{\m}{4\,\alpha\,\varphi\,{\rm T}}\ .
\ee
Over this time span, the corrections (\ref{8.6.2})
can only change the initial value $\varphi=\varphi_0$
if the typical time $\tau=(\Delta m)^{-1}({\rm T})\sim(\alpha\,{\rm T})^{-1}$
is less than the age of the universe $t$, whereupon we
obtain the condition
\be
\label{8.6.4}
\varphi_0\la\frac{\m}{3}\ .
\ee
High-temperature corrections can thus affect the initial value of
the field $\varphi$ only if the latter is less than
$\displaystyle \frac{\m}{3}$.  Meanwhile, theories with
$\vf\sim\varphi^n$ place no constraints on the initial value of
the field $\varphi$ except to require that $\vf\la\m^4$.  For
example, in the Shafi--Vilenkin theory (as in a
$\displaystyle \frac{\lambda\,\varphi^4}{4}$
theory with $\lambda\sim10^{-14}$),
the constraint ${\rm V}(\chi)\la\m^4$ implies that the field
$\chi$ can initially take on any value in the range
\be
\label{8.6.5}
-10^4\,\m\la\chi\la10^4\,\m\ ,
\ee
and only less than $10^{-4}$ of this interval is pertinent to
values of $\chi$ for which high-temperature corrections can play
any role.

For the case $\displaystyle \varphi\la\frac{\m}{3}$ one can make
another estimate.  In a hot universe with N different particle
species,
\be
\label{8.6.6}
t\la\frac{1}{4\,\pi}\,\sqrt{\frac{45}{\pi\,{\rm N}}}\,\frac{\m}{{\rm T}^2}\ ,
\ee
(see (\ref{1.3.20})).  Comparing $t$ from (\ref{8.6.6}) with
$\tau\sim(\alpha\,{\rm T})^{-1}$, we see that high-temperature effects
only begin to change the field $\varphi$ at a temperature
\be
\label{8.6.7}
{\rm T}\la{\rm T}_1\sim
\frac{\alpha\,\m\,\sqrt{45}}{4\,\pi\,\sqrt{\pi\,{\rm N}}}
\sim\frac{\alpha\,\m}{50}\ ,
\ee
when the overall energy density of hot matter is
\be
\label{8.6.8}
\rho({\rm T}_1)\sim\frac{\pi^2}{30}\,{\rm N}\,{\rm T}_1^4\la
\alpha^4\,\m^4\sim
\frac{3\cdot 10^{-3}\,\alpha^4\,\m^4}{{\rm N}^2}\sim
10^{-7}\,\alpha^4\,\m^4
\ee
for ${\rm N} \sim 200$ (as would be the case in grand unified
theories).  Notice, however, that the process whereby the field
$\varphi$ decreases can continue only so long as the total
density $\rho({\rm T})$ is not comparable to ${\rm V}(0)$, since
soon afterward, high-temperature effects become exponentially
small due to inflation.  This leads to the constraint
\be
\label{8.6.9}
10^{-7}\,\alpha^4\,\m^4>{\rm V}(0)\ .
\ee

In the theory of a field $\varphi$ that interacts only with
itself (with a coupling constant $\lambda\sim10^{-14}$), and in
primordial inflation models, the parameter $\alpha^2$ is of order
$10^{-14}$, so that (\ref{8.6.9}) then becomes
\be
\label{8.6.10}
{\rm V}(0)\la10^{-35}\,\m^4\ .
\ee
This value is much smaller than the actual value of ${\rm V}(0)$
in all realistic models of new inflation.

The situation is somewhat better in the
Shafi--Vilenkin model.\linebreak[10000]
There, $\alpha^2$ is of order $10^{-7}$, and (\ref{8.6.9})
remains valid.  We must ascertain, however, whether
high-temperature effects can make $\chi$ smaller than $5\cdot 10^{15}$
GeV (\ref{8.5.15}), which would be necessary for the
universe to inflate by a factor of $e^{60}$--$e^{70}$.

For this to happen, the field $\chi$ must be reduced to $5\cdot
10^{15}$ GeV at the time when the quantity
$$
\frac{d\Delta{\rm V}(\chi,{\rm T})}{d\chi}\approx\alpha^2\,{\rm T}^2\,\chi
$$
becomes less than
$$
\frac{d{\rm V}(\chi)}{d\chi}\sim4\,{\rm A}\,\chi^3\,\ln\frac{\chi_0}{\chi}\ .
$$
This occurs at a temperature
\be
\label{8.6.11}
{\rm T}_2\sim10^{12}\;\mbox{GeV}\ .
\ee

While the temperature T drops from ${\rm T}_1$ to ${\rm T}_2$, the field
$\chi$ oscillates in the potential
$\displaystyle \Delta{\rm V}(\chi,{\rm T})\sim
\frac{\alpha^2}{2}\,\chi^2\,{\rm T}^2$
at a frequency $m_\chi\sim\alpha\,{\rm T}$.  The rate of production of
pairs by this field is very low (\ref{8.5.20}), so that its
oscillation amplitude in the early universe decreases mainly by
virtue of the expansion of the universe.  It is readily shown
that in the present case (with
$\displaystyle \Delta {\rm V}(\chi,{\rm T})\sim
\frac{\alpha^2}{2}\,\chi^2\,{\rm T}^2$),
the falloff in the field $\chi$ is proportional to the
temperature T.  As the temperature drops from ${\rm T}_1\sim10^{14}$
GeV to ${\rm T}_2\sim10^{12}$ GeV, the original amplitude of the field
$\chi$ is reduced by a factor of $10^2$, and it becomes less than
$\sim5\cdot 10^{15}$ GeV only if the initial field $\chi$ was
less than $5\cdot 10^{17}$ GeV.

Thus, in order to implement the new inflationary universe
scenario in the Shafi--Vilenkin model, the field $\chi$ must
originally be a factor of 20 less than $\m$, a requirement that
is quite unnatural.

It must be understood that the foregoing estimates are
model-depen\-dent.  There do exist theories in which the effective
potential ${\rm V}(\varphi)$ rises so rapidly with increasing
field $\varphi$ that it becomes greater than $\m^4$ for
$\varphi\ga\m$.  In that event, the condition $\varphi_0\la\m$
may be warranted.  In general, it is possible to suggest
mechanisms whereby a field $\varphi\la\m$ in the early universe
rapidly drops to $\varphi\ll\m$.  But the examples considered
above show that it is indeed difficult to obtain consistency
among all the requirements needed for a successful implementation
of the new inflationary universe scenario.  As a result, a
consistent implementation of this scenario within the framework
of a realistic elementary particle theory is still lacking.

Of course, one cannot rule out the possibility that some future
elementary particle theory will automatically satisfy all the
necessary conditions.  But for now, there is no need to insist
that all of these conditions be satisfied, as there is another
scenario amenable to realization over a much wider class of
theories, namely the chaotic inflation scenario.
\index{Inflationary universe scenario!new!problems and prospects|)}%
\index{Inflationary universe scenario!new|)}%


\chapter{\label{c9}The Chaotic Inflation Scenario}
\index{Chaotic inflation scenario|(}

\section[Introduction.  Basic features of chaotic inflation]%
{\label{s9.1}Introduction.  Basic features of the scenario.\protect\\
The question of initial
conditions}\index{Chaotic inflation scenario!basic features of|(}

The general underpinnings of the chaotic inflation scenario were
described in some detail in Chapter \ref{c1}.  Rather than
reiterating what has already been said, we shall attempt here to
review the basic features of this scenario, which may perhaps
stand out in better relief against the backdrop of the preceding
discussion of the new inflationary universe scenario.

The basic idea behind this scenario is simply that one need no
longer assume that the field $\varphi$ lies at a minimum of its
effective potential ${\rm V}(\varphi)$ or ${\rm V}(\varphi, {\rm T})$
from the very outset in the early universe.  Instead, it is
only necessary to study the evolution of $\varphi$ for a variety
of fairly natural initial conditions, and check to see whether or
not inflation sets in.

If one requires in addition that a solution to the flatness
problem be accessible within the context of this scenario even
when the universe is closed, then it becomes necessary that
inflation be able to start with $\vf\sim\m^4$.  As demonstrated
in Chapter \ref{c1}, this requirement is satisfied by a broad
class of theories in which the effective potential ${\rm
V}(\varphi)$ increases no faster than a power of the field
$\varphi$ in the limit $\varphi\gg\m$.  In principle, inflation
can also take place in theories for which $\displaystyle
\vf\sim\exp\left(\alpha\,\frac{\varphi}{\m}\right)$ when
$\varphi\gg\m$ if $\alpha$ is sufficiently small $\alpha\la5$.
The general criterion for the onset of inflation follows from the
condition $\displaystyle \dot{\rm H}\ll{\rm H}^2=
\frac{8\,\pi\,{\rm V}}{3\,\m^2}$ and (\ref{1.7.16}):

\be \label{9.1.1} \frac{d\ln{\rm
V}}{d\varphi}\ll\frac{4\,\sqrt{\pi}}{\m}\ . \ee
 As we stated in
Chapter \ref{c1}, the most natural initial conditions for the
field $\varphi$ on a scale $l\sim{\rm H}^{-1}\sim\m^{-1}$ are that
$\partial_0\varphi\,\partial^0\varphi \sim
\partial_i\varphi\,\partial^i\varphi \sim \vf \sim \m^4$.  The
probability of formation of an inflationary region of the universe
then becomes significant --- we might estimate it to be perhaps
$1/2$ or $1/10$.  For our purposes, the only important
consideration is that the probability is not reduced by a factor
like $\exp(-1/\lambda)$ [\cite{118}].  Meanwhile, it has been
argued by some authors (see [\cite{258}, \cite{278}], for example)
that the probability of inflation in the $\displaystyle
\frac{\lambda\,\varphi^4}{4}$ theory might actually be suppressed
by a factor of this kind.  A thorough investigation of this
question is necessary for a proper understanding of those changes
that the inflationary universe scenario has introduced into our
conception of the world about us.  We will discuss this question
here, following the arguments put forth in [\cite{118}].

1.  First, let us try to understand whether it is actually
necessary to assume that $\displaystyle \frac{\dot\varphi}{2}\ll\vf$
from the very outset.  For
simplicity, we consider Eqs. (\ref{1.7.12}) and (\ref{1.7.13}) in
a flat universe ($k=0$) with a uniform field $\varphi$, and with
$\dot\varphi^2\gg\vf$.  Then (\ref{1.7.12}) and (\ref{1.7.13})
imply that $\ddot\varphi\gg{\rm V}'(\varphi)$ and\vspace{-3pt}
\be \label{9.1.2}
\ddot\varphi=\frac{2\,\sqrt{3\,\pi}}{\m}\,\dot\varphi^2\ ,
\vspace{-3pt}
\ee
so that\vspace{-3pt}
\ba
\label{9.1.3}
\dot\varphi&=&-|\dot\varphi_0|\,\left(1+\frac{2\,\sqrt{3\,\pi}}{\m}\,
|\dot\varphi_0|\,t\right)^{-1}\ ,\\
\label{9.1.4}
\varphi&=&\varphi_0-\frac{\m}{2\,\sqrt{3\,\pi}}\,
\ln\left(1+\frac{2\,\sqrt{3\,\pi}}{\m}\,|\dot\varphi_0|\,t\right)\ .
\vspace{-3pt}
\ea This means that when $\displaystyle t\ga{\rm
H}^{-1}\sim\frac{\m}{|\dot\varphi_0|}$, the kinetic energy of the
field $\varphi$ falls off according to a power law, as
$\dot\varphi^2\sim t^{-2}$, while the magnitude of the field
itself (and thus of $\vf\sim\varphi''$) falls off only
logarithmically.  The kinetic energy of the field $\varphi$
therefore drops rapidly, and after a short time (just several
times the value of ${\rm H}^{-1}$), the field enters the
asymptotic regime $\dot\varphi^2\ll\vf$ [\cite{118}, \cite{110}].

The thrust of this result, a more general form of which was
derived in [\cite{279}, \cite{280}] for both an open and closed
universe, is quite simple. When $\dot\varphi^2>\vf$, the
energy-momentum tensor has the same form as the energy-momentum
tensor of matter whose equation of state is $p=\rho$.  The energy
density of such matter rapidly decreases as the universe expands,
while the value of a sufficiently flat potential ${\rm
V}(\varphi)$ changes very slowly.

Let us estimate the fraction of initial values of $\dot\varphi$
for which the universe fails to enter the inflationary regime in
the $\displaystyle \frac{\lambda\,\varphi^4}{4}$ theory.  This
requires that $\dot\varphi^2$ remains larger than ${\rm
V}(\varphi)$ until $\varphi$ becomes smaller than $\displaystyle
\frac{\m}{3}$.  The initial value of $\displaystyle
\frac{\dot\varphi^2}{2}$ is of order $\m^4$  (prior to that point,
it is impossible to describe the universe classically), and the
initial value $\varphi_0$  of the field $\varphi$ can take on any
value in the range
$-\lambda^{1/4}\,\m\la\varphi\la\lambda^{1/4}\,\m$.  In that
event, we see from (\ref{9.1.4}) that the total time needed for
the field $\varphi$ to decrease from $\varphi_0$ to
$\varphi\sim\m$ is of order $\displaystyle
\frac{1}{2\sqrt{6\pi}\m}
\exp\left(\frac{2\sqrt{3\pi}\,\varphi_0}{\m}\right)$.  In this
time span, $\dot\varphi$ is reduced in magnitude by approximately
a factor $\displaystyle
\exp\left(\frac{2\,\sqrt{3\,\pi}\,\varphi_0}{\m}\right)$.  We then
find that when $\lambda\ll1$, $\dot\varphi^2$ can remain larger
than ${\rm V}(\varphi)$ during the whole process only if
$\varphi_0\sim\m$.  The probability that the field $\varphi$,
which initially can take any value in the range from
$-\lambda^{1/4}\,\m$ to $\lambda^{1/4}\,\m$, winds up being of
order $\m$ can be estimated to be $\lambda^{1/4}\sim3\cdot
10^{-4}$ for $\lambda\sim10^{-14}$.  It is therefore practically
inevitable that a homogeneous, flat universe passes through the
inflationary regime in the $\displaystyle
\frac{\lambda\,\varphi^4}{4}$ theory [\cite{280}, \cite{110},
\cite{118}].  One comes to the same conclusion for an open
universe as well.  For a closed universe, the corresponding
probability is of order $1/4$ [\cite{280}]. The reason that the
probability is smaller for a closed universe is that it may
collapse before $\dot\varphi^2$ becomes smaller than ${\rm
V}(\varphi)$.  In any event, the probability of occurrence of an
inflationary regime turns out to be quite significant, as we
expected.

2.  Next, let us discuss a more general situation in which the
field $\varphi$ is inhomogeneous.  If the universe is closed,
then its overall initial size $l$ is ${\rm O}(\m^{-1})$ (when
$l\ll\m^{-1}$, the universe cannot be described in terms of
classical space-time, and in particular one cannot say that its
size $l$ is much less than $\m^{-1}$).  If
$\partial_0\varphi\,\partial^0\varphi$  and
$\partial_i\varphi\,\partial\varphi^i$ are both severalfold
smaller than ${\rm V}(\varphi)$, which is not improbable, then
the universe begins to inflate, and the gradients
$\partial_i\varphi$  soon become exponentially small.  Thus, the
probability of formation of a closed inflationary universe
remains almost as large as in the case considered above, even
when possible inhomogeneity of the field $\varphi$ is taken into
account.

If the universe is infinite, then the probability that the
\index{Infinite universe}%
\index{Universe!infinite}%
conditions necessary for inflation actually come to pass would
seem at first glance to be extremely low [\cite{258}].  In fact,
if a typical initial value of the field $\varphi$ in a
$\displaystyle \frac{\lambda\,\varphi^4}{4}$ theory is, as we
have said, of order $\varphi_0\sim\lambda^{-1/4}\,\m\sim3000\,\m$,
then the condition $\partial_i\varphi\,\,\partial^i\varphi\la\m^4$
might lead one to conclude that $\varphi$ should remain larger
than $\sim\lambda^{-1/4}\,\m$ on a scale
$l\ga\lambda^{-1/4}\,\m\sim3000\,\m^{-1}$.  But this is highly
improbable, since initially (i.e., at the Planck time $t_{\rm
P}\sim\m^{-1}$) there can be no correlation whatever between
values of the field $\varphi$ in different regions of the
universe separated from one another by distances greater than
$\m^{-1}$.  The existence of such correlation would violate
causality (see the discussion of the horizon problem in Chapter
\ref{c1}).

The response to this objection is very simple [\cite{118},
\cite{78}, \cite{79}].  We have absolutely no reason to expect
that the overall energy density $\rho$ will {\it simultaneously} become
less than the Planck energy $\m^4$ in all causally disconnected
regions of an infinite universe, since {\it that} would imply the
existence of an acausal correlation between values of $\rho$ in
different domains of size ${\rm O}(\m^{-1})$.  Each such domain
looks like an isolated island of classical space-time, which
emerges from the space-time foam independently of other such
islands.  During inflation, each of these islands acquires
dimensions many orders of magnitude larger than the size of the
observable part of the universe.  If some gradually join up with
others through connecting necks of classical space-time, then in
the final analysis, the universe as a whole will begin to look
like a cluster (or several independent clusters) of topologically
connected mini-universes.  However, such a structure may only
come into being later (see Chapter \ref{c10} in this regard), and
a typical {\it initial} size of a domain of classical space-time with
$\rho\la\m^4$ is extremely small --- of the order of the Planck
length $l_{\rm P}\sim\m^{-1}$.  Outside each of these domains the
condition $\partial_i\varphi\,\partial^i\varphi\la\m^4$ no longer
holds, and there is no correlation at all between values of the
field $\varphi$ in different disconnected regions of classical
space-time of size ${\rm O}(\m^{-1})$.  But such correlation is
not really necessary for the realization of the inflationary
universe scenario --- according to the ``no hair'' theorem for de
Sitter space, a sufficient condition for the existence of an
inflationary region of the universe is that inflation take place
{\it inside} a region whose size is of order ${\rm H}^{-1}\sim\m^{-1}$,
which in our example is actually the case.

We wish to emphasize once again (and this will subsequently be of
some importance) that the confusion that we have analyzed above,
involving the correlation between values of the field $\varphi$
in different causally disconnected regions of the universe, is
rooted in the familiar notion of a universe that is {\it instantaneously}
created from a singular state with $\rho \rightarrow\infty$,
and {\it instantaneously} passes through a state with the Planck
density $\rho\sim\m^4$.  The lack of justification for such a
notion is the very essence of the horizon problem;  see Section
\ref{s1.5}.  Now, having disposed of the horizon problem with the
aid of the inflationary universe scenario, we may possibly manage
to familiarize ourselves with a different picture of the
universe, a picture whose specific features are gradually
becoming clear.  We shall return to this question in the next
chapter.

Evidently, the condition
$\partial_i\varphi\,\partial^i\varphi<\vf$ invoked above can also
be relaxed, in the same way that we relaxed the requirement
$\partial_0\varphi\,\partial^0\varphi<\vf$.  The basic idea here
is that if the effective potential ${\rm V}(\varphi)$ is flat
enough, then during the expansion of the universe (in all regions
which neither drop out of the general expansion process nor
collapse), the gradients of the field $\varphi$ fall off rapidly,
whereas the mean value of $\varphi$ decreases relatively slowly.
The net result is that just as in the case of kinetic energy
$\displaystyle \frac{\dot\varphi^2}{2}$, we ought to arrive at a
situation in which the energy density associated with gradients
of the field $\varphi$ in a significant part of the universe will
have fallen to much less than ${\rm V}(\varphi)$;  in other
words, the conditions necessary for inflation appear.  We shall
not discuss this possibility any further, as the results obtained
above suffice for our purposes.

To conclude this section, let us note that the foregoing question
involving acausal correlation does not arise in realistic
theories, in general, where apart from a ``light'' field
$\varphi$ with $\lambda\sim10^{-14}$, there is at least one
scalar field $\Phi$ with a bigger coupling constant
$\lambda_\Phi\ga10^{-2}$.  In such theories, the
\index{``Acausal correlation length''}``acausal
correlation length'' between values of $\Phi$ in different
regions is only marginally bigger than the distance to the
horizon, so that even if the aforementioned arguments concerning
the acausal correlation between values of $\Phi$ were true, the
probability that inflation would be driven by the field $\Phi$
would not be noticeably suppressed.  As demonstrated in
[\cite{281}], long-wave fluctuations of the light field $\varphi$
that are generated at the time of inflation bring about
self-reproducing inflationary regions (see Section \ref{s1.8})
filled with a quasihomogeneous field $\varphi$ for which
$\vf\la\m^4$.  The heavy field $\Phi$ rapidly decreases in these
regions, so that the last stages of inflation are governed by the
field $\varphi$ with $\lambda\sim10^{-14}$, as before.

Our principal conclusion, then, is that there exists a broad
class of elementary particle theories within the scope of which
inflation sets in under natural initial
conditions.\index{Chaotic inflation scenario!basic features of|)}

\section[The simplest SU(5) model]%
{\label{s9.2}The simplest model based on the SU(5) theory}
\index{Chaotic inflation scenario!simplest model based on SU(5) theory|(}%
\index{SU(5) theory!simplest model based on|(}%

The chaotic inflation scenario can be implemented within the
framework of many models (and in particular the Shafi--Vilenkin
model, which was originally designed to implement the new
inflationary universe scenario).  But one can achieve the same
end using simpler models, since we no longer need to satisfy the
numerous constraints imposed on the theory by the new
inflationary universe scenario.  To be specific, there is no need
to invoke the Coleman--Weinberg mechanism; the superheavy $\Phi$
and ${\rm H}_5^{\vphantom{+}}$ field sector in SU(5) theory can
be cast in a standard form;  we can omit interactions between the
$\chi$-field and $\Phi$-fields, and so on.

Consider, for example, a theory with the effective potential
\ba
\label{9.2.1}
{\rm V}&=&\frac{1}{4}\,a\,\tr(\Phi^2)^2+\frac{1}{2}\,b\,\tr\Phi^4-
\frac{{\rm M}_\Phi^2}{2}\,\tr\Phi^2-
a\,({\rm H}_5^+\,{\rm H}^{\vphantom{+}}_5)\,\tr\Phi^2\nonumber \\
&+&\frac{\lambda}{4}\,({\rm H}^+_5\,{\rm H}^{\vphantom{+}}_5)^2
-\beta\,{\rm H}^+_5\,\Phi^2\,{\rm H}^{\vphantom{+}}_5
+m_5^2\,{\rm H}^+_5\,{\rm H}^{\vphantom{+}}_5\nonumber \\
&-&\frac{m^2}{2}\,\chi^2+\frac{\lambda_1}{4}\,\chi^4
+\frac{\lambda_2}{2}\,\chi^2\,{\rm H}^+_5\,{\rm H}^{\vphantom{+}}_5\ ,
\ea
and assume that $a\sim b\sim\alpha\sim g^2$, $\lambda_1\gg
\lambda_2^2$ so that quantum corrections to $\lambda_1$  may be
neglected.  In this theory, in contrast to Eq. (\ref{8.5.3}) and
(\ref{8.5.4}), the masses $m_2$ and $m_3$ are given by
\ba
\label{9.2.2}
m_2^2&=&m_5^2+\lambda_2\,\chi^2-(\alpha+0.3\,\beta)\,\varphi^2\ ,\\
\label{9.2.3}
m_3^2&=&m_2^2+\frac{\beta}{6}\,\varphi^2\ .
\ea
Inflation takes place during the time that the $\chi$-field rolls
down from $\chi\sim\lambda_1^{-1/4}\,\m$ to the minimum of ${\rm V}(\chi)$
at $\displaystyle \chi_0=\frac{m}{\sqrt{\lambda_1}}$.
We will assume for simplicity that $\chi_0\la\m$;  then
$\displaystyle \frac{\delta\rho}{\rho}\sim10^{-5}$  for
$\lambda_1\sim10^{-14}$.  The universe is reheated much more
efficiently than in the Shafi--Vilenkin model, as the terms in
the Lagrangian responsible for decay of the field $\chi$ are now
of the form $\sim\lambda_2\,\chi^2\,{\rm H}^+_5\,{\rm H}^{\vphantom{+}}_5$
(there is no additional coefficient $\beta\sim10^{-6}$ resulting
from the simultaneity of oscillations of the $\varphi$ and $\chi$
fields).  This effect and additional energy transfer during
oscillations of the ${\rm H}_1$ and ${\rm H}_2$ fields (which
result from sign changes in $m_2^2$ as the field $\chi$
oscillates in the neighborhood of $\chi_0$) lead to rapid
reheating of the universe.  This is also facilitated by an
increase in oscillation frequency of the field $\chi$.  If, for
example, one takes $m\sim10^{12}$ GeV, then $\chi_0\sim\m$.  The
oscillation frequency of the field $\chi$ then becomes
$\sqrt{2}\,m=1.5\cdot 10^{12}$ GeV.  The reheating temperature
${\rm T}_{\rm R}$ in this model can reach $10^{12}$--$10^{13}$
GeV.  The decay $\chi\,\chi\rightarrow{\rm H}^+_3\,{\rm
H}^{\vphantom{+}}_3$ takes place at $m_3\la10^{12}$ GeV;  the
particular difficulties with proton decay which are related to
the low mass of the $m_3$ do not appear in this model, and the
temperature ${\rm T}_{\rm R}$ is large enough to facilitate the
standard baryogenesis mechanism based on the decay of
${\rm H}^{\vphantom{+}}_3$ particles.

The model presented here admits of a great many generalizations.
For example, one can delete the terms $\displaystyle -
\frac{m^2}{2}\,\chi^2$ and $\displaystyle
\frac{\lambda_1}{4}\,\chi^4$ from (\ref{9.2.1}), leaving only the
last term $\displaystyle
\frac{\lambda_2}{2}\,\chi^2\,{\rm H}^+_5\,{\rm H}^{\vphantom{+}}_5$.
Then due to radiative corrections, an induced term like
$\displaystyle {\rm C}\,\frac{\lambda_2^2\,\chi^4}{64\,\pi^2}\,
\left(\ln\frac{\chi}{\chi_0}-\frac{1}{4}\right)$  will be
responsible for inflation.  When $\lambda_2\sim10^{-6}$, this
term gives rise to density inhomogeneities $\displaystyle
\frac{\delta\rho}{\rho}\sim10^{-5}$.  This model is analogous to
the Shafi--Vilenkin model, but it is much simpler, and it is also
shares none of the latter's problems with baryogenesis.
Likewise, the extremely small coupling constant $\lambda_1\sim10^{-14}$
once again need not be introduced beforehand --- the constant
$\lambda_2\sim10^{-6}$ is sufficient, which seems more natural,
inasmuch as similar constants do appear in such popular models as
\index{Glashow--Weinberg--Salam theory}%
the Glashow--Weinberg--Salam theory.
\index{Chaotic inflation scenario!simplest model based on SU(5) theory|)}%
\index{SU(5) theory!simplest model based on|)}%

\section{\label{s9.3}Chaotic inflation in
\index{Chaotic inflation scenario!supergravity and|(}%
\index{Supergravity theory!chaotic inflation in|(}%
supergravity}

There are currently several different models that describe
chaotic inflation in the context of supergravity [\cite{273},
\cite{274}, \cite{282}].  Here we examine one of these that seems
to us to be particularly simple, a model related to
\index{SU$(n,1)$ supergravity}%
\index{Supergravity theory!SU$(n,1)$}%
$\mbox{SU}(n,1)$ supergravity [\cite{283}], several versions of
which arise in the low-energy limit of superstring theory
[\cite{17}].

One of the major problems that comes up in constructing realistic
models based on supergravity theory is how to make the effective
potential ${\rm V}(z)$ vanish at its minimum $z_0$.  As a first
step toward such a theory, one can attempt to find a general form
of the function ${\rm G}(z,z^*)$  for which the potential ${\rm V}(z,z^*)$
in (\ref{8.4.2}) is identically zero.  This occurs when
[\cite{284}]
\be
\label{9.3.1}
{\rm G}(z,z^*)=-\frac{3}{2}\,\ln(g(z)+g^*(z))^2\ ,
\ee
where $g(z)$ is some arbitrary function.  In that case, the
Lagrangian is
\be
\label{9.3.2}
{\rm L}={\rm G}_{zz^*}\,\partial_\mu z\,\partial^\mu z
=3\,\frac{\partial_\mu g\,\partial^\mu g}{(g+g^*)^2}\ .
\ee
All such theories with different $g(z)$ are equivalent to one
another after the transformation  $g(z) \rightarrow z$.  The
Lagrangian
\be
\label{9.3.3}
{\rm L}=3\,\frac{\partial_\mu z\,\partial^\mu z}{(z+z^*)^2}
\ee
is invariant under the group of $\mbox{SU}(1, 1)$ transformations
\be
\label{9.3.4}
z\rightarrow\frac{\alpha\,z+i\,\beta}{i\,\gamma\,z+\delta}
\ee
with real parameters $\alpha$, $\beta$, $\gamma$, $\delta$
such that $\alpha\,\delta+\beta\,\gamma=1$ [\cite{284}]. Such theories
have been called $\mbox{SU}(1,1)$ supergravity for that reason.

One possible generalization of the function ${\rm G}(z,z^*)$ of
(\ref{9.3.1}) that leads to a potential
${\rm V}(z,z^*,\varphi,\varphi^*)\ge0$, where $\varphi$ is the scalar
(inflation) field responsible for inflation, is
\be
\label{9.3.5}
{\rm G}=-\frac{3}{2}\,\ln(z+z^*+h(\varphi,\varphi^*))^2
+g(\varphi,\varphi^*)\ ,
\ee
where $h$ and $g$
are arbitrary real-valued functions of $\varphi$ and $\varphi^*$.
In the theory (\ref{9.3.5}),
\be
\label{9.3.6}
{\rm V}=\frac{1}{|z+z^*|^2}\,e^g\,
\frac{|g_\varphi|^2}{{\rm G}_{\varphi\varphi^*}}\ ,
\ee
where ${\rm G}_{\varphi\varphi^*}= g_{\varphi\varphi^*}+
{\rm G}_z\,h_{\varphi\varphi^*}\ge0$ if the kinetic term for the field
$\varphi$ has the correct (positive) sign.

Cast in terms of the variables $z$ and $\varphi$, the theory
(\ref{9.3.5}) looks rather complicated, but it can be simplified
considerably by diagonalizing the kinetic part of the Lagrangian,
reducing it to the form [\cite{285}]
\be
\label{9.3.7}
{\rm L}_{kin}=\frac{1}{12}\,\partial_\mu\zeta\,\partial^\mu\zeta
+\frac{3}{4}e^{\frac{2}{3}\,\zeta}\,{\rm I}_\mu^2
+{\rm G}_{\varphi\varphi^*}\,\partial_\mu\varphi^*\,\partial^\mu\varphi\ ,
\ee
where
\ba
\label{9.3.8}
\zeta&=&-\frac{3}{2}\,\ln(z+z^*+h(\varphi,\varphi^*))^2\ ,\nonumber \\
{\rm I}_\mu&=&i\,[\partial_\mu(z-z^*)+h_\varphi\,\partial_\mu\varphi
-h_{\varphi^*}\,\partial_\mu\varphi^*]\ ,\nonumber \\
{\rm G}_{\varphi\varphi^*}&=&
g_{\varphi\varphi^*}+{\rm G}_z\,h_{\varphi\varphi^*}
=g_{\varphi\varphi^*}-3\,e^{\zeta/3}\,h_{\varphi\varphi^*}\ .
\ea
In terms of $\zeta$, the potential becomes
\be
\label{9.3.9}
{\rm V}=e^{\zeta+g}\,\frac{|g_\varphi|^2}{{\rm G}_{\varphi\varphi^*}}\ .
\ee
As the simplest realization of the chaotic inflation scenario in
this model [\cite{274}], one can consider the theory
(\ref{9.3.5}) with
\be
\label{9.3.10}
g(\varphi,\varphi^*)=(\varphi-\varphi^*)^2
+\ln|f(\varphi)|^2\ ,
\ee
while $h(\varphi,\varphi^*)$ satisfies
\be
\label{9.3.11}
h_{\varphi\varphi^*}=(2\,a)^{-1}\,g_{\varphi\varphi^*}=-a^{-1}\ ,
\ee
where $a$ is a positive constant.  Then
\be
\label{9.3.12}
{\rm V}_\zeta\equiv\frac{\partial{\rm V}}{\partial\zeta}
={\rm V}\frac{a-e^{\zeta/3}}{\displaystyle a-\frac{3}{2}\,e^{\zeta/3}}\ ,
\ee
that is, ${\rm V}_\zeta=0$  when $e^{\zeta/3}=a$.
Notice that at the extremum of V (i.e., at ${\rm V}_\zeta=0$), the field
$\varphi$ has the canonical kinetic term
\be
\label{9.3.13}
{\rm G}_{\varphi\varphi^*}=-\frac{1}{2}\,g_{\varphi\varphi^*}=1\ ,
\ee
and
\be
\label{9.3.14}
{\rm V}_{\zeta\zeta^*}=\frac{2}{3}\,{\rm V}>0\ .
\ee
This means that the potential ${\rm V}(\varphi, \zeta)$ has a
hollow located at $\zeta = 3\, \ln a$, $-\infty < \varphi <
\infty$.  At the bottom of the hollow, the potential ${\rm
V}(\varphi, \zeta)$ is
\be
\label{9.3.15}
{\rm V}(\varphi)=a^3\,e^g\,|g_\varphi|^2=a^3\,e^{-4\,\eta^2}\,
|f_\varphi+4\,i\,\eta|^2\ ,
\ee
where $\varphi=\xi+i\,\eta$.
Equation (\ref{9.3.15}) implies that for all real $\varphi$,
\be
\label{9.3.16}
{\rm V}=a^3\,|f_\varphi|^2\ .
\ee
This resembles the expression for the effective potential in a
globally supersymmetric theory with the superpotential
$f(\varphi)$.  Inflation takes place in this theory for a broad
class of superpotentials, such as those with
$f(\varphi)\sim\varphi^n$, $n >1$.  A complete description of
inflation in this theory is quite complicated, particularly on
account of the presence of non-minimal kinetic energy terms in
(\ref{9.3.7}).  The third term in (\ref{9.3.7}), for example,
leads to an extra term $\sim
a^{-1}\,e^{\zeta/3}\,|\partial_\mu\varphi|^2$  in ${\rm V}_\zeta$
(\ref{9.3.12}).  Fortunately, $|\partial_\mu\varphi|^2\ll{\rm V}$
during inflation, and the corresponding correction turns out to
be negligible.

In order to study the evolution of the universe in this model, we
assume that $\varphi$ is originally a fairly large field,
$|\varphi|\gg1$  (or $\displaystyle |\varphi|\gg\frac{\m}{\sqrt{8\,\pi}}$
in conventional units).  Then both the curvature
${\rm V}_{\eta\eta}\sim a^3\,|f|^2$ and the curvature
${\rm V}_{\zeta\zeta}\sim  a^3\,|f_\varphi|^2$ are much greater
than the curvature ${\rm V}_{\xi\xi}$, which in the theory under
consideration is of order $a^3\,|f_\varphi|^2\,\varphi^{-2}$.  If
$\zeta\neq3\,\ln a$ ($\displaystyle \zeta>3\,\ln\frac{2\,a}{3}$)
from the outset and $\eta\neq0$ ($|\eta|\la 1$), then the fields
$\zeta$ and $\varphi$ quickly roll down to the bottom of the
hollow, where $\zeta=3\,\ln a$ and $\eta=0$, and the effective
potential is given by (\ref{9.3.16}).  The field $\varphi$ then
has the usual kinetic energy term (\ref{9.3.13}), and for
$f=\mu^3\,\varphi^n$,
\be
\label{9.3.17}
\vf=n^2\,a^3\,\mu^6\,\varphi^{2\,n-2}\ .
\ee
In particular, if  $f=\mu^3\,\varphi^3$,
\be
\label{9.3.18}
\vf=9\,a^3\,\mu^6\,\varphi^4\ .
\ee
The universe undergoes inflation as the field $\varphi$ rolls
down from $\varphi\gg1$ to $\varphi\la1$.  The density
inhomogeneities that are produced in the theory (\ref{9.3.18})
are of order $\displaystyle \frac{\delta\rho}{\rho}\sim10^{-5}$
when $\sqrt{\alpha}\,\mu\sim10^{-2}$--$10^{-3}$.  There is thus
no need to introduce any anomalously small coupling constants
like $\lambda\sim10^{-14}$; in this scenario, the combination
$a^3\,\mu^6$ takes on that role.  A typical inflation factor for
the universe in this model is of order $10^{10^7}$.  The process
whereby the universe is reheated depends on the manner in which
the field $\varphi$ interacts with the matter fields.  As a rule,
reheating to a temperature ${\rm T}_{\rm R}\ga10^8$ GeV is
readily accomplished [\cite{286}], enabling baryon asymmetry
production by the mechanisms described in Chapter
\ref{c7}.
\index{Chaotic inflation scenario!supergravity and|)}%
\index{Supergravity theory!chaotic inflation in|)}%

\section[modified Starobinsky model and combined scenario]%
{\label{s9.4}The modified Starobinsky model and the combined\protect\\
scenario}
\index{Chaotic inflation scenario!combined|(}%
\index{Starobinsky model!modified|(}%

{\looseness=1
In all of the models discussed thus far, inflation is driven by
an elementary scalar field $\varphi$.  However, the role of this
field can also be played by a condensate of fermions
$\langle\bar\psi\,\psi\rangle $ or vector particles
$\langle {\rm G}_{\mu\nu}^a\,{\rm G}_{\mu\nu}^a\rangle $, or
simply by the curvature scalar R itself.  The latter possibility
provided the basis for the Starobinsky model [\cite{52}], which
could be considered the first version of the inflationary
universe scenario, and which predated even the model of Guth.  In
its original form, this model was based on the observation of
Dowker and Critchley [\cite{106}] that when the conformal anomaly
of the energy-momentum tensor is taken into account, de Sitter
space with energy density approaching the Planck density turns
out to be a self-consistent solution of the Einstein equations
with quantum corrections.  Starobinsky showed that the
corresponding solution is unstable;  the curvature scalar starts
to decrease slowly at some moment, and this decrease accelerates.
Finally, after the oscillatory stage, the universe is reheated,
and is then described by the standard hot universe model.

}

The formal description of the decay of the initial de Sitter
space in the Starobinsky model bears a close resemblance to the
theory of the decay of the unstable state $\varphi=0$ in the new
inflationary universe scenario.  When this model was proposed, it
elicited an enormous amount of interest from cosmologists
[\cite{287}].  But the origin of the unstable de Sitter state in
the Starobinsky model remained somewhat enigmatic --- the
conventional wisdom was that either such a state came into being
as a result of an asymmetric collapse of a previously existing
universe [\cite{288}], or that the universe appeared ``from
nothing'' in an unstable vacuum-like state [\cite{289},
\cite{290}].  These suggestions seemed rather more complicated
than the principles underlying the new inflationary universe
scenario.  Furthermore, the original Starobinsky model, like the
first versions of the new inflationary universe scenario, leaves
us with density inhomogeneities $\displaystyle
\frac{\delta\rho}{\rho}$ after inflation that are too large
[\cite{107}], and it fails to provide a solution to the
primordial monopole problem.

Subsequently, however, it proved to be possible to modify this
model and implement it in a manner that was similar in spirit to
the chaotic inflation scenario [\cite{108}--\cite{110}].  The
crux of this modification entailed replacing the study of
one-loop corrections to the energy-momentum tensor ${\rm T}_\mu^\nu$
with an examination of a gravitational theory in which terms
quadratic in the curvature tensor ${\rm R}_{\mu\nu\alpha\beta}$
are added to the Einstein Lagrangian $\displaystyle
\frac{{\rm R}}{16\,\pi\,{\rm G}}$.

In general, this is far from an innocuous procedure, inasmuch as
metric perturbations then turn out to be described by
fourth-order equations, and this frequently leads to particles
\index{Indefinite metric}%
\index{Metric!indefinite}%
\index{Tachyon}%
having imaginary mass (tachyons) or negative energy (indefinite
metric) [\cite{291}].  Fortunately, these problems do not appear
if just one term
$\displaystyle \frac{{\rm R}^2\,\m^2}{96\,\pi^2\,{\rm M}^2}$ is
added, with ${\rm M}^2\ll\m^2$.  When the sign in front of ${\rm R}^2$
is correctly chosen, this term leads to the emergence of a
\index{Scalaron}%
scalar excitation (scalaron) corresponding to a particle with
positive energy and mass ${\rm M}^2>0$.  Taking the term
$\sim{\rm R}^2$ into consideration, the Einstein equations are
then modified.  Specifically, in a flat Friedmann space ($k=0$),
Eq.  (\ref{1.7.12}) for the universe filled by a uniform field
$\varphi$ is replaced by
\be
\label{9.4.1}
{\rm H}^2=\frac{8\,\pi}{3\,\m^2}\left[\frac{1}{2}\,\dot\varphi^2+\vf\right]
-\frac{{\rm H}^2}{{\rm M}^2}\,\left[\dot{\rm H}
+2\,\frac{\ddot{\rm H}}{{\rm H}}-\left(\frac{\dot{\rm H}}{{\rm H}}\right)^2
\right]\ .
\ee
Let us first neglect the contribution to (\ref{9.4.1}) from the
field $\varphi$, and consider solutions of the modified Einstein
equations in the absence of matter fields.  Equation
(\ref{9.4.1}) will then admit of a solution satisfying the
conditions $|\dot{\rm H}|\ll{\rm H}^2$,
$|\ddot{\rm H}|\ll|\dot{\rm H}\,{\rm H}|$, or in other words, a
solution that describes an inflationary universe with a slowly
changing parameter H [\cite{108}, \cite{109}]:
\ba
\label{9.4.2}
{\rm H}&=&\frac{1}{6}\,{\rm M}^2\,(t_1-t)\ ,\\
\label{9.4.3}
a(t)&=&a_0\,\exp\left(\frac{{\rm M}^2}{12}\,(t_1-t)^2\right)\ .
\ea
These conditions prevail until H becomes smaller than M;  after
that, the stage of inflation ends, and H starts to oscillate
about some mean value $\displaystyle {\rm H}_0(t)\sim\frac{1}{t}$.
The universe then heats up, and it can
subsequently be described by the familiar hot universe theory.

Strictly speaking, Eqs. (\ref{9.4.2}) and (\ref{9.4.3}) are only
applicable if ${\rm R}^2$, ${\rm R}_{\mu\nu}\,{\rm R}^{\mu\nu}\ll\m^4$.
Moreover, depending on the initial value of
H, inflation may start much later than the Planck time, at which
${\rm R}^2$ and ${\rm R}_{\mu\nu}\,{\rm R}^{\mu\nu}$ become of
the same order as $\m^4$.  We thus arrive once again at the
problem of the evolution of a universe in which inflation takes
place only in regions with suitable initial conditions (in no way
associated with high-temperature phase transitions).  In other
words, we again wind up with the chaotic inflation scenario,
where the curvature scalar R (equal to $12\,{\rm H}^2$ at the
time of inflation) takes on the role of the scalar inflaton
field.  In the more general case in which scalar fields $\varphi$
(\ref{9.4.1}) are also present, several different stages of
inflation are possible, where the dominant effects are either
associated with the scalar fields or they are purely
gravitational, as described above [\cite{110}].

The succession of different stages is governed by the
\index{Scalaron}%
relationship between the mass M of the scalaron and the mass $m$
of the scalar field $\varphi$ when $\varphi\sim\m$
($m\sim\sqrt{\lambda}\,\m$ in the $\displaystyle
\frac{\lambda}{4}\,\varphi^4$ theory).  When $m\gg{\rm M}$, the
stage in which the field $\varphi$ dominates rapidly comes to an
end, and the next stage of inflation is associated with purely
gravitational effects.  During that stage, as usual, density
inhomogeneities $\displaystyle \frac{\delta\rho}{\rho}$ are
produced which on a galactic scale are given to order of
magnitude by [\cite{107}, \cite{221}]
\be
\label{9.4.4}
\frac{\delta\rho}{\rho}\sim10^3\,\frac{{\rm M}}{\m}\ ,
\ee
or in other words, $\displaystyle \frac{\delta\rho}{\rho}\sim10^{-5}$ when
\be
\label{9.4.5}
{\rm M}\sim10^{11}\;\mbox{GeV}\ .
\ee
Reheating of the universe in this instance is also driven by
purely gravitational effects [\cite{52}, \cite{134}].  According
to (\ref{7.9.17}),
\be
\label{9.4.6}
{\rm T}_{\rm R}\sim10^{-1}\,\sqrt{\Gamma\,\m}\sim
10^{-1}\,\sqrt{\frac{{\rm M}^3}{\m}}\sim10^6\;\mbox{GeV}\ .
\ee
To account for baryogenesis at a temperature
${\rm T}\la{\rm T}_{\rm R}\sim10^6$ GeV as in the case of the
Shafi--Vilenkin model and a number of other models based on
supergravity, it is necessary to invoke the nonstandard
mechanisms described in Section \ref{s7.10}.  Note, however, that
terms like $\displaystyle \frac{{\rm R}^2\,\m^2}{96\,\pi^2\,{\rm M}^2}$,
if they do appear in elementary particle theories or
superstring theories, will do so, as a rule, with ${\rm M}\sim\m$
rather than ${\rm M}\sim10^{-8}\,\m$.  It therefore seems more
likely that realistic theories will yield $m\ll {\rm M}$.   The
modified Starobinsky model may then turn out to be responsible
for the description of the earliest stages of inflation, while
the formation of the observable structure of the universe and its
reheating take place during the stage when the scalar field
$\varphi$ dominates.  Reference \cite{110} describes a more
detailed investigation of the combined model (\ref{9.4.1}),
effects related to the scalar field $\varphi$, and effects
associated with quadratic corrections to the Einstein
Lagrangian.
\index{Chaotic inflation scenario!combined|)}%
\index{Starobinsky model!modified|)}%

\section{\label{s9.5}Inflation in
\index{Inflation!in Kaluza--Klein and superstring theories|(}%
\index{Kaluza--Klein theories!inflation in|(}%
\index{Superstring theory!inflation in|(}%
Kaluza--Klein and superstring theories}

It was noted in Chapter \ref{c1} that our fondest hopes for
constructing a unified theory of all fundamental interactions
have been tied in recent years to Kaluza--Klein and superstring
theories.  One feature common to both of these theories is the
proposition that original space-time has a dimensionality
$d\gg4$.  Theories with $d = 10$ [\cite{17}], $d = 11$
[\cite{16}], $d = 26$ [\cite{94}], and even $d = 506$ [\cite{95},
\cite{96}] have all been entertained.  The assumption is that
$d-4$ dimensions are compactified,
\index{Compactified dimensions}\index{Dimensions!compactified}and
that space takes on
dimensions of order $\m^{-1}$ in the corresponding directions, so
that we are actually able to move only in the one remaining time
and three remaining space directions.  It is usually assumed that
the compactified directions are spatial, but the possibility of
compactifying multidimensional time has also aroused some
interest [\cite{292}, \cite{293}].  The properties of a
compactified space, in the final analysis, determine the basic
properties of the elementary particle theory that it engenders.

Unfortunately, neither specific elementary particle models based
on Kaluza--Klein and superstring theories nor associated
cosmological models have yet come close to fruition.
Nevertheless, it does make sense to examine the results that have
been obtained in this field.

One of the most interesting and detailed models of inflation
\index{Shafi--Wetterich model}%
based on Kaluza--Klein models is that of Shafi and Wetterich
[\cite{237}].  The basis for this model is the Einstein action
with corrections that are quadratic in the $d$-dimensional
curvature:
\ba
\label{9.5.1}
{\rm S}&=&-\frac{1}{{\rm V}_{\cal D}}\,\int d^dx\,\sqrt{g_d}\,\nonumber \\
&\times&
\left\{\alpha\,\hat{\rm R}^2
+\beta\,\hat{\rm R}_{\hat\mu\hat\nu}\,\hat{\rm R}^{\hat\mu\hat\nu}
+\gamma\,\hat{\rm R}_{\hat\mu\hat\nu\hat\sigma\hat\lambda}\,
\hat{\rm R}^{\hat\mu\hat\nu\hat\sigma\hat\lambda}+\delta\cdot \hat{\rm R}
+\varepsilon\right\}\ .\;\mbox{\hspace{5pt}}
\ea
Here $\hat\mu,\hat\nu,\ldots=0,1,2,\ldots\,,d - 1$;
$\hat{\rm R}_{\hat\mu\hat\nu\hat\sigma\hat\lambda}$
is the curvature tensor in $d$-dimensional space; ${\rm V}_{\cal D}$
is a volume in ${\cal D}$-dimensional  compactified space,
${\cal D}=d-4$;  and $\alpha$, $\beta$, and $\gamma$ are
dimensionless parameters.  With
\ba
\label{9.5.2}
\zeta&=&{\cal D}\,({\cal D}-1)\,\alpha+
({\cal D}-1)\,\beta+2\,\gamma>0\ ,\\
\label{9.5.3}
\delta&>&0\ ,\\
\label{9.5.4}
\varepsilon&=&\frac{1}{4}\,\delta^2\,{\cal D}\,({\cal D}-1)\,\zeta^{-1}\ ,
\ea
the equations for the $d$-dimensional metric have a solution of
the form ${\rm M}^4\times{\rm S}^{\cal D}$, where ${\rm M}^4$ is
a Minkowski space, and ${\rm S}^{\cal D}$ is a sphere of radius
\be
\label{9.5.5}
{\rm L}_0^2=\frac{2\,\zeta}{\delta}\ .
\ee
For
\be
\label{9.5.6}
\chi=({\cal D}-1)\,\beta+2\,\gamma>0\ ,
\ee
the effective gravitational constant describing the interaction
at large distances in ${\rm M}^4$ space is positive:
\be
\label{9.5.7}
{\rm G}^{-1}=\m^2=16\,\pi\,\frac{\chi}{\zeta}\,\delta\ .
\ee
A question remains concerning the stability of the solution ${\rm
M}^4\times{\rm S}^{\cal D}$, but it has at least been proven that
with certain constraints on the parameters of the theory,
compactification is stable against variations of the radius of
the sphere ${\rm S}^{\cal D}$ [\cite{294}].

To describe cosmological evolution in this model, it is
convenient to introduce the four-dimensional scalar field
\be
\label{9.5.8}
\varphi(x)=\ln\frac{{\rm L}(x)}{{\rm L}_0}\ .
\ee
After a suitable change of scale of the metric
$g_{\hat\mu\hat\nu}(x)$, the effective action in four-dimensional
space can be expressed as
\ba
\label{9.5.9}
{\rm S}&=&-\int d^4x\:\sqrt{g_4}\nonumber \\
&\times&\biggl[\frac{\m^2\,{\rm R}}{16\,\pi}+\exp{\cal D}\,\varphi\cdot
(\alpha\,{\rm R}^2+\beta\,{\rm R}_{\mu\nu}\,{\rm R}^{\mu\nu}
+\gamma\,{\rm R}_{\mu\nu\sigma\lambda}\,{\rm R}^{\mu\nu\sigma\lambda})
\nonumber \\
&-&\frac{1}{2}\,f^2(\varphi)\,\partial_\mu\varphi\,\partial^\mu\varphi
-\frac{1}{2}\,f_{\rm R}(\varphi)\cdot
{\rm R}\,\partial_\mu\varphi\,\partial^\mu\varphi\nonumber \\
&-&\tilde h(\varphi)\,\partial_\mu\varphi\,\partial^\mu{\rm R}+\vf
+\Delta{\rm L}_{kin}\biggr]\ .
\ea
Here $\mu, \nu, \ldots = 0, 1, 2, 3$, and $\Delta{\rm L}_{kin}$
takes in terms that comprise many derivatives of the field
$\varphi$, like $\partial_\mu\partial_\nu\varphi\cdot
\partial^\mu\partial^\nu\varphi$, etc.  The potential
${\rm V}(\varphi)$ takes the form
\be
\label{9.5.10}
\vf=\left(\frac{\m^2}{16\,\pi}\right)^2\,
\frac{{\cal D}\,({\cal D}-1)}{4\,\zeta}e^{-{\cal D}\,\varphi}
\left(\frac{1-e^{-2\,\varphi}}{1-\sigma\,e^{-2\,\varphi}}\right)^2\ ,
\ee
where $\displaystyle \sigma=\frac{\chi}{\zeta}-1$.  The functions
$f^2(\varphi)$, $f_{\rm R}(\varphi)$, and $\tilde h(\varphi)$ in
(\ref{9.5.9}) depend on $\alpha$, $\beta$, $\gamma$, and {\cal
D}.  From (\ref{9.5.10}), it follows that ${\rm V}(\varphi) \ge0$;
${\rm V}(\varphi)$ tends to zero only when $\varphi=0$ ---
that is, when ${\rm L}(x)={\rm L}_0$.  When
${\rm R}_{\mu\nu\sigma\lambda}\neq0$, however, the term
\be
\label{9.5.11}
e^{{\cal D}\,\varphi}\,{\cal K}_\varphi=e^{{\cal D}\,\varphi}\,
(\alpha\,{\rm R}^2+\beta\,{\rm R}_{\mu\nu}\,{\rm R}^{\mu\nu}
+\gamma\,{\rm R}_{\mu\nu\sigma\lambda}\,{\rm R}^{\mu\nu\sigma\lambda})
\ee
also contributes to the equation of motion of the field $\varphi$,
and the function
\be
\label{9.5.12}
{\rm W}(\varphi)=\vf+e^{{\cal D}\,\varphi}\,{\cal K}_\varphi
\ee
then plays the role of the potential energy of the field $\varphi$.

On the other hand, it is readily demonstrated that at the stage
of inflation, the term (\ref{9.5.11}) makes a contribution
$\sim e^{{\cal D}\,\varphi}\,{\rm H}^2\,\dot{\rm H}$ to the
Einstein equations in four-dimensional space-time, and with
${\rm H}=\mbox{const}$, this contribution can be neglected.  In
that approximation, then, the rate of inflation of the universe
does not depend on the fact that the additional term
(\ref{9.5.11}) is present, and is governed solely by the
potential ${\rm V}(\varphi)$,
\be
\label{9.5.13}
{\rm H}^2=\frac{8\,\pi}{3\,\m^2}\,\vf\ ,
\ee
while at the same time, the evolution of the field $\varphi$
depends on the form of the potential ${\rm W}(\varphi)$
(\ref{9.5.12}):
\be
\label{9.5.14}
3\,{\rm H}\,h^2(\varphi)\,\dot\varphi=
-\frac{\partial{\rm W}}{\partial\varphi}=
-\frac{\partial{\rm V}}{\partial\varphi}-{\cal D}\,e^{{\cal D}\,\varphi}\,
{\cal K}_\varphi({\rm H}(\varphi))\ .
\ee
The function $h^2(\varphi)$ appears in (\ref{9.5.14}) by virtue
of the non-minimal nature of the kinetic energy terms pertaining
to the field $\varphi$ in (\ref{9.5.9}).  This function varies
slowly as $\varphi$ changes, and goes to a constant as
$\varphi\rightarrow\infty$.  The function
$\displaystyle \frac{\partial{\rm W}}{\partial\varphi}$ at large
$\varphi$ behaves as follows:
\be
\label{9.5.15}
\lim_{\varphi\rightarrow\infty}\frac{\partial{\rm W}}{\partial\varphi}=
\left(\frac{\m^2}{16\,\pi}\right)^2\,
\frac{{\cal D}^2\,({\cal D}-1)}{4\,\zeta}\,(\mu-1)\,e^{-{\cal D}\,\varphi}\ ,
\ee
where
\be
\label{9.5.16}
\mu-1=\frac{{\cal D}-4}{12\,\zeta}\,
[3\,({\cal D}-1)\,\beta+2\,({\cal D}+3)\,\gamma]\ .
\ee
For $\mu > 1$, the potential ${\rm W}(\varphi)$ approaches some
constant from below, with the corresponding difference becoming
exponentially small.  This implies that the field $\varphi$ rolls
down to the minimum of ${\rm W}(\varphi)$ at $\varphi=0$
exponentially slowly.  On the other hand, the potential
${\rm V}(\varphi)$, which determines the rate of expansion of the
universe, is also exponentially small at large $\varphi$.
Nonetheless, with an initial value of $\varphi\ga{\rm O}(1)$ and
a reasonable choice of the constants $\alpha$, $\beta$, and
$\gamma$, it is possible simultaneously to obtain both a high
degree of inflation and small density perturbations.  In
particular, the duration of the inflationary stage in this model
is given approximately [\cite{237}] by
\be
\label{9.5.17}
\Delta t\sim {\rm H}^{-1}\,\frac{2\,{\cal K}_\infty}{\mu-1}\,\varphi\ ,
\ee
where ${\cal K}_\infty$ is defined by
\be
\label{9.5.18}
\lim_{\varphi\rightarrow\infty} h^2(\varphi)=
h^2_\infty=\frac{\m^2}{16\,\pi}\,
\frac{{\cal D}}{4\,\zeta}\,{\cal K}_\infty\ .
\ee
The quantity ${\cal K}_\infty$ may be expressed in terms of
$\alpha$, $\beta$, and $\gamma$, and is usually of order unity.
It can easily be shown that by choosing $\alpha, \beta,\gamma \sim 1$,
which is quite natural, one can obtain $\Delta
t\sim70\,{\rm H}^{-1}$, assuming an initial value $\varphi\sim3$
[\cite{237}].

In this model, the quantity $\displaystyle \frac{\delta\rho}{\rho}$
is given by an expression like (\ref{7.5.21}),
the only difference being that $\dot\varphi\,h(\varphi)$
appears instead of $\dot\varphi$:
\be
\label{9.5.19}
\frac{\delta\rho}{\rho}\sim0.2\,\frac{{\rm H}^2}{\dot\varphi\,h(\varphi)}
\sim0.2\,\frac{{\rm H}^2\,\Delta t}{h_\infty\,\varphi}
\sim\frac{2\,{\rm H}}{\m}\,
\left(\frac{\pi\,\varphi}{{\cal D}\,{\cal K}_\infty}\right)^{1/2}\,
\frac{{\rm H}\,\Delta t}{\varphi}\ .
\ee
At the epoch of interest, ${\rm H}\,\Delta t\sim70$, and $\varphi\sim3$ so
\be
\label{9.5.20}
\frac{\delta\rho}{\rho}\sim{\rm C}\frac{{\rm H}}{\m}\ ,
\ee
where ${\rm C} = {\rm O}(1)$. In particular,
$\displaystyle \frac{\delta\rho}{\rho}\sim10^{-5}$ when
\be
\label{9.5.21}
\frac{\rm {}H}{\m}\sim\frac{1}{8}\,
\left(\frac{{\cal D}\,({\cal D}-1)}{6\,\pi\,\zeta}\right)^{1/2}\,
\exp\left(-\frac{{\cal D}}{2}\,\varphi\right)\sim10^{-5}\ .
\ee
For $\varphi\sim3$, (\ref{9.5.20}) is satisfied in theories with
$d={\cal D}+4={\rm O}(10)$.  One interesting feature of the
perturbation spectrum obtained under these circumstances is its
decline at large $\varphi$ --- that is, at long wavelengths.
This is related to the fact that the behavior of the field and
the rate of expansion of the universe in this model are
determined by the two different functions ${\rm W}(\varphi)$ and
${\rm V}(\varphi)$, respectively, rather than the single function
${\rm V}(\varphi)$.

The Shafi--Wetterich model is also interesting in that the
\index{Shafi--Wetterich model}%
curvature of the effective potential ${\rm W}(\varphi)$ is quite
large for $\varphi \ll 1$.  After inflation, $\varphi$ oscillates
in the vicinity of $\varphi = 0$ at close to the Planck
frequency, and the universe is reheated very rapidly and
efficiently.  In this model, the temperature can reach
${\rm T}_{\rm R}\sim10^{17}$ GeV after reheating [\cite{295}].

The main problem here is related to the initial conditions
required for inflation.  Indeed, it would be unnatural to assume,
within the framework of Kaluza--Klein theories, that
three-dimensional space has been infinite from the very
beginning, since that would mean that the distinction between
compactified and non-compactified dimensions would have to have
been inserted into the theory from the outset, rather than
arising spontaneously.  It would be more natural to suppose that
the universe has been compact since its birth, but that it has
expanded at different rates in different directions:  in three
dimensions, it has grown exponentially, while in $d-4$
dimensions, it has gradually acquired a size of approximately
${\rm L}_0\sim\m^{-1}$ (\ref{9.5.5}), (\ref{9.5.7}).  To phrase
it differently, we are dealing with a compact (closed, for
example) universe governed by an expansion law that is asymmetric
in different directions.

In Chapter \ref{c1}, it was noted that a
\index{Closed universe}\index{Universe!closed}%
closed universe has a
typical total lifetime of order $\m^{-1}$, and the only thing
that can rescue it from collapse is inflation that begins
immediately after it has emerged from a state of Planck energy
density.  In the Shafi--Wetterich model, however, inflation ought
to begin when $\varphi\ga3$, ${\rm H}\la10^{-5}\,\m$
(\ref{9.5.21}), that is, when $\vf\ll\m^4$.  In that event,
inflation cannot save the universe from a premature death.  In
order to circumvent this problem, it was suggested in
[\cite{237}] that the entire universe came into being as a result
of a quantum jump from the space-time foam (from ``nothing'')
into a state with $\varphi\ga3$, ${\rm H}\la10^{-5}\,\m$, a
possibility we shall discuss in the next chapter.  Unfortunately,
however, estimates of the probability for such processes
[\cite{296}] lead to an expression of the form $\displaystyle
{\rm P}\sim\exp\left(-\frac{\m^4}{\vf}\right)$, giving
${\rm P}\sim\exp(-10^{10})$ in the present case.  Thus, the
outlook for a natural implementation of the inflationary universe
scenario in the context of the Shafi--Wetterich model is not very
good.  In fact, we are the victims here of difficulties of the
same type as those that prevent a successful implementation of
the new inflationary universe scenario.

It might be hoped that these problems will all go away when we
make the transition to a superstring theory.  Such theories
present several different candidates for the inflaton field
responsible for the inflation of the universe --- it may be some
combination of the dilaton field that appears in superstring
theory and the logarithm of the compactification radius.
Regrettably, our current understanding of the phenomenological
and cosmological aspects of superstring theory are still not
entirely satisfactory.  Existing models of inflation based on
superstring theories [\cite{297}] rest on various assumptions
about the structure of these theories, and these assumptions are
not always well-founded.  But it is the initial conditions, as
before, that are the main problem.  Our view is that the initial
conditions prerequisite to the onset of inflation in most of the
models based on superstring theories that have been proposed thus
far are unnatural.

Does this mean that we are headed down the wrong road?  At the
moment, that is a very difficult question to answer.  It is quite
\index{Superstring theory!future development of}%
possible that with the future development of superstring theory,
the inflationary universe scenario will be implemented in the
context of the latter in some nontrivial way (see, for example,
Ref. \cite{353}).  On the other hand, one should recall that over
the past decade, three palace revolts have taken place in the
land of elementary particles.  In place of grand unified theories
came theories based on supergravity, followed by Kaluza--Klein
theories, and finally the presently reigning superstring theory.
The inflationary universe scenario can be successfully
implemented in some of these theories;  in some, this has not yet
been accomplished, but there are no ``no-go'' theorems that say
it is impossible.  In our opinion, what we have encountered here
is a somewhat nonstandard aspect of a standard situation.  A
theory should be constructed in such a way that it describes
experimental data, but this cannot always be done, and the theory
must then be changed.  Until recently, however, cosmological data
have not been counted among the most important experimental
facts.  This situation has now been radically altered, and it
might just be that models in which inflation of the universe
cannot be implemented in a natural way will be rejected as being
inconsistent with the experimental data (if, of course, we find
no alternative solution for all of the cosmological problems
outlined in Chapter \ref{c1} that is not based on the
inflationary universe scenario).  In analyzing the current state
of affairs in this field, it must also be borne in mind that our
understanding of the inflationary universe scenario, and in
particular the most important question of initial conditions, is
far from complete.  In recent years, our conception of the
initial-condition problem in cosmology and our ideas about the
global structure of the inflationary universe have undergone a
significant change.  Progress in this field depends primarily on
\index{Quantum cosmology}%
the development of quantum cosmology, the topic to which we now
turn.
\index{Chaotic inflation scenario|)}%
\index{Inflation!in Kaluza--Klein and superstring theories|)}%
\index{Kaluza--Klein theories!inflation in|)}%
\index{Superstring theory!inflation in|)}%


\epigraph{%
If a man will begin with certainties, he shall end\\
in doubts;  but if he will be content to begin\\
with doubts he shall end in certainties.}%
{Francis Bacon (1561--1626)\\
The Advancement of Learning, Book V}
\chapter{\label{c10}Inflation and Quantum Cosmology}

\section{\label{s10.1}The wave
\index{Quantum cosmology|(}%
\index{Wave function of universe|(}%
\index{Universe!wave function of|(}%
\index{Wheeler--DeWitt equation|(}%
function of the universe}

Quantum cosmology is conceptually one of the most difficult
branches of theoretical physics.  This is due not just to such
difficulties as the ultraviolet divergences encountered in the
quantum theory of gravitation, but also in large measure to the
fact that the very formulation of the problems studied in quantum
cosmology is not at all a trivial matter.  The results of
research often appear paradoxical, and it requires an especially
open mind not to dismiss them outright.

The foundations of quantum cosmology were laid at the end of the
1960's by Wheeler and DeWitt [\cite{298}, \cite{299}].  But prior
to the advent of the inflationary universe scenario, a
description of the universe as a whole within the framework of
quantum mechanics seemed to most scientists to be an unnecessary
luxury.  When one describes macroscopic objects using quantum
mechanics, the results are usually the same as those given by
classical mechanics.  If the universe is indeed the largest
macroscopic entity that exists, then why bother to describe it
using quantum theory?

In the standard hot universe theory, this was a perfectly
legitimate question, since according to that theory the
observable part of the universe resulted from the expansion of a
region containing a total of perhaps $10^{87}$ elementary
particles.  But in the inflationary universe scenario, the entire
observable part of the universe (and possibly the entire universe
itself) was formed by virtue of the rapid expansion of a region
of size $l\la\m^{-1}\sim10^{-33}$ cm  containing perhaps not a
single elementary particle!  Quantum effects could thus have
played a pivotal role in events during the earliest stages of
expansion of the universe.

Until recently, the fundamental working tool in quantum cosmology
has been the Wheeler--DeWitt equation for the wave function of
the universe $\Psi(h_{ij},\varphi)$, where $h_{ij}$ is the
\index{Metric!three-dimensional spatial}%
\index{Three-dimensional spatial metric}%
three-dimensional spatial metric, and $\varphi$ is the matter
\index{Matter field}%
field.  The Wheeler--DeWitt equation is essentially the
Schr\"odinger equation for the wave function in the stationary
state given by $\displaystyle \frac{\partial\Psi}{\partial t}=0$
(see below).  It describes the behavior of the quantity $\Psi$ in
\index{Superspace}%
so-called superspace --- the space of all three-dimensional
metrics $h_{ij}$ (not to be confused with the superspace used to
describe supersymmetric theories!).  A detailed exposition of the
corresponding theory may be found in [\cite{298}--\cite{301}].
But the most interesting results in this sphere were obtained
using a simplified approach in which only a portion of the full
\index{Minisuperspace}%
\index{Superspace!minisuperspace}%
superspace, known as a minisuperspace, was considered, giving a
description of a homogeneous Friedmann universe;  the scale
factor of the universe $a$ took up the role of all the quantities
$h_{ij}$.  In this section, we will therefore illustrate the
basic problems relating to the calculation and interpretation of
the wave function of the universe with an example of the
minisuperspace approach.  In subsequent sections, we will discuss
the limits of applicability of this approach, the results
obtained via recently developed stochastic methods for describing
an inflationary universe [\cite{134}, \cite{135}, \cite{57},
\cite{132}, \cite{133}], and a number of other questions with a
bearing on quantum cosmology.

Let us consider, then, the theory of a scalar field $\varphi$
with the Lagrangian
\be
\label{10.1.1}
{\cal L}(g_{\mu\nu},\varphi)=-\frac{{\rm R}\,\m^2}{16\,\pi}
+\frac{1}{2}\,\partial_\mu\varphi\,\partial^\mu\varphi-\vf
\ee
in a closed Friedmann universe whose metric can be represented in
the form
\be
\label{10.1.2}
ds^2={\rm N}^2(t)\,dt^2-a^2(t)\,d\Omega_3^2\ ,
\ee
where ${\rm N}(t)$ is an auxiliary function that defines the
scale on which the time $t$ is measured, and
$d\Omega_3^2=d\chi^2+\sin^2\chi\,(d\theta^2+\sin^2\theta\,d\varphi^2)$
is the element of length on a three-dimensional sphere of unit
radius.  To obtain an effective Lagrangian that depends on $a(t)$
and $\varphi(t)$, one must integrate over angular variables in
the expression for the action ${\rm S}(g,\varphi)$, which with
the factor $\sqrt{g}$ taken into account gives $2\,\pi^2\,a^3$.
Then, making use of the fact that the universe is closed (has no
boundaries), one obtains the action in a form that depends only
on $a$ and $\dot a$, but not on $\ddot a$:
\be
\label{10.1.3}
{\rm L}(a,\varphi)=-\frac{3\,\m^2\,\pi}{4}\,
\left(\frac{\dot a^2\,a}{{\rm N}}-{\rm N}\,a\right)
+2\,\pi^2\,a^3\,{\rm N}\,
\left(\frac{\dot\varphi^2}{2\,{\rm N}^2}-\vf\right)\ .
\ee
The canonical momenta are
\ba
\label{10.1.4}
\pi_\varphi&=&\frac{\partial{\rm L}}{\partial\dot\varphi}
=\frac{2\,\pi^2\,a^3}{{\rm N}}\,\dot\varphi\ ,\\
\label{10.1.5}
\pi_a&=&\frac{\partial{\rm L}}{\partial\dot a}
=-\frac{3\,\m^2\,\pi}{2\,{\rm N}}\,\dot a\,a\ ,\\
\label{10.1.6}
\pi_{\rm N}&=&\frac{\partial{\rm L}}{\partial\dot{\rm N}}=0\ ,
\ea
and the Hamiltonian is
\ba
\label{10.1.7}
{\cal H}&=&\pi_\varphi\,\dot\varphi+\pi_a\,\dot a-{\rm L}(a,\varphi)
\nonumber \\
&=&-\frac{{\rm N}}{a}\,\left(\frac{\pi_a^2}{3\,\pi\,\m^2}
+\frac{3\,\pi\,\m^2}{4}\,a^2\right)
+\frac{{\rm N}}{a}\,\left(\frac{\pi^2_\varphi}{4\,\pi^2\,a^2}
+2\,\pi^2\,a^4\ \vf\right)\nonumber \\
&=&{\cal H}_a+{\rm H}_\varphi\ .
\ea
Here ${\cal H}_a$ and ${\cal H}_\varphi$ are the effective
Hamiltonians of the scale factor $a$ and scalar field $\varphi$
in the Friedmann universe.  The equation relating the canonical
variables $\pi_a$, $\pi_\varphi$, $a$,  and $\varphi$ follows
from (\ref{10.1.6}):
\ba
\label{10.1.8}
0=\frac{\partial{\cal H}}{\partial{\rm N}}=
\frac{{\cal H}}{{\rm N}}
&=&-\frac{1}{a}\,\left(\frac{\pi_a^2}{3\,\pi\,\m^2}
+\frac{3\,\pi\,\m^2}{4}\,a^{2}\right)\nonumber \\
&+&\frac{1}{a}\,\left(\frac{\pi_\varphi^2}{4\,\pi^2\,a^2}
+2\,\pi^2\,a^4\,\vf\right)\ .
\ea
Upon quantization, Eq. (\ref{10.1.8}) gives the relation that
governs the wave function of the universe:
\be
\label{10.1.9}
i\,\frac{\partial\Psi(a,\varphi)}{\partial t}={\cal H}\,\Psi=0\ .
\ee
In the usual fashion, the canonical variables are replaced with
the operators
\ba
\label{10.1.10}
\varphi&\rightarrow&\varphi,\qquad \pi_\varphi\rightarrow\frac{1}{i}\,
\frac{\partial}{\partial\varphi}\nonumber \\
a&\rightarrow& a,\qquad \pi_a\rightarrow\frac{1}{i}\,
\frac{\partial}{\partial a}\ ,
\ea
and Eq. (\ref{10.1.9}) takes the form
\be
\label{10.1.11}
\left(-\frac{1}{3\pi\m^2}\,\frac{\partial^2}{\partial a^2}
+\frac{3\pi\m^2}{4}\,a^2
+\frac{1}{4\pi^2a^2}\,\frac{\partial^2}{\partial\varphi^2}
-2\pi^2a^4\vf\right)\,\Psi(a,\varphi)=0\; .
\ee
\index{Superspace!minisuperspace}%
\index{Minisuperspace}%
This then is the Wheeler--DeWitt equation in minisuperspace.

Strictly speaking, it should be pointed out that ambiguities
relating to the commutation properties of $a$ and $\pi_a$ can
arise in the derivation of Eq. (\ref{10.1.11}). Instead of the
term $\displaystyle -\frac{\partial^2}{\partial a^2}$ in
(\ref{10.1.11}), one sometimes writes $\displaystyle
-\frac{1}{a^p}\,\frac{\partial}{\partial a}\,
a^p\,\frac{\partial}{\partial a}$, where the parameter $p$ can
take on various values.  In the semiclassical approximation,
which will be of particular interest to us later on, the actual
value of this parameter is unimportant, and in particular, one
can take $p=0$ and seek a solution of (\ref{10.1.11}).

Clearly, however, Eq. (\ref{10.1.11}) has many different
solutions, and one of the most fundamental questions facing us is
which of these solutions actually describes our universe.  Before
launching into a discussion of this question, we make several
remarks of a general nature that bear upon the interpretation of
the wave function of the universe.

First of all, we point out that the wave function of the universe
depends on the scale factor $a$ but, according to (\ref{10.1.9}),
{\it it is time-independent}.  How can this be reconciled with
\index{Time!in universe}%
\index{Universe!time in}%
the fact that our observable universe does depend on time?

Here we encounter one of the principal paradoxes of quantum
cosmology, a proper understanding of which is exceedingly
important.  The universe {\it as a whole} does not depend on time
because the very concept of such a change presumes the existence
of some immutable reference that does not appertain to the
universe, but relative to which the latter evolves.  If by ``the
universe'' we mean ``everything,'' then there remains no
{\it external observer} according to whose clocks the universe could
develop.  But in actuality, we are not asking why the universe is
developing, we are inquiring as to why we {\it perceive} it to be
developing.  We have thereby separated the universe into two
parts:  a macroscopic observer with clocks, and all the rest.
The latter can perfectly well evolve in time (according to the
clocks of the observer), despite the fact that the wave function
\index{Time!in universe}%
\index{Universe!time in}%
of the {\it entire} universe is time-independent [\cite{299}].

In other words, we arrive at our customary picture of a world
that evolves in time only after the universe has been divided
into two macroscopic parts, each of which develops
semiclassical.  The situation that ensues is reminiscent of
the theory of tunneling through a barrier:  the wave function is
defined inside the barrier, but it yields the probability
amplitude of finding a particle propagating in real time only
outside the barrier, where classical motion is allowed.  By
analogy, the universe too exists in its own right, in a certain
\index{Time!in universe}%
\index{Universe!time in}%
sense, but one can only speak of its {\it temporal} existence in
the context of the semiclassical evolution of the part that
remains after a macroscopic observer with clocks has emerged.

Thus, by the very fact of his existence, an observer somehow
reduces the overall wave function of the universe to that part
which describes the world that is observable to him.  This is
exactly the point of view espoused in the standard Copenhagen
\index{Copenhagen interpretation of quantum mechanics}%
\index{Quantum mechanics!Copenhagen interpretation of}%
interpretation of quantum mechanics --- the observer becomes not
just a passive viewer, but something more like a participant in the
\index{Creation!of universe}%
\index{Universe!creation of}%
creation of the universe [\cite{302}].

The situation is somewhat different in the
many-worlds
\index{Quantum mechanics!many-worlds interpretation of}%
\index{Many-worlds interpretation of quantum mechanics}%
interpretation of quantum mechanics [\cite{303}--\cite{309}],
which presently enjoys a sizable following among quantum
cosmologists.  In this interpretation, the wave function
$\Psi(h_{ij},\varphi)$ simultaneously describes all possible
universes together with the observers (of all possible kinds)
that inhabit them.  In performing a measurement, rather than
reducing the wave function of all of these universes to the wave
function of one of them (or a part of one of them), an observer
simply refines the issue of who he is and in which of these
universes he resides.  The same results are then obtained as in
the standard approach, but without recourse to the somewhat
ill-founded assumption of the reduction of the wave function at
the instant of measurement.

We shall not engage here in a detailed discussion of the
interpretation of quantum mechanics, a problem which becomes
particularly acute in the context of quantum cosmology
[\cite{302}, \cite{309}];  instead, we return to our discussion
\index{Universe!evolution of}%
of the evolution of the universe.

Another manifestation of the fact that the universe as a whole
does not change in time is that the wave function
$\Psi(a,\varphi)$ depends only on the quantities $a$ and
$\varphi$, and not on whether the universe is contracting or
expanding.  One might interpret this to mean that at the point of
maximum expansion of a closed universe,
the
\index{Arrow of time}%
\index{Time!arrow of}%
arrow of time is
somehow reversed, the total entropy of the universe begins to
decrease --- and observers are rejuvenated [\cite{310}].
However, to determine the direction of the arrow of time, one
must first divide the universe into two semiclassical
\index{Time!in universe}%
\index{Universe!time in}%
subsystems, one of which contains an observer with clocks.  In
general, the wave function of each such subsystem will certainly
not be symmetric under a change in the sign of $\dot a$.  After
the division of the universe into two semiclassical subsystems,
one can make use of the usual classical description of the
universe, according to which the total entropy of the universe
can only increase with time, and there is no way in which the
direction of the arrow of time can ever be reversed at the
instant of maximum expansion [\cite{311}].

We have discussed all these problems in such detail here in order
to demonstrate that not just the solution but even the
formulation of problems in the context of quantum cosmology is a
nontrivial matter.  The question of whether entropy can decrease
in a contracting universe, whether the arrow of time can be
reversed in a singularity or at the point of maximum expansion of
a closed universe, whether the universe can oscillate, has thus
far bothered many experts in quantum cosmology;  see, for
example, [\cite{312}, \cite{313}].  Above, we enunciated our own
viewpoint in this regard, but it should be understood that the
comprehensive investigation of these questions is only just
beginning.

The Wheeler--DeWitt equation (\ref{10.1.11}) has many different
solutions, and it is very difficult to ascertain which of them
actually describes our universe.  One of the most interesting
suggestions here was advanced by Hartle and Hawking [\cite{314}],
who proposed that the universe possesses a ground state, or a
\index{Vacuum}%
state of least excitation, similar to the vacuum state in quantum
field theory in Minkowski space.  By carrying out short-time
measurements in Minkowski space, one can see that the vacuum is
\index{Vacuum}%
\index{Virtual particles}%
\index{Particles!virtual}%
not empty, but is instead filled with virtual particles.
Similarly, the universe that we observe might be a virtual state
(but with a very long lifetime, due to inflation), and the
probability of winding up in such a state might be determinable
if the wave function of the ground state of the universe were
\index{Hartle--Hawking wave function}%
\index{Wave function!Hartle--Hawking}%
known.  According to the Hartle and Hawking hypothesis, the wave
function $\Psi(a,\varphi)$ of the ground state of the universe
with scale factor $a$ which is filled with a homogeneous field
$\varphi$ is given in the semiclassical approximation by
\be
\label{10.1.12}
\Psi(a,\varphi)\sim{\rm N}\,e^{-{\rm S}_{\rm E}(a,\varphi)}\ .
\ee
Here N is a normalizing factor, and ${\rm S}_{\rm E}(a,\varphi)$
is the Euclidean action corresponding to solutions of the
equations of motion for $a(\varphi(\tau),\tau)$ and
$\varphi(\tau)$ with boundary conditions
$a(\varphi(0),0)=a(\varphi)$, $\varphi(0)=\varphi$ in space with
a metric that has  Euclidean signature.

The reason for choosing this particular solution of the
Wheeler--DeWitt equation was explained as follows.  Consider the
Green's function of a particle that moves from the point $(0,t')$
to $({\bf x}, 0)$:
\ba
\label{10.1.13}
\langle {\bf x},0|0,t'\rangle &=&
\sum_n\Psi_n({\bf x})\,\Psi_n(0)\,e^{i\,{\rm E}_n\,t'}\nonumber \\
&=&\int d{\bf x}(t)\:\exp\{i\,{\rm S}[{\bf x}(t)]\}\ ,
\ea
where $\Psi_n({\bf x})$ is a time-independent eigenfunction of
the energy operator with eigenvalue ${\rm E}_n\ge0$.  Let us now
perform a rotation $t\rightarrow-i\,\tau$  and take the limit as
$\tau'\rightarrow-\infty$.  The only term that survives in the
sum (\ref{10.1.13}) is the one corresponding to the smallest
eigenvalue ${\rm E}_n$ (normalized to zero).  This implies that
\be
\label{10.1.14}
\Psi_0({\bf x})\sim{\rm N}\,\int d{\bf x}\:
\exp\{-{\rm S}_{\rm E}[{\bf x}(\tau)]\}\ .
\ee
Hartle and Hawking have argued that the generalization of this
\index{Hartle--Hawking wave function}%
\index{Wave function!Hartle--Hawking}%
result to quantum cosmology in the semiclassical approximation
will yield (\ref{10.1.13}).  For a slowly varying field $\varphi$
(and this is precisely the most interesting case from the
standpoint of implementing the inflationary universe scenario),
the solution of the Euclidean version of the Einstein equations
for $a(\varphi,\tau)$ is
\be
\label{10.1.15}
a(\varphi,\tau)\approx{\rm H}^{-1}(\varphi)\,
\cos[{\rm H}(\varphi)\,\tau]\equiv
a(\varphi)\,\cos[{\rm H}(\varphi)\,\tau]\ ,
\ee
where $\displaystyle {\rm H}(\varphi)=\sqrt{\frac{8\,\pi\,\vf}{3\,\m^2}}$,
and the corresponding Euclidean action is
\be
\label{10.1.16}
{\rm S}_{\rm E}(a,\varphi)=-\frac{3\,\m^4}{16\,\vf}\ ,
\ee
whereupon
\ba
\label{10.1.17}
\Psi[a(\varphi),\varphi]&\sim&{\rm N}\,
\exp\left(\frac{3\,\m^4}{16\,\vf}\right)=
{\rm N}\,
\exp\left(\frac{\pi\,\m^2}{2\,{\rm H}^2(\varphi)}\right)\nonumber \\
&=&{\rm N}\,\exp\left(\frac{\pi\,\m^2\,a^2(\varphi)}{2}\right)\ .
\ea
Hence, it should follow that the probability of detecting a
closed universe in a state with field $\varphi$ and scale factor
$a(\varphi)={\rm H}^{-1}(\varphi)$ is
\ba
\label{10.1.18}
{\rm P}[a(\varphi),\varphi]&\sim&{\rm N}^2\,|\Psi[a(\varphi),\varphi)]|^2
\sim{\rm N}^2\,\exp\left(\frac{3\,\m^4}{8\,\vf}\right)\nonumber \\
&=&{\rm N}^2\,\exp[\pi\,\m^2\,a^2(\varphi)]\ .
\ea
If the ground state of the universe were a state with
$\varphi=\varphi_0$ and $0<{\rm V}(\varphi_0)\ll\m^4$, then the
normalization factor ${\rm N}^2$ that ensures a total probability
of all realizations being unity would have to be
\be
\label{10.1.19}
{\rm N}\sim\exp[-\pi\,\m^2\,a_0^2]
=\exp\left(-\frac{3\,\m^4}{8\,{\rm V}(\varphi_0)}\right)\ ,
\ee
where $a_0={\rm H}^{-1}(\varphi_0)$.  From Eqs. (\ref{10.1.18})
and (\ref{10.1.19}), it follows that
\be
\label{10.1.20}
{\rm P}[a(\varphi),\varphi]\sim
\exp\left[\frac{3\,\m^4}{8}\,
\left(\frac{1}{\vf}-\frac{1}{{\rm V}(\varphi_0})\right)\right]\ .
\ee
To calculate the probability that $a\ll a_0={\rm H}^{-1}(\varphi_0)$
or $a\gg a_0={\rm H}^{-1}(\varphi_0)$ when
$\varphi=\varphi_0$, we must venture outside the confines of the
semiclassical approximation (\ref{10.1.12}) or solve Eq.
\index{WKB approximation}%
(\ref{10.1.11}) directly in the WKB approximation.  According to
[\cite{314}],
\ba
\label{10.1.21}
\Psi(a\ll a_0)&\sim&
\exp\left[\frac{\pi}{2}\,\m^2(a^2-a_0^2)\right]\ ,\\
\label{10.1.22}
\Psi(a\gg a_0)&\sim&
\exp\left[\frac{i\,{\rm H}(\varphi_0)\,\m^2\,a^3}{3}\right]\nonumber \\
&+&\exp\left[-\frac{i\,{\rm H}(\varphi_0)\,\m^2\,a^3}{3}\right]\ .
\ea

Unfortunately, the arguments used by Hartle and Hawking to
justify (\ref{10.1.12}) are far from universally applicable.  In
\index{Hartle--Hawking wave function}%
\index{Wave function!Hartle--Hawking}%
fact, the Euclidean rotation alluded to above can be used to
eliminate all but the zeroth-order term from (\ref{10.1.13}) only
if ${\rm E}_n>0$ for all $n>0$.  While the energy of excitations
of the scalar field $\varphi$ is positive, the energy of the
scale factor $a$ is negative, so that these sum to zero;  see
(\ref{10.1.7}) and (\ref{10.1.9}).  In such a situation, there is
no general prescription for extracting the ground state $\Psi_0$
from the sum (\ref{10.1.13}) by rotation to Euclidean space.  To
investigate the properties of the field $\varphi$ on scales much
smaller than the size of the closed universe, this is an
unimportant issue, and we can simply quantize the field $\varphi$
against the background of the classical gravitational field and
perform the standard Euclidean rotation $t\rightarrow-i\,\tau$.
\index{Probability density function}%
This is exactly the reason why the probability density function
(\ref{10.1.20}) is the same as the distribution (\ref{7.4.7}),
which was derived using more conventional methods.  On the other
hand, in those situations where the scale factor $a$ itself must
be quantized (for instance, in a description of the quantum
creation\index{Creation!of universe}\index{Universe!creation of}
of the universe from a state with $a = 0$, i.e., from
``nothing'' [\cite{315}--\cite{317}, \cite{289}, \cite{290},
\cite{318}]), the corresponding problem becomes much more
serious.

Fortunately, this can be avoided if the quantum properties of the
field $\varphi$ are unimportant for our purposes at the epoch of
interest --- for example, if $\varphi$ is a classical slowly
varying field whose sole role is to produce a nonzero vacuum
energy ${\rm V}(\varphi)$ (cosmological term).  One can then
neglect quantum effects associated with the scalar field, and to
isolate the ground state $\Psi(a,\varphi)$, corresponding to the
lowest excitation state of the scale factor $a$, one need only
carry out the rotation $t\rightarrow+i\,\tau$ in order to
suppress the contribution to (\ref{10.1.13}) from negative-energy
excitations.\footnote{It should be borne in mind that there are
no {\it physical} excitations of the gravitational field with
negative energy. Therefore, for a consistent quantization of the
scale factor $a$ one should introduce Faddeev--Popov ghosts.
\index{Faddeev--Popov ghosts}%
\index{Ghosts, Faddeev--Popov}%
However, as usual, ghosts do not contribute to $\Psi(a,\varphi)$
in the semiclassical approximation.}
This gives contribution to
(\ref{10.1.13}) from negative-energy
excitations.\footnotemark[1] This gives
\be
\label{10.1.23}
\Psi(a,\varphi)\sim{\rm N}\,e^{{\rm S}_{\rm E}(a,\varphi)}
\sim{\rm N}\,\exp\left(-\frac{3\,\m^4}{16\,\vf}\right)\ ,
\ee
and the probability that the universe appears in a state with
field $\varphi$ is
\be
\label{10.1.24}
{\rm P}[a(\varphi),\varphi)]\sim|\Psi|^2
\sim{\rm N}^2\,\exp\left(-\frac{3\,\m^4}{8\,\vf}\right)\ .
\ee

Equation (\ref{10.1.23}) was first obtained using the method
described above [\cite{319}];  it was later derived by Zeldovich
and Starobinsky [\cite{320}], Rubakov [\cite{321}], and Vilenkin
[\cite{322}] using a different method.  For the reasons to be
\index{Tunneling wave function}%
\index{Wave function!tunneling}%
discussed soon, we will call (\ref{10.1.23}) {\it tunneling wave
function}.

The obvious difference between Eqs. (\ref{10.1.24}) and
(\ref{10.1.18})--\linebreak[10000]
(\ref{10.1.21}) is in the sign of the argument
of the exponential.  This difference is extremely important,
since according to (\ref{10.1.18}) and (\ref{10.1.20}), the
probability of {\it detecting} the universe in a state with a
large value of ${\rm V}(\varphi)$ is exponentially small.  In
contrast, Eq. (\ref{10.1.24}) tells us that the universe is most
likely {\it created} in a state with ${\rm V}(\varphi)\sim\m^4$.
This is consistent with our previous expectations, and leads to a
natural implementation of the chaotic inflation scenario
[\cite{319}].

In order to comprehend the physical meaning of the
Hartle--Hawking wave function (\ref{10.1.12}), let us compare the
\index{Hartle--Hawking wave function}%
\index{Wave function!Hartle--Hawking}%
solutions of Eqs. (\ref{10.1.21}) and (\ref{10.1.22}) with
solutions for the scalar field (\ref{1.1.3}).  One possible
interpretation of the solution (\ref{10.1.21}) is that the wave function
$\displaystyle\exp\left[\frac{i\,{\rm H}(\varphi_0)\,\m^2\,a^3}{3}\right]$
describes a universe with decreasing scale factor $a$ (compare
with the wave function of a particle with momentum $p$, $\psi\sim
e^{-i\,p\,x}$), while the wave function
$\displaystyle\exp\left[-\frac{i\,{\rm H}(\varphi_0)\,\m^2\,a^3}{3}\right]$
corresponds to a universe with increasing scale factor.  Bearing
in mind, then, that the corresponding motion takes place,
according to (\ref{10.1.11}), in a theory for which the effective
potential of the scale factor is
\be
\label{10.1.25}
{\rm V}(a)=\frac{3\,\pi\,\m^2}{4}\,a^2-2\,\pi^2\,a^4\,\vf\ ,
\ee
the interpretation of the solutions (\ref{10.1.21}) and
(\ref{10.1.22}) becomes quite straightforward (although different
authors are still not in complete agreement on this point).  The
wave function (\ref{10.1.22}) describes a wave incident upon the
barrier ${\rm V}(a)$ from the large-$a$ side and a wave reflected
from the barrier;  when $a<{\rm H}^{-1}$ (i.e., incidence below
the barrier height), the wave is exponentially damped in
\begin{figure}[t]\label{f35}
\centering \leavevmode\epsfysize=6cm \epsfbox{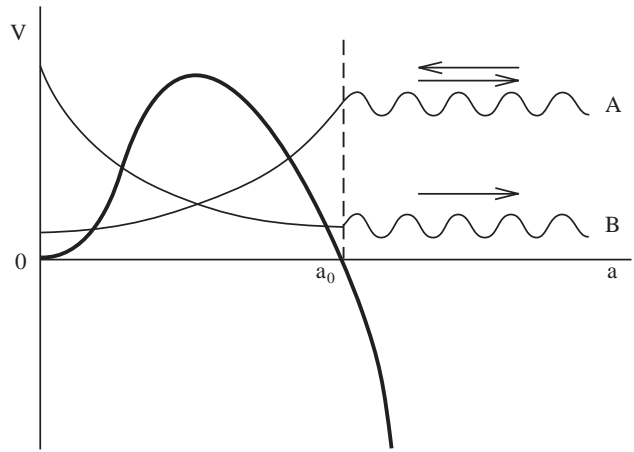} \caption{The
effective potential ${\rm V}(a)$ for the scale factor $a$ of
(\ref{10.1.25}).  This figure also gives a
\index{Hartle--Hawking wave function}%
\index{Wave function!Hartle--Hawking}%
somewhat tentative representation of the Hartle--Hawking wave
function (\ref{10.1.22}) (curve A), and of the wave function
\index{Quantum birth of universe}%
\index{Universe!quantum birth of}%
(\ref{10.1.23}), which describes the quantum birth of the
universe from the state $a =0$ (curve B).}
\end{figure}
accordance with (\ref{10.1.21}) (see Fig. 10.1).  The
physical meaning of this solution is most easily grasped if one
recalls that a closed de Sitter space with ${\rm V}(\varphi_0)>0$
first contracts and then expands:  $a(t)={\rm H}^{-1}\cosh({\rm H}\,t)$.
\index{Hartle--Hawking wave function}%
\index{Wave function!Hartle--Hawking}%
The Hartle--Hawking wave function (\ref{10.1.21}) accounts for
the ``broadening'' of this semiclassical trajectory, and allows
for the fact that at the quantum level, the scale factor can
become less than ${\rm H}^{-1}$ at the point of maximum
contraction.  The absence of exponential suppression for $a>{\rm H}^{-1}$
(\ref{10.1.22}) is related to the fact that values $a>{\rm H}^{-1}$
are classically allowed [\cite{314}].  Observational cosmological
data put the present-day energy density of the vacuum
${\rm V}(\varphi_0)$ at no more than $10^{-29}$  g/cm$^3$, which
corresponds to ${\rm H}^{-1}\ga10^{28}$ cm.  The evolution of a
de Sitter space whose minimal size exceeds $10^{28}$ cm has
\index{Universe!evolution of}%
nothing in common with the evolution of the universe in which we
now live.  Therefore, within the scope of the foregoing
\index{Hartle--Hawking wave function}%
\index{Wave function!Hartle--Hawking}%
interpretation, the Hartle--Hawking wave function does not give a
proper description of our universe in the minisuperspace
approximation studied above.  We face a similar difficulty if we
attempt (without justification) to use this wave function instead
of the function (\ref{10.1.23}) to account for the very earliest
\index{Universe!evolution of}%
stages in the evolution of the universe, since according to
(\ref{10.1.18}) and (\ref{10.1.20}), the likelihood of a
prolonged inflationary stage would in that case be exponentially
small.

One might suggest another possible interpretation of the
\index{Hartle--Hawking wave function}%
\index{Wave function!Hartle--Hawking}%
Hartle--Hawking wave function, namely that its square gives the
probability density function for an observer to detect that he is
in a universe of a given type not at the instant of its creation,
but at the instant of his first measurement, prior to which he
\index{Universe!evolution of}%
cannot say anything about the evolution of the
universe.\footnote{ From this standpoint, one might tentatively
say that the wave function (\ref{10.1.23}) is associated with the
creation of the universe, while (\ref{10.1.12}) is associated with
the creation of an observer.}
Such an interpretation may turn out to be eminently reasonable
(and, in the final analysis, independent of the choice of
observer) if, as originally supposed by Hartle and Hawking, a
ground state actually exists for the system in question, so that
the probability distribution under consideration turns out to be
stationary, like the vacuum state or ground state of an
equilibrium thermodynamic system.  And in fact, as we have
\index{Hartle--Hawking wave function}%
\index{Wave function!Hartle--Hawking}%
already noted, the Hartle--Hawking wave function provides a good
description of the quasistationary distribution of the field
$\varphi$ in an intermediate metastable state (see
(\ref{10.1.20}) and (\ref{7.4.7})).

On the other hand, a stationary distribution (\ref{10.1.20}) of
the field $\varphi$ is only possible if
$\displaystyle m^2=
\frac{d^2{\rm V}}{d\varphi^2}\ll{\rm H}^2\sim\frac{\vf}{\m^2}$
in the vicinity of the absolute minimum of $\vf$. Not a single
realistic model of the inflationary universe satisfies this
requirement.  The only stationary distribution of a
(quasi)classical field $\varphi$ in realistic models that we are
presently aware of (see the discussion of this question in
Sections \ref{s7.4}, \ref{s10.2}, and \ref{s10.3}) is the trivial
delta-function distribution, with the field totally concentrated
at the minimum of $\vf$.  But this is not at all the result that
researchers in quantum cosmology are trying to obtain when they
\index{Hartle--Hawking wave function}%
\index{Wave function!Hartle--Hawking}%
discuss the Hartle--Hawking wave function and assume that the
appropriate probability distribution is given by Eq.
(\ref{10.1.20}).

All these caveats notwithstanding, we would rather not draw any
hasty conclusions, which would be a particularly dangerous thing
to do in a science whose ultimate foundations have yet to be
laid.  The mathematical structure proposed by Hartle and Hawking
is quite elegant in and of itself, and quite possibly we may yet
find a way to take advantage of it.  Our fundamental objection to
the possibility of a stationary distribution of the field
$\varphi$ in an inflationary universe is based upon a study of a
(typical) situation in which the field possesses one (or a few)
absolutely stable vacuum states.  But instances are known in
which theories are characterized by the value of some
time-independent field, topological invariant, or other parameter
characterizing properties of the vacuum state which might govern,
say, the strength of CP violation, the energy of the vacuum, and
so forth.

One such parameter is the angle $\theta$ which characterizes the
vacuum properties in quantum chromodynamics [\cite{183}].  It is
possible that the cosmological term and many coupling constants
in elementary particle theory [\cite{345}, \cite{346},
\cite{349}] are also vacuum parameters of this type.  Their
\index{Superselection rules}%
time-independence may be guaranteed by superselection rules of
some sort [\cite{346}].  But in the context of the
many-worlds
\index{Quantum mechanics!many-worlds interpretation of}%
\index{Many-worlds interpretation of quantum mechanics}%
interpretation of quantum mechanics, the question of what the
properties of the world (of the vacuum state) are in which the
observer finds himself at the instant of his first observation is
a perfectly reasonable one.  The suggestion that the appropriate
probability distribution will be given by the square of the
\index{Hartle--Hawking wave function}%
\index{Wave function!Hartle--Hawking}%
Hartle--Hawking wave function [\cite{346}] seems to us worthy of
serious consideration.  At the same time, the question of
choosing between the Hartle--Hawking function and the function
(\ref{10.1.23}) under these circumstances becomes especially
important.  As we pointed out earlier, the Hartle--Hawking wave
function actually gives the proper results when one considers the
(quasi)stationary distribution of a scalar field $\varphi$ with
positive energy  in a classical de Sitter background (see
(\ref{10.1.20})).  On the other hand, if the evolution of matter
fields (and vacuum parameters) is insignificant, then the wave
function may possibly be determined by an expression like
(\ref{10.1.23}).  We will return to the discussion of this
question in Section \ref{s10.7}.

A possible interpretation (and an alternative derivation) of the
wave function (\ref{10.1.23}) can also be obtained by studying
\index{Tunneling wave function}%
\index{Wave function!tunneling}%
tunneling through the barrier (\ref{10.1.25}), but from the
direction of small $a$ rather than large $a$
[\cite{320}--\cite{322}].  Indeed, one can readily show that
(with $\varphi \approx \mbox{const}$) a solution of Eq.
(\ref{10.1.11}) exists,  which behaves as $\displaystyle
\exp\left(-\frac{\pi}{2}\,\m^2\,a^2\right)$  for
$$
a<a_0={\rm H}^{-1}(\varphi)\gg\m^{-1}
$$
(cf. (\ref{10.1.21})), while for $a\gg{\rm H}^{-1}(\varphi)$,
it is represented by a wave
$$
\sim\exp\left(-\frac{\pi\,\m^2\,a_0^2}{2}
-\frac{i\,{\rm H}\,\m^2\,a^3}{3}\right)
$$
emerging from the barrier and moving off toward large $a$; see
Fig. 10.1.  Damping of the wave as it emerges from the
barrier is of order
$$
\exp\left(-\frac{\pi\,\m^2\,a_0^2}{2}\right)\sim
\exp\left(-\frac{3\,\m^4}{16\,\vf}\right)\ ,
$$
which exactly corresponds to Eq. (\ref{10.1.23}) above.  The wave
function (\ref{10.1.23}) thus describes the process of quantum
creation of a closed inflationary universe filled with a
\index{Tunneling wave function}%
\index{Wave function!tunneling}%
homogeneous field $\varphi$ due to tunneling from a state with
scale factor $a=0$, or in other words, from ``nothing''
[\cite{319}--\cite{322}].

We now attempt to provide a plausible interpretation of this
result, and to clarify the reason why the probability of the
quantum creation of a closed universe only becomes large when
$\vf\sim\m^4$.  To this end we examine a closed de Sitter space
with energy density $\vf$.  Its volume at the epoch of maximum
contraction ($t=0$) is of order
${\rm H}^{-3}(\varphi)\sim\m^3\,[\vf]^{-3/2}$,  and the total energy of
the scalar field contained in de Sitter space at that instant is
approximately $\displaystyle
{\rm E}\sim\vf\,{\rm H}^{-3}(\varphi)\sim\frac{\m^3}{\sqrt{\vf}}$.
When $\vf\sim\m^4$, the total energy of the scalar field is ${\rm E}\sim\m$.
By the uncertainty principle, one cannot rule out the
possibility of quantum fluctuations of energy E lasting for a
time $\Delta t\sim{\rm E}^{-1}\sim\m^{-1}$.  However, within a
time of this order (or slightly longer), de Sitter space of
initial size $\sim{\rm H}^{-1}\sim\m^{-1}$ becomes exponentially
large, and one can consider it to be an inflationary universe
emerging from ``nothing'' (or from the space-time foam).  For
small $\vf$, the probability that this process will come to pass
should be extremely low, since as $\vf$ decreases, the minimum
energy E of a scalar field in de Sitter space increases, rather
than falling off, and the typical lifetime of the corresponding
quantum fluctuation becomes much shorter than the Planck time.

It is important to note here that we are dealing with the
creation of a compact universe with no boundaries, so that no
supplementary conditions such as $\varphi=0$ are required at the
boundary of the bubble that is formed (compare this with the
discussion of the ``budding'' of the universe from  Minkowski
space in Section \ref{s10.3}).  If the effective potential
${\rm V}(\varphi)$ is flat enough, so that the field $\varphi$ rolls
down to its minimum in a time much longer than ${\rm H}^{-1}$,
then at the instant of its creation, the universe will ``know
nothing'' of the location of the minimum of $\vf$ or how far the
initial field $\varphi$ is displaced from it.  To a first
approximation, the probability of creation of the universe is
given solely by the magnitude of $\vf$, in accordance with
(\ref{10.1.24}).

Generally speaking, the extent to which the probability of
creation of a universe with $\vf\ll\m^4$ is suppressed may be
lessened somewhat if particle creation at the time of tunneling
is taken into account [\cite{321}].  Furthermore, exponential
suppression may be absent, by and large, during the creation of a
compact (flat) universe with nontrivial topology [\cite{320}].
For us, the only important thing is that, as expected, there is
no exponential suppression of the probability for creation of an
inflationary universe with $\vf\sim\m^4$ --- that is, from the
present point of view, the initial conditions for implementation
of the chaotic inflation scenario are also found to be quite
natural.

Note that the distinction between the creation of the universe
from a singularity and quantum creation from ``nothing'' at
Planck\index{Density!Planck}\index{Planck density}
density is rather tentative.  In either case, one is
dealing with the emergence of a region of classical space-time
from the space-time foam.  The terminological distinction
consists of the fact that by  creation from ``nothing,'' one
usually means that a description of the evolution of the universe
according to the classical equations of motion begins only at
large enough $a$.  But due to large quantum fluctuations of the
metric at $\rho\ga\m^4$, it also turns out to be impossible to
describe the universe near the singularity (i.e., at small $a$).
An important feature of either case is that as $a\rightarrow0$,
only quantum cosmology can provide a description of the evolution
of the universe, a circumstance that can lead to surprising
consequences.

Consider, for example, a possible model for the evolution of a
closed inflationary universe.  This model will be incomplete, and
\index{Inflationary universe!closed evolution of}%
aspects of its interpretation will be open to argument, but on
the whole it furnishes a good illustration of some novel
possibilities being discussed within the context of quantum
cosmology.

Thus, suppose that the universe was originally in a state $a=0$.
Quantum fluctuations of the metric at that time were extremely
large, and there were neither clocks nor rulers.  Any
observations made by an imaginary observer at that epoch would
have been uncorrelated with one another, and one could not even
have said which of those observations came first or last.  The
results of measurements could not be remembered, which implies
that with each new measurement the observer would effectively
find himself in a completely new space.  If during one of these
measurements he found himself inside a hot universe that was not
passing through an inflationary stage, then the characteristic
lifetime of such a universe would turn out to be of order
$\m^{-1}$  and its total energy ${\rm E}\sim\m$; it would
therefore be essentially indistinguishable from a quantum
fluctuation.  But if the observer detected that he was in an
inflationary universe, he would then be able to make clocks and
rulers, and over an exponentially long period of time he could
describe the evolution of the universe with the aid of the
classical Einstein equations.  After a certain time, the universe
would be reheated, following which it would proceed through a
state of maximum expansion and begin to contract.  When it had
reached a state of
Planck\index{Density!Planck}\index{Planck density}
density (which would occur when
$a\gg\m^{-1}$), it would become impossible to use clocks and
rulers and thereby introduce any meaningful concept of time,
entropy density, and so forth, due to large quantum fluctuations
of the metric.  One could say that quantum fluctuations near the
singularity in effect erase from the memory of the universe any
information about the properties it had during its period of
semiclassical evolution.  Consequently, after Planck densities
have been attained, subsequent observations again become
disordered; it is even dubious, strictly speaking, that one could
say they are {\it subsequent}.  At some point, the observer
detects that he is in an inflationary universe, and everything
begins anew.  Here the parameters of the inflationary universe
depend only on the value of the wave function $\Psi(a,\varphi)$,
and not on its history, which has been ``forgotten'' during the
passage through the purgatory of
Planck\index{Density!Planck}\index{Planck density}
densities.  We thus
obtain a model for an oscillating universe in which there is no
increase of entropy during each successive cycle [\cite{298},
\cite{323}].  Other versions of this model exist, and are based
on a hypothesized limiting density $\rho\sim\m^4$ [\cite{313}] or
gravitational confinement at $\rho\ga\m^4$ [\cite{116}].

The examples considered in this section indicate how interesting
the investigation of solutions of the Wheeler--DeWitt equation
is,  and by the same token, how difficult it is to choose and
interpret a satisfactory solution.  Studies of this question are
only just beginning [\cite{324}].  Some of the problems
encountered are related to the minisuperspace approximation
employed, and some to the fact that we wish to derive (or guess)
the correct solution to the full quantum mechanical problem
without an adequate understanding of the properties of the global
structure of the inflationary universe at a more elementary
level.  To fill this gap, it will prove very useful to study the
properties of the inflationary universe via the stochastic
approach to inflation, which occupies an intermediate position
between the classical description of the inflationary universe
and an approach based on solving the Wheeler--DeWitt equation.
\index{Quantum cosmology|)}%
\index{Wave function of universe|)}%
\index{Universe!wave function of|)}%
\index{Wheeler--DeWitt equation|)}%

\section[Quantum cosmology and global structure]%
{\label{s10.2}Quantum cosmology and the global
\index{Global structure of inflationary universe|(}%
\index{Quantum cosmology!global structure of inflationary universe and|(}%
structure of the\protect\\ inflationary universe}

One of the principal shortcomings of an approach based on
minisuperspace is the initial presumption of global homogeneity
of the universe.  The only explanation for the homogeneity of the
universe which is known at present is based on the inflationary
universe scenario.  However, as we showed in Section \ref{s1.8},
effects related to long-wave fluctuations of the scalar field
prevent the geometry of an inflationary universe from having
anything in common with that of a homogeneous Friedmann space on
\index{Universe!homogeneous}%
much larger scales (at $l\ga l^*$).  Instead of a homogeneous
universe that comes into being as a whole at some instant of time
$t=0$, we must deal with a globally inhomogeneous
self-reproducing inflationary universe, whose evolution has no
end and quite possibly may not have had a unique beginning.
Thus, many of the most important properties of the inflationary
universe cannot be understood or studied within the context of
the minisuperspace approach.

In Chapter \ref{c1}, we presented the simplest description of the
mechanism of self-regeneration of inflationary domains of the
universe in the chaotic inflation scenario [\cite{57}].  Below,
we engage in a more detailed investigation of this problem
[\cite{132}, \cite{133}].

One could carry out this investigation in the coordinate system
(\ref{7.5.8}), which is especially convenient for analyzing
density inhomogeneities in an inflationary universe [\cite{218},
\cite{220}].  If, however, one is interested in describing the
evolution of the universe from the point of view of a comoving
observer, then it is more convenient to go to a synchronous
coordinate system, which can be so chosen that the metric of an
inflationary universe on scales much larger than ${\rm H}^{-1}$
may be written in the form [\cite{135}, \cite{133}]
\be
\label{10.2.1}
ds^2\approx dt^2-a^2({\bf x},t)\,d{\bf x}^2\ ,
\ee
where
\be
\label{10.2.2}
a({\bf x},t)\sim
\exp\left\{\int^t_0{\rm H}[\varphi({\bf x},t)]\:dt\right\}\ .
\ee
What these expressions mean is that an inflationary universe in a
neighborhood of size $l\ga{\rm H}^{-1}$ about every point x looks
\index{Hubble ``constant''}%
like a homogeneous inflationary universe with Hubble parameter
${\rm H}[\varphi({\bf x},t)]$.  To study the global structure of
the inflationary universe in this approximation, it turns out to
be sufficient to study the independent local evolution (in
accordance with the ``no hair'' theorem in de Sitter space) of
the field $\varphi$ within each individual region of the
inflationary universe of size $l\sim{\rm H}^{-1}$ (or with {\it initial}
size $l_0\sim{\rm H}^{-1}$), and then attempt to discern
the overall picture using Eqs. (\ref{10.2.1}) and (\ref{10.2.2}).
The local evolution of the field $\varphi$ in regions whose size
is of order ${\rm H}^{-1}$ is governed by the diffusion equations
(\ref{7.3.22}), (\ref{7.4.4}), and (\ref{7.4.5}), taking into
consideration the dependence of the diffusion and mobility
coefficients $\displaystyle {\rm D}=\frac{{\rm H}^3}{8\,\pi^2}$
on the magnitude of the field $\varphi$ [\cite{135}, \cite{133}].

The simplest possibility would be to study stationary solutions
of Eqs. (\ref{7.4.4}) and (\ref{7.4.5}). The corresponding
solution in the general case might also depend on the stationary
probability flux $j_c=\mbox{const}$, and for $\vf\ll\m^4$, it is
given by [\cite{135}]
\be
\label{10.2.3}
{\rm P}_c(\varphi)\sim
\mbox{const}\cdot \exp\left[\frac{3\,\m^4}{8\,\vf}\right]-
2\,j_c\,\frac{\sqrt{6\,\pi\,\vf}}{\m\,{\rm V}'(\varphi)}\ .
\ee
Unfortunately, attempts to provide a physical interpretation of
this solution meet with all sorts of difficulties.  As in Section
\ref{s7.4}, let us first consider the case $j_c=0$.  One can
readily show that Eq. (\ref{10.2.3}) is identical to the square
\index{Hartle--Hawking wave function}%
\index{Wave function!Hartle--Hawking}%
of the Hartle--Hawking wave function (\ref{10.1.17}), namely
(\ref{10.1.18}).  But since the effective potential $\vf$
vanishes at its minimum, which corresponds to the vacuum state in
the observable part of the universe, the distribution
(\ref{10.2.3}) is non-normalizable.  This is an especially easy
problem to comprehend in the chaotic inflation scenario for
theories with $\vf\sim\varphi^{2\,n}$; in these theories,
inflation takes place only when $\varphi\ga\m$.  In such
theories, then, there is no diffusion flux out of the region with
$\varphi\la\m$ and into the region with $\varphi\ga\m$.  But such
a flux would provide the only way to compensate for the classical
rolling down of the field to the minimum of $\vf$, which is
necessary for the existence of a stationary distribution
${\rm P}_c(\varphi)$ when $j_c=0$.

The issue of the interpretation of the second term in
(\ref{10.2.3}) is even more complicated.  As we noted in Section
\ref{s7.4}, formally, in theories with $\vf\sim\varphi^{2\,n}$,
this solution does not exist in general, as it is odd in
$\varphi$ while ${\rm P}_c(\varphi)$ must always be positive.  We
can avert this problem to a certain extent by recalling that in
these theories, Eq. (\ref{7.4.5}) itself holds only on the
segment $\m\la\varphi\la\varphi_{\rm P}$, where
${\rm V}(\varphi_{\rm P})\sim\m^4$.  But this is not an entirely
satisfactory response.  It is actually straightforward to show
that the second term in (\ref{10.2.3}) is a solution of Eq.
(\ref{7.3.22}) in which the first (diffusion) term is omitted.
Thus, we are simply dealing with a classical ``rolling'' of the
field $\varphi$ away from a region with $\vf\gg\m^4$, where the
diffusion equation is not valid.  In this instance, a stationary
distribution ${\rm P}_c(\varphi,t)$ can only be sustained through
a constant flux $j_c$ out of a region with
${\rm V}(\varphi_{\rm P})\gg\m^4$.  One could attempt to
interpret this flux as a current corresponding to the probability
of quantum creation of new domains of the universe with
${\rm V}(\varphi_{\rm P})\ga\m^4$, per unit initial coordinate volume.
But as Starobinsky has already emphasized in [\cite{135}], where
a solution of the type (\ref{10.2.3}) was first derived, at
present we can neither give a rigorous existence proof for such a
solution, nor can we say anything definite about the magnitude of
$j_c$ if it is nonzero.  The feasibility of the interpretation
suggested above does not follow from the derivation of Eqs.
(\ref{7.4.4}) and (\ref{7.4.5}) given in
[\cite{132}--\cite{135}].  Moreover, since the vast preponderance
\index{Universe!volume of}%
\index{Volume of universe}%
of the initial coordinate volume of the inflationary universe
eventually transforms into a state with $\varphi\la\m$,
${\rm V}(\varphi_{\rm P})\ll\m^4$, the validity of the assumed
constancy of the probability current for quantum creation of the
new domains of the universe per unit {\it initial} volume seems
\index{Universe!volume of}%
\index{Volume of universe}%
not to be well-founded.  It is not entirely clear, for example,
why it is necessary to require a stationary distribution
${\rm P}_c(\varphi,t)$ rather than a probability distribution for
finding the field $\varphi$ at time $t$, per unit {\it physical}
volume, and taking into account the increase in volume due to the
quasiexponential expansion of the universe (\ref{10.2.2}), which
proceeds at different rates in regions filled with different
fields $\varphi$.

One can attain a proper understanding of the situation for
stationary solutions only through a comprehensive analysis of
nonstationary solutions of the diffusion equation under the most
general initial conditions for the distribution ${\rm
P}_c(\varphi,t)$.  As stated above, we are interested in the
probability ${\rm P}_c(\varphi,t)$ of finding the field $\varphi$
at time $t$ in a region of initial size ${\rm O}({\rm H}^{-1})$.
Due to inflation, the initial inhomogeneities of the field
$\varphi$ on this scale become exponentially small, while the
amplitude of semiclassical perturbations $\delta\varphi$ with
wavelengths $l\ga{\rm H}^{-1}$ on this scale do not exceed H;
see (\ref{7.3.12}).  Bearing in mind that inflation occurs in
theories with $\vf\sim\varphi^{2\,n}$ when $\varphi\ga\m$,  the
condition $\vf\ll\m^4$ implies that $\delta\varphi\sim{\rm
H}\ll\varphi$, or in other words the field $\varphi$ is, to a
very good approximation, homogeneous on a scale $l\sim{\rm H}^{-1}$.

We may therefore assume without loss of generality that at time
$t=0$, the field $\varphi$ is equal to some constant $\varphi_0$
in the region under consideration --- the size of which is
${\rm O}({\rm H}^{-1})$ --- or in other words,
${\rm P}_c(\varphi,t=0)=\varphi(\varphi-\varphi_0)$.  Solutions
of Eq.  (\ref{7.3.22}) with  such initial conditions were
investigated in [\cite{132}, \cite{133}], and there it was found
that these solutions are all nonstationary.  The distribution
${\rm P}_c(\varphi,t)$ first broadens, and its center is then
displaced toward small $\varphi$, being governed by the same law
as the classical field $\varphi(t)$.  Meanwhile, the distribution
${\rm P}_p(\varphi,t)$ of the physical volume occupied by the
field $\varphi$ behaves differently depending on the value of the
initial field $\varphi=\varphi_0$.  For small $\varphi_0$,
${\rm P}_p(\varphi,t)$ behaves in almost the same way as
${\rm P}_c(\varphi,t)$, but for sufficiently large $\varphi_0$, the
distribution ${\rm P}_p(\varphi,t)$ starts to edge toward large
values of the field $\varphi$ as $t$ increases, which leads to
the onset of the self-reproducing inflationary regime discussed
in Section \ref{s1.8}.

Referring the reader to [\cite{132}, \cite{133}] for details, let
us elucidate the behavior of the distributions ${\rm
P}_c(\varphi,t)$ and ${\rm P}_p(\varphi,t)$ using the theory
$\displaystyle \vf=\frac{\lambda}{4}\,\varphi^4$ as an example.
In order to do so, we divide the semiclassical field $\varphi$
into a homogeneous classical field $\varphi(t)$ and
inhomogeneities $\delta\varphi({\bf x},t)$ with wavelengths
$l\ga{\rm H}^{-1}$ (see (\ref{7.5.7})):
\be
\label{10.2.4}
\varphi({\bf x},t)=\varphi(t)+\delta\varphi({\bf x},t)\ .
\ee
It can readily be shown that during the inflationary stage, the equations of
motion for $\varphi(t)$ and $\delta\varphi$ in the metric
(\ref{10.2.1}), (\ref{10.2.2}) are given, to terms linear in
$\delta\varphi$,  by
\ba
\label{10.2.5}
3\,{\rm H}\,\dot\varphi&=&-\frac{d{\rm V}}{d\varphi}=-\lambda\,\varphi^3\ ,\\
\label{10.2.6}
3\,{\rm H}\,\delta\dot\varphi-\frac{1}{a^2}\,\Delta\delta\varphi&=&
\left[{\rm V}''-\frac{({\rm V}')^2}{2\,{\rm V}}\right]\,\delta\varphi
=-\frac{5}{2}\,\lambda\,\varphi^2\,\delta\varphi\ .\;\mbox{\hspace{8pt}}
\ea
The term $\displaystyle \frac{({\rm V}')^2}{2\,{\rm V}}\,\delta\varphi$
in (\ref{10.2.6}) appears by virtue of the dependence of the
\index{Hubble ``constant''}%
Hubble parameter H on $\varphi$.  Equations (\ref{10.2.5}) and
(\ref{10.2.6}) make it clear that to lowest order in
$\delta\varphi$, an investigation of the evolution of the field
$\varphi({\bf x},t)$ reduces to an investigation of the motion of
the homogeneous field $\varphi(t)$ as governed by the classical
equation of motion (\ref{10.2.5}), and to a subsequent
investigation of the evolution of the distribution
${\rm P}_c(\delta\varphi,t)$ subject to the initial condition
${\rm P}_c(\delta\varphi,0)\sim\delta(\delta\varphi)$.

It is important that when $\varphi\gg\m$ (that is, during inflation),
the effective
\index{Mass!squared}%
mass squared of the field $\delta\varphi$,
\be
\label{10.2.7}
m^2_{\delta\varphi}={\rm V}''-\frac{({\rm V}')^2}{2\,{\rm V}}
=\frac{5}{2}\,\lambda\,\varphi^2\ ,
\ee
be much less than the square of the Hubble parameter:
\index{Hubble ``constant''}%
$m^2_{\delta\varphi}\ll{\rm H}^2$.  This means that during the
first stage of ``broadening'' of the delta
functional\index{Delta functional}
distribution ${\rm P}_c(\delta\varphi,0)\sim\delta(\delta\varphi)$,
right up to the time
$$
t_1\sim\frac{3\,{\rm H}}{2\,m^2_{\delta\varphi}}\sim
(2\,\sqrt{\lambda}\,\m)^{-1}\ ,
$$
the dispersion squared of fluctuations  grows linearly with time,
(\ref{7.3.12}):
\be
\label{10.2.8}
\langle \delta\varphi^2\rangle =\frac{{\rm H}^3(\varphi)\,t}{4\,\pi^2}
=\frac{\lambda\,\sqrt{\lambda}\,\varphi^6}{3\,\sqrt{6\,\pi}\,\m^3}\,t\ .
\ee
The growth of $\langle \delta\varphi^2\rangle $ then slows down
(see (\ref{7.3.13})), and by
$$
t_2\sim\frac{\sqrt{6\,\pi}}{\sqrt{\lambda}\,\m}\sim10\,t_1\ ,
$$
the dispersion of fluctuations $\delta\varphi$ will have
essentially reached its asymptotic value (\ref{7.3.3}):
\be
\label{10.2.9}
\Delta_0=\sqrt{\langle \delta\varphi^2\rangle }
={\rm C}\,\sqrt{\frac{3\,{\rm H}^4}{8\,\pi^2\,m^2_{\delta\varphi}}}
\approx\sqrt{\frac{\lambda}{15}}\,\frac{\varphi^3}{\m^2}\ ,
\ee
where ${\rm C} \approx 1$.  Equation (\ref{1.7.22}) tells us that
at this stage (with $t\ll t_2$), the mean field $\varphi(t)$
hardly decreases at all.  For $t>t_2$, both the field
$\varphi(t)$ and the quantity ${\rm H}(\varphi)$ start to fall
off rapidly, and that is why fluctuations produced at $t\gg t_2$
make a negligible contribution to the total dispersion
$\Delta(t)=\sqrt{\langle \delta\varphi^2\rangle }$.  The latter
quantity is basically determined by fluctuations that make their
appearance when $t\la t_2$.  To analyze the behavior of
$\Delta(t)$ for $t>t_2$, it is sufficient to note that the
amplitude of fluctuations $\delta\varphi$ that arise when $t<t_2$
subsequently behaves in the same way as the magnitude of
$\dot\varphi$ [\cite{114}] (the reason being that, as one can
readily prove, $\displaystyle \dot\varphi=\frac{d\varphi}{dt}$
obeys the same equation of motion (\ref{10.2.6}) as
$\delta\varphi$).  This implies that for $t\gg t_2$,
\be
\label{10.2.10}
\Delta(t)\approx\Delta_0\,\frac{\dot\varphi(t)}{\dot\varphi(t_2)}
\approx\Delta_0\,\frac{\dot\varphi(t)}{\dot\varphi(0)}\ .
\ee
In the theory with $\displaystyle \vf=\frac{\lambda}{4}\,\varphi^4$,
Eq. (\ref{1.7.22}) implies that $\dot\varphi\sim\varphi(t)$;
that is, for $t\gg  t_2$,
\be
\label{10.2.11}
\Delta(t)={\rm C}\,\sqrt{\frac{\lambda}{15}}\,
\frac{\varphi(t)\,\varphi_0^2}{\m^2}\ ,
\ee
with ${\rm C}\approx  1$.

The foregoing is an elementary derivation of the expressions
(\ref{10.2.8})--(\ref{10.2.11}) for the dispersion $\Delta(t)$,
as an attempt to elucidate the physical nature of the phenomena
taking place [\cite{57}, \cite{78}].  The same results can be
obtained in a more formal manner by solving the diffusion
equation (\ref{7.3.22}) directly for the distribution
${\rm P}_c(\varphi,t)$ with the initial condition
${\rm P}_c(\varphi,t)\sim\delta(\varphi-\varphi_0)$.  This
problem has been solved in [\cite{132}, \cite{133}];  here we
simply present the final result for $\Delta(t)$  in theories with
$\displaystyle \vf=\frac{\lambda\,\varphi^n}{n\,\m^{n-4}}$:
\be
\label{10.2.12}
\Delta^2(t)=\frac{4\,\lambda\,\varphi^{n-2}(t)}{3\,n^2\,\m^n}\,
[\varphi_0^4-\varphi^4(t)]\ .
\ee
In particular, in a theory with
$\displaystyle \vf=\frac{\lambda}{4}\,\varphi^4$,
\be
\label{10.2.13}
\Delta(t)=\frac{1}{2}\,\sqrt{\frac{\lambda}{3}}\,\frac{\varphi(t)}{\m^2}\,
[\varphi_0^4-\varphi^4(t)]^{1/2}\ .
\ee
Using (\ref{1.7.21}), one can easily show that this result is
consistent with Eqs. (\ref{10.2.8})--(\ref{10.2.11}), which were
obtained by a simpler method.

In what follows, it will be especially important to analyze the
evolution of the scalar field $\varphi$ during the initial stage
of the process (for a time $t\la t_2$), during which time the
field $\varphi(t)$ changes by an amount
$\Delta\varphi\la\varphi_0$.  We already know from
(\ref{10.2.12}) that at that stage $\Delta(t)\ll\varphi(t)$  if
${\rm V}(\varphi_0)\ll\m^4$.  Thus, if the initial energy density
is much less than the Planck density, the dispersion of the
scalar field distribution during the stage in question will
always be much less than the mean value of the field $\varphi(t)$
--- that is, we are justified in investigating the evolution of
the field $\varphi(x,t)$ to first order in $\delta\varphi(x,t)$
(\ref{10.2.5}), (\ref{10.2.6}).  On the other hand, when
$\Delta(t)\ll\varphi(t)$, ${\rm P}_c(\varphi,t)$ is a Gaussian
distribution in the vicinity of its maximum at
$\varphi=\varphi(t)$, i.e.,
\ba
\label{10.2.14}
{\rm P}_c(\varphi,t)&\sim&
\exp\left(-\frac{[\varphi-\varphi(t)]^2}{2\,\Delta^2}\right)\nonumber \\
&=&\exp\left(
-\frac{3\,n^2\,[\varphi-\varphi(t)]^2\,\m^n}{8\,\lambda\,\varphi^{n-2}(t)\,
[\varphi_0^4-\varphi^4(t)]}\right) ,
\ea
where $\varphi(t)$ is a solution of Eq. (\ref{10.2.5}); see
(\ref{1.7.21}), (\ref{1.7.22}).  In particular,
\be
\label{10.2.15}
\varphi(t)=\varphi_0\,\exp\left(
-\sqrt{\frac{\lambda}{6\,\pi}}\,\m\,t\right)
\ee
in the theory with $\displaystyle \vf=\frac{\lambda}{4}\,\varphi^4$, and
\be
\label{10.2.16}
\varphi^{2-\frac{n}{2}}(t)=\varphi_0^{2-\frac{n}{2}}-t\,
\left(2-\frac{n}{2}\right)\,
\sqrt{\frac{n\,\lambda}{24\,\pi}}\,\m^{3-\frac{n}{2}}
\ee
in the theory with $\displaystyle
\vf=\frac{\lambda\,\varphi^n}{n\,\m^{n-4}}$ for $n\neq4$.

In any given domain of the universe, then, the distribution
${\rm P}_c(\varphi,t)$ is nonstationary, and in the course of time, the
probability of detecting a large field $\varphi$ at any given
point of space becomes exponentially low.

If, however, one would like to know the distribution
\index{Universe!volume of}%
\index{Volume of universe}%
${\rm P}_p(\varphi,t)$  of the physical volume of the universe,
taking account of expansion proportional to
$$
\exp\left(\int^t_0{\rm H}({\bf x},t)\:dt\right)\ ,
$$
that contains the field $\varphi$ at time $t$, the answer will be
completely different.  To attack this question, let us consider
for definiteness the evolution of the distribution ${\rm
P}_c(\varphi,t)$ for a time $\Delta t$, during which the mean
field $\varphi(t)$ decreases by $\displaystyle
\Delta\varphi=\frac{\varphi_0}{{\rm N}}\ll\varphi_0$, where N is
some number satisfying ${\rm N} \gg 1$.  According to
(\ref{10.2.14}),
\be
\label{10.2.17}
{\rm P}_c(\varphi,\Delta t)\approx
\exp\left(-\frac{
\displaystyle 3\,n^2\,{\rm N}\left[\varphi-\varphi_0\,
\left(1-\frac{1}{{\rm N}}\right)\right]^2\,\m^n}{32\,\lambda\,
\varphi_0^{n+2}}\right)\ .
\ee
We see from this result that the fraction of the original coordinate volume
remaining in a state with $\varphi=\varphi_0$ after a time $\Delta t$ is
\be
\label{10.2.18}
{\rm P}_c(\varphi_0,\Delta t)\approx
\exp\left(-\frac{3\,n^2\,\m^n}{32\,\lambda\,{\rm N}\,\varphi_0^n}\right)
=\exp\left(-\frac{3\,n\,\m^4}{32\,{\rm N}\,{\rm V}(\varphi_0)}\right)\ .
\ee
Notice that ${\rm P}_c(\varphi_0,\Delta t)\ll1$ when
${\rm V}(\varphi_0)\ll\m^4$.  In other words, the dispersion
$\Delta$ is much less than the difference $\varphi_0-\varphi(t)$.
In a time $\Delta t$, the volume of a region with
$\varphi=\varphi_0$ expands by a factor
$e^{3\,{\rm H}(\varphi_0)\,\delta t}$, on the average.  It then
follows from (\ref{10.2.15}) and (\ref{10.2.16}) that
\be
\label{10.2.19}
\Delta t=\frac{2}{{\rm N}}\,\sqrt{\frac{6\,\pi}{n\,\lambda}}\,
\frac{\m^{\frac{n}{2}-3}}{\varphi^{\frac{n}{2}-2}}\ .
\ee
The original volume occupied by the field $\varphi_0$ thus
changes in a time $\Delta t$ given by (\ref{10.2.19}) by a factor
${\rm P}_p (\varphi_0,\Delta t)$, where
\ba
\label{10.2.20}
{\rm P}_p (\varphi_0,\Delta t)&\approx&{\rm P}_c(\varphi_0,\Delta t)\,
\exp[3\,{\rm H}(\varphi_0)\,\Delta t]\nonumber \\
&=&\exp\left(-\frac{3\,n^2}{32\,\lambda\,{\rm N}}\,\frac{\m^n}{\varphi_0^n}+
\frac{24\,\pi}{{\rm N}\,n}\,\frac{\varphi_0^2}{\m^2}\right)\ .
\ea
Clearly, then, when $\varphi_0\gg\alpha\,\varphi^*$, where
\be
\label{10.2.21}
\varphi^*=\lambda^{-\frac{1}{n+2}}\,\m\ ,\qquad
\alpha=\left(\frac{n^3}{2^8\,\pi}\right)^{\left(\frac{1}{n+2}\right)}=
{\rm O}(1)\ ,
\ee
the volume occupied by the field $\varphi_0$ will grow during the
time $\Delta t$ rather than decrease.  The same thing will be
repeated during the next time interval $\Delta t$, and so on.
This means that during inflation, regions of the inflationary
universe with $\varphi>\varphi^*$ reproduce themselves endlessly:
the process of inflation, once having started, continues forever
\index{Universe!volume of}%
\index{Volume of universe}%
unabated, and the volume of the inflationary part of the universe
grows without bound.

Behavior of the distribution ${\rm P}_p (\varphi,\Delta t)$
for $\displaystyle \varphi-\varphi_0\gg\Delta
\varphi=\frac{\varphi_0}{{\rm N}}$ is even more interesting.  In
point of fact, a field $\varphi$ that occasionally jumps to
values with $\displaystyle
\varphi-\varphi_0\gg\frac{\varphi_0}{{\rm N}}$ due to quantum
fluctuations in some domain of the inflationary universe cannot
be significantly reduced in a time $\sim \Delta t$, either by
classical rolling (by $\Delta\varphi\ll\varphi-\varphi_0$) or by
diffusion (by $\sim\Delta\ll\Delta \varphi$).  The volume of all
regions occupied by the field $\varphi$ increases in a time
$\Delta t$ (\ref{10.2.19}) by a factor $\exp[3\,{\rm H}(\varphi)\,\Delta t]$,
whereupon
\ba
\label{10.2.22}
{\rm P}_p (\varphi,\Delta t)\!\!\!\!&\approx&\!\!\!\!
{\rm P}_c(\varphi,\Delta t)
\exp[3\,{\rm H}(\varphi)\,\Delta t]\nonumber \\
\!\!\!\!&=&\!\!\!\!
\exp\left\{-\frac{3\,n^2\,{\rm N}\,[\varphi-\varphi(t)]^2}{32\,\lambda}\,
\frac{\m^n}{\varphi_0^{n+2}}+\frac{24\,\pi}{{\rm N}\,n}\,
\frac{\varphi_0^2}{\m^2}\,\left(\frac{\varphi}{\varphi_0}\right)^{n/2}
\right\}\; .\nonumber \\
\ea
It can readily be shown that when $\varphi>\beta\,\varphi^*$,
where $\displaystyle
\beta=\left(\frac{n^2\,{\rm N}}{2^6}\right)^{\frac{1}{n+2}}=
{\rm O}(1)$, the maximum of the distribution
${\rm P}_p (\varphi,\Delta t)$ is shifted not towards
$\varphi<\varphi_0$, like the maximum of ${\rm P}_c(\varphi,\Delta t)$,
but towards $\varphi>\varphi_0$.

This means that when $\varphi\gg\varphi^*$, the universe not only
continually regenerates itself, but in the process, most of the
physical volume of the universe gradually fills with a larger and
larger field $\varphi$ [\cite{132}, \cite{133}].  This result is
entirely consistent with the result obtained in Section
\ref{s1.8} by more elementary methods [\cite{57}].
\index{Global structure of inflationary universe|)}%
\index{Quantum cosmology!global structure of inflationary universe and|)}%

\section[self-reproducing universe and quantum cosmology]%
{\label{s10.3}The self-reproducing inflationary universe
\index{Quantum cosmology!self-reproducing inflationary universe and|(}%
\index{Self-reproducing inflationary universe!quantum cosmology and|(}%
and quantum cosmology}

The possibility of an eternally existing, self-reproducing
universe is one of the most important and surprising consequences
of the theory of the inflationary universe, and it merits
detailed discussion (see also Section \ref{s1.8}).  Let us first
of all provide a more accurate interpretation of the results
obtained in Section \ref{s10.2}.

The physical meaning of the distributions ${\rm P}_c(\varphi,t)$
and ${\rm P}_p(\varphi,t)$ is as follows.  Consider a domain of
the inflationary universe having initial size $l\ga{\rm H}^{-1}$,
and let us assume that initially (at $t=0$) it is uniformly
filled throughout its entire volume by observers with identical
clocks, which are synchronized at $t=0$.  In that event, the
quantity ${\rm P}_c(\varphi,t)$ determines the fraction of all
observers who at time $t$ as measured by their own clocks (i.e.,
in a synchronous coordinate system) are located in a region
filled with a practically homogeneous (on a scale
$l\ga{\rm H}^{-1}(\varphi)$) semiclassical field $\varphi$.  The
distribution ${\rm P}_p(\varphi,t)$ determines how much of the
physical volume of the universe is occupied by observers who, at
time $t$ as measured by their own clocks, live in regions filled
with the field $\varphi$ that is homogeneous on a scale larger
than ${\rm H}^{-1}(\varphi)$.

The results obtained in the preceding section imply that in no
particular region of the inflationary universe can the
distribution ${\rm P}_c(\varphi,t)$ be stationary.  It can be
quasistationary  during tunneling from a metastable vacuum state
\index{Hawking--Moss equation}%
\index{Tunneling!Hawking--Moss}%
to a stable state, as in the case of Hawking--Moss tunneling in
the new inflationary universe scenario.  But in any model in
which the universe becomes hot after inflation and the field
$\varphi$ rolls down to its minimum at $\vf=0$, the function
${\rm P}_c(\varphi,t)$ cannot (and should not) be a nontrivial
stationary distribution (at least within the range of validity of
the approximation that we have employed;  see below).  In other
words, the fraction of all observers initially situated in an
unstable state away from the absolute minimum of the effective
potential $\vf$ should decrease with time.  This conclusion is
confirmed by results obtained above --- for example, see Eq.
(\ref{10.2.14}), which shows that the probability of remaining in
an unstable state $\varphi\ga\m$ in a theory with $\displaystyle
\vf=\frac{\lambda}{4}\,\varphi^4$ after a time $\displaystyle
t\ga\frac{\sqrt{6\,\pi}}{\sqrt{\lambda}\,\m}\ln\frac{\varphi_0}{{\rm M}}$
becomes exponentially small.

At the same time, when $\varphi_0\gg\varphi^*$ (see
(\ref{10.2.21})), the distribution ${\rm P}_p(\varphi,t)$ begins
to increase with increasing $\varphi$ in a region with
$\varphi\ga\varphi^*$, i.e., the fraction of the volume of the
inflationary universe occupied by observers who at time $t$ by
their clocks find themselves located in an unstable state
$\varphi\ga\varphi^*$ increases at large $\varphi$ and $t$, and
consequently the total volume of the inflationary regions of the
universe continues to grow without bound.  It follows from
(\ref{10.2.22}) that at large $t$, most of the volume of the
universe should be in a state with the very largest possible
value of the field $\varphi$, such that $\vf\sim\m^4$.

Here, to be sure, we must state an important reservation.  The
fraction of the volume of the universe in a state with a given
field $\varphi$ {\it at a given time} depends on what we mean by
\index{Time!defined}%
the word {\it time}.  The results obtained above refer to the
proper time $t$ of comoving observers whose clocks were
synchronized at some time $t=0$, when they were all quite close
to one another.  The same phenomena can be described using
another coordinate system, namely the coordinates (\ref{7.5.8}),
which are especially convenient in investigating the density
inhomogeneities produced during inflation.  In order to
distinguish between the proper time $t$ in the synchronous
\index{Time!defined}%
coordinates and the ``time'' in the coordinates (\ref{7.5.8}), we
denote the latter here as $\tau$. Investigation of diffusion in
the coordinate system (\ref{7.5.8}) also shows that the total
volume of the universe filled with the field $\varphi>\varphi^*$
increases exponentially with time $\tau$ [\cite{133}].  But due
to the specific way in which the ``time'' $\tau$ is defined, the
rate of exponential expansion of the universe, which is $\sim
e^{{\rm H}\,\tau}$ in the coordinates (\ref{7.5.8}), is the same
everywhere, regardless of any local increases or decreases in the
field $\varphi$.  The fraction of the physical volume of the
universe filled with a large field $\varphi$ on a hypersurface of
constant $\tau$ therefore falls off in almost the same way as
${\rm P}_c(\varphi,t)$, and thus the fraction of the physical
volume of a self-reproducing universe that is transformed over
time into a state with the largest possible value of the field
\index{Time!defined}%
$\varphi$ depends on what exactly one means by the word {\it
time}.  It is for just this reason that we have engaged here in a
more detailed discussion of this question, the answer to which
turns out to depend on exactly how the question is formulated.
Fortunately, however, our basic conclusion about
self-regeneration and exponential expansion of regions of the
universe filled with a field $\varphi>\varphi^*$ is
coordinate-independent [\cite{133}].

{\looseness=-1
It is worthwhile to examine these results from yet another
standpoint.  If the universe is self-reproducing, then the
standard question about the initial conditions {\it over the
whole universe} may be irrelevant, since the universe may turn
out not to have had a global initial spacelike singular
hypersurface to play the role of a global Cauchy hypersurface.
At present, we do not have sufficient reason to believe that the
universe as a whole was created approximately $10^{10}$ years ago
in a singular state, prior to which classical space-time did not
exist at all.  Inflation could begin and end at different times
in different domains of the universe, and this would be
completely consistent with the existing observational data.

}

Accordingly, the
\index{Density!matter}%
\index{Matter density}%
matter density in different regions of the
universe will drop to $\rho_0\sim10^{-29}$ g/cm$^3$ at different
times, approximately $10^{10}$ years after inflation ends in each
of these regions.  It is just after this point in each of these
regions that the conditions required for the emergence of
observers like ourselves will first appear.  The number of such
observers should plainly be proportional to the volume of the
universe at the density hypersurface(s) with
\index{Density hypersurface}%
\index{Hypersurface!density}%
$\rho=\rho_0\sim10^{-29}$ g/cm$^3$.  Therefore, having
investigated the question of what processes take part in the
creation of most of the volume of the universe at the density
hypersurface $\rho=\rho_0\sim10^{-29}$ g/cm$^3$  (i.e., $10^{10}$
years after inflation ceases in each particular domain), we have
prepared ourselves to assess the most likely history of the part
of the universe that we are able to observe.

In order to look into this question, one should take into account
that the universe expands by a factor of approximately $10^{30}$
over the $10^{10}$ years after the end of inflation, and that
during inflation in the theory with
$\displaystyle \vf=\frac{\lambda}{4}\,\varphi^4$, the universe
typically expands by a factor
$\displaystyle \exp\left(\frac{\pi\,\varphi_0^2}{\m^2}\right)$,
where $\varphi_0$ is the original value of the field $\varphi$.
However, for $\varphi_0\ga\varphi^*\sim\lambda^{-1/6}\,\m$, this
result is modified.

Indeed, as at the end of Section \ref{s10.2}, let us consider a
domain of the inflationary universe in which the scalar field,
due to its long-wave quantum fluctuations during the time $\Delta t$,
jumps from $\varphi=\varphi_0$ up to some value $\varphi$
such that $\varphi-\varphi_0\gg|\Delta\varphi|$, where
$\Delta\varphi$ is the value of the classical decrease in the
field $\varphi$ during the time $\Delta t$.  If the jump in the
field $\varphi$ is large enough, its mean value in this domain
will return to the original value $\varphi=\varphi_0$, mainly due
to classical rolling.  According to (\ref{1.7.25}), during
classical rolling, the domain under consideration inflates by an
additional factor of $\displaystyle
\exp\left[\frac{\pi}{\m^2}(\varphi^2-\varphi_0^2)\right]$.

The probability of a large jump in the field $\varphi$ is
exponentially suppressed (see (\ref{10.2.14}), (\ref{10.2.22})),
but it is not hard to prove that when $\varphi\gg\varphi^*$, this
suppression pays us back with interest on account of the
aforementioned additional inflation of the region filled with the
field $\varphi$ making the jump.  This means that most of the
volume of the universe after inflation (for example, on the
hypersurface $\rho=\rho_0$) results from the evolution of those
relatively rare but additionally inflated regions in which the
field $\varphi$ has jumped upward as a result of long-wavelength
quantum fluctuations.  To continue this line of reasoning, it can
be shown that the overwhelming preponderance of the physical
volume of the universe in a state with given density $\rho=\rho_0$
is formed as a result of the inflation of regions in which the
field $\varphi$, over the longest possible times, has been
fluctuating about its maximum possible values, such that
$\vf\sim\m^4$.  In that sense, a state with potential energy
density close to the Planck value (i.e., the space-time foam) can
be considered to be a source, continuously producing the greater
part of the physical volume of the universe.  We shall return to
this point subsequently, but for the moment we wish to compare our
conclusions with the basic expectations and assumptions that have
been made in analyzes of the wave function of the universe.

In deriving the expression for the wave function (\ref{10.1.12}),
(\ref{10.1.17}) proposed by Hartle and Hawking, it was assumed
that the universe has a stationary ground state, or a state of
least excitation (vacuum), the wave function $\Psi(a,\varphi)$ of
which they attempted to determine;  see Section \ref{s10.1}.  The
square of this wave function $|\Psi(a,\varphi)|^2$ should then
give the stationary distribution for the probability of detecting
the universe in a state with the homogeneous scalar field
$\varphi$ and scale factor $a$ (\ref{10.1.18}), (\ref{10.1.20}).
The fact that the quasistationary distribution ${\rm P}_c(\varphi)$
(\ref{10.2.3}) is proportional to the square of
\index{Hartle--Hawking wave function}%
\index{Wave function!Hartle--Hawking}%
the Hartle--Hawking wave function could be taken as a important
indication of the validity of this assumption.  The results
obtained in the preceding section, however, show that with fairly
general initial conditions, the distribution ${\rm P}_c(\varphi,t)$
in the inflationary universe scenario does not
wind up in the stationary regime (\ref{10.2.3}).  Nevertheless,
another type of stationary regime is possible, described in part
by the distribution ${\rm P}_p(\varphi,t)$.  In this regime, the
universe continually produces exponentially expanding regions
\index{Mini-universes}%
\index{Universe!mini-universes}%
(mini-universes) containing the large field $\varphi$  (with
$\varphi^*\la\varphi\la\varphi_{\rm P}$), where
${\rm V}(\varphi_{\rm P})\sim\m^4$, and the properties of the
universe within such regions do not depend on the properties of
neighboring regions (by virtue of the ``no hair'' theorem for de
Sitter space), nor do they depend on the history or epoch of
formation of those regions.  Here we can speak of stationarity in
the sense that regions of the inflationary universe containing a
field $\varphi\ga\varphi^*$ constantly come into being, and in an
exponentially large neighborhood of each such region, the average
properties of the universe are the same and do not depend on the
epoch at which the region was formed.  This implies that the
inflationary universe has a fractal structure [\cite{133},
\cite{325}].

Thus, the Hartle--Hawking wave function may be useful in
describing the intermediate stages of inflation, during which the
distribution ${\rm P}_c(\varphi,t)$ can sometimes (in the
presence of metastable vacuum states) turn out to be
quasistationary.  This function may also turn out to be quite
useful in the study of certain other important problems of
quantum cosmology --- see, for example, a discussion of this
possibility in Section \ref{s10.7}.  At the present time,
however, we are unable to ascribe to this wave function the
fundamental significance sometimes assigned to it in the
literature.

What then can we say about the tunneling wave function
\index{Tunneling wave function}%
\index{Wave function!tunneling}%
\index{Quantum creation of universe}%
\index{Universe!quantum creation of}%
(\ref{10.1.23}) used to describe the quantum creation of the
universe?

To answer this question, let us study in somewhat more detail the
first (diffusive) stage of spreading of the initial distribution
${\rm P}_c(\varphi,0)=\delta(\varphi-\varphi_0)$ when the
magnitude of the classical field $\varphi$ is almost constant,
$\varphi(t)\approx\varphi_0$.  Equation (\ref{10.2.14}) holds
when the dispersion $\Delta$ and the difference between $\varphi$
and $\varphi_0$ are much less than $\varphi_0$ itself.  At the
same time, when $\varphi-\varphi_0\sim\varphi_0$  the
distribution ${\rm P}_c(\varphi,t)$ is far from Gaussian.  In
order to calculate  in that case, one should bear in mind that
the classical rolling of the field $\varphi$, i.e., the last term
in the diffusion equation (\ref{7.3.22}), can be neglected during
the first stage:
\be
\label{10.3.1}
\frac{\partial{\rm P}_c(\varphi,t)}{\partial t}=
\frac{2\,\sqrt{2}}{3\,\sqrt{3\,\pi}\,\m^3}\,
\frac{\partial^2}{\partial\varphi^2}\,
[\vf\,{\rm P}_c(\varphi,t)]\ .
\ee
It is convenient to seek a solution of this equation in the form
$$
{\rm P}_c(\varphi,t)\sim{\rm A}(\varphi,t)\cdot
\exp\left[-\frac{{\rm S}(\varphi)}{t}\right]\ ,
$$
where ${\rm A}(\varphi,t)$ and ${\rm S}(\varphi)$
are relatively slowly varying functions of $\varphi$ and $t$.

It can readily be shown that in the theory $\displaystyle
\vf=\frac{\lambda\,\varphi^n}{n\,\m^{n-4}}$ with
$\varphi\ll\varphi_0$, the corresponding solution is
\be
\label{10.3.2}
{\rm P}_c(\varphi,t)={\rm A}\,
\exp\left[-\frac{3\,\sqrt{6\,\pi}}{t\,\lambda\,\sqrt{\lambda}\,(3\,n-4)^2}\,
\left(\frac{\m}{\varphi}\right)^{\frac{3n}{2}-1}\right]\ ,
\ee
and neglect of the last term in (\ref{7.3.22}) is warranted when
\be
\label{10.3.3}
t\la\Delta t(\varphi)=
\sqrt{\frac{6\,\pi}{n\,\lambda}}\,\m^{-1}\,
\left(\frac{\m}{\varphi}\right)^{\frac{n}{2}-2}
\ee
(compare with (\ref{10.2.19})).  If the effective potential
${\rm V}(\varphi)$ is not too steep ($n \le 4$), then the diffusion
approximation first ceases to work at small $\varphi$, and
subsequently at $\varphi\sim\varphi_0$.  One can say in that
event that regions of space with a small field $\varphi$, in
which classical motion prevails over quantum fluctuations, are
formed by virtue of a quantum diffusion process that operates for
a time $t\la\Delta t(\varphi)$, and the probability distribution
for the creation of a region (mini-universe) with a given field
$\varphi$ when quantum diffusion ceases to dominate
($t\sim\Delta t(\varphi)$) is, according to (\ref{10.3.2}) and
(\ref{10.3.3}),
\be
\label{10.3.4}
{\rm P}_c(\varphi,\Delta t(\varphi))\sim
\exp\left[-{\rm C}\,\frac{\m^4}{\vf}\right]\ ,
\ee
where ${\rm C}={\rm O}(1)$.  This formula holds for $n\le4$,
$\varphi\ll\varphi_0$ regardless of the initial value of the
field $\varphi_0$.  In particular, when ${\rm V}(\varphi_0)\ga\m^4$,
it can be interpreted as the probability for the quantum creation
\index{Quantum creation of universe}%
\index{Universe!quantum creation of}%
of a (mini-)universe from the space-time foam with $\vf\ga\m^4$.
It is easily seen that up to a factor ${\rm C}={\rm O}(1)$,
(\ref{10.3.4}) and the probability (\ref{10.1.24}) of quantum
creation of a universe from ``nothing'' are identical.

Is this merely formal consistency between these equations, or is
there more to it than that?  An answer requires additional
investigation, but there are some ideas on this score that may be
enunciated right now.  First of all, note that rather than
describing the creation of the entire universe from ``nothing,''
Eq. (\ref{10.3.4}) describes the creation of only a part of the
universe of size greater than ${\rm H}^{-1}(\varphi)$ due to
quantum diffusion from a previously existing region of the
inflationary universe.  Moreover, Eq. (\ref{10.3.4}) holds in
theories with $\vf\sim\varphi^n$ only when $n\le4$;  for $n>4$,
it can be shown that
\be
\label{10.3.5}
{\rm P}_c(\varphi,\Delta t(\varphi))={\rm P}_c(\varphi,\Delta t(\varphi_0))
\sim\exp\left[
-\frac{{\rm C}\,\m^4}{\vf}\,
\left(\frac{\varphi_0}{\varphi}\right)^{\frac{n}{2}-2}\right]\ .
\ee
In theories in which the interval between $\varphi$ and
$\varphi_0$ contains segments where the field $\varphi$ rolls
down rapidly and the universe is not undergoing inflation,
equations like (\ref{10.3.4}) and (\ref{10.3.5}) will generally
not be valid; that is, the diffusion equation that we have used
will not be applicable to such segments.  More specifically, one
cannot derive equations like (\ref{10.3.4}) for the probability
of diffusion from a space-time foam with ${\rm V}(\varphi_0)\sim\m^4$
onto the top of the effective potential at $\varphi=0$ in the new
inflationary universe scenario.  Meanwhile, it is usually assumed
that Eq. (\ref{10.1.23}) (perhaps somewhat modified to take
account of the effects of quantum creation of particles during
tunneling [\cite{321}]) can describe the quantum creation of the
universe as a whole, even if a continuous diffusive transition
between $\varphi_0$ and $\varphi$ is not possible.

This indicates that we are dealing here with two distinct
complementary or competing processes described by Eqs.
(\ref{10.3.4}) and (\ref{10.1.24}), respectively.  However,
experience with the Hawking--Moss theory of tunneling
(\ref{7.4.1}) engenders a certain amount of caution in this
regard.  Recall that Eq. (\ref{7.4.1}), which was originally
derived via the Euclidean approach to tunneling theory, was
interpreted as giving the probability of uniform tunneling over
the entire universe [\cite{121}].   However, a rigorous derivation
of Eq. (\ref{7.4.1}) and a justification for this interpretation
were lacking.  In our opinion, a rigorous derivation of Eq.
(\ref{7.4.1}) was first provided by solving the diffusion equation
(\ref{7.3.22}), and its interpretation was different from the
original one based on the Euclidean approach [\cite{134},
\cite{135}], though consistent with the interpretation proposed in
[\cite{209}].  Likewise, neither approach to deriving Eq.
(\ref{10.1.24}) (using the (anti-)Wick rotation $t\rightarrow
i\,\tau$ or considering tunneling from the point $a=0$) is
sufficiently rigorous, and the interpretation of (\ref{10.1.24})
as the probability of tunneling from ``nothing'' also falls
somewhere on the borderline between physics and poetry.  One of
the fundamental questions to emerge here had to do with what {\it
exactly} was tunneling, if there were no {\it incoming} wave.  A
plausible response is that one simply cannot identify the incoming
wave within the framework of the minisuperspace approach.  Indeed,
by solving the diffusion equation in the theory of chaotic
inflation, it was shown that during inflation there is a steady
process of creation of inflationary domains whose density is close
to the Planck density, and whose size is $l\sim l_{\rm
P}\sim\m^{-1}$. Tunneling (or diffusion) involving an increase in
the size of each such region and a change in the magnitude of the
scalar field within each region can be (approximately) associated
with the process of quantum creation of the universe.  The process
of formation of such Planck-size inflationary domains (the
``incoming wave'') cannot be described in the context of the
minisuperspace approach, but it has a simple interpretation within
the scope of the stochastic approach to inflation.

We have thus come closer to
substantiating the validity of Eq.\linebreak[10000]
(\ref{10.1.24}) as a probability of quantum creation of the
universe from ``nothing.'' It is nevertheless still not entirely
clear whether in this expression there is anything that is both
true and at the same time different from Eq. (\ref{10.3.4}),
which was derived with the aid of the stochastic approach to
inflation, and which has a much more definite physical meaning.
This is a particularly important question, as it pertains to the
theory of the quantum creation of the universe in the state
$\varphi=0$, corresponding to a local maximum of $\vf$ located at
$\vf\ll\m^4$, since diffusion into this state from a space-time
foam with $\vf\sim\m^4$ is impossible.

To conclude this section, let us examine one more question, relating
\index{Creation!of inflationary universe from Minkowski space}%
\index{Inflationary universe!creation of, from Minkowski space}%
\index{Minkowski space, creation of inflationary universe from}%
to the possibility of creating an inflationary universe
from Minkowski space.  The issue here is that quantum
fluctuations in the latter can bring into being an inflationary
domain of size $l\ga{\rm H}^{-1}(\varphi)$, where $\varphi$ is a
scalar field produced by quantum fluctuations in this domain.
The ``no hair'' theorem for de Sitter space implies that such a
domain inflates in an entirely self-contained manner, independent
of what occurs in the surrounding space.  We could then conceive
of a ceaseless process of creation of inflationary mini-universes
that could take place even at the very latest stages of
development of the part of the universe that surrounds us.

A description of the process whereby a region of the inflationary
universe is produced as a result of quantum fluctuations could
proceed in a manner similar to that for the formation of regions
of the inflationary universe with a large field $\varphi$ through
the buildup of long-wave quantum fluctuations $\delta\varphi$.
The basic difference here is that long-wave fluctuations
$\delta\varphi$ of the massive scalar field $\varphi$ at the time
of inflation with $m\ll{\rm H}$  are ``frozen'' in amplitude,
while there is no such effect in Minkowski space.  But if the
buildup of quantum fluctuations in some region of Minkowski space
were to engender the creation of a fairly large and homogeneous
field $\varphi$, then that region in and of itself could start to
inflate, and such a process could stabilize (``freeze in'') the
fluctuations $\delta\varphi$ that led to its onset.  In that
event, one could sensibly speak of a self-consistent process of
formation of inflationary domains of the universe due to quantum
fluctuations in Minkowski
space.
\index{Creation!of inflationary universe from Minkowski space}%
\index{Inflationary universe!creation of, from Minkowski space}%
\index{Minkowski space, creation of inflationary universe from}%

Without pretending to provide a complete description of such a
process, let us attempt to estimate its probability in theories
with $\displaystyle \vf=\frac{\lambda\,\varphi^n}{n\,\m^{n-4}}$.
A domain formed with a large field $\varphi$ will only be a part
of de Sitter space if in its interior
$(\partial_\mu\varphi)^2\ll\vf$.  This means that the size of the
domain must exceed $l\sim\varphi\,\vf^{-1/2}$, and the field
inside must be greater than $\m$.  Such a domain could arise
through the buildup of quantum fluctuations $\delta\varphi$ with
a wavelength
$$
k^{-1}\ga l\sim\varphi\,{\rm V}^{-1/2}(\varphi)\sim m^{-1}(\varphi)\ .
$$
One can estimate the dispersion $\langle \varphi^2\rangle_{k<m}$
of such fluctuations using the simple formula
\be
\label{10.3.6}
\langle \varphi^2\rangle_{k<m}\sim\frac{1}{2\,\pi^2}\,
\int^{m(\varphi)}_0\frac{k^2\:dk}{\sqrt{k^2+m^2(\varphi)}}
\sim \frac{m^2}{\pi^2}\sim\frac{\vf}{\pi^2\,\varphi^2}\ ,
\ee
and for a Gaussian distribution ${\rm P}(\varphi)$ for the
appearance of a field $\varphi$ which is sufficiently homogeneous
on a scale $l$, one has [\cite{133}]
\be
\label{10.3.7}
{\rm P}(\varphi)\sim
\exp\left[-{\rm C}\,\frac{\pi^2\,\varphi^4}{\vf}\right]\ ,
\ee
where ${\rm C}={\rm O}(1)$.  In particular, for a theory with
$\displaystyle \vf=\frac{\lambda}{4}\,\varphi^4$,
\be
\label{10.3.8}
{\rm P}(\varphi)\sim\exp\left[-{\rm C}\,\frac{4\,\pi^2}{\lambda}\right]\ .
\ee
Naturally, this method is rather crude;  nevertheless, the
estimates that it provides are quite reasonable, to order of
magnitude.  For example, practically the same lines of reasoning
could be employed in assessing the probability of tunneling from
the point $\varphi=0$ in a theory with $\displaystyle
\vf=-\frac{\lambda}{4}\,\varphi^4$; see Chapter \ref{c5}.  The
estimate that one obtains for the formation of a bubble of the
field $\varphi$ is also given by Eq. (\ref{10.3.8}). This result
is in complete accord with the equation
$\displaystyle {\rm P}\sim\exp\left(-\frac{8\,\pi^2}{3\,\lambda}\right)$
(\ref{5.3.12}), which was derived using Euclidean methods.  In
fact, one can easily verify that {\it all} results concerning tunneling
which were obtained in Chapter \ref{c5} can be reproduced (up to
a numerical factor ${\rm C}={\rm O}(1)$ in the exponent) by using
the simple method suggested above.  This makes the validity of
the estimates (\ref{10.3.7}), (\ref{10.3.8}) quite plausible.

The main objection to the possibility of quantum creation of an
inflationary
\index{Creation!of inflationary universe from Minkowski space}%
\index{Inflationary universe!creation of, from Minkowski space}%
\index{Minkowski space, creation of inflationary universe from}%
universe in Minkowski space is that energy
conservation forbids the production of an object with positive
energy out of the vacuum in this space.  Within the scope of
classical field theory, in which the energy is everywhere
positive, such a process would therefore be impossible (a related
problem is discussed in [\cite{213}, \cite{326}]).  But at the
quantum level, the energy density of the vacuum is zero by virtue
of the cancellation between the positive energy density of
classical scalar fields, along with their quantum fluctuations,
and the negative energy associated with quantum fluctuations of
fermions, or the bare negative energy of the vacuum.  The
creation of a positive energy-density domain through the buildup
of long-wave fluctuations of the field $\varphi$ is inevitably
accompanied by the creation of a region surrounding that domain
in which the long-wave fluctuations of the field $\varphi$ are
suppressed, and the
\index{Density!vacuum energy}%
\index{Vacuum energy density}%
\index{Energy density, vacuum}%
vacuum energy density is consequently
negative.  Here we are dealing with the familiar quantum
fluctuations of the vacuum energy density about its zero point.
It is important here that from the point of view of an external
observer, the total energy of the inflationary region of the
universe (and indeed the total energy of the closed inflationary
universe) does not grow exponentially;  the region that emerges
forms a universe distinct from ours, to which it is joined only
\index{Wormholes}%
by a connecting throat (wormhole) which, like a black hole, can
disappear by virtue of the Hawking effect [\cite{327},
\cite{213}].  At the same time, the shortfall of long-wave
fluctuations of the field $\varphi$ surrounding that region is
quickly replenished by fluctuations arriving from neighboring
regions, so the negative energy of the region near the throat can
be be rapidly spread over a large volume around the inflationary
domain.

Our discussion of the creation of
an inflationary universe in\linebreak[10000]
Minkowski
\index{Creation!of inflationary universe from Minkowski space}%
\index{Inflationary universe!creation of, from Minkowski space}%
\index{Minkowski space, creation of inflationary universe from}%
space is highly speculative, and is only intended to
illustrate the basic feasibility of such a process;  this is
clearly a problem that requires closer study.  If this process
can actually transpire, and if it is accompanied by burnout of
\index{Wormholes}%
the wormhole connecting the parent (Minkowski) space with the
inflationary universe that is its offspring, then the theory will
have one more mode for the stationary production of regions of an
inflationary universe.  We wish to emphasize, however, that in
our approach, the likelihood of this regime being realized is in
no way related to the distribution ${\rm P}_c(\varphi)$, which is
proportional to the square of the wave function (\ref{10.1.17}).
The Euclidean approach to the theory of baby-universe formation
has been developed in [\cite{350}--\cite{352}], and will be
discussed  in Section \ref{s10.7}.
\index{Quantum cosmology!self-reproducing inflationary universe and|)}%
\index{Self-reproducing inflationary universe!quantum cosmology and|)}%

\section[global structure of the inflationary universe]%
{\label{s10.4}The global structure of the inflationary universe
\index{Cosmological singularity!global structure of
inflationary universe and|(}%
\index{Global structure of inflationary universe!and
general cosmological singularity|(}
and the problem of the general cosmological singularity}

One of the most important consequences of the inflationary
universe scenario is that under certain conditions, once a
universe has come into being it can never again collapse as a
whole and disappear completely.  Even if it initially resembles a
homogeneous closed Friedmann universe, it will most likely cease
to be locally homogeneous and become markedly inhomogeneous on
the largest scales, and there will then be no global end of the
world such as that which occurs in a homogeneous closed Friedmann
universe.

There are versions of the theory of the inflationary universe in
which self-regeneration of the universe does not take place, such
as the Shafi--Wetterich model [\cite{237}], which is based on a
study of inflation in a particular version of Kaluza--Klein
theory --- see Section \ref{s9.5}.  But for most of the
inflationary models studied thus far, the evolution of the
universe and the process of inflation has no end.  In the old
\index{Guth scenario}%
Guth scenario, for example, when the probability of forming
bubbles of a new phase with $\varphi\neq0$ is low enough, these
bubbles will never fill the entire physical volume of the
universe, since the distance separating any two of them increases
exponentially, and the resulting increase in the volume of the
universe in the state $\varphi=0$ is greater than the decrease of
this volume due to the creation of new bubbles [\cite{53},
\cite{113}, \cite{327}, \cite{328}].  One encounters a similar
effect in the new inflationary universe scenario [\cite{266},
\cite{267}], and a detailed theory of this process [\cite{204}]
is quite similar to the corresponding theory in the chaotic
inflation scenario [\cite{57}, \cite{133}], the basic difference
being that in both the old and new inflationary universe
scenarios one is dealing with the production of regions
containing a field $\varphi$ close to zero and with $\vf\ll\m^4$,
while in the chaotic inflation scenario there can be a steady
output of regions with very high values of $\vf$, right up to
$\vf\sim\m^4$.  We shall subsequently find this to be a very
important circumstance.

The possibility of the ceaseless regeneration of inflationary
regions of the universe, which implies the absence of a general
cosmological singularity (i.e., a global spacelike singular
hypersurface) {\it in the future}, compels us to reconsider the
problem of the
\index{Cosmological singularity!initial}%
{\it initial} cosmological singularity as well.
At present, it seems unnecessary to assume that there was a
unique beginning to this endless production of inflationary
\index{Nonsingular universe}%
\index{Universe!nonsingular}%
regions.  Models of a nonsingular universe based on this idea
have been proposed in the context of both the old [\cite{327},
\cite{328}] and new [\cite{267}] inflationary universe scenarios.
According to these models, most of the physical volume of the
universe remains forever in a state of exponential expansion with
$\varphi\approx0$, engendering ever newer exemplars of our type
of mini-universe.

Unfortunately, it is not yet entirely clear how one would go
about implementing this possibility.  To understand where the
main difficulty lies, recall that exponentially expanding (flat)
de Sitter space is not geodesically complete --- it comprises
only a part of the closed de Sitter space, which originally
contracts (at $t<0$) rather than expanding:
$$
a(t)={\rm H}^{-1}\,\cosh{\rm H}\,t
$$
(see Section \ref{s7.2}).  In
\index{de Sitter space!contracting}%
de Sitter space contracting at an
exponential rate, a phase transition from the state $\varphi=0$
can in principle take place in a finite time over the entire
volume, and there would then remain no regions that could lead to
an infinite expansion of the universe for $t>0$.  This question,
and the problem of the geodesic completeness of a
self-reproducing inflationary universe, requires further study,
both because the theory of phase transitions in an exponentially
contracting space is not well understood, and because the {\it global}
geometry of a self-reproducing universe differs from the
geometry of de Sitter space.  At present, therefore, we cannot
say with absolute certainty that the new inflationary universe
scenario with no initial singularities is impossible.  That there
be {\it no} singularities in the past, however, is probably too
strong a requirement.

A more natural possibility is suggested by the chaotic inflation
scenario, where most of the physical volume of the universe at a
hypersurface of given density is comprised of regions which, by
virtue of fluctuations in the field $\varphi$, passed through a
stage with $\vf\sim\m^4$.  In this scenario, classical space-time
behaves as if it were in a state of dynamic equilibrium with the
\index{Space-time!space-time foam and}%
\index{Space-time foam!space-time and}%
space-time foam:  regions of classical space are continually
created out of the space-time foam, and some of these are
\index{Space-time foam}%
reconverted to the foam with $\vf\ga\m^4$.  In that sense, the
occurrence of spatial ``singularities'' is part and parcel of
this scenario.  At the same time, what is graphically clear about
this scenario is that instead of dealing with the problem of
creation of an {\it entire} universe from a singularity, prior to
which {\it nothing} existed, and its subsequent d\'enouement into
{\it nothingness}, we are simply concerned with an endless
process of interconversion of phases in which quantum
fluctuations of the metric are either large or small.  Our
results imply that once it has arisen, classical space-time --- a
phase in which quantum fluctuations of the metric are small ---
will never again disappear.  Even more than in the new
inflationary universe scenario, the global geometric properties
of a region filled with this phase\footnote{We traditionally
refer to this particular region as the universe, although
\index{Universe}%
strictly speaking, {\it the universe} (i.e., all that exists)
includes those regions occupied by the space-time foam as well.}
differ from those of de Sitter space.  If this region turns out
to be geodesically complete, then one can plausibly discuss a
model in which the universe has neither a unique beginning nor a
unique end.

Actually, however, as we have already pointed out, this
possibility arises in the chaotic inflation scenario without even
taking the self-regeneration process into consideration.
Specifically, if the universe is finite and initially no larger
than the Planck size, $l\la\m^{-1}$, then it is not unreasonable
to suppose that at some initial time $t=0$ (to within perhaps
$\Delta t\sim\m^{-1}$) the entire universe came into being as a
whole out of the space-time foam (in classical language, it
appeared from a singularity).  If, however, the universe is
infinite, then the possibility that an infinity of causally
disconnected regions of classical space will simultaneously
\index{Space-time foam}%
appear out of the space-time foam seems totally
unlikely.\footnote{From this standpoint, open and flat Friedmann
models, which are quite useful in the description of local
properties of our universe at any stage of its existence. At the
same time, the model of a closed Friedmann universe {\it can}
describe the global properties of the universe, but only during
the earliest stages of its evolution, until diffusion of the
field $\varphi$ results in a large distortion of the original
metric.}

To avoid confusion, it should again be emphasized that the
existence of an initial general (global) spacelike cosmological
singularity is not, in and of itself, a necessary consequence of
the general topological theorems on singularities.  This
conclusion is based primarily on the assumption of global
homogeneity of the universe.  Within the framework of the hot
universe theory, such an assumption, even though it had no
fundamental justification to back it up, nevertheless seemed
unavoidable, since in that theory the observable part of the
universe arose by virtue of the expansion of an enormous number
of causally disconnected regions in which for some unknown reason
the matter\index{Density!matter}\index{Matter density}
density was virtually the same (see the discussion of
the homogeneity and horizon problems in Chapter \ref{c1}).  On
the other hand, in the inflationary universe scenario, the
assumption that the universe is globally homogeneous is
unnecessary, and in many cases it is simply wrong.  Therefore, in
the context of inflationary cosmology, the conventional statement
that in the very early stages of evolution of the universe there
was some instant of time before which there was no time at all
(see Section \ref{s1.5}) is,  at the very least, not
well-founded.\index{Cosmological singularity!global structure of
inflationary universe and|)}\index{Global structure of inflationary
universe!and general cosmological singularity|)}

\section{\label{s10.5}Inflation and the Anthropic
\index{Inflation!Anthropic Principle and|(}%
\index{Anthropic Principle!inflation and|(}%
Principle}

One of the main desires of physicists is to construct a theory
that unambiguously predicts the observed values for all
parameters of all the elementary particles that populate our
\index{Elementary particles!parameters of}%
universe.  The noble idealism of the researcher compels many to
believe that the correct theory describing our world should be
both beautiful and unique.  This does not at all imply, however,
that all parameters of elementary particles in such a theory must
be uniquely calculable.  For example, in supersymmetric SU(5)
theory, the effective potential ${\rm V}(\Phi, {\rm H})$ for the
Higgs fields $\Phi$ and H that figure in this theory has several
different minima, and without taking gravitational effects into
account, the
\index{Density!vacuum energy}%
\index{Vacuum energy density}%
\index{Energy density, vacuum}%
vacuum energy ${\rm V}(\Phi, {\rm H})$ would be the
same at all of these minima.  Each of the minima corresponds to a
different type of symmetry breaking in this theory, i.e., to
different properties of elementary particles. Gravitational
interactions remove the energy degeneracy between these minima.
But the lifetime of the universe in a state corresponding to any
such minimum turns out either to be infinite or at least many
orders of magnitude greater than $10^{10}$ years [\cite{329}].
This means that prescribing a specific grand unified theory will
not always enable one to uniquely determine properties of
elementary particles in our universe.  An even richer spectrum of
possibilities comes to the fore in Kaluza--Klein and superstring
theories, where an exponentially large variety of
\index{Compactification schemes}%
compactification schemes is available for the original
multidimensional space; the type of compactification determines
the coupling constants, the vacuum energy, the symmetry breaking
properties in low-energy elementary particle physics, and
finally, the effective dimensionality of the space we live in
(see Chapter \ref{c1}).  Under these circumstances, the most
\index{Elementary particles!parameters of}%
varied sets of elementary-particle parameters (mass, charge,
etc.) can appear.  It is conceivable that this is the very reason
why we have not yet been able to identify any particular
regularity in comparisons of the electron, muon, proton, W-meson,
and Planck masses.  Most of the parameters of elementary
particles look more like a collection of random numbers than a
unique manifestation of some hidden harmony of Nature.
Meanwhile, it was pointed out long ago that a minor change (by a
\index{Electron mass}%
\index{Mass!electron}%
factor of two or three) in the mass of the electron, the
fine-structure constant $\alpha_e$, the strong-interaction
constant $\alpha_s$, or the gravitational constant would lead to
a universe in which life as we know it could never have arisen.
For example, increasing the mass of the electron by a factor of
two and one-half would make it impossible for atoms to exist;
multiplying $\alpha_e$ by one and one-half would cause protons
and nuclei to be unstable; and more than a ten percent increase
in $\alpha_s$ would lead to a universe devoid of hydrogen.
Adding or subtracting even a single spatial dimension
\index{Dimensions}would make
planetary systems impossible, since in space-time with
dimensionality $d > 4$, gravitational forces between distant
bodies fall off faster than $r^{-2}$ [\cite{330}], and in
\index{General theory of relativity}%
\index{Relativity!general theory of}%
space-time with $d<4$, the general theory of relativity tells us
that such forces are absent altogether.

Furthermore, in order for life as we know it to exist, it is
necessary that the universe be sufficiently large, flat,
homogeneous, and isotropic.  These facts, as well as a number of
other observations and remarks, lie at the foundation of the
so-called Anthropic Principle in cosmology [\cite{77}].
According to this principle, we observe the universe to be as it
is because only in such a universe could observers like ourselves
exist.  There are presently several versions of this principle
extant (the
\index{Weak Anthropic Principle}%
\index{Anthropic Principle!Weak}%
Weak Anthropic Principle, the Strong Anthropic
Principle, the Final Anthropic Principle, etc.) --- see
[\cite{331}].  All versions, formulated in markedly different
ways, in one way or another interrelate the properties of the
universe, the properties of elementary particles, and the very
fact that mankind exists in this universe.

{\looseness=1
At first glance, this formulation of the problem looks to be
faulty, inasmuch as mankind, having appeared $10^{10}$ years
after the basic features of our universe were laid down, could in
no way influence either the structure of the universe or the
properties of the elementary particles within it.  In reality,
however, the issue is not one of cause and effect, but just of
correlation between the properties of the observer and the
\index{Observer}%
properties of the universe that he observes (in the same sense as
in the Einstein--Podolsky--Rosen experiment [\cite{332}], where
\index{Einstein--Podolsky--Rosen experiment}%
there is a correlation between the states of two different
particles but no interaction between them).  In other words, what
is at issue is the {\it conditional probability} that the
universe will have the properties that we observe, with the
obvious and apparently trivial condition that observers like
ourselves, who take an interest in the structure of the universe,
do indeed exist.

}

All this discussion can make sense only if one can actually
compare the probabilities of winding up in different universes
having different properties of space and matter, but that is
possible only if such universes do in fact exist.  If it is not
so, any talk of altering the mass of the electron, the fine
structure constant, and so forth is perfectly meaningless.

One possible response to this objection is that the wave function
of the universe describes both the observer and the rest of the
universe in all its possible states, including all feasible
variants of compactification and spontaneous symmetry breaking
(see Section \ref{s10.1}).  By making a measurement that improves
our knowledge of the properties of the observer, one
simultaneously obtains information about the rest of the
universe, just as by measuring the spin of one particle in the
\index{Einstein--Podolsky--Rosen experiment}%
Einstein--Podolsky--Rosen experiment one promptly obtains
information about the spin of the other [\cite{302}, \cite{304},
\cite{359}].

This answer, in our view, is correct and entirely sufficient.
Nevertheless, it would be more satisfying to have an alternative
reply to the foregoing objection, one that is conceptually
simpler and that does not require an analysis of the somewhat
obscure foundations of quantum cosmology for its justification.
Moreover, we would like to obtain an answer to another (and in
our opinion, the most important) objection to the Anthropic
Principle, namely that it does not seem at all necessary for the
existence of life as we know it to have identical conditions
(homogeneity, isotropy, ratios
$\displaystyle \frac{n_{\rm B}}{n_\gamma}\sim10^{-9}$,
$\displaystyle \frac{\delta\rho}{\rho}\sim10^{-5}$, etc.) over
the whole observable part of the universe.  The random occurrence
of such uniformity seems completely unlikely.

As noted in Chapter \ref{c1}, both of these objections can be
dispatched particularly simply by the theory of the
self-reproducing inflationary universe.  Specifically, long-wave
fluctuations are generated during inflation not only of the
inflaton field $\varphi$, which drives inflation, but of all
other scalar fields $\Phi$ with mass $m_\Phi\ll{\rm H}$ as well
(and having a small coupling constant $\xi$ in possible
interactions of the type $\xi\,{\rm R}\,\Phi^2$).  In the chaotic
inflation scenario, this means that in certain regions of the
universe where the potential energy density of the scalar field
$\varphi$ permanently fluctuates near $\vf\sim\m^4$  (by virtue
of the self-regeneration process taking place in such regions),
long-wave fluctuations of practically every scalar field $\Phi$
grow until the mean value of the potential energy density of each
of these fields is no longer of order $\m^4$.  This follows
\index{Hawking--Moss equation}%
simply from an analysis of the Hawking--Moss distribution
(\ref{7.4.1}) for the field $\Phi$ with ${\rm V}(\Phi=0,\varphi)\sim\m^4$.

The upshot of this process is that a distribution of the scalar
fields $\varphi$ and $\Phi$ is set up that is quite uniform on
exponentially large scales because of inflation;  but on the
scale of the universe as a whole, on the other hand, the fields
can take on practically any value for which their potential
energy density does not exceed the Planck value.  In those
regions of the universe where inflation has ended, the fields
$\varphi$ and $\Phi$ roll down to different {\it local} minima of
${\rm V}(\varphi,\Phi)$, and since all feasible initial
conditions for rolling are realized in different regions of the
universe, the universe becomes divided into different
exponentially large domains containing the fields $\varphi$ and
$\Phi$ corresponding to all local minima of ${\rm
V}(\varphi,\Phi)$, i.e., all possible types of symmetry breaking
in the theory.

At the stage of strong fluctuations with ${\rm V}(\varphi,\Phi)\sim\m^4$,
not only can the magnitudes of the
scalar fields vary, but large metric fluctuations can also be
generated, leading to local compactification or
decompactification of space in Kaluza--Klein or superstring
theories.  If a region of space with changing compactification is
inflating and has an initial size ${\rm H}^{-1}\sim\m^{-1}$  (at
Planck densities, the probability of this happening should not be
too low), then as a result of inflation this region will turn
into an exponentially large domain with a new type of
compactification (for example, a different dimensionality)
[\cite{78}].

Thus, the universe will be partitioned into enormous regions
(mini-universes), which will manifest all possible types of
compactification and all possible types of spontaneous symmetry
breaking compatible with the process of inflation, leading to an
exponential growth in the size of these regions.  Reference
[\cite{333}] contains an implementation of this scenario in the
context of certain specific Kaluza--Klein theories.

It should be emphasized that due to the unlimited temporal extent
of the process of inflation in a self-reproducing universe, such
a universe will support an unbounded collection of mini-universes
of all types, even if the probability of creation of some of them
is extremely small.  But this is just what is needed in support
of the so-called
\index{Weak Anthropic Principle}%
\index{Anthropic Principle!Weak}%
Weak Anthropic Principle:  we reside in regions
with certain space-time and matter properties not because other
regions are impossible, but because regions of the required type
exist, and life as we know it (or, to be more precise, the carbon
based life of human observers of our type) would not be possible
in others.\footnote{ Zelmanov proposed a similar formulation of
the Anthropic Principle [\cite{77}], saying that we are witnesses
only to certain definite kinds of physical processes because
other processes take place without witnesses.}  It is important
here that the total volume of regions in which we could live be
unbounded, so that life as we know it will come into being even
if the probability of its spontaneous appearance is vanishingly
small.  This does not mean, of course, that we can choose the
laws of physics at random.  The issue simply involves choosing
one type of compactification and symmetry breaking or another, as
allowed by the theory at hand.  The search for theories in which
{\it the part of the world that surrounds us} can have the
properties that we observe is still a difficult problem, but it
is much easier than a search for theories in which {\it the whole
world} is not permitted to have properties different from those
in the part where we now live.

Naturally, most of what we have said would remain true if we
simply considered an infinitely large universe with chaotic
initial conditions.  But when inflation is not taken into
account, the Anthropic Principle is incapable of explaining the
uniformity of properties of the observable part of the universe
(see Chapter \ref{c1}).  Furthermore, the mechanism for
self-regeneration of the inflationary universe enables one to put
the Anthropic Principle on a firm footing, given the most natural
initial conditions in the universe, and regardless of whether it
is finite or infinite.

We now consider several examples that demonstrate the various
ways in which the Anthropic Principle can be applied in
inflationary cosmology.

1)  Consider first the symmetry breaking process in
supersymmetric SU(5) theory.  After inflation, the universe
breaks up into exponentially large domains containing the fields
$\Phi$ and H, corresponding to all possible types of symmetry
breaking.  Among these domains, there will be some in the
SU(5)-symmetric phase and some in the
$\mbox{SU}(3) \times {\rm U}(1)$-symmetric phase, corresponding
to the type of symmetry breaking that we observe.  Within each of
these domains, the vacuum state will have a lifetime many orders
of magnitude longer than the $10^{10}$ years which have passed
since the end of inflation in our part of the universe.  We live
inside a domain with $\mbox{SU}(3) \times {\rm U}(1)$ symmetry,
within which there are strong, weak, and electromagnetic
interactions of the type actually observed.  This occurs not
because there are no other regions whose properties differ from
ours, and not because life is totally impossible in other
regions, but because life {\it as we know it} is only possible in
a region with $\mbox{SU}(3) \times {\rm U}(1)$ symmetry.

2)  Consider next the theory of the\index{Axion fields}
axion field $\theta$ with a
potential of the form (\ref{7.7.22}):
\be
\label{10.5.1}
{\rm V}(\theta)\sim m^4_\pi
\left(1-\cos\frac{\theta}{\sqrt{2}\,\Phi_0}\right)\ .
\ee
The field $\theta$ can take any value in the range
$-\sqrt{2}\,\pi\,\Phi_0$ to $\sqrt{2}\,\pi\,\Phi_0$.  A natural
estimate for the initial value of
the\index{Axion fields} axion field would therefore
be $\theta={\rm O}(\Phi_0)$, and the initial value of ${\rm V}(\theta)$
should be of order $m_\pi^4$.  An investigation of
the rate at which the energy of
the\index{Axion fields} axion field falls off as the
universe expands shows that for $\Phi_0\ga10^{12}$ GeV, most of
the energy density would presently be contributed by axions,
while the baryon energy density would be considerably lower than
its presently observed value of $\rho_{\rm B}\ga2\cdot 10^{-31}$
g/cm$^3$.  (Since the universe becomes almost flat after
inflation, the overall matter density in the universe should now
be $\rho_c\sim2\cdot 10^{-29}$ g/cm$^3$, independent of the value
of the parameter $\Phi_0$.) This information was used to derive
the strong constraint $\Phi_0\la10^{11}$--$10^{12}$ GeV
[\cite{49}].  This is not a terribly felicitous result, as axion
fields with $\Phi_0\sim10^{15}$--$10^{17}$ GeV show up in a
natural way in many models based on superstring theory
[\cite{50}].

Let us now take a somewhat closer look at whether one can
actually obtain the constraint $\Phi_0\la10^{11}$--$10^{12}$ GeV
in the context of inflationary cosmology.

As we remarked in Section \ref{s7.7}, long-wave fluctuations of
the axion field $\theta$ are generated at the time of inflation
(if Peccei--Quinn symmetry breaking, resulting in the potential
\index{Peccei--Quinn symmetry breaking}%
\index{Symmetry breaking!Peccei--Quinn}%
(\ref{10.5.1}), takes place before the end of inflation).  By the
end of inflation, therefore, a quasihomogeneous distribution of
the field $\theta$ will have appeared in the universe, with the
field taking on all values from $-\sqrt{2}\,\pi\,\Phi_0$ to
$\sqrt{2}\,\pi\,\Phi_0$  at different points in space with a
probability that is almost independent of $\theta_0$ [\cite{276},
\cite{224}].  This means that one can always find exponentially
large regions of space within which $\theta\ll\Phi_0$.  The energy
of the axion field always remains relatively low in such regions,
and there is no conflict with the observational data.

In and of itself, this fact does not suffice to remove the
constraint $\Phi_0\la10^{12}$ GeV, since when $\Phi_0\gg10^{12}$
GeV, only within a very small fraction of the volume of the
universe is the axion field energy density small enough by
comparison with the baryon density.  It might therefore seem
unlikely that we just so happened make our appearance in one of
these particular regions.  Consider, for example, those regions
initially containing a field $\theta_0\ll\Phi_0$, for which the
present-day ratio of the energy density of the axion field to the
baryon density is consistent with the observational data.  It can
be shown that the total number of baryons in regions with
$\theta\sim10\,\theta_0$ should be ten times the number in
regions with $\theta\sim\theta_0$.  One might therefore expect
the probability of randomly winding up in a region with
$\theta\sim10\,\theta_0$ (contradicting the observational data)
to be ten times that of winding up in a region with
$\theta\sim\theta_0$.  Closer examination of this problem
indicates, however, that the mean matter density in galaxies at
time $t\sim10^{10}$ years is proportional to $\theta^8$, and in
regions with $\theta\sim10\,\theta_0$  it should be $10^8$ times
higher than in regions like our own, with $\theta\sim\theta_0$
\index{Star formation process}%
[\cite{334}].  A preliminary study of the star formation process
in galaxies with $\theta\sim10\,\theta_0$ indicates that
solar-type stars are most likely not formed in such galaxies.  If
that is really the case, then the conditions required for the
appearance of life as we know it can only be realized when
$\theta\sim\theta_0$, and that is exactly why we find ourselves
to be in one such region rather than in a typical region with
$\theta\gg\theta_0$.  Generally speaking, then, the observational
data do not imply that $\Phi_0\la10^{12}$ GeV.  In any event,
since regions with $\theta\sim\theta_0$ most assuredly exist, a
derivation of the constraint $\Phi_0\la10^{12}$ GeV would require
that one show that it is much more improbable for life as we know
it to arise in regions with $\theta\sim\theta_0$ than in regions
with $\theta\gg\theta_0$.  As we have already said, an
investigation of this question indicates just the opposite.

{\looseness=-1
The preceding discussions are quite general in nature, and can be
applied not just to the theory of axions, but to the theory of
any other light, weakly interacting scalar fields as well ---
dilatons, for example [\cite{335}].  In principle, by applying
the Anthropic Principle to axion cosmology, one might attempt to
explain why the present-day baryon density\index{Baryon matter density}
\index{Density!baryon matter}$\rho_{\rm B}$ is
$10^{-1}$--$10^{-2}$ times the total matter density
$\rho_0\approx\rho_c$ in the universe.  Actually, for
$\theta\ll\theta_0$, the energy density of the axion field would
be low ($\rho_a\sim\theta^2$), so the main contribution to
$\rho_0$ would come from baryons,
$\rho_0\approx\rho_c\approx\rho_{\rm B}$.  But only a small
fraction of the baryons in the universe (proportional to
$\displaystyle \frac{\theta}{\Phi_0}$) are to be found in regions
with $\theta\ll\theta_0$.  At the same time, for
$\theta\gg\theta_0$, the conditions of life would be markedly
different from our own, and it is most unlikely that one would
find himself inside such a region of the universe.  The location
of the maximum probability for the existence of life as we know
it, as a function of $\theta$, depends on the value of $\Phi_0$,
and for a particular value $\Phi_0\gg10^{12}$ GeV, the maximum
may be attained precisely in a state with initial value
$\theta\sim \theta_0$, in which now
$\rho_{\rm B}\sim10^{-1}$--$10^{-2}\,\rho_0$.  Consequently, a
\index{Galaxy formation}%
\index{Star formation process}%
study of the theory of star and galaxy formation together with a
detailed study of the conditions necessary for the existence of
life as we know it may actually enable us to determine the most
likely value of the parameter $\Phi_0$ in the theory of axions. }

3)  The preceding results can be employed practically unchanged
to avoid one of the principal difficulties encountered in using
the mechanism for generating the
\index{Baryon asymmetry}%
\index{Asymmetry baryon}%
\index{Universe!baryon asymmetry of}%
baryon asymmetry of the universe
proposed by Affleck and Dine [\cite{97}, \cite{98}].  Recall that
the baryon asymmetry produced by this mechanism is typically too
large:  the value of $\displaystyle \frac{n_{\rm B}}{n_\gamma}$
ranges from $-{\rm O}(1)$ to $+{\rm O}(1)$, depending on the
magnitude of the angle $\theta$ in isotopic spin space between
the initial values of two different scalar fields.  According to
the inflationary universe scenario, one can always find
exponentially large regions in which this angle is small and
$\displaystyle \frac{n_{\rm B}}{n_\gamma}\sim10^{-9}$.  Such
regions occupy a very small fraction of the total volume of the
universe.  But in regions, say, with
$\displaystyle \frac{n_{\rm B}}{n_\gamma}\sim10^{-7}$, the
density of matter in galaxies will be eight orders of magnitude
greater than in our own, and life as we know it will either be
impossible or extremely unlikely.  Naturally, there are a number
of other ways to get rid of the excess
\index{Asymmetry baryon}%
\index{Baryon asymmetry}%
\index{Universe!baryon asymmetry of}%
baryon asymmetry of
the universe (see Section \ref{s7.10}), but interestingly enough, the
Anthropic Principle as applied within the scope of the theory of
the inflationary universe may turn out to be sufficient to
resolve this problem all by itself.

4)  The last example that we present here is somewhat different
from its predecessors.  We know that in the standard
Friedmann\index{Cosmology!Friedmann}\index{Friedmann cosmology}
cosmology, the universe, if it is closed, will spend
approximately half its life in a state of expansion, and the
other half in a state of contraction.  A similar phenomenon can
also occur locally in an inflationary universe on scales at which
density inhomogeneities that came into being during the
inflationary stage become large, with
$\displaystyle \frac{\delta\rho}{\rho}\sim1$ [\cite{336}].  The
question that arises is then `Why is the observable part of the
universe expanding?' Do we live in an expanding part of the
universe by accident, or is there some special reason for this
circumstance?

The answer to this question is related to the fact that in the
simplest $\displaystyle \frac{\lambda}{4}\,\varphi^4$ theory with
$\lambda\sim10^{-14}$, for example, the size of a homogeneous
locally Friedmann part of the universe is of order
$l\sim\m^{-1}\,\exp(\pi\,\lambda^{-1/3})\sim\m^{-1}\cdot10^{6\cdot 10^4}$
(\ref{1.8.8}), and a typical mass concentrated
in such a region is of order ${\rm M}\sim\m\cdot 10^{2\cdot 10^5}$,
so according to (\ref{1.3.15}), the typical time until
the beginning of contraction within such a region is
$t\sim10^{2\cdot 10^5}$ years [\cite{336}].  In a
self-reproducing universe, inasmuch as it exists without end,
there should be regions that are much older and regions that are
much younger.  We happen to live in a relatively young region,
which has existed a total of $10^{10}$ years since the end of
inflation (in this region).  This is related to the fact that
life {\it as we know} it exists near solar-type stars, whose
maximum lifetime is $10^{10}$--$10^{11}$ years.  This is
precisely why the part of the universe that surrounds us is still
in the initial stage of its expansion, and that expansion (within
the framework of the simple model considered here) should last at
least another $10^{2\cdot 10^5}$ years.

Taken by itself, the foregoing certainly does not mean that no
life at all is possible during the contraction stage
[\cite{336}];  the issue is simply that at the current rate of
evolution of living organisms (and also taking into account the
probable decay of baryons after $10^{35}$--$10^{40}$ years),
observers $10^{2\cdot 10^5}$ years hence will scarcely resemble
the way we are now.

We wish to emphasize once again that the
so-called
\index{Weak Anthropic Principle}%
\index{Anthropic Principle!Weak}%
Weak Anthropic
Principle, as formulated and used above, is conceptually quite
simple.  It involves an assessment of the probability of
observing a region of the universe with given properties, under
the condition that the fundamental properties of the observer
himself also be known.  The preceding discussions require no
philosophical sophistication, and their import is of a trivial,
mundane nature --- for instance, we live on the surface of the
Earth not because there is more room here than in interstellar
space, but simply because in interstellar space we could never
breathe.

Furthermore, the richness and heuristic value of the results
obtained via the
\index{Weak Anthropic Principle}%
\index{Anthropic Principle!Weak}%
Weak Anthropic Principle have impelled many
authors to try to expand and generalize it as much as possible,
even if such a generalization is presently not entirely justified
(see [\cite{331}]).  The possibility of such a generalization is
suggested by the unusually important role played by the concept
of the observer in the construction and interpretation of quantum
cosmology.  Most of the time, when discussing quantum cosmology,
one can remain entirely within the bounds set by purely physical
categories, regarding the observer simply as an automaton, and
not dealing with questions of whether he has
\index{Consciousness}%
consciousness or
feels anything during the process of observation [\cite{305}].
This we have done in all of the preceding discussions.  But we
cannot rule out the possibility {\it a priori} that carefully
avoiding the concept of
\index{Consciousness}%
\index{Quantum cosmology!consciousness and}%
consciousness in quantum cosmology
constitutes an artificial narrowing of one's outlook.  A number
of authors have underscored the complexity of the situation,
\index{Participant}%
replacing the word {\it observer} with the word {\it participant},
and introducing such terms as ``self-observing
\index{``Self-observing universe''}%
\index{Universe!``self-observing''}%
universe'' (see, for example, [\cite{302}, \cite{323}]).  In
fact, the question may come down to whether standard physical
theory is actually a closed system with regard to its description
of the universe as a whole at the quantum level:  is it really
possible to fully understand what the universe is without first
understanding what life is?

Leaving aside the question of how well motivated such a statement
of the problem is, let us note that similar problems often appear
in the development of science.  We know, for example, that
classical electrodynamics is incomplete, an example being the
problem of the self-acceleration of an electron, requiring the
use of quantum theory for a solution [\cite{65}].  Quantum
\index{Quantum electrodynamics}%
\index{Electrodynamics, quantum}%
electrodynamics likely suffers from the zero-charge problem
[\cite{156}, \cite{157}], which can be circumvented by including
electrodynamics in a unified nonabelian gauge theory [\cite{3}].
The quantum theory of gravitation presents even greater
conceptual difficulties, and attempts have been made to overcome
these by significantly broadening and generalizing the original
theory [\cite{14}--\cite{17}].  We do not know whether it is
possible to assign an exact meaning to many of the concepts
employed in quantum cosmology (probability of creation of the
universe from ``nothing,'' splitting of the universe in the
many-worlds
\index{Many-worlds interpretation of quantum mechanics}%
\index{Quantum mechanics!many-worlds interpretation of}%
interpretation of quantum mechanics, etc.) without
stepping outside the confines of the existing theory, and
possible ways of generalizing this theory are still far from
clear.  The only thing we can do at this point is to attempt to
draw upon analogies from the history of science which may prove
to be instructive.

Prior to the advent of the special theory of relativity, space,
\index{Special theory of relativity}%
\index{Relativity!special theory of}%
time, and matter seemed to be three fundamentally different
entities.  In fact, space was thought to be a kind of
three-dimensional coordinate grid which, when supplemented by
clocks, could be used to describe the motion of matter.  Special
relativity did away with the insuperable distinction between
space and time, combining them into a unified whole.  But
space-time nevertheless remained something of a fixed arena in
which the properties of matter became manifest.  As before, space
itself possessed no intrinsic degrees of freedom, and it
continued to play a secondary, subservient role as a backdrop for
the description of the truly substantial material world.

{\looseness=1
The general theory of relativity brought with it a decisive
\index{General theory of relativity}%
\index{Relativity!general theory of}%
change in this point of view.  Space-time and matter were found
\index{Space-time}%
\index{Matter}%
to be interdependent, and there was no longer any question of
which was the more fundamental of the two.  Space-time was also
found to have its own inherent degrees of freedom, associated
\index{Gravitational waves}%
\index{Waves, gravitational}%
with perturbations of the metric --- gravitational waves.  Thus,
\index{Space}%
space can exist and change with time in the absence of electrons,
protons, photons, etc.;  in other words, in the absence of
anything that had {\it previously} (i.e., prior to general
relativity) been subsumed by the term {\it matter}.  (Note that
because of the weakness with which they interact, gravitational
waves are exceedingly difficult to detect experimentally, an
as-yet unsolved problem.)

}

{\looseness=1
A more recent trend, finally, has been toward a unified geometric
theory of all fundamental interactions, including gravitation.
Prior to the end of the 1970's, such a program --- a dream of
Einstein's --- seemed unrealizable;  rigorous theorems were
proven on the impossibility of unifying spatial symmetries with
the internal symmetries of elementary particle theory
[\cite{337}].  Fortunately, these theorems were sidestepped after
the discovery of supersymmetric theories [\cite{85}].  In
principle, with the help of supergravity, Kaluza--Klein, and
superstring theories, one may hope to construct a theory in which
all matter fields will be interpreted in terms of the geometric
properties of some multidimensional superspace
[\cite{13}--\cite{17}].  Space would then cease to be simply a
requisite mathematical adjunct for the description of the real
world, and would instead take on greater and greater independent
significance, gradually encompassing all the material particles
under the guise of its own intrinsic degrees of freedom.  Of
course, this does not at all mean that the concept of matter
becomes useless.  The issue at hand is simply the revelation of
the fundamental unity of space, time, and matter, which is hidden
from us in much the same way that the unity of the weak and
electromagnetic interactions was hidden until recently.

}

According to standard materialistic doctrine,
\index{Consciousness}%
consciousness, like\linebreak[10000]
space-time before the invention of general relativity, plays a
secondary, subservient role, being considered just a function of
matter and a tool for the description of the truly existing
material world.  It is certainly possible that nothing similar to
the modification and generalization of the concept of space-time
will occur with the concept of consciousness in the coming
decades.  But the thrust of research in quantum cosmology has
taught us that the mere statement of a problem which might at
first glance seem entirely metaphysical can sometimes, upon
further reflection, take on real meaning and become highly
significant for the further development of science.  We should
therefore like to take a certain risk and formulate several
questions to which we do not yet have the answers.

{\looseness=1
Is it not possible that
\index{Consciousness}%
consciousness, like space-time, has its
own intrinsic degrees of freedom, and that neglecting these will
lead to a description of the universe that is fundamentally
incomplete?  Will it not turn out, with the further development
of science, that the study of the universe and the study of
consciousness will be inseparably linked, and that ultimate
progress in the one will be impossible without progress in the
other?  After the development of a unified geometrical
description of the weak, strong, electromagnetic, and
gravitational interactions, will the next important step not be
the development of a unified approach to our entire world,
including the world of consciousness?
\index{Consciousness}%

}

All of these questions might seem somewhat naive and out of place
in a serious scientific publication, but to work in the field of
quantum cosmology without an answer to these, and without even
trying to discuss them, gradually becomes as difficult as working
on the hot universe theory without knowing why there are so many
different things in the universe, why nobody has ever seen
parallel lines intersect, why the universe is almost homogeneous
and looks approximately the same at different locations, why
space-time is four-dimensional, and so on (see Section
\ref{s1.5}).  Now, with plausible answers to these questions in
hand, one can only be surprised that prior to the 1980's, it was
sometimes taken to be bad form even to discuss them.  The reason
is really very simple:  by asking such questions, one confesses
one's own ignorance of the simplest facts of daily life, and
moreover encroaches upon a realm which may seem not to belong to
the world of positive knowledge.  It is much easier to convince
oneself that such questions do not exist, that they are somehow
not legitimate, or that someone answered them long ago.

It would probably be best then not to repeat old mistakes, but
instead to forthrightly acknowledge that the problem of
consciousness and the related problem of human life and death are
not only unsolved, but at a fundamental level they are virtually
completely unexamined.  It is tempting to seek connections and
analogies of some kind, even if they are shallow and superficial
ones at first, in studying one more great problem --- that of the
birth, life, and death of the universe.  It may conceivably
become clear at some future time that these two problems are not
so disparate as they might seem.
\index{Inflation!Anthropic Principle and|)}%
\index{Anthropic Principle!inflation and|)}%

\section{\label{s10.6}Quantum cosmology and the
\index{Metric!signature of|(}%
\index{Signature of space-time|(}%
\index{Space-time!signature of|(}%
\index{Quantum cosmology!signature of space-time and|(}%
signature of space-time}

The most significant modification to the concept of
four-dimensional space-time that we have discussed so far is that
of a space with one temporal and $d-1$ spatial coordinates, some
of the latter being compactified.  This construction, however, is
clearly not the most general.  Our intuitive ideas about
space-time are linked to our study of the dynamics of objects
whose dimensions may be arbitrarily small, but in the quantum
theory of gravitation it is difficult (or impossible) to consider
objects smaller than $\m^{-1}$.  If the theory is to be based on
the study of extended objects like strings or membranes, many of
our intuitive ideas about the geometrical objects associated with
them (points, straight lines, etc.) will be found to be largely
inadequate [\cite{17}].

Unanswered questions arise, however, even at a simpler level.
For example, why are there many spatial coordinates, but only one
temporal coordinate?  That is, why does our space have the
signature $(+ ---\ldots-)$?  Why could it not be Euclidean,  with
signature $(+ + \ldots +)$ or $(--\ldots-)$?  Why are just the
spatial dimensions compactified, and not the temporal?  Are
transformations which change the signature of the metric possible
[\cite{292}]?

In the context of a model of the universe consisting of large
regions with differing properties, these may all prove to be
sensible questions.  It is therefore worth considering, if only
briefly, how the properties of the universe would be altered
under a change in the signature of the metric.  There are many
aspects to this question, some of which stand out particularly
clearly in supergravity and the theory of superstrings.  For
instance, the 16-component Majorana--Weyl spinors required for a
\index{Majorana--Weyl spinors}%
\index{Spinors, Majorana--Weyl}%
formulation of supergravity in a space with $d=10$ only exist for
three different signatures of the metric:  $1 + 9$ (one temporal
and nine spatial dimensions), $5 + 5$, and $9 + 1$ [\cite{338}];
a supersymmetric theory has only been formulated in the first
case.

There exists one additional more general problem that arises in a
very broad class of theories when space contains more than one
time dimension.  The problem is most readily understood by
studying a scalar field in a flat space with signature $(+ + - -)$
as an example.  The usual dispersion relation for the field
$\varphi$, which takes the form $k_0^2={\bf k}^2+m^2$  in
Minkowski space, then becomes
\be
\label{10.6.1}
k_0^2=k_2^2+k_3^2+m^2-k_1^2\ .
\ee
Here the momentum $k_1$ can clearly change the sign of the
effective mass squared in (\ref{10.6.1}), or in other words, it
can induce exponentially rapid growth of fluctuations of the
field $\varphi$ when $k_1^2>k_2^2+k_3^2+m^2$:
\be
\label{10.6.2}
\delta\varphi\sim
\exp(\sqrt{k_1^2-k_2^2-k_3^2-m^2}\,t)\ .
\ee
This effect is analogous to the instability of the vacuum state
\index{Instability of vacuum state}%
\index{Vacuum state, instability of}%
with $\varphi=0$ in the theory of a scalar field with negative
mass squared (see (\ref{1.1.5}), (\ref{1.1.6})).  In the theory
(\ref{1.1.5}), however, the development of the instability came
to a halt when the sign of the effective squared mass
$m^2(\varphi)$ changed as the field $\varphi$ increased.  But in
the present example, the instability grows without bound, since
there are exponentially growing modes with arbitrary values of
$m^2$ for sufficiently large momenta $k_1$.  Since the
instability is associated precisely with the very largest momenta
(shortest wavelengths), the existence of such an instability will
most likely be a general feature of theories in a space with more
than one time dimension, regardless of either the topology of the
space or whether the additional time dimensions are compactified.
In some theories it proves to be possible to avoid instability in
modes corresponding to particles that have relatively low mass
after compactification [\cite{293}], but there still remains an
instability due to heavy particles with masses $m$ of the order
of the reciprocal of the compactification radius ${\rm R}_c^{-1}$.
From (\ref{10.6.2}), it follows that this instability is not the
least bit less dangerous.  One might hold out hope that for some
reason there might be a cutoff at $k_0$, $k_1\sim{\rm R}_c^{-1}$
in this theory;  modes with $m\ga{\rm R}_c^{-1}$ might then not
appear.  But if there were a cutoff at momenta of order ${\rm R}_c^{-1}$,
it would become impossible even to discuss compactification in
conventional semiclassical language.  To put it differently,
until such time as we are capable of describing a classical space
containing more than one time dimension, instability seems
unavoidable.

In Euclidean space there is no instability, but there is also no
\index{Euclidean space}%
\index{Space!Euclidean}%
evolution over the course of time, which is necessary for the
existence of life as we know it.  Furthermore, Euclidean space
also lacks the requisite instability with respect to exponential
growth of the universe, which leads to inflation and makes the
universe so large.

To summarize this section, we might say that where there is no
time, there is neither evolution nor life, and where there is too
much time, instability is rampant and life is short.  From this
perspective, the familiar signature of the metric seems to be a
necessary condition for progress to take place within a
relatively orderly framework.
\index{Metric!signature of|)}%
\index{Signature of space-time|)}%
\index{Space-time!signature of|)}%
\index{Quantum cosmology!signature of space-time and|)}%

\section[Cosmological constant and the Anthropic Principle]
{\label{s10.7}The cosmological constant, the Anthropic Principle,
and reduplication of the universe and
\index{Anthropic Principle!cosmological constant and|(}%
\index{Cosmological constant!Anthropic Principle and|(}%
\index{Anthropic Principle}%
\index{Energies!vacuum|(}%
\index{Reduplication of universe|(}%
\index{Universe!reduplication of|(}%
life after inflation}

{\looseness=1%
\noindent{}As noted in Section \ref{s1.5}, one of the most difficult
problems in modern physics is the problem of the vacuum energy,
or the cosmological constant.  There have been a great many
interesting suggestions as to how this problem might be solved
--- for example, see [\cite{17}, \cite{78}, \cite{116},
\cite{292}, \cite{335}, \cite{339}--\cite{359}].  This multitude
of proposals can be divided into two main groups.  The most
attractive possibility is that due to some mechanism related to a
symmetry of the theory, for example, the vacuum energy must be
exactly zero.  The second possibility, presently an active topic
of discussion among experts in the theory of the formation of the
large-scale structure of the universe, is that there is some sort
of mechanism that may engender a
\index{Density!vacuum energy}%
\index{Vacuum energy density}%
\index{Energy density, vacuum}%
vacuum energy density $\rho_\nu$
at the present time of the same order of magnitude as the present
total matter density $\rho_0\sim\rho_c\sim2\cdot 10^{-29}$ g/cm$^3$.
While deep-seated reasons may exist for the vanishing
of the vacuum energy, however, ensuring the equality of
$\rho_\nu$  and $\rho_0$ at the present epoch (if only to order
of magnitude) is difficult without placing unnatural constraints
on the parameters of the theory.

}

{\looseness=1%
The
\index{Anthropic Principle}%
Anthropic Principle suggests one possible escape from this
situation.  To illustrate the basic idea behind this approach to
the problem of the cosmological constant, consider the theory of
a scalar field $\Phi$ with effective potential ${\rm
V}(\Phi,\varphi)=\alpha\,\m^3\,\Phi+\vf$ [\cite{78}, \cite{341}].
Here $\vf$ is the potential of the field $\varphi$ responsible
for inflation, with a minimum at the point $\varphi_0$.  We shall
assume that the constant $\alpha$ is very small,
$\alpha\la10^{-120}$.  Fluctuations of the field $\Phi$ that set
in at the time of inflation result in space being partitioned
into regions with all possible values of ${\rm
V}(\Phi,\varphi_0)$, ranging from $-\m^4$ to $+\m^4$.  In those
regions where ${\rm V}(\Phi,\varphi_0)\ll-10^{-29}$ g/cm$^3$ at
present, the universe looks locally like de Sitter space with
negative vacuum energy.  In such regions, all structures come
into being and pass out of existence in much less than $10^{10}$
years, and life as we know it cannot emerge.  In regions with
${\rm V}(\Phi,\varphi)>2\cdot 10^{-29}$ g/cm$^3$, inflation is
still ongoing, and if the potential ${\rm V}(\Phi,\varphi_0)$ is
very flat ($\alpha\la10^{-120}$) the field $\Phi$ will vary quite
slowly;  the time needed for ${\rm V}(\Phi,\varphi_0)$ to
decrease to $10^{-29}$ g/cm$^3$ will then be more than $10^{10}$
years.  In regions with ${\rm V}(\Phi,\varphi_0)\ga10^{-27}$
\index{Galaxy formation}%
g/cm$^3$, the standard mechanism for galaxy formation is altered
significantly, and for ${\rm V}(\Phi,\varphi_0)\gg10^{-27}$
g/cm$^3$, galaxies and stars like our own are hardly formed at
all [\cite{348}].  This still does not tell us why presently
${\rm V}(\Phi,\varphi_0)\la10^{-29}$ g/cm$^3$, but the fact that
the observational constraints on the vacuum energy density must
be satisfied in at least a few percent of the ``habitable''
reaches of the universe makes the problem of the cosmological
constant much less acute.  An even better model would be one in
which the spectrum of possible values of the vacuum energy
$\rho_{\rm v}$ is discrete rather than continuous, and includes
states with $\rho_{\rm v}=0$ but not those with energy density
less than $10^{-27}$ g/cm$^3$.  With the enormous number of
possible types of compactification in Kaluza--Klein theories,
such a possibility can actually be realized [\cite{292}], and a
similar possibility emerges in superstring theory [\cite{353}].
In any case, the very fact that the
\index{Anthropic Principle}%
Anthropic Principle may
enable us to narrow the range of possible values of $\rho_{\rm v}$
in the observable part of the universe from
$-10^{94}\;\mbox{g/cm$^3$}\la\rho_{\rm v}\la10^{94}\;\mbox{g/cm$^3$}$ to
$-10^{-29}\;\mbox{g/cm$^3$}\la\rho_{\rm v}\la10^{-27}\;\mbox{g/cm$^3$}$
(i.e., to reduce the range by a factor of $10^{121}$) seems
worthy of note.

}

We will soon return to a discussion of using the
\index{Anthropic Principle}%
Anthropic Principle to solve the problem of the cosmological constant.
Here we consider the possibility that the cosmological constant
vanishes
\index{Cosmological constant!vanishing}%
because of some hidden symmetry.  Several suggestions
have been made on this score.  One of the most interesting and
promising ideas has to do with the application of supersymmetric
theories, and superstring theories in particular [\cite{17}].  In
certain versions of such theories, the vacuum energy in the
absence of supersymmetry breaking vanishes to all orders of
perturbation theory [\cite{17}, \cite{353}].  In the real world,
however, supersymmetry is broken, and it is still unclear whether
the vacuum energy remains at zero after supersymmetry breaking.
Another possibility has to do with so-called dilatation
\index{Dilatation invariance}%
invariance, which (if certain rather strong assumptions are made)
might be of some help despite the fact that it is also broken
[\cite{335}] (see also [\cite{354}]).  Below, we discuss one more
possibility with a direct bearing on quantum cosmology that shows
how many surprises this science may hold in store [\cite{344}].

Consider a model that simultaneously describes two different
universes X and $\bar {\rm X}$, with coordinates $x_\mu$ and
$\bar x_\alpha$ respectively ($\mu$, $\alpha = 0$, 1, 2, 3) and
metrics $g_{\mu\nu}(x)$  and $\bar g_{\alpha\beta}(\bar x)$,
containing the fields $\varphi(x)$ and $\bar\varphi(\bar x)$ with
action of the following unusual type
[\cite{344}]:\footnote{Somewhat different (but similar) models
have also been considered elsewhere [\cite{116}, \cite{293}].}
\ba
\label{10.7.1}
{\rm S}&=&{\rm N}\,\int d^4x\,d^4\bar x\:\sqrt{g(x)}\,\sqrt{\bar g(\bar x)}
\nonumber \\
&\times&\left\{\frac{\m^2}{16\,\pi}\,{\rm R}(x)+{\rm L}[\varphi(x)]
-\frac{\m^2}{16\,\pi}\,{\rm R}(\bar x)-
{\rm L}[\bar\varphi(\bar x)]\right\}\ .
\ea
Here N is a normalizing constant.  The action (\ref{10.7.1}) is
invariant under general covariant transformations in each of the
universes individually.  A novel symmetry of the action
(\ref{10.7.1}) is the symmetry under the transformations
$\varphi(x)\rightarrow\bar\varphi(x)$,
$g_{\mu\nu}(x)\rightarrow\bar g_{\alpha\beta}(x)$,
$\bar\varphi(\bar x)\rightarrow\varphi(\bar x)$,
$\bar g_{\alpha\beta}(\bar x)\rightarrow g_{\mu\nu}(\bar x)$
with a subsequent change of sign, ${\rm S}\rightarrow-{\rm S}$.
For reasons that will soon become clear, we call
this
\index{Antipodal symmetry}%
\index{Symmetry!antipodal}%
antipodal symmetry.  (In principle, other terms that do not violate this
symmetry could be added to the integrand of Eq. (\ref{10.7.1}),
such as any odd function of $\varphi(x)-\bar\varphi(\bar x)$;
this would not affect our basic result.)

One immediate consequence of antipodal symmetry is invariance
under a shift in the values of the effective potentials
$\vf\rightarrow\vf+{\rm C}$,
${\rm V}(\bar\varphi)\rightarrow{\rm V}(\bar\varphi)+{\rm C}$,
where C is an arbitrary constant.  Thus, nothing in the theory
depends on the value of the potentials $\vf$ and
${\rm V}(\bar\varphi)$ at their absolute minima $\varphi_0$ and
$\bar\varphi_0$ (note that $\varphi_0=\bar\varphi_0$ and
${\rm V}(\varphi_0)={\rm V}(\bar\varphi_0)$ by virtue of the same
symmetry).  This is precisely why it proves possible to solve the
cosmological constant problem in the theory (\ref{10.7.1}).

However, the main reason for invoking this new symmetry was not
just to solve the cosmological constant problem.  Just as the
theory of mirror particles was originally proposed in order to
make the theory CP-symmetric while maintaining CP-asymmetry in
its observable sector, the theory (\ref{10.7.1}) is proposed in
order to make the theory symmetric with respect to the choice of
the sign of energy.  This removes the old prejudice that even
though the overall change of sign of the Lagrangian (i.e., of
both its kinetic and potential terms) does not change the
solutions of the theory, one {\it must say} that the energy of
all particles is positive.  This prejudice was so strong that
several years ago people preferred to quantize {\it particles}
with {\it negative energy} as {\it antiparticles} with {\it positive energy},
which resulted in the appearance of such
meaningless concepts as negative probability.  We wish to
emphasize that there is no problem with performing a consistent
quantization of theories which describe particles with negative
energy.  Difficulties appear only when there exist interacting
species with both signs of energy.  (As noted in Section
\ref{s10.1}, this is one of the main problems of quantum
cosmology, where one must quantize fields with positive energy
and the scalar factor $a$ with negative energy.)  In the present
case no such problem exists, just as there is no problem of
antipodes falling off the opposite side of the Earth.  The reason
is that the fields $\bar\varphi(\bar x)$ do not interact with the
fields $\varphi(x)$, and the equations of motion for the fields
$\bar\varphi(\bar x)$ are the same as for the fields $\varphi(x)$
(the overall minus sign in front of ${\rm L}[\bar\varphi(\bar x)]$
does not change the Lagrange equations).  In other words, in
spite of the fact that from the standpoint of the sign of the
energy of matter, universe $\bar {\rm X}$ is an antipodal world
where everything is upside-down, there is no instability there,
and particles of the field $\bar\varphi(\bar x)$ are completely
unaware that they have energy of the ``wrong'' sign, just as our
antipodal counterparts living on the other side of the globe are
completely unruffled by the fact that they are upside-down from
our point of view.

Similarly, gravitons from different universes do not interact
\index{Gravitons}%
with each other.  However, some interaction between the two
universes does exist.  In the theory (\ref{10.7.1}), the Einstein
equations take the form
\ba
\label{10.7.2}
{\rm R}_{\mu\nu}(x)-
\frac{1}{2}\,g_{\mu\nu}(x)\,{\rm R}(x)&=&
-8\,\pi\,{\rm G}\,{\rm T}_{\mu\nu}(x)\nonumber \\
&-&g_{\mu\nu}(x)\,\left\langle \frac{{\rm R}(\bar x)}{2}
+8\,\pi\,{\rm G}\,{\rm L}[\bar\varphi(\bar x)]\right\rangle \ ,\nonumber \\
\\
\label{10.7.3}
{\rm R}_{\alpha\beta}(\bar x)-
\frac{1}{2}\,g_{\alpha\beta}(\bar x)\,{\rm R}(\bar x)&=&
-8\,\pi\,{\rm G}\,{\rm T}_{\alpha\beta}(\bar x)\nonumber \\
&-&g_{\alpha\beta}(\bar x)\,\left\langle \frac{{\rm R}(x)}{2}
+8\,\pi\,{\rm G}\,{\rm L}[\varphi(x)]\right\rangle\ . \nonumber \\
\ea
Here ${\rm G}=\m^{-2}$, ${\rm T}_{\mu\nu}$ is the energy-momentum
tensor of the field $\varphi(x)$, ${\rm T}_{\alpha\beta}$ is the
energy-momentum tensor of the field $\bar\varphi(\bar x)$, and
the meaning of the angle-bracket notation
is\index{Angle-bracket notation}
\ba
\label{10.7.4}
\langle R(x)\rangle =\frac{\displaystyle \int
d^4x\:\sqrt{g(x)}\,{\rm R}(x)}%
{\displaystyle \int
d^4x\:\sqrt{g(x)}}\ ,\\
\label{10.7.5}
\langle R(\bar x)\rangle =\frac{\displaystyle \int
d^4\bar x\:\sqrt{\bar g(\bar x)}\,{\rm R}(\bar x)}%
{\displaystyle \int
d^4\bar x\:\sqrt{\bar g(\bar x)}}\ ,
\ea
and similarly for $\langle {\rm L}[\varphi(x)]\rangle $ and
$\langle {\rm L}[\bar \varphi(\bar x)]\rangle $.
Thus, although particles in universes X and $\bar{\rm X}$ do not
interact with each other, the universes themselves {\it do}
interact, but only globally:  each one makes a time-independent
contribution to the average vacuum energy density of the other,
with averaging taking place over the entire history of the
universe.  At the quantum cosmological level, when one writes
down the equations for universe X, for example, averaging should
take place over {\it all} possible states of universe $\bar{\rm
X}$ --- that is, the result should not depend on the initial
conditions in each of the two universes.

Generally speaking, it is extremely difficult to calculate the
averages (\ref{10.7.4}) and (\ref{10.7.5}), but in the
inflationary universe scenario (at least at the classical level),
everything turns out to be quite simple.  Indeed, after inflation
the universe becomes almost flat, and its lifetime becomes
exponentially long (or even infinite, if it is open or flat).  In
that event, the dominant contribution to the mean values $\langle
{\rm R}\rangle $  and $\langle {\rm L}\rangle$ comes from the late
stages of the evolution of the universe, when the fields
$\varphi(x)$ and $\bar \varphi(\bar x)$ relax near the global
minima of $\vf$ and ${\rm V}(\bar \varphi)$.  As a consequence,
the mean value of $-{\rm L}[\varphi(x)]$ is the same, to
exponentially high accuracy, as the value of the potential $\vf$
at its global minimum at $\varphi=\varphi_0$, and the mean value
of the curvature scalar ${\rm R}(x)$ is identical to its value
during the late stages of evolution of the universe X, when the
universe transforms to the state $\varphi=\varphi_0$,
corresponding to the global minimum of $\vf$.  The analogous
statement also holds true for $\langle {\rm L}[\bar \varphi(\bar
x)]\rangle $ and $\langle {\rm R}(\bar x)\rangle $.  For that
reason, Eqs. (\ref{10.7.2}) and (\ref{10.7.3}) take the following
form in the late stages of evolution of the universes X and $\bar
{\rm X}$:
 \ba \label{10.7.6} {\rm R}_{\mu\nu}(x)-
\frac{1}{2}\,g_{\mu\nu}(x)\,{\rm R}(x)&=& 8\,\pi\,{\rm
G}\,g_{\mu\nu}(x)\, [{\rm V}(\bar\varphi_0)-{\rm V}(\varphi_0)]
\nonumber \\
&-&\frac{1}{2}\,g_{\mu\nu}(x)\,{\rm R}(\bar x)\ ,\\
\label{10.7.7}
{\rm R}_{\alpha\beta}(\bar x)-
\frac{1}{2}\,g_{\alpha\beta}(\bar x)\,{\rm R}(\bar x)&=&
8\,\pi\,{\rm G}\,g_{\alpha\beta}(\bar x)\,
[{\rm V}(\varphi_0)-{\rm V}(\bar\varphi_0)]
\nonumber \\
&-&\frac{1}{2}\,g_{\alpha\beta}(\bar x)\,{\rm R}(x)\ ,
\ea
yielding
\ba
\label{10.7.8}
{\rm R}(x)&=&2\,{\rm R}(\bar x)
+32\,\pi\,{\rm G}\,[{\rm V}(\varphi_0)-{\rm V}(\bar\varphi_0)]\ ,\\
\label{10.7.9}
{\rm R}(\bar x)&=&2\,{\rm R}(x)
+32\,\pi\,{\rm G}\,[{\rm V}(\bar\varphi_0)-{\rm V}(\varphi_0)]\ .
\ea
Recall from our earlier discussion that $\varphi_0 =
\bar\varphi_0$ and ${\rm V}(\varphi_0) = {\rm V}(\bar\varphi_0)$
by virtue of
\index{Antipodal symmetry}%
\index{Symmetry!antipodal}%
antipodal symmetry.
This implies that in the late stages of evolution of the universe X,
\be
\label{10.7.10}
{\rm R}(x)=-{\rm R}(\bar x)=
\frac{32}{3}\,{\rm G}\,[{\rm V}(\varphi_0)-{\rm V}(\bar\varphi_0)]=0\ .
\ee

We emphasize that the contribution made by universe $\bar{\rm X}$
to the effective vacuum energy of universe X does not depend on
the time $t$ in the latter.  The cancellation represented by
(\ref{10.7.10}) therefore takes place only in the late stages of
evolution of universe X, and its sole effect is to add a constant
term to $\vf$ in such a way as to obtain ${\rm V}(\varphi_0)=0$.
Thus, the mechanism considered here does not alter the standard
inflationary scenario at all.

Note that this model differs from the conventional Kaluza--Klein
theory in which, as we have already stated, the introduction of
two time coordinates immediately leads to instability.  If would
be straightforward to generalize the theory (\ref{10.7.1}), for
example, by writing the action as an integral over universes
${\rm X}_1, {\rm X}_2, \ldots$, and taking the Lagrangian to be a
sum of Lagrangians from the various fields
$\varphi_1(x_1),\varphi_2(x_2),\ldots$, each of which resides in
only one of these universes.  With such a scheme, our world would
consist of arbitrarily many different universes interacting with
each other only globally, the inhabitants of each having their
own time and their own physical laws.  This approach would
provide a basis for the\index{Anthropic Principle}
Anthropic Principle in its strongest form.

Of course there are shortcomings to be dealt with.  This scheme
could be generalized to supersymmetric theories, but it would be
difficult to do so in a way that ensures that the cosmological
constant\index{Cosmological constant!vanishing}
vanishes automatically.  If the universe is
self-reproducing, one could encounter difficulties in
calculating the averages (\ref{10.7.4}), (\ref{10.7.5}), since
they may become infrared-divergent and the result may depend on
the cutoff.  This question has yet to be thoroughly examined, due
to the very complicated large-scale structure of the
self-reproducing universe.  However, one can easily avoid such
questions in theories in which $\vf$ grows rapidly enough at
$\varphi\ge\varphi^*$, since there will be no self-regeneration
of the universe in such theories.  Another problem is that the
integral over $d^4\bar x$ in (\ref{10.7.1}) renormalizes the
effective Planck constant in the universe X, and one should take
a very small normalization factor
(${\rm N}\sim\exp(-\lambda^{1/3})$ in the $\lambda\,\varphi^4$
theory) in order to compensate this renormalization.  A further
possibility is that in constructing the quantum theory in a
reduplicated universe, one should only do so in each of the
noninteracting universes individually, without taking the
foregoing renormalization of N into consideration.  Note also
that the mechanism for cancellation of the cosmological term that
was suggested above works independently of the value of N.

In any case, the very fact that (at least at the classical level)
there is a large class of models within which the cosmological
constant automatically vanishes, regardless of the detailed
structure of the theory, seems noteworthy.  Moreover, the
possibility of building a consistent theory of many universes
that interact with one another only globally may be of interest
in its own right.

A very interesting and nontrivial generalization of the ideas we
have discussed here has been proposed quite recently in papers by
Coleman [\cite{345}, \cite{346}], Giddings and Strominger
[\cite{349}], and Banks [\cite{347}]; these are based on earlier
work by Hawking [\cite{350}], Lavrelashvili, Rubakov, and
Tinyakov [\cite{351}], and Giddings and Strominger [\cite{352}]
\index{Wormholes}%
dealing with wormholes and the loss of coherence in quantum
gravitation, as well as on a paper by Hawking [\cite{340}] that
treated a possible mechanism for the vanishing of the
cosmological constant in the context of quantum cosmology.

The basic idea in Refs. \cite{345}--\cite{347} and \cite{349} is
that because of quantum effects, the universe can split into
several topologically disjoint but globally interacting parts.
Such processes can take place at any location in our universe
(see [\cite{350}--\cite{352}], [\cite{133}], and Section
\ref{s10.3}).  The
\index{Baby universes}%
\index{Universe!baby}%
baby universes can carry off electron-positron
pairs or some other combination of particles and fields if they
are not prevented from doing so by conservation laws.  The
simplest way to describe this effect is to say that the existence
of baby universes leads to a modification of the effective
Hamiltonian that describes particles and fields in our own
universe [\cite{345}, \cite{349}]:
\be
\label{10.7.11}
{\cal H}(x)={\cal H}_0(\varphi(x),\psi(x),\ldots)
+\sum{\cal H}_i(\varphi(x),\psi(x),\ldots)\,{\rm A}_i\ .
\ee
This Hamiltonian describes the fields $\varphi, \psi, \ldots$ in
our universe on scales exceeding $\m^{-1}$.  In (\ref{10.7.11}),
${\cal H}_0$ is the part of the Hamiltonian unrelated to
topological fluctuations;  the ${\cal H}_i$ are certain local
functions of the fields $\varphi, \psi, \ldots$; the ${\rm A}_i$
represent a combination of creation and annihilation operators in
the
\index{Baby universes}%
\index{Universe!baby}%
baby universes. Thus, for example, a term like ${\cal
H}_1\,{\rm A}_1$, where ${\cal H}_1$ is constant, is associated
with the possibility of a change in the vacuum energy density due
to an interaction with baby universes, the term $\bar e(x)\,e(x)\,{\rm A}_2$
is associated with possible electron-positron pair exchange, and
so on.  The operators ${\rm A}_i$ are independent of the position
$x$, since baby universes cannot carry off energy or momentum.
According to [\cite{345}, \cite{346}], the requirement that the
Hamiltonian in our universe be local,
\be
\label{10.7.12}
[{\cal H}(x),{\cal H}(y)]=0
\ee
for spacelike $x-y$, implies that all of the operators ${\rm
A}_i$ commute with each other.  They can therefore all be
simultaneously diagonalized by the ``$\alpha$-states''
$|\alpha_i\rangle $, so that
\be
\label{10.7.13}
{\rm A}_i\,|\alpha_i\rangle =\alpha_i\,|\alpha_i\rangle \ .
\ee
If the quantum state of the universe is an eigenstate of the
operators ${\rm A}_i$, then one consequence of the complicated
structure of the vacuum (\ref{10.7.13}) will be the introduction
of an infinitude of {\it a priori} undetermined parameters  into
the effective Hamiltonian:  in (\ref{10.7.11}), one simply
replaces the operators ${\rm A}_i$ by their eigenvalues
$\alpha_i$ in the given state.  If the universe is not originally
in an eigenstate of the  ${\rm A}_i$, its wave function, after a
series of measurements, will nonetheless quickly be reduced to
such an eigenstate [\cite{345}].

This makes it possible to consider anew many of the fundamental
problems of physics.  It is often supposed that the basic goal of
theoretical physics is to find exactly what Lagrangian or
Hamiltonian correctly describes our entire world.  But one could
well ask the following question:  if we assume that there was a
time when our universe (or the part in which we live) did not
exist (at least as a classical space-time), in what sense can one
speak of the existence of laws {\it at that time} which would
have governed its birth and evolution?  We know, for example,
that the laws that control biological evolution are recorded in
our genetic code.  But where were the laws of physics recorded if
there was no universe?

One possible answer is that the final structure of the effective
Hamiltonian, including whatever values are taken on by the
constants $\alpha_i$, is not fixed until a series of measurements
have been made, determining with finite accuracy which of the
possible quantum states of the universe $|\alpha_i\rangle $ we
live in [\cite{355}].  This implies that the concept of an
observer may play an important role not just in discussions of
the various characteristics of our universe, but in the very laws
by which it is governed.

In general, the wave function of the universe can depend on the
parameters $\alpha_i$.  This possibility lies at the root of
Coleman's explanation for the vanishing of the cosmological
constant
$$
\Lambda=\frac{8\,\pi\,{\rm V}(\varphi_0)}{\m^2}\ ,
$$
where ${\rm V}(\varphi_0)$ is the present value of the vacuum
energy density.  The basic idea goes back to a paper by Hawking
[\cite{340}], who made use of the Hartle--Hawking wave function
(\ref{10.1.12}), (\ref{10.1.17}).  According to Hawking, if the
cosmological constant could for some reason take on arbitrary
values, the probability of winding up in a universe with a given
value of $\Lambda$ is
\be
\label{10.7.14}
{\rm P}(\Lambda)\sim\exp[-{\rm S}_{\rm E}(\Lambda)]
=\exp\frac{3\,\m^4}{8\,{\rm V}}=\exp\frac{3\,\pi\,\m^2}{\Lambda}
\ee
(compare with (\ref{10.1.18})).  In the context of the approach
that we are considering, which is based on the theory
(\ref{10.7.11}), (\ref{10.7.13}), the cosmological constant, like
any other constant, could actually take on different values,
depending on exactly what quantum state we happen to be in.  In
calculating ${\rm P}(\Lambda)$ for this case, however, one must
also sum over all topologically disconnected configurations of the
\index{Baby universes}%
\index{Universe!baby}%
baby universes, which leads to a modified expression for
${\rm P}(\Lambda)$ [\cite{346}]:
\be
\label{10.7.15}
{\rm P}(\Lambda)\sim\exp\left(\exp\frac{3\,\pi\,\m^2}{\Lambda}\right)\ .
\ee
From (\ref{10.7.14}) and (\ref{10.7.15}), we see that ${\rm
P}(\Lambda)$ is sharply peaked at  $\Lambda=0$, i.e. among all
possible universes, the most probable are those with a
vanishingly small cosmological constant.

Just how reliable is this conclusion?  At the moment, it is
difficult to say.  In fact, the probability of formation of Baby
\index{Baby universes}%
\index{Universe!baby}%
universes, resulting in a gravitational vacuum with highly
complicated structure, has yet to be firmly established.  The
description of this process using Euclidean methods
[\cite{350}--\cite{352}] differs from that obtained via the
stochastic approach [\cite{133}] (see Section \ref{s10.3}, Eq.
(\ref{10.3.7})).  Furthermore, as noted in Section \ref{s10.2},
the use of the Hartle--Hawking wave function in the inflationary
cosmology is justified only when there exists a stationary
distribution of the field $\varphi$, and therefore of ${\rm V}(\varphi)$
and $\Lambda(\varphi)$.  Thus far, it has not been possible to
find any inflationary models in which such a stationary
distribution might actually exist.  The probability distribution
for the quantity $\Lambda(\alpha_i)$ ought to have been
stationary, but this is the probability distribution for finding
a cosmological constant equal to $\Lambda$ {\it in different
universes} (or to be more precise, in different quantum states of
the universe), rather than in different parts of a single
universe.  The stochastic approach is not capable of justifying
expressions like (\ref{10.7.14}) and (\ref{10.7.15}) under these
circumstances, and validity of Euclidean methods used for the
derivation of Eq. (\ref{10.7.15}) is  not quite clear.

Some authors argued that a correct distribution of probability
for the universe to be in a given quantum state $|\alpha_i\rangle $
does not have a peak at $\Lambda=0$, in contrast with Eq.
(\ref{10.7.15}) [\cite{356}, \cite{357}]. It may be important
also that actually we are not asking why the universe lives in a
given quantum state $|\alpha_i\rangle $.  We are just trying to
explain the experimental fact that {\it we} live in the universe
in the quantum state $|\alpha_i\rangle $ corresponding to
$\rho_{\rm v}=|{\rm V}(\varphi_0)|\la10^{-29}$ g/cm$^3$
[\cite{358}, \cite{359}].

In this regard, recall that the application of the Anthropic Principle
based on an analysis of the galaxy formation process enables one to place
constraints on the
\index{Energy density, vacuum}%
\index{Density!vacuum energy}\index{Vacuum energy density}%
vacuum energy density [\cite{348}],
$$
-10^{-29}\;\mbox{g/cm$^3$}\la{\rm V}(\varphi_0)\la
10^{-27}\;\mbox{g/cm$^3$}\ ,
$$
and these constraints are quite close to the experimental figure,
$$
|{\rm V}(\varphi_0)|\la10^{-29}\;\mbox{g/cm$^3$}\ .
$$
This is an especially interesting result when viewed in the
context of the
\index{Baby universes}%
\index{Universe!baby}%
baby universe theory which makes it possible to
choose among the various $\Lambda$ [\cite{345}].

Can the Anthropic constraint on the
\index{Energy density, vacuum}%
\index{Density!vacuum energy}%
\index{Vacuum energy density}%
vacuum energy density be made stronger, so as to make the inequality
${\rm V}(\varphi_0)\la10^{-29}$ g/cm$^3$ a necessary consequence of the
Anthropic Principle?  There is still no final answer, but there
are a few inklings of how one might go about solving this
problem.

As follows from the results obtained in Sections
\ref{s10.2}--\ref{s10.4}, life in all its possible forms will
appear again and again in different domains of the
self-reproducing inflationary universe.  This does not mean,
however, that one can be very optimistic about the fate of
mankind.  An investigation of this question reveals that within
the presently observable part of the universe, life as we know it
cannot endure indefinitely, due to the decay of baryons and the
local collapse of matter [\cite{336}].  The only possibility that
we are presently aware of for the perpetual promulgation of life
is that in the scenario under consideration, for example in the
$\displaystyle \frac{\lambda\,\varphi^4}{4}$ theory, there must
\index{Domains!large number of}%
now exist a large number of domains in every region of size
\ba
\label{10.7.16}
l\ga l^*&\sim&
10^{30}\,\m^{-1}\,\exp\left(\frac{\pi\,(\varphi^*)^2}{\m^2}\right)
\nonumber \\
&\sim&10^{30}\,\m^{-1}\,\exp\left(\frac{\pi\,(\varphi^*)^2}{\m^2}\right)
\ea
in which the process of inflation continues unabated, and will do
so forever.  There will always be sufficiently dense regions
(like our own) near such domains, in which inflation came to an
end relatively recently and in which baryons have not yet had a
chance to decay.  One possible survival strategy for mankind
might consist of continual spaceflight bound for such regions.
In the worst case, if we will be unable to travel to such distant
places ourselves, we can try to send some information about us,
our life and our knowledge, and maybe even stimulate development
of such kinks of life there, which would be able to receive and
use this information.  In such a case one would have a comforting
thought that even though life in our part of the universe will
disappear, we will have some inheritors, and in this sense our
existence is not entirely meaningless.  (At least it would be not
worse than what we have here now.)

Leaving aside the question of the optimal strategy for the
\index{Mankind, survival of}%
\index{Survival of mankind}%
survival of mankind, we wish to note that the appropriate process
is necessarily impossible if the
\index{Energy density, vacuum}%
\index{Density!vacuum energy}%
\index{Vacuum energy density}%
vacuum energy density
${\rm V}(\varphi_0)$ is greater than
\be
\label{10.7.17}
{\rm V}^*\sim\rho_0\cdot 10^{200}\,\exp(-6\,\pi\,\lambda^{-1/3})\ .
\ee
In the theory with $\lambda\sim10^{-14}$,  ${\rm V}^*$ is
vanishingly small:
\be
\label{10.7.18}
{\rm V}^*\sim10^{-5\cdot 10^6}\;\mbox{g/cm$^3$}\ll \rho_0\ .
\ee
The reason there is a critical value ${\rm V}^*$ is that when
${\rm V}(\varphi_0)>{\rm V}^*$, the size ${\rm
H}^{-1}(\varphi_0)$ of the event horizon in a world with vacuum
energy density ${\rm V}(\varphi_0)$  turns out to be less than
the typical distance between domains in which the process of
self-regeneration of the universe is taking place.  (This
distance is presently $l^*$ of (\ref{10.7.16}); by the time the
vacuum energy ${\rm V}(\varphi_0)$ begins to dominate, the
distance will have grown by a factor of approximately
$10^{-60}\,\exp(2\,\pi\,\lambda^{-1/3})$.) Under such
circumstances, it would be impossible in principle either to fly
or to send a signal from our region of the universe to regions in
the vicinity of self-reproducing domains;  see Section
\ref{s1.4}.

Thus, in the present model, any quantum state of the universe
\index{Energy density, vacuum}%
\index{Density!vacuum energy}%
\index{Vacuum energy density}%
$|\alpha_i\rangle$ with vacuum energy density ${\rm
V}(\varphi_0)\ga10^{-5\cdot 10^6}$ g/cm$^3$ is a sort of cosmic
prison, and life within the universe in such a state is
inescapably condemned to extinction as a result of proton decay
and the exponential falloff of the density of matter when the
vacuum energy ${\rm V}(\varphi_0)$ becomes dominant.  There is
still no consensus on the probability of the spontaneous emergence
of complex life forms on Earth through just a single evolutionary
chain.  If, as some believe, this probability is extremely low,
and  if some mechanism of indefinitely long reproduction of life
at ${\rm V}(\varphi_0) < {\rm V}^*$ does actually exist (and this
is not at all obvious {\it a priori} [\cite{336}]), then the
existence of such a mechanism  can drastically increase the
fraction of the ``habitable'' universes in a quantum state with
${\rm V}(\varphi_0)<10^{-5\cdot 10^6}$ g/cm$^3$  as compared with
the fraction in a state with ${\rm V}(\varphi_0)>10^{-5\cdot
10^6}$ g/cm$^3$. The net result could then be [\cite{359}] that
any observer like ourselves, who is capable of inquiring about the
vacuum energy
\index{Energy density, vacuum}%
\index{Density!vacuum energy}\index{Vacuum energy density}%
density, would very likely find himself in a universe
corresponding to a quantum state $|\alpha_i\rangle $ with
$$
{\rm V}(\varphi_0)\ll 10^{-29}\;\mbox{g/cm$^3$}\ .
$$

Our treatment of this problem has merely provided a sketch of the
course of future research, illustrating the novel possibilities
which have emerged in recent years in elementary particle theory
and cosmology.  If we are to develop a successful strategy for
\index{Mankind, survival of}%
\index{Survival of mankind}%
the survival of mankind (if such a strategy exists), we will need
to undertake a much deeper study of the global structure of the
inflationary universe and the conditions required for the
emergence and/or propagation of life therein.  In any event,
however, the possibility that there is a the correlation between
\index{Energy density, vacuum}%
\index{Density!vacuum energy}%
\index{Vacuum energy density}%
\index{Energy density, vacuum}%
the value of the vacuum energy density and the existence of a
mechanism of eternal reproduction of life in the universe seems
\index{Universe!eternal reproduction of life in}%
\index{Eternal reproduction of life in universe}%
\index{Life in universe, eternal reproduction}%
\index{Universe!eternal reproduction of life in}%
to us to be noteworthy.
\index{Anthropic Principle!cosmological constant and|)}%
\index{Cosmological constant!Anthropic Principle and|)}%
\index{Energies!vacuum|)}%
\index{Reduplication of universe|)}%
\index{Universe!reduplication of|)}%


\pagestyle{myheadings}
\markboth{CONCLUSION}{\bktit}
\chapter*{Conclusion}

Elementary particle theorists and cosmologists might be compared
to two teams tunneling toward each other through the enormous
mountain of the unknown.  The analogy, however, is not entirely
accurate.  If the two teams of construction workers miss each
other, they will simply have built two tunnels instead of one.
But in our case, if the particle theorists fail to meet the
cosmologists, we wind up without any complete theory at all.
Furthermore, even if they do meet, and manage to build an
internally consistent theory of all processes in the micro- and
macro-worlds, that still does not mean that their theory is
correct.

In the face of the now familiar (and inevitable) dearth of
experimental data on particle interactions at energies
approaching $10^{19}$ GeV, and on the structure of the universe
on scales $l\gg10^{28}$ cm, it becomes especially important to
guess, if only in broad outline, the true direction that this
science will take, a direction that should remain valid even if
many specific details of the theory under construction should
change.  This explains the recent emergence of such unusual terms
as {\it scenario} and {\it paradigm} in the physicists' lexicon.

The major developments in elementary particle physics over the
past two decades can be characterized by a few key words, such as
{\it gauge invariance, unified theories with spontaneous symmetry
\index{Inflation}%
breaking, supersymmetry}, and {\it strings}.  The term inflation
has become such a word in the cosmology of the 1980's.

{\looseness=1
The inflationary universe scenario could only
\index{Inflationary universe!evolution of}%
have been created\linebreak[10000]
through the joint efforts of cosmologists and elementary particle
theorists.  The need for and fruitfulness of such a collaboration
\index{Inflation}%
is now obvious.  It should be pointed out that {\it inflation} is
certainly not a magic word that will automatically solve all our
problems and open all doors.  In some theories of elementary
particles, it is difficult to implement the inflationary universe
\index{Inflationary universe!evolution of}%
scenario, whereas many other theories fail to lead to a good
\index{Inflation}%
cosmology, even with the help of inflation.  The road to a
consistent cosmological theory may yet prove to be a very long
one, and we may still find that many details of our present
scenarios will be cast off as inessential excess baggage.  At the
moment, however, it does seem necessary to have something like
inflation to obtain a consistent cosmology at peace with particle
physics.

}

Inflationary cosmology continues to develop rapidly.  We are
\index{Inflationary universe!evolution of}%
witnessing a gradual change in our most general concepts about
the evolution of the universe.  Just a few years ago, most
authorities had virtually no doubt that the universe was born in
a unique Big Bang approximately 10--15 billion years in the past.
It seemed obvious that space-time was four-dimensional from the
very outset, and that it remains four-dimensional all over the
universe.  It was believed that if the universe were closed, its
total size could hardly exceed that of its observable part, $l
\sim 10^{28}$ cm, and that in no more than $10^{11}$ years such a
universe would collapse and disappear.  If, on the other hand,
the universe were open or flat, then it would be infinite, and
the general conviction was that it would then exhibit properties
everywhere that were almost the same as those in its observable
part.  Such a universe would exist forever, but after its protons
had decayed, as predicted by unified theories of the weak,
strong, and electromagnetic interactions, no baryon matter would
remain to support life.  The only alternatives, then, were a
``hot end'' with the expected collapse of the universe, and a
``cold end'' in the infinite void of space.

It now seems more likely that the universe as a whole exists
eternally, endlessly spawning ever newer exponentially large
\index{Life in universe, eternal reproduction}%
\index{Universe!eternal reproduction of life in}%
\index{Eternal reproduction of life in universe}%
\index{Universe!eternal reproduction of life in}%
regions in which the low-energy laws governing elementary
particle interactions, and even the effective dimensionality of
space-time itself, may be different.  We do not know whether life
can evolve forever in each such region , but we do know for
certain that life will appear again and again in different
regions of the universe, taking on all possible forms.  This
change in our notions of the global structure of the universe and
our place within it is one of the most important consequences to
\index{Inflationary universe!evolution of}%
come out of inflationary cosmology.

We have finally come to a newfound appreciation of why it was
necessary to write the scenario even though the performance was
already over ---  the show is still going on, and most likely
will continue forever.  In different parts of the universe,
different audiences will observe it in its infinite variations.
We cannot witness the whole play in all its grandeur, but we can
try to imagine its most essential features --- and ultimately
perhaps even grasp its meaning.


\newpage

\pagestyle{refheadings}

\end{document}